# Scalable String and Suffix Sorting

## Algorithms, Techniques, and Tools

Timo Bingmann

# Scalable String and Suffix Sorting: Algorithms, Techniques, and Tools

zur Erlangung des akademischen Grades eines

Doktors der Naturwissenschaften

von der Fakultät für Informatik
des Karlsruher Instituts für Technologie (KIT)

**genehmigte**

**Dissertation**

von

**Timo Fabian Horst Bingmann**

aus Mainz

*To my parents.*

# Abstract

This dissertation focuses on two fundamental sorting problems: *string sorting* and *suffix sorting*. The first part considers parallel string sorting on shared-memory multi-core machines, the second part external memory suffix sorting using the induced sorting principle, and the third part distributed external memory suffix sorting with a new distributed algorithmic big data framework named *Thrill*.

Sorting strings or vectors is a basic algorithmic challenge different from integer sorting because it is important to access components of the keys to avoid repeated operations on the entire string. We focus on sorting large inputs which fit into the RAM of a shared-memory machine. String sorting is needed for instance in database index construction, suffix sorting algorithms, and to order high-dimensional geometric data.

We first survey engineered variants of basic sequential string sorting algorithms and perform an extensive experimental evaluation to measure their performance. Furthermore, we perform experiments to quantify parallel memory bandwidth and latency experiments as preliminary work for designing parallel string sorting algorithms.

We then propose *string sample sort* as an adaptation of sample sort to string objects and present its engineered version *Super Scalar String Sample Sort*. This parallel-ready algorithm runs in $\mathcal{O}(\frac{D}{w} + n \log n)$ expected time, makes effective use of the cache hierarchy, uses word- and instruction-level parallelism, and avoids branch mispredictions. Our parallelization named *Parallel Super Scalar String Sample Sort* (pS$^5$) employs voluntary work sharing for load balancing and is the overall best performing algorithm on single-socket multi-core machines in our experiments.

For platforms with non-uniform memory access (NUMA) we propose to run pS$^5$ on each NUMA node independently and then merge the sorted string sequences. To accelerate the merge with longest common prefix (LCP) values we present a new *LCP-aware multiway merge* algorithm using a tournament tree. The merge algorithm is also used to construct a stand-alone *LCP-aware $K$-way mergesort*, which runs in $\mathcal{O}(D + n \log n + \frac{n}{K})$ time and benefits from long common prefixes in the input.

Broadly speaking, we propose both *multiway distribution-based* with string sample sort and *multiway merge-based* string sorting with LCP-aware merge and mergesort, and engineer and parallelize both approaches. We also present parallelizations of multikey quicksort and radix sort, and perform an extensive experimental evaluation using six machines and seven inputs. For all input instances, except random strings and URLs, pS$^5$ achieves higher speedups on modern *single-socket* multi-core machines than our own parallel multikey quicksort and radix sort implementations, which are already better than any previous ones. On *multi-socket* NUMA machines $\text{pS}^5$ combined with the LCP-aware top-level multiway merging was fastest on most inputs.

We then turn our focus to *suffix sorting*, which is equivalent to *suffix array construction*. The suffix array is one of the most popular text indexes and can be used for fast substring search in DNA or text corpora, in compression applications, and is the basis for many string algorithms. When augmented with the *LCP array* and additional tables, the suffix array can emulate the suffix tree in a myriad of stringology algorithms. Our goal is to create fast and scalable suffix sorting algorithms to generate large suffix arrays for real-world inputs. As introduction to suffix array construction, we first present a brief survey of their principles and history.

Our initial contribution to this field is *eSAIS*, the first external memory suffix sorting algorithm which uses the *induced sorting* principle. Its central loop is an elegant reformulation of this principle using an external memory priority queue, and our theoretical analysis shows that eSAIS requires at most $\textsc{Sort}(17n) + \textsc{Scan}(9n)$ I/O volume. We then extend eSAIS to also construct the LCP array while suffix sorting, which yields the first implementation of fully external memory suffix and LCP array construction in the literature. Our experiments demonstrate that eSAIS is a factor two faster than DC3, the previously best external memory suffix sorting implementation. After our initial publication of eSAIS, many authors showed interest in the topic and we review their contributions and improvements over eSAIS.

For scaling to even larger inputs, we then consider suffix sorting on a distributed cluster machine. To harness the computational power of a such a system in a convenient data-flow style functional programming paradigm, we propose the new high-performance distributed big data processing framework *Thrill*. Thrill's central concept is a *distributed immutable array* (DIA), which is a virtual array of C++ objects distributed onto the cluster. Such arrays can be manipulated using a small set of scalable primitives, such as mapping, reducing, and sorting. These are implemented using pipelined distributed external memory algorithms encapsulated as C++ template classes, which can be efficiently coupled to form large complex applications. Our Thrill prototype is evaluated using five micro benchmarks against the popular frameworks Apache Spark and Flink on up to 16 hosts in the AWS Elastic Compute Cloud. Thrill consistently outperforms the other frameworks in all benchmarks and on all numbers of hosts.

Using Thrill we then implement five suffix sorting algorithms as a *case study*. Three are based on prefix doubling and two are variants of the linear-time difference cover algorithm DC. The implementation of these complex algorithms demonstrates the *expressiveness* of the scalable primitives provided by Thrill. They also are the first distributed external memory suffix sorters presented in the literature. We compare them experimentally against two hand-coded MPI implementations and the fastest non-distributed sequential suffix sorters. Our results show that algorithms implemented using Thrill are *competitive to MPI programs*, but scale to larger inputs due to automatic usage of external memory. In the future, these implementations can benefit from improvements of Thrill such as fault tolerance or specialized sorting algorithms.

# Deutsche Zusammenfassung

Die vorliegende Dissertation behandelt zwei grundlegende Sortierprobleme: Sortieren von *Zeichenketten* und Sortieren aller *Suffixe* eines Texts. Der erste Teil betrachtet paralleles Sortieren von Zeichenketten auf Mehrkernrechnern mit gemeinsam genutztem Speicher, der zweite Teil ein neues Verfahren zum Sortieren von Suffixen im Externspeicher und der dritte Teil Sortieren von Suffixen auf verteilen Parallelrechnersysteme mit dem neuen algorithmischen Framework *Thrill*.

Das Sortieren von Zeichenketten oder Vektoren unterschiedet sich von Sortieren von Zahlen durch die zusätzliche Komponentenstruktur der Schlüssel, die systematisch ausgenutzt werden muss um teure Operationen auf den ganzen Objekten zu vermeiden. Wir betrachten dabei Eingaben die in den gemeinsamen Speicher einer modernen Mehrkernmaschine passen. Große Mengen von Zeichenketten werden beispielsweise sortiert bei der Konstruktion von Datenbankindices, beim Sortieren von Suffixen, oder um hochdimensionale geometrische Daten anzuordnen.

Als vorbereitende Arbeit für den Entwurf von parallelen Sortieralgorithmen für Zeichenketten diskutieren wir zuerst hochentwickelte Varianten der bestehenden sequentiellen Basisalgorithmen und führen eine umfangreiche experimentelle Auswertung dieser durch. Darüber hinaus berichten wir von einer quantitativen Untersuchung der parallelen Speicherbandbreiten und -latenz in modernen Mehrkernsystemen.

Mit dem Wissen aus dieser Vorarbeit entwickeln wir als ersten Algorithmus *String Sample Sort*, der eine Anpassung von Samplesort für Zeichenketten ist und präsentieren dessen optimierte Version *Super Scalar String Sample Sort*. Dieser neue Algorithmus ist leicht zu parallelisieren, benötigt $\mathcal{O}(\frac{D}{w} + n \log n)$ erwartete Zeit, nutzt die Cache-Hierarchie effektiv, verwendet Parallelität auf Wort- und Anweisungsebene und vermeidet teure Fehlvorhersagen von Verzweigungen. Seine Parallelisierung namens *Parallel Super Scalar String Sort* (pS$^5$) verwendet ein freiwilliges Lastbalanceverfahren und ist in unseren Experimenten der insgesamt leistungsfähigste Algorithmus für Mehrkernrechner mit einem Sockel.

Für Plattformen mit non-uniform memory access (NUMA) entwerfen wir einen Hybridansatz, in dem zuerst pS$^5$ auf jedem NUMA-Knoten unabhängig voneinander ausgeführt wird und dann gemeinsam die vorsortierten Zeichenkettenfolgen zusammen gemischt werden. Um die Zusammenführung durch ein Array der längsten gemeinsamen Präfixe (LCP) zu beschleunigen, präsentieren wir einen neuen *LCP-beschleunigten*

*Mehrwege-Mischalgorithmus* (multiway merge), der auf einem Turnierbaum basiert. Der Mischalgorithmus wird darüber hinaus auch verwendet, um einen eigenständigen LCP-beschleunigten $K$-Wege-Mischsortieralgorithmus (multiway mergesort) zu entwerfen. Dieser läuft in $\mathcal{O}(D + n \log n + \frac{n}{K})$ Zeit und profitiert von langen gemeinsamen Präfixen in der Eingabe.

Kurz gesagt, schlagen wir sowohl Sortieralgorithmen auf Basis von *Mehrwege-Verteilen* mit String Sample Sort als auch von *Mehrwege-Mischen* mit LCP-beschleunigten Merge und Mergesort vor und optimieren und parallelisieren beide Ansätze. Darüber hinaus entwickeln wir auch Parallelisierungen von Multikey Quicksort und Radix Sort und führen eine umfangreiche experimentelle Analyse auf sechs Maschinen und sieben Eingaben durch. Auf allen Instanzen, außer zufälligen Zeichenketten und URLs, erreicht $\text{pS}^5$ höhere Geschwindigkeiten auf modernen Mehrkernrechnern mit *einem Sockel* als unsere Multikey-Quicksort- und Radix-Sort-Parallelisierungen, die bereits besser sind als alle bestehenden Verfahren. Auf *Mehrsockel*-NUMA-Rechnern war der Hybridansatz bestehend aus $\text{pS}^5$ und LCP-beschleunigten Mehrwege-Mischen auf den meisten Instanzen am schnellsten.

Danach konzentrieren wir uns auf das Sortieren der *Suffixe eines Text*, welches auch *Suffix-Array-Konstruktion* genannt wird. Das Suffix-Array ist eines der beliebtesten Textindizes und dient zur Beschleunigung des Suchen nach Teilzeichenfolgen in DNA- oder Textkorpora, wird in Kompressionsverfahren verwendet werden und ist die Grundlage für viele komplexe String-Algorithmen. Wenn das Suffix-Array um das *LCP-Array* und weitere zusätzliche Tabellen ergänzt wird, kann diese Kombination den Suffix-Tree in einer Vielzahl von String-Algorithmen ersetzen. Unser Ziel sind schnelle und skalierbare Suffix-Sortieralgorithmen, um große Suffix-Arrays für reale Eingaben zu generieren. Als Einführung präsentieren wir zunächst einen kurzen Überblick über die Prinzipien und Geschichte von Suffix-Sortieralgorithmen.

Unser erster Beitrag zu diesem Gebiet ist *eSAIS*, der erste Suffix-Sortieralgorithmus für Externspeicher, der das *induzierte Sortierprinzip* (induced sorting) verwendet. Seine zentrale Schleife ist eine elegante Neuformulierung dieses Prinzips mittels einer Prioritätswarteschlange für Externspeicher. Unsere theoretische Analyse zeigt, dass eSAIS höchstens $\textsc{Sort}(17n) + \textsc{Scan}(9n)$ I/O-Volumen erfordert. eSAIS wird daraufhin um die gleichzeitige Konstruktion des LCP-Array während der Suffix-Sortierung erweitert. Dies ergibt die erste Implementierung eines vollständig externen Suffix- und LCP-Array-Konstruktionsalgorithmus in der Literatur. Unsere Experimente zeigen, dass eSAIS um einen Faktor zwei schneller ist als DC3, dem bisher besten Suffix-Sortierverfahren für Externspeicher. Nach unserer ersten Veröffentlichung von eSAIS zeigten viele weitere Autoren Interesse an dem Thema, und wir besprechen ihre Beiträge und Verbesserungen.

Um die Verfahren auf noch größere Eingaben zu skalieren, betrachten wir dann Suffix-Sortierverfahren für verteilten Parallelrechnersysteme. Hierzu präsentieren wir zuerst das neue verteilte Big-Data-Framework *Thrill*, mit dessen Hilfe komplexe Algorithmen für solch hochleistungsfähige Systeme leichter entworfen und programmiert werden können. Das zentrale Konzept von Thrill ist ein *verteiltes unveränderbares Array* (DIA), das nahezu beliebige C++ Objekte enthalten kann und transparent auf dem Cluster verteilt liegt. Es ist jedoch kein direkter Zugriff möglich. Statt dessen können DIAs nur mittels eines kleinen Satzes von skalierbaren Primitiven wie *Map*, *Reduce* und *Sort* manipuliert werden. Diese werden als verteilt-externe Basisalgorithmen implementiert und in C++ template Klassen gekapselt. Die Basisalgorithmen können mit anwendungsspezifischen Funktoren parametrisiert und effizient zu größeren Anwendungen gekoppelt werden. Unser Thrill-Prototyp wird anhand von fünf Mikro-Benchmarks mit den populären Frameworks Apache Spark und Apache Flink auf bis zu 16 Maschinen in der AWS Elastic Compute Cloud evaluiert. Thrill ist schneller als die anderen Frameworks in allen Benchmarks und für jede Anzahl von Maschinen.

Als Fallstudie implementieren wir dann fünf Suffix-Sortieralgorithmen mit Thrill. Drei basieren auf Präfixverdopplung und zwei sind Varianten des linearen difference cover Algorithmus DC. Die Implementierung dieser komplexen Algorithmen demonstriert die *Ausdruckskraft* der von Thrill bereitgestellten skalierbaren Primitiven. Darüber hinaus sind sie die ersten verteilten externen Suffix-Sortierer, die in der Literatur vorgestellt werden. Wir vergleichen sie experimentell mit zwei von Hand erstellten MPI-Implementierungen und mit den schnellsten nicht verteilten sequentiellen Suffix-Sortierern. Unsere Ergebnisse zeigen, dass mit Thrill implementierte Algorithmen *mit MPI-Programmen konkurrieren* können und dass sie aufgrund der automatischen Verwendung von externem Speicher auf größere Eingaben skalieren. Darüber hinaus können diese Implementierungen von zukünftigen Verbesserungen in Thrill so wie Fehlertoleranz oder spezialisierten Sortieralgorithmen profitieren.

# Acknowledgements

*First and foremost I would like to thank Peter Sanders for enabling, supporting, and guiding my research. He is truly an extraordinary computer science scholar with a wide field of interests and amazing in depth knowledge in virtually all of them.*

*Thanks go also to my first two coauthors, Johannes Fischer and Vitaly Osipov, who helped me through the perils of writing and publishing the first research paper.*

*The academic research in this dissertation is the result of fruitful interactions with many people through the years. I would like to thank Michael Axtmann, Christian Schulz, and Peter Sanders for their close collaboration on distributed integer sorting problems.*

*With Johannes Fischer, Simon Gog, Dominik Kempa, and Florian Kurpicz I share a liking for string algorithms, which are often intricate and unfathomable at first sight but develop a charm when studied in detail.*

*Andreas Eberle, Daniel Feist, Florian Gauger, Thomas Keh, and Alexander Noe were extraordinary students who I had the pleasure to guide for their bachelor or master thesis. Each tackled a challenging and interesting algorithmic problem with vigor.*

*The big data framework Thrill is the outcome of an ambitious collaboration of doctoral researchers and students. Robert Hangu, Emanuel Jöbstl, Sebastian Lamm, Huyen Chau Nguyen, Alexander Noe, Matthias Stumpp, and Tobias Sturm worked together with Michael Axtmann, Sebastian Schlag, Peter Sanders, and myself for over a year on the project. So many things could have gone wrong but didn't.*

*And then I would like to thank my coworkers and office mates: Yaroslav Akhremtsev for first occupying my flat, then the office sofa, all while talking about probabilistic proofs, Michael Axtmann for extensive discussions on sampling and sample sort, Veit Batz for pleasant talks in our shared office, ethics discussions, and bass guitar music, Daniel Funke for his clearheaded view of the world, Simon Gog for asking me lots of C++ questions and having a more hands-on approach to string algorithms than myself, Demian Hespe for faultlessly increasing the office volume and playing SNES, Lorenz Hübschle-Schneider for twittering in my office all the time, keeping me from working, enjoying life, and having missed his calling to be a pizza baker, Moritz Kobitzsch for baking a cake, Sebastian Lamm for always having an alternative, quicker approach to solving a coding problem in Thrill, Dennis Luxen for ridiculing nerds of all kind and*

*being the most lovable Star Wars nerd himself, Tobias Maier for showing me what kind of hobbies "normal" people have, Dennis Schieferdecker for an example of how to calmly complete a dissertation and wonderful days in southern France, Sebastian Schlag for keeping the difficulty of our lecture exercises in check, lots of whiskey, taking the student's side in Thrill meetings, and letting me put up pictures on his side of our office, Christian Schulz for demonstrating academic efficiency and working together on algorithm lectures and exercises, Jochen Speck for pursing his life goal of being a Wikipedia polymath, and Sascha Witt for keeping the peace in our group. More thanks go to Anja Blancani for gracefully abiding with all the peculiarities considered perfectly normal in a group of computer scientists and to Norbert Berger for steadfast and long-term organization of our computing systems.*

*Furthermore, I would like to thank the Amazon Cloud Credits for Research program for making the experiments with Thrill possible on the Amazon Web Services and the Steinbuch Computing Center (SCC) at the KIT for computing resources for external memory suffix sorting.*

*And then again more thanks goes to my enduring proof readers, Yaroslav Akhremtsev, Michael Axtmann, Demian Hespe, Lorenz Hübschle-Schneider, and Sebastian Schlag, for all their well-minded corrections.*

*Last but not least, I would like to thank my parents and siblings for everlasting loving emotional support.*

# Table of Contents

Chapter 1

# 1 Introduction and Overview of the Dissertation

*In this dissertation we consider how to scale two important basic sorting problems:* string sorting *and* suffix sorting. *Our focus lies on multi-core machines for string sorting, and external memory and distributed systems for suffix sorting. In the context of scaling suffix sorting to distributed machines, we also introduce our new high-performance C++ framework* Thrill *for general purpose parallel and distributed external memory data processing.*

*Before diving into the two challenges in parts I to III, we introduce the reader in this chapter to our motivation and methodology. We discuss why developing scalable algorithms is an important current and future research topic, and how* algorithm engineering *and appropriate* theoretical machine models *can be of aid on this quest.*

Around the year 2004 there was a turning point in CPU technology: while Moore's law [Moo65], an observation and prediction that the number of transistors doubles every two years, has continued to be valid, the clock frequency of individual processors no longer increased at the same rate due to physical limitations. Instead, the number of cores per socket sharing a common memory system has increased from one to more than 64. Anecdotally, this observation can be attested by considering that today one cannot buy a smartphone without at least four cores. Figure 1.1 presents a more founded chart of the number of transistors per socket, the clock frequency, and the number of cores per socket from a list of microprocessor models dating back to 1970. Two regression lines were added: the red line clearly shows the exponential increase in transistors, which is described by Moore's law, while the blue line shows that the increase in clock frequency has clearly stalled. At the same time, the number of cores has been increasing, starting at around 2004 and continuing on to this day.

Hence, looking into the future, barring a revolution in CPU technology, *parallelism* is now the only way to extract performance gains from Moore's law. This means that any performance critical algorithm and implementation needs to be parallelized efficiently. This has been known for many years, yet many algorithmicists still think sequentially, and most algorithms are still developed for the RAM model. This is clearly because parallel programming, both in theory and practice, is much harder [McK17] due to concurrency, asynchronicity, and synchronization. However, both the available programming tools and university education focusing on parallelization have improved greatly in the last ten years and will hopefully lead to a more wide-spread adoption of this technology.

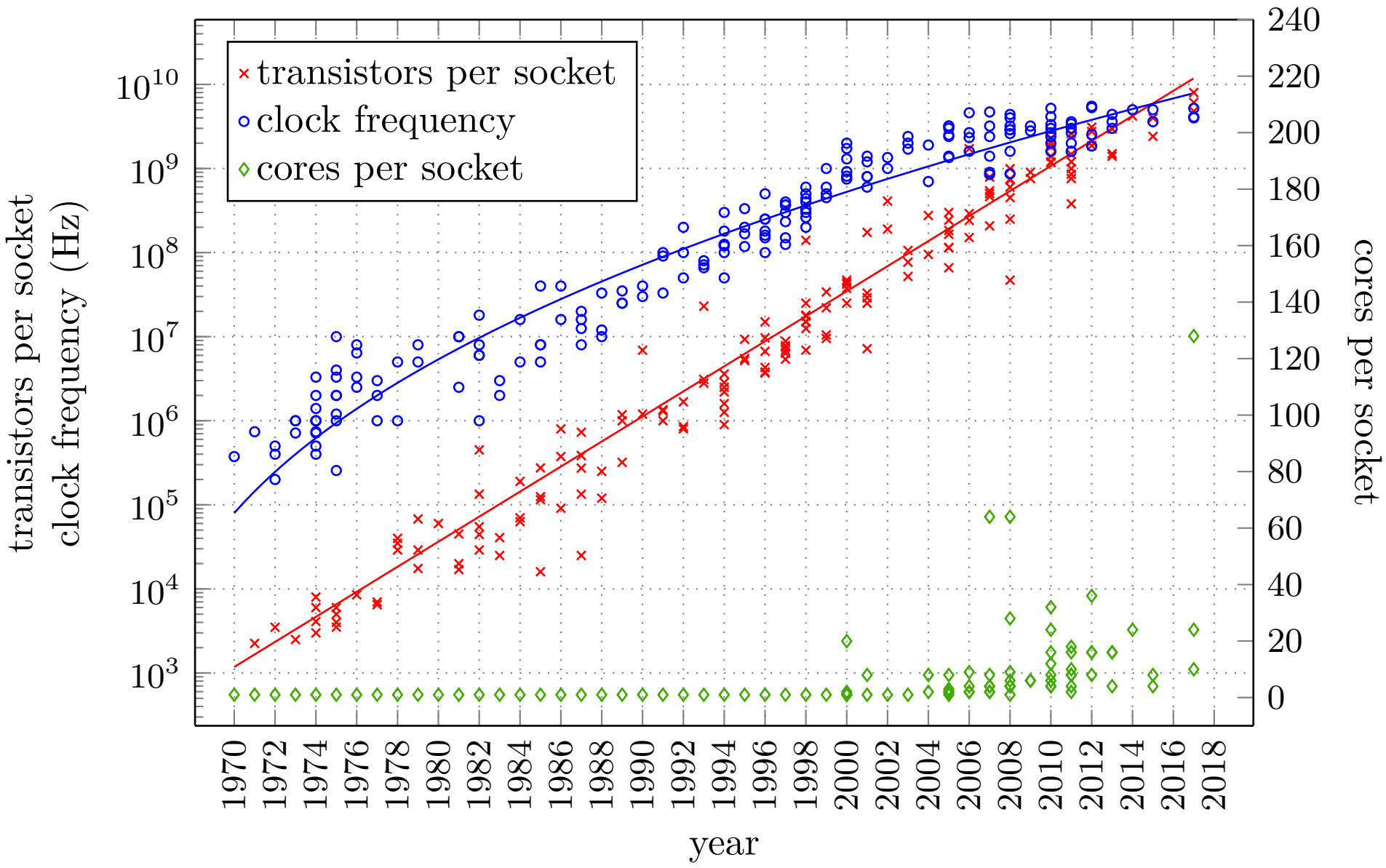

**Figure 1.1:** Increase of transistors per socket, clock frequency, and cores per socket in microprocessors from 1970–2017. Data from Wikipedia's microprocessor chronology*.

In the same time span, 2004 to 2018, rotational disk memory and RAM have become cheaper by a factor of about 29 (figure 1.2), or even 36 if one adjusts for inflation. While the price ratio of RAM to rotational disk has always stayed in a range of 50–400, external disk memory has become so cheap in absolute terms that the amount of stored data has skyrocketed in recent years. As storage space is readily available, (parallel) external memory algorithms have become an increasingly important field for processing huge amounts of information. This phenomenon has been termed the *big data* revolution, as larger datasets require better algorithms and tools to process the volume, velocity, and variety of these emerging information age resources [ABC+06; Hil16].

To tackle the software engineering aspects of these tasks, distributed processing frameworks such as Hadoop MapReduce, Apache Spark, and Apache Flink have been developed for commodity hardware and gained enormous popularity. They promise automatic parallelization, automatic data management, automatic load balancing,

*`https://en.wikipedia.org/wiki/Microprocessor_chronology` (accessed February 2018)

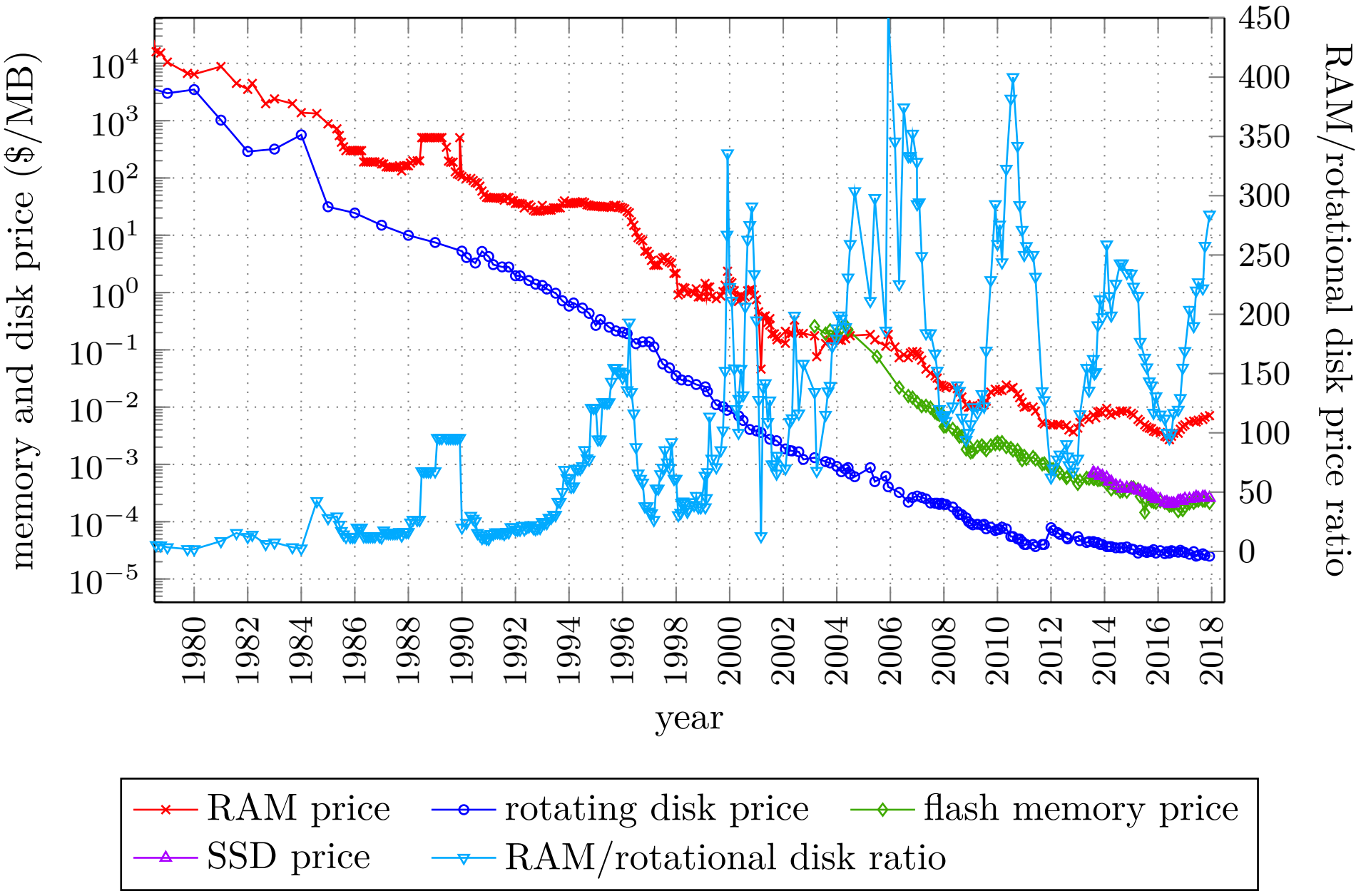

**Figure 1.2:** History of the cost of memory (1980–2017): RAM, rotational disks, flash memory, and SSDs (not adjusted for inflation). Data collected by John C. McCallum at `http://www.jcmit.com/`.

and automatic fault tolerance, but notoriously lack in absolute performance [MIM15]. While we believe that the number of truly massive data sets, such as Google's web search indices, ad click streams, and particle accelerator sensor data, will remain small, we believe that the number of *medium-sized* data sets from industrial and scientific applications will increase dramatically. These can be stored and processed on a small cluster of machines.

## 1.1 Algorithm Engineering and Machine Models

This dissertation adopts the *algorithm engineering* methodology [San09; San10]. Traditionally, algorithm theory relies strongly on mathematics and logic for the analysis of algorithms. Undoubtedly, proving correctness of algorithms for all possible inputs is an essential task. Likewise, estimating and comparing the running time or other costs of algorithms is important for determining their applicability.

However, provable worst-case or average-case performance guarantees for all possible inputs are often difficult to attain due to the complexity of advanced algorithms. On the one hand this has restricted the exploration space of theoretical algorithmics, while on the other hand it has fostered a culture of combining known building block algorithms to ever more complex super-algorithms that are often never implemented. In extreme cases, promising simple algorithmic approaches have been neglected due to missing theoretical analysis or some pathological inputs which result in outlier worst-case behavior [DSSW09]. Nevertheless, theoretical analyses of algorithms which exhibit insights into their properties must always remain a prime objective of computer science.

At the center of such theoretical analyses lie abstract models of the processing machine an algorithm is to run on. These models must strike a balance between being simple enough for theoretical tractability and adequately representing the performance characteristics of real machines. However, these real machines have become increasingly complex in the pursuit of performance, and hence simple models have progressively become less and less justifiable.

Therefore, *experimental algorithmics* has become increasingly popular. However, just like theory, experimental analysis has to be performed with scientific rigor. The first step of actually implementing an algorithm *well* is equally difficult and time consuming, and the second step of designing and running experiments such that they produce meaningful results is a challenge and can be very costly. The gold standard of *reproducible* experiments is hard to obtain and maintain in the fast moving world of computer hardware. Furthermore, actually implementing and evaluating algorithms often leads to a better understanding of the underlying challenges, and hence, to better algorithm designs and new approaches for analysis.

The *algorithm engineering cycle* (figure 1.3) ties together this scientific process into a feedback loop of *design*, *analysis*, *implementation*, and *experiments*. This feedback loop is centered on falsifiable hypotheses about an algorithm and may be run through several times in the search for incremental improvements. However, the engineering loop must not remain isolated and self-entertaining. Instead, it is embedded into the framework of *real-world applications*, which supply realistic settings, specific inputs, and maybe even concrete experiments. Part of algorithm engineering is then to correctly model these settings, and to collect or generate sufficient real-world inputs for experiments.

While the developed implementations can directly improve applications, the actual step of transferring knowledge from academic insights to industry-grade applications is an entire separate *application engineering* process. Another important avenue to improve applications is to collect and develop *algorithm libraries* containing well-engineered implementations with precise interfaces, which can then be used by applications. This both cleanly separates the algorithm engineering efforts from actual applications and puts the burden of designing, developing, and documenting a library on the algorithm engineer. A third route is to gain tighter performance guarantees through analysis of algorithms, which in turn may come from better design or experimental observations.

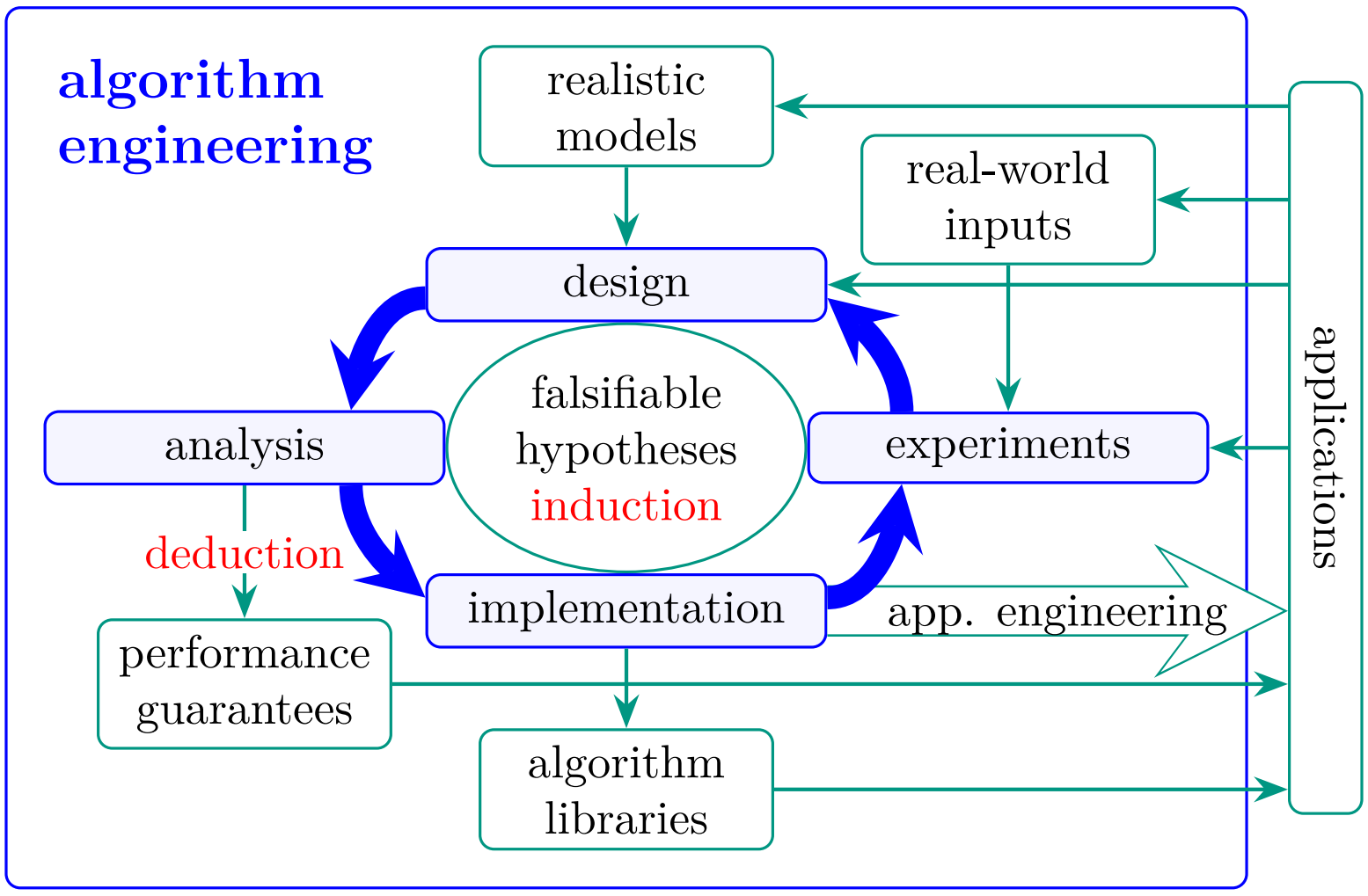

**Figure 1.3:** Algorithm engineering as a cycle of design, analysis, implementation, and experiments. Adapted from [San10].

In all three parts of this dissertation, the algorithm engineering cycle is clearly visible. Part I focuses on *string sorting*, and we first perform an extensive study of the design and experimental performance of existing sequential string sorters in chapter 2, and of multi-core machines in chapter 3. The gained knowledge is then used to design new parallel string sorters in chapter 4, which are analyzed theoretically in sections 4.2.6 and 4.3.2 and evaluated with practical real-world inputs in section 4.5. Part II then turns to *suffix sorting* and first reviews existing algorithms and their history in chapter 5. With these insights, an external memory algorithm is designed in section 6.2, analyzed theoretically using the external memory model in section 6.2.3, and experimentally in section 6.3. For distributed external memory suffix sorting in part III, a different route is chosen: first a new *algorithm library*, Thrill, is introduced in chapter 7 and then used to implement and evaluate suffix sorting algorithms in chapter 8.

The following four subsections summarize the theoretical models used for analysis in this dissertation. We discuss their weaknesses and how to judge the gap between theory and practical results. The final subsection then critically focuses on experimental algorithmics.

### 1.1.1 The Random Access Machine (RAM)

Theoretical models of computing machines have to incorporate enough detail to be able to predict the actual performance, while at the same time remain simple enough to reason about using mathematical arguments. The most widely used model is the *random access machine (RAM) model* (figure 1.4 (a)) [Neu45; SS63]. It assumes one processor containing multiple registers, which can perform elementary operations, and a finite amount of memory to load or store information from or to. Both elementary operations and load/store operations incur one unit cost.

The RAM model is still the fundamental model for analysis of algorithms mainly due to its simplicity and universal applicability. However, considering the complexity of today's mainstream processors, basically none of the assumptions of the RAM model remain valid [Dre07; HP12].

The RAM model assumes equal cost for all operations and memory accesses. This is clearly not true: today's machines have a *cache hierarchy*, consisting of in-processor level 1 (L1) and level 2 (L2) caches, and sometimes larger on-chip level 3 (L3) caches, followed by larger dynamic RAM (DRAM) chips (see section 3.1). A random access to DRAM is around 100 times slower than an access to L1 cache (see figure 3.11 on page 73). If one considers NUMA architectures, the divide widens further: even the name "non-uniform memory access" is in clear opposition to the RAM model.

While cache effects are a well-known and studied subject [San99; San00], modern processors contain many more transparent features which improve performance. Less known effects are caused by *prediction of branches*, in which a processor guesses the outcome of a conditional jump to avoid pipeline stalls. The penalties of mispredictions were studied in the context of quicksort [KS06a; EW16]. The recent revelation of the Meltdown [LSG+18] and Spectre [KGG+18] security flaws in Intel and AMD chips have brought *speculative execution* of instructions to the headlines of many computer magazines. In essence, processors today not only try to predict branches, they even execute them prior to checking all the conditions and security restrictions. While the effects of the executed instruction remain hidden until the checks are completed, the two vulnerabilities were able to extract information from *side effects* this speculative execution causes despite it being rolled back in case of security violations.

Another less known and understood feature of modern processors is *superscalar processing*: again in the name of performance, processors today can dispatch and execute not only a single instruction per clock cycle, but multiple ones in parallel. This is called *instruction-level parallelism* and requires multiple arithmetic units, instruction decoders, spare hidden registers, and more on the processor chip. But all these features are the norm on present-day processors. To organize superscalar processing, they contain a unit which analyzes the data dependencies of the next few instructions and dispatches operations in parallel, all completely transparent to the running code. These look-ahead and planning capabilities are implemented in hardware for speed, and hence must be limited in their scope and in the complexity of the detected constructs. In section 4.2 we carefully design a string sorting algorithm based on previous work [SW04], which takes advantage of this implicit parallelism by drawing data dependencies apart.

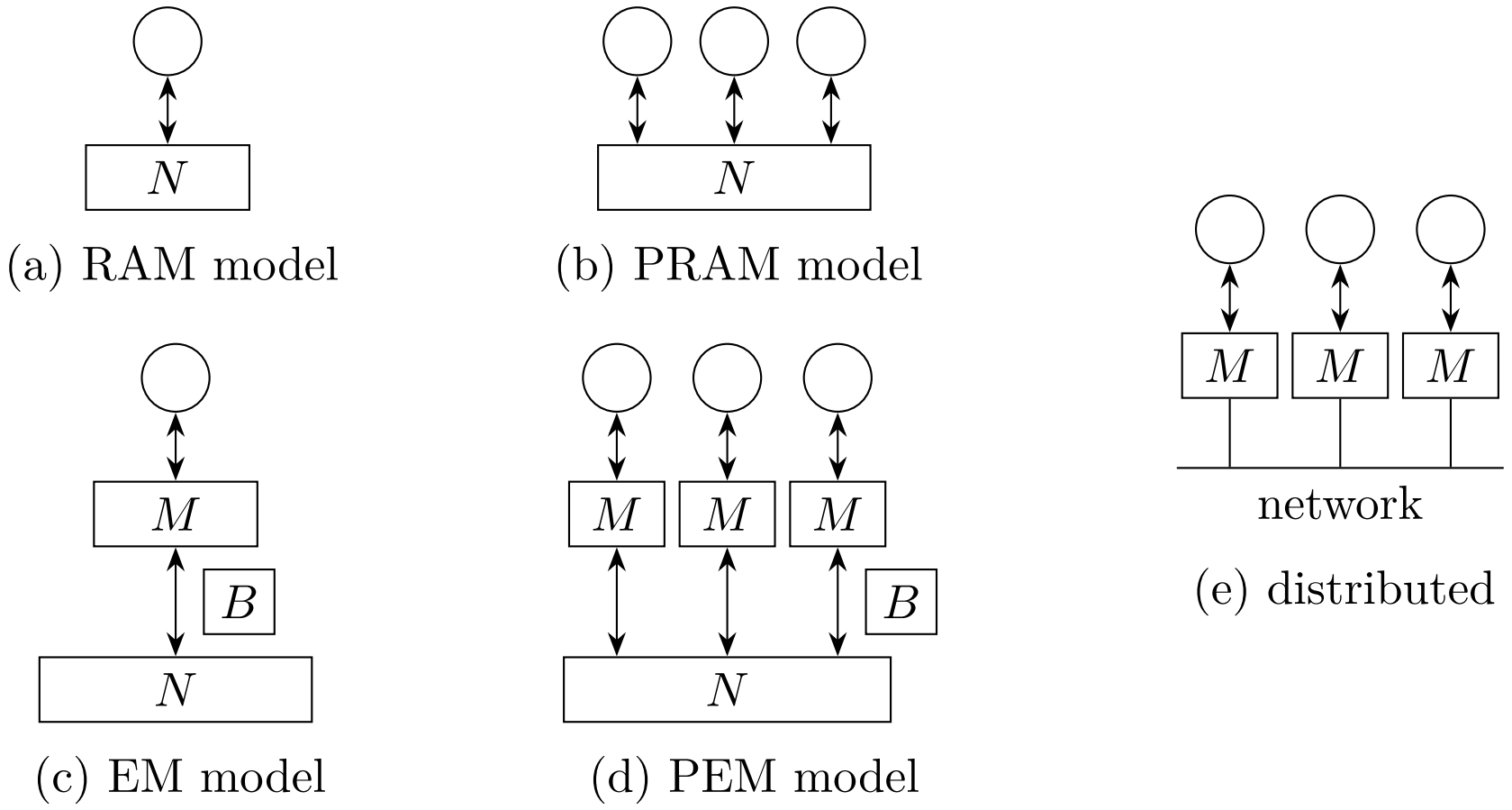

**Figure 1.4:** Illustrations of RAM [Neu45; SS63], PRAM [FW78; Gol78; SS79], EM [Vit98], PEM [AGNS08], and a distributed machine model.

As heat is one of the main problems in modern processors, chip manufacturers developed many *power saving* and *temperature control* techniques. These allow software (the operating system) to control the clock frequency with increasingly fine granularity. For example, if only one or a few cores on a multi-core system are used, then their clock frequency can be increased as long as the remaining cores are idle, such that the total heat generated does not exceed specifications. More generally, processors allow *dynamic frequency scaling* depending on their temperature, available power (e.g. from battery), future expected work load, the work load of other cores, and many other factors. These enhancements are very useful for common computer workstations, but algorithms analysis usually assumes *one fixed* processor speed. In particular, it can happen that algorithms start with a faster (turbo) frequency, and then are forced to slow down due to the internal temperature rising.

All performance optimizations described above maintain the *appearance* of one sequential processor close to the RAM model. Interestingly, the performance optimizations tend to stem from violating some aspect of a sequential RAM, and then reestablishing compatibility by hiding the acceleration. Since CPU clock frequencies are no longer increasing, performance gains must be realized via other means of optimization. Hence, we believe this trend is likely to continue.

While theoreticians will argument away all these performance features in the $\mathcal{O}(\cdot)$ notation by attributing them with constant factors, we consider this unsatisfactory because these factors can be very large. While algorithm theory may celebrate removing a $\log^* n$ factor from the asymptotic running time of an algorithm, $\log^* n \leq 5$ for any countable entity in the known universe. Even $\log_2 n \leq 50$, except on the very largest scale applications. Considering these numbers, we have to keep in mind that the difference between a single access to cache memory and to DRAM can be a factor of 100, as stated above. Hence, an $\mathcal{O}(n \log n)$ algorithm can be as fast as an $\mathcal{O}(n)$ algorithm which incurs a random DRAM access per operation. For example, $\mathcal{O}(n)$ pointer chasing can be slower than an $\mathcal{O}(n \log n)$ sorting algorithm for any realistic number of items. Thus, in this day and age asymptotic analysis using the RAM model can only be used to determine an *expected input range* an algorithm can cope with in a reasonable time span. An algorithm with time complexity $\mathcal{O}(n^2)$ for example can operate on different input sizes than an $\mathcal{O}(n^3)$ algorithm. Predictions of running time using the RAM model, even extrapolations thereof, have become difficult due to the increased complexity of our computation platforms. Thus theoretical analysis has to be paired with experimental results.

### 1.1.2 The External Memory Model (EM)

Another very successful model is the *external memory (EM) model* (figure 1.4 (c)) [AV88; VS94]. Its purpose is to model input/output (I/O) transfers between a main memory of size $M$ and a secondary memory of size $N$, which is presumed to be larger and slower. I/O transfers are always in terms of whole *blocks* of size $B$, and *I/Os* are the main cost measure of the model. The model can be extended to the *parallel disk model (PDM)* to account for multiple disks, which allow $P$ parallel block transfers of size $B$ in one unit of cost. The model assumes $1 \leq B \leq M < N$, and $1 \leq P \leq \frac{M}{B}$.

The EM model was designed to model transfers from RAM to rotational disks, and specifically for developing large scale sorting algorithms with data being written/read to/from disks. Accessing a block on a rotational disk, however, is actually a multi-step mechanical process: first the arm containing the read/write head has to be moved to the right track, then the head has to wait until the disks have rotated to the right position, after which the actual transfer can be started. The unit cost of block transfers in the EM model *amortizes* these steps by assuming averages for the non-constant mechanical access times.

Due to its simplicity, the EM model has been widely used for analysis of algorithms processing large amounts of data [Vit98; Vit01; MSS03; DSSS04; AMO07; MO09]. A linear scan through $N$ items takes $\text{scan}(N) = \Theta(\frac{N}{PB})$ I/Os, and sorting $N$ items takes $\text{sort}(N) = \Theta(\frac{N}{PB} \log_{\frac{M}{B}} \frac{N}{M})$ I/Os. Comparing these to a naive random access of $N$ items, which takes $\mathcal{O}(N)$ I/Os, one can see that efficient external memory algorithms can really accelerate applications and are often indispensable.

However, actually *implementing* EM algorithms efficiently is a technical challenge. For this reason, two high quality open-source libraries containing many well-implemented EM containers and algorithms have been developed. *TPIE*, previously short for a *Transparent Parallel I/O Environment*, but now the *Templated Portable I/O Environment* [APV02; ARST17], is the older library and was started in 1994 at the Duke University in North Carolina, U.S.A. In 2003, nine years later, *STXXL*, the *Standard*

*Template library for XXL data sets* [DKS05; Dem06; DKS08; BDS09], was initiated in Germany. Both are C++ template libraries which implement I/O layers, containers, algorithms, and many other concepts to make working with external memory easier. During the preparation time of this dissertation, the author took over maintenance of the STXXL and released version 1.4.0 in December 2013 and 1.4.1 in October 2014. The goals of these two releases were to incorporate all work since the prior release of 1.3.1 in March 2011, and to bring the build and development processes of STXXL to modern standards. These updates made it easier for other people and ourselves to work with STXXL, thus hopefully increasing STXXL's usage.

While the EM model has proven to be a good approximation for working with rotational disks, it does have its weaknesses. As discussed above, modern machines actually have a multi-level memory hierarchy, of which one can view external memory as the last level. Hence, a significant amount of the I/O transfers could be answered by the higher level caches, which are considerably faster. The EM and PDM models only represent two levels in this hierarchy. The *cache-oblivious* model [FLPR99; FLPR12] was proposed to design algorithms which would be efficient on *all levels* of a cache hierarchy by not using knowledge of the block size $B$ and memory size $M$. Despite these restrictions, researchers have developed efficient sorting algorithms [FLPR99; FLPR12; BFV04; BFV08], priority queues [ABD+02; BF02], graph algorithms [ABD+02; BFMZ04], dictionaries [BDFC00; BDFC05], spatial data structures [ABH05; ABH09], and many more. However, the drawback of cache-oblivious algorithms is that they are often only *asymptotically* optimal, and the hidden constants in the $\mathcal{O}(\cdot)$ notation of their I/O complexity are rather large. A well-engineered algorithm which knows $B$ and $M$ obviously has an advantage.

Besides representing only two levels in a hierarchy, the EM model also oversimplifies the rotational disk hardware: data on the outside of the disk platter can be read faster than on the inside due to the amount of memory surface passing under the read/write head per unit of time. Furthermore, if one writes/reads blocks sequentially on a disk, these will often be stored in sequence on the memory platter, hence, there is near-zero seek time between two blocks instead of the amortized average time. Writing in sequence, however, may not be possible due to the file system rearranging blocks as necessary. But generally, file systems and algorithms are optimized to work sequentially, hence the amortized time is probably overstated.

*Solid-state drives* (SSDs), which store data using flash chips instead of rotational platters, have become increasingly cheap and popular in recent years (compare also figure 1.2 on page 3). Each SSD contains a memory controller which maps disk sectors to flash memory cells. Contrary to rotational disks, block access on SSDs does not incur a seek time as there are no mechanical moving parts. However, even SSDs have non-zero per-access overhead due to the memory controller; but this overhead is less than 0.1 ms and hence much lower than the 1 ms of current rotational disks. In 2015, a typical rotational disk achieved sequential data transfer speeds of about 125 MiB/s on our systems, while even a single consumer-grade SATA SSD reached 500 MiB/s read and 400 MiB/s write speeds [BKS15a]. The newer *non-volatile memory express* (NVMe) SSDs now commonly used in medium- to high-end consumer devices and in our experiments in section 8.4 reach effective sequential read speeds of 2.1 GiB/s and write speeds of 800 MiB/s.

### 1.1.3 The Parallel Random Access Machine (PRAM)

Due to the increased prevalence of multi-core machines, a good machine model for parallel computation is needed. Interestingly, the most commonly used model, the *parallel random access machine* (PRAM) (figure 1.4 (b)) [FW78; Gol78; SS79; JáJ92], was proposed and well-studied by theoretical computer scientists in the late 70s and 80s, when parallel machines were exotic and existed only in very small numbers.

A PRAM consists of several sequential processors, which each have individual registers for computation and may run different programs. The processors can all access a *global shared memory*, which is also the only method of communication between processors. The processors are clocked asynchronously or synchronously, meaning access to the common memory is synchronized or unsynchronized. The most common variant is to assume synchronized access, which then opens the question of how to deal with simultaneous read and write operations. This issue branches the machine model into different submodels: the EREW PRAM only allows *exclusive* read and *exclusive* write access to memory cells in each time step, thus disallowing any collisions. The CREW PRAM allows *concurrent* read and *exclusive* write operations, which is maybe most relevant because it disallows problematic write collisions but allows common reads. Lastly, the CRCW PRAM allows both *concurrent* read and *concurrent* write access, and one has to state how write conflicts are resolved: common strategies are to assume that an *arbitrary* processor succeeds in the operation or to assume that the smallest processor id is given *priority*. The CRCW PRAM is the most powerful, but can be simulated by the EREW PRAM at the cost of an $\mathcal{O}(\log p)$ slowdown [JáJ92]. There are many more variants of PRAM and simulation theorems between them, but these do not differ too much. The main issue with the PRAM model is that it is considered unrealistic by many researchers.

A common objection, for example, is that authors frequently assume a PRAM with lots of processors, sometimes up to $n$ (one per input element), or even more. These large, "unrealistic" machine models, however, can in theory be simulated by PRAMs with priority write and a lower number of processors without incurring any slowdown [Bre74]. In reality, however, switching between tasks has non-negligible overhead, which can have significant impact on fine-grained PRAM algorithms.

Another objection is that PRAMs are impractical and uneconomical to build. While some prototypes of PRAMs have actually been built [ADK+93], it is improbable that machines with these theoretical assumptions will ever become wide-spread. The main problem of the model is that *synchronization* via the shared global memory is assumed to have little or zero cost. This is contrary to reality, where ensuring a common shared memory view by maintaining *cache coherence* [Ste90; Mar08] among the private caches of processors imposes the largest costs when scaling to more and more cores. Cache coherence can be seen as a sophisticated automated hardware message passing system, but it scales only to a limited number of processors. This limit will be pushed out farther in the next years, with hundreds and maybe even thousands of cores on a single shared-memory system [MHS12]. For higher scalability, however, one must focus on shared-nothing machines with explicit communication via a network.

Recently, the *parallel external memory (PEM) model* (figure 1.4 (d)) [AGNS08; AGS10] was proposed as a combination of the PRAM and the EM model. It models multiple parallel processors and a two-level memory hierarchy: each processor has a private internal memory (cache) and together they share a large external memory (main memory). Like in the EM model, transfers between internal and external memory are performed in blocks of size $B$, and the number of I/Os performed is the main cost metric. The model is quite versatile and by adapting parameters can model the cache/RAM memory level pair, the main/external memory pair, and also matches some GPU architectures. At the same time this versatility makes designing general algorithms difficult, and it remains to be seen how popular this model will become.

### 1.1.4 Distributed Models

Shared-memory machines retain an easy programming interface via hardware cache coherence. But this automated, algorithm agnostic communication protocol can only scale so far. For larger scale machines, explicit communication becomes imperative as bandwidth and latency of an interconnection network will always be the ultimate bottleneck of scalability. Cluster computers and supercomputers are traditionally "shared-nothing" machines (figure 1.4 (e)) in which the processors are independent (multi-core) machines with distributed memory and an explicit communication network such as Ethernet or Infiniband. Figure 1.5 shows the exponential increase in the number of processors in the most powerful supercomputer systems in the world over the last 25 years.

Maybe the most well-known and popular programming model for high-performance "shared-nothing" machines is the *message passing interface* (MPI) [MPI95; MPI97]. This standard defines an elaborate set of point-to-point communication primitives, collective operations, and supporting utility functions which facilitate programming distributed applications without restricting them too much to any particular communication pattern. The standard was implemented by multiple open-source projects [GLDS96; GFB+04] and vendors of high-performance supercomputers and interconnection networks, which often provide performance-tuned implementations of communications primitives with an MPI interface. On the other hand, MPI was developed in the 90s, focused on the C, C++, and Fortran languages, and was designed with the software engineering practices of the time.

While the fine-grained communication primitives provided by MPI are still necessary for developing high-performance algorithms with explicit communication patterns, MPI has become increasingly too cumbersome for complex applications. In chapter 7 we propose *Thrill*, which provides a higher level abstraction for writing distributed algorithms by combining primitive operations on virtual distributed arrays. In this regard Thrill follows popular distributed processing frameworks such as Apache Hadoop/MapReduce, Apache Spark [ZCF+10], and Apache Flink [ABE+14], which allow defining applications and algorithms using "data-flow" style functional programming.

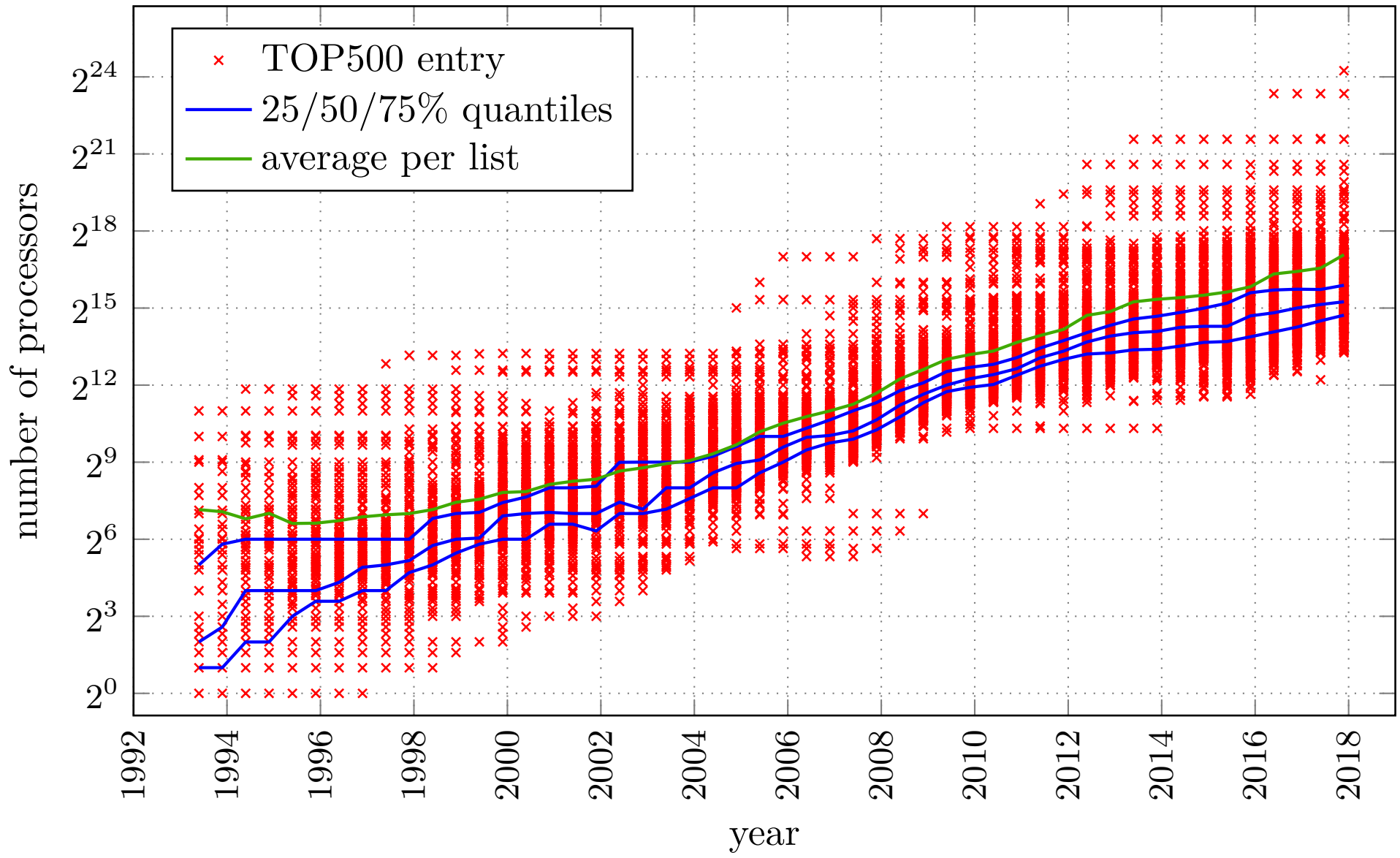

**Figure 1.5:** Graph of all biannual TOP500 lists of the most powerful commercially available computer systems in the world. Data from `https://top500.org`.

Modeling distributed machines for theoretical analysis is notoriously difficult due to the complexity of all the components in the system combined. Thus researchers usually pick only the most relevant aspect of the distributed system and model it. For example, when modeling collective operations such as *Broadcast*, *Reduce*, or *Alltoall*, authors often abstract the network topology using a network model such as a lattice, torus, or hypercube, and assume that sending a message of $k$ bytes between linked nodes costs $T_{\text{start}} + k \cdot T_{\text{byte}}$, where $T_{\text{start}}$ is a per message startup time for routing and other overhead and $T_{\text{byte}}$ is the time for transferring one byte across the link. This network model is good for analyzing short collective interactions, but is too detailed for large applications.

A much broader view of parallel algorithms is established by the *bulk synchronous parallel* (BSP) machine model (figure 1.6) [Val90; GV92; GV94]. A BSP computer consists of a number of processing components (processors), a network that routes one-sided messages between them, and a facility for synchronization of all or a subset of the processors. A BSP algorithm proceeds in a sequence of global *supersteps*, which consist of three phases. In the first phase all processors concurrently perform *local work* independent from another. The second phase allows *communication* between processors, which consists only of one-sided unacknowledged messages. The first and second phase can be intertwined, since the messages can be sent at any time during the local work. Synchronization between processors only happens in the third phase, which consists of a *barrier synchronization.* When local work is finished, all processors must wait until the global barrier is reached. After the barrier, all messages become available to the receiving processors and can be processed in the next local work phase.

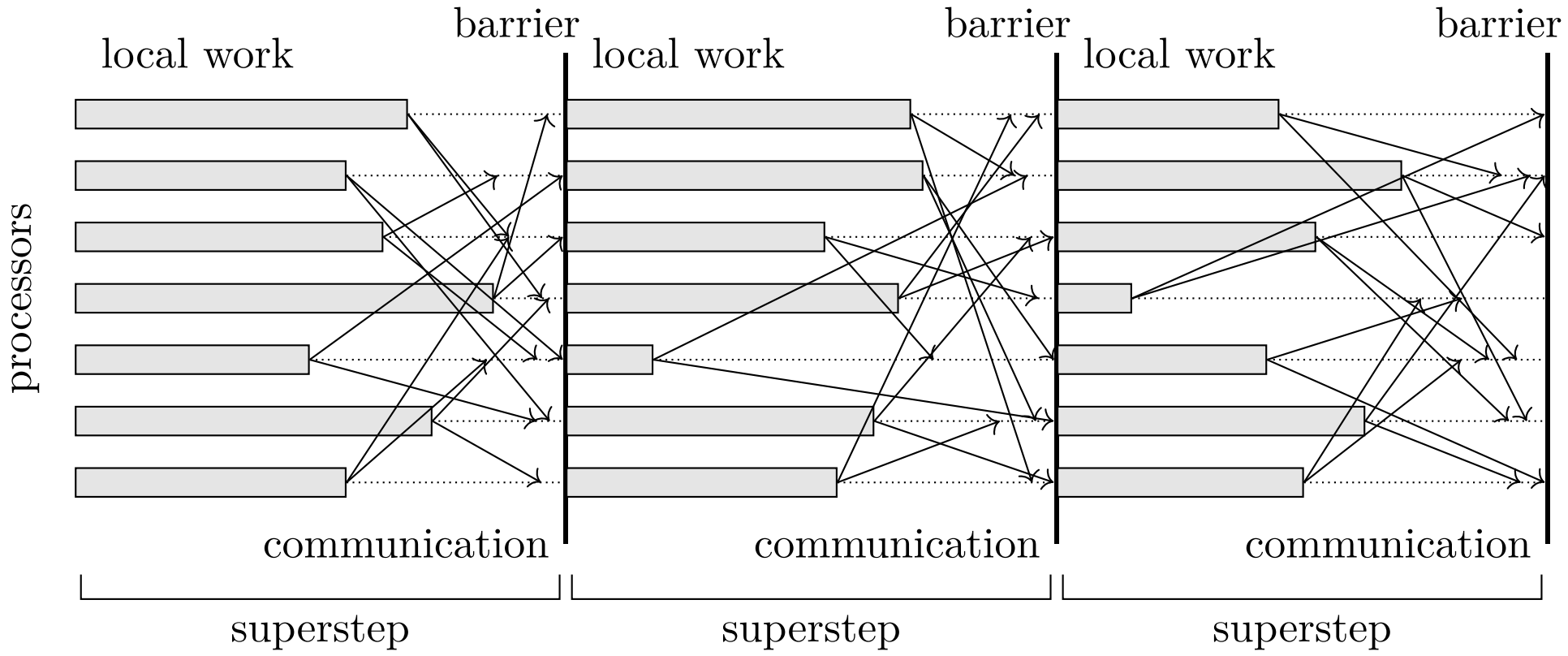

**Figure 1.6:** The bulk synchronous parallel (BSP) machine model.

As implied by its name, the BSP model considers *bulk* messaging. In the original model the communication network is said to simply incur a cost $g$ per message, regardless of size. A processor sending out $h$ messages incurs a communication cost of $gh$. An entire superstep is then said to cost the *maximum* communication of all processors, plus a cost $s$ for the synchronization barrier. While this model definitely has its drawbacks, maybe the most important take-away is to focus on the *number* of supersteps in an algorithm. In our evaluation of algorithms in Thrill (chapter 7), it turns out that barrier synchronizations are some of the most costly aspects: not only do processors have to synchronize, but they have to *wait* for the others to finish. This idle time is wasted computation time and money, and it increases with the number of processors due to statistical fluctuations in local running time. On the other hand *balancing* the amount of local work is very difficult as it means estimating the required time with as little overhead as possible and/or redistributing work online.

## 1.2 Challenges of Algorithmic Experiments

Experimental algorithmics has become an active field of research [DI00; Joh02; McG12], which until recently has taken a back seat to theoretical algorithms. However, conducting experimental algorithmic research may be even more challenging than pure theory due to the higher level of complexity and many pitfalls in producing tangible results. In this section we discuss what we consider the most important hazards and how to stay clear of them.

Experiments and empirical evaluations play a prominent role in many natural sciences, such as biology, chemistry, and physics. There they validate, complement, and interplay with pure theory. In computer science however the approach has long been neglected and only recently gained popularity. On the reasons one can only speculate. Maybe the astounding speed of improvement in microprocessors made it impossible to extrapolate experimental results from one generation of machines to the next. Maybe the field of algorithms was not as well explored in the early days of computer science, and it presented a large amount of low hanging fruit that could be harvested easily and more quickly with theoretical analysis. Maybe the focus on theory is due to the mathematical passion of a small group of influential computer science pioneers. However, since the early days, our computing platforms have become far more complex and their performance much more difficult to predict. At the same time, real-world applications of algorithms have become economically significant and hence their real-world performance much more important. While theoretical analysis will stay a major hallmark of informatics due to its abstract nature, experimental analysis is gaining practical impact. However, the gold standard of scientific rigor is difficult to achieve in experimental algorithmics.

Maybe the aspect of rigorous science most challenging for experimental algorithmics is to ensure *reproducibility* of experiments. The blessing of fast innovation cycles in processor technology which delivered breathtaking performance increases in the last few decades are simultaneously a curse for reproducibility of experiments. Every few years the performance of computing systems doubled and their architecture has been continuously changing. However, in recent years the hardware cycles appear to have become longer, and while computing hardware will continue to improve, the rate of change seems to have decreased.

At the same time, papers in experimental algorithmics are being scrutinized more thoroughly. The renowned ACM Journal of Experimental Algorithmics (JEA) has joined the Replicated Computational Results Initiative, which invites authors to apply for a certificate that their experimental results have been replicated independently. While this initiative is definitely worthwhile, computing platforms will remain diverse and not every researcher will have access to the necessary hardware to reproduce experiments. One has to note that this is also not the case in the natural sciences. However, with cloud computing providers like Amazon Web Services, and ubiquitous and cheap platforms such as the Raspberry Pi, access to common hardware has become more readily available.

Besides the experiment hardware, selection of the algorithms and their *implementations* also has potential hidden hazards which directly affect the reported results' validity. Maybe most difficult to gauge is the *expertise* of the implementer of the algorithms in question. Depending on the skill level, experimental results can vary greatly, and it is nearly impossible to determine the level from reading just the experimental results. Even further, it is also difficult for the authors to gauge their own skill level [KD99].

The gold standard should be to apply the *same* amount of time and expertise to optimizing each implementation in an experimental comparison. The reality of many experimental papers, however, is that the authors implement some new algorithm very well and compare it to only a few prior implementations (not by the authors) on a limited input set. And quite often, the source code or implementations of prior work is no longer available or functional.

Scientific rigor must also extend to the selection of experimental *inputs*. It is all too easy to focus only on those inputs which show the desired outcome, and ignore inopportune outlier results. This has been countered in some communities by agreeing on a certain set of *benchmark instances*. But this in turn opens the possibility of overfitting algorithms and heuristics to this limited benchmark set. Thus designing these benchmark sets must be done with care and foresight.

In general, planning and conducting experiments is hard. Test scenarios must be designed to precisely capture specific aspects such as performance and other metrics, which ultimately confirm or contradict an experimental assumption. Both novice and experienced experimenters are prone to mistaking experimental artifacts as a newsworthy signal. Due to the complexity of our computing machines, the variety of factors which may influence an experiment's outcome and validity is so large that probably no experimenter can grasp the entire system in depth. Experiments may be influenced by random noise, kernel version peculiarities, compiler optimization effects, cache or NUMA effects, alignment of code, transient failures in RAM or disk, delays due to memory fragmentation, or even the weather due to temperature and atmospheric pressure. In our experiments we noticed that even something as simple as rebooting a machine, which resets it into a pristine state, can have an effect on the performance of experiments. In some sense, this aspect of experimental algorithmics is very similar to a natural science, in which the full detail of the object of study will possibly never be known.

Reporting experimental results can also be easily skewed by formatting the results in a way which accentuates aspects advantageous to the authors and hides less fortunate results. Extracting statistically significant results from experimental data is an entire branch of mathematics [FW86].

Targeting a high standard of experimental algorithmics in this dissertation, we try to avoid many of these pitfalls. In the worst case, purely experimental research delivers little more than random numbers for some very specific combinations of machines and algorithms, which carry no meaning beyond the single experiment.

Our first ground rule is to present *precise pseudocode* for our algorithms. The pseudocode we use in this dissertation is written in “concrete implementable detail”, which means a novice programmer should be able to transcribe it into any programming language without having to solve guessing games. While this makes our pseudocode representation rather technical and demanding to read, we believe this level of precision is necessary since the algorithm pseudocode will remain valid even in the case of a revolution in computer architecture.

To reduce the impact of experimental artifacts, we try to perform our experiments on as many inputs and machines as possible: six machines in part I, and two machines in part II. Only in part III were we limited by time and money to one specific instance type on Amazon Web Services. But for those experiments, the machine hardware carried less significance than software quality.

To improve credibility and the ability to reproduce our experimental results, we published *all our implementations and test frameworks as open-source software*. Furthermore, we developed *SqlPlotTools* [Bin14] for this dissertation, which defines a workflow for producing, analyzing, plotting, and formatting experimental results. SqlPlotTools is a tool to extract data series from algorithm experiment logs, convert, process, or aggregated them using SQL statements, and embed the results in gnuplot datafiles or pgfplots LaTeX files.

The input to SqlPlotTools are simple text logs from experiment runs which contain special “RESULT” lines containing key-value metrics. The remaining experiment log is ignored but can kept for future reference or debugging. The result rows are collected into an SQL table from which one can then directly generate tables and plots in LaTeX documents. All tables and figures in this dissertation were created using this process, which precludes most human errors when transcribing or calculating numbers. This process also makes it easy to include many pages of result data, which enrich this dissertation. Including data tables from experiments is common practice in natural sciences, where these may even be the only contribution of a paper.

## 1.3 Preliminaries and Pseudocode

Throughout this dissertation we distinguish between the decimal prefixes of the International System of Units (SI), such as $\mathrm{k} = 10^3$, $\mathrm{M} = 10^6$, and $\mathrm{G} = 10^9$, and the binary prefixes defined by the International Electrotechnical Commission (IEC), such as $\mathrm{Ki} = 2^{10} = 1\,024$, $\mathrm{Mi} = 2^{20} = 1\,048\,576$, and $\mathrm{Gi} = 2^{30} = 1\,073\,741\,824$. Furthermore, to symbolize *bytes* we use a uppercase “B” as unit, and for *bits* the lowercase “b”, as recommended by the IEEE 1541 standard.

The algorithms in this dissertation are written in a tuple pseudocode language, which mixes Pascal-like control flow with array manipulation and mathematical set notation. This enables powerful expressions like $A := [\,(i^2 \bmod 7, i) \mid i \in [0\,..\,5)\,]$, which assigns the array of pairs $[\,(0,0),(1,1),(4,2),(2,3),(2,4)\,]$ to the symbol $A$.

Ordered sequences like arrays are written using square brackets $[\,\ldots]$, '+' can be used to concatenate arrays, and $[a\,..\,b] := [\,a,\ldots,b\,]$, $[a\,..\,b) := [\,a,\ldots,b-1\,]$, and $]a\,..\,b[ := [\,a+1,\ldots,b-1\,]$ are defined as ranges of integers. To make array operations more concise, we assume both $A_i$ and $A[i]$ to be the $i$-th element in the array $A$. Arrays and variables are usually not allocated or declared beforehand, so $A_i := 1$ also implicitly defines an array $A$. The unary operators "++" and "−−" increment and decrement integer variables by one. The symbol $\mathbb{1}_{cond} \in \{0,1\}$ is short for a boolean variable indicating the truthfulness of condition *cond*.

The individual operations in the tuple pseudocode are implementable in the RAM model, with external memory, and even on distributed machines with appropriate algorithms: for example $(i,j) \in A$ can be implemented as a scan over the array $A$, and $A' := \text{Sort}(A)$ calls an appropriate sorting algorithm, which by default sorts tuples lexicographically.

## 1.4 Contributions

This dissertation is composed of three parts. The first part focuses on parallel string sorting on shared-memory multi-core machines, the second part on external memory suffix sorting using the induced sorting principle, and the third part on distributed external memory suffix sorting with the new algorithmic framework Thrill.

Part I addresses parallel string sorting and begins with a comprehensive survey and evaluation of *sequential* string sorting algorithms in chapter 2. In section 2.2 all basic string sorting algorithms in the literature such as multikey quicksort, most significant digit radix sort, burstsort, and LCP-merge sort are reviewed in detail. However, the survey is not limited to the basic algorithms. As preparatory research we analyze how previous authors engineered the string sorting algorithms and discuss optimizations from which we can draw when designing parallel string sorters. In section 2.3 we contribute the most comprehensive experimental evaluation of sequential string sorting algorithms to date in the literature, which compares 39 implementations on six different machines and seven real-world inputs. Our analysis of the results shows that engineered radix sort variants with character caching perform best across all instances.

In chapter 3 further preliminary research for designing parallel string sorting is discussed. A new benchmark tool called *pmbw* is presented, which measures sequential and parallel memory bandwidth and latency to better understand their quantitative characteristics when executing algorithms in parallel. The pmbw tool contains a small set of experimental loops coded in assembly which perform simple memory operations such as scanning an array and walking a random permutation. While sequential memory bandwidth has been measured previously by many authors, to the best of our knowledge pmbw is the first to focus on *parallel* memory bandwidth. Additionally, pmbw is also used to quantify NUMA effects in section 3.2 by measuring access time on remote NUMA nodes. The insights gained with pmbw are interesting not only for parallel string sorting but generally for designing any shared-memory parallel algorithm.

Chapter 4 then concentrates on parallel string sorting algorithms. In section 4.2 we propose *string sample sort*, which is an adaptation of sample sort to string objects. To avoid long string comparisons, splitters in string sample sort are limited in length. By sampling the string set and selecting splitters to balance bucket sizes, string sample sort combines the advantages of both multikey quicksort and radix sort. With *Super Scalar String Sample Sort* we present an engineered version which avoids branch mispredictions, exploits instruction-level superscalar parallelism, and is optimized to keep the classification data structure in cache. The algorithm is designed to be parallelized efficiently and with *Parallel Super Scalar String Sample Sort* ($\mathrm{pS}^5$) we propose our first engineered parallel string sorter. In section 4.2.6 we show that Super Scalar String Sample Sort with word size $w$ has expected sequential running time $\mathcal{O}(\frac{D}{w} + n \log n)$ for $n$ strings with distinguishing prefix $D$ when comparing for equality at each splitter and $\mathcal{O}((\frac{D}{w} + n) \log v + n \log n)$ when unrolling the classification. Furthermore, in section 4.2.7 we show that one step of Parallel Super Scalar String Sample Sort runs on a CREW PRAM with $p < \frac{n}{v}$ processors in $\mathcal{O}(\frac{n}{p} \log v + \log p)$ time and $\mathcal{O}(n \log v + pv)$ work where $v$ is the number of splitters. More details on the engineered parallel implementation are given in sections 4.2.3 and 4.2.8.

With $\mathrm{pS}^5$ we focus on *distribution-based* string sorting by splitting string sets into smaller sets with longer distinct common prefixes. In section 4.3 we turn to *merge-based* string sorting and propose *LCP-aware multiway merge*, which can be used to merge multiple sorted string sequences and use known LCP values to accelerate the operation. Besides the sorted string array, the algorithm can also output the corresponding LCP array. We show that LCP-aware $K$-way merge requires at most $\Delta L + n \log_2 K + K$ character comparisons to merge $K$ sorted sequences containing $n$ strings in total, where $\Delta L$ is the difference of the sum of all LCP array entries of the presorted sequences and the sum of the output LCP array. LCP-aware merge is then used to propose *multiway LCP-Mergesort* as a stand-alone sequential merge-based string sorting algorithm, which requires at most $L + n\lceil \log_K n \rceil \log_2 K + (n-1)\frac{K}{K-1}$ character comparisons and runs in $\mathcal{O}(D + n \log n + \frac{n}{K})$ time, where $L$ is the sum of all entries in the LCP array. We parallelize multiway LCP-Mergesort and explore different splitting heuristics to balance work onto processors. Furthermore, we adapt our parallel algorithms to NUMA machines by designing a hybrid algorithm which first runs $\mathrm{pS}^5$ on each NUMA node independently and then uses LCP-aware multiway merge as *top-level* algorithm to combine the presorted sequences. For this hybrid algorithm, $\mathrm{pS}^5$ is extended to also save the LCP values when sorting strings, and *LCP-insertion sort* is proposed in section 4.3.5 as an LCP-aware base-case sorter for pS[5] which runs in $\mathcal{O}(D + n^2)$ time and reuses information while building the LCP array.

In section 4.5 we compare all newly proposed parallel algorithms and the few existing competitors in a large experimental evaluation. We use six different machines and seven inputs to determine the practical performance of the algorithms across a large set of instances. Overall our parallel string sorting implementations yield high speedups, reaching a factor of three on a 4-core desktop machine and a factor of 20 on a 32-core server-class machine. We believe that future applications which sort large string sets on multi-core systems will benefit greatly from the algorithms we designed.

With part II we turn our focus to the problem of suffix sorting. Chapter 5 introduces the reader to suffix arrays and the versatile history of suffix and LCP array construction algorithms. This survey brings context and an overview to the development of suffix sorting algorithms, which are largely derived from three basic suffix sorting principles: *prefix doubling*, *induced sorting*, and *recursion*.

Chapter 6 presents our initial contribution to the field of suffix sorting: the first external memory algorithm based on the induced sorting principle, called *eSAIS*. Induced sorting is used by the fastest RAM-based suffix sorters and we were able to accelerate external memory suffix sorting by a factor of two using this principle. The exposition of eSAIS begins in section 6.1 by presenting induced suffix sorting and LCP array construction in *main memory*. In section 6.2 the algorithm is then transferred to external memory using an elegant reformulation of the central loop of induced sorting with an external memory *priority queue*. Our analysis shows that eSAIS requires at most $\textsc{Sort}(17n) + \textsc{Scan}(9n)$ I/O volume for a string of length $n$. We then extend eSAIS and integrate external memory LCP array construction using range minimum queries. Our implementation of this algorithm using STXXL is the first fully external suffix and LCP array construction published in the literature. We demonstrate the scalability of eSAIS with experiments on real-world inputs such as 80 GiB of Wikipedia text using machines with only 4 GiB of main memory. eSAIS was first published in 2013 and since then the interest in the area of external memory suffix and LCP array construction algorithms has increased. Chapter 6 closes with a review of algorithms published after eSAIS, which greatly enhanced our work and resulted in the currently best external memory suffix sorters.

In part III we turn to larger distributed memory cluster systems to harness scalable computation resources for suffix sorting and other algorithms. Chapter 7 introduces our new high-performance C++ big data framework *Thrill*, which attempts to bridge the gap between MPI and MapReduce-like frameworks such as Apache Spark and Apache Flink. Thrill's central concept is the *distributed immutable array* (DIA) which is a virtual distributed array of C++ objects. These arrays can be manipulated using a small set of primitive operations such as sorting, mapping, merging, reducing, prefix sums, and scanning. Large complex algorithms can be implemented using Thrill by parameterizing and combining this small set of scalable operations using data-flow-style functional programming. Thrill uses C++ template meta-programming to efficiently couple DIA operations without intermediate buffering and with minimal indirections. The individual DIA operations are implemented as pipelined distributed external memory algorithms, which transparently use disk memory when the processed data exceeds main memory. More details on the prototype implementation are given in section 7.3. In section 7.4 an experimental study is conducted comparing Thrill with Apache Spark and Apache Flink using five micro benchmarks: WordCount, PageRank, Terasort, KMeans, and Sleep. Thrill consistently outperforms the other frameworks on all instances run on a cluster of up to 16 hosts in the AWS Elastic Compute Cloud. Thrill is free and open-source software which other researchers can use to develop their own algorithms and solve their big data processing needs.

In chapter 8, Thrill is applied to suffix sorting as a case study. Using Thrill we present the first *distributed external memory* suffix sorting implementations in the literature. We propose five suffix sorters: three based on prefix doubling and two based on the difference cover algorithm [KS03; KSB06]. These are the most complex algorithms implemented using Thrill to date and demonstrate the expressiveness of combining the small set of scalable primitives provided by Thrill. Using these primitives also makes the pseudocode in the chapter much more precise than relying on prose. In section 8.4 we run the five implementations on up to 32 hosts with fast NVMe SSDs and RAM limited to 8 GiB in the AWS Elastic Compute Cloud (EC2). We compare them against two MPI implementations and against the fastest non-distributed sequential suffix sorters. Our experimental results show that algorithms implemented in Thrill are competitive with hand-coded MPI implementations. By using the Thrill framework the algorithms automatically benefit from possible future enhancements, such as fault tolerance and faster sorting implementations. Using 32 hosts, we can suffix sort 16 GiB of Wikipedia text in 30 min, or 16 GiB of digits of $\pi$ in 15 min. Our Thrill implementations scale higher than the MPI implementations which are constrained by RAM. Compared to the fastest sequential suffix sorters (divsufsort and sais), our best Thrill implementations are faster on digits of $\pi$ when run with 2 hosts (32 cores), and on Wikipedia when run with 4 hosts (64 cores).

In total, we present and analyze three novel sequential string sorting algorithms, engineer more than five parallel string sorting algorithms variants, propose a new external memory suffix sorting algorithm, present a new distributed computing framework in C++, and five suffix sorting algorithm implementations in the framework as a case study.

# I

# Engineering Parallel String Sorting

*We discuss how string sorting algorithms can be parallelized on modern multi-core shared-memory machines. As a synthesis of the best sequential string sorting algorithms and successful parallel sorting algorithms for atomic objects, we propose* string sample sort, *and its engineered parallelization, Parallel Super Scalar String Sample Sort (p$S^5$), in section 4.2. In section 4.3 we turn our focus to NUMA architectures and contribute parallel LCP-aware multiway merging both as a stand-alone string sorter and as a top-level algorithm for combining presorted sequences. Broadly speaking, we propose both* multiway distribution-based *string sorting with $S^5$ and* multiway merge-based *string sorting with LCP-aware merge(-sort), and parallelize both approaches. Additionally, we develop parallelizations of caching multikey quicksort and radix sort in section 4.4.*

*Preliminary research on the properties of sequential string sorting algorithms and on parallel memory bandwidth and latency are discussed in chapters 2 and 3. The insights gained therein are reflected in the design of our parallel string sorting algorithms and are of independent interest for future parallel algorithm development.*

*We compare all our parallel string sorting algorithms experimentally in section 4.5 on six modern multi-core machines using seven inputs. In all our experiments, our new parallel sorting implementations show very good speedups, which are much higher than those of all previous implementations.*

Chapter 2

# Overview of Sequential String Sorting Algorithms

2

*Sorting is perhaps the most thoroughly studied algorithmic challenge in computer science. While the simplest model for sorting assumes* atomic keys*, an important class of keys are* strings *or* vectors *to be sorted lexicographically. Here, it is important to exploit the structure of the keys to avoid costly repeated operations on the entire string. String sorting is used for example in database index construction, some suffix sorting algorithms, or when sorting high-dimensional geometric data.*

*In this chapter we give an overview of pre-existing basic sequential string sorting algorithms, acceleration techniques, and further related work as the basis for developing parallel algorithms. Section 2.3 presents an experimental evaluation of many sequential string sorting algorithms and a discussion of their techniques.*

While the main topic of this dissertation part are parallel string sorting algorithms, we first review the basic sequential string sorting algorithms in section 2.2: multikey quicksort, most significant digit (MSD) radix sort, burstsort, LCP-mergesort, insertion sort, and more. These well-known algorithms form the basis for any development of new string sorting algorithms.

Whereas these basic algorithms are common textbook knowledge, we also dive deeper into engineered variants thereof. These are maybe even more important, because such acceleration techniques turn out to be the key to fast practical implementations: engineered versions employ *character caching*, *loop fission*, and *in-place* or *out-of-place pointer redistribution*.

In section 2.3 we then present a large experimental evaluation of virtually every string sorting algorithm available. We believe this comparison to be the most comprehensive evaluation of sequential string sorting to date. Our final recommendations are given at the end of section 2.3.4.

Part I of this dissertation is based on our papers on engineering parallel string sorting [BS13a; BES17], of which we are the main author. This chapter on *sequential* string sorting, however, is composed largely of unpublished exploratory research prior to developing parallel algorithms. These precursory findings are of independent interest, as we are not aware of any broad evaluation of sequential string sorting algorithms.

## 2.1 Notation and Preliminaries

Our input is an array $\mathcal{S} = [\, s_0, \dots, s_{n-1} \,]$ of $n$ strings with total length $N$. A *string* $s$ is a zero-based array of characters from the *alphabet* $\Sigma = \{1, \dots, \sigma\}$. The canonical lexicographic ordering relation '$<$' is assumed when comparing strings, and our goal is to sort $\mathcal{S}$ lexicographically. For the implementation and pseudocode, we require that strings are zero-terminated, i.e. $s[|s|-1] = 0 \notin \Sigma$, where $|s|$ is the total number of characters including the additional terminating zero. This convention can be replaced using other end-of-string indicators, like an explicit string length.

Let $D(\mathcal{S})$ or just $D$ denote the *distinguishing prefix size* of $\mathcal{S}$, i.e. the total number of characters that need to be inspected in order to establish the lexicographic ordering of $\mathcal{S}$. $D$ is a natural lower bound for the execution time of sequential string sorting. If, moreover, sorting is based on character comparisons, we get a lower bound of $\Omega(D + n \log n)$.

Arrays of strings are usually represented as arrays of *pointers* to the beginning of each string. This indirection means that, in general, every access to the characters of a string incurs a cache fault even if we are scanning an array of strings. This is a major difference to sorting algorithms for atomic keys where scanning is very cache efficient. Our target machine is a shared-memory system supporting $p$ hardware threads or processing elements (PEs), on $\Theta(p)$ cores.

To avoid special cases in the algorithm descriptions, we use the following sentinels: '$\varepsilon$' is the empty string, which is lexicographically smaller than any other string, '$\infty$' is a character or string larger than any other character or string, and '$\perp$' is an undefined value.

For two arrays $s$ and $t$, let $\textsc{lcp}(s,t)$ denote the length of the *longest common prefix* (LCP) of $s$ and $t$. This function is symmetric, and for zero-based arrays the LCP value denotes the first index where $s$ and $t$ mismatch, while all positions up to and including $\textsc{lcp}(s,t) - 1$ match in $s$ and $t$. In a sequence $x$ let $\textsc{lcp}_x(i)$ denote $\textsc{lcp}(x_{i-1}, x_i)$. For a sorted sequence of strings $\mathcal{S} = [\, s_0, \dots, s_{n-1} \,]$ the *associated LCP array* $H(\mathcal{S})$ or just $H$ is $[\, \perp, h_1, \dots, h_{n-1} \,]$ with $h_i = \textsc{lcp}_{\mathcal{S}}(i) = \textsc{lcp}(s_{i-1}, s_i)$. For the empty string $\varepsilon$, let $\textsc{lcp}(\varepsilon, s) = 0$ for any string $s$.

We will often need the sum over all values in the LCP array $H(\mathcal{S})$ (excluding the first), and denote this as $L(\mathcal{S}) := \sum_{i=1}^{n-1} H(\mathcal{S})[i]$, or just $L$ if $\mathcal{S}$ is clear from the context. The distinguishing prefix size $D$ and $L$ are related but not identical. While $D$ includes all characters counted in $L$, additionally, $D$ also accounts for the distinguishing characters, some string terminators and characters of the first string (see figure 2.1). In general, we have:

**Lemma 2.1 (Relationship of Distinguishing Prefix $D$ and LCP Sum $L$)**
*For any string set $\mathcal{S}$,*

$$n + L \leq D \leq 2L + n \,.$$

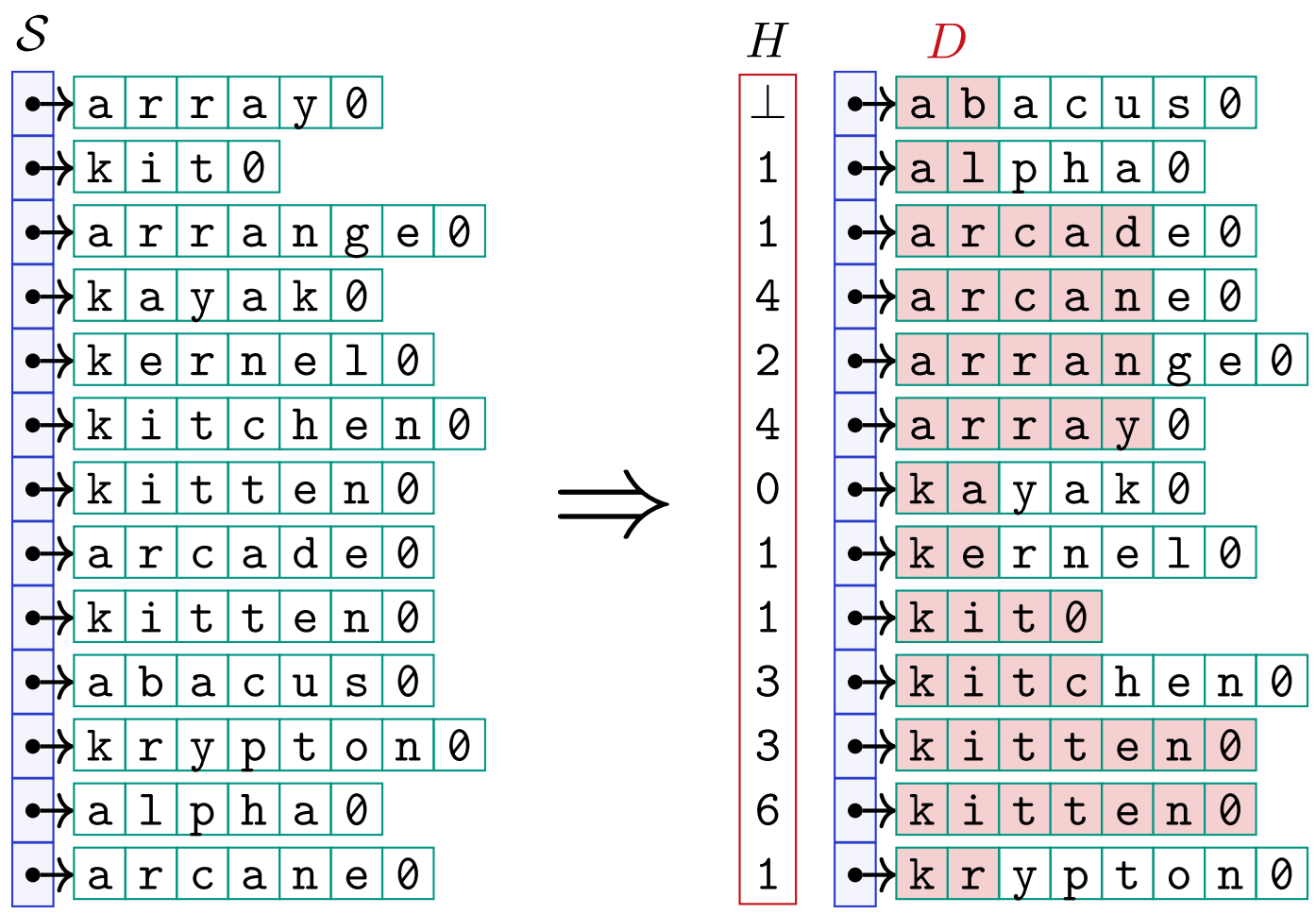

**Figure 2.1:** Example of an array of strings $\mathcal{S}$, the array after sorting, the distinguishing prefix $D(\mathcal{S}) = 52$, and the LCP array $H(\mathcal{S})$ with $L(\mathcal{S}) = 27$.

*Proof.* Consider the distinguishing prefix $d_i$ of $s_i$ from the *sorted* string array $\mathcal{S} = [\,s_0, \ldots, s_{n-1}\,]$. We have $d_i = \max\{\textsc{lcp}_\mathcal{S}(i) + 1,\ \textsc{lcp}_\mathcal{S}(i+1) + 1\}$, since the maximum number of letters needed to determine the order is exactly one more than the longest common prefix with either of its neighbors in the sorted array. To handle the corner cases, we can assume $\textsc{lcp}_\mathcal{S}(0) = 0$ and $\textsc{lcp}_\mathcal{S}(n) = 0$. Hence, we have

$$\begin{aligned} D = \textstyle\sum_{i=0}^{n-1} d_i &= n + \textsc{lcp}_\mathcal{S}(1) + \textstyle\sum_{i=1}^{n-2} \max\{\textsc{lcp}_\mathcal{S}(i),\ \textsc{lcp}_\mathcal{S}(i+1)\} + \textsc{lcp}_\mathcal{S}(n-1) \\ &\le n + \textsc{lcp}_\mathcal{S}(1) + \textstyle\sum_{i=1}^{n-2} \textsc{lcp}_\mathcal{S}(i) + \textstyle\sum_{i=1}^{n-2} \textsc{lcp}_\mathcal{S}(i+1) + \textsc{lcp}_\mathcal{S}(n-1) \\ &= n + \textstyle\sum_{i=1}^{n-1} \textsc{lcp}_\mathcal{S}(i) + \textstyle\sum_{i=1}^{n-1} \textsc{lcp}_\mathcal{S}(i)\,, \end{aligned}$$

from which the inequality easily follows due to $L = \sum_{i=1}^{n-1} \textsc{lcp}_\mathcal{S}(i)$. □

Note that $D \le 2L + n$ is a pathological bound, but necessary to show $D = \Theta(L)$. An example of a string set with $D = 2L + n$ is $[\,\texttt{a}, \texttt{ab}, \texttt{b}, \texttt{bb}\,]$, with both $D = 8$ and $2L + n = 2 \cdot 2 + 4 = 8$.

## 2.2 Basic Sequential String Sorting Algorithms

In the following subsections we give an overview of most efficient sequential string sorting algorithms. Nearly all algorithms split the original string array $\mathcal{S}$ into smaller subarrays with a distinct common prefix of length $h$. The smaller arrays are then sorted recursively, until only a single item remains or a base-case string sorter is called.

In principle, one could switch algorithms at each level of recursion, and the following pseudocode implementations generically invoke the function "StringSort($\mathcal{S}, h$)" as a placeholder for any algorithm. This function can select the algorithm by string set size and available resources. The sorting procedures are initially started with the whole string set and $h = 0$.

### 2.2.1 Multikey Quicksort

Bentley and Sedgewick [BS97] proposed a simple but effective adaptation of quicksort to strings (which they call multikey data). When all strings in $\mathcal{S}$ have a common prefix of length $h$, the algorithm uses character $x = s[h]$ of a pivot string $s \in \mathcal{S}$ (e.g. a pseudo-median or random string) as a *splitter* character. $\mathcal{S}$ is then partitioned into $\mathcal{S}_<$, $\mathcal{S}_=$, and $\mathcal{S}_>$ depending on comparisons of the $(h+1)$-th character with $x$. Recursion is done on all three subproblems, with the exception of $\mathcal{S}_=$ if $x = 0$ is the zero-termination. Algorithm 2.1 shows concrete pseudocode for multikey quicksort.

The key observation is that the strings in $\mathcal{S}_=$ have common prefix length $h + 1$ which means that compared characters found to be equal with $x$ will never be considered again. Insertion sort (section 2.2.5) is used as a base case for constant size inputs. This leads to a total expected execution time of $\mathcal{O}(D + n \log n)$. Multikey quicksort works well in practice in particular for inputs which fit into cache, and it very efficiently handles string sets with large common prefixes.

In fact, one of the overall best sequential string sorting algorithm in our experiments was a *variant* of multikey quicksort from Rantala's extensive algorithm collection [Ran07], which was also described but not implemented by Ng and Kakehi [NK07]. Two simple enhancements were added: First, instead of comparing single characters of the pivot string, a whole machine word of $w = 8$ characters is compared with each string. And second, instead of repeating random access in each level of recursion, the compared $w$ characters are fetched and *cached* in an additional array. The cache array is aligned with the current string pointer array, hence, an additional $nw$ bytes are required. A *parallel* version of caching multikey quicksort is presented in section 4.4.2.

### 2.2.2 Most Significant Digit (MSD) Radix Sort

Given a string array with common prefix length $h$, most significant digit (MSD) radix sort looks at the $(h + 1)$-th character and produces $\sigma$ subproblems which are then sorted recursively with common prefix length $h + 1$. This natural approach is a good algorithm for large inputs and small alphabets as it uses the maximum amount of information within a single character. Many very important improvements to the base radix sort algorithm have been proposed in the literature.

The first obvious but significant improvement is that for input sizes $o(\sigma)$ MSD radix sort is no longer efficient and one has to switch to a different algorithm for the base case.

**Algorithm 2.1 :** Sequential Multikey Quicksort, adapted from [BS97]

**Function** MultikeyQuicksort($\mathcal{S}, h$)
  **Input :** $\mathcal{S} = [\, s_0, \ldots, s_{n-1} \,]$ an array of $n$ strings with common prefix $h$.
  **swap**($s_0, s_{\text{SelectPivot}(0,\ldots,n-1)}$)   *// Select pivot string, and set up iterators:*
  $a := b := 1, \quad c := d := n-1, \quad x := s_0[h]$   //

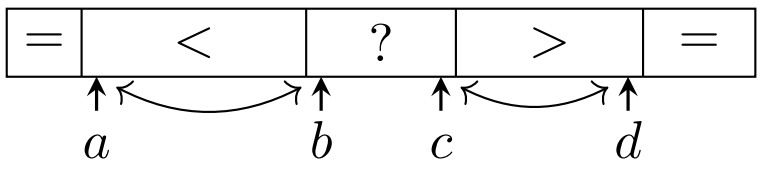

  **while** *true* **do**
    **while** $b \le c$ **and** $(r := s_b[h] - x) \le 0$ **do**
      **if** $r = 0$ **then** **swap**($s_a, s_b$), $a{+}{+}$   *// Swap equal elements to a,*
      $b{+}{+}$   *// advance b while strings are* $< s_0$.
    **while** $b \le c$ **and** $(r := s_c[h] - x) \ge 0$ **do**
      **if** $r = 0$ **then** **swap**($s_c, s_d$), $d{-}{-}$   *// Swap equal elements to d,*
      $c{-}{-}$   *// advance c while strings are* $> s_0$.
    **if** $b > c$ **then** **break**
    **swap**($s_b, s_c$), $b{+}{+}$, $c{-}{-}$   *// Swap* $s_b$*, which is* $> s_0$*, and* $s_c$*, which is* $< s_0$.
  $p := \min(a, b-a), \quad q := \min(d-c, n-1-d)$   *// Swap equal element to center,*
  **swap**($\mathcal{S}[0 \mathinner{..} p), \mathcal{S}[b-p \mathinner{..} b)$)
  **swap**($\mathcal{S}[b \mathinner{..} b+q), \mathcal{S}[n-q \mathinner{..} n)$)   //

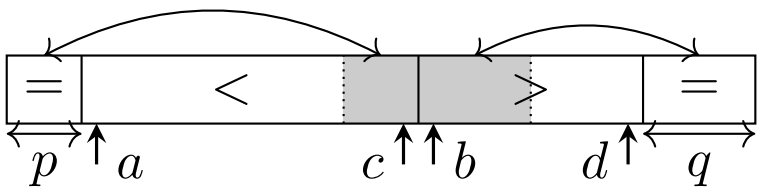

  StringSort($\mathcal{S}[0 \mathinner{..} b-a), h$)
  **if** $s_{b-a}[h] \ne 0$ **then**   *// then recurse on* $\mathcal{S}_<$ *part, on* $\mathcal{S}_=$ *part if*
    StringSort($\mathcal{S}[b-a \mathinner{..} (n-1-(d-b))), h+1$)   *// not zero-terminated,*
  StringSort($\mathcal{S}[n+c-d \mathinner{..} n), h$)   *// and on* $\mathcal{S}_>$ *part.*
  **Output :** The array $\mathcal{S}$ is fully sorted lexicographically.

The running time is $\mathcal{O}(D)$ plus the time for solving the base cases. Using multikey quicksort for the base case yields an algorithm with running time $\mathcal{O}(D + n \log \sigma)$.

McIlroy, Bostic, and McIlroy [MBM93] were the first to systematically engineer different variants of radix sort and propose good concrete practical implementations. They first recognized the problem of “bookkeeping piles”: either one scans over the strings once and requires dynamically growing lists or arrays for each of the $\sigma$ subproblems (variant “D” for *dynamic*), or one scans the strings twice, first to count the occurrences of each character, and then to distribute the subproblems into the predetermined buckets of a continuous array (variant “C” for *counting*). Despite scanning twice, the best variants of the second approach are faster in their (and our) experiments due to the high cost of managing dynamic lists.

But also for variant “C” they propose two different solutions: one with a second temporary pointer array, into which the strings are distributed in $\sigma$ buckets by $(h+1)$-st character, and a second called “American flag sort” which permutes the pointers *in-place.* While the first variant yields a stable string sorter, the in-place version is not stable. The difference is discussed in detail later in this section.

**Algorithm 2.2 :** Sequential Radix Sort "CE0", adapted from [KR08; Ran07]

```
Function RadixSortCE0(S, h)
    Input : S = [ s_0, ..., s_{n-1} ] an array of n strings with common prefix h.
    c := [ 0, ..., 0 ]                      // Allocate |Σ| integer counters initialized with zero,
    for i = 0, ..., n − 1 do  c[s_i[h]]++              // and count character occurrences.
    b := [ 0, ⊥, ..., ⊥ ]                     // Calculate exclusive prefix sum of counters
    for i = 1, ..., n − 1 do  b[i] := b[i − 1] + c[i − 1]          // as bucket pointers.
    T := allocate(n, string pointer)      // Allocate temporary array for sorted output.
    for i = 0, ..., n − 1 do                // Reorder
        T[b[s_i[h]]] := move(s_i)           // into
        b[s_i[h]]++                         // buckets
    copy(T → S),   deallocate(T, b).
    x := c[0]                                       // Track beginning of bucket as x,
    for i = 1, ..., |Σ| − 1 do                  // recurse into every unfinished bucket,
        StringSort(S[x .. x + c[i]), h + 1)      // except for the first (zero-termination),
        x := x + c[i]                            // which contains all fully sorted strings.
    Output : The array S is fully sorted lexicographically.
```

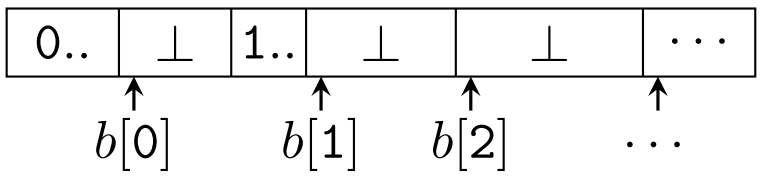

Paige and Tarjan [PT87] previously presented theoretical considerations on sorting binary strings by iteratively refining them by prefix. They presented the first $\mathcal{O}(D+\sigma)$ radix sorting algorithm, but did not consider any implementation. Andersson and Nilsson [AN94] improved these previous theoretical results from $\mathcal{O}(n(\frac{\bar{D}}{\log n}+1))$ to $\mathcal{O}(n\log(\frac{\bar{D}}{\log n}+2))$ on a word-RAM under the assumption that a machine word $w=\Omega(\bar{D})$ with $\bar{D}=\frac{D}{n}$ the average number of bits in a distinguishing prefix. While these theoretical considerations did not yield a practical algorithm, the same authors presented implementations of "Forward Radix Sort" and "Adaptive Radix Sort" [AN98], which promised asymptotically good theoretical properties by using linked lists to store the buckets. Their experiments exhibited acceptable performance, however, we were unable to reproduce such good results with their source code in our own experiments on modern hardware.

The collection of radix string sorting implementations was extended by Ng and Kakehi [NK07] with a *caching* variant. Instead of fetching just one character from a string, they fetch $z$ characters at once and store them in a "cache" buffer aligned with the string pointer array. The next $z-1$ recursion steps then no longer suffer a cache miss for the character lookup on the string. This caching radix sort, named "CRadix sort", outperformed all others in their experiments. As mentioned previously, Ng and Kakehi [NK07] also describe a *caching* variant of multikey quicksort, but they do not go into details and concentrate on radix sort.

Kärkkäinen and Rantala [KR08] presented an up-to-date experimental study of many radix string sorter variants. Besides also incorporating caching of characters and

**Algorithm 2.3 :** Sequential Radix Sort "CE1", adapted from [KR08; Ran07]

1 **Function** RadixSortCE1($\mathcal{S}, h$)
  **Input :** $\mathcal{S} = [\, s_0, \dots, s_{n-1} \,]$ an array of $n$ strings with common prefix $h$.
2 $o :=$ allocate($n$, character) *// Allocate temporary "oracle" array for caching chars*
3 $c := [\, 0, \dots, 0 \,]$ *// and $|\Sigma|$ integer counters initialized with zero.*
4 **for** $i = 0, \dots, n-1$ **do** *// Cache characters and count occurrences.*
5   $o[i] := s_i[h], \quad c[o[i]]{+}{+}$ *// See figure 2.2 for an improvement.*
6 $b := [\, 0, \bot, \dots, \bot \,]$ *// Calculate exclusive prefix sum of counters*
7 **for** $i = 1, \dots, n-1$ **do** $b[i] := b[i-1] + c[i-1]$ *// as bucket pointers.*
8 $\mathcal{T} :=$ allocate($n$, string pointer) *// Allocate temporary array for sorted output.*
9 **for** $i = 0, \dots, n-1$ **do** *// Reorder:*
10   $\mathcal{T}[b[o[i]]] :=$ **move**$(s_i), \quad b[o[i]]{+}{+}$
11 copy($\mathcal{T} \to \mathcal{S}$), deallocate($\mathcal{T}, o, b$)
12 ... perform recursion as in lines 11 to 14 of algorithm 2.2 (page 30) ...
  **Output :** The array $\mathcal{S}$ is fully sorted lexicographically.

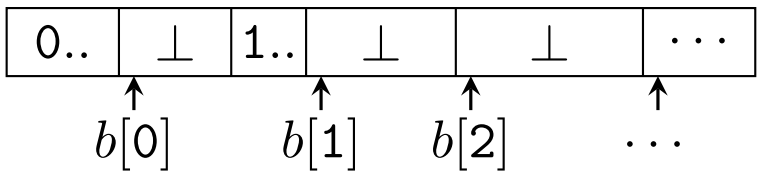

adding adaptive 16-bit radix sorts, they highlighted *counter-intuitive* modifications to the inner loops which take advantage of modern processors' super-scalar accelerations and memory latency hiding. Beyond their paper, Rantala [Ran07] implemented an entire collection of other string sorting algorithms, all of high quality, and many of them are compared in section 2.3.

Kärkkäinen and Rantala's [KR08] radix sort algorithms are the *fastest practical string sorters*, and we discuss the steps in their evolution in great detail in the next paragraphs, since we have to incorporate these accelerations into our parallel string sorting algorithms. Reaching ahead to section 2.3, we also mention some experimental results of the radix sort implementations from table 2.3 (page 46).

**CE0:** Algorithm 2.2 shows their baseline radix sort which sorts by the $h$-th character by performing two scans over the strings. The first scan (line 3) is used to count the occurrences of each character in an array $c$. Then the exclusive prefix sum of $c$ is calculated (lines 4 to 5), which serves as an array $b$ of indices into a temporary string array $\mathcal{T}$. The strings are then scanned a second time, and each string $s_i$ is inserted into the bucket $b[s_i[h]]$ matching its character $s_i[h]$ (lines 7 to 8). This redistribution of strings is performed *out-of-place* by writing into the temporary array $\mathcal{T}$. After the scan, the temporary array swaps names with $\mathcal{S}$, and the buckets are recursively sorted with common prefix $h+1$ (lines 11 to 14). This algorithm, called variant CE0, requires $|\Sigma|$ counters per recursion level of stack memory and $n$ pointers of temporary extra memory.

**CE1:** Their first enhancement, variant CE1, shown in algorithm 2.3, adds caching of characters in an additional temporary array $o$ (line 5). This array costs $n$ characters of extra temporary memory, but greatly accelerates the second scan during the **for** $i = 0, \dots, n-1$ **do**
  ⌊ $o[i] := s_i[h], \quad c[o[i]]$++

$\Rightarrow$

**for** $i = 0, \dots, n-1$ **do**
  ⌊ $o[i] := s_i[h]$
**for** $i = 0, \dots, n-1$ **do**
  ⌊ $c[o[i]]$++

(a) fused loop (b) fissioned loop

**Figure 2.2:** Loop fission in radix sort [KR08]. Splitting the loop improves performance due to memory latency hiding and superscalar parallelism.

redistribution of strings in line 10 by cutting the number of random access to characters (cache misses) in half. In our experiments the CE1 variant was about 33% faster than CE0.

**CE2:** As second enhancement step in variant CE2, they separate the loop in lines 4 to 5 into two passes, as shown in figure 2.2. This acceleration technique is called *loop fission* and is counter-intuitive, because more work is performed. Instead of reading each character $s_i[h]$, storing it in $o[i]$, and incrementing $c[o[i]]$, the fissioned loop touches each character *twice*: first read $s_i[h]$ and store it in $o[i]$ for each string, and then retrieve $o[i]$ and increment $c[o[i]]$ in a second loop. In our experimental evaluation this variant is about 40% faster than with the fused loop. Kärkkäinen and Rantala [KR08] attribute this speedup to a hardware acceleration called *memory latency hiding*, which is triggered by the processor when it detects loops of a certain kind and issues multiple independent requests for memory simultaneously. Since random access to the string's characters is one of the main bottlenecks, exploiting this hardware feature is crucial.

**CE3:** The third enhancement step CE3 adaptively performs 16-bit radix sorting for string sets larger than $2^{16}$. This requires a larger character cache of $2n$ characters, and a larger bucket array of $|\Sigma|^2$ counters per recursion level, but yields another small performance improvement in practice. One problem with such large alphabets is that one will incur many cache faults during redistribution of the string pointers if the cache cannot support $|\Sigma|^2$ concurrent output streams (see [MS03] for details). For $n < 2^{16}$ the variant CE3 falls back to running 8-bit radix sort variant CE2. On average, CE3 is about 14% faster than CE2 in our experiments.

The extra memory requirement of CE3 is large: $2n$ characters for the cache, $n$ string pointers for the out-of-place reordering, and $|\Sigma|^2$ counters per level of recursion. An attractive method to reduce this requirement is to permute the string pointers *in-place* using a very elegant but tricky technique already developed by McIlroy, Bostic, and McIlroy [MBM93], which they called "American flag sort".

**CI2:** Algorithm 2.4 presents our adaptation of this algorithm, as variant CI2, based on the version by Rantala [Ran07]. The basic idea is that every permutation is composed of cycles. Counting characters with caching (lines 2 to 5), and calculating an *inclusive* prefix sum (lines 6 to 9) is done as in previous algorithms. In line 12 the

**Algorithm 2.4 :** Sequential Radix Sort "CI2", adapted from [KR08; Ran07]

```
1  Function RadixSortCI2(S, h)
     Input : S = [ s_0, ..., s_{n-1} ] an array of n strings with common prefix h.
2    o := allocate(n, character)  // Allocate temporary "oracle" array for caching chars
3    c := [ 0, ..., 0 ]            // and |Σ| integer counters initialized with zero.
4    for i = 0, ..., n − 1 do  o[i] := s_i[h]   // Fissioned loop: first cache characters,
5    for i = 0, ..., n − 1 do  c[o[i]]++         // then count occurrences.
6    b := [ c[0], ⊥, ..., ⊥ ],   ℓ := c[0]
7    for i = 1, ..., n − 1 do                    // Calculate inclusive prefix sum
8        b[i] := b[i − 1] + c[i − 1]             // of counters as bucket pointers.
9        if c[i] ≠ 0 then  ℓ := c[i]             // Save last non-empty bucket's size.
10   i := 0                                      // Start at front, take out string s_i.
11   while i < n − ℓ do                          // Reorder:
12       s̄ := move(s_i),   ō := move(o[i])
13       while (j := −−b[ō]) > i do
14           swap(s̄, s_j),   swap(ō, o[j])       // Walk cycles of the permutation, swap to
15       s_i := move(s̄),   i := i + c[ō]         // back of buckets, until current bucket is done.
16   deallocate(o, b)
17   ... perform recursion as in lines 11 to 14 of algorithm 2.2 (page 30) ...
     Output : The array S is fully sorted lexicographically.
```

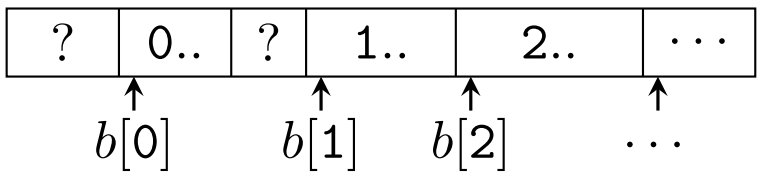

first remaining unsorted string in the current bucket is taken out of the set and stored as $\bar{s}$. Correspondingly, the cached character is stored as $\bar{o}$. In the loop, lines 13 to 14, the correct location for $\bar{s}$ is determined as $j = b[\bar{o}] - 1$, and $\bar{s}$ and $\bar{o}$ are swapped with the unsorted item at this position. This corrects one transposition in the current cycle. The remaining trick of the algorithm is that buckets are filled from back to front, by decrementing $b[\bar{o}]$ in the inclusive prefix sum, and breaking the inner loop when $i = j$, which occurs exactly when the current bucket is fully sorted. The outer loop, lines 11 to 15, hence processes one bucket at a time, which is why $i$ is advanced by $c[\bar{o}]$ after placing $\bar{s}$ into the last remaining free slot in the bucket ($\bar{o}$ is no longer needed). The outer loop is terminated when $i$ reaches the beginning of the *last* non-empty bucket at $n - \ell$, since at that time this last bucket is already correctly ordered. Note that this in-place reordering does not yield a stable sorting algorithm.

The in-place variant CI2 only needs a bucket array of $|\Sigma|$ counters per recursion level, which makes it much more memory efficient than CE2. Amazingly, despite the complex string pointer exchange pattern, CI2 was only 7.5% slower than CE2 in our experiments. In section 2.3 we will highlight the relative performance of all radix sort variants further.

In addition to CI2, we also implemented the corresponding baseline in-place radix sort variant CI0, the first improvement with only character caching CI1, and added adaptive 16-bit radix sorting to CI2 to gain variant CI3 (analogously to the step from CE2 to CE3).

Besides the "C" counting variants above, Rantala's [Ran07] library also contains many "D" variants with *dynamic* lists. In these radix sort variants, the string set is scanned only once and each string is immediately inserted into one of $\sigma$ dynamic lists or arrays. In Rantala's implementations the dynamic lists are then read to create a continuous string pointer array, on which recursion is performed. Combined with the loop fission optimization, "D" variants are competitive with simpler "C" variants in Kärkkäinen and Rantala's [KR08] experiments. They also tried different dynamic list data structures like arrays and arrays of arrays.

Conventional wisdom is that managing dynamically growing data structures always comes at a cost. This makes the approach of Wassenberg and Sanders [WS11] even more surprising: by using the virtual memory system they are able to bring the extra cost down to almost zero. In their integer radix sort implementation they allocate a huge amount of virtual memory for the radix sort buckets. The trick is that this special area is provided by the memory system without actually backing the area with physical memory. Their radix sort can then distribute the integers into buckets, whose underlying memory is dynamically filled by the memory system as needed. The "bookkeeping of piles" is passed on to lower layers in the systems, and hence their management is at virtually zero extra cost. Their highly-tuned implementation is specifically for 32-bit integer keys, and we are not aware of any efforts to attempt this approach for string sorting.

As radix sort yields very fast string sorting implementations, we consider *parallel* radix sorting in section 4.4.1.

### 2.2.3 Burstsort

*Burstsort* [SZ03a; SZ03b; SZ04a] dynamically builds a *burst trie* data structure (see figure 2.3) for the input strings which implicitly sorts the string set. The burst trie consists of an *access trie*, an ordinary trie or compressed trie, with *containers* at the leaves, which are unordered data structures holding all strings with the common prefix distinguished by the path to the leaf. In order to reduce the involved work to the distinguishing prefix and to become cache efficient, a burst trie is built *lazily*: only when the number of strings accumulated in a container exceeds a threshold, the trie is expanded at this container. Once all strings are inserted, the trie is traversed to deliver the sorted string set. During the final traversal, the relatively small containers stored at the leaves of the trie are sorted individually, without need of further expansion. The burst trie [HZW02] was originally designed for vocabulary accumulation of a large text corpus.

Crucial factors for the performance of burstsort are the implementation of the trie, the containers, the threshold when to burst, and the algorithm used to sort the containers during the final traversal. Sinha and Zobel [SZ03a] use an array of size $|\Sigma|$ for each trie node, and keep unordered dynamic arrays of string pointers at the leaf containers. They empirically select 8 192 as threshold at which to burst a container, and use multikey quicksort to sort the leaves. These choices yield a total expected running time of $\mathcal{O}(D + n \log \sigma)$, equal to radix sort. In their experiments, they show that burstsort outperforms all other basic string sorting algorithms, though they do not compare with the caching variants mentioned in previous sections. They attribute their results to burstsort being more cache efficient than the other basic string sorters, and show this experimentally using a cache simulator.

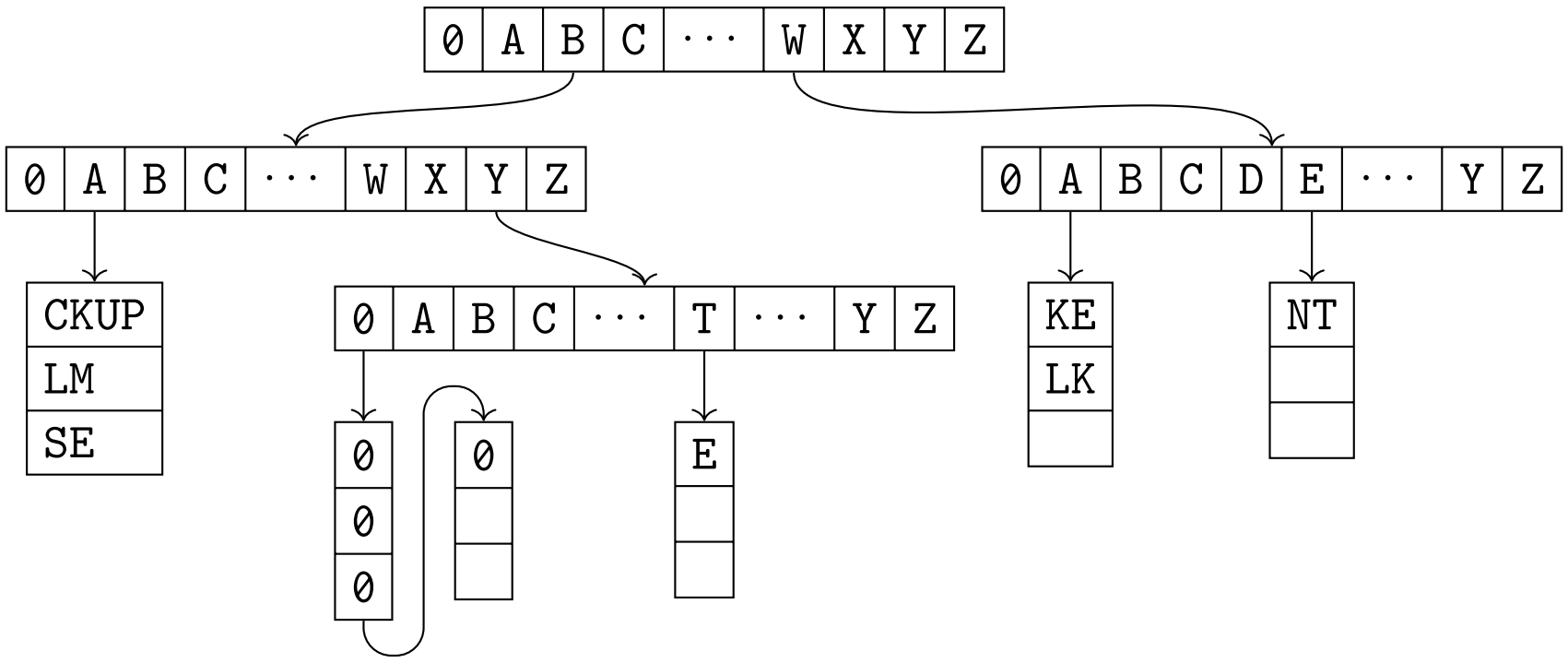

**Figure 2.3:** Burst trie containing `backup`, `balm`, `base`, `by`, `by`, `by`, `by`, `byte`, `wake`, `walk`, and `went` (adapted from [SZ04a]).

To further reduce cache misses, the same authors proposed in follow-up work [SZ04b; SZ05] to pre-initialize the access trie using a random sample of the input. The initialization method they designed chooses $R$ strings uniformly at random, and uses them to initialize an access trie with one container for each distinct string in the sample. Empirically they select $R = n/S$ with $S = 8\,192$ (to match the burst threshold), and show experimentally that despite the additional work the pre-initialization yields a faster practical algorithm for some inputs.

In the first version of burstsort, also called *P-burstsort*, the leaf containers store pointers to strings (contrary to what figure 2.3 suggests). Alternatively, to optimize access to strings when bursting a container or during the final sort, the authors propose *copying* the entire string tails into a container and discarding the original string [SZR07]. During traversal the string set is reconstructed in sorted order. This variant is named *C-burstsort* or *copy-burstsort* and is closer to the illustration in figure 2.3. While this approach is good for short string sets, its time complexity is no longer bounded by the distinguishing prefix size $D$, but instead by the total number of characters $N$. Furthermore, the algorithm's interface of discarding and recreating the underlying string set is unusual and incompatible with many applications.

Sinha, Zobel, and Ring [SZR07] implemented three variants of copy-burstsort: *C-burstsort* which copies all characters in the tails into the container, *CP-burstsort* which does not discard the original string and additionally stores a pointer to it in order to maintain stability, and *CPL-burstsort*, which is not mentioned in the paper and copies only a limited number of characters into the container (by default, 80). Furthermore, these variants are configured with a *free bursts* tuning parameter, which specifies a simpler pre-initialization strategy than sampling: while free bursts are available, containers are burst when they contain only two strings. This builds up the trie faster with the first strings as samples.

In the final paper on burstsort [SW08; SW10], the authors make attempts to bring down the memory requirements of P-burstsort by adding a more space-efficient container. This container basically is a dynamically growing array of arrays, which saves space but adds a costly layer of indirection for each access to a string pointer. While the paper's title suggests they are attempting to bring burstsort down to $\mathcal{O}(\log n)$ in-place sorting, this is not the case: they settle for $2n$ extra space. The source code described in the paper is not publicly available.

The algorithm collection of Rantala [Ran07] contains an large number of *independent* burstsort implementations, which are included in our experimental survey in section 2.3.

It must be noted that compared to the previous basic string sorters, burstsort has a large number of tuning parameters, e.g., the growth strategy of containers, the burst threshold, number of free bursts, and length of tails copied into the container. For our experiments we kept the values the original authors supplied, but depending on the input, better values could be determined.

### 2.2.4 LCP-Mergesort

LCP-Mergesort is an adaptation of mergesort to strings that saves and reuses the LCPs of consecutive strings in the sorted subproblems [NK08], which yields an algorithm with $\mathcal{O}(D + n \log n)$ worst-case time complexity.

Consider the basic comparison of two strings $s_a$ and $s_b$. If there is no additional LCP information, the strings must be compared characterwise. However, if one has the LCP of $s_a$ and $s_b$ to another smaller or equal string $p$, namely $\textsc{lcp}(p, s_a)$ and $\textsc{lcp}(p, s_b)$ with $p \le s_a$ and $p \le s_b$, then one can first compare these LCP values. If $\textsc{lcp}(p, s_a) < \textsc{lcp}(p, s_b)$, then $s_a > s_b$, and symmetrically if $\textsc{lcp}(p, s_a) > \textsc{lcp}(p, s_b)$, then $s_a < s_b$. Hence, their order can be determined without additional character comparisons.

Ng and Kakehi [NK08] developed a binary LCP-Mergesort implementation which interleaves string pointers and LCP values for fast cache-efficient merging. In their experiments they show that their LCP-Mergesort implementation is only a factor 1.3

**Algorithm 2.5 :** String Insertion Sort

**Function** InsertionSort($\mathcal{S}, h$)
  **Input :** $\mathcal{S} = [\, s_0, \ldots, s_{n-1} \,]$ an array of $n$ strings with common prefix $h$.
  **for** $i = 1, \ldots, n-1$ **do** *// Insert $x = s_i$ into sorted sequence $[\, s_0, \ldots, s_{i-1} \,]$.*
    $j := i, \quad x :=$ **move**$(s_j)$ *// Take $x$ out of $\mathcal{S}$, and*
    **while** $j > 0$ **do** *// set up iterators:*

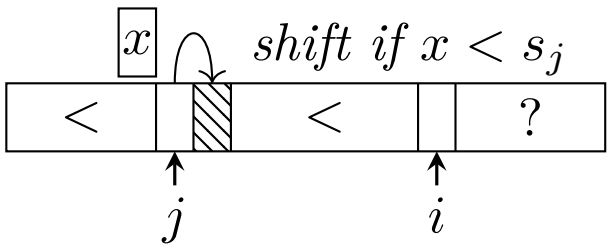

      $j{-}{-}, \quad p := h$
      **while** $s_j[p] = x[p]$ **do** $p{+}{+}$
      **if** $s_j[p] \le x[p]$ **then** *// Compare characters in $x$ and $s_j$. If $x \ge s_j$,*
        $j{+}{+}$, **break** *// then move $j$ back to vacant position,*
      $s_{j+1} :=$ **move**$(s_j)$ *// else shift $s_j$ to $s_{j+1}$ because $x < s_j$.*
    $s_j :=$ **move**$(x)$ *// Insert $x$ at correct free position $j$.*
  **Output :** The array $\mathcal{S}$ is fully sorted lexicographically.

slower than CRadix sort, while the other (non-caching) algorithms they tested were considerably slower.

In section 4.3, we develop a parallel multiway variant of LCP-merge, which is used to improve performance on NUMA machines. Our multiway LCP-merge is also interesting for merging string arrays stored in external memory.

Rantala's algorithm collection [Ran07] contains an independent implementation of binary LCP merge sort. It also contains a multiway non-LCP implementation based on a tournament tree, and an implementation of funnelsort [BF02], which is a cache-oblivious merge sort.

### 2.2.5 Insertion Sort

Insertion sort [Knu98] is important even for string sorting because it is commonly used as base case sorter for small string sets. It keeps an ordered array into which unsorted items are inserted by linearly scanning for their correct position. As strings are considered atomic, full string comparisons are done during the linear scan. This leads to an $\mathcal{O}(nD)$ worst-case time complexity, which is prohibitive for large string sets, but provides good performance for small $n$ and $D$ due to cache-efficient scanning of the strings. Algorithm 2.5 shows pseudocode for basic naive string insertion sort.

If one additionally keeps the associated LCP array, the number of character comparisons can be decreased, trading them for integer comparisons of LCPs. An LCP-aware insertion sort is presented in section 4.3.5 as a base case sorter for LCP-accelerated string sorting algorithms.

### 2.2.6 Other Algorithms and Models

Besides the basic algorithms presented in the previous sections, the string sorting algorithm set by Rantala [Ran07] contains more algorithms and variants: (non-LCP) merge sort with full string comparisons, multikey quicksort with multiple pivots, multikey quicksort with SIMD parallelism, least-significant-digit (LSD) radix sort, and many more. While these are interesting examples of the variety of solutions to this basic problem, the ones we reviewed in the preceding sections 2.2.1 to 2.2.5 are the most important genres.

Despite the wealth of string sorting algorithms designed for the RAM model, very few are efficient for external memory due to the high cost of random access. The only exceptions are variants of merge sort, e.g. funnelsort [BF02], which is a showcase example of a cache-oblivious algorithm.

Since sorting stringsis an important application in external memory databases, special external memory string sorting algorithms were designed [AFGV97; FPP06]. These utilize techniques to reduce the working set such as advanced buffering data structures and hashing of string prefixes. While these external memory techniques can be applied to the cache-RAM hierarchy step as well, we are not aware of any optimized implementations of the approaches.

## 2.3 Empirical Performance of Sequential Algorithms

The practical performance of string sorting algorithms is impossible to determine using theoretical asymptotic analysis alone due to the large impact of hardware acceleration effects like caching and memory latency hiding. We therefore collected and integrated many sequential string sorting algorithm implementations in a test framework and present an experimental evaluation in this chapter as preliminary work for designing parallel string sorting algorithms.

All implementations in the framework are in C or C++. We collected most of the original source code by the authors mentioned in section 2.2:

The algorithm library by Rantala [Ran07] contains 37 versions of radix sort (in-place, out-of-place, and one-pass with various dynamic memory allocation schemes), 26 variants of multikey quicksort (with caching, block-based, different dynamic memory allocation strategy, and SIMD instructions), 10 different funnelsorts, 38 implementations of burstsort (again with different dynamic memory managements), and 29 mergesorts (with losertree and LCP caching variants). In total these are 140 original implementation variants, all of high quality.

The other main source of string sorting implementations are the publications of Sinha [SZ03a; SZ03b; SZ04a; SZ04b; SZ05; SZR07; SW08; SW10]. We included the original burstsort implementations (one with dynamically growing arrays and one with linked lists), and 9 versions of copy-burstsort. The original copy-burstsort code was written for 32-bit machines, and we modified it to work with 64-bit pointers.

We also incorporated the implementations of CRadix sort [NK07] and LCP-Mergesort [NK08] by Ng and Kakehi, the original multikey quicksort code by Bentley and Sedgewick, the original in-place American flag radix sort by McIlroy, Bostic, and McIlroy [MBM93], and Forward and Adaptive Radix Sort by Andersson and Nilsson [AN98].

We believe the test framework contains virtually every efficient string sorting implementation publicly available, and our experimental evaluation to be the most comprehensive in the literature to date. Our implementations, the test framework, and most input sets are available from `http://panthema.net/2013/parallel-string-sorting`. It also contains our new parallel string sorters which are discussed in sections 4.2 to 4.4 and evaluated using the same framework in section 4.5.

### 2.3.1 Experimental Platforms and Setup

We tested our implementations and those by other authors on six different platforms. All platforms run Linux and their main properties are listed in table 2.1. We compiled all programs using gcc 4.8 or gcc 5 with optimizations `-O3 -march=native`. The six platforms were chosen to encompass a wide variety of multi-core systems, which exhibit different characteristics in their memory system and also cover today's most popular hardware. By experimenting on a large number of systems (and inputs), we aspire to present robust experimental results.

The test framework sets up a separate environment for every run. To isolate heap fragmentation, it was very important to `fork()` a child process for each run. The string data is loaded before the `fork()`, allocating exactly the required amount of RAM, and shared read-only with the child processes. No precaution to lock the program's memory into RAM was taken. The Linux CPU frequency scaling governor "performance" was activated to disable power-saving mechanisms.

Before an algorithm is called, the string pointer array is generated inside the child process by scanning the string data for zero characters, thus flushing caches and TLB entries. Time measurement is done with `clock_gettime()` and encompasses only the sorting algorithm. Because many algorithms have a deep recursion stack for our large inputs, we increased the stack size limit to 64 MiB.

The output of each string sorting algorithm was verified by first checking that the resulting pointer list is a permutation of the input set, and then checking that strings are in non-descending order. The underlying characters are shared read-only with the child process running the algorithm and thus cannot be modified.

Methodologically we have to discuss whether measuring only the algorithm's running time is a good decision. The issue is that cleanup work like deallocation and defragmentation in both heap allocators and kernel page tables is done *lazily*. This was most notable when running two algorithms consecutively without precautions: the second run is generally much slower due to heap memory fragmentation. The `fork()` process isolation excludes both variables from the experimental results by creating a pristine process environment, and tearing it down after the clock stops. However, when considering that the algorithms should run in a real program context these costs should not be ignored. These issues must be discussed in greater detail in future work for sound results with big data applications in RAM. We briefly considered HugePages, but these did not yield a change in performance. This is probably due to random accesses (cache misses) being the main time cost of string sorting, while the number of TLB misses is not a bottleneck.

**Table 2.1:** Hard- and software characteristics of experimental platforms. L1, L2, and L3 cache are per socket, RAM is for the whole machine.

| Name | Processor | Clock (GHz) | Micro-architecture | Introduction Date |
|---|---|---|---|---|
| A.Intel-1×8 | 1 × Intel Core i7 920 | 2.67 | Bloomfield | Nov. 2008 |
| B.AMD-4×4 | 4 × AMD Opteron 8350 | 2.0 | Barcelona | Sep. 2007 |
| C.AMD-4×12 | 4 × AMD Opteron 6168 | 1.9 | Magny-Cours | Mar. 2010 |
| D.Intel-4×8 | 4 × Intel Xeon E5-4640 | 2.4 | Sandy Bridge | May 2012 |
| E.Intel-2×16 | 2 × Intel Xeon E5-2683 v4 | 2.1 | Broadwell | Mar. 2016 |
| F.AMD-1×16 | 1 × AMD Ryzen 7 1800X | 3.6 | Zen | Mar. 2017 |

| Name | Sockets × Cores × SMT | Cache: L1 (I+D) (KiB) | L2 (KiB) | L3 (MiB) | RAM (GiB) |
|---|---|---|---|---|---|
| A.Intel-1×8 | $1\times\ 4\times 2 = 8$ | $4\times(32+32)$ | $4\times 256$ | 8 | 12 |
| B.AMD-4×4 | $4\times\ 4 = 16$ | $4\times(64+64)$ | $4\times 512$ | 2 | 64 |
| C.AMD-4×12 | $4\times 12 = 48$ | $12\times(64+64)$ | $12\times 512$ | $2\times 6$ | 256 |
| D.Intel-4×8 | $4\times\ 8\times 2 = 64$ | $8\times(32+32)$ | $8\times 256$ | 20 | 512 |
| E.Intel-2×16 | $2\times 16\times 2 = 64$ | $16\times(32+32)$ | $16\times 256$ | 40 | 512 |
| F.AMD-1×16 | $1\times\ 8\times 2 = 16$ | $8\times(64+32)$ | $8\times 512$ | $2\times 8$ | 64 |

| Name | Memory Channels | NUMA Nodes | Interconnect | Ubuntu Linux/Kernel/ gcc Versions |
|---|---|---|---|---|
| A.Intel-1×8 | 3 × DDR3-800 | 1 | 1 × 4.8 GT/s QPI | 14.04.5/3.13.0/4.8.5 |
| B.AMD-4×4 | 2 × DDR2-533 | 4 | 3 × 1.0 GHz HT | 14.04.5/3.13.0/4.8.4 |
| C.AMD-4×12 | 4 × DDR3-667 | 8 | 4 × 3.2 GHz HT | 14.04.5/3.13.0/4.8.4 |
| D.Intel-4×8 | 4 × DDR3-1600 | 4 | 2 × 8.0 GT/s QPI | 14.04.5/3.13.0/4.8.4 |
| E.Intel-2×16 | 4 × DDR4-2133 | 2 | 2 × 9.6 GT/s QPI | 16.04.2/4.4.0/5.4.0 |
| F.AMD-1×16 | 2 × DDR4-2133 | 1 | – | 16.04.2/4.4.0/5.4.0 |

**Table 2.2:** Characteristics of the selected input instances.

| Name | $n$ | $N$ | $\frac{D}{N}$ $(D)$ | $\frac{L}{n}$ | $\lvert\Sigma\rvert$ | avg. $\lvert s\rvert$ |
|---|---:|---:|---:|---:|---:|---:|
| URLs | 1.11 G | 70.7 Gi | 93.5 % | 62.0 | 84 | 68.4 |
| Random | $\infty$ | $\infty$ | – | – | 94 | 10.5 |
| GOV2 | 11.3 G | 425 Gi | 84.7 % | 32.0 | 255 | 40.3 |
| Wikip | 83.3 G | $\frac{1}{2}n(n+1)$ | (79.56 T) | 954.7 | 213 | $\frac{1}{2}(n+1)$ |
| Sinha URLs | 10 M | 304 Mi | 97.5 % | 29.4 | 114 | 31.9 |
| Sinha DNA | 31.6 M | 302 Mi | 100 % | 9.0 | 4 | 10.0 |
| Sinha NoDup | 31.6 M | 382 Mi | 73.4 % | 7.7 | 62 | 12.7 |

### 2.3.2 Inputs

We selected the following seven datasets, all with 8-bit characters. The most important characteristics of these instances are shown in table 2.2.

**URLs** contains all URLs found on a set of web pages which were crawled breadth-first from the author's institute website. They include the protocol name, e.g. `http://`.

**Random** (from [SZ04a]) are strings of length $[0..20)$ over the ASCII alphabet $[33..127)$, with both length and characters chosen uniformly at random.

**GOV2** is a TREC test collection consisting of 25 million HTML pages, PDF and other documents retrieved from websites under the .gov domain. We consider the whole corpus for *line-based* string sorting, concatenated in order of document identifier.

**Wikip** is an XML dump of all pages (only the most recent version, excluding history) in the English Wikipedia. Our dump is dated `enwiki-20120601`, and was obtained from `http://dumps.wikimedia.org/`. Since the XML data is not line-based, we perform *suffix sorting* on this input.

We also include the three largest inputs **Sinha** [SZ04a] tested burstsort on: a set of **URLs** excluding the protocol name, a sequence of genomic strings of length nine over a **DNA** alphabet, and a list of non-duplicate English words called **NoDup**. The "largest" among these is NoDup with only 382 MiB, which is why we consider these inputs more as reference datasets than as our target.

The inputs were chosen to represent both real-world datasets and to exhibit extreme results when sorting. Random has a very low average LCP, while URLs have a high average LCP. GOV2 is a general text file with all possible ASCII characters, and Sinha DNA has a small alphabet size. By taking suffixes of Wikipedia we have a very large sorting problem instance, which needs little memory for characters.

Our inputs are very large, one infinite, and most of our platforms did not have enough RAM to process them. For each platform, we determined a large prefix $[0..n)$, which can be processed with the available RAM and time, and leave sorting of the remainder to future work.

### 2.3.3 Algorithm List

Of the 197 different sequential string sorting variants in our experimental framework, we selected the following 39 implementations to represent both the fastest ones in a preliminary test and each of the basic algorithms from section 2.2. Furthermore, for the sake of completeness, we also already include experimental results for sequential versions of the parallel string sorters we propose in sections 4.2 and 4.3. The 39 algorithms were run on all six test platforms on up to 4 GiB portions of the test instances described in section 2.3.4.

**std::sort** is `gcc` 4.9's standard comparison-based atomic introsort [Mus97] with naive lexicographic string comparisons.

**BS.mkqs** is the original multikey quicksort by Bentley and Sedgewick [BS97] as discussed in section 2.2.1.

**R.mkqs-cache8** is a modified multikey quicksort with a pivot and cache size of eight characters by Rantala [Ran07], slightly improved by us.

**MBM.radixsort** is the original 8-bit American Flag radix sort by McIlroy, Bostic, and McIlroy [MBM93].

**KR.radixsort-CE0** is the baseline 8-bit out-of-place radix sort by Kärkkäinen and Rantala [KR08], as shown in algorithm 2.2.

**KR.radixsort-CE1** is the 8-bit out-of-place radix sort with character caching to optimize memory access by Kärkkäinen and Rantala [KR08], as shown in algorithm 2.3.

**KR.radixsort-CE2** is the 8-bit out-of-place radix sort with character caching and fissioned loops by Kärkkäinen and Rantala [KR08], shown in algorithm 2.3 combined with figure 2.2.

**KRB.radixsort-CE3s** is the adaptive 16-/8-bit out-of-place radix sort with character caching to optimize memory latency by Kärkkäinen and Rantala [KR08], completely rewritten by us to use an explicit recursion stack to avoid stack overflow.

**KR.radixsort-CI0** is the baseline 8-bit in-place radix sort by Kärkkäinen and Rantala [KR08].

**KR.radixsort-CI1** is the 8-bit in-place radix sort with character caching to optimize memory latency by Kärkkäinen and Rantala [KR08].

**KR.radixsort-CI2** is the 8-bit in-place radix sort with character caching and fissioned loops by Kärkkäinen and Rantala [KR08], as shown in algorithm 2.4.

**KRB.radixsort-CI3s** is the adaptive 16-/8-bit in-place radix sort with character caching to optimize memory latency by Kärkkäinen and Rantala [KR08], completely rewritten by us to use an explicit recursion stack to avoid stack overflow. This implementation is used as base case sorter for our parallel algorithms.

**KR.radixsort-CE6** is an improvement of radix sort CE3 by Kärkkäinen and Rantala [KR08], which unrolls fetching of characters into the cache array.

**KR.radixsort-CE7** is another improvement of CE6 by Kärkkäinen and Rantala [KR08], which checks for sorted string array sequences to avoid out-of-place reordering steps.

**KR.radixsort-D-vec** is an 8-bit radix sort using `std::vector` as dynamic data structure for buckets, implemented by Kärkkäinen and Rantala [KR08].

**KR.radixsort-D-vecblk** is an 8-bit radix sort using a `std::vector` of string pointer arrays as dynamic data structure for buckets, implemented by Kärkkäinen and Rantala [KR08].

**KR.radixsort-DB** is an 8-bit radix sort variant using a custom dynamic block-list data structure with in-place reordering by Kärkkäinen and Rantala [KR08].

**AN.AdaptiveRadix** is the original Adaptive Radix Sort code by Andersson and Nilsson [AN98].

**AN.ForwardRadix8** is the original 8-bit Forward Radix Sort code by Andersson and Nilsson [AN98], slightly modified to work correctly with 64-bit pointers.

**AN.ForwardRadix16** is the original 16-bit Forward Radix Sort code by Andersson and Nilsson [AN98], slightly modified to work correctly with 64-bit pointers.

**NK.CRadix** is the original CRadix sort code by Ng and Kakehi [NK07], unmodified.

**B.Seq-S$^5$-U** is our sequential Super Scalar String Sample Sort (S$^5$) implementation with unrolling of tree descents, as discussed in section 4.2, with KRB.radixsort-CI3s as base case.

**B.Seq-S$^5$-UI** is our sequential Super Scalar String Sample Sort (S$^5$) implementation with unrolling and interleaving of four tree descents, as discussed in section 4.2, with KRB.radixsort-CI3s as base case sorter.

**B.Seq-S$^5$-E** is our sequential Super Scalar String Sample Sort (S$^5$) implementation with equality checking after each comparison, as discussed in section 4.2, with KRB.radixsort-CI3s as base case sorter.

**B.Seq-S$^5$-UC** is our sequential Super Scalar String Sample Sort (S$^5$) implementation with unrolling of tree descents and the pre-order/level-order calculation trick, as discussed in section 4.2, with KRB.radixsort-CI3s as base case sorter.

**B.Seq-S$^5$-UIC** is our sequential Super Scalar String Sample Sort (S$^5$) implementation with unrolling and interleaving of four tree descents and the pre-order/level-order calculation trick, as discussed in section 4.2, with KRB.radixsort-CI3s as base case sorter.

**NK.LCP-Mergesort** is the original binary LCP mergesort code by Ng and Kakehi [NK08], unmodified.

**B.LCP-MS-2way** is our binary LCP mergesort, as discussed in section 4.3.

**B.LCP-MS-16way** is our 16-way LCP mergesort with an LCP tournament tree, as discussed in section 4.3.

**R.funnelsort-32way** is a 32-way cache-oblivious funnelsort [BFV04; BFV08] implemented by Rantala [Ran07].

**SZ.burstsortA** is the original burstsort implementation using dynamic arrays by Sinha and Zobel [SZ04a], from Rantala's [Ran07] implementation library.

**SZ.burstsortL** is the original burstsort implementation using dynamic linked lists by Sinha and Zobel [SZ04a], from Rantala's [Ran07] implementation library.

**SZR.C-burstsort** is the original copy-burstsort implementation by Sinha, Zobel, and Ring [SZR07] with fully copied string tails in the containers and free bursts for initializing the trie (see section 2.2.3), heavily repaired and modified to work correctly with 64-bit pointers.

**SZR.CP-burstsort** is the original copy-burstsort implementation by Sinha, Zobel, and Ring [SZR07] with pointers in the containers and free bursts for initializing the trie, also heavily repaired.

**SZR.CPL-burstsort** is the original copy-burstsort implementation by Sinha, Zobel, and Ring [SZR07] with limited string tails in the containers and free bursts for initializing the trie, also heavily repaired.

**R.burstsort-vec** is an independent burstsort implementation by Rantala [Ran07], which stores trie nodes as arrays of size $|\Sigma|$ with a bit set indicating whether the children are leaves or inner trie nodes. In this variant leaves are `std::vector`s of string pointers.

**R.burstsort2-vec** is a slight modification of the independent burstsort implementation by Rantala [Ran07] which uses the least significant bit of a pointer to distinguish leaves and inner trie nodes. In this variant leaves are `std::vector`s of string pointers.

**R.burstsort-vecblk** is an independent burstsort implementation by Rantala [Ran07], identical to R.burstsort-vec, but using a `std::vector` of string pointer array blocks as containers.

**R.burstsort2-vecblk** is an independent burstsort implementation by Rantala [Ran07], identical to R.burstsort2-vec, but using a `std::vector` of string pointer array blocks as containers.

### 2.3.4 Experimental Results

Tables 2.6 to 2.11 on pages 52 to 57 present our detailed experimental results for the 39 selected algorithms run with seven inputs on six machines. Every execution was repeated three times, and the median result is shown in the detailed tables. For each input and machine, the fastest algorithm's running time is marked in bold.

For a better comparison of the algorithms *across all inputs* on an individual machine, the detailed tables show in column "GeoM" the geometric mean of each algorithm's *slowdown factor* relative to the fastest algorithm on the same input and machine. The column "Rank" shows the rank of each algorithm ordered by "GeoM". This evaluation method was chosen to rank algorithms since the geometric mean of the normalized performance factors correctly emphasizes the small relative differences of the fastest algorithms [FW86].

For example, on A.Intel-1×8, algorithm BS.mkqs (second row in table 2.6) runs a factor $\frac{26.4}{14.8} = 1.78$ slower on URLs, $\frac{56.1}{12.5} = 4.49$ on Random, $\frac{25.3}{13.9} = 1.82$ on GOV2, $\frac{89.4}{56.4} = 1.59$ on Wikip, $\frac{3.03}{1.86} = 1.63$ on Sinha URLs, $\frac{6.70}{2.58} = 2.60$ on Sinha DNA, and $\frac{7.50}{3.03} = 2.48$ on Sinha NoDup over the fastest algorithm on the same input and machine. The geometric mean of these seven slowdown factors is 2.19, which has rank 24 among all other algorithms on this machine.

A summary of the results, aggregated *over all machines*, is presented in table 2.4 on page 50. The summary table shows the geometric mean of the slowdown factors for each algorithm and input across all machines, and in the column "GeoM" the overall geometric mean of the slowdown factors across all experiments. The "Rank" column then presents the overall final rank of each algorithm in our experiment.

Empty cells in the tables indicate out-of-memory exceptions or extremely long running times ($> 1\,\text{h}$). This occurred for the copy-burstsort variants mostly on the GOV2 and Wikipedia inputs because they perform excessive caching of characters. Additionally, on A.Intel-1×8 copy-burstsort required more memory than the available 12 GiB to sort a 4 GiB prefix of URLs. In such cases when algorithms did not finish, the *maximum* runtime among all other algorithms is used in the geometric mean calculation, such as not to skew the result. These entries are marked with an asterisk in the tables.

Regarding "GeoM" in summary table 2.4, one can see that no single algorithm is *always* fastest for all machines and inputs. In a sense this is to be expected, since the inputs were selected to have different characteristics with regard to their distinguishing prefix. On the other hand, the algorithms are designed to exploit different structural aspects and utilize hardware acceleration tricks in different ways, which in turn may or may not be supported on each machine.

The fastest algorithm across all our experiments was KR.radixsort-CE6, which however was only concretely fastest on Wikip. The second fastest KRB.radixsort-CE3s was less than one percent overall slower than KR.radixsort-CE6, and was fastest in the geometric mean on Random, Sinha DNA, and Sinha NoDup. The difference between these two implementations is not large, KR.radixsort-CE6 contains only few additional optimizations, and it remains unclear if they are worth it. In particular, KR.radixsort-CE7 adds even more optimizations, but this implementation comes in third.

**Table 2.3:** Percentage change of geometric mean of slowdown factors over the best radix sorter for the different variants in our experiments, and their additional space requirements over $|\Sigma|$ bucket pointers per level of recursion.

| Radix Sort | Relative Performance | | | | |
|---|---|---|---|---|---|
| Variant | GeoM | Over Prev. | Over Next | Additional Space | Features |
| KR-CI0 | 4.122 | | -30.7 % | none | baseline in-place |
| KR-CE0 | 3.154 | 23.5 % | -20.1 % | $n$ pointers | out-of-place |
| KR-CI1 | 2.626 | 16.7 % | -25.6 % | $n$ chars | character cache |
| KR-CE1 | 2.091 | 20.4 % | -54.9 % | $n$ chars and $n$ pointers | cache out-of-place |
| KR-CI2 | 1.350 | 35.4 % | -8.0 % | $n$ chars | loop fission |
| KR-CE2 | 1.250 | 7.4 % | -12.1 % | $n$ chars and $n$ pointers | loop fission o-o-p. |
| KRB-CI3s | 1.115 | 10.8 % | -4.0 % | $2n$ chars | 16-bit in-place |
| KR-CE7 | 1.072 | 3.8 % | -0.1 % | $2n$ chars and $n$ pointers | sortedness |
| KRB-CE3s | 1.071 | 0.1 % | -0.2 % | $2n$ chars and $n$ pointers | 16-bit adaptive |
| KR-CE6 | 1.068 | 0.2 % | | $2n$ chars and $n$ pointers | unroll cache |

Remarkably, KRB.radixsort-CI3s ranks fourth, which is surprising due to the more complex in-place exchange pattern described in section 2.2.2. To better understand the performance differences of the radix sort variants, consider table 2.3. In this table, all variants are ordered descending by running time. The fastest variant, KR.radixsort-CE6 is on the bottom, but it is only 0.2 % faster than KRB.radixsort-CE3s, which again in turn is only 0.1% faster than KR.radixsort-CE7. These margins are very small, and can also be attributed to measurement noise. Most interesting is that KRB.radixsort-CI3s is only 4 % slower than KR.radixsort-CE7, and it uses a lot less memory: $n$ fewer pointers, which are $8n$ bytes on a 64-bit machine.

In fifth place ranks R.mkqs-cache8, Rantala's caching multikey quicksort, which was the fastest algorithm for GOV2 and Sinha URLs. Again, the margin to the best radix sort is relatively small, less than 6 %. We attribute this speedup on the two inputs to the difference in how well radix sort and R.mkqs-cache8 handle large string sets with long common prefixes, e.g. in case of URLs, the protocol and domain names, and in case of GOV2, text with relatively large distinct words. For such inputs, radix sort performs more iterations over the strings than R.mkqs-cache8, which will read chunks of eight characters.

Our own, much larger URL dataset is handled even better by R.burstsort2-vecblk. Sorting URLs with long common prefixes is really a showcase example for burstsort. Interestingly, Rantala's independent (and much simpler) burstsort implementations generally perform better than Sinha's original versions. Sinha's burstsort implementations do not perform particularly well on our inputs.

Our own *sequential* implementations of $S^5$ (to be discussed in section 4.2) were never the fastest, but they consistently fall in the middle field, without any outlier. This is expected, since $S^5$ is mainly designed to be used as an efficient top-level parallel algorithm, and to be conservative with memory bandwidth, which is the limiting factor for data-intensive multi-core applications.

All mergesort implementations, with and without LCP acceleration, rank among the slower implementations, except for URLs, where they are among the fastest due to the long common prefixes.

The classical string sorting implementations, such as AN.AdaptiveRadix, AN.Forward-Radix8, AN.ForwardRadix16, MBM.radixsort, and BS.mkqs also are among the slower implementations. These older source codes clearly were not written with today's hardware accelerations in mind.

However, running time is not the only important metric for string sorting algorithms. We also measured the peak memory usage of all sequential implementations using our heap and stack profiling tool `malloc_count` [Bin13b]. The results are presented in table 2.5 in MiB, excluding the string data array and the string pointer array (we only have 64-bit systems, so pointers are eight bytes). We must note that the profiler considers *allocated virtual memory*, which may not be identical to the amount of physical memory actually used. However, none of the algorithms perform allocation tricks which would result in a large discrepancy.

From the table we plainly see that the more *caching* an implementation does, the higher its peak memory requirements. However, the memory usage of SZR.C-burstsort and SZR.CP-burstsort are extreme: they require nearly eight times the input character set size for our URL dataset. This is clearly excessive, even if one considers that the implementation can deallocate and recreate the string data from the burst trie. The SZR.CPL-burstsort variant brings the memory requirement down somewhat for URLs and Random, but it is still excessive for the GOV2 input.

High memory requirements seem to correlate with slower running time. There is, however, one important exception: R.mkqs-cache8 requires a lot of cache memory, but is among the best performing algorithm. Furthermore, the fast counting radix sorts clearly benefit from using the additional memory.

Generally, string sorters with lots of dynamic data structures, such as SZR.C-burstsort and AN.ForwardRadix8/16, can require very large memory allocations for some inputs due to overhead and fragmentation of the bookkeeping information. Furthermore, complex bookkeeping is also slow performance-wise, which makes these algorithms less attractive in both metrics.

Our sequential $S^5$ implementations, B.Seq-$S^5$, fare well with respect to memory consumption because they do not use caching and permute the string pointers in-place (note that KRB.radixsort-CI3s is used for small string subsets in sequential $S^5$).

To summarize our sequential string sorting performance experiment, depending only on the amount of additional memory available, without knowledge of the input, one should use the following algorithms:

(i) If memory for $2n$ characters and $n$ pointers is available, use the fastest out-of-place radix sort, KR.radixsort-CE7.

(ii) The $n$ additional pointers can be conserved by using KRB.radixsort-CI3s, which is only 6 % slower than the fastest out-of-place radix sorts.

(iii) While counting radix sorts clearly outperform radix sorts with dynamic data structures, KR.radixsort-DB stands out by requiring very little memory, being nearly in-place, and still providing good performance.

(iv) For sorting with next to zero extra memory, plain multikey quicksort (BS.mkqs) is still a good choice.

## 2.4 Conclusion: Draw on Architecture Specific Enhancements

To conclude our empirical performance evaluation of basic sequential string sorting algorithms, we want to highlight some of most important optimization principles employed by the fastest implementations.

**Memory access time** varies greatly in modern systems. While the RAM model considers all memory accesses to take unit time, current architectures have multiple levels of cache, require additional memory access on TLB misses, and may have to request data from "remote" nodes on NUMA systems. While there are few hard guarantees, one can still expect recently used memory to be in cache and use these assumptions to design cache-efficient algorithms. Furthermore, predicable memory access patterns, such as linear scanning or scanning with regular skips, are an order of magnitude faster than random access, because they can be accelerated by the hardware memory prefetcher.

**Caching of characters** is very important for modern memory hierarchies as it reduces the number of cache misses due to random access on strings, which is the most costly operation in string sorting. When performing character lookups, a caching algorithm copies successive characters of the string into a more convenient memory area. Subsequent sorting steps can then avoid random access until the cache needs to be refreshed. This technique has successfully been applied to radix sort [NK07], multikey quicksort [Ran07], and in its extreme to burstsort [SZR07], and much of the performance gain of these implementations can be attributed to this technique. However, caching comes at the cost of increased space requirements and memory accesses, hence a good trade-off must be found.

**Unrolling, fission, and vectorization of loops** are methods to exploit out-of-order execution and superscalar parallelism now standard in modern CPUs. The processor's instruction scheduler automatically analyzes the machine code, detects data dependencies, and can dispatch multiple parallel operations. However, only specific, simple data independencies can be detected and thus inner loops must be designed with care (e.g. for radix sort [KR08]). The performance increase by reorganizing loops is most difficult to predict and is highly dependent on the particular hardware. The fastest radix sorts have fissioned loops which fetch the characters from strings into the character cache. This enables hardware-based parallelization and hiding of the memory latency of character accesses and greatly accelerates these implementations.

**Super-Alphabets** can be used to accelerate string sorting algorithms which originally look only at single characters. Instead, multiple characters are grouped as one and sorted together. However, most algorithms are very sensitive to large alphabets, thus the group size must be chosen carefully. This approach results in 16-bit MSD radix sort and fast sorters for DNA strings. If the grouping is done to fit many characters into a machine word for processing as a whole block using arithmetic instructions, then this is also called *word parallelism*.

Overall, the radix sort implementations by Kärkkäinen and Rantala [KR08] outperform other more complex algorithms by employing sophisticated architecture specific enhancements. The next chapter focuses on the memory hardware in modern multi-core machines.

**Table 2.4:** Geometric mean of slowdown factors of sequential algorithms over the best algorithm for each instance across all machines.

| | Overall | | Our Datasets | | | | Sinha's | | |
|---|---|---|---|---|---|---|---|---|---|
| | Rank | GeoM | URLs | Random | GOV2 | Wikip | URLs | DNA | NoDup |
| std::sort | 38 | 5.45 | 5.77 | 8.73 | 4.30 | 3.77 | 4.26 | 7.20 | 5.70 |
| BS.mkqs | 23 | 2.36 | 2.37 | 5.13 | 1.70 | 1.70 | 1.58 | 2.74 | 2.64 |
| R.mkqs-cache8 | 5 | 1.34 | 1.20 | 2.08 | **1.00** | 1.26 | **1.02** | 1.42 | 1.67 |
| MBM.radixsort | 34 | 4.82 | 6.53 | 5.98 | 4.59 | 2.98 | 4.58 | 5.83 | 4.24 |
| KR.radixsort-CE0 | 30 | 3.76 | 6.38 | 3.95 | 4.29 | 2.11 | 3.89 | 3.95 | 3.01 |
| KR.radixsort-CE1 | 26 | 2.49 | 4.13 | 2.54 | 2.69 | 1.54 | 2.67 | 2.67 | 1.91 |
| KR.radixsort-CE2 | 6 | 1.49 | 2.89 | 1.17 | 1.95 | 1.11 | 1.43 | 1.40 | 1.10 |
| KRB.radixsort-CE3s | 2 | 1.28 | 2.11 | **1.04** | 1.75 | 1.06 | 1.28 | **1.03** | **1.03** |
| KR.radixsort-CI0 | 36 | 4.91 | 6.57 | 6.07 | 5.17 | 2.88 | 4.39 | 6.71 | 3.93 |
| KR.radixsort-CI1 | 29 | 3.13 | 4.17 | 3.23 | 3.30 | 2.17 | 3.04 | 3.72 | 2.68 |
| KR.radixsort-CI2 | 9 | 1.61 | 2.71 | 1.24 | 2.18 | 1.30 | 1.47 | 1.53 | 1.29 |
| KRB.radixsort-CI3s | 4 | 1.33 | 1.87 | 1.26 | 1.61 | 1.12 | 1.22 | 1.20 | 1.16 |
| KR.radixsort-CE6 | 1 | 1.27 | 2.04 | 1.14 | 1.59 | **1.01** | 1.30 | 1.07 | 1.04 |
| KR.radixsort-CE7 | 3 | 1.28 | 2.02 | 1.14 | 1.63 | 1.03 | 1.28 | 1.08 | 1.04 |
| KR.radixsort-D-vec | 11 | 1.70 | 3.00 | 1.46 | 2.06 | 1.28 | 1.66 | 1.50 | 1.41 |
| KR.radixsort-D-vecblk | 17 | 1.80 | 3.08 | 1.60 | 2.19 | 1.40 | 1.52 | 1.59 | 1.65 |
| KR.radixsort-DB | 7 | 1.55 | 2.92 | 1.26 | 1.93 | 1.11 | 1.53 | 1.48 | 1.23 |
| AN.AdaptiveRadix | 27 | 2.82 | 2.25 | 4.21 | 2.28 | 2.26 | 2.55 | 3.87 | 2.97 |
| AN.ForwardRadix8 | 39 | 7.95 | 6.76 | 7.61 | 8.35 | 6.92 | 9.31 | 8.95 | 8.07 |
| AN.ForwardRadix16 | 35 | 4.88 | 4.24 | 5.03 | 4.82 | 4.48 | 5.41 | 5.11 | 5.18 |
| NK.CRadix | 15 | 1.76 | 2.68 | 1.37 | 1.87 | 1.36 | 1.69 | 2.10 | 1.58 |
| B.Seq-$S^5$-E | 18 | 1.81 | 1.64 | 3.26 | 1.84 | 1.70 | 1.17 | 1.73 | 1.92 |
| B.Seq-$S^5$-U | 20 | 1.97 | 1.95 | 3.41 | 1.92 | 1.71 | 1.38 | 1.80 | 2.13 |
| B.Seq-$S^5$-UI | 10 | 1.67 | 1.68 | 2.74 | 1.75 | 1.51 | 1.23 | 1.32 | 1.83 |
| B.Seq-$S^5$-UC | 21 | 2.05 | 2.03 | 3.57 | 1.98 | 1.76 | 1.45 | 1.86 | 2.22 |
| B.Seq-$S^5$-UIC | 13 | 1.72 | 1.72 | 2.76 | 1.78 | 1.53 | 1.27 | 1.44 | 1.86 |
| NK.LCP-Mergesort | 28 | 2.96 | 1.32 | 8.73 | 1.89 | 2.50 | 1.89 | 4.13 | 4.67 |
| B.LCP-MS-2way | 24 | 2.45 | 1.11 | 5.86 | 1.55 | 2.35 | 1.51 | 3.61 | 4.07 |
| B.LCP-MS-16way | 25 | 2.46 | 1.14 | 6.16 | 1.56 | 2.21 | 1.49 | 3.79 | 3.97 |
| R.funnelsort-32way | 31 | 4.03 | 3.06 | 9.80 | 2.77 | 2.79 | 2.72 | 5.65 | 4.87 |
| SZ.burstsortA | 16 | 1.78 | 1.22 | 3.75 | 1.43 | 1.73 | 1.39 | 1.83 | 1.97 |
| SZ.burstsortL | 22 | 2.08 | 1.24 | 3.32 | 1.64 | 1.88 | 1.82 | 2.76 | 2.68 |
| SZR.C-burstsort | 33* | 4.17 | 3.56 | 2.75 | | | 5.43 | 1.38 | 4.69 |
| SZR.CP-burstsort | 37* | 4.94 | 3.26 | 2.99 | | | 5.26 | 2.80 | 7.75 |
| SZR.CPL-burstsort | 32* | 4.16 | 2.48 | 3.92 | 2.15 | | 4.20 | 4.64 | 7.75 |
| R.burstsort-vec | 14 | 1.72 | 1.10 | 4.43 | 1.33 | 1.80 | 1.27 | 1.62 | 1.86 |
| R.burstsort2-vec | 19 | 1.95 | 1.36 | 4.87 | 1.52 | 2.11 | 1.46 | 1.73 | 2.01 |
| R.burstsort-vecblk | 8 | 1.57 | **1.06** | 3.80 | 1.23 | 1.57 | 1.16 | 1.53 | 1.70 |
| R.burstsort2-vecblk | 12 | 1.70 | 1.22 | 3.83 | 1.35 | 1.77 | 1.22 | 1.64 | 1.82 |

* The SZR.burstsort variants sometimes require too much memory. In the geometric mean empty cells are replaced with the maximum running time of any other algorithm for such instances.

**Table 2.5:** Memory usage of sequential algorithms on F.AMD-1×16 in MiB, excluding input and string pointer array.

| | Speed | Our Datasets | | | | Sinha's | | |
|---|---|---|---|---|---|---|---|---|
| | Rank | URLs | Random | GOV2 | Wikip | URLs | DNA | NoDup |
| $n$ | | 66 M | 409 M | 80.2 M | 256 Mi | 10 M | 31.5 M | 31.6 M |
| $N$ | | 4 Gi | 4 Gi | 4 Gi | 32 Pi | 304 Mi | 302 Mi | 382 Mi |
| $D/N$ $(D)$ | | 92.7 % | 43.0 % | 69.7 % | (13.6 G) | 97.5 % | 100 % | 73.4 % |
| $L/n$ | | 57.9 | 3.3 | 34.1 | 33.0 | 29.4 | 9.0 | 7.7 |
| std::sort | 38 | 0 | 0 | 0 | 0 | 0 | 0 | 0 |
| BS.mkqs | 23 | 0 | 0 | 1 | 0 | 0 | 0 | 0 |
| R.mkqs-cache8 | 5 | 1002 | 6242 | 1225 | 4096 | 153 | 483 | 483 |
| MBM.radixsort | 34 | 0 | 0 | 0 | 0 | 0 | 0 | 0 |
| KR.radixsort-CE0 | 30 | 505 | 3121 | 618 | 2052 | 77 | 241 | 241 |
| KR.radixsort-CE1 | 26 | 568 | 3511 | 695 | 2308 | 86 | 271 | 272 |
| KR.radixsort-CE2 | 6 | 568 | 3511 | 695 | 2308 | 86 | 271 | 272 |
| KRB.radixsort-CE3s | 2 | 723 | 3902 | 830 | 2573 | 120 | 305 | 305 |
| KR.radixsort-CI0 | 36 | 4 | 0 | 6 | 4 | 1 | 0 | 0 |
| KR.radixsort-CI1 | 29 | 65 | 390 | 104 | 258 | 10 | 30 | 30 |
| KR.radixsort-CI2 | 9 | 65 | 390 | 104 | 258 | 10 | 30 | 30 |
| KRB.radixsort-CI3s | 4 | 222 | 781 | 218 | 525 | 44 | 64 | 64 |
| KR.radixsort-CE6 | 1 | 669 | 3902 | 786 | 2566 | 111 | 303 | 303 |
| KR.radixsort-CE7 | 3 | 669 | 3902 | 786 | 2566 | 111 | 303 | 303 |
| KR.radixsort-D-vec | 11 | 4895 | 3392 | 1206 | 2958 | 267 | 288 | 370 |
| KR.radixsort-D-vecblk | 17 | 504 | 3125 | 621 | 2054 | 77 | 242 | 242 |
| KR.radixsort-DB | 7 | 8 | 12 | 9 | 9 | 1 | 1 | 1 |
| AN.AdaptiveRadix | 27 | 1533 | 9550 | 1874 | 6267 | 233 | 738 | 738 |
| AN.ForwardRadix8 | 39 | 2204 | 13732 | 2695 | 9011 | 336 | 1062 | 1062 |
| AN.ForwardRadix16 | 35 | 2204 | 13732 | 2695 | 9011 | 336 | 1062 | 1062 |
| NK.CRadix | 15 | 752 | 4681 | 919 | 3072 | 114 | 362 | 362 |
| B.Seq-$S^5$-E | 18 | 126 | 780 | 154 | 512 | 19 | 60 | 60 |
| B.Seq-$S^5$-U | 20 | 126 | 780 | 154 | 512 | 19 | 60 | 60 |
| B.Seq-$S^5$-UI | 10 | 126 | 780 | 154 | 512 | 19 | 60 | 60 |
| B.Seq-$S^5$-UC | 21 | 126 | 780 | 154 | 512 | 19 | 60 | 60 |
| B.Seq-$S^5$-UIC | 13 | 126 | 780 | 154 | 512 | 19 | 60 | 60 |
| NK.LCP-Mergesort | 28 | 1002 | 6242 | 1225 | 4096 | 153 | 483 | 483 |
| B.LCP-MS-2way | 24 | 1628 | 10143 | 1990 | 6656 | 248 | 784 | 784 |
| B.LCP-MS-16way | 25 | 1628 | 10143 | 1990 | 6656 | 248 | 784 | 784 |
| R.funnelsort-32way | 31 | 501 | 3121 | 612 | 2048 | 76 | 241 | 241 |
| SZ.burstsortA | 16 | 1466 | 7384 | 1436 | 5868 | 200 | 531 | 792 |
| SZ.burstsortL | 22 | 1089 | 6268 | 1273 | 4200 | 162 | 491 | 489 |
| SZR.C-burstsort | 33 | 31962 | 6200 | | | 2875 | 436 | 4182 |
| SZR.CP-burstsort | 37 | 39426 | 7372 | | | 3159 | 1400 | 9403 |
| SZR.CPL-burstsort | 32 | 9948 | 7262 | 21406 | | 987 | 1471 | 6654 |
| R.burstsort-vec | 14 | 768 | 4002 | 894 | 2989 | 115 | 342 | 347 |
| R.burstsort2-vec | 19 | 734 | 3997 | 874 | 2956 | 111 | 337 | 344 |
| R.burstsort-vecblk | 8 | 924 | 7066 | 1114 | 5400 | 135 | 262 | 585 |
| R.burstsort2-vecblk | 12 | 902 | 7061 | 1097 | 5333 | 132 | 261 | 573 |

**Table 2.6:** Running time of sequential string sorting algorithms on A.Intel-1×8 in seconds.

| | Overall | | Our Datasets | | | | Sinha's | | |
|---|---|---|---|---|---|---|---|---|---|
| | Rank | GeoM | URLs | Random | GOV2 | Wikip | URLs | DNA | NoDup |
| $n$ | | | 66 M | 205 M | 80.2 M | 256 Mi | 10 M | 31.5 M | 31.6 M |
| $N$ | | | 4 Gi | 2 Gi | 4 Gi | 32 Pi | 304 Mi | 302 Mi | 382 Mi |
| $D/N$ $(D)$ | | | 92.7 % | 42.0 % | 69.8 % | (14.4 G) | 97.5 % | 100 % | 73.4 % |
| $L/n$ | | | 57.9 | 3.2 | 34.1 | 34.8 | 29.4 | 9.0 | 7.7 |
| | A.Intel-1×8 (2008) | | | | | | | | |
| std::sort | 36 | 4.90 | 79.5 | 93.9 | 58.1 | 180.1 | 7.42 | 16.46 | 15.24 |
| BS.mkqs | 24 | 2.19 | 26.4 | 56.1 | 25.3 | 89.4 | 3.03 | 6.70 | 7.50 |
| R.mkqs-cache8 | 4 | 1.33 | **14.8** | 30.0 | **13.9** | 65.9 | **1.86** | 3.87 | 5.21 |
| MBM.radixsort | 33 | 4.48 | 63.9 | 64.1 | 73.4 | 146.5 | 8.72 | 16.99 | 11.80 |
| KR.radixsort-CE0 | 25 | 2.60 | 50.0 | 33.4 | 59.5 | 85.8 | 5.38 | 6.45 | 5.83 |
| KR.radixsort-CE1 | 16 | 1.82 | 36.6 | 20.9 | 40.7 | 67.9 | 3.59 | 4.25 | 4.31 |
| KR.radixsort-CE2 | 5 | 1.43 | 37.9 | **12.5** | 31.0 | 57.8 | 2.95 | 3.07 | 3.32 |
| KRB.radixsort-CE3s | 2 | 1.27 | 28.7 | 12.6 | 27.7 | **56.4** | 2.55 | 2.59 | 3.13 |
| KR.radixsort-CI0 | 34 | 4.48 | 63.5 | 62.5 | 76.0 | 153.0 | 8.59 | 16.67 | 11.73 |
| KR.radixsort-CI1 | 29 | 3.00 | 40.4 | 40.5 | 55.9 | 115.3 | 6.52 | 9.16 | 7.42 |
| KR.radixsort-CI2 | 10 | 1.68 | 36.6 | 16.1 | 35.3 | 71.2 | 3.28 | 3.98 | 4.15 |
| KRB.radixsort-CI3s | 6 | 1.45 | 27.0 | 17.0 | 28.0 | 64.8 | 2.76 | 3.34 | 3.82 |
| KR.radixsort-CE6 | 3 | 1.28 | 28.3 | 13.5 | 26.9 | 58.4 | 2.51 | 2.58 | **3.03** |
| KR.radixsort-CE7 | 1 | 1.27 | 27.4 | 13.5 | 26.8 | 58.3 | 2.45 | **2.58** | 3.05 |
| KR.radixsort-D-vec | 9 | 1.61 | 37.3 | 20.4 | 32.0 | 67.5 | 3.04 | 3.21 | 3.79 |
| KR.radixsort-D-vecblk | 15 | 1.80 | 39.3 | 22.4 | 35.7 | 78.6 | 3.12 | 3.59 | 4.81 |
| KR.radixsort-DB | 7 | 1.46 | 36.5 | 13.5 | 30.4 | 62.6 | 2.94 | 3.25 | 3.36 |
| AN.AdaptiveRadix | 27 | 2.70 | 28.8 | 41.3 | 36.4 | 118.1 | 5.15 | 9.90 | 8.58 |
| AN.ForwardRadix8 | 39 | 7.44 | 94.7 | 80.5 | 124.4 | 327.7 | 18.84 | 21.56 | 21.38 |
| AN.ForwardRadix16 | 32 | 4.38 | 53.7 | 54.4 | 69.0 | 203.2 | 10.05 | 11.85 | 13.52 |
| NK.CRadix | 11 | 1.72 | 36.5 | 13.8 | 30.9 | 69.9 | 3.33 | 5.36 | 4.83 |
| B.Seq-$S^5$-E | 19 | 1.87 | 23.5 | 39.4 | 30.4 | 88.1 | 2.72 | 4.27 | 5.97 |
| B.Seq-$S^5$-U | 21 | 2.06 | 29.2 | 42.2 | 32.2 | 89.7 | 3.21 | 4.60 | 6.31 |
| B.Seq-$S^5$-UI | 12 | 1.73 | 24.9 | 32.7 | 29.6 | 78.6 | 2.84 | 3.48 | 5.33 |
| B.Seq-$S^5$-UC | 23 | 2.14 | 30.6 | 44.3 | 33.0 | 92.8 | 3.30 | 4.85 | 6.58 |
| B.Seq-$S^5$-UIC | 14 | 1.78 | 25.4 | 34.3 | 30.0 | 80.5 | 2.87 | 3.63 | 5.54 |
| NK.LCP-Mergesort | 30 | 3.13 | 21.6 | 99.5 | 30.2 | 137.4 | 4.12 | 11.88 | 14.20 |
| B.LCP-MS-2way | 26 | 2.66 | 19.4 | 71.2 | 25.1 | 131.6 | 3.36 | 10.59 | 12.25 |
| B.LCP-MS-16way | 28 | 2.76 | 20.0 | 79.9 | 25.7 | 129.2 | 3.42 | 11.46 | 12.47 |
| R.funnelsort-32way | 37 | 4.92 | 56.7 | 144.6 | 48.6 | 173.0 | 6.90 | 18.00 | 17.39 |
| SZ.burstsortA | 18 | 1.84 | 20.5 | 44.9 | 22.6 | 94.7 | 2.92 | 4.70 | 5.59 |
| SZ.burstsortL | 20 | 2.04 | 20.3 | 41.3 | 24.1 | 98.2 | 3.41 | 6.54 | 7.14 |
| SZR.C-burstsort | 35* | 4.52 | | 38.6 | | | 10.27 | 3.89 | 13.88 |
| SZR.CP-burstsort | 38* | 5.15 | | 32.8 | | | 9.75 | 7.69 | 21.69 |
| SZR.CPL-burstsort | 31* | 3.94 | 39.1 | 40.1 | 33.1 | | 8.00 | 11.19 | 20.78 |
| R.burstsort-vec | 17 | 1.84 | 17.4 | 51.0 | 22.0 | 102.7 | 2.76 | 4.60 | 5.90 |
| R.burstsort2-vec | 22 | 2.07 | 22.6 | 54.7 | 24.9 | 116.9 | 3.14 | 4.90 | 6.26 |
| R.burstsort-vecblk | 8 | 1.60 | 16.5 | 35.5 | 20.6 | 85.9 | 2.35 | 4.42 | 5.41 |
| R.burstsort2-vecblk | 13 | 1.75 | 20.0 | 37.1 | 22.3 | 99.6 | 2.52 | 4.62 | 5.58 |

* mark incomplete result, see table 2.4.

**Table 2.7:** Running time of sequential string sorting algorithms on B.AMD-4×4 in seconds.

| | Overall | | Our Datasets | | | | Sinha's | | |
|---|---|---|---|---|---|---|---|---|---|
| | Rank | GeoM | URLs | Random | GOV2 | Wikip | URLs | DNA | NoDup |
| $n$ | | | 66 M | 409 M | 80.2 M | 256 Mi | 10 M | 31.5 M | 31.6 M |
| $N$ | | | 4 Gi | 4 Gi | 4 Gi | 32 Pi | 304 Mi | 302 Mi | 382 Mi |
| $D/N$ $(D)$ | | | 92.7 % | 43.0 % | 69.8 % | (14.4 G) | 97.5 % | 100 % | 73.4 % |
| $L/n$ | | | 57.9 | 3.3 | 34.1 | 34.8 | 29.4 | 9.0 | 7.7 |
| | B.AMD-4×4 (2007) | | | | | | | | |
| std::sort | 38 | 5.76 | 240 | 567 | 180 | 451 | 23.4 | 53.2 | 39.4 |
| BS.mkqs | 25 | 2.60 | 115 | 369 | 77 | 206 | 9.4 | 18.5 | 17.5 |
| R.mkqs-cache8 | 3 | 1.27 | 54 | 108 | **40** | 135 | **5.0** | 8.8 | 9.7 |
| MBM.radixsort | 34 | 4.88 | 331 | 332 | 190 | 319 | 24.5 | 41.6 | 24.8 |
| KR.radixsort-CE0 | 37 | 5.45 | 434 | 309 | 231 | 332 | 30.1 | 44.2 | 26.7 |
| KR.radixsort-CE1 | 28 | 3.42 | 249 | 199 | 137 | 235 | 18.0 | 26.9 | 18.1 |
| KR.radixsort-CE2 | 7 | 1.48 | 115 | 71 | 73 | 126 | 7.5 | 9.1 | 7.7 |
| KRB.radixsort-CE3s | 2 | 1.25 | 84 | **62** | 64 | 117 | 6.4 | **7.4** | 6.7 |
| KR.radixsort-CI0 | 36 | 5.20 | 338 | 390 | 208 | 336 | 24.8 | 45.2 | 25.5 |
| KR.radixsort-CI1 | 29 | 3.52 | 241 | 206 | 151 | 257 | 19.5 | 27.1 | 17.0 |
| KR.radixsort-CI2 | 6 | 1.45 | 100 | 70 | 76 | 136 | 6.8 | 8.6 | 8.1 |
| KRB.radixsort-CI3s | 1 | 1.24 | 72 | 70 | 59 | 120 | 5.7 | 7.8 | 7.1 |
| KR.radixsort-CE6 | 5 | 1.38 | 87 | 116 | 61 | **114** | 6.6 | 8.0 | 6.5 |
| KR.radixsort-CE7 | 4 | 1.35 | 84 | 105 | 61 | 115 | 6.4 | 8.0 | **6.5** |
| KR.radixsort-D-vec | 18 | 1.88 | 129 | 142 | 81 | 154 | 9.3 | 10.4 | 9.7 |
| KR.radixsort-D-vecblk | 21 | 1.96 | 135 | 140 | 90 | 164 | 9.1 | 10.9 | 10.4 |
| KR.radixsort-DB | 14 | 1.69 | 122 | 129 | 74 | 132 | 7.9 | 10.5 | 7.9 |
| AN.AdaptiveRadix | 27 | 2.81 | 105 | 232 | 92 | 242 | 13.7 | 24.6 | 19.2 |
| AN.ForwardRadix8 | 39 | 7.10 | 261 | 416 | 296 | 722 | 39.7 | 51.0 | 49.5 |
| AN.ForwardRadix16 | 33 | 4.48 | 183 | 288 | 182 | 467 | 22.8 | 27.1 | 33.3 |
| NK.CRadix | 19 | 1.89 | 117 | 86 | 74 | 172 | 8.9 | 16.9 | 11.5 |
| B.Seq-$S^5$-E | 8 | 1.52 | 61 | 165 | 62 | 162 | 5.2 | 8.4 | 11.2 |
| B.Seq-$S^5$-U | 13 | 1.64 | 68 | 182 | 64 | 171 | 5.7 | 8.7 | 12.1 |
| B.Seq-$S^5$-UI | 9 | 1.53 | 64 | 164 | 62 | 163 | 5.4 | 7.8 | 11.2 |
| B.Seq-$S^5$-UC | 15 | 1.72 | 71 | 190 | 67 | 178 | 5.9 | 9.6 | 12.6 |
| B.Seq-$S^5$-UIC | 11 | 1.56 | 65 | 167 | 62 | 163 | 5.5 | 8.4 | 11.3 |
| NK.LCP-Mergesort | 26 | 2.75 | 51 | 525 | 72 | 281 | 8.2 | 25.0 | 27.2 |
| B.LCP-MS-2way | 24 | 2.06 | 39 | 308 | 51 | 235 | 6.2 | 20.6 | 21.8 |
| B.LCP-MS-16way | 22 | 2.00 | 38 | 315 | 49 | 214 | 6.0 | 19.8 | 22.1 |
| R.funnelsort-32way | 31 | 4.10 | 125 | 695 | 111 | 340 | 13.1 | 35.5 | 32.8 |
| SZ.burstsortA | 17 | 1.85 | 43 | 275 | 64 | 223 | 7.0 | 12.3 | 13.5 |
| SZ.burstsortL | 23 | 2.03 | 46 | 210 | 71 | 217 | 8.7 | 17.4 | 16.3 |
| SZR.C-burstsort | 30* | 3.83 | 133 | 158 | | | 27.4 | 7.7 | 32.9 |
| SZR.CP-burstsort | 35* | 4.94 | 130 | 188 | | | 31.4 | 19.8 | 56.1 |
| SZR.CPL-burstsort | 32* | 4.21 | 86 | 296 | 85 | | 21.4 | 32.3 | 55.7 |
| R.burstsort-vec | 16 | 1.72 | 39 | 308 | 52 | 230 | 6.0 | 10.6 | 12.6 |
| R.burstsort2-vec | 20 | 1.89 | 45 | 339 | 58 | 268 | 6.5 | 11.2 | 13.1 |
| R.burstsort-vecblk | 10 | 1.55 | **38** | 262 | 48 | 193 | 5.5 | 9.4 | 11.9 |
| R.burstsort2-vecblk | 12 | 1.63 | 41 | 269 | 51 | 217 | 5.3 | 10.2 | 12.1 |

* mark incomplete result, see table 2.4.

**Table 2.8:** Running time of sequential string sorting algorithms on C.AMD-4×12 in seconds.

| | Overall | | Our Datasets | | | | Sinha's | | |
|---|---|---|---|---|---|---|---|---|---|
| | Rank | GeoM | URLs | Random | GOV2 | Wikip | URLs | DNA | NoDup |
| $n$ | | | 66 M | 409 M | 80.2 M | 256 Mi | 10 M | 31.5 M | 31.6 M |
| $N$ | | | 4 Gi | 4 Gi | 4 Gi | 32 Pi | 304 Mi | 302 Mi | 382 Mi |
| $D/N$ $(D)$ | | | 92.7 % | 43.0 % | 69.8 % | (14.4 G) | 97.5 % | 100 % | 73.4 % |
| $L/n$ | | | 57.9 | 3.3 | 34.1 | 34.8 | 29.4 | 9.0 | 7.7 |
| | C.AMD-4×12 (2010) | | | | | | | | |
| std::sort | 39 | 5.21 | 238.9 | 428.8 | 127.3 | 380 | 18.79 | 39.06 | 29.71 |
| BS.mkqs | 26 | 2.55 | 113.9 | 283.1 | 53.2 | 200 | 7.80 | 16.49 | 16.53 |
| R.mkqs-cache8 | 1 | 1.27 | 50.4 | 95.4 | **33.2** | 130 | **4.28** | 7.08 | 9.07 |
| MBM.radixsort | 34 | 4.00 | 315.2 | 186.8 | 110.8 | 233 | 22.75 | 25.39 | 19.29 |
| KR.radixsort-CE0 | 37 | 4.13 | 399.6 | 138.7 | 131.0 | 244 | 20.54 | 27.98 | 20.97 |
| KR.radixsort-CE1 | 28 | 2.99 | 228.3 | 98.8 | 95.9 | 218 | 18.37 | 21.61 | 11.91 |
| KR.radixsort-CE2 | 6 | 1.48 | 119.1 | 50.5 | 58.9 | 110 | 6.99 | 8.78 | 6.64 |
| KRB.radixsort-CE3s | 3 | 1.28 | 81.1 | 49.9 | 59.2 | 113 | 5.39 | **6.13** | 6.48 |
| KR.radixsort-CI0 | 36 | 4.07 | 320.6 | 183.3 | 120.3 | 230 | 20.89 | 28.65 | 19.57 |
| KR.radixsort-CI1 | 29 | 3.12 | 227.5 | 117.8 | 105.6 | 238 | 11.44 | 20.68 | 18.69 |
| KR.radixsort-CI2 | 8 | 1.54 | 110.5 | 60.1 | 69.9 | 130 | 6.09 | 8.14 | 7.24 |
| KRB.radixsort-CI3s | 4 | 1.28 | 72.3 | 63.6 | 50.0 | 119 | 4.95 | 6.78 | **6.44** |
| KR.radixsort-CE6 | 2 | 1.28 | 83.6 | 49.6 | 51.1 | **104** | 5.95 | 6.72 | 6.48 |
| KR.radixsort-CE7 | 5 | 1.31 | 82.2 | **49.6** | 51.0 | 112 | 6.34 | 7.20 | 6.44 |
| KR.radixsort-D-vec | 16 | 1.75 | 131.1 | 62.7 | 67.1 | 139 | 7.84 | 9.59 | 8.86 |
| KR.radixsort-D-vecblk | 22 | 1.91 | 137.5 | 69.3 | 74.5 | 158 | 7.86 | 10.31 | 10.64 |
| KR.radixsort-DB | 9 | 1.56 | 124.2 | 52.8 | 60.8 | 116 | 7.44 | 9.25 | 7.26 |
| AN.AdaptiveRadix | 25 | 2.24 | 86.9 | 156.2 | 63.0 | 180 | 8.62 | 15.36 | 14.38 |
| AN.ForwardRadix8 | 38 | 5.15 | 184.4 | 241.8 | 173.7 | 478 | 24.43 | 30.66 | 35.93 |
| AN.ForwardRadix16 | 32 | 3.66 | 154.5 | 174.3 | 105.4 | 362 | 19.09 | 20.04 | 23.14 |
| NK.CRadix | 20 | 1.85 | 109.7 | 60.4 | 56.7 | 166 | 8.38 | 15.44 | 9.55 |
| B.Seq-$S^5$-E | 10 | 1.58 | 61.5 | 129.3 | 53.2 | 167 | 4.40 | 8.01 | 10.25 |
| B.Seq-$S^5$-U | 15 | 1.71 | 68.8 | 145.3 | 57.0 | 162 | 5.26 | 8.75 | 10.43 |
| B.Seq-$S^5$-UI | 11 | 1.60 | 65.2 | 129.0 | 54.3 | 167 | 4.94 | 6.61 | 10.99 |
| B.Seq-$S^5$-UC | 18 | 1.78 | 71.9 | 153.1 | 58.5 | 168 | 5.92 | 8.23 | 11.01 |
| B.Seq-$S^5$-UIC | 14 | 1.67 | 65.7 | 132.2 | 54.8 | 168 | 5.55 | 7.98 | 10.80 |
| NK.LCP-Mergesort | 27 | 2.76 | 47.2 | 495.7 | 54.6 | 234 | 7.38 | 20.26 | 28.04 |
| B.LCP-MS-2way | 23 | 2.05 | **35.8** | 239.9 | 41.2 | 204 | 5.78 | 16.89 | 22.48 |
| B.LCP-MS-16way | 24 | 2.11 | 37.7 | 255.2 | 46.9 | 198 | 5.76 | 18.22 | 20.79 |
| R.funnelsort-32way | 31 | 3.60 | 114.5 | 436.8 | 81.7 | 258 | 9.98 | 29.33 | 26.53 |
| SZ.burstsortA | 19 | 1.78 | 42.7 | 176.0 | 43.5 | 188 | 6.73 | 10.91 | 13.15 |
| SZ.burstsortL | 17 | 1.77 | 41.7 | 133.4 | 44.9 | 169 | 7.86 | 13.52 | 12.57 |
| SZR.C-burstsort | 30* | 3.40 | 108.9 | 137.9 | | | 23.04 | 7.43 | 25.51 |
| SZR.CP-burstsort | 35* | 4.04 | 102.4 | 159.5 | | | 20.75 | 15.90 | 40.62 |
| SZR.CPL-burstsort | 33* | 3.80 | 84.4 | 196.5 | 61.8 | | 19.22 | 27.36 | 46.01 |
| R.burstsort-vec | 13 | 1.67 | 38.3 | 202.0 | 42.1 | 180 | 5.67 | 9.71 | 11.71 |
| R.burstsort2-vec | 21 | 1.87 | 45.4 | 227.2 | 47.7 | 217 | 6.08 | 10.54 | 12.20 |
| R.burstsort-vecblk | 7 | 1.53 | 37.6 | 171.3 | 39.8 | 163 | 5.27 | 8.77 | 10.56 |
| R.burstsort2-vecblk | 12 | 1.63 | 41.0 | 181.5 | 42.7 | 181 | 5.29 | 9.67 | 10.62 |

* mark incomplete result, see table 2.4.

**Table 2.9:** Running time of sequential string sorting algorithms on D.Intel-4×8 in seconds.

| | Overall | | Our Datasets | | | | Sinha's | | |
|---|---|---|---|---|---|---|---|---|---|
| | Rank | GeoM | URLs | Random | GOV2 | Wikip | URLs | DNA | NoDup |
| $n$ | | | 66 M | 409 M | 80.2 M | 256 Mi | 10 M | 31.5 M | 31.6 M |
| $N$ | | | 4 Gi | 4 Gi | 4 Gi | 32 Pi | 304 Mi | 302 Mi | 382 Mi |
| $D/N$ ($D$) | | | 92.7 % | 43.0 % | 69.8 % | (14.4 G) | 97.5 % | 100 % | 73.4 % |
| $L/n$ | | | 57.9 | 3.3 | 34.1 | 34.8 | 29.4 | 9.0 | 7.7 |
| | D.Intel-4×8 (2012) | | | | | | | | |
| std::sort | 37 | 6.03 | 136.8 | 367.0 | 112.9 | 284.9 | 14.98 | 32.98 | 29.70 |
| BS.mkqs | 27 | 2.73 | 76.0 | 220.5 | 49.4 | 125.0 | 5.51 | 12.92 | 12.65 |
| R.mkqs-cache8 | 1 | 1.25 | 28.6 | 60.5 | **21.8** | 91.8 | 3.36 | 5.04 | 6.84 |
| MBM.radixsort | 36 | 5.87 | 191.2 | 438.6 | 109.4 | 319.1 | 13.98 | 18.97 | 25.50 |
| KR.radixsort-CE0 | 32 | 3.79 | 135.2 | 260.1 | 107.6 | 168.3 | 9.12 | 10.84 | 14.76 |
| KR.radixsort-CE1 | 21 | 2.19 | 87.5 | 105.6 | 45.8 | 97.7 | 5.89 | 9.64 | 8.38 |
| KR.radixsort-CE2 | 7 | 1.57 | 77.3 | 62.9 | 48.4 | 90.9 | **3.29** | 7.05 | **3.81** |
| KRB.radixsort-CE3s | 2 | 1.30 | 55.8 | **43.7** | 42.7 | 76.9 | 4.14 | **4.01** | 3.84 |
| KR.radixsort-CI0 | 38 | 6.12 | 187.1 | 439.7 | 168.5 | 259.3 | 12.07 | 36.98 | 16.53 |
| KR.radixsort-CI1 | 31 | 3.49 | 88.0 | 181.4 | 68.5 | 186.2 | 9.46 | 20.28 | 13.15 |
| KR.radixsort-CI2 | 13 | 1.78 | 74.6 | 58.3 | 54.1 | 101.6 | 5.09 | 7.82 | 4.88 |
| KRB.radixsort-CI3s | 5 | 1.41 | 49.3 | 47.9 | 40.7 | 81.3 | 4.29 | 5.90 | 4.55 |
| KR.radixsort-CE6 | 3 | 1.33 | 50.4 | 44.1 | 36.4 | 75.1 | 4.00 | 4.98 | 4.89 |
| KR.radixsort-CE7 | 4 | 1.34 | 52.8 | 44.1 | 39.1 | **75.1** | 3.92 | 4.98 | 4.88 |
| KR.radixsort-D-vec | 19 | 1.94 | 87.9 | 54.4 | 53.9 | 103.3 | 5.57 | 7.85 | 7.20 |
| KR.radixsort-D-vecblk | 14 | 1.79 | 87.1 | 54.9 | 49.1 | 96.0 | 3.77 | 7.78 | 7.29 |
| KR.radixsort-DB | 10 | 1.73 | 83.5 | 56.3 | 48.9 | 80.0 | 4.68 | 7.27 | 6.06 |
| AN.AdaptiveRadix | 26 | 2.72 | 53.4 | 210.9 | 50.2 | 157.7 | 6.09 | 14.36 | 11.52 |
| AN.ForwardRadix8 | 39 | 8.10 | 160.1 | 348.8 | 184.1 | 521.6 | 24.94 | 39.11 | 35.96 |
| AN.ForwardRadix16 | 35 | 4.97 | 96.4 | 200.8 | 109.7 | 354.8 | 13.92 | 24.22 | 24.12 |
| NK.CRadix | 12 | 1.76 | 53.7 | 63.9 | 43.1 | 98.9 | 5.01 | 8.48 | 7.02 |
| B.Seq-$S^5$-E | 20 | 2.13 | 44.0 | 166.3 | 50.5 | 142.6 | 4.05 | 10.12 | 7.61 |
| B.Seq-$S^5$-U | 23 | 2.37 | 52.5 | 150.2 | 48.3 | 159.1 | 5.19 | 9.87 | 11.30 |
| B.Seq-$S^5$-UI | 17 | 1.89 | 42.1 | 130.7 | 42.6 | 124.4 | 4.39 | 6.50 | 8.61 |
| B.Seq-$S^5$-UC | 24 | 2.40 | 53.4 | 155.5 | 51.0 | 138.9 | 5.31 | 10.25 | 11.67 |
| B.Seq-$S^5$-UIC | 15 | 1.80 | 40.7 | 108.0 | 41.7 | 110.8 | 4.40 | 6.54 | 8.59 |
| NK.LCP-Mergesort | 28 | 2.83 | 28.5 | 353.8 | 37.9 | 170.3 | 5.88 | 15.41 | 20.18 |
| B.LCP-MS-2way | 25 | 2.45 | 24.2 | 249.7 | 35.7 | 184.2 | 4.40 | 13.34 | 18.74 |
| B.LCP-MS-16way | 22 | 2.33 | 24.7 | 240.1 | 33.8 | 154.3 | 4.28 | 13.65 | 16.92 |
| R.funnelsort-32way | 29 | 3.15 | 55.9 | 300.9 | 51.2 | 161.4 | 6.50 | 17.12 | 16.27 |
| SZ.burstsortA | 11 | 1.74 | 27.6 | 118.2 | 36.0 | 116.6 | 3.68 | 8.66 | 9.13 |
| SZ.burstsortL | 18 | 1.92 | **22.8** | 112.3 | 37.0 | 123.1 | 4.65 | 10.43 | 13.83 |
| SZR.C-burstsort | 30* | 3.41 | 66.8 | 72.2 | | | 14.11 | 4.31 | 15.52 |
| SZR.CP-burstsort | 34* | 4.03 | 62.7 | 84.5 | | | 14.63 | 7.52 | 25.55 |
| SZR.CPL-burstsort | 33* | 3.98 | 52.6 | 149.0 | 45.2 | | 12.21 | 17.99 | 31.97 |
| R.burstsort-vec | 8 | 1.58 | 25.2 | 146.9 | 31.5 | 113.3 | 3.56 | 5.82 | 7.35 |
| R.burstsort2-vec | 16 | 1.81 | 30.5 | 156.4 | 34.9 | 135.7 | 4.51 | 5.92 | 8.82 |
| R.burstsort-vecblk | 6 | 1.49 | 23.9 | 137.7 | 29.7 | 109.2 | 3.73 | 5.54 | 6.08 |
| R.burstsort2-vecblk | 9 | 1.62 | 27.5 | 125.2 | 33.3 | 114.3 | 3.90 | 6.02 | 7.88 |

* mark incomplete result, see table 2.4.

**Table 2.10:** Running time of sequential string sorting algorithms on E.Intel-2×16 in seconds.

| | Overall | | Our Datasets | | | | Sinha's | | |
|---|---|---|---|---|---|---|---|---|---|
| | Rank | GeoM | URLs | Random | GOV2 | Wikip | URLs | DNA | NoDup |
| $n$ | | | 66 M | 409 M | 80.2 M | 256 Mi | 10 M | 31.5 M | 31.6 M |
| $N$ | | | 4 Gi | 4 Gi | 4 Gi | 32 Pi | 304 Mi | 302 Mi | 382 Mi |
| $D/N$ $(D)$ | | | 92.7 % | 43.0 % | 69.8 % | (14.4 G) | 97.5 % | 100 % | 73.4 % |
| $L/n$ | | | 57.9 | 3.3 | 34.1 | 34.8 | 29.4 | 9.0 | 7.7 |
| | E.Intel-2×16 (2016) | | | | | | | | |
| std::sort | 37 | 4.88 | 87.0 | 219.9 | 63.5 | 229.0 | 8.60 | 19.03 | 18.26 |
| BS.mkqs | 22 | 1.89 | 29.0 | 117.3 | 22.2 | 90.4 | 2.81 | 6.89 | 8.29 |
| R.mkqs-cache8 | 8 | 1.41 | **18.4** | 74.5 | **16.7** | 77.2 | **2.40** | 5.09 | 6.51 |
| MBM.radixsort | 31 | 3.03 | 64.4 | 107.4 | 53.7 | 129.0 | 5.34 | 11.31 | 10.20 |
| KR.radixsort-CE0 | 14 | 1.58 | 46.5 | 45.5 | 33.7 | 63.5 | 3.35 | 4.49 | 4.58 |
| KR.radixsort-CE1 | 6 | 1.36 | 37.8 | 41.3 | 29.1 | 59.5 | 2.89 | 3.44 | 4.04 |
| KR.radixsort-CE2 | 9 | 1.43 | 44.9 | 33.9 | 30.4 | 66.1 | 3.36 | 3.55 | 4.18 |
| KRB.radixsort-CE3s | 3 | 1.25 | 33.1 | 31.6 | 27.0 | 65.9 | 3.04 | 2.77 | 3.89 |
| KR.radixsort-CI0 | 30 | 2.91 | 63.8 | 104.4 | 55.3 | 113.0 | 5.14 | 11.09 | 9.46 |
| KR.radixsort-CI1 | 10 | 1.44 | 35.9 | 41.4 | 29.7 | 71.8 | 2.71 | 4.04 | 4.53 |
| KR.radixsort-CI2 | 12 | 1.56 | 45.1 | 32.0 | 35.9 | 80.5 | 3.28 | 4.21 | 4.82 |
| KRB.radixsort-CI3s | 4 | 1.29 | 31.0 | 36.5 | 25.5 | 68.2 | 2.72 | 3.19 | 4.30 |
| KR.radixsort-CE6 | 1 | 1.19 | 31.5 | 31.4 | 24.8 | **58.9** | 2.91 | **2.68** | **3.83** |
| KR.radixsort-CE7 | 2 | 1.19 | 30.6 | 31.4 | 24.7 | 62.4 | 2.83 | 2.68 | 3.84 |
| KR.radixsort-D-vec | 11 | 1.48 | 41.5 | 36.1 | 29.8 | 71.4 | 3.48 | 3.72 | 4.68 |
| KR.radixsort-D-vecblk | 16 | 1.65 | 44.0 | 42.3 | 33.1 | 83.1 | 3.35 | 4.19 | 5.89 |
| KR.radixsort-DB | 5 | 1.33 | 40.0 | **28.2** | 28.4 | 60.7 | 3.17 | 3.66 | 4.14 |
| AN.AdaptiveRadix | 27 | 2.46 | 31.6 | 109.3 | 32.8 | 129.9 | 5.26 | 9.89 | 8.95 |
| AN.ForwardRadix8 | 39 | 7.81 | 113.3 | 212.6 | 142.6 | 416.0 | 23.49 | 26.14 | 25.27 |
| AN.ForwardRadix16 | 36 | 4.53 | 62.1 | 143.1 | 78.1 | 246.8 | 12.90 | 14.22 | 15.54 |
| NK.CRadix | 7 | 1.41 | 37.7 | 37.4 | 26.2 | 68.3 | 3.05 | 3.75 | 4.77 |
| B.Seq-$S^5$-E | 21 | 1.81 | 27.6 | 83.3 | 30.2 | 108.5 | 2.69 | 5.41 | 7.38 |
| B.Seq-$S^5$-U | 23 | 1.93 | 34.0 | 90.3 | 31.6 | 95.4 | 3.04 | 5.72 | 7.67 |
| B.Seq-$S^5$-UI | 13 | 1.56 | 27.8 | 65.8 | 28.0 | 82.1 | 2.66 | 4.04 | 6.16 |
| B.Seq-$S^5$-UC | 26 | 2.06 | 36.0 | 97.1 | 32.8 | 114.7 | 3.17 | 5.85 | 8.04 |
| B.Seq-$S^5$-UIC | 17 | 1.68 | 30.0 | 71.8 | 29.4 | 92.0 | 2.80 | 4.49 | 6.48 |
| NK.LCP-Mergesort | 32 | 3.03 | 23.2 | 222.4 | 33.1 | 159.4 | 4.79 | 13.64 | 16.56 |
| B.LCP-MS-2way | 28 | 2.76 | 22.0 | 180.5 | 29.8 | 154.7 | 4.22 | 12.71 | 15.60 |
| B.LCP-MS-16way | 29 | 2.82 | 23.0 | 192.5 | 29.9 | 152.2 | 4.15 | 13.47 | 15.71 |
| R.funnelsort-32way | 33 | 3.49 | 44.0 | 218.7 | 40.0 | 146.7 | 5.78 | 15.92 | 15.30 |
| SZ.burstsortA | 18 | 1.71 | 23.7 | 100.0 | 21.0 | 95.7 | 3.43 | 4.99 | 6.54 |
| SZ.burstsortL | 25 | 2.02 | 23.5 | 87.3 | 24.6 | 111.4 | 4.52 | 7.63 | 8.91 |
| SZR.C-burstsort | 35* | 4.42 | 73.2 | 85.9 | | | 13.17 | 4.83 | 17.41 |
| SZR.CP-burstsort | 38* | 5.11 | 63.9 | 93.9 | | | 11.86 | 8.59 | 31.41 |
| SZR.CPL-burstsort | 34* | 3.84 | 44.6 | 98.0 | 32.2 | | 9.04 | 12.02 | 24.13 |
| R.burstsort-vec | 19 | 1.73 | 21.9 | 114.7 | 20.5 | 104.8 | 3.33 | 5.00 | 6.55 |
| R.burstsort2-vec | 24 | 1.96 | 28.9 | 121.1 | 24.7 | 115.2 | 3.84 | 5.34 | 6.91 |
| R.burstsort-vecblk | 15 | 1.62 | 21.0 | 107.4 | 18.9 | 91.9 | 2.97 | 5.02 | 6.39 |
| R.burstsort2-vecblk | 20 | 1.76 | 25.9 | 103.3 | 21.9 | 104.1 | 3.26 | 5.07 | 6.58 |

* mark incomplete result, see table 2.4.

**Table 2.11:** Running time of sequential string sorting algorithms on F.AMD-1×16 in seconds.

| | Overall | | Our Datasets | | | | Sinha's | | |
|---|---|---|---|---|---|---|---|---|---|
| | Rank | GeoM | URLs | Random | GOV2 | Wikip | URLs | DNA | NoDup |
| $n$ | | | 66 M | 409 M | 80.2 M | 256 Mi | 10 M | 31.5 M | 31.6 M |
| $N$ | | | 4 Gi | 4 Gi | 4 Gi | 32 Pi | 304 Mi | 302 Mi | 382 Mi |
| $D/N$ $(D)$ | | | 92.7 % | 43.0 % | 69.8 % | (14.4 G) | 97.5 % | 100 % | 73.4 % |
| $L/n$ | | | 57.9 | 3.3 | 34.1 | 34.8 | 29.4 | 9.0 | 7.7 |
| | F.AMD-1×16 (2017) | | | | | | | | |
| std::sort | 33 | 6.04 | 64.0 | 162.0 | 48.6 | 139.7 | 6.52 | 13.46 | 12.61 |
| BS.mkqs | 21 | 2.28 | 21.7 | 81.1 | 15.9 | 57.8 | 2.08 | 4.86 | 5.25 |
| R.mkqs-cache8 | 5 | 1.50 | 12.9 | 39.4 | **11.0** | 47.5 | 1.63 | 2.76 | 3.77 |
| MBM.radixsort | 36 | 8.07 | 89.0 | 164.0 | 75.7 | 152.1 | 11.69 | 21.18 | 14.18 |
| KR.radixsort-CE0 | 35 | 7.99 | 116.4 | 151.3 | 70.8 | 117.0 | 15.53 | 20.40 | 11.97 |
| KR.radixsort-CE1 | 27 | 4.31 | 67.9 | 100.0 | 39.4 | 65.2 | 8.26 | 9.16 | 5.57 |
| KR.radixsort-CE2 | 7 | 1.57 | 30.6 | 17.8 | 20.5 | 39.0 | 2.30 | 2.60 | 2.37 |
| KRB.radixsort-CE3s | 3 | 1.30 | 23.0 | 15.5 | 17.6 | 36.0 | 1.85 | 1.83 | 2.14 |
| KR.radixsort-CI0 | 37 | 8.28 | 92.5 | 164.6 | 79.4 | 160.4 | 11.89 | 21.38 | 14.38 |
| KR.radixsort-CI1 | 31 | 5.69 | 71.2 | 104.5 | 52.9 | 108.2 | 9.76 | 13.00 | 9.43 |
| KR.radixsort-CI2 | 10 | 1.66 | 27.5 | 19.5 | 21.9 | 45.8 | 2.06 | 2.87 | 2.91 |
| KRB.radixsort-CI3s | 4 | 1.31 | 18.7 | 20.6 | 15.3 | 38.2 | 1.66 | 1.90 | 2.48 |
| KR.radixsort-CE6 | 1 | 1.20 | 21.0 | **14.1** | 15.6 | 33.4 | 1.92 | **1.63** | 1.93 |
| KR.radixsort-CE7 | 2 | 1.21 | 21.1 | 15.0 | 16.9 | **33.2** | 1.83 | 1.64 | **1.92** |
| KR.radixsort-D-vec | 6 | 1.57 | 30.1 | 17.9 | 20.2 | 39.5 | 2.20 | 2.55 | 2.53 |
| KR.radixsort-D-vecblk | 11 | 1.68 | 28.9 | 22.6 | 20.9 | 45.3 | 2.04 | 2.53 | 3.11 |
| KR.radixsort-DB | 8 | 1.60 | 31.5 | 18.5 | 20.4 | 39.6 | 2.27 | 2.61 | 2.51 |
| AN.AdaptiveRadix | 28 | 4.47 | 27.9 | 108.3 | 29.6 | 125.0 | 6.48 | 13.06 | 9.95 |
| AN.ForwardRadix8 | 39 | 14.64 | 109.0 | 217.8 | 149.3 | 441.3 | 27.76 | 30.80 | 28.59 |
| AN.ForwardRadix16 | 38 | 8.40 | 60.5 | 139.3 | 81.7 | 264.8 | 14.53 | 16.61 | 17.74 |
| NK.CRadix | 17 | 2.00 | 37.1 | 25.7 | 21.5 | 46.2 | 2.84 | 3.94 | 3.15 |
| B.Seq-S$^5$-E | 18 | 2.04 | 16.9 | 68.8 | 19.0 | 63.9 | 1.68 | 3.50 | 4.73 |
| B.Seq-S$^5$-U | 20 | 2.22 | 21.0 | 69.8 | 20.2 | 64.7 | 2.14 | 3.54 | 4.84 |
| B.Seq-S$^5$-UI | 12 | 1.74 | 17.6 | 48.3 | 18.0 | 52.7 | 1.77 | 2.35 | 3.80 |
| B.Seq-S$^5$-UC | 22 | 2.30 | 21.8 | 71.5 | 20.9 | 66.6 | 2.20 | 3.76 | 5.00 |
| B.Seq-S$^5$-UIC | 16 | 1.82 | 18.4 | 50.8 | 18.6 | 54.8 | 1.82 | 2.53 | 3.96 |
| NK.LCP-Mergesort | 26 | 3.28 | 14.0 | 146.1 | 22.4 | 98.6 | 2.91 | 8.16 | 10.17 |
| B.LCP-MS-2way | 23 | 2.84 | 12.5 | 115.4 | 18.4 | 91.4 | 2.38 | 7.36 | 9.28 |
| B.LCP-MS-16way | 25 | 2.89 | 12.9 | 123.3 | 17.9 | 89.4 | 2.39 | 8.04 | 9.11 |
| R.funnelsort-32way | 29 | 5.39 | 39.0 | 206.0 | 36.5 | 129.2 | 5.50 | 13.20 | 12.67 |
| SZ.burstsortA | 13 | 1.76 | 12.4 | 73.3 | 13.1 | 60.0 | 1.94 | 2.89 | 3.52 |
| SZ.burstsortL | 24 | 2.87 | 15.9 | 79.2 | 20.5 | 87.9 | 3.40 | 7.44 | 7.39 |
| SZR.C-burstsort | 32* | 5.93 | 50.9 | 56.1 | | | 9.73 | 3.09 | 12.06 |
| SZR.CP-burstsort | 34* | 6.87 | 43.5 | 61.1 | | | 8.81 | 6.39 | 19.47 |
| SZR.CPL-burstsort | 30* | 5.42 | 32.7 | 71.4 | 28.4 | | 6.90 | 9.62 | 18.75 |
| R.burstsort-vec | 15 | 1.80 | 12.0 | 95.4 | 12.8 | 66.6 | 1.72 | 2.73 | 3.54 |
| R.burstsort2-vec | 19 | 2.12 | 15.0 | 112.5 | 15.2 | 81.2 | 2.06 | 3.08 | 3.86 |
| R.burstsort-vecblk | 9 | 1.61 | **11.2** | 86.2 | 11.3 | 55.0 | **1.46** | 2.60 | 3.28 |
| R.burstsort2-vecblk | 14 | 1.79 | 13.0 | 90.8 | 12.4 | 64.5 | 1.66 | 2.91 | 3.45 |

* mark incomplete result, see table 2.4.

Chapter 3

# pmbw – Measuring Parallel Memory Bandwidth and Latency

3

*Jurkiewicz and Mehlhorn [JM13; JM15] phrase it perfectly: "Modern computers are not random access machines (RAMs)." Furthermore, modern* multi-core *machines are not* parallel *random access machines (PRAMs). In this chapter we investigate the properties of modern multi-core machines with regard to memory performance when working with* one *processor and with* many *processors in parallel.*

Memory in modern machines is a complex hierarchical system [Dre07; HP12]. Current machines have four to five levels in their memory hierarchy (compare also figure 3.1). The registers are the fastest level and attached to the CPU's logic circuits themselves. Level 1 (L1) and level 2 (L2) caches are usually also immediately close to the processor registers on the silicon die and very fast. Many machines have a level 3 (L3) cache which is shared among cores on a single socket but inside the CPU enclosure. And the final level is composed of RAM chips accessible from the CPUs via memory controllers. Furthermore, depending on one's perspective, the memory hierarchy can be extended beyond volatile RAM to external memory such as hard disks or SSDs.

This complex multi-level hierarchy is hidden from a user program behind the concept of virtual shared memory. Each memory access is mapped by the virtual address translation system to a physical memory cell. The system kernel sets up the virtual-to-physical address translation by maintaining *page tables*, and changes these when the user program requests more memory. This enables sharing physical memory between multiple isolated tasks. As each and every memory access goes through this system, modern systems have a *translation look-aside buffer* (TLB) on-chip which caches the mapping for better performance. All these intricate details are ignored by the RAM model, and while specialized *virtual address translation (VAT) models* have been proposed [JM13; JM15], they have not become popular yet due to their complexity.

The previous paragraphs focused on the hierarchy of only a single processor. To enable parallel processing with a common (shared) view of the memory on a multi-*core* machine the caches of all processors have to cooperate and maintain *cache coherence*. This common view is established using a hardware protocol [Ste90; Mar08] which communicates via the memory bus among the cores on a single socket, but also across a fast interconnect network on multi-*socket* machines.

To develop parallel string sorting algorithms, we are interested in the speed of different memory access patterns when they are *performed in parallel* by all processors in a system. Section 3.1 introduces our new pmbw benchmark tool which measures the parallel memory bandwidth and latency, and discuss the results from our experimental platforms. Section 3.2 considers the same access patterns when memory on a remote memory bank (NUMA node) is accessed across the socket interconnection.

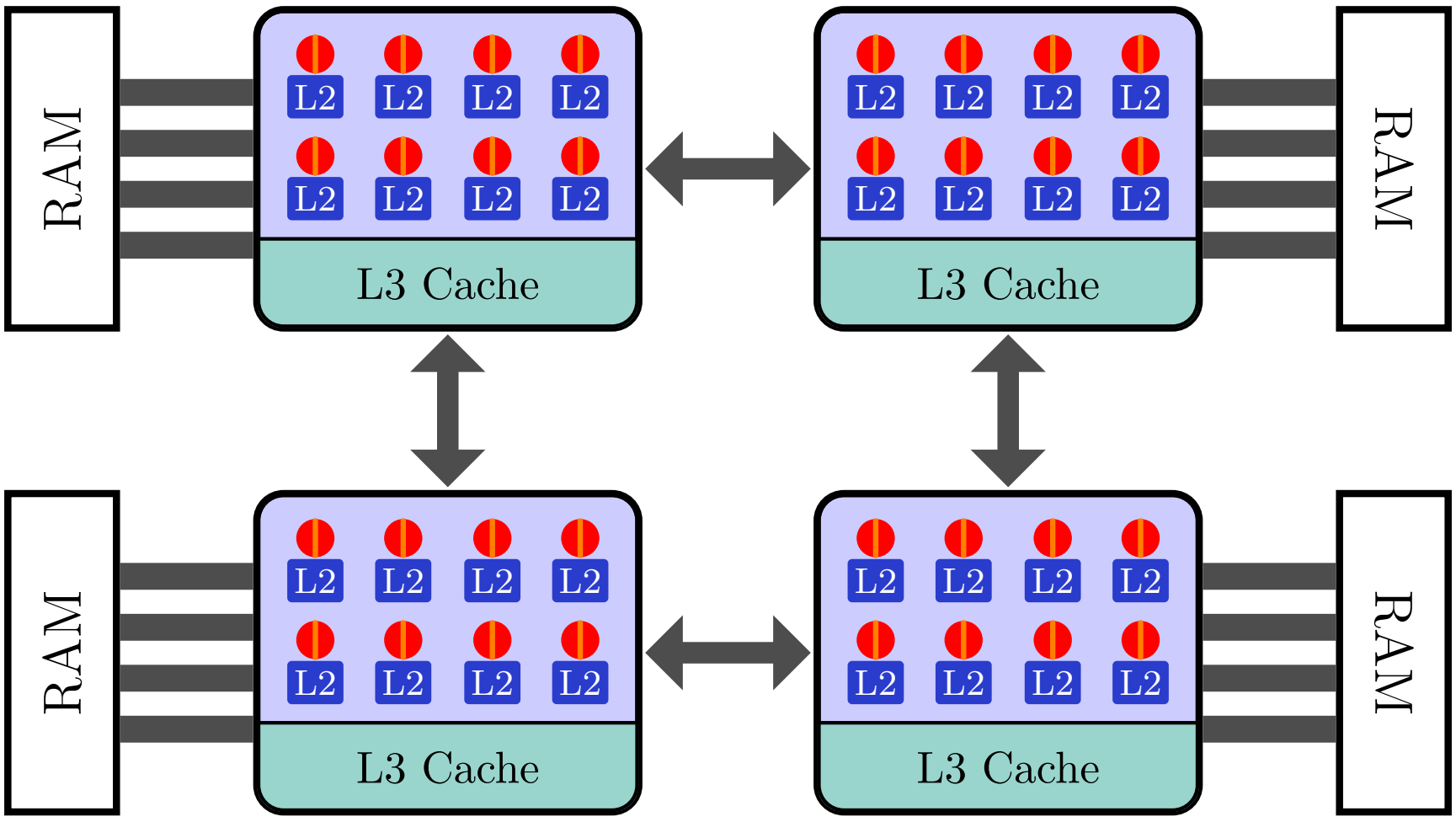

**Figure 3.1:** Schematic drawing of a modern multi-core architecture. The machine has 32 physical cores, each exposing two hardware threads, and each with an associated L2 cache. The 32 cores are distributed over four sockets with eight cores per socket, which share a larger L3 cache. Each socket is connected to a RAM bank via four memory channels, and the sockets form a coherent shared-memory system using the interconnect.

This chapter was newly written for this dissertation, but pmbw was developed while researching parallel string sorting.

## 3.1 The pmbw Benchmark Tool

Memory bandwidth has been measured by many experimental computer scientists in other contexts. McCalpin [McC95a; McC95b] introduced the STREAM benchmark for high performance supercomputers, which measures sustainable memory bandwidth and the corresponding computation rate for a small set of simple vector kernels like addition. Smith [Smi05] maintains a simple memory bandwidth benchmark for `x86`, `x86_64`, and recently also for ARM computers.

However, none of them focus on *parallel* memory operations by all cores at once, which is why we created *pmbw* [Bin13a], short for "parallel memory bandwidth benchmark".

(a) `ScanRead64IndexSimpleLoop`

```
for (size_t i = 0; i < n; ++i) {
    uint64_t x = array[i];
}
```

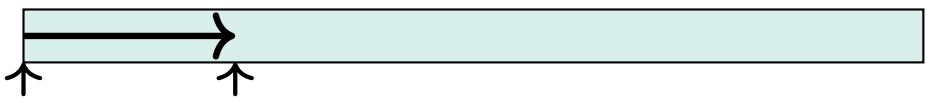

(b) `ScanRead64IndexUnrollLoop`

```
for (size_t i=0; i < n; i += 16) {
    uint64_t x0 = array[i + 0];
    // ... 14 more times ...
    uint64_t x15 = array[i + 15];
}
```

(c) `ScanRead64PtrSimpleLoop`

```
for (uint64_t* p = array;
             p < array + n; ++p) {
    uint64_t x = *p;
}
```

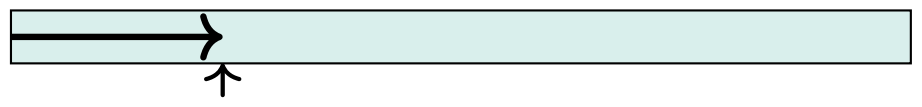

(d) `ScanRead64PtrUnrollLoop`

```
for (uint64_t* p = array;
         p < array + n; p += 16) {
    uint64_t x0 = *(p + 0);
    // ... 14 more times ...
    uint64_t x15 = *(p + 15);
}
```

(e) `PermRead64SimpleLoop`

```
uint64_t p = *array;
while((uint64_t*)p != array)
    p = *(uint64_t*)p;
```

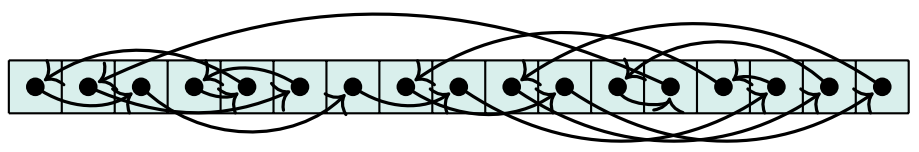

(f) `PermRead64UnrollLoop`

```
uint64_t p = *array;
while((uint64_t*)p != array) {
    p = *(uint64_t*)p;
    // ... 14 more times ...
    p = *(uint64_t*)p;
}
```

**Figure 3.2:** The pmbw memory access pattern loops expressed in C/C++ code.

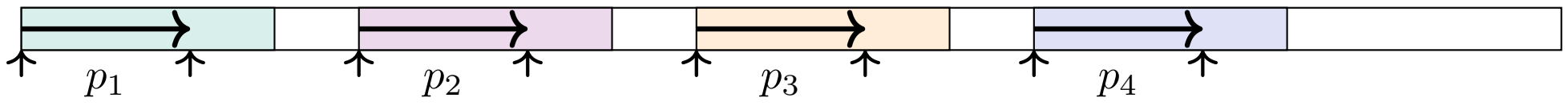

(a) Scanning an array in parallel.

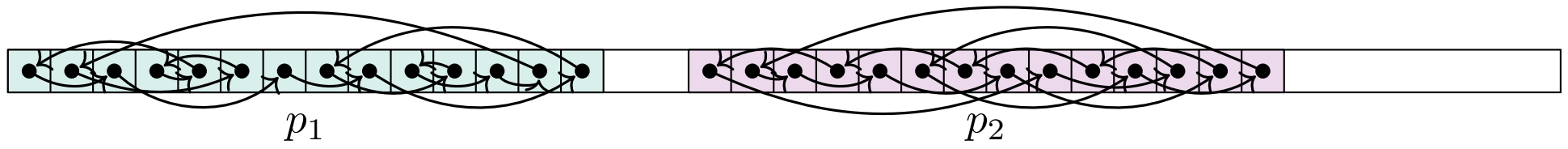

(b) Random access on a permutation in parallel.

**Figure 3.3:** Segmentation of an array when running pmbw's experiment loops *in parallel*. The spacing is required to avoid cache trashing.

It is available online as open-source software at `http://panthema.net/2013/pmbw/`, and we have collected a large number of experimental results on the website.

The goal of pmbw is to measure the single-thread and multi-thread performance properties of two memory access patterns: sequential *scan*-based (streaming) access and *random* memory access. These two patterns were chosen as extremes in terms of predictability: scan access can be accelerated by the hardware memory prefetcher and should reach the maximum memory bandwidth speed, while the random access pattern is impossible to predict and thus measures the full memory request round-trip time (latency). At the same time, these patterns resemble the extremes of the access patterns used in any program or basic algorithm, and especially in sorting and string sorting. Again: *any* program performs a mixture of these two memory access patterns.

Additionally, we are interested in the impact of the different ways one can program the access pattern loops on the resulting memory performance: Is array-index-based access slower than iteration with pointers? Is loop unrolling necessary? How much slower are 32-bit transfers compared to 256-bit?

Each experiment in pmbw is composed of a loop of a very basic memory operation (read or write) performed on a continuously allocated array of increasing size $n$. The experiment loops themselves were programmed *in assembly* to take all compiler optimizations *out of the equation*. Read/write operations to memory are performed using simple 32-, 64-, 128-, or 256-bit wide register-to-memory or memory-to-register `MOV` instructions. No extra calculations aside from the address registers and loop counters are performed.

The *scan* memory access pattern resembles unit-by-unit sequential reading or writing of a memory area. We only consider forward (increasing) address scanning. To benchmark *random* accesses we pre-generate a random *1-cycle permutation* of 64-bit pointers in an array using Sattolo's [Sat86] algorithm. The experiment loop then walks the 1-cycle, hence the location of each memory access is determined by the result of the previous one, which precludes any acceleration techniques.

Figure 3.2 shows the experiment loops of pmbw expressed in C/C++ code. The experiment names are concatenations of the following components:

(i) the array access pattern: `Scan` for sequential forward scanning, and `Perm` for walking a pre-generated random 1-cycle permutation,

(ii) the operation and register size used for memory transfers: `Read64` for reading 64-bit values, `Write64` for writing 64-bit values, `Read32`/`Write32` for 32-bit values, while `Read128`, `Write128`, `Read256`, and `Write256` work with 128-bit and 256-bit wide MMX/SSE/AVX registers when available,

(iii) the array indexing method: `Index` accesses the array via `array[i]`, whereas `Ptr` uses a pointer iterator such as `*p++`, and

(iv) the loop type: `SimpleLoop` for a single-step loop, and `UnrollLoop` for a loop containing 16 step manually unrolled in assembly.

For example, `ScanRead64PtrUnrollLoop` specifies an experiment loop for the *scan* pattern, reading 64-bit values using a pointer iterator in an unrolled loop. `ScanRead64PtrUnrollLoop` is also maybe the *most interesting* experiment loop as 64-bit is the default access size and unrolled pointer loops the most common way to write fast loops.

All experiment loops are run on arrays ranging from 1024 B to at most 1 TiB in size to measure the performance of all levels of the cache and RAM hierarchy. Our pmbw starts by determining the amount of physical memory available, then allocating the largest available power of two, and filling it with data to force the virtual memory system to actually assign physical memory.

The allocated memory area is then split among $p$ threads which run the experiment loops on independent segments (see figure 3.3). The *sum* over all *bytes* in the test working set is labeled as the area size $n$, hence each thread works on $\frac{n}{p}$ bytes. We selected a list of 63 *target* area sizes. This list contains common L1, L2, and L3 cache sizes and nearby values to distinguish the levels of the hierarchy.

Given a target area size, the available size is divided by $p$ and the result is rounded down to the largest array accepted by the experiment loop such that all threads operate on a segment of $\frac{n}{p}$ bytes. For example, for an experiment loop running on 32 threads, which unrolls 16 accesses to 64-bit of data, the smallest $n$ possible is 4 KiB such that $\frac{n}{p} = 16 \cdot 8$. The independent segments are spaced out in the available memory area with a *non-regular* amount of bytes between the segments. In our experiments we noticed bad performance when using a *regular* spacing (e.g. by 1 MiB). This is probably due to the way address bits are hashed by an unknown hardware hash function [CKD+10] in the cache coherence tables, which leads to a large number of collisions in the cache directories. Using a non-regular spacing (4 MiB plus 16 KiB) avoids this phenomenon.

As many of the experiment loops run for a very short time for small array sizes, pmbw performs a number of repetitions of the loop. This number is adaptively set such that each experiment instance runs for an expected 1.5 seconds (calculated from previous experiments), and always at least 1.0 second (otherwise it is repeated). The resulting bandwidth and latency are then calculated by dividing by the number of repetitions. This adaptive method is required because the range of experiment durations in pmbw is quite large: while experiments with small arrays which fit into L1 cache require a very high repetition count, walking a 128 GiB permutation with a single thread can take hours.

All threads are pinned by pmbw to a specific processor number. For NUMA machines, the memory area is allocated as separate arrays on all NUMA nodes, and the threads are allocated round-robin to the NUMA nodes. All pinned threads operate exclusively on the local NUMA memory. No changes were done to the standard Linux scheduling settings.

### 3.1.1 Single-Thread Memory Performance

We are first going to consider *single-thread* performance of various experiment loops. Figures 3.4 to 3.9 show the single-thread memory *bandwidth* of the loops on the top in GiB/s and the memory *latency* in nanoseconds per access on the bottom. Top and bottom plots show different metrics from the same experiment runs.

Focus first on the results from A.Intel-1×8 in figure 3.4, which is an older single-socket desktop-style machine. The top plot clearly shows four levels of memory hierarchy: the highest bandwidth of up to 44 GiB/s was attained with array size $n$ from $2^{10}$ to $2^{15}$, which clearly corresponds to the 32 KiB of L1 data cache in the processor (see table 2.1, page 40). While the transition at $2^{15}$ from L1 to L2 cache is sharp, the next drop in bandwidth around $2^{18}$ is more gradual (L2 cache is 256 KiB). The reasons behind this are the various optimizations of cache performance [HP12]: the hardware tries to predict access patterns and preload memory it expects will be needed. Notice that bandwidth degrades even for $2^{17.5}$ due to prefetching of unneeded cache lines which evict needed lines from the working set. The same gradual performance drop occurs around $2^{23}$ due to the L3 cache size of 8 MiB.

On A.Intel-1×8, 128-bit *scan* read operations deliver the highest bandwidth: 44 GiB/s from L1 cache, around 30 GiB/s from L2 cache, 21 GiB/s from L3 cache, and 13 GiB/s from RAM. Notice that 64-bit operations deliver half of the bandwidth from L1 cache, and 32-bit again half of 64-bit operations from L1 cache. This is because any access to L1 cache costs the same 0.33 nanoseconds per access regardless of the datatype, as one can see in the bottom plot of figure 3.4. These ratios do not hold true for the L2 cache, L3 cache, and RAM: larger operations take longer per access, but also deliver higher throughput per byte.

Considering single-step simple loops and unrolled loops: unrolling loops is clearly necessary to get the full performance from the fast caches. `ScanRead64Ptr`*`Simple`*`Loop` delivers only 11 GiB/s from L1, L2, and L3 cache, and around 9 GiB/s from RAM, while the unrolled variant `ScanRead64Ptr`*`Unroll`*`Loop` achieves 22 GiB/s from L1 cache, about 19 GiB/s from L2 cache, 17 GiB/s from L3 cache, and 11 GiB/s from RAM.

On the topic of different loop implementations: surprisingly, `ScanRead64`*`Index`*`Unroll-Loop` and `ScanRead64`*`Ptr`*`UnrollLoop` achieve almost identical performance, despite the longer and more complex assembly instructions. Apparently, modern processors can perform the additional offset calculations needed in the *Index* variant at no extra cost in the same memory operation.

While we focused mostly on *Read* operations, the plots also contain two *Write* operations. Writes to arrays in L1 cache are just as fast as reads. Beyond L1 cache, however, write operations are consistently slower than reads, which is likely caused by the additional overhead of guaranteeing cache coherence. Interestingly, 128-bit writes are faster than 64-bit writes in L2 cache, but converge to exactly the same bandwidth for L3 cache and RAM. This phenomenon is probably due to *write combining*, where multiple write operations in a cache line are combined prior to shipping the data out of the processor chip.

The results from E.Intel-2×16 are shown in figure 3.8. These are similar to those from A.Intel-1×8, even though the processors' release dates are almost a decade apart. The newer processor additionally supports 256 bit wide registers and memory transfers. As in A.Intel-1×8 the L1 cache bandwidth doubles from 128-bit operations, because each operation has a fixed latency of only 0.24 nanoseconds. As before, wider data transfers achieve higher bandwidth from L2 and L3 cache, though when reading from RAM, 256-bit, 128-bit, and 64-bit operations all deliver the same bandwidth of 10 GiB/s. 32-bit operations achieve slightly lower performance. The same occurs with write operations, though these reach only about 7 GiB/s.

The other single-thread memory bandwidth and latency results in figures 3.4 to 3.9 have the same overall shape. The interesting discussion of comparing the *ratio* of cache and RAM access on the different platform is postponed to the next subsection.

Figures 3.10 and 3.11 present all results of the 64-bit unrolled scan benchmark and the permutation walking benchmark across all platforms. In the top plots with logarithmic time scale one can see the levels of memory hierarchy on the whole, while the bottom plots show bandwidth and access latency of small and large arrays on separate linear scales, split at 256 KiB. In figure 3.10 one can see that newer machines generally have a higher bandwidth: the improvement from the oldest machine, B.AMD-4×4, to the newest, F.AMD-1×16, is about a factor for 2.3 for L1 cache, and even 5.7 for RAM bandwidth. Figure 3.11 shows the results of the permutation walking benchmark which exhibits random access memory latency without any acceleration across all experimental platforms. Random L1 cache access on all machines takes 1–2 nanoseconds, while random L2 cache access 1.5–19 nanoseconds. Maybe the most important number is the access time to RAM for random cache misses: depending on the machine, this time is between 70 and 150 nanoseconds for moderately sized arrays. On E.Intel-2×16 the access time increases up to 260 nanoseconds for $n = 2^{37}$.

Notice that the access time increases steadily even for $n \geq 2^{30} = 1\,\text{GiB}$, which is definitely beyond L3 cache size. This plainly contradicts the RAM model's assumptions that memory access time is constant, independent of the input size $n$. The explanation for this phenomenon lies with virtual memory mapping, which is defined by the page tables. These page tables are a tree-like data structure which map virtual to physical addresses. Larger arrays require more space in the memory page tables, and modern machines have three or four level of hierarchy in the page tables due to the large amounts of RAM. For small arrays, all page table entries are in one of the high-level caches. However, in the extreme, a single random access to RAM can cause a cache fault on every level of the page table hierarchy. The larger the random permutation array in our experiment is, the higher the probability that a random access causes an additional cache fault on one of the page tables. This is the reason for the gradual steady rise in access time for large arrays and also the sharper ascent on E.Intel-2×16.

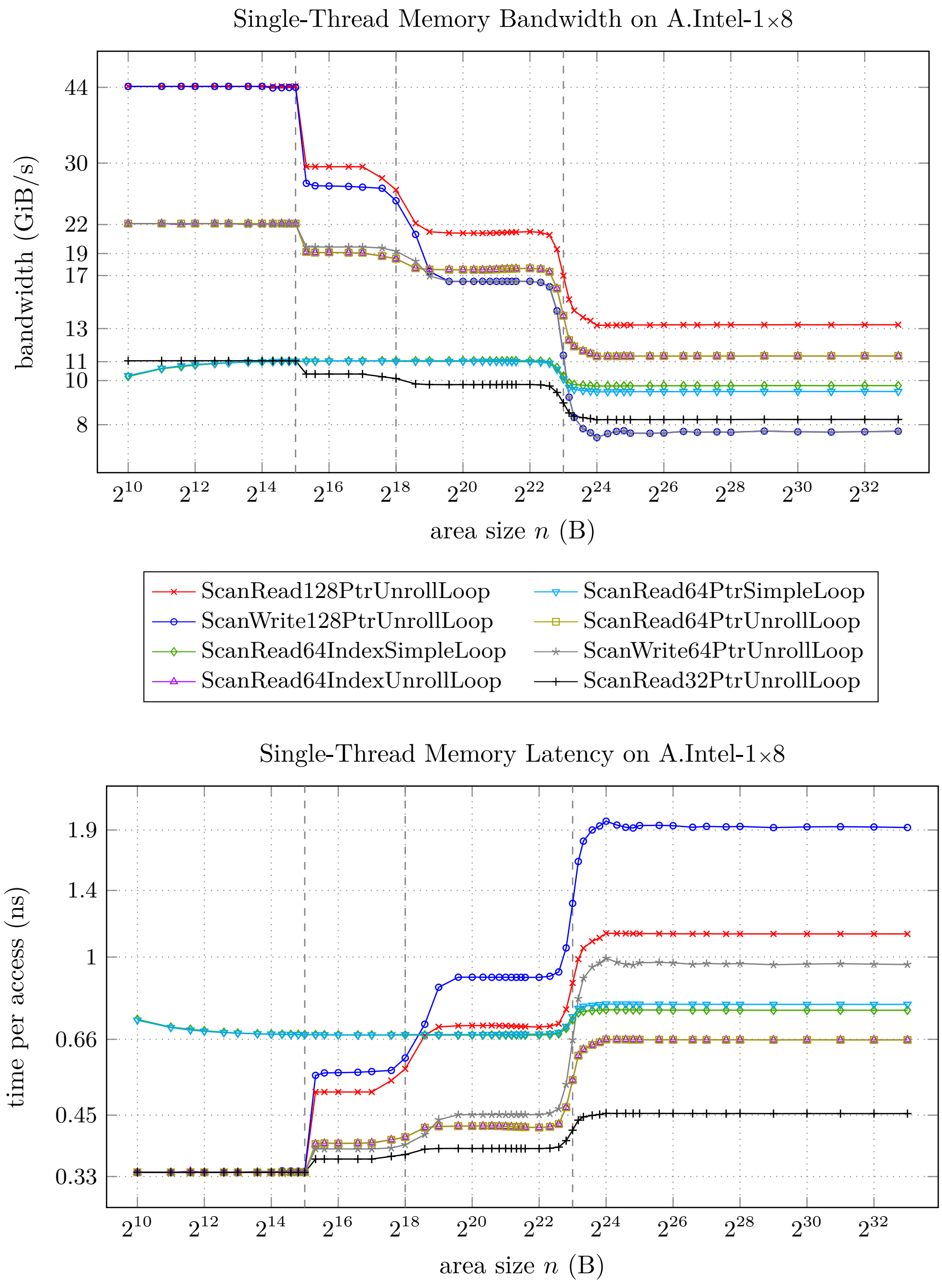

**Figure 3.4:** Single-thread scan memory bandwidth and latency on A.Intel-1×8.

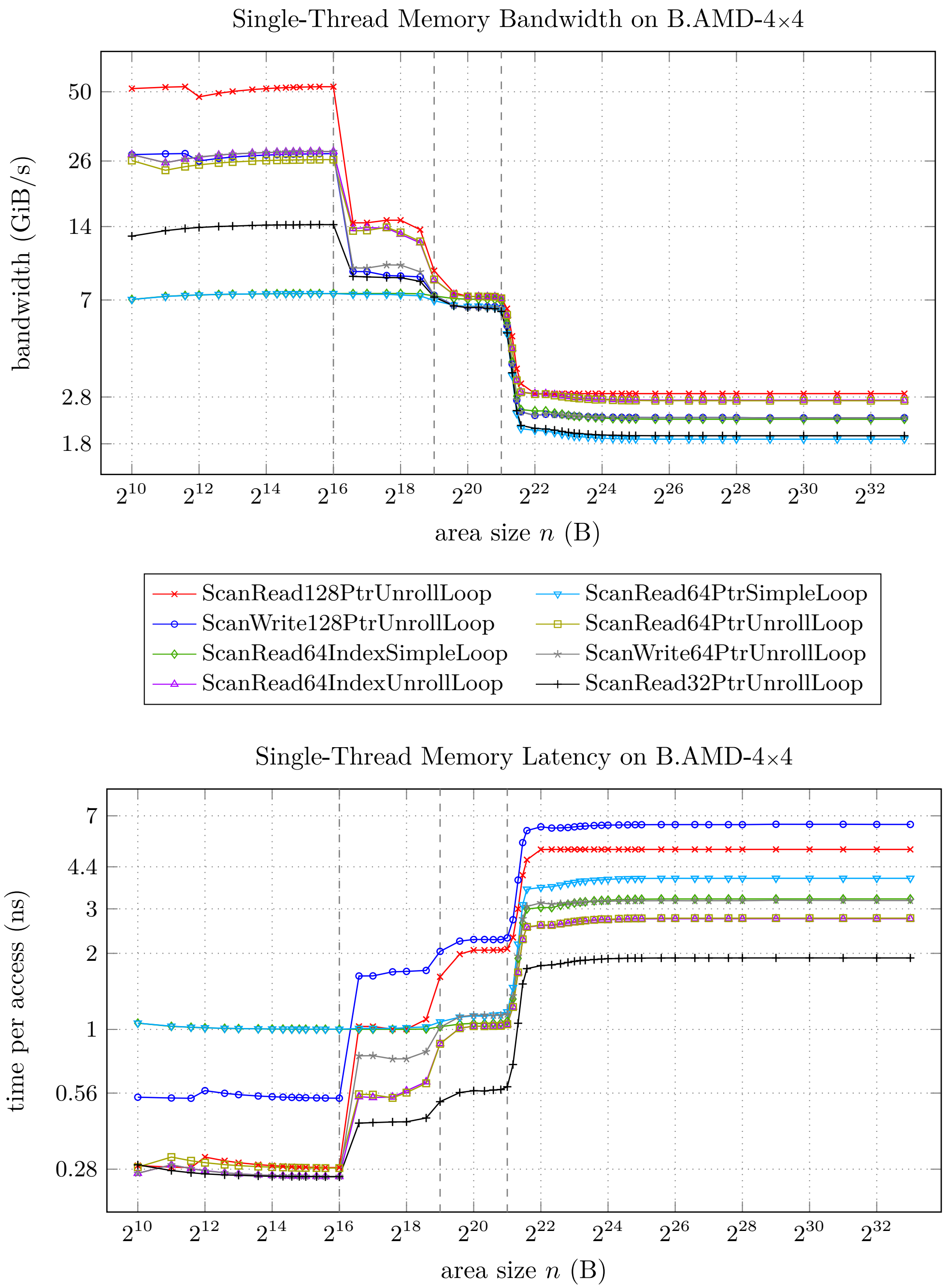

**Figure 3.5:** Single-thread scan memory bandwidth and latency on B.AMD-4×4.

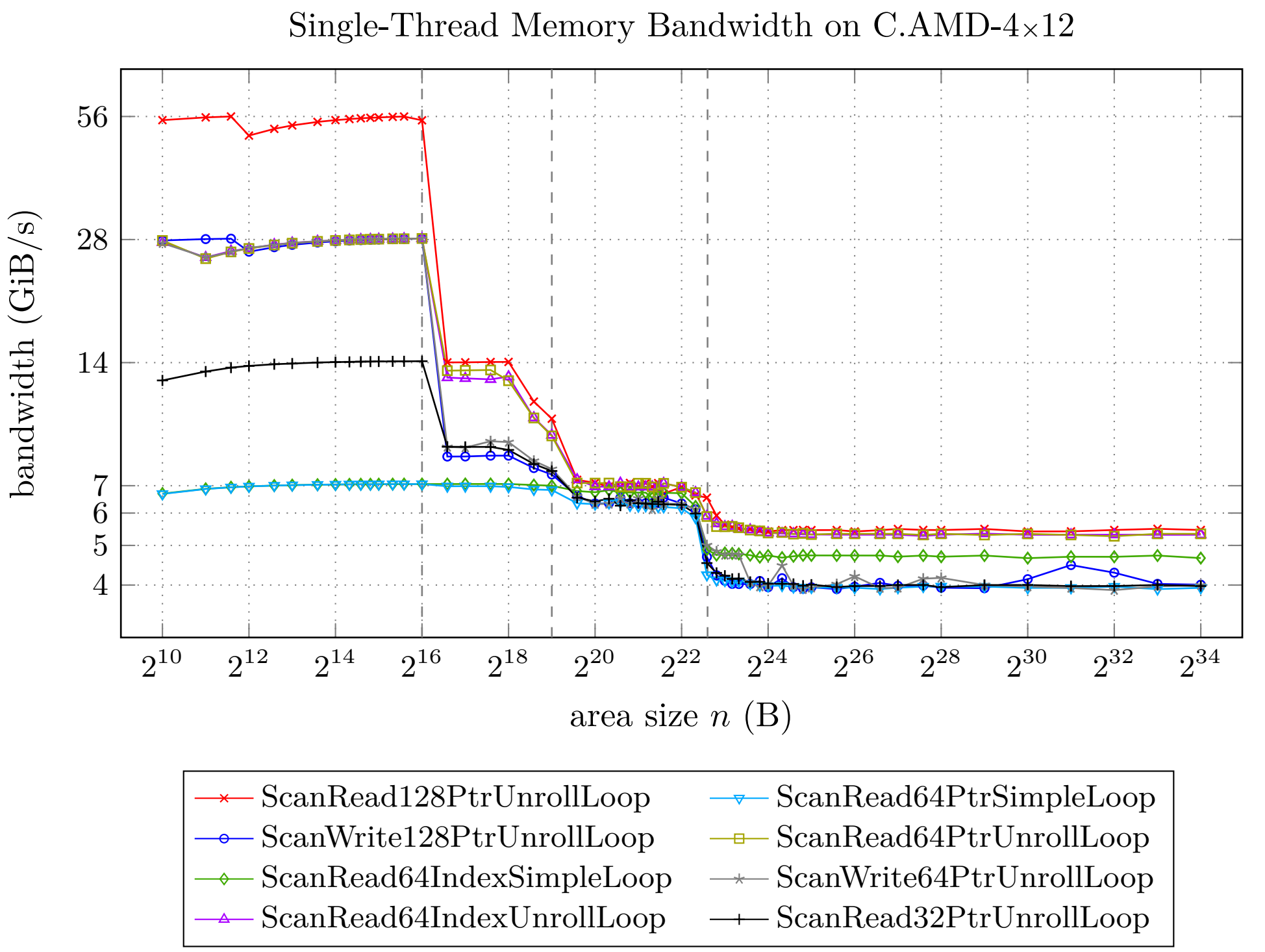

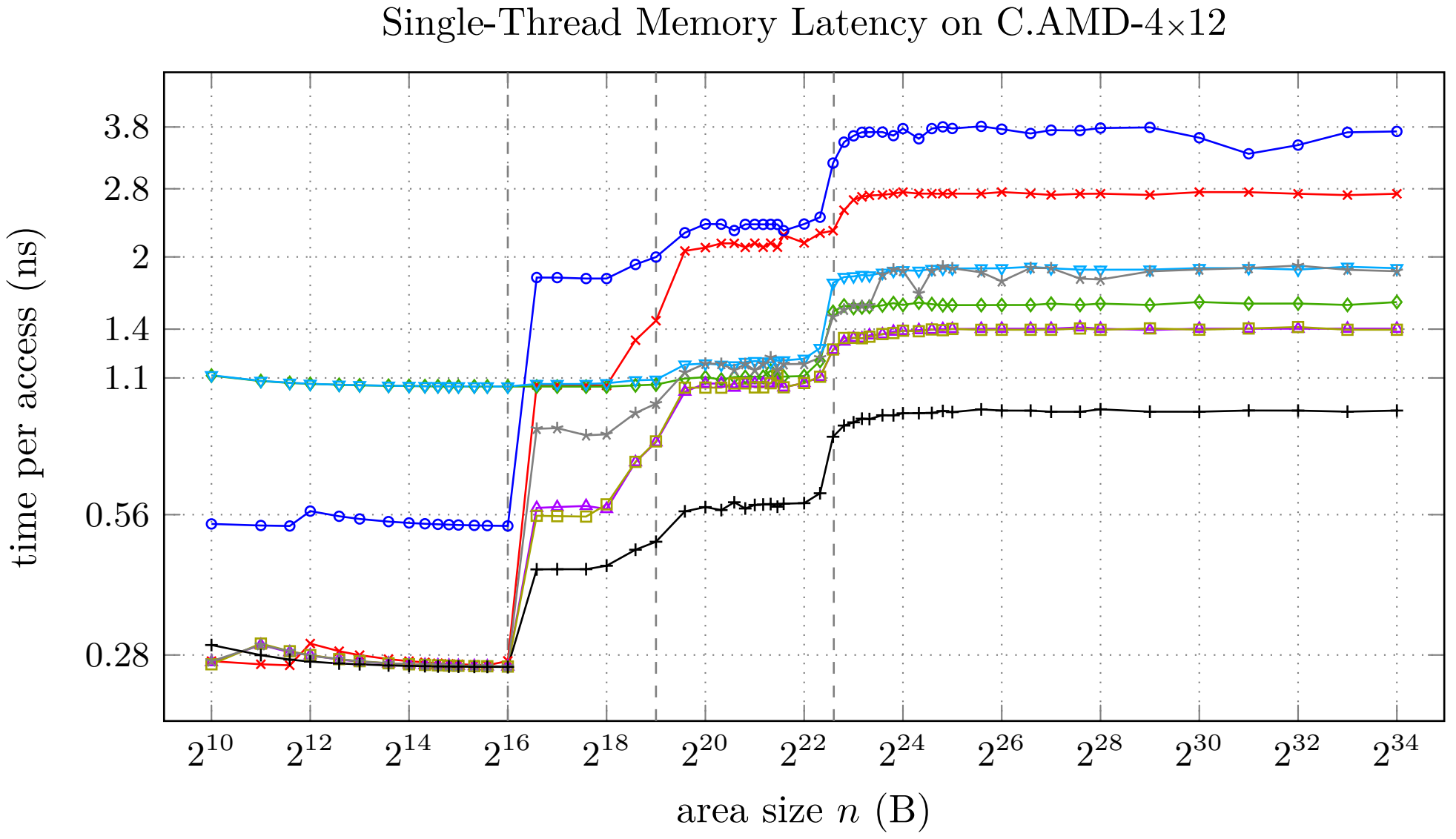

**Figure 3.6:** Single-thread scan memory bandwidth and latency on C.AMD-4×12.

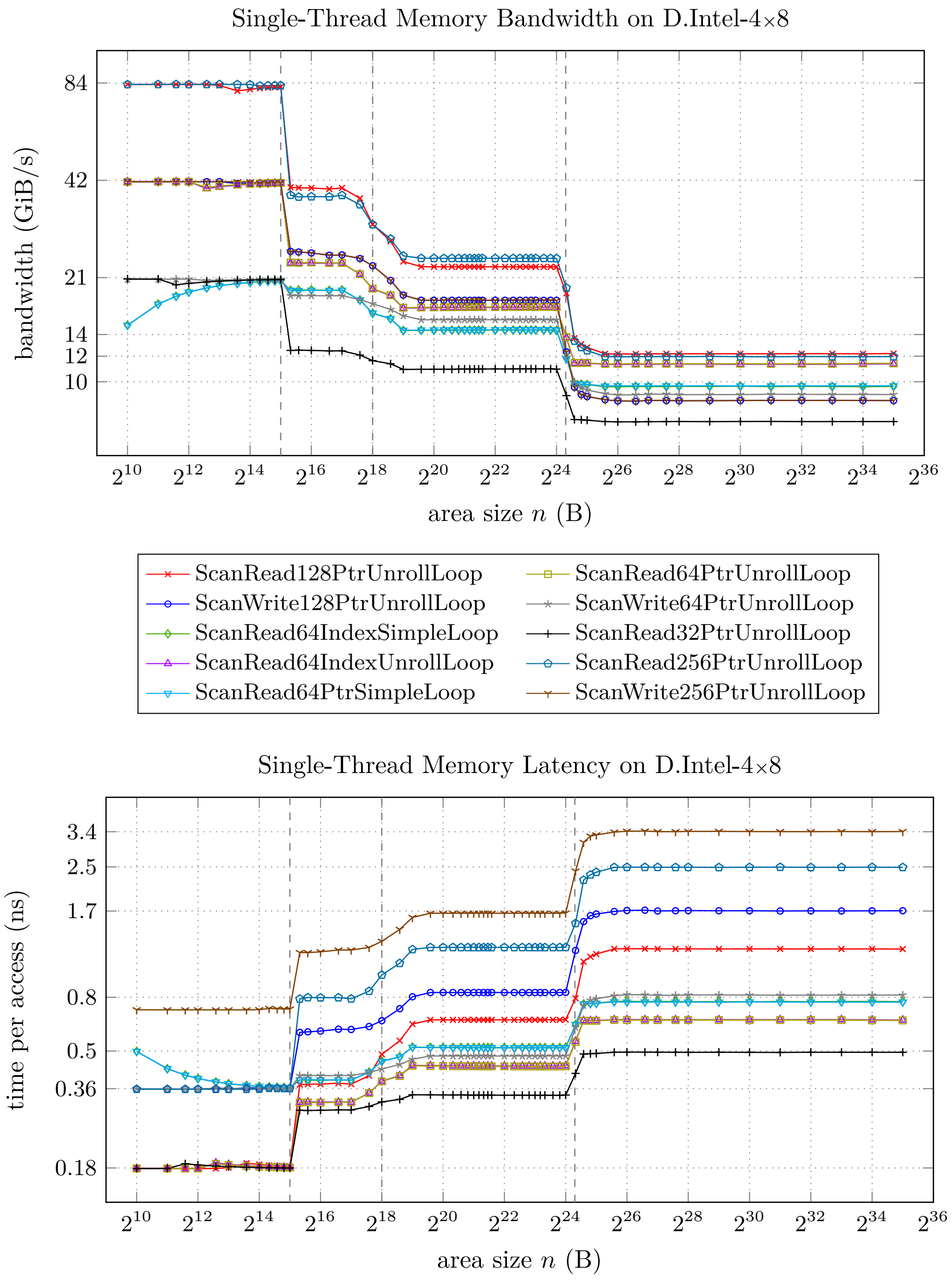

**Figure 3.7:** Single-thread scan memory bandwidth and latency on D.Intel-4×8.

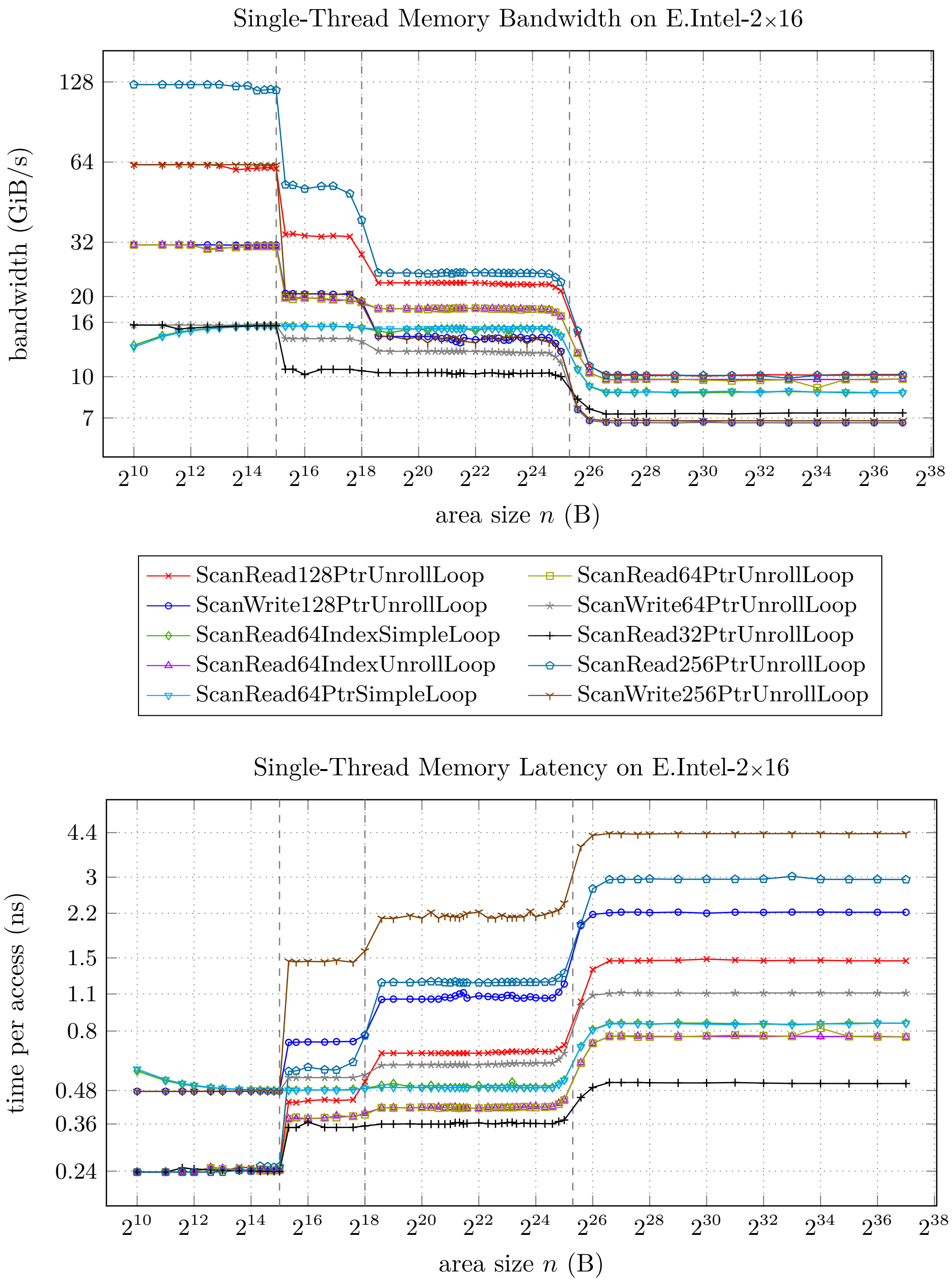

**Figure 3.8:** Single-thread scan memory bandwidth and latency on E.Intel-2×16.

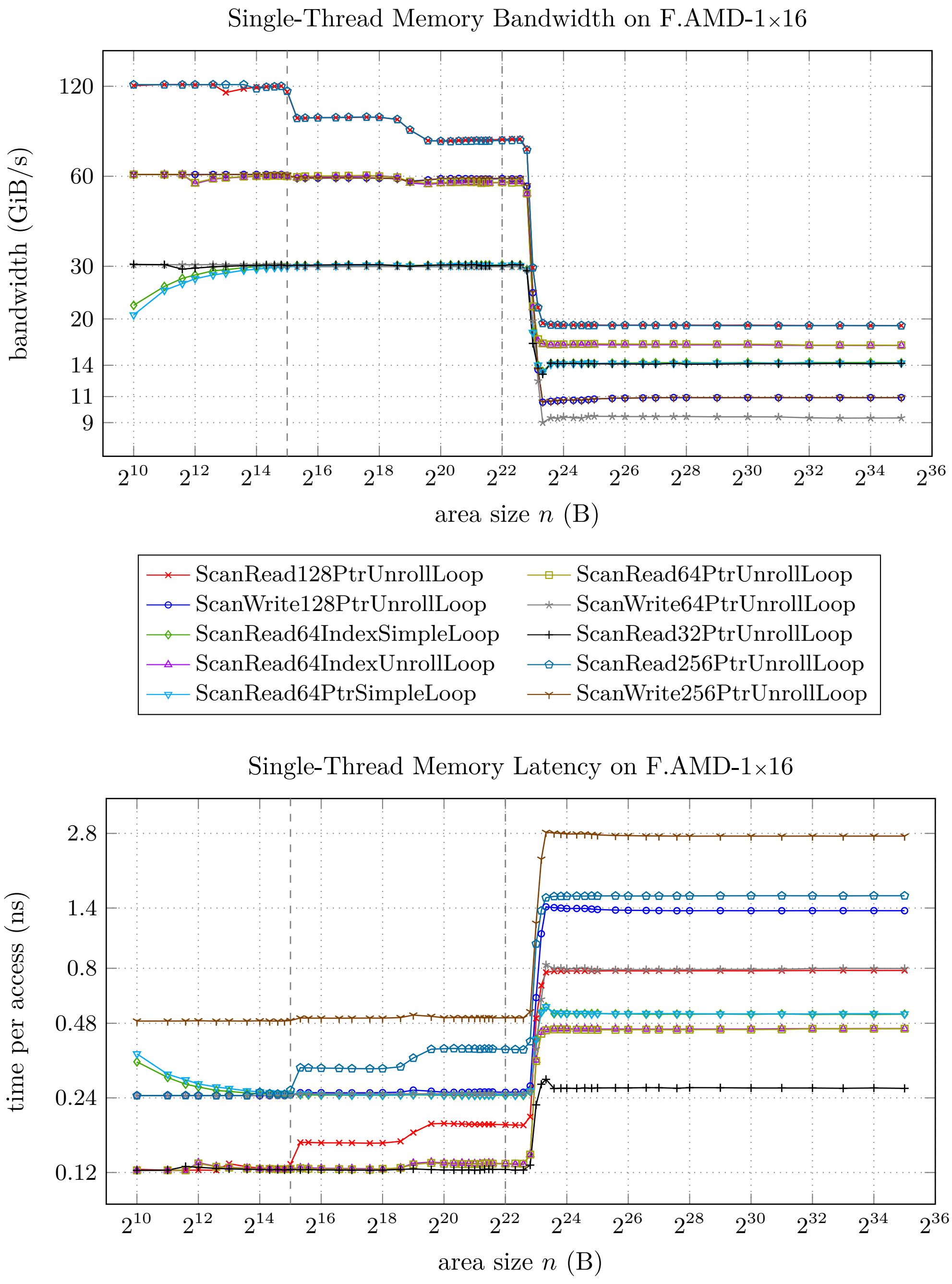

**Figure 3.9:** Single-thread scan memory bandwidth and latency on F.AMD-1×16.

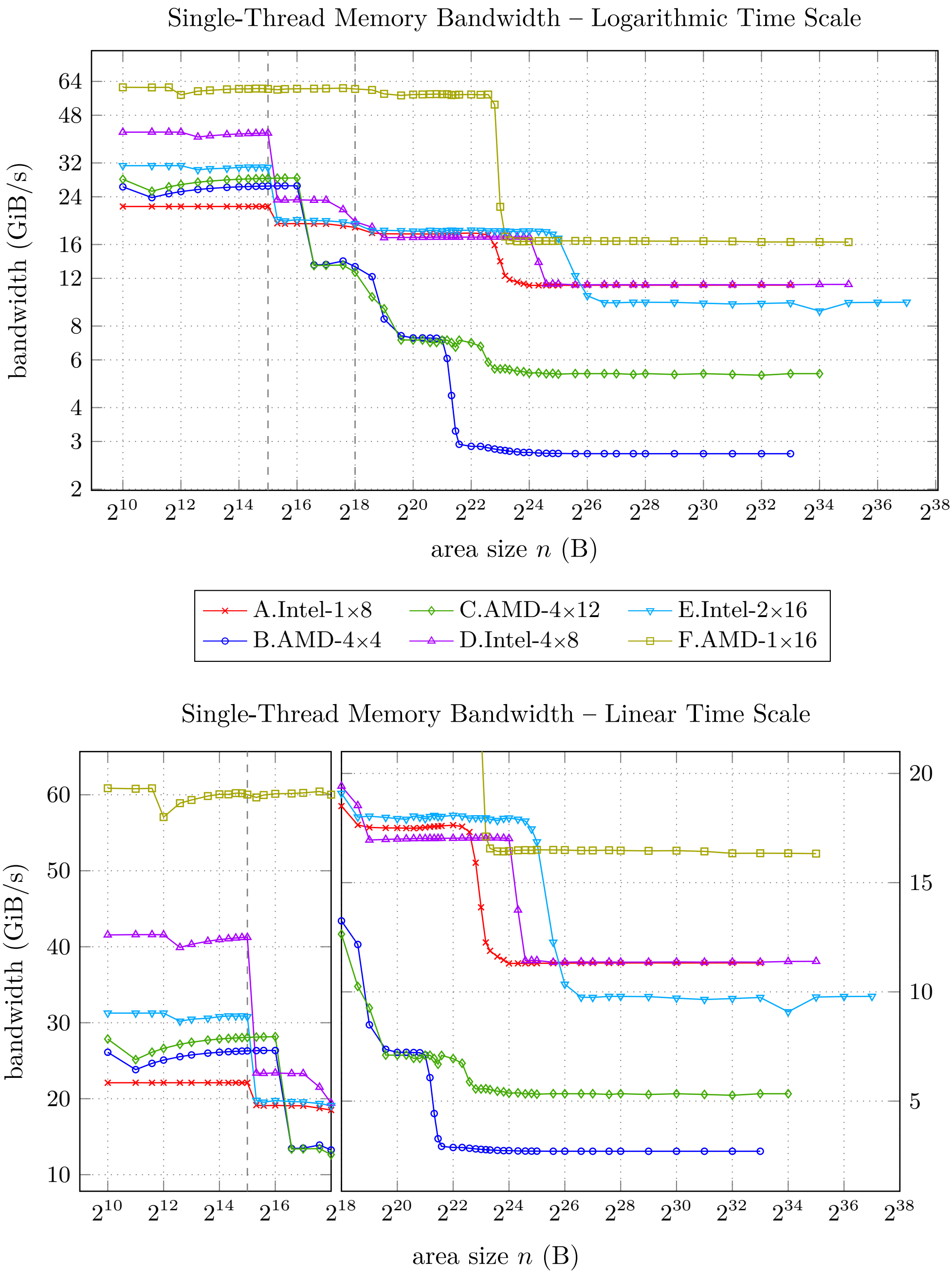

**Figure 3.10:** Single-thread scan bandwidth experiment ScanRead64PtrUnroll-Loop across all machines.

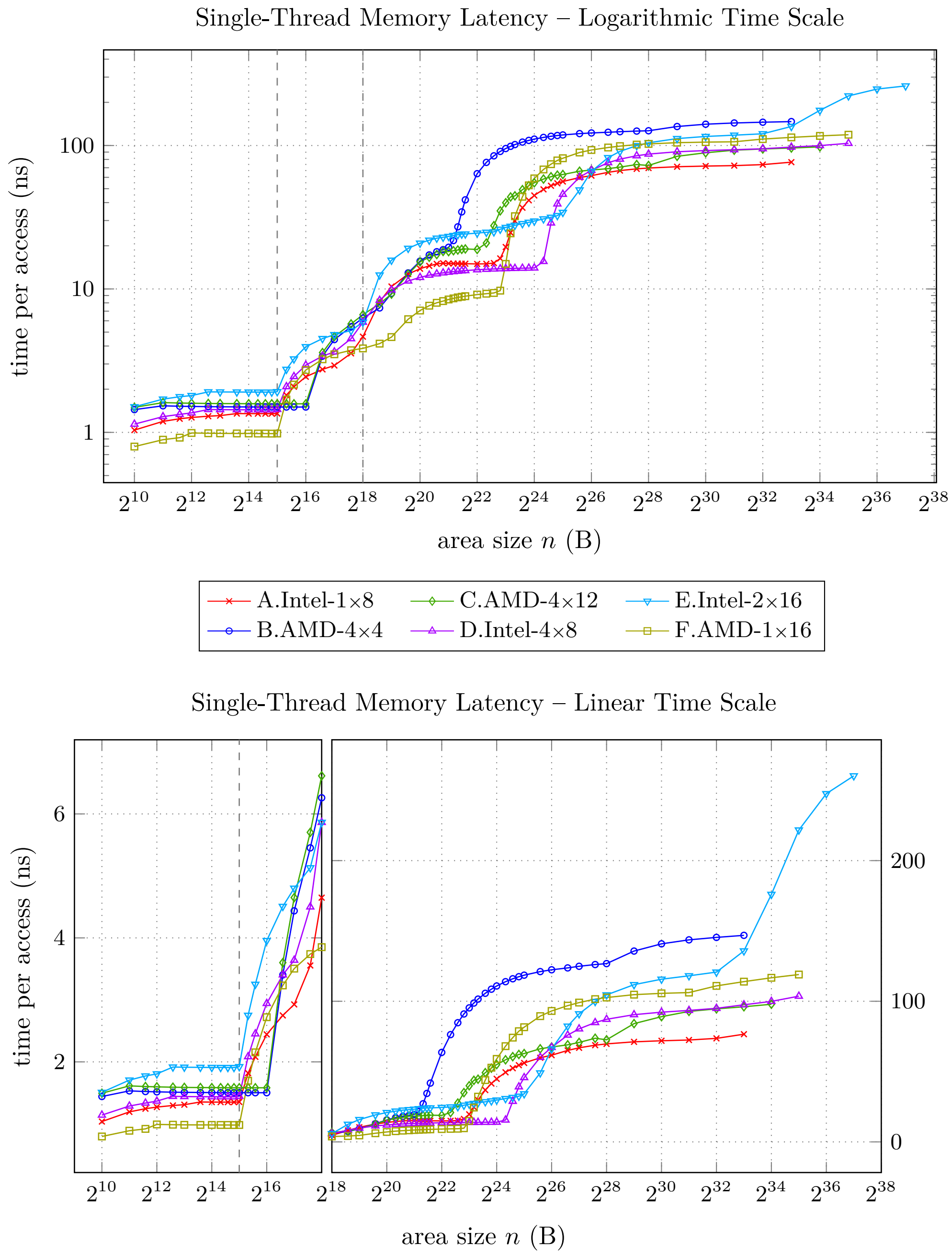

**Figure 3.11:** Single-thread random cyclic permutation walking experiment PermRead64UnrollLoop across all machines.

### 3.1.2 Multi-Thread Memory Performance

We now turn to multi-threaded parallel memory performance. Figures 3.12 to 3.14 show the 64-bit parallel scan read *bandwidth* performance while figures 3.15 to 3.17 show the 64-bit parallel memory *latency* as measured with our random permutation walk.

In the experiments we increase the number of processors, usually doubling in each step, until the number of physical cores is reached. For machines with a core count that is a not a power of two, we try to align the steps with the number's factors. On some experiments, we then perform additional runs with more threads than physical processors to measure the overcommitment penalty.

In principle, we would expect the memory bandwidth shown in figures 3.12 to 3.14 to double for each step in which the number of threads is doubled. On E.Intel-2×16, the top panel of figure 3.13, we can see exactly this doubling for the bandwidth from L1 and L2 cache: for $p$ threads with $1 \leq p \leq 32$ the system achieves approximately $32p$ GiB/s from L1 cache, $20p$ GiB/s from L2 cache, and $18p$ GiB/s from L3 cache. Simultaneous multithreading (SMT) (Intel's Hyperthreading) with $p = 64$ achieves the same bandwidth as $p = 32$ from L1 cache, however, for L2 cache the performance increases from 500 GiB/s to about 780 GiB/s with SMT.

But due to limited bandwidth of the memory controller and memory channels, the linear scaling of bandwidth with the number of processors does not continue into RAM. This is commonly referred to as the *memory wall* [WM95]: on E.Intel-2×16 one thread achieves 10 GiB/s from RAM, two threads about 18 GiB/s, four threads 32 GiB/s, eight threads 62 GiB/s, sixteen threads 110 GiB/s, and thirty-two threads 120 GiB/s. Hence, one thread can read 10 GiB/s from RAM, while thirty-two can read only 3.75 GiB/s *per thread* from RAM. Concerning the ratios of how well bandwidth to RAM scales on our various experiment platforms, E.Intel-2×16 is actually the best multi-core machine. On F.AMD-1×16, one thread can read 16.5 GiB/s from RAM, while sixteen threads achieve only 1.69 GiB/s from RAM per thread.

Table 3.1 on page 75 shows a summary of the sequential and parallel bandwidth measurements across our platforms. The L1 cache columns shows the highest bandwidth with one thread and all threads for $n \leq 2^{14}$, and the RAM columns the highest bandwidth for $n \geq 2^{30}$. We are most interested in the ratio columns, which shows the ratio of cache bandwidth over RAM bandwidth in two scenarios: first for a single threaded program ($c_1/r_1$), and second for a multi-threaded program using all cores of the machine ($c_p/r_p$). In a sense, these ratios determine *how many access to cache* are allowed *per access to RAM* in an algorithm which wishes to effectively utilize the cache.

While, by this reasoning, a multi-threaded parallel algorithm should perform 4.95 accesses to cache for one access to RAM to saturate bandwidth on A.Intel-1×8, on B.AMD-4×4 the ratio rises to 25.0 and on F.AMD-1×16 to 16.3. This is simply due to the limited bandwidth to RAM over cache.

**Table 3.1:** L1 cache single-thread bandwidth $c_1$, L1 cache multi-thread bandwidth $c_p$ with all cores, RAM single-thread bandwidth $r_1$, and RAM multi-thread bandwidth $r_p$ with all cores, and ratios between single- and multi-thread performance.

| | L1 Cache | | RAM | | Ratios | | |
|---|---|---|---|---|---|---|---|
| Host | $c_1$ (GiB/s) | $c_p$ (GiB/s) | $r_1$ (GiB/s) | $r_p$ (GiB/s) | $c_1/r_1$ | $c_p/r_p$ | $\frac{c_p/r_p}{c_1/r_1}$ |
| A.Intel-1×8 | 22.1 | 84 | 11.3 | 17.1 | 2.0 | 4.9 | 2.5 |
| B.AMD-4×4 | 26.1 | 421 | 2.7 | 16.8 | 9.7 | 25.0 | 2.6 |
| C.AMD-4×12 | 27.9 | 1 347 | 5.3 | 91.3 | 5.2 | 14.7 | 2.8 |
| D.Intel-4×8 | 41.6 | 1 182 | 11.4 | 110.0 | 3.6 | 10.7 | 2.9 |
| E.Intel-2×16 | 31.3 | 1 000 | 9.8 | 116.9 | 3.2 | 8.5 | 2.7 |
| F.AMD-1×16 | 60.9 | 439 | 16.5 | 27.0 | 3.7 | 16.3 | 4.4 |

The takeaway is that bandwidth-saturating parallel algorithms have to be *even more* cache-efficient than sequential algorithms to scale well on multi-core machines. Depending on the machine the factor by which they have to be more cache efficient ranges from 2.53 to 4.40.

When comparing parallel bandwidth across machines, as shown in figure 3.18 (top panel), we can see that processors' bandwidth to cache has continuously improved, while RAM bandwidth has grown at a lower pace and stagnated at around 100 GiB/s in the last few processor generations.

Contrary to bandwidth, walking a random permutation *scales nearly perfectly* with the number of threads used. Figures 3.15 to 3.17 show the rate per access with increasing number of threads. One can see that the rates basically do not change with more threads. Instead, the plot series shift to the right due to more fast cache being available, as more threads are allocated onto the cores. The reason why walking permutation scales is simply because *waiting scales*. Each thread is mostly waiting for the memory system to return the next value. While walking a permutation scales perfectly due to the waiting, *batches* of parallel random accesses by a single core, e.g. as performed in some string sorters, may still incur bottlenecks in the memory system.

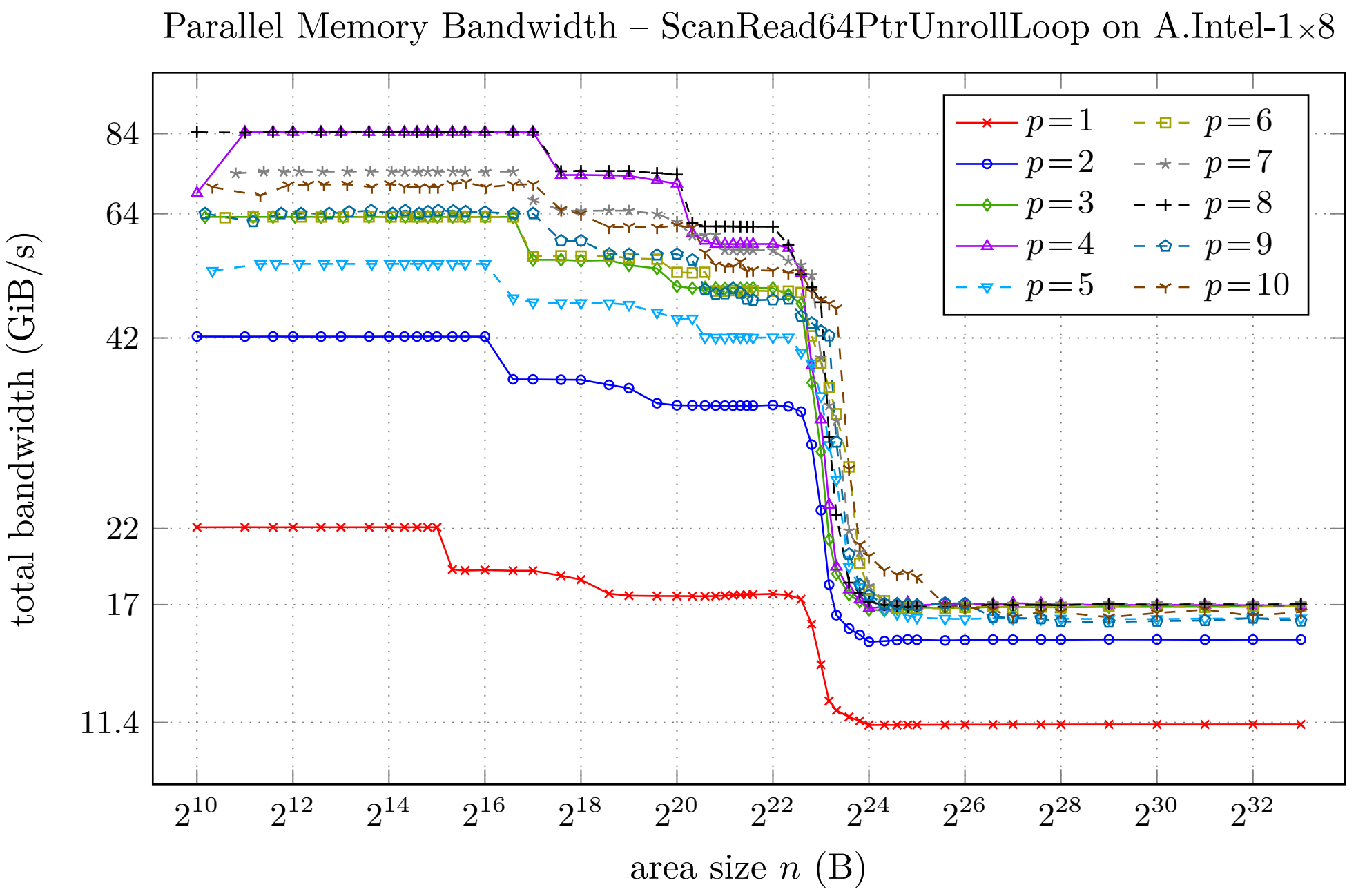

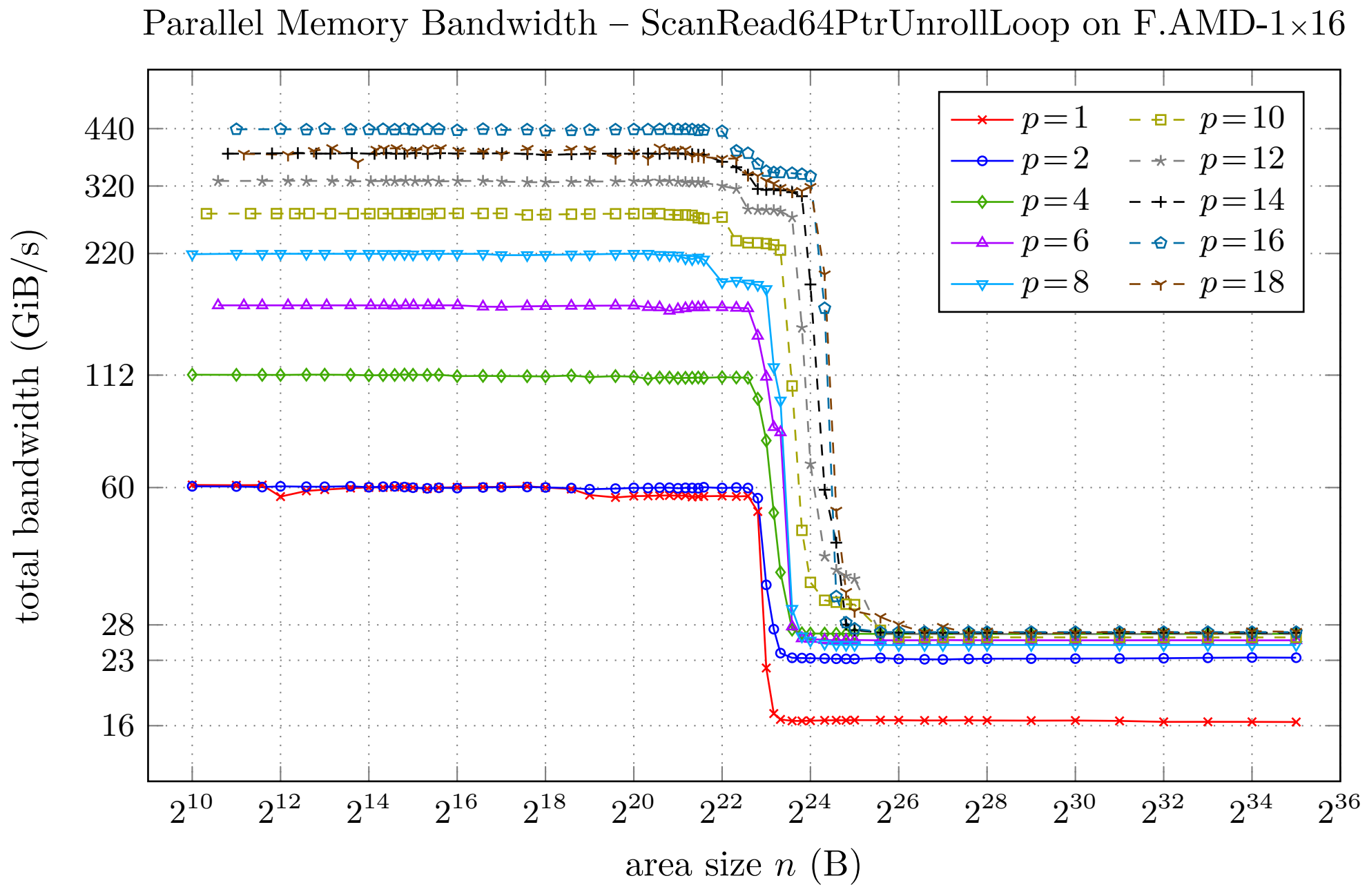

**Figure 3.12:** 64-bit parallel memory bandwidth on A.Intel-1×8 and F.AMD-1×16.

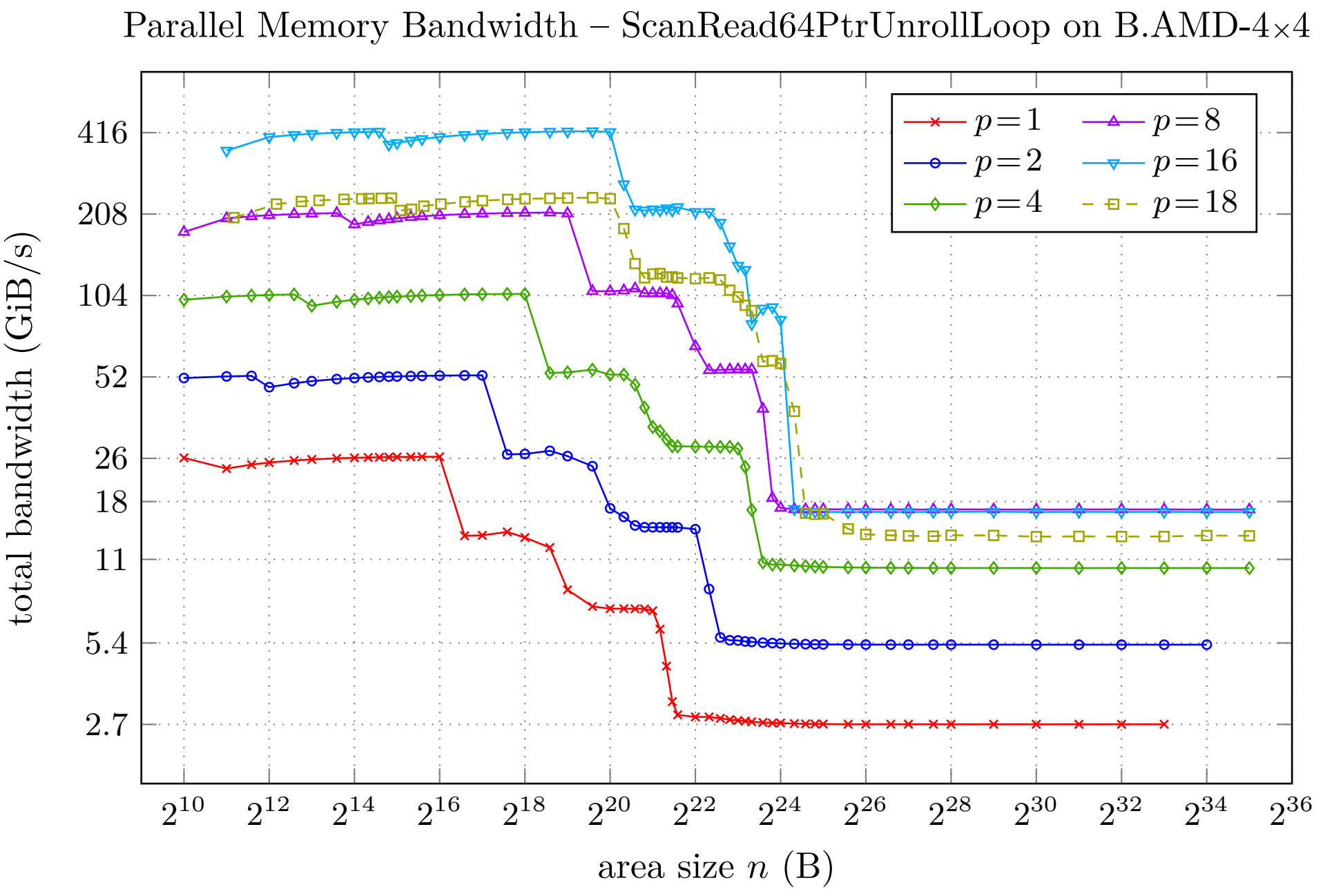

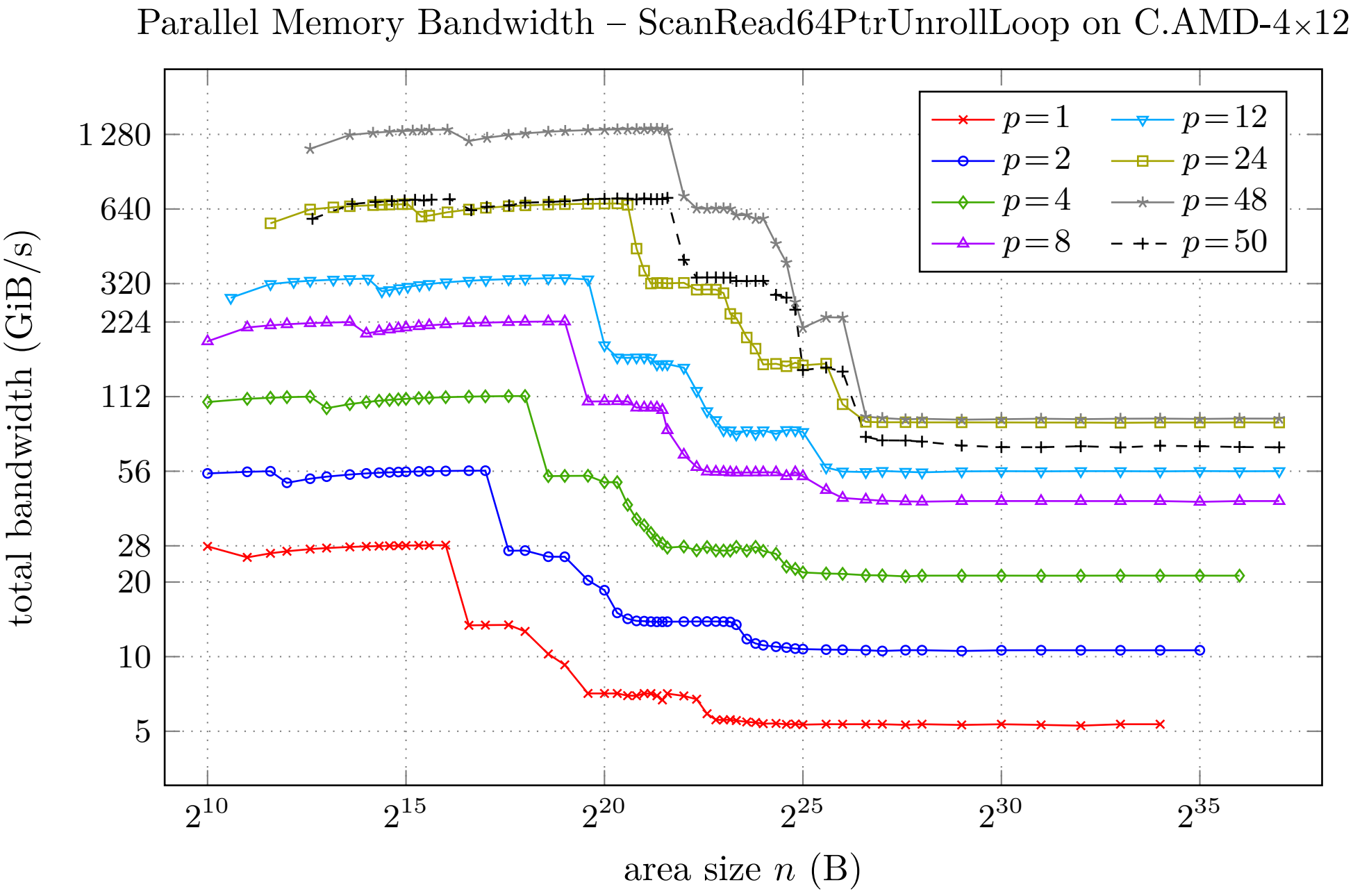

**Figure 3.13:** 64-bit parallel memory bandwidth on B.AMD-4×4 and C.AMD-4×12.

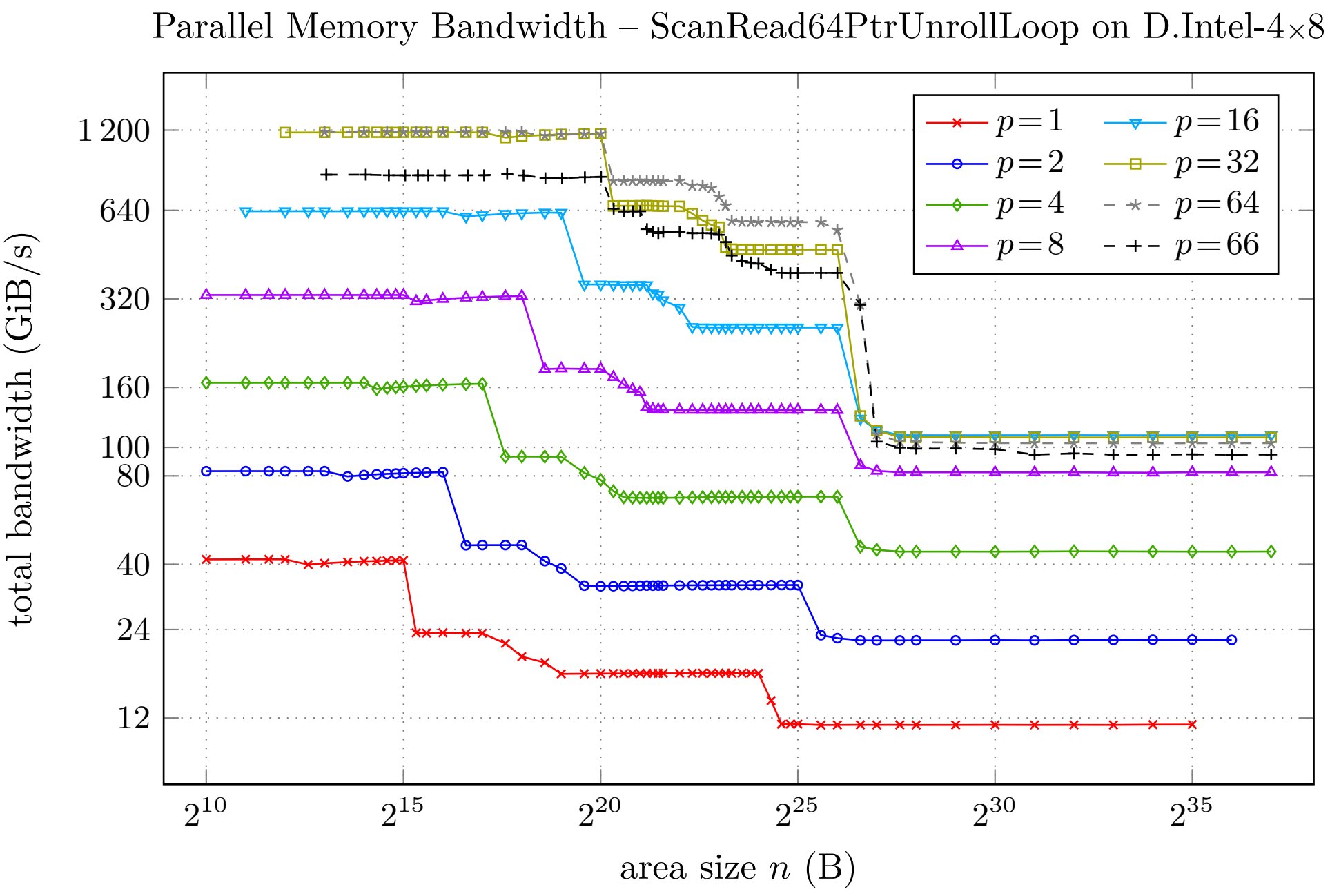

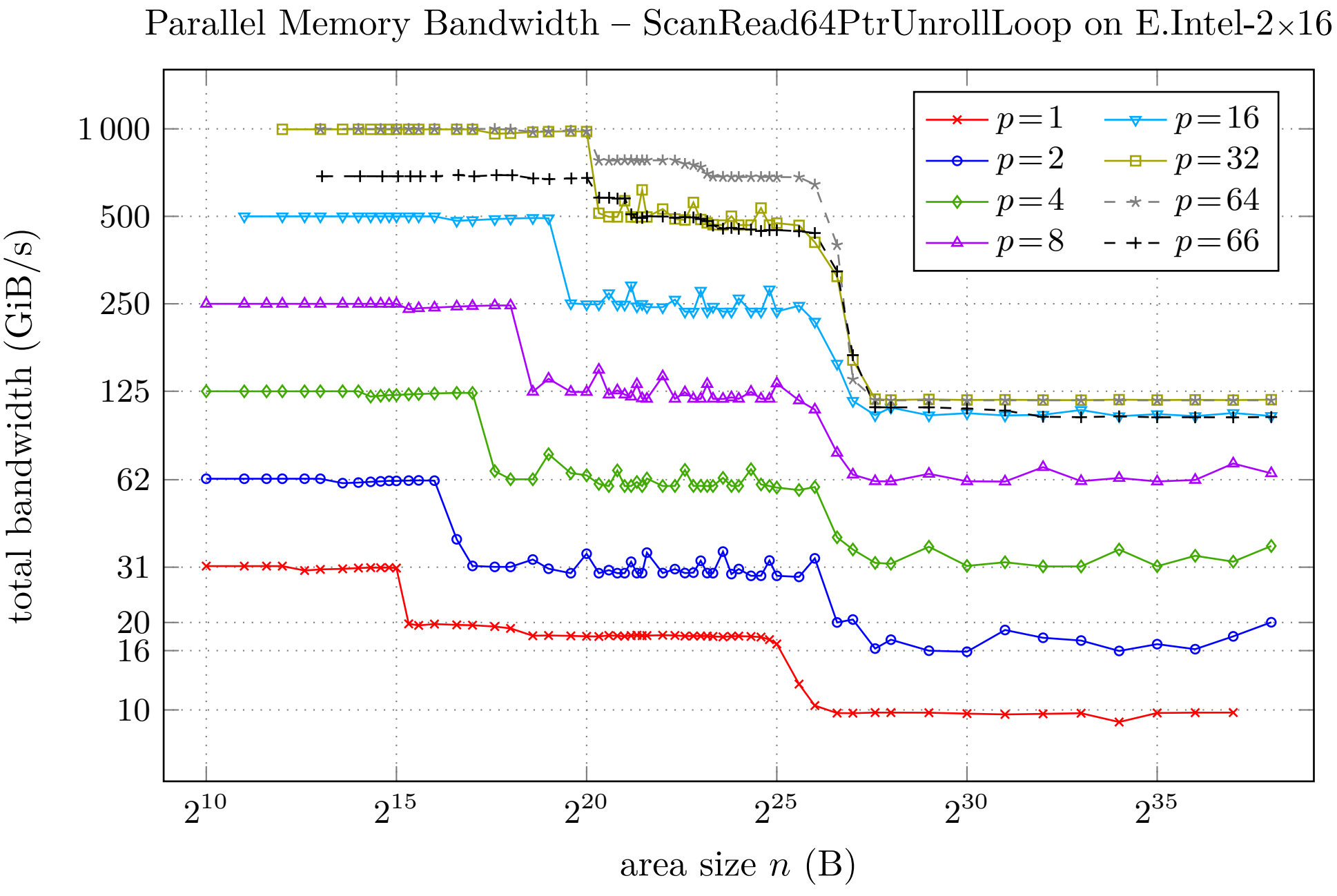

**Figure 3.14:** 64-bit parallel memory bandwidth on D.Intel-4×8 and E.Intel-2×16.

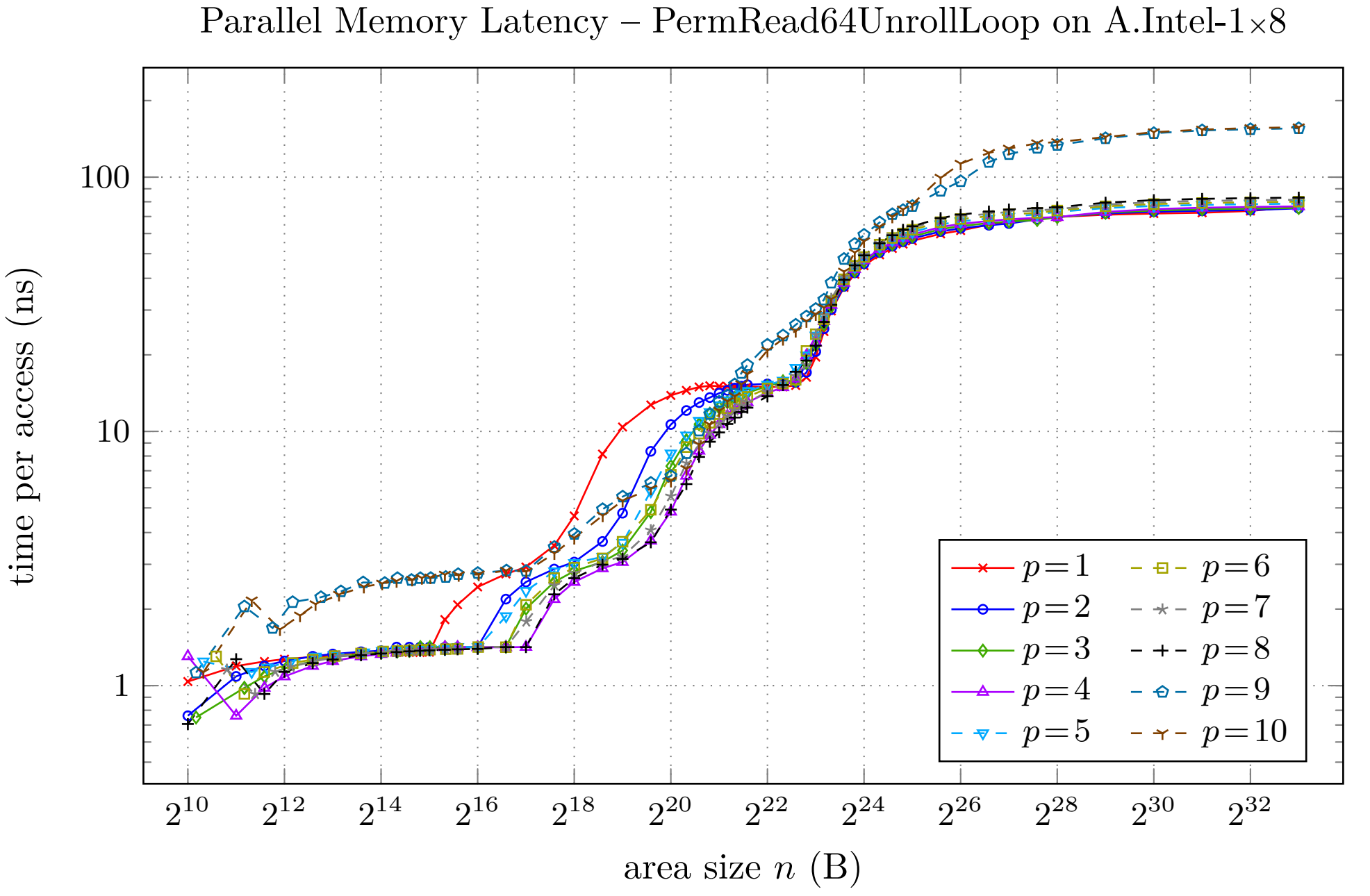

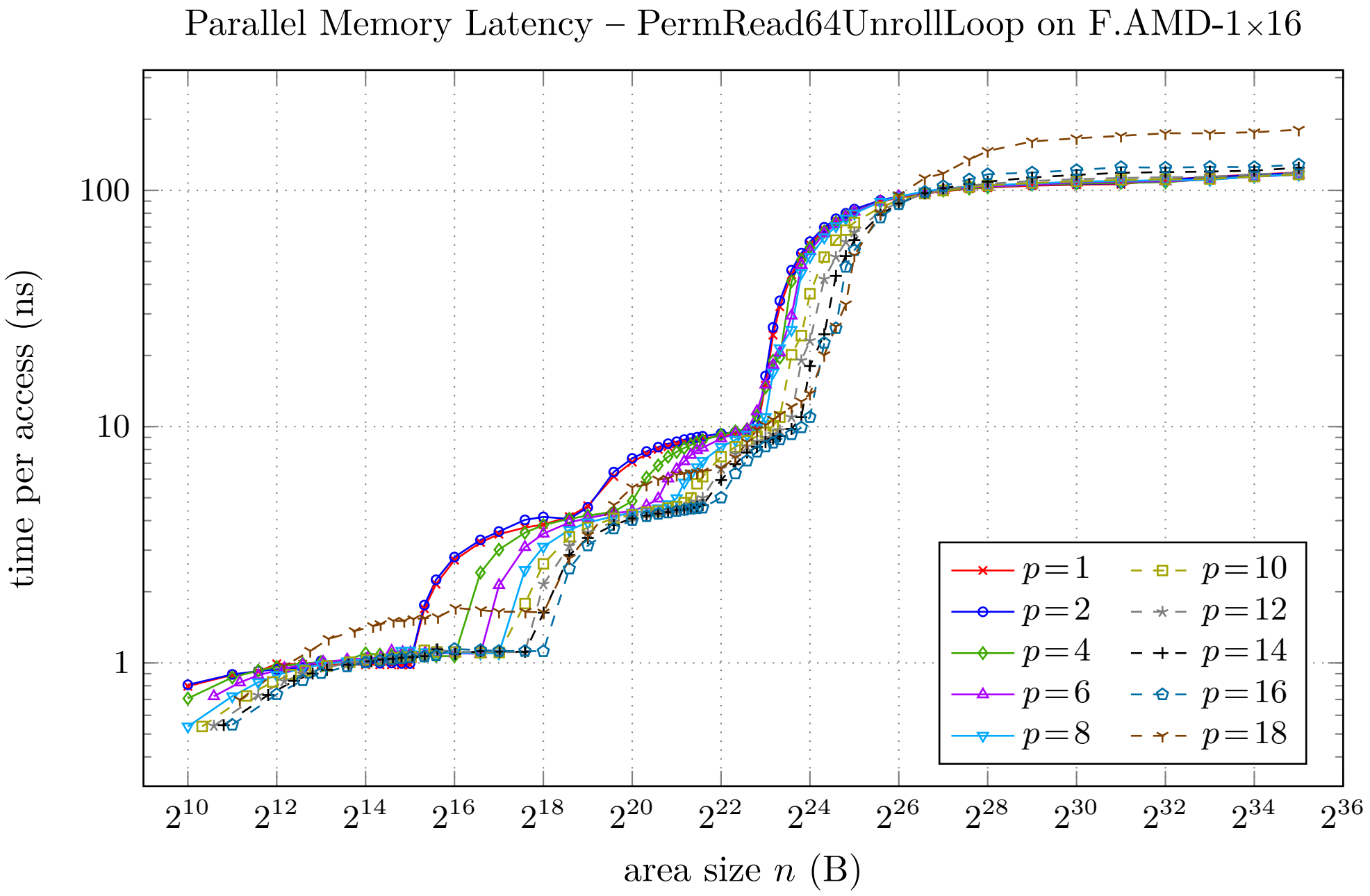

**Figure 3.15:** 64-bit parallel memory latency on A.Intel-1×8 and F.AMD-1×16.

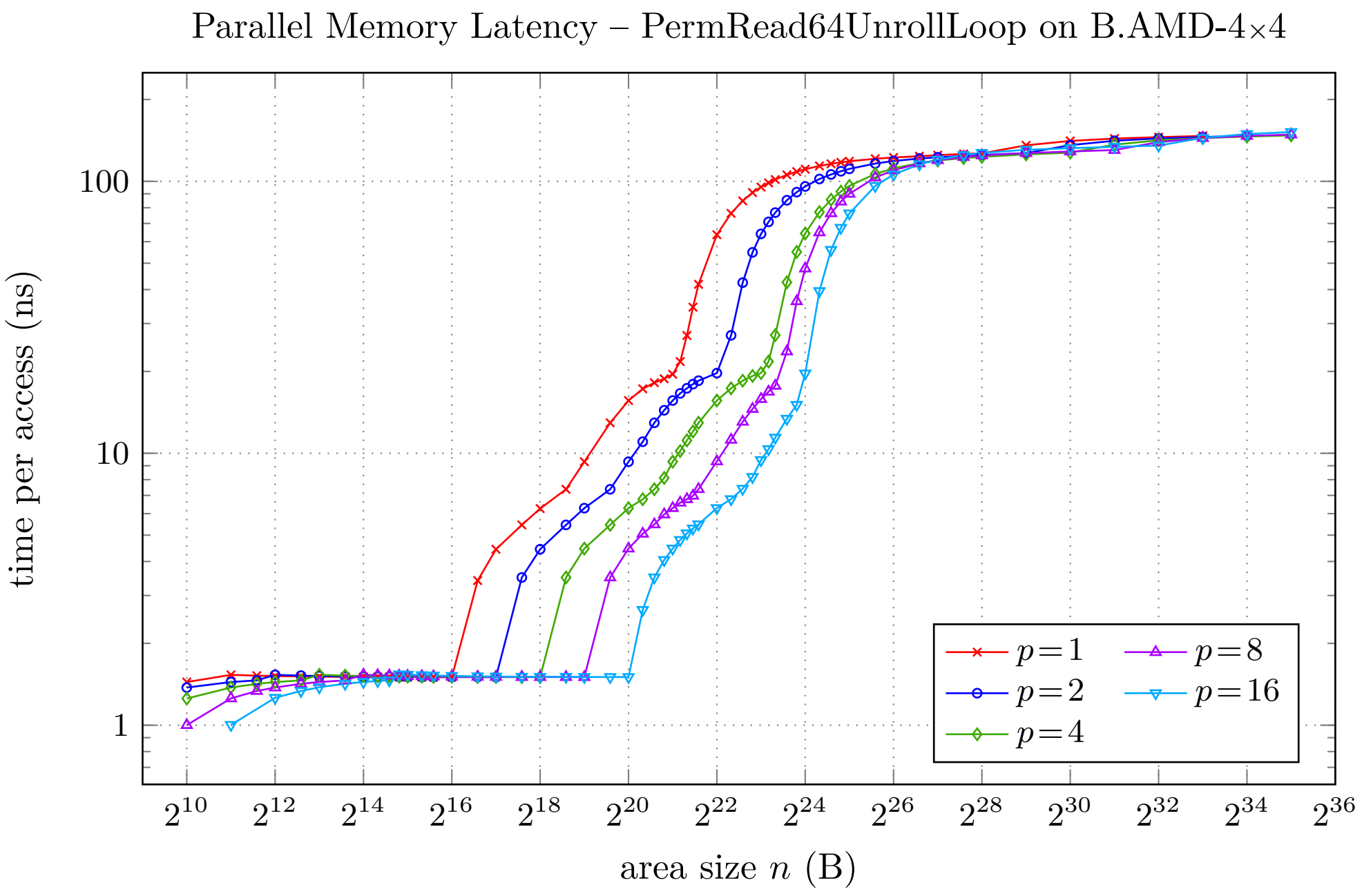

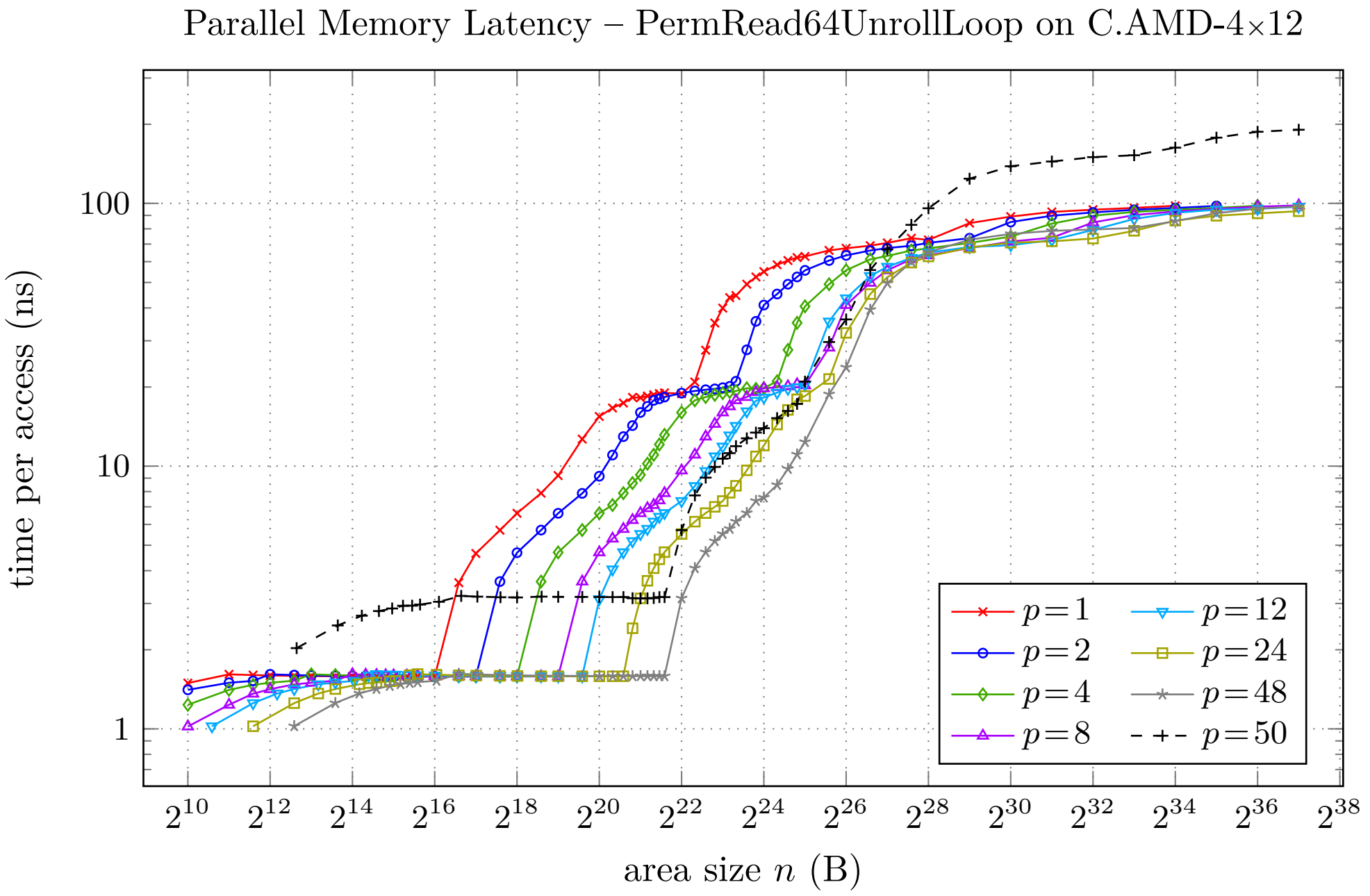

**Figure 3.16:** 64-bit parallel memory latency on B.AMD-4×4 and C.AMD-4×12.

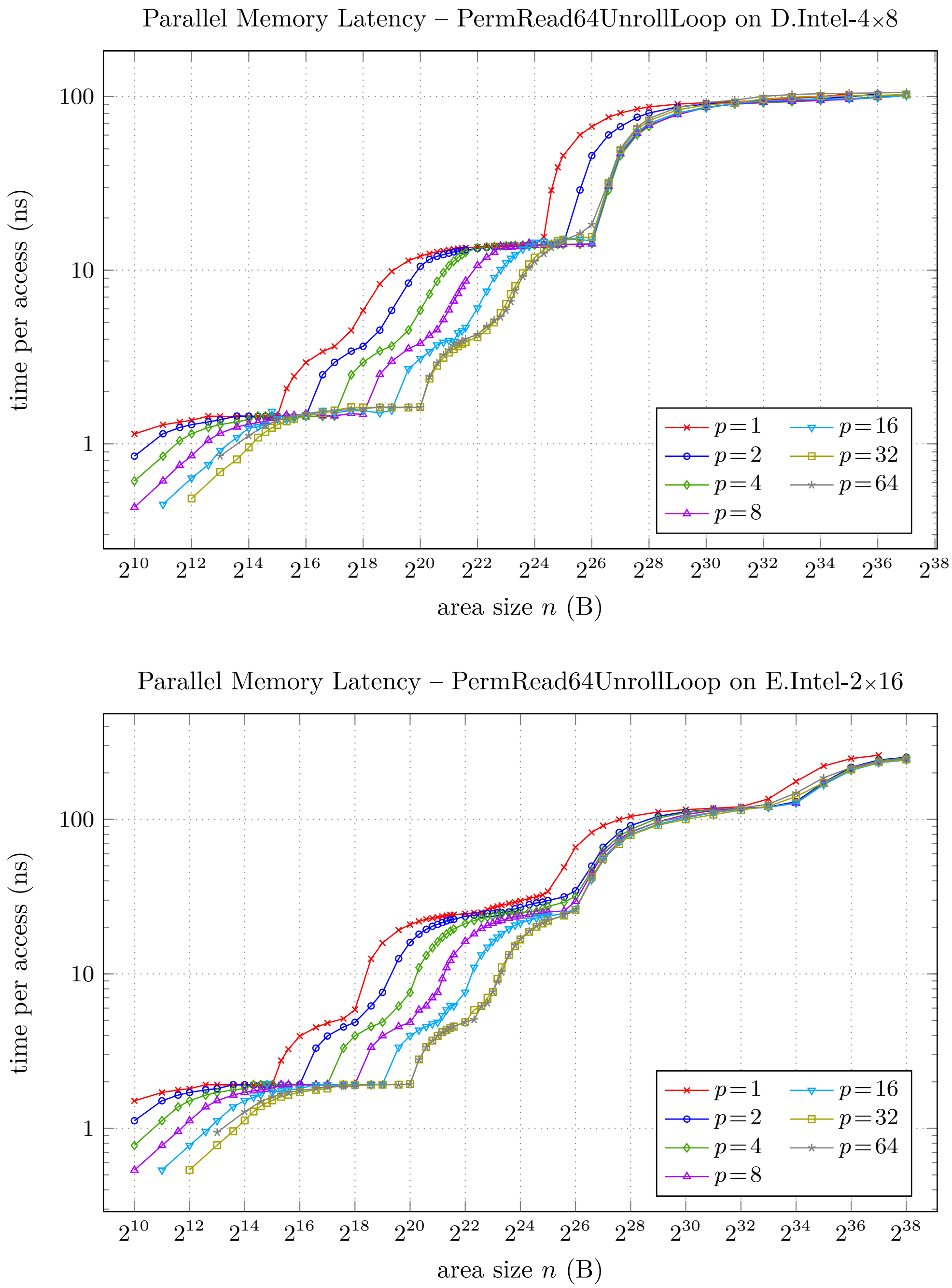

**Figure 3.17:** 64-bit parallel memory latency on D.Intel-4×8 and E.Intel-2×16.

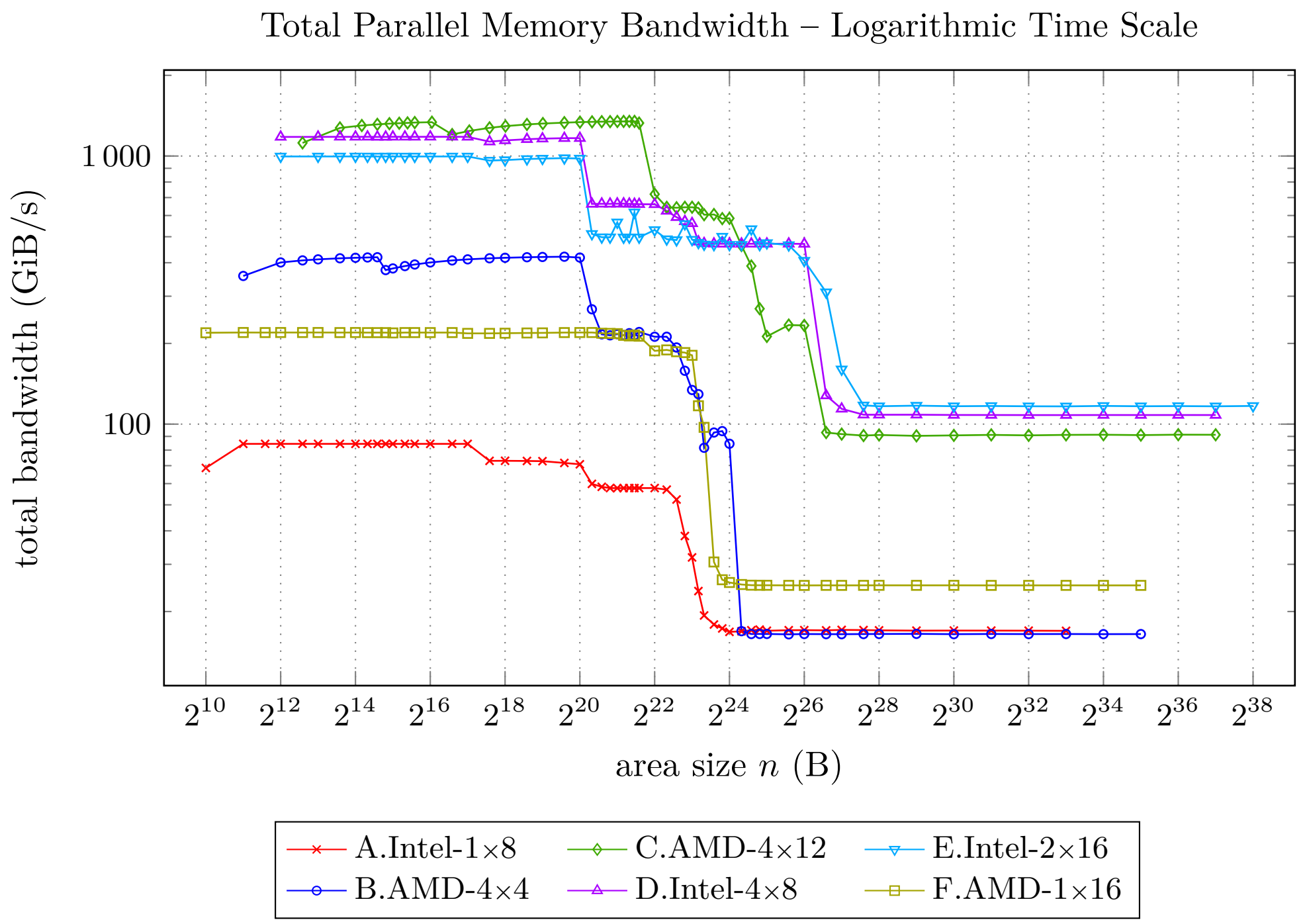

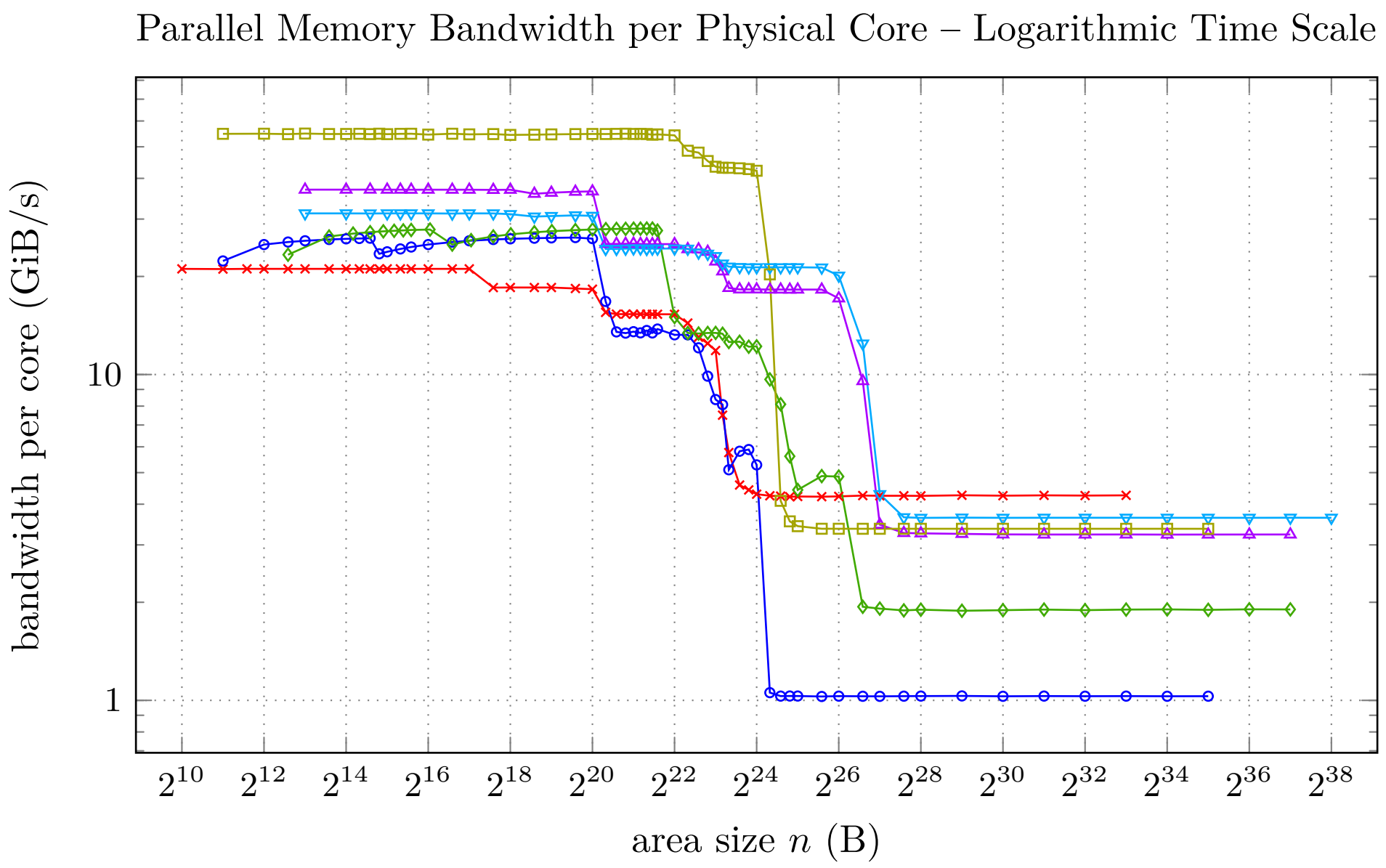

**Figure 3.18:** Parallel scan bandwidth on all machines.

## 3.2 Cross-NUMA Node Bandwidth and Latency

The previous section considered memory bandwidth and latency only for the local NUMA node. We now focus on how high the performance penalty is when accessing memory on a *remote* NUMA node. Let us first turn back to figure 3.1 (page 60), which shows a schematic drawing of a four socket machine. Memory on remote banks (NUMA nodes) can be accessed only via the interconnection network between the processor sockets. Depending on the network architecture, sockets may or may not have direct connections; in the example in figure 3.1, the machine only has a ring interconnect and not a full mesh. Nevertheless, the whole memory is visible to a program as shared memory, albeit with *non-uniform* performance characteristics.

For our cross-NUMA-node experiments we modified pmbw to execute the experiment loops with memory on remote nodes. The new parameter $m$ specifies the distance in "NUMA hops" which each thread's memory is located. A thread on socket $k$ performs a memory access pattern on NUMA node $(k + m)$ modulo the number of nodes on the machine. In essence, the thread accesses memory on the $m$ next socket, with a cyclic wrap-around.

We performed only the two fundamental experiments: `ScanRead64PtrUnrollLoop`, high-bandwidth 64-bit scan reading in an unrolled loop, and `PermRead64UnrollLoop`, high-latency 64-bit random permutation walking. These two experiments encompass the most common memory access patterns in applications. Both experiments are run with a fixed target array size of $2^{30}\,\text{B} = 1\,\text{GiB}$, and with an increasing number of threads.

Figures 3.22 to 3.19 show results of the four NUMA systems in our platform portfolio (see table 2.1, page 40). Maybe the most interesting system is C.AMD-4×12 because it has 8 NUMA nodes (even though the machine has only two physical sockets). The top of figure 3.20 shows the bandwidth results from `ScanRead64PtrUnrollLoop` for all possible NUMA hops. With $m = 0$, all operations are performed on the local RAM, and hence the bandwidth results match those shown at $n = 2^{30}$ in figure 3.6. The NUMA bandwidth series exhibit a "sawtooth" character, for C.AMD-4×12 with period 8, due to the round-robin style allocation of threads onto the NUMA sockets: imbalance in the number of threads per socket reduces the overall performance.

On C.AMD-4×12, local memory can be read at about 80 GiB/s, while remote memory can be read with 10–16 GiB/s across the interconnect. Notice that eight threads, one thread per socket, are sufficient to saturate the interconnection network when reading memory with the high-bandwidth scan access pattern. While $m \in \{2, 4, 6\}$ is faster than the experiments with odd hop numbers, the difference is not large relative to local NUMA node bandwidth.

Cross-NUMA latency results in the lower half of figure 3.20 are similar to the bandwidth results, except that the odd NUMA hop shifts show a higher latency than even shifts. This discrepancy is probably due to the interconnect's internal structure. The latency diagram exhibits a “steps” character, since adding more threads on the cores requires more coordination within one socket.

**Table 3.2:** Local NUMA-node bandwidth $b_l$, remote NUMA-node bandwith $b_r$, local NUMA-node latency $l_l$, and remote NUMA-node latency $l_r$, averaged for each host over all experiment instances with at least half of all cores active. Additionally, shows the ratios of bandwidth and latency of remote and local memory access.

| | Bandwidth | | | Latency | | |
|---|---|---|---|---|---|---|
| Host | $b_l$ (GiB/s) | $b_r$ (GiB/s) | $b_l/b_r$ | $l_l$ (ns) | $l_r$ (ns) | $l_r/l_l$ |
| B.AMD-4×4 | 15.7 | 7.5 | 2.10 | 153 | 218 | 1.43 |
| C.AMD-4×12 | 82.4 | 13.8 | 5.98 | 79 | 173 | 2.18 |
| D.Intel-4×8 | 101.6 | 14.2 | 7.16 | 93 | 364 | 3.89 |
| E.Intel-2×16 | 113.7 | 50.8 | 2.24 | 103 | 153 | 1.48 |

Platforms D.Intel-4×8 and E.Intel-2×16 show only a small difference between odd and even NUMA hop shifts, while B.AMD-4×4 again show a large discrepancy. On all machines it is possible to nearly saturate the interconnect bandwidth with *one thread per socket*, with the exception of E.Intel-2×16, the newest NUMA machine.

Table 3.2 shows the ratios between local and remote RAM bandwidth and latency. Machine D.Intel-4×8 exhibits the highest ratios: local RAM bandwidth is a factor 7.2 faster than remote access, and random access latency is a factor 3.9 smaller. These ratios make this machine the most interesting for NUMA experiments.

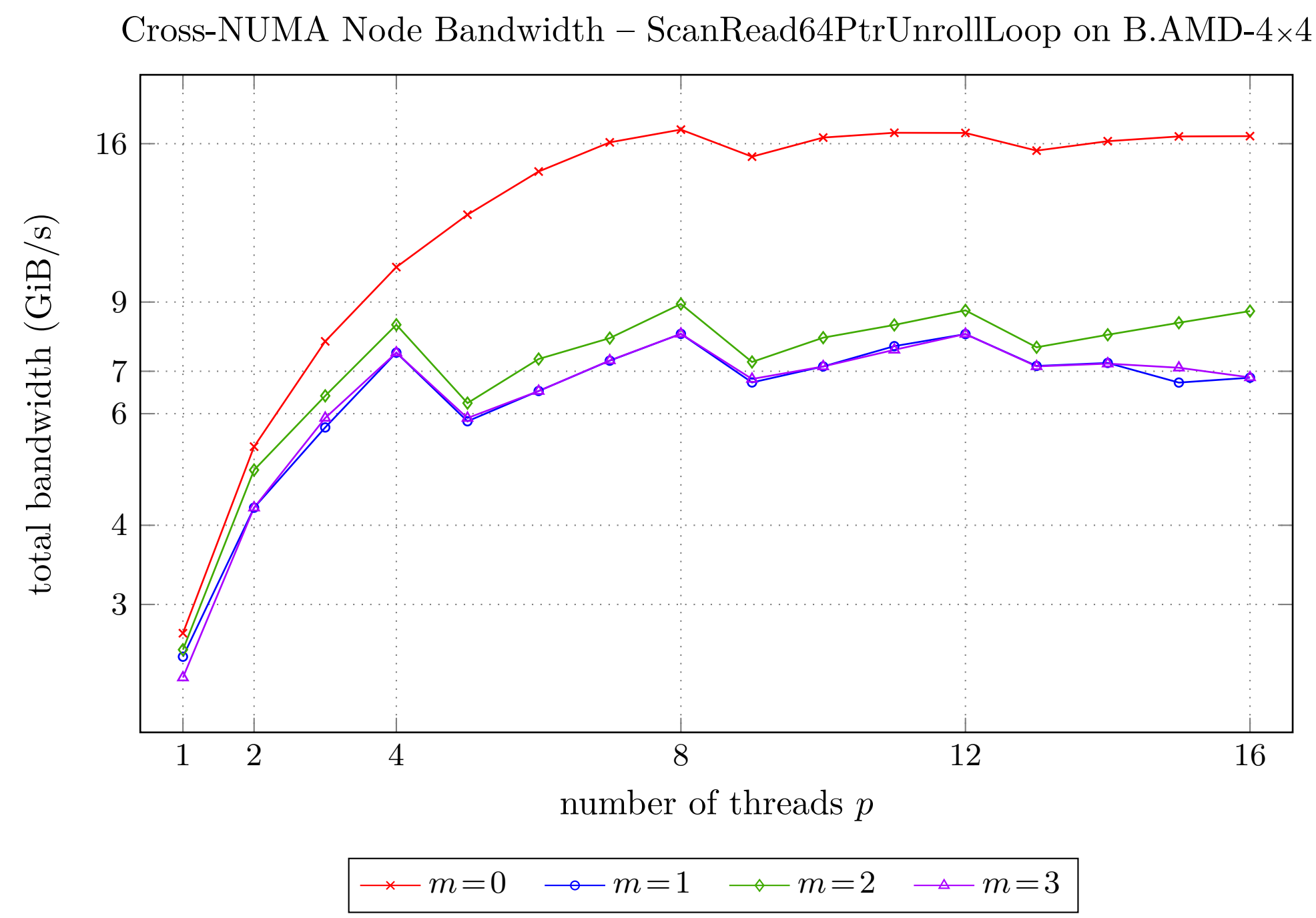

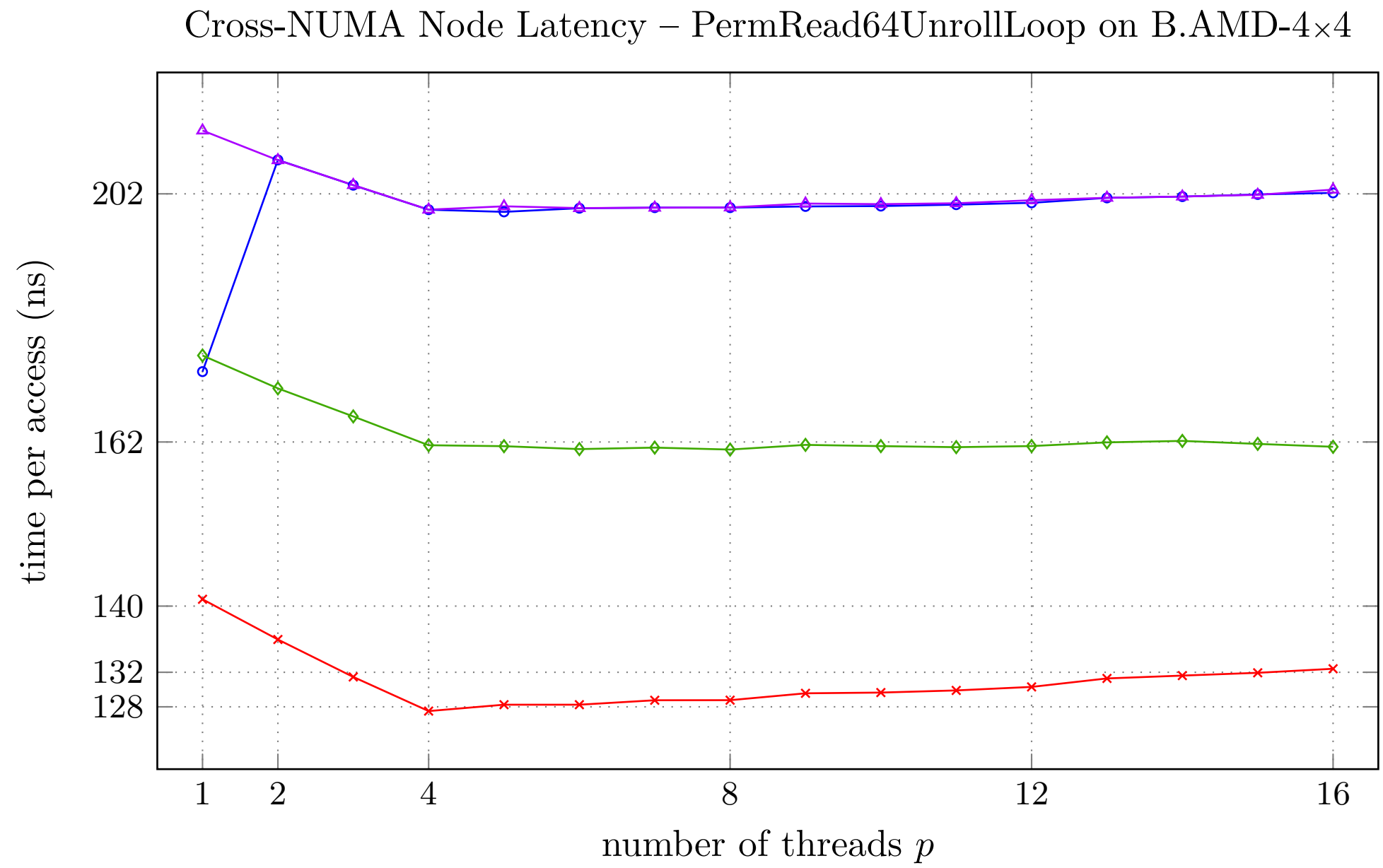

**Figure 3.19:** Cross-NUMA node scan bandwidth and random access latency on B.AMD-4×4. The parameter $m$ is the number of hops to the remote memory.

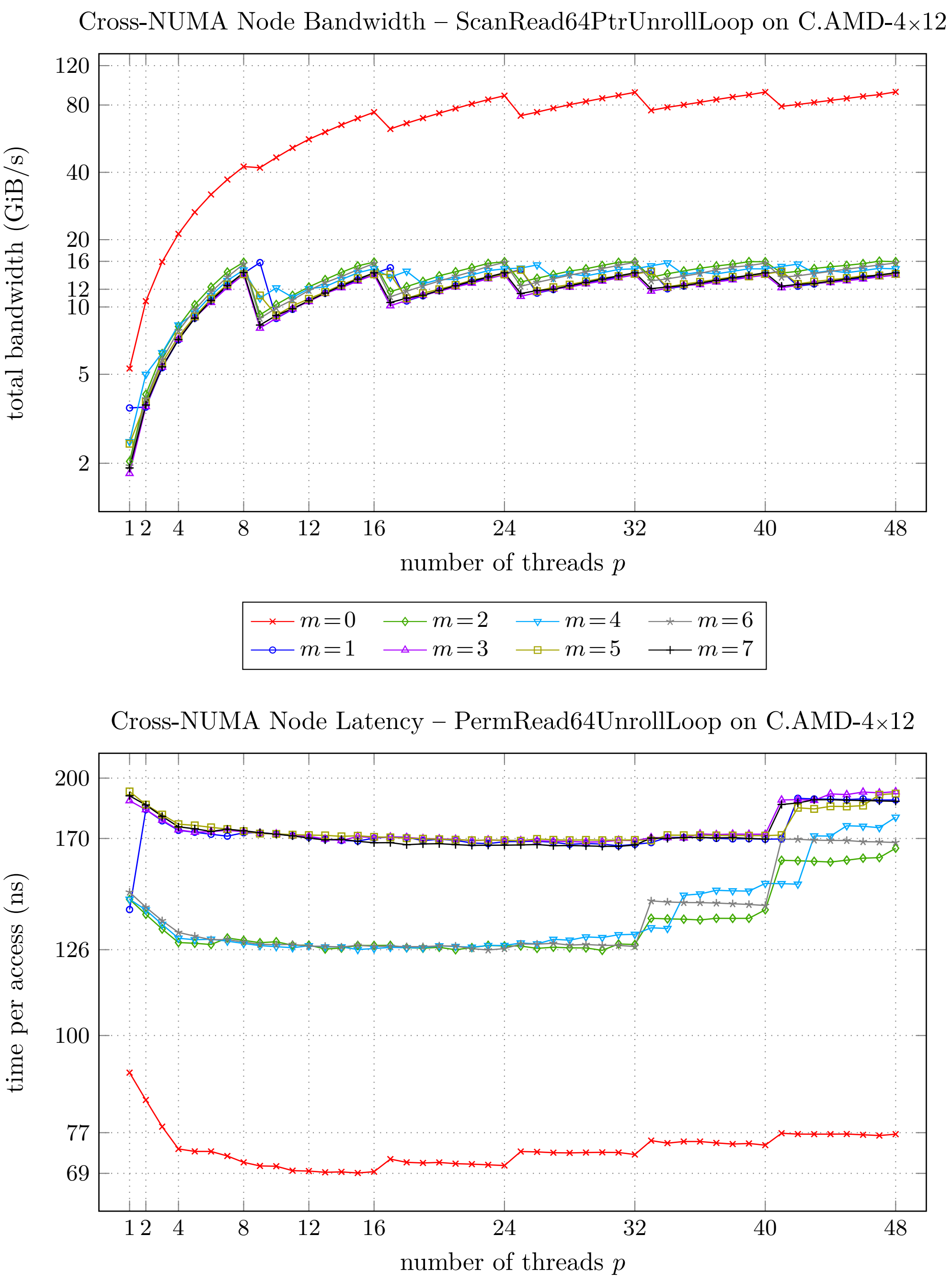

**Figure 3.20:** Cross-NUMA node scan bandwidth and random access latency on C.AMD-4×12. The parameter $m$ is the number of hops to the remote memory.

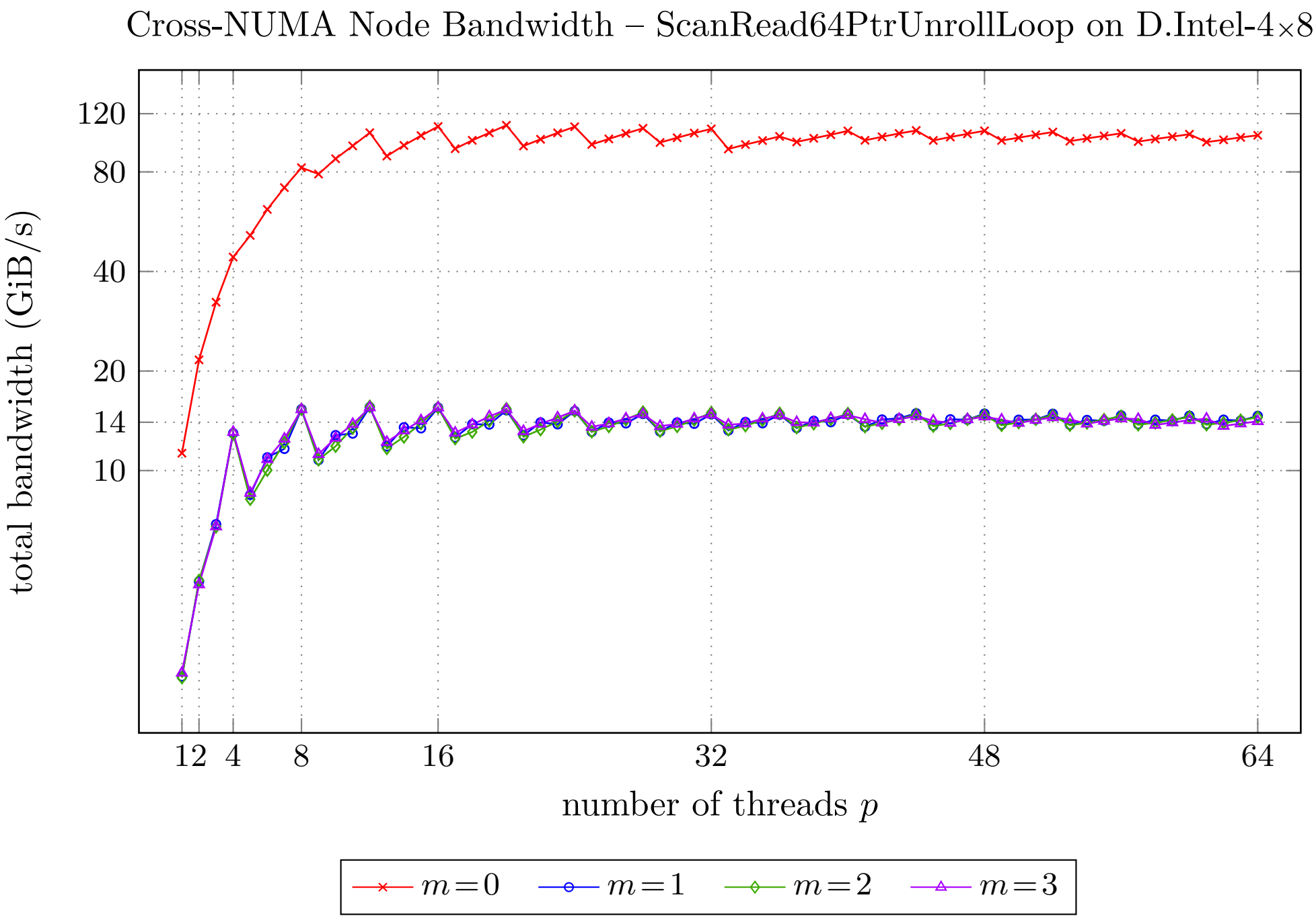

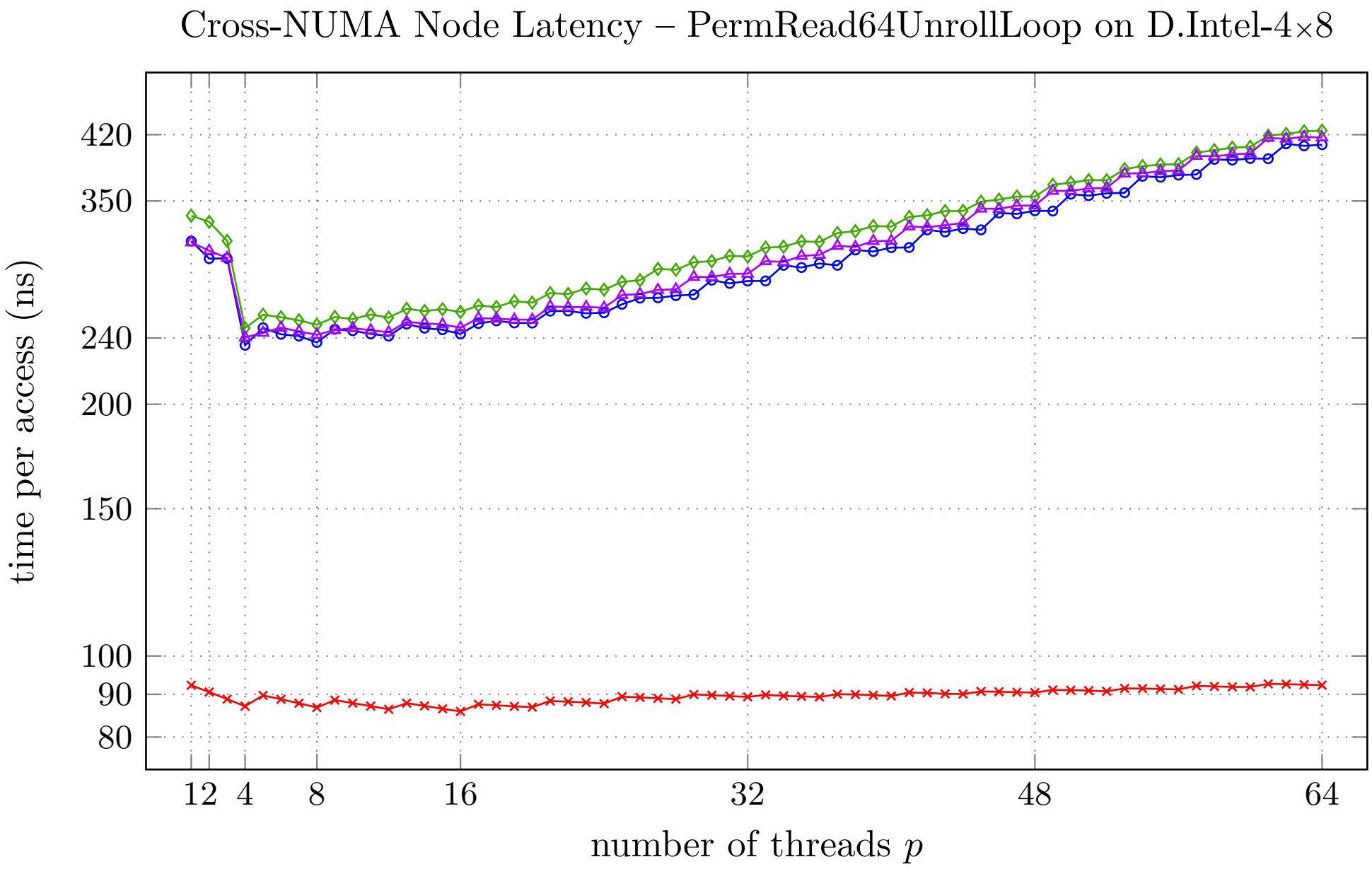

**Figure 3.21:** Cross-NUMA node scan bandwidth and random access latency on D.Intel-4×8. The parameter $m$ is the number of hops to the remote memory.

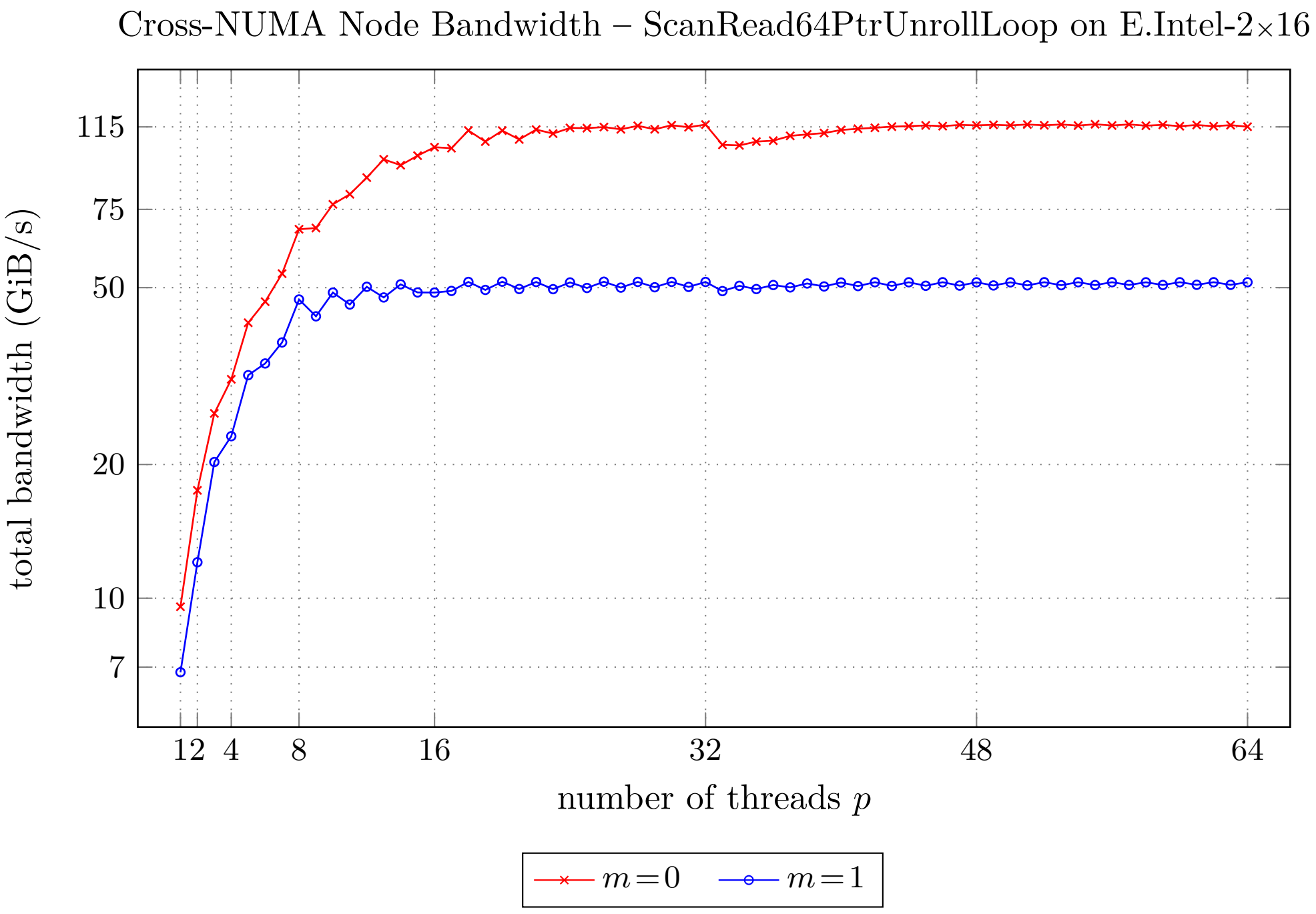

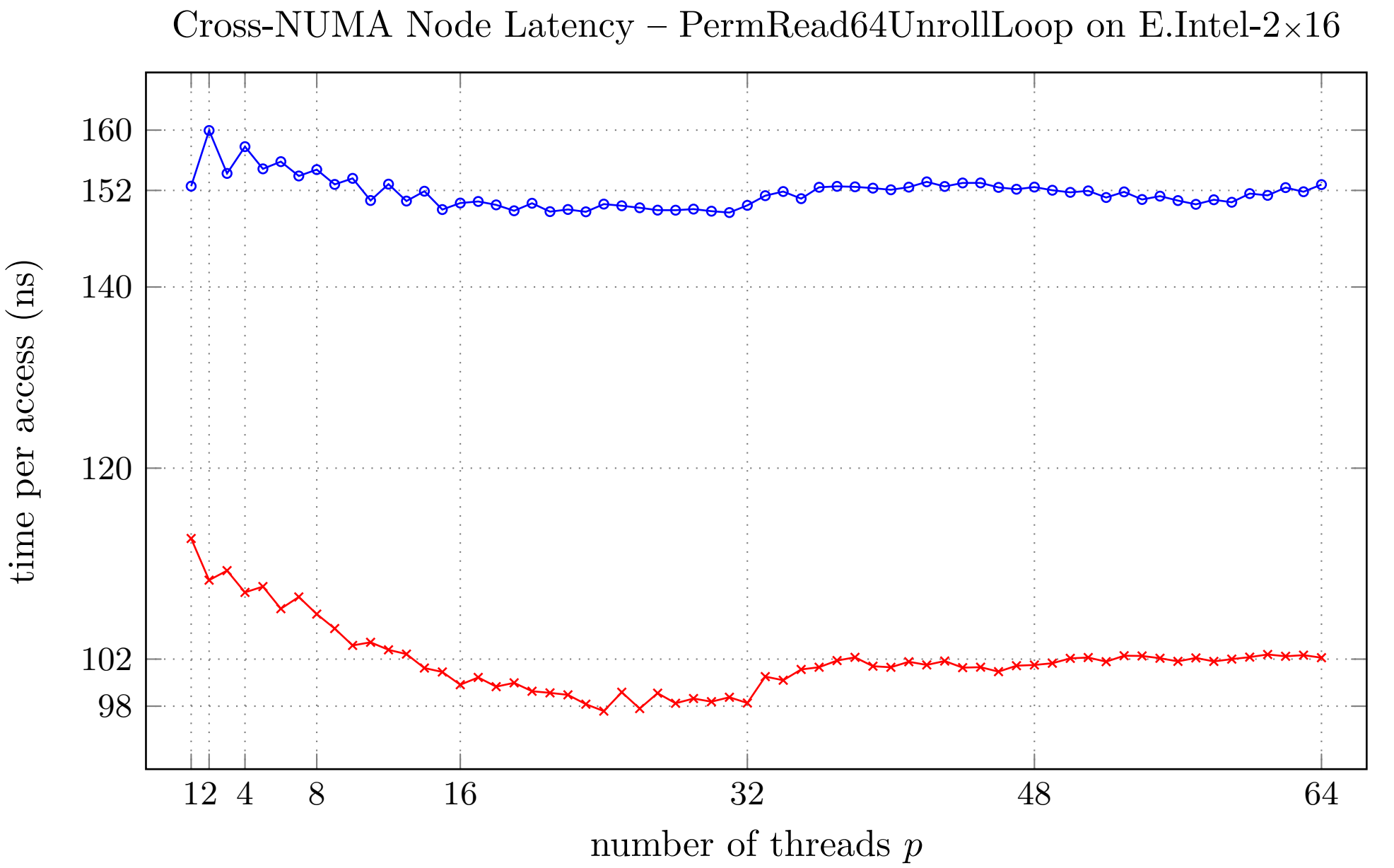

**Figure 3.22:** Cross-NUMA node scan bandwidth and random access latency on E.Intel-2×16. The parameter $m$ is the number of hops to the remote memory.

Chapter 4

# 4 Parallel String Sorting Algorithms

*While there is a long history of work on sequential string sorting, very little work exists on* parallel string sorting*. This is very surprising since parallelism is now the only way to get performance out of Moore's law so that any performance critical algorithm needs to be parallelized. We therefore consider practical parallel string sorting algorithms for modern multi-core shared-memory machines. Our focus is on large inputs which fit into RAM. This means that besides parallelization and load balancing we have to take the memory hierarchy, memory layout, and processor features like word parallelism, superscalar processing, and the high cost of branch mispredictions into account. Looking beyond single-socket multi-core architectures, we also consider multi-core machines with multiple sockets and non-uniform memory access (NUMA).*

In chapter 2 we reviewed the most important sequential string sorting algorithms and discussed a comprehensive performance evaluation. Then, in chapter 3, we focused on parallel memory bandwidth, which will guide our algorithmic design decisions when developing parallel string sorting algorithms in the following sections.

In section 4.2, we propose our first new string sorting algorithm: Super Scalar String Sample Sort ($S^5$), and its parallelization: Parallel Super Scalar String Sample Sort ($pS^5$). The algorithm makes effective use of the memory hierarchy, uses word-level parallelism, and largely avoids branch mispredictions.

Thereafter, we turn our focus to NUMA architectures in section 4.3, and contribute parallel LCP-aware multiway merging as a top-level algorithm for combining presorted sequences, but also as a stand-alone algorithm for string sets with large common LCPs such as URL lists. As base-case sorter for LCP-aware string sorting we describe sequential LCP-insertion sort which calculates the LCP array and accelerates its insertions using the information in the array.

Broadly speaking, we propose both *multiway distribution-based* string sorting with $S^5$ and *multiway merge-based* string sorting with LCP-aware mergesort, and *parallelize* both approaches. Table 4.1 summarizes the theoretical running time bounds we show for our new sequential string sorting algorithms and compares them with existing algorithms from chapter 2. In section 4.4 we develop parallelizations of caching multikey quicksort and radix sort, which are two additional competitors in the parallel string sorting field, and discuss possible parallelizations of burstsort.

**Table 4.1:** Sequential Running Time Bounds on Our New Parallel-Ready String Sorting Algorithms and Existing Prior Work.

| Algorithm | Running Time | Page |
|---|---|---|
| String Sample Sort with Equality Checks | $\mathcal{O}(\frac{D}{w} + n \log n)$ exp. | 108 |
| String Sample Sort with Unrolled Trees | $\mathcal{O}((\frac{D}{w} + n) \log v + n \log n)$ exp. | 108 |
| Multiway LCP-Mergesort | $\mathcal{O}(D + n \log n + \frac{n}{K})$ | 121 |
| LCP-Insertion Sort | $\mathcal{O}(D + n^2)$ | 128 |
| Multikey Quicksort [BS97] | $\mathcal{O}(D + n \log n)$ exp. | 28 |
| MSD Radix Sort [MBM93] | $\mathcal{O}(D + n \log \sigma)$ | 28 |
| Burstsort [SZ03a] | $\mathcal{O}(D + n \log \sigma)$ exp. | 34 |
| Binary LCP-Mergesort [NK08] | $\mathcal{O}(D + n \log n)$ | 36, 116 |
| Insertion Sort [Knu98] | $\mathcal{O}(nD)$ | 37 |

We then compare our parallel string sorting algorithms experimentally in section 4.5 on six current multi-core platforms with seven real-world inputs. For all our input instances, except random strings and URLs, parallel $S^5$ achieves higher speedups on modern *single-socket* multi-core machines than our own parallel multikey quicksort and radix sort implementations, which are already better than any previous ones. For the URLs input with large common LCPs, our parallel $K$-way LCP-mergesort outperforms $pS^5$ on a machine with few memory channels.

For our Intel and AMD *multi-socket* NUMA machines the picture is more complex: parallel multiway LCP-merge with node-local parallel $S^5$ achieves the highest speedups on 11 of the 16 input/platforms combinations with large real-world inputs. For random strings on a current-generation NUMA machine, our parallel radix sort is faster than $pS^5$ with $K$-way parallel LCP-merging. For large URLs on an older NUMA machine, our parallel $K$-way LCP-mergesort also outperforms, and for suffix sorting Wikipedia articles, regular $pS^5$ is faster than the combination with parallel $K$-way LCP-merging. Overall, our parallel string sorters are much faster than any previous implementations.

This chapter is based on joint work with Peter Sanders [BS13a; BES17] and five students: Florian Drews, Andreas Eberle, Michael Hamann, Christian Käser, and Sascha Denis Knöpfle. The idea to consider parallel string sorting algorithms was initiated by Peter Sanders, who was surprised at the lack of work in this area. String sample sort was our first choice due to Peter Sanders' prior work in this area [SW04]. Sascha Denis Knöpfle performed a first investigation of Super Scalar String Sample Sort in his bachelor thesis, which we supervised. Additional work lead to the first conference paper on Parallel String Sample Sort [BS13a] with Peter Sanders. As part of a practical lab course, Florian Drews, Andreas Eberle, Michael Hamann, and Christian Käser later implemented prototypes of parallel multikey quicksort, parallel radix sort, parallel LCP-merge sort, and parallel burstsort. Together with Andreas

Eberle we then studied parallel multiway LCP-merge and LCP-merge sort in depth, on which he wrote his bachelor thesis [Ebe14].

All our work on parallel string sorting was then consolidated in a journal article [BES17], which was primarily written by the author of this dissertation. Peter Sanders contributed some parts of the theoretical string sample sort description and many of the possible future improvements ideas (section 4.6). Many parts of this chapter were copied from our journal article verbatim. However, this dissertation extends the previous material on string sample sort with a proof of its theoretical expected running time (section 4.2.6), a more detailed analysis of the best tree size and number of interleaved descents (section 4.2.4), how to use a bit mapping operation to calculate level-order from pre-order indices (lemma 4.1), an experiment running sequential string sorters independently in parallel on multiple cores (section 4.2.5), and a more detailed description of the internals of $pS^5$'s parallelization framework (section 4.2.8). The section on LCP merge sort additionally contains more details on our splitting techniques (section 4.3.3), and all experiments in section 4.5 were rerun for the dissertation using newer hardware and an up-to-date software stack.

## 4.1 Related Work On Parallel String Sorting

There is a huge amount of work on parallel sorting of *atomic* objects such that we can only discuss the most relevant results. Cole's celebrated merge sort [Col88] runs on a CREW or EREW PRAM with $n$ processors in $\mathcal{O}(\log n)$ time, but it is mostly of theoretical interest. For analysis of parallel algorithms we will use the CREW PRAM in which $p$ independent RAM processors can perform operations in parallel on a shared memory, as long as write operations are without conflict (see also section 1.1.3). Besides simpler versions of (multiway) mergesort [AS87; SSP07], perhaps the most practical parallel sorting algorithms are parallelizations of radix sort (e.g. [WS11]) and quicksort [TZ03] as well as sample sort [BLM+91; AWFS17].

There is some work on PRAM algorithms for string sorting [HP92; JRV94; Hag94; JRV96]. The fastest among these [Hag94] recursively combines pairs of adjacent characters into single characters and applies fast PRAM integer sorting to them. One obtains algorithms with work $\mathcal{O}(N \log N)$ and time $\mathcal{O}(\log N / \log\log N)$. Compared to the sequential algorithms this is suboptimal unless the distinguishing prefix $D = \mathcal{O}(N) = \mathcal{O}(n)$, since all characters in the strings are touched. It is unclear how this parallelization approach can be modified to avoid work on characters outside the distinguishing prefixes.

Aside from our own work [BS13a; BES17] and work by our students [Knö12; Ebe14], we found no academic publications on practical parallel string sorting. However, Akiba has implemented a parallel radix sort [Aki11], Rantala's library [Ran07] contains multiple parallel mergesorts and a parallel SIMD variant of multikey quicksort, and Shamsundar [Sha09] also parallelized Ng and Kakehi's LCP-mergesort [NK08]. Of all these implementations, only the radix sort by Akiba scales reasonably well on multi-core architectures. We discuss the scalability issues of these implementations in the experimental section 4.5.

## 4.2 Parallel Super Scalar String Sample Sort (pS$^5$)

Already in a sequential setting, theoretical considerations and experiments (see section 2.3) indicate that *the* best string sorting algorithm does not exist. Rather, it depends at least on $n$, $D$, $\sigma$, and the hardware. Therefore we decided to parallelize several algorithms taking care that components like data distribution, load balancing, or base case sorter can be reused. Remarkably, most algorithms in section 2.2 can be parallelized rather easily and we will discuss straight-forward parallel versions in section 4.4. However, none of these parallelizations make use of the striking new attribute of modern multi-core systems discussed in chapter 3: many multi-core processors with individual cache levels but relatively few and slow memory channels to shared RAM. Therefore we decided to design a new string sorting algorithm based on *sample sort* [FM70], which exploits these properties.

### 4.2.1 Traditional (Parallel) Atomic Sample Sort

Sample sort [FM70; BLM+91] is a generalization of quicksort using $k-1$ pivots at the same time. For small inputs, sample sort uses a sequential base case sorter. Larger inputs are split into $k$ *buckets* $b_1, \ldots, b_k$ by determining $k-1$ splitter keys $x_1 \leq \cdots \leq x_{k-1}$ and then classifying the input elements — element $s$ goes to bucket $b_i$ if $x_{i-1} < s \leq x_i$ (where $x_0$ and $x_k$ are defined as sentinel elements — $x_0$ being smaller than all possible input elements and $x_k$ being larger). Splitters can be determined by drawing a random sample of size $\alpha k - 1$ from the input, sorting it, and then taking every $\alpha$-th element as a splitter. Integer parameter $\alpha \geq 1$ is the *oversampling* factor. The buckets are then sorted recursively and concatenated.

The original sample sort paper [FM70] focuses on determining the sample size and number of splitters $k$ depending on $n$ such that the overall algorithm retains the *expected* $\mathcal{O}(n \log n)$ runtime. In the original algorithm, sample sort is run once and then each bucket is individually sorted using a quicksort variant.

"Traditional" *distributed* parallel sample sort [BLM+91; GV92; GV94] for "shared-nothing", message passing machines chooses $k = p$ and uses a sample big enough to ensure that all buckets have approximately equal size with sufficiently high probability. As the number of processors $p$ increases, maintaining load balance becomes the main concern for sample sort. This issue has been addressed using multiple rounds of sample selection [HBJ98; HJB98], and by adaptively performing multi-level sample sort with multiple data exchanges [ABSS15; AS17].

Sample sort is also attractive as a sequential algorithm since it is more cache efficient than quicksort and since it is particularly easy to avoid branch mispredictions. Super Scalar Sample Sort (S$^4$) [SW04] describes an elegant technique to take advantage of these optimizations. In this case, $k$ is chosen in such a way that classification and data distribution can be done cache-efficiently.

Recently, Super Scalar Sample Sort has been engineered into an *almost in-place* fast shared-memory parallel algorithm for sorting atomic objects using comparisons [AWFS17]. Their implementation, called IPS$^4$o, outperforms all other shared-memory parallel sorters, including the previously fastest ones [PSS07; SSP07; Rei07; SK08; SBF+12].

### 4.2.2 Super Scalar String Sample Sort (S$^5$)

In order to adapt the atomic sample sort from the previous section to strings, we have to devise an efficient classification algorithm. Most importantly, we want to avoid comparing whole strings as atomic objects, and thus focus on character comparisons. Also, in order to approach total work $\mathcal{O}(D + n \log n)$, we have to use the information gained during classification in the recursive calls. This can be done by observing that strings in buckets have a common prefix depending on the LCP of the two splitters: assuming bucket $b$ contains all strings $s$ with $x_{i-1} < s \leq x_i$, we have

$$\forall s, t \in b : \textsc{lcp}(s, t) \geq \textsc{lcp}_x(i) \,. \tag{4.1}$$

Thus, we know how to advance the common depth when recursively sorting a bucket. Hence, it remains to select the length and number of splitters $x_i$.

Rather than using whole strings as arbitrarily long splitters, or all characters of the alphabet as in radix sort, we design the splitter keys to consist of *as many characters as fit into a machine word* and choose the number of splitters such that the classification data structure is kept in fast cache memory.

We adapt the implicit binary search tree data structure used in (atomic) Super Scalar Sample Sort (S$^4$) [SW04] to strings, and call the algorithm "Super Scalar String Sample Sort" (S$^5$). Algorithm 4.1 shows pseudocode of the main variant of S$^5$ as a guideline through the following discussion, and figure 4.1 illustrates the classification tree with buckets $b_i$ and splitters $x_i$.

In the following let $w$ denote the number of characters that fit into one machine word, though other values are also possible. For 8-bit characters and 64-bit machine words this yields $w = 8$. We choose $v = 2^d - 1$ splitters $x_0, \dots, x_{v-1}$ for some integer $d$ from a sorted sample to construct a *perfect binary search tree* with $d - 1$ levels (lines 1 to 5), which is used to classify an array of strings based on the next $w$ characters at common prefix $h$. The main disadvantage of this limited character depth over comparing entire strings is that we may have many input strings whose next $w$ characters are identical. For these strings, the classification does not reveal much information. We make the best out of such inputs by adding *equality buckets* for strings whose next $w$ characters exactly match $x_i$. For equality buckets, we can increase the common prefix length by $w$ in the recursive calls, i.e., these characters will never be inspected again. In total, we have $k = 2v + 1$ different buckets $b_0, \ldots, b_{2v}$ for a ternary search tree (see figure 4.1).

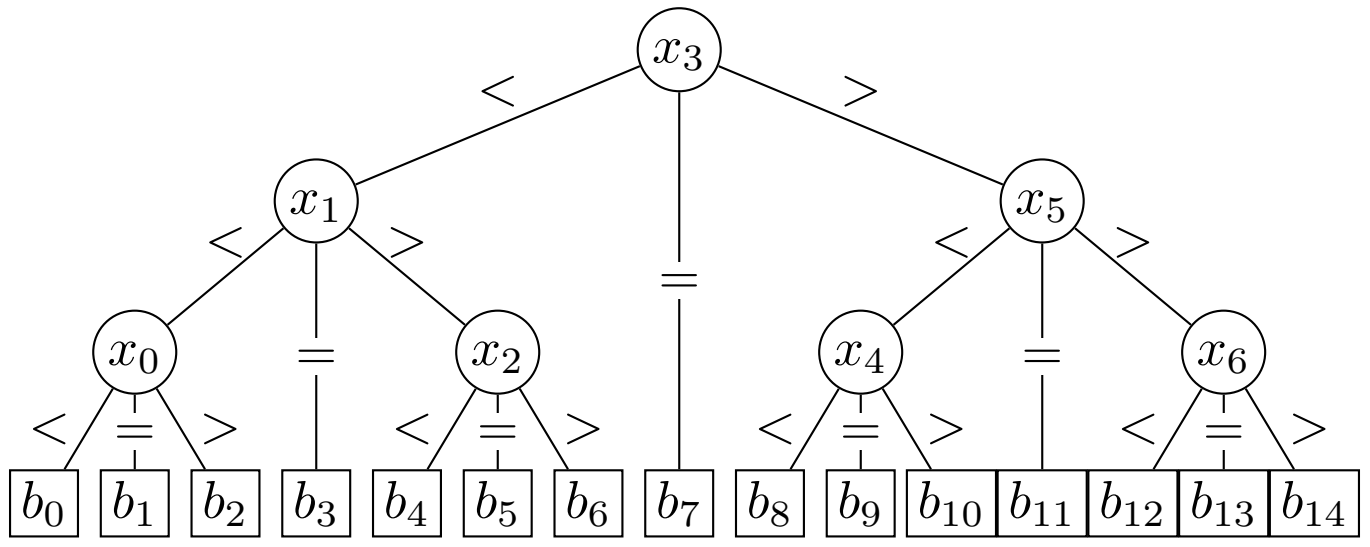

**Figure 4.1:** Ternary classification tree for $v = 7$ splitters and $k = 15$ buckets.

Testing for equality can either be implemented by explicit equality checks at each node of the search tree (which saves time when most elements end up in a few large equality buckets) or by going down the search tree all the way to a bucket $b_i$ ($i$ even) performing only $\leq$-comparisons, followed by a single equality test with $x_{\frac{i}{2}}$, unless $i = 2v$. The latter variant is shown in algorithm 4.1, and the equality test is done in line 11.

After reordering, the strings in the "$< x_0$" and "$> x_{v-1}$" buckets $b_0$ and $b_{2v}$ have to be recursively sorted with common prefix length $h$. For *even* buckets $b_i$ the common prefix length is increased by $\text{LCP}_x(\frac{i}{2})$ due to equation (4.1), while sorting depth in *odd* (equality) buckets $b_i$ advances by the full splitter width $w$.

### 4.2.3 Discussion of Implementation Details

One goal of $S^5$ is to have a common classification data structure that fits into the cache on all cores. Using this data structure, all processors can independently classify a subset of the strings into buckets in parallel. The overall $S^5$ algorithm follows the classic distribution-based sorting steps: we first classify strings (lines 6 to 12), counting how many fall into each bucket (line 14), then calculate a prefix sum (line 15) and redistribute the string pointers accordingly (line 16). To avoid traversing the tree twice, the bucket index of each string is stored in a cache array $o$ (sometimes called oracle) (lines 12, 14 and 16). Additionally, to make better use of superscalar parallelism and memory latency hiding, we separate the classification loop (lines 6 to 12) from the counting loop (line 14), as discussed in the case of radix sort [KR08] in section 2.2.2.

**Tree Traversal.** Like in $S^4$, the binary tree of splitters is stored in *level-order* as a *one-based* array $t$ (line 4), e.g. $t = [\,x_3, x_1, x_5, x_0, x_2, x_4, x_6\,]$ for the tree in figure 4.1.

**Algorithm 4.1 :** Sequential Super Scalar String Sample Sort – a single step

**Input :** $\mathcal{S} = [\, s_0, \dots, s_{n-1} \,]$ an array of $n$ strings with common prefix $h$, $v = 2^d - 1$ a number of splitters, $v' := v + 1$, and $\alpha \geq 1$ an oversampling factor.

1 $p_i := \text{chars}_h(s_{\text{random}(0,\dots,n-1)}) \quad \forall\, i = 1, \dots, v\alpha + \alpha - 1$ *// Read sample p of $\mathcal{S}$,*

2 $\text{sort}([\, p_1, \dots, p_{v\alpha+\alpha-1} \,])$ *// sort it, and select*

3 $[\, x_1, x_2, \dots, x_{v-1}, x_v \,] := [\, p_\alpha, p_{2\alpha}, p_{3\alpha}, \dots, p_{v\alpha} \,]$ *// equidistant splitters.*

4 $t = [\, t_1, \dots, t_v \,] := [\, x_{\frac{v'}{2}}, x_{\frac{v'}{4}}, x_{\frac{3v'}{4}}, x_{\frac{v'}{8}}, x_{\frac{3v'}{8}}, x_{\frac{5v'}{8}}, x_{\frac{7v'}{8}}, \dots \,]$ *// Construct tree, save*

5 $[\, h'_0, \dots, h'_v \,] := [\, 0 \,] + [\, \textsc{lcp}(x_{i-1}, x_i) \mid i = 1, \dots, v - 1 \,] + [\, 0 \,]$ *// LCPs of splitters.*

6 **for** $j := 0, \dots, n-1$ **do** *// Process strings (interleavable loop).*

7     $i := 1, \quad c := \text{chars}_h(s_j)$ *// Start at root, get w chars from $s_j$,*

8     **for** $1, \dots, \log_2(v+1)$ **do** *// and traverse tree (unrollable loop)*

9         $i := 2i + \mathbb{1}_{(c \leq t_i)}$ *// without branches using $\mathbb{1}_{(c \leq t_i)} \in \{0, 1\}$.*

10     $i := i - (v + 1), \quad m := 2i$ *// Calculate matching non-equality bucket.*

11     **if** $i < v$ **and** $x_i = c$ **then** $m{+}{+}$ *// Test for equality with next splitter.*

12     $o_j := m$ *// Save final bucket number for string $s_j$ in an oracle.*

13 $b_i := 0 \quad \forall\, i = 0, \dots, 2v$ *// Calculate inclusive prefix sum*

14 **for** $i := 0, \dots, n-1$ **do** $(b_{o_i}){+}{+}$ *// over bucket sizes*

15 $[\, b_0, \dots, b_{2v}, b_{2v+1} \,] := [\, \sum_{j \leq i} b_j \mid i = 0, \dots, 2v \,] + [\, n \,]$ *// as fissioned loops.*

16 **for** $i := 0, \dots, n-1$ **do** $s'_{(b_{o_i})--} := \mathbf{move}(s_i)$ *// Reorder strings into new subarrays.*

**Output :** $\mathcal{S}'_i = \{ s'_j \mid j = b_i, \dots, b_{i+1} - 1 \text{ if } b_i < b_{i+1} \}$ for $i = 0, \dots, 2v$ are string subarrays with $\mathcal{S}'_i < \mathcal{S}'_{i+1}$. The subarrays have common prefix $h + h'_{i/2}$ for even values of $i$, and common prefix $h + w$ for odd values.

This allows efficient traversal using $i := 2i + \{0, 1\}$ without branch mispredictions in line 9. The pseudocode "$\mathbb{1}_{(c \leq t_i)}$" yields 0 or 1, and can be implemented using different non-branching machine instructions. On `x86`, one method is to use the instruction `SETA`, which sets a register to 0 or 1 depending on a preceding comparison. Alternatively, newer processors have predicated instructions like `CMOVA` to conditionally move one register to another, again depending on a preceding comparison's outcome. We noticed that `CMOVA` was slightly faster than flag arithmetic.

**Character Reordering.** While traversing the classification tree, we compare $w$ characters using one arithmetic comparison. However, we need to make sure that these comparisons have the desired outcome, e.g., that the most significant bits of the register hold the first character. For little-endian `x86` machines and 8-bit characters, which are used in all of our experiments, we need to *swap the byte order* when loading characters from a string. In our implementation, we do this using the `BSWAP` machine instruction. In the pseudocode (algorithm 4.1) this operation is symbolized by "$\text{chars}_h(s_i)$", which fetches $w$ characters from $s_i$ at depth $h$, and swaps them appropriately.

**From Level-Order to Tree-Order and Back.** For performing the equality check, already mentioned in the previous section we want to discuss three alternatives in more technical detail in the next paragraph on $\mathrm{S}^5$ variants.

Depending on the variant, the splitter tree is stored in *level-order*, e.g. for seven splitters as $t = [\,x_3, x_1, x_5, x_0, x_2, x_4, x_6\,]$, in *pre-order*, e.g. $x = [\,x_0, x_1, x_2, x_3, x_4, x_5, x_6\,]$, or in both. While implementing the variants, we found the following $\mathcal{O}(1)$ transformation between the two, involving only bit operations. This transformation is probably known as folklore in the literature, but we are unaware of any explicit reference.

**Lemma 4.1 (Bit Mapping of Level-Order to/from Pre-Order Indices)**
*Given a vertex $v$ in a perfect binary tree containing $2^w - 1$ vertices in $w$ levels.*

*(i) If $l = l_1 l_2 \cdots l_w$ is the binary representation of $v$'s level-order one-based index in the tree ($l_i \in \{0,1\}$), wherein $l_x$ is the first one bit ($l_x = 1$ and $l_{x'} = 0$ for all $1 \leq x' < x$), then $p = l_{x+1} \cdots l_w l_x l_1 \ldots l_{x-1}$ is the binary representation of $v$'s pre-order index.*

*(ii) Vice versa, if $p = p_1 p_2 \cdots p_w$ is the binary representation of $v$'s pre-order one-based index, wherein $p_y$ is the last one bit ($p_y = 1$ and $p_{y'} = 0$ for all $y < y' \leq w$), then $l = p_{y+1} \cdots p_w p_y p_1 \ldots p_{y-1}$ is the binary representation of $v$'s level-order index.*

For example, consider the level-order index $0101_l$ of a vertex in a perfect binary tree with 15 vertices (shown in figure 4.2). The highest one bit is $l_2 = 1$, hence we have the prefix $l_1 l_2 = 01$, and the suffix $l_3 l_4 = 01$. Rearranging the bits as $l_3 l_4 l_1 l_2$ yields the pre-order index $0110_p$ of the same vertex.

*Proof (lemma 4.1).* Let $l = l_1 l_2 \cdots l_w$ be the binary representation of $v$'s level-order index and $x$ as described in the lemma. As levels are enumerated zero-based (the root is on level 0), $v$ is the $l_{x+1} \cdots l_w$-th vertex on level $w - x$. Consider the pre-order vertex indices in the leaves (level $w - 1$): all are odd, because every other vertex in pre-order is a leaf. More generally, all pre-order indices on level $k$ have $10 \cdots 0$ as their $w - k + 1$ least significant bits (0 repeated $w - k$ times) because every $(w - k + 1)$-th vertex is on the same level. Furthermore, the remaining $k - 1$ bits represent the rank of the vertex among those on the same level. Hence, if we denote $v$'s pre-order index $p = p_1 \cdots p_{y-1} p_y \cdots p_w$, as in the lemma with $p_y = 1$ and $p_{y'} = 0$ for $y < y' \leq w$, we have $p_1 \cdots p_{y-1} = l_{x+1} \cdots l_w$. Together with $l_x = 1$ and $l_{x'} = 0$ for $1 \leq x' < x$ this shows the first part of the lemma. The second part follows analogously by reversing the steps. □

**The Equality Check – $\mathbf{S}^5$ Variants.** Lemma 4.1 enables us to calculate the pre-order index from the level-order index and vice versa, hence, in algorithm 4.1 we can optionally discard the pre-order splitter array after line 5, and calculate $x_i$ by mapping $i$ from pre-order to level-order and using array $t$. Array $t$ is required for the branchless tree traversal regardless.

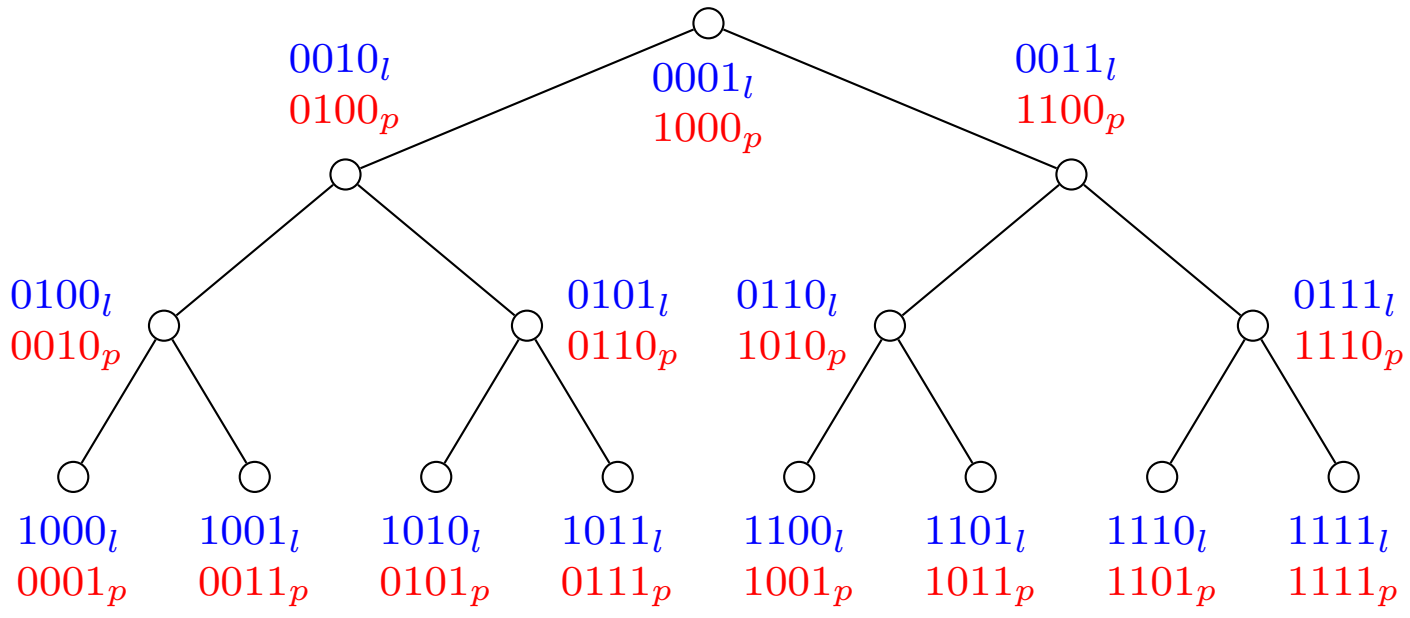

**Figure 4.2:** Binary representation of level-order (subscript $l$) and pre-order (subscript $p$) one-based enumeration of a perfect binary tree.

We developed the following three variants of the classification kernel of S$^5$:

(i) The basic naive variant of S$^5$, named **S$^5$-E** for "Equal", checks for equality after each splitter comparison. This requires only one additional jump-if-equal (`JE`) instruction and no extra comparison (`CMP`) in the inner-most loop in line 9. The branch misprediction cost of the `JE` is counter-balanced by skipping comparisons in the remaining levels of the tree.

As $i$ is a level-order index when exiting the inner loop, we need to apply lemma 4.1 (i) to $i$ in any case, since the correct output is simply the equality bucket number in pre-order. Thus no additional pre-order splitter array is needed in this variant.

(ii) In the main variant of S$^5$, named **S$^5$-UI** for "Unroll/Interleave", the tree is traversed using only $\leq$-comparisons and all checks for equality are postponed to the end, as shown in algorithm 4.1. This allows us to completely *unroll* the loop descending the search tree (line 8). Additionally, we can *interleave* independent tree descents, since there is no exit condition (contrary to S$^5$-E). As in [SW04], this is an important optimization since it allows the instruction scheduler in a superscalar processor to parallelize the operations by drawing data dependencies apart. The number of interleaved descents is limited in practice by the number of registers to hold local variables like $i$ and $c$. We will further investigate the best parameters for tree size $x$ and interleaved tree descents $y$ in section 4.2.4, and label the variant **S$^5$-U$_x$I$_y$** given the two parameters.

In the S$^5$-UI variant, we keep the splitters $x_i$ in a pre-order array, in addition to the classification tree $t$ which contains them in level-order for the traversal. Duplicating the splitters avoids additional work in line 11, where $i$ is a pre-order index.

(iii) The additional pre-order array from the previous variant can be removed using lemma 4.1 (ii): after the descent, $i$ is in pre-order in line 10. With the lemma, we can transform $i$ back to level-order and reuse the classification tree $t$ for the equality check. We call this variant $\mathbf{S^5\text{-}U_{\boldsymbol{x}}I_{\boldsymbol{y}}C}$ for “Calculate”, where $x$ is the number of tree levels, and $y$ the number of interleaved tree descents. Depending on the processor hardware, this variant is sometimes slightly faster.

**Sampling.** The sample is drawn pseudo-randomly with an oversampling factor $\alpha = 2$ to keep it in cache when sorting with STL's `std::sort` (introsort [Mus97] in the current gcc) and building the search tree. Instead of using the straight-forward equidistant method to draw splitters from the sample, as shown in algorithm 4.1 (line 3), we developed a simple recursive scheme that tries to avoid using the same splitter repeatedly: Select the middle sample $m$ of a range $a..b$ (initially the whole sample) as the middle splitter $\bar{x}$. Find new boundaries $b'$ and $a'$ by scanning left and right from $m$ *skipping* samples equal to $\bar{x}$. Recurse on $a..b'$ and $a'..b$. The splitter tree selected by this heuristic was never slower than equidistant selection, but slightly faster for inputs with many equal common prefixes. It is used in all our experiments.

**LCP via XOR.** The LCP of two consecutive splitters in line 5 of algorithm 4.1 can be calculated without a loop using just two machine instructions: `XOR` and `BSR` (to count the number of leading zero bits in the result of `XOR`). In our implementation, these calculations are done while selecting splitters. Similarly, we need to check if splitters contain end-of-string terminators, and skip the recursion in this case.

**Fitting Splitters into Cache.** For current 64-bit machines with 256 KiB L2 cache, we could use $v = 8191$. Note that the limiting data structure which must fit into L2 cache is not the splitter tree $t$, which is only 64 KiB for this $v$, but the bucket counter array $b$ containing $2v + 1$ counters, each of which is eight bytes long. So while theory suggests $v = 8191 = 2^{13} - 1$, in the following subsection, we perform an experimental study to determine a good tree size for practice.

### 4.2.4 Experimental Evaluation of $S^5$ Variants

In the previous section, we proposed three different classification kernels for $S^5$: $S^5$-E with an explicit equality check, $S^5$-$U_xI_y$ with $x$ unrolled $\leq$-comparisons prior to the equality check performed as $y$ simultaneous interleaved unrolled descents, and $S^5$-$U_xI_yC$ which modifies the previous version using lemma 4.1 to remove the pre-order splitter array.

We are interested in how the variants fare against each other and in determining good values for $x$, the number of levels in the tree, and $y$, the number of interleaved descents. So we performed a preliminary parameter optimization experiment on two inputs and two machines. We designed a new input, named **Random2**, to require $S^5$ to perform a large number of tree descents. Random2 consists of random ASCII strings with only two characters, '`0`' and '`1`'.

We ran the $S^5$ variants on 2 GiB of Random2 and on a 128 MiB suffix dataset from Wikipedia (Wikip) on A.Intel-1×8 and F.AMD-1×16. Note that $S^5$ uses KRB.radixsort-CI3s as base case sorter for subsets containing less than 32 Ki strings, which actually handles a large portion of the sorting work. Nevertheless, the inputs are large enough to expose differences in the $S^5$ variants.

Figure 4.3 shows the absolute running time of our $S^5$ variants depending on the number of levels $x$ in the splitter tree. For example, for $x = 10$ there are $2^{10} - 1$ splitters, each eight bytes, such that the classification tree has a total of about 8 KiB. The plot shows the running time for $y \in \{1, 2, 4\}$ interleaved tree descents: Random2 are solid lines, and Wikip suffixes are dashed lines.

On both A.Intel-1×8 and F.AMD-1×16, $S^5$-$U_xI_4C$ consistently outperformed the other variants in figure 4.3, despite the extra calculations for lemma 4.1. On F.AMD-1×16 the variant $S^5$-$U_xI_4C$ was faster than $S^5$-$U_xI_4$ by a larger margin than on A.Intel-1×8, possibly because L1/L2 cache latency rises faster for F.AMD-1×16 than for A.Intel-1×8 (see figure 3.11, page 73). The plot highlights that *interleaving* of tree descents is *crucial* for performance, which rules out $S^5$-E. The optimal tree size appears to be around $x = 10$ for Random2, and $x = 14$ for Wikip suffixes. This is surprising because theoretically $x = 13$ is optimal for 256 KiB L2 caches. We decided to use $x = 10$ for all our main experiments.

Since interleaving proved pivotal for performance, we performed another set of preliminary experiments to determine a good parameter $y$. The plot in figure 4.4 shows absolute running time depending on the number of interleaved tree descents $y$ for classification trees with $x \in \{6, 8, 10, 12, 14\}$ levels. Again the lower plot series are for the Random2 input, and the upper plot series for Wikip suffixes.

Clearly, interleaving is very important, but the running time does not improve beyond 3–4 interleaved tree descents. We therefore chose $y = 4$ for our main experiments.

### 4.2.5 Running Sequential String Sorters in Parallel

Super Scalar String Sample Sort is designed with the intent to extract *more information* per (random) access to the characters of a string than 8- or 16-bit radix sort. This is the idea behind the classification tree. $S^5$ is proposed as a good parallel string sorting algorithm because memory bandwidth is more limited when using multiple cores in parallel, as we saw in chapter 3. Hence, high bandwidth algorithms, such as R.mkqs-cache8, should fare worse in a parallel setting.

Another preliminary experiment attempts to highlight this by simply running the sequential algorithms independently on all cores in parallel, and comparing their parallel speed with running the algorithm on just a single core. First, a large string pointer array is allocated with $p \cdot m$ strings. For the parallel run, $p$ threads independently sort $m$ strings each; for the single core run, one thread sorts $m$ strings. The slowdown of the algorithms when run simultaneously over the algorithm run on a single core should give a good idea how efficiently the algorithms utilize the hardware resources. Note that in the parallel run $p$ times more strings are sorted, hence, a lot more work is performed. However, on a perfectly scaling machine, the $p$ processors *should* actually perform $p$ times more work in the same amount of time. But due to the shared memory hardware, machines obviously cannot scale perfectly.

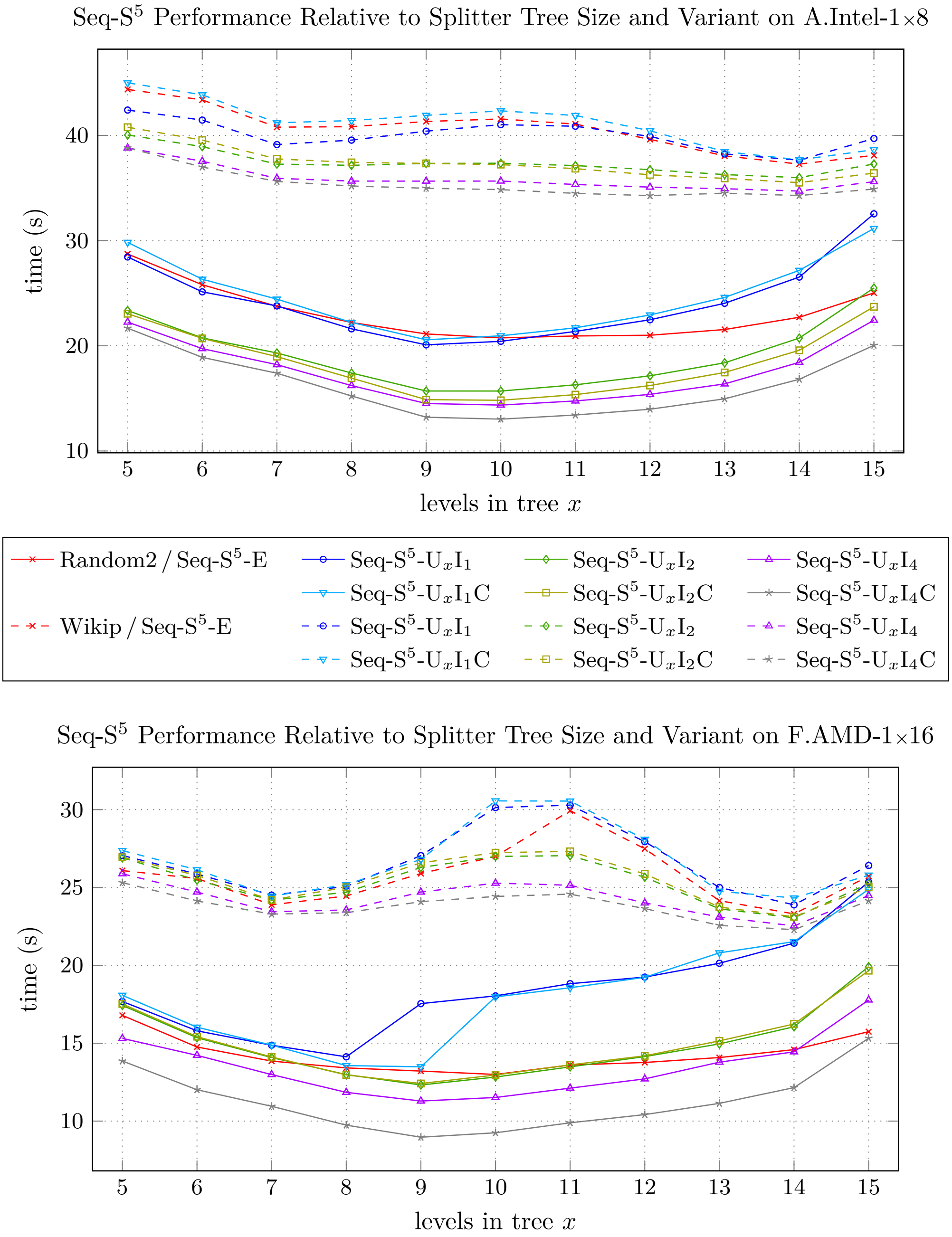

**Figure 4.3:** Sequential $S^5$ running time depending on splitter tree depth and traversal variant on A.Intel-1×8 and F.AMD-1×16.

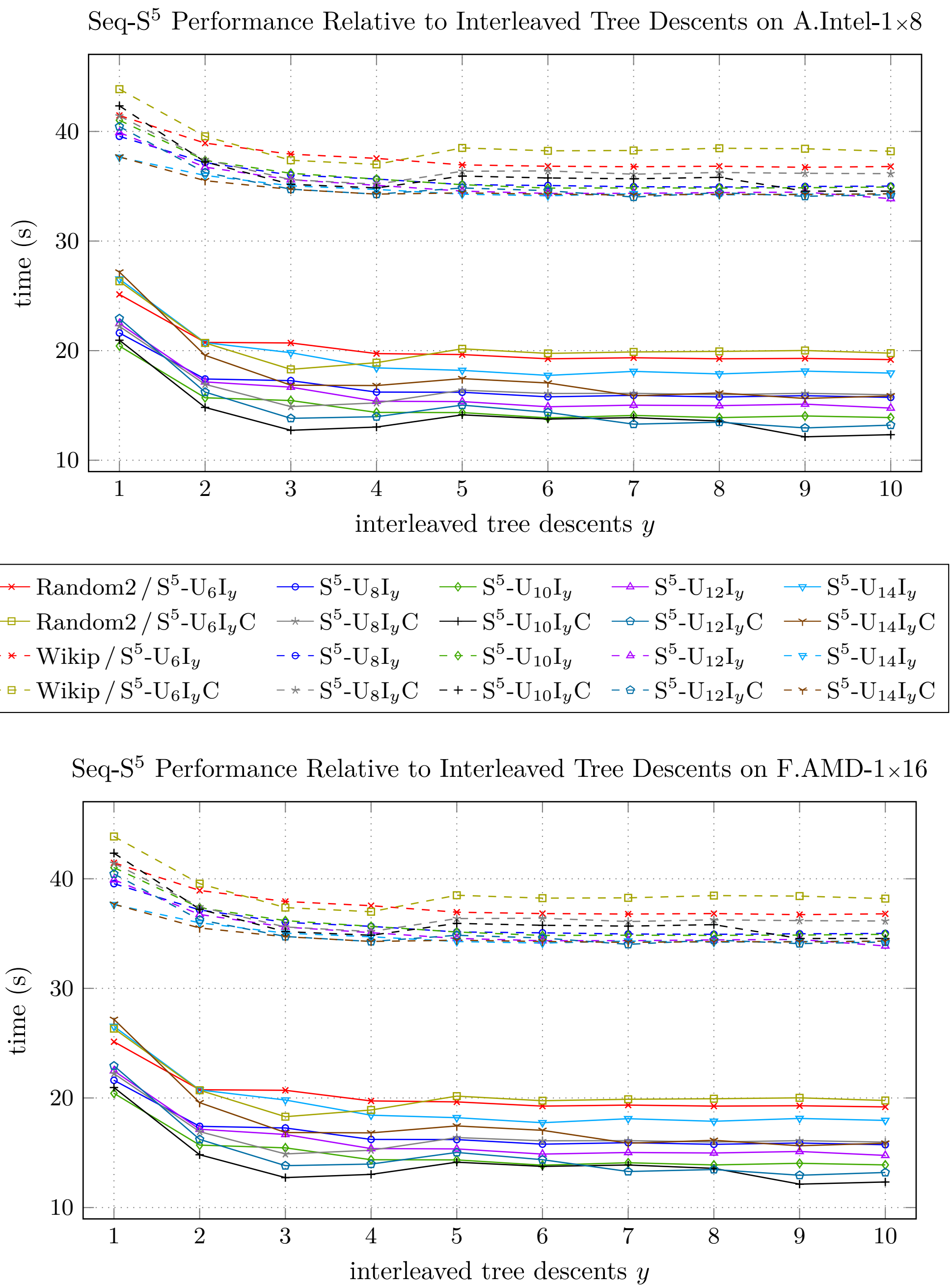

**Figure 4.4:** Sequential S$^5$ running time depending on splitter tree depth and number of interleaved tree descents on A.Intel-1×8 and F.AMD-1×16.

**Table 4.2:** Slowdown factor of running sequential string sorters independently in parallel on all cores (eight on A.Intel-1×8, and sixteen on F.AMD-1×16) over running the sorter on only a single core.

| | Overall | | Our Datasets | | | | Sinha's | | |
|---|---|---|---|---|---|---|---|---|---|
| | Rank | AriM | URLs | Random | GOV2 | Wikip | URLs | DNA | NoDup |
| | | | | A.Intel-1×8 (2008) | | | | | |
| R.mkqs-cache8 | 5 | 2.40 | 3.75 | 1.88 | 2.39 | 2.42 | 2.42 | 2.04 | 1.89 |
| KRB.radixsort-CI3s | 4 | 2.18 | 2.70 | 2.18 | 1.86 | 2.47 | 2.05 | 2.18 | 1.82 |
| KR.radixsort-CE6 | 7 | 2.56 | 3.09 | 2.49 | 2.04 | 2.90 | 2.27 | 2.98 | 2.14 |
| KR.radixsort-CE7 | 6 | 2.53 | 2.94 | 2.46 | 2.06 | 2.88 | 2.21 | 3.01 | 2.12 |
| B.Seq-$S^5$-E | 1 | **1.74** | **1.75** | **1.59** | **1.77** | **2.23** | **1.70** | **1.57** | 1.58 |
| B.Seq-$S^5$-UI | 3 | 1.82 | 1.80 | 1.79 | 1.80 | 2.37 | 1.77 | 1.65 | **1.58** |
| B.Seq-$S^5$-UIC | 2 | 1.80 | 1.81 | 1.74 | 1.78 | 2.35 | 1.73 | 1.60 | 1.59 |
| | | | | F.AMD-1×16 (2017) | | | | | |
| R.mkqs-cache8 | 5 | 3.59 | 5.53 | 3.49 | 3.56 | 3.61 | 2.96 | 3.18 | 2.80 |
| KRB.radixsort-CI3s | 6 | 3.65 | 4.83 | 4.88 | 3.20 | 3.57 | 2.53 | 3.96 | 2.56 |
| KR.radixsort-CE6 | 7 | 3.66 | 4.61 | 4.36 | 3.08 | 3.48 | 2.51 | 4.60 | 2.98 |
| KR.radixsort-CE7 | 4 | 3.59 | 4.38 | 4.39 | 3.19 | 3.32 | 2.27 | 4.58 | 2.98 |
| B.Seq-$S^5$-E | 1 | **2.25** | 2.45 | **2.12** | **2.02** | 2.87 | **1.92** | **2.26** | 2.13 |
| B.Seq-$S^5$-UI | 3 | 2.35 | 2.43 | 2.40 | 2.06 | **2.75** | 2.29 | 2.30 | 2.22 |
| B.Seq-$S^5$-UIC | 2 | 2.30 | **2.38** | 2.35 | 2.03 | 2.83 | 2.07 | 2.37 | **2.08** |

Table 4.2 shows the results of this scaling experiment on A.Intel-1×8 and F.AMD-1×16 with the same inputs as in section 2.3. The table shows only the slowdown factors of running the sequential string sorters in parallel, independently on $p$ cores, over running only a single thread on one core (see tables 4.16 to 4.17 on pages 158 to 159 for the absolute running times). Clearly, B.Seq-$S^5$-E performs best, which means it has the least congestion on common resources. B.Seq-$S^5$-UI and B.Seq-$S^5$-UIC are marginally slower. The fast radix sorts and multikey quicksort have higher slowdown factors, hence, will prospectively deliver worse scaling parallel algorithms on shared-memory machines. R.mkqs-cache8, however, performs better than its high bandwidth requirements suggest. The slowdown effect is more pronounced on F.AMD-1×16 because the machine has fewer memory channels than A.Intel-1×8. In summary, the results match our expectation of sample sort outperforming radix sort in a relative speedup test.

### 4.2.6 Theoretical Run-time Analysis

In this section we show that Sequential Super Scalar String Sample Sort runs in expected $\mathcal{O}(\frac{D}{w} + n\log n)$ time when using equality checks with each splitter, and in expected $\mathcal{O}((\frac{D}{w} + n)\log v + n\log n)$ time when using unrolled tree descents. To achieve this goal, we first show that sample sort for atomic keys still takes $\mathcal{O}(n\log n)$ running time with high probability when run *recursively* with a *constant* number of samples and a *constant* number of splitters. Using this result, we then show the running time bounds for string sample sort by combining the result of recursive sample sort with the analysis of multikey quicksort [BS97].

The work of both algorithms can be viewed as a decision tree. In such a tree, edges correspond to the possible outcomes of comparisons with a splitter and can be labeled with $<$, $=$, or $>$. In multikey quicksort each inner node $z$ of the tree has exactly three children. Each character comparison during partitioning at node $z$ can be associated with the edge determined thereby. By selecting pivots randomly or using a sample median, the expected number of $<$ and $>$ edges in all paths from the root is in $\mathcal{O}(\log n)$ [Hoa62; BS97] since this approach is identical to atomic quicksort. Thus the expected time spent on all comparisons accounted for by these edges is in $\mathcal{O}(n\log n)$. All comparisons associated with $=$ edges correspond to characters from the distinguishing prefix, and are thus bounded by $D$. In total we have $\mathcal{O}(D + n\log n)$ work in the multikey quicksort tree.

We can view string sample sort as a multikey quicksort using multiple pivots in the classification tree, as shown in figure 4.1. However, to further adapt this proof we have to consider the asymptotic running time of sample sort. The issue is that to the best of our knowledge, all analyses in the literature fall into two categories. The first type is for a sequential one-level scenario [FM70; Ape78; Mah00], in which a large number of samples and splitters is determined such that the resulting buckets are small enough to finish with quicksort (or another sorting algorithm). The second type is for distributed scenarios [YHC87; BLM+91; GV92; GV94; HBJ98; HJB98; ABSS15], in which only one round of sample sort is performed to distribute elements roughly evenly onto $p$ processors. Our Sequential Super Scalar String Sample Sort however is different because it is *recursively* applied to unfinished buckets. This is essential for sorting strings, because the depth has to be increased stepwise and because sampling a large number of strings can take a long time in practice. Since it is run repeatedly and recursively it is not necessary to draw a large $\Omega(\log n)$ sample. We are thus first going to prove that a *constant* number of splitters is sufficient for a *constant* number of buckets provided sample sort is run recursively. We call this variant *recursive sample sort.* IPS$^4$o proposed by Axtmann, Witt, Ferizovic, and Sanders [AWFS17] also uses this recursive sample sort approach for atomic keys, but they present an I/O analysis in the PEM model instead of one considering comparisons.

The following proof requires mathematical notation and techniques from probability theory and randomized algorithms. We adopt the notation from standard text-books [MU05; HP18]: $\mathrm{P}[\mathcal{E}]$ is the *probability* of an event $\mathcal{E}$ occurring and $\mathbb{E}(X)$ is the *expected value* of a random variable $X$.

To bound tail probabilities, we will use the following *Chernoff bound* [MU05, theorem 4.5]: given $n$ independent Bernoulli random variables $X_1, \dots, X_n$ with $\mathrm{P}[X_i = 1] = p_i$, $\mathrm{P}[X_i = 0] = 1 - p_i$, and $X = \sum_{i=1}^{n} X_i$, then

$$\mathrm{P}[X \leq (1-\delta)\mathbb{E}(X)] \leq \exp\left(-\frac{\delta^2}{2}\mathbb{E}(X)\right) \tag{4.2}$$

for any $0 < \delta < 1$.

We now present a proof of the running time of recursive sample sort for atomic keys with equality buckets. To simplify the challenge, we assume that all keys are distinct. This assumption does not restrict the proof, because equal keys can be made unique by including their memory address as a second component. Moreover, duplicate keys are handled well by sample sort and hence not a concern of the proof. The following analysis is a combination of a proof of high probability for quicksort's $\mathcal{O}(n \log n)$ running time by Har-Peled [HP14, chapter 10] and a proof of one-level sample sort by Sanders [San18, p. 120–122].

**Theorem 4.2 (Running Time of Recursive (Atomic) Sample Sort)**
*Recursive sample sort with $k$ buckets and oversampling factor $S$ sorts $n$ elements in $\mathcal{O}(n \log n)$ time with high probability (i.e. with probability at least $1 - \frac{1}{n^2}$) if $k$ and $S$ are constants.*

*Proof.* We first focus on one round of sample sort. To partition the input into $k$ buckets, the algorithm selects $k-1$ splitters from $Sk$ randomly picked sample elements $\{s_1, \dots, s_{Sk}\}$ of the input. We will analyze the easier case of selecting samples *with replacement*, such that each element has the fixed chance of $\frac{1}{n}$ to be the sample $s_i$ regardless of other sample choices.

For the analysis, we fix an arbitrary individual element $e_j$ of the *sorted* input $[\, e_1, \dots, e_n \,]$. We are interested in the probability that a bad event occurs for element $e_j$. For one round of sample sort, this is the probability that $e_j$ lands in a "big" bucket due to the samples being picked inconveniently. To formalize this, let $\mathcal{E}_j$ be the event that element $e_j$ falls into a bucket of size $\geq (1+\varepsilon)\frac{n}{k}$ with imbalance factor $\varepsilon > 0$. We will call $\mathcal{E}_j$ an *unlucky round.* A sufficient condition for this event to occur is that there is a sequence of $(1+\varepsilon)\frac{n}{k}$ elements in the sorted input containing $e_j$ and at most $S$ samples. In this case the two splitters defining the bucket will be too far apart. So consider an indicator variable $X_i$ for each sample $s_i$ defined as

$$X_i = \begin{cases} 1 & s_i \in [\, e_j, \dots, e_{j+(1+\varepsilon)\frac{n}{k}-1} \,]\,, \\ 0 & \text{else}\,. \end{cases}$$

As each input element is equally likely to be the sample $s_i$, we have $\mathrm{P}[X_i = 1] = (1+\varepsilon)\frac{n}{k} \cdot \frac{1}{n} = \frac{1+\varepsilon}{k}$. The probabilities of the indicator variable obviously do not depend on $e_j$ or $i$, since only the number of elements in the range is important and all samples are drawn with replacement.

To determine $\mathrm{P}[\mathcal{E}_j]$, let $X = \sum_{i=1}^{Sk} X_i$ be the sum of the indicators. The expected value is simply $\mathbb{E}(X) = Sk \cdot \frac{1+\varepsilon}{k} = S(1+\varepsilon)$. Using the expected value, we can now bound $\mathrm{P}[\mathcal{E}_j]$, the probability of an unlucky round, as $\mathrm{P}[\mathcal{E}_j] \le \mathrm{P}[X < S] = \mathrm{P}[X < \frac{\mathbb{E}(X)}{(1+\varepsilon)}] = \mathrm{P}[X < (1-\frac{\varepsilon}{1+\varepsilon})\mathbb{E}(X)]$. The $X_i$ are independent 0/1 random variables, so the Chernoff bound (equation (4.2)) can be applied: $\mathrm{P}[\mathcal{E}_j] < \exp(-\frac{\varepsilon^2}{2(1+\varepsilon)^2}\mathbb{E}(X)) = \exp(-\frac{S\varepsilon^2}{2(1+\varepsilon)})$. Let $p := \exp(-\frac{S\varepsilon^2}{2(1+\varepsilon)})$ be this constant depending only on the allowed imbalance $\varepsilon$ and oversampling factor $S$.

Since we have bounded the probability of an individual element $e_j$ encountering an unlucky round in a single run of sample sort, we now consider the probability of sufficiently many lucky rounds occurring over multiple runs. As samples are drawn independently in each run, the probability of a *lucky round* (i.e. an unlucky round *not* occurring) is simply $\mathrm{P}[\text{not } \mathcal{E}_j] > 1-p$ in each case.

Let $Y_\ell$ be a Bernoulli random variable that element $e_j$ has a lucky round in level $\ell$ of recursive sample sort. From the previous analysis, we know that $\mathrm{P}[Y_\ell = 1]$ is at least $1-p$. Let $r$ be the maximum number of lucky rounds that an element can participate in before its bucket contains at most one element. This parameter can be determined by considering $n((1+\varepsilon)\frac{1}{k})^r \le 1$ because in each lucky round the bucket containing $e_j$ shrinks by at least a factor of $(1+\varepsilon)\frac{1}{k}$. Solving for $r$ yields $r \ge \log_{\frac{k}{1+\varepsilon}} n$. As $\mathbb{E}(Y_\ell) \ge 1-p$, we thus need $\frac{r}{1-p}$ rounds *in expectation* until a bucket contains only the element $e_j$.

For a bound with high probability, we can consider running $m$ rounds of recursive sample sort and are going to study the tail probability of "too many" unlucky rounds occurring within those $m$ rounds. Let $Y = \sum_{\ell=1}^{m} Y_\ell$ be the sum of the indicator variables for lucky rounds. From the further analysis it becomes clear that the number of rounds $m$ has to be at least $\frac{16r}{1-p} \cdot \ln(\frac{k}{1+\varepsilon}) \ge \frac{16 \ln n}{1-p}$. As all $Y_\ell$ are independent random variables, $\mathbb{E}(Y) \ge m(1-p)$, and we can apply the Chernoff bound again to determine $\mathrm{P}[Y \le (1-\delta)\mathbb{E}(Y)] \le \exp(-\frac{\delta^2}{2}\mathbb{E}(Y)) \le \exp(-\frac{\delta^2 m(1-p)}{2}) \le \exp(-8\delta^2 \ln n) = (\frac{1}{n})^{8\delta^2}$ for any $0 < \delta < 1$. This is the tail probability that an individual element $e_j$ participates in an improbably high number of rounds of recursive sample sort. There are $n$ input elements and their number of rounds are not independent due to the samples. We can however still calculate the union bound of all $n$ being in buckets of size one after $m$ rounds. Hence, the probability that recursive sample sort exceeds $m$ rounds is $1 - n(\frac{1}{n})^{8\delta^2} = 1 - (\frac{1}{n})^{8\delta^2-1}$. For any $(\frac{3}{8})^{\frac{1}{2}} < \frac{3}{4} < \delta < 1$ this probability is less than $1 - \frac{1}{n^2}$.

The total running time of recursive sample sort is proportional to the number of comparisons performed by it. Each comparison corresponds to the decision on which edge to take in the classification tree, except the comparisons needed to sort the sample in each round. We have shown that the number of comparisons needed for sorting $n$ elements is $\mathcal{O}(n \log n)$ with high probability because the maximum depth needed is $m = \mathcal{O}(\log n)$. In each round only a constant number of samples is sorted, and all other comparisons can be attributed to the elements being classified in the tree. Hence in total, recursive sample sort runs in $\mathcal{O}(n \log n)$ time with high probability. □

Before focusing on string sample sort, let us reconsider the previous proof from a practical standpoint. While recursive sample sort was shown to be *asymptotically* optimal, the constant factors (due to $k$ and $S$) for small inputs are definitely larger than for simpler sorting algorithms such as quicksort and mergesort. (Recursive) sample sort is only worthwhile for large inputs, and for moderate and small input sizes other algorithms should be used in the recursion. Furthermore, due to the absolute running time of each round of sample sort, it may be worthwhile to sample more elements in the initial rounds and fewer in lower rounds.

For the context of this dissertation, the main result of the previous proof is that each element is fully classified by recursive sample sort after $\mathcal{O}(\log n)$ steps in expectation. Using this knowledge, we can transfer the reasoning of the proof of multikey quicksort to string sample sort.

**Theorem 4.3 (Complexity of String Sample Sort)**
*String sample sort with implicit binary trees, word parallelism, and equality checking at each splitter node can be implemented to run in expected time* $\mathcal{O}(\frac{D}{w} + n \log n)$*. String sample sort with unrolled tree descents runs in expected time* $\mathcal{O}((\frac{D}{w} + n) \log v + n \log n)$*.*

*Proof.* The sorting operation of string sample sort corresponds to a ternary tree. An $=$ edge matches $w$ characters, of which at least one is from the distinguishing prefix $D$. If any of the $w$ characters is not in the distinguishing prefix, then the $=$ edge leads to a leaf, which corresponds to the final bucket of an element. There are at most $n$ such comparisons leading to leaves, all other $=$ edges match $w$ characters. Thus we have at most $\frac{D}{w} + n$ comparisons attributed to $=$ edges.

From the previous proof, we know that for recursive sample sort on atomic keys the expected number of $<$ and $>$ edges on all paths from the root is $\mathcal{O}(\log n)$. This result transfers to string sample sort: while each $=$ edge increases the depth by $w$, $<$ and $>$ edges partition the string set. Due to the way sample sort classifies string sets into smaller subsets, the number of $<$ and $>$ edges is limited: in total there are only $\mathcal{O}(n \log n)$ $<$ and $>$ edges in expectation. Thus the overall work attributed to them is $\mathcal{O}(n \log n)$ in expectation due to the fixed sample size in each iteration (theorem 4.2). By using the additional LCP information gained at $<$ and $>$ edges from equation (4.1) one could decrease the expected path length from the root further, though probably not asymptotically.

If the $=$ edges are taken immediately, as done in the variant with explicit equality checking at each node, then in total string sample sort takes expected $\mathcal{O}(\frac{D}{w} + n \log n)$ time. However, if we choose to unroll descents of the tree, then the splitter at the root may match and the $\Theta(\log v)$ additional steps down the tree are superfluous. This happens when many strings are identical, and the corresponding splitters are high up in the tree. We thus have to attribute $\mathcal{O}((\frac{D}{w} + n)\log v)$ time to the $=$ edges. Together with the expected cost of $<$ and $>$ edges, this yields in total an expected $\mathcal{O}((\frac{D}{w} + n)\log v + n \log n)$ bound. □

### 4.2.7 Parallelization of S$^5$

For parallelization of S$^5$ we have to reconcile the load balancing perspective of traditional parallel sample sort with the cache efficiency of Super Scalar Sample Sort, where the splitters are designed to fit into cache. We do this by using *dynamic* load balancing which includes parallel execution of recursive calls as in parallel quicksort. Dynamic load balancing is very important and probably unavoidable for parallel string sorting because any algorithm must adapt to the input string array's characteristics.

String sample sort is particularly easy to parallelize for large string sets and current multi-core architectures where $n \gg pv$, and we can state the following theorem. We only consider a single step here, and thus cannot use the distinguishing prefix $D$ to bound the overall work.

**Theorem 4.4 (Parallel Runtime of One Step of pS$^5$)**
*A single step of Super Scalar String Sample Sort (algorithm 4.1) can be implemented to run on a CREW PRAM with $p < \frac{n}{v}$ processors in $\mathcal{O}(\frac{n}{p}\log v + \log p)$ time and $\mathcal{O}(n \log v + pv)$ work.*

*Proof.* Sorting the sample requires $\mathcal{O}(\frac{a \log a}{p} + \log p)$ time and $\mathcal{O}(a \log a)$ work [Col88; Bre74], where $a := \alpha v + \alpha - 1 \ll n$ is the sample size. Selecting the sample, picking splitters, constructing the tree, and saving the LCP of splitters is all $\mathcal{O}(\frac{a}{p})$ time and $\mathcal{O}(a)$ work. Each processors gets $\frac{n}{p}$ strings and in the worst case runs all $\log v$ steps down the classification tree, which takes $\mathcal{O}(\frac{n}{p}\log v)$ time and $\mathcal{O}(n \log v)$ work. Departing from lines 13 to 16 in algorithm 4.1 (page 97), each processor keeps its own bucket array $b_i$ of size $2v+1$, initializes it in $\mathcal{O}(v)$ time, and classifies only those strings in its string set. Then, an interleaved global prefix sum over the $p(2v+1)$ bucket counters yields the boundaries in which each processor can independently redistribute its strings. The prefix sum runs in $\mathcal{O}(\log pv)$ time and $\mathcal{O}(pv)$ work [KS73], while counting and redistribution runs in $\mathcal{O}(\frac{n}{p})$ time and $\mathcal{O}(n)$ work. Summing all time and work yields our result. □

### 4.2.8 Practical Parallelization of S$^5$

While the previous section delivered a theoretical parallelization result, we now focus on a practical implementation. Our parallel S$^5$ (pS$^5$) is composed of four subalgorithms for differently sized subsets of strings. For a string subset $\mathcal{S}'$ from $\mathcal{S}$ with $|\mathcal{S}'| \geq \frac{n}{p}$, a *fully parallel version* of $S^5$ is run, for large sizes $\frac{n}{p} > |\mathcal{S}'| \geq t_m$ a sequential version of $S^5$ is used, and for sizes $t_m > |\mathcal{S}'| \geq t_i$ the fastest sequential algorithm for medium-size inputs (caching multikey quicksort from section 2.2.1) is called, which internally uses insertion sort when $|\mathcal{S}'| < t_i$. We empirically determined $t_m = 2^{20} = 1\,\text{Mi}$ and $t_i = 32$ as good thresholds to switch subalgorithms.

**Fully Parallel $\mathbf{S^5}$.** The *fully* parallel version of $S^5$ uses $p' = \Theta(\frac{p}{n}|\mathcal{S}'|)$ threads for a subset $\mathcal{S}'$ of $\mathcal{S}$ with $|\mathcal{S}| = n$. It consists of four stages: selecting samples and generating a splitter tree, parallel classification and counting, global prefix sum, and redistribution into buckets. Selecting the sample and constructing the search tree are done sequentially, as these steps have negligible running time. Classification is done independently, dividing the string set evenly among the $p'$ threads. The prefix sum is done sequentially once all threads finish counting.

**Out-of-place Redistribution.** In both the sequential and parallel versions of $S^5$ we permute the string pointer array using out-of-place redistribution into an extra array. In principle, we could do an in-place permutation in the sequential version by walking cycles of the permutation [MBM93] (see also section 2.2.2). Compared to out-of-place copying, the in-place algorithm uses fewer input/output streams and requires no extra space. However, we found that modern processors optimize the sequential reading and writing pattern of the out-of-place version better than the random access pattern of the in-place walking. Furthermore, for fully parallel $S^5$, an in-place permutation cannot be done in the same manner. Therefore, the extra string pointer array of size $n$ is needed anyway, and hence we always use *out-of-place redistribution.* For recursive calls, the role of the extra array and original array are swapped, which saves superfluous copying work.

**Dynamic Load-Balancing and Voluntary Work Sharing.** All work in parallel $S^5$ is dynamically load balanced via a central job queue. We use the lock-free queue implementation from Intel's Thread Building Blocks (TBB) [Rei07] and threads initiated by OpenMP to create a *light-weight thread pool.*

To make work balancing most efficient, we modified all sequential subalgorithms of parallel $S^5$ to use an *explicit* recursion stack. The traditional way to implement dynamic load balancing would be to use work stealing among the sequentially working threads [BS81; ABP98; BL99]. This would require the operations on the local recursion stacks to be synchronized or atomic. However, for our application fast stack operations are crucial for performance as they are very frequent. We therefore choose a different method: *voluntary work sharing* (similar to lazy task creation [MKH91]). If the global job queue is empty and a thread is idle, then a global atomic counter is incremented to indicate that other threads should share their work. These then free the stack level with the *largest subproblems* from their local recursion stack and enqueue these as separate, independent jobs (see figure 4.5). This method avoids costly atomic operations on the local stacks, replacing it by a faster counter check, which itself *need not be synchronized*. The short wait of an idle thread for new work does not occur often because the largest recursive subproblems are shared. Furthermore, the global job queue never gets large because most subproblems are kept on local stacks. Storage of the subproblems is also more compact on the stacks than in the global job queue.

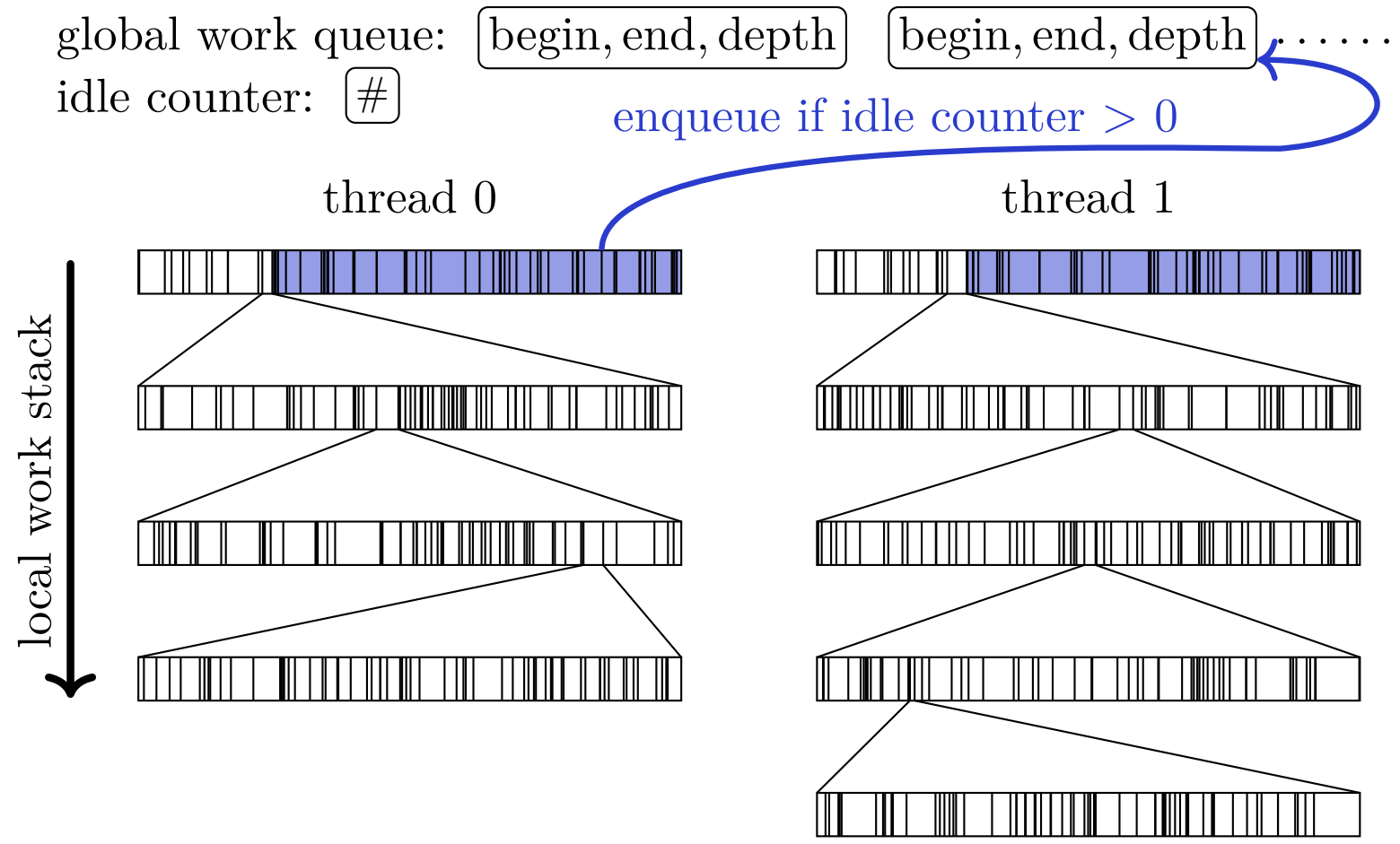

**Figure 4.5:** Threads work on independent parts of the string set from the global work queue. Each thread keeps its own recursion stack, but frees the remaining *top-most* level if any thread is idle.

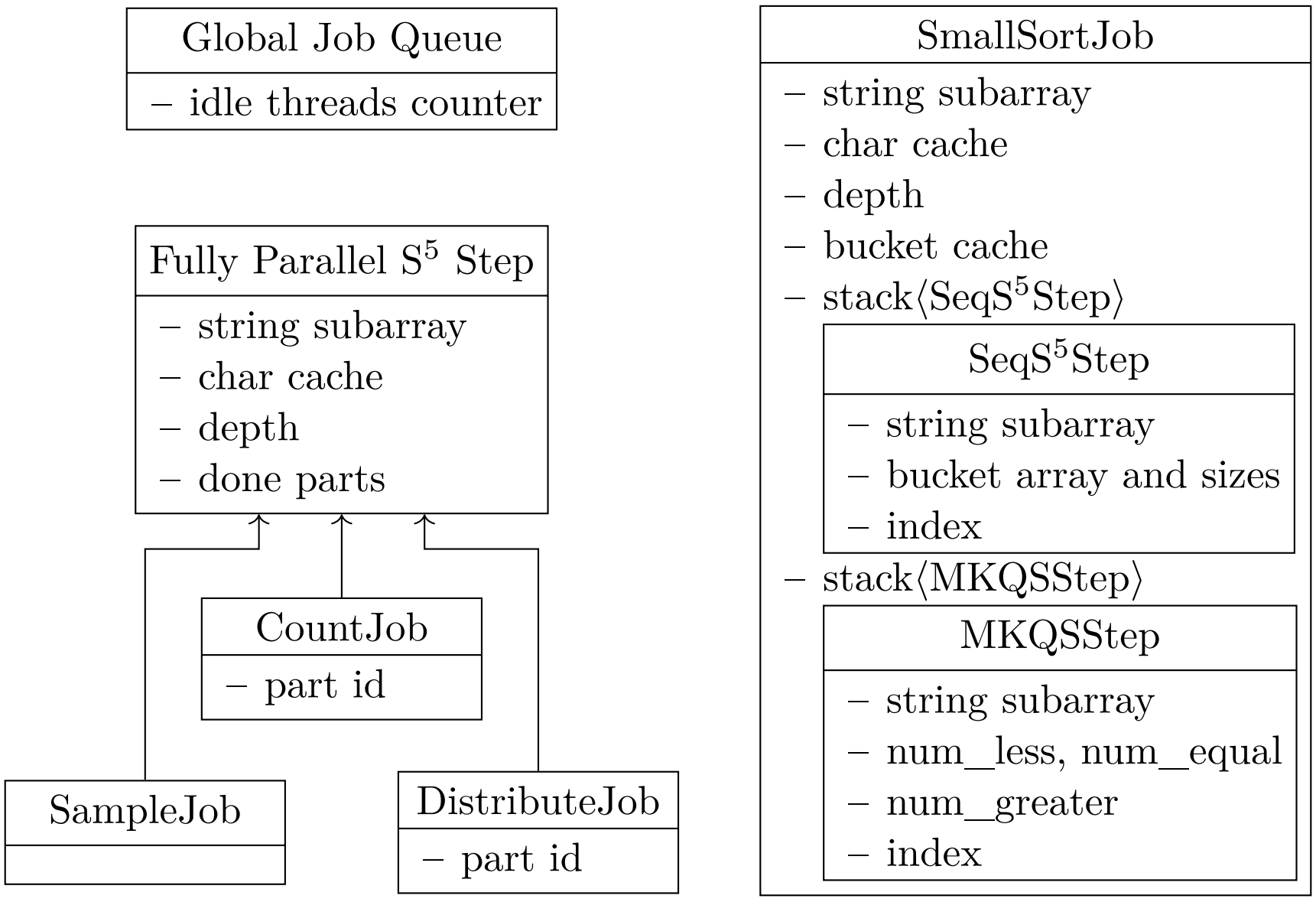

**Figure 4.6:** Schematic breakup of parallel S$^5$ into concurrently running sorting *jobs* dispatched via a central job queue, and recursive sorting *steps* containing the bookkeeping information

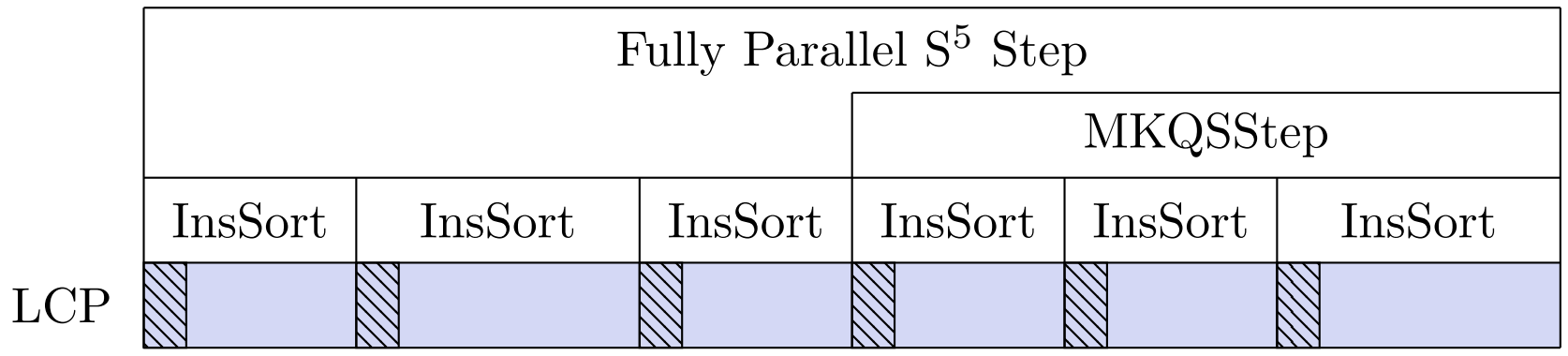

**Figure 4.7:** Hierarchy of sorting steps and LCP array partitions highlighting the first entries. These entries at the bucket boundaries must be calculated when all subarrays of a sorting step have recursive been completed.

**Steps and Jobs.** All work of parallel $S^5$ is broken down into sorting *steps* and sorting *jobs* (see figure 4.6), both coordinated via the central job queue. The goal of a sorting step is to further classify a string set with a known common prefix. Depending on the sorting algorithm, the sorting step data structure contains different fields such as a splitter or bucket array. A sorting job is a piece of code, run by any thread from the central job queue, furthering the completion of a sorting step.

Fully parallel $S^5$ is decomposed into smaller independent jobs, such as sampling the strings (SampleJob, run by one thread), classifying $\frac{n}{p}$ strings (CountJob, run by $p'$ threads), or reordering strings in parallel out-of-place into the target array (DistributeJob, run by $p'$ threads). All fully parallel $S^5$ sorting jobs reference the same sorting step. In this structure we also count how many of the jobs have been completed and spawn new ones. For example, once all $p'$ CountJobs have completed, the interleaved prefix sum can be calculated, and $p'$ DistributeJobs are created. This way jobs from all subalgorithms can run simultaneously, which is necessary for skewed datasets.

Once $|\mathcal{S}'| < \frac{n}{p}$, sorting work is performed by SmallSortJobs which contain a stack of sorting steps. The SmallSortJobs periodically check the idle threads counter of the job queue and voluntarily free their largest sorting step by pushing more SmallSortJobs into the global work queue for better load balancing.

**LCP Calculation.** For running parallel $S^5$ as a subalgorithm of parallel multiway LCP-merge, which we are going to focus on in the next chapter, we extended it to also calculate the LCP array of the sorted string set. While most LCP entries are computed in the base case sorter, for which we use insertion sort, the difficulty is determining the correct LCP values at the boundaries between base case sorting instances (see figure 4.7). These LCP values at the boundaries are undefined with regard to the smaller string subset (it is the first), but valid inside the LCP array for the larger string set.

For multikey quicksort, two of the boundaries values have to be corrected: between the $<$ and $=$ and the $=$ and $>$ parts. We calculate these values after the two insertion sort subproblems are sorted by taking $w$ characters from largest string in the $>$ part and the smallest string in the $>$ part, and calculating their LCP with the $w$ pivot characters.

For string sample sorting, $k - 1$ values between the $k$ buckets must be correctly calculated after each of the $k$ buckets is sorted recursively. Due to the alternating equal- and non-equal buckets of string sample sort, we can calculate the LCP values *directly* from the splitters.

In our experiments, string LCP calculation has little to no overhead when performed integrated into the sorting algorithm as described above. In section 4.5 we present an extensive experimental comparison of all our parallel string sorting algorithm implementations, including $pS^5$.

## 4.3 Parallel Multiway LCP-Mergesort

When designing $pS^5$ we considered L2 cache sizes, word parallelism, superscalar parallelism, and other modern processor features. However, new architectures with large amounts of RAM are now commonly *non-uniform memory access* (NUMA) systems, and the RAM chips are distributed onto different memory banks, called *NUMA nodes*. In preliminary synthetic experiments in section 3.2, access to memory on "remote" nodes was 2–7 times slower than memory on the local socket because the requests must pass over an additional interconnection bus. This latency and throughput disparity brings algorithms for external and distributed memory to mind, but the divide is much less pronounced and block sizes are smaller.

In light of this disparity, we propose *independent string sorters* to be used on each NUMA node, and then *merge* the sorted results. During merging, the amount of information per transmission unit passed via the interconnect (64-byte cache lines) should be maximized. Thus, besides the sorted string pointers, we also want to use LCP information to skip over known common prefixes, and cache the distinguishing characters.

While merging sorted sequences of strings with associated LCP information is a very intuitive idea, remarkably, only one very recent paper by Ng and Kakehi [NK08] fully considers LCP-aware mergesort for strings. They describe *binary* LCP-mergesort and perform an average case analysis yielding estimates for the number of comparisons needed. For the NUMA scenario, however, we need a practical *parallel $K$-way LCP-merge*, where $K$ is the number of NUMA nodes. Furthermore, we also needed to extend our existing string sorting algorithms to store the LCP array.

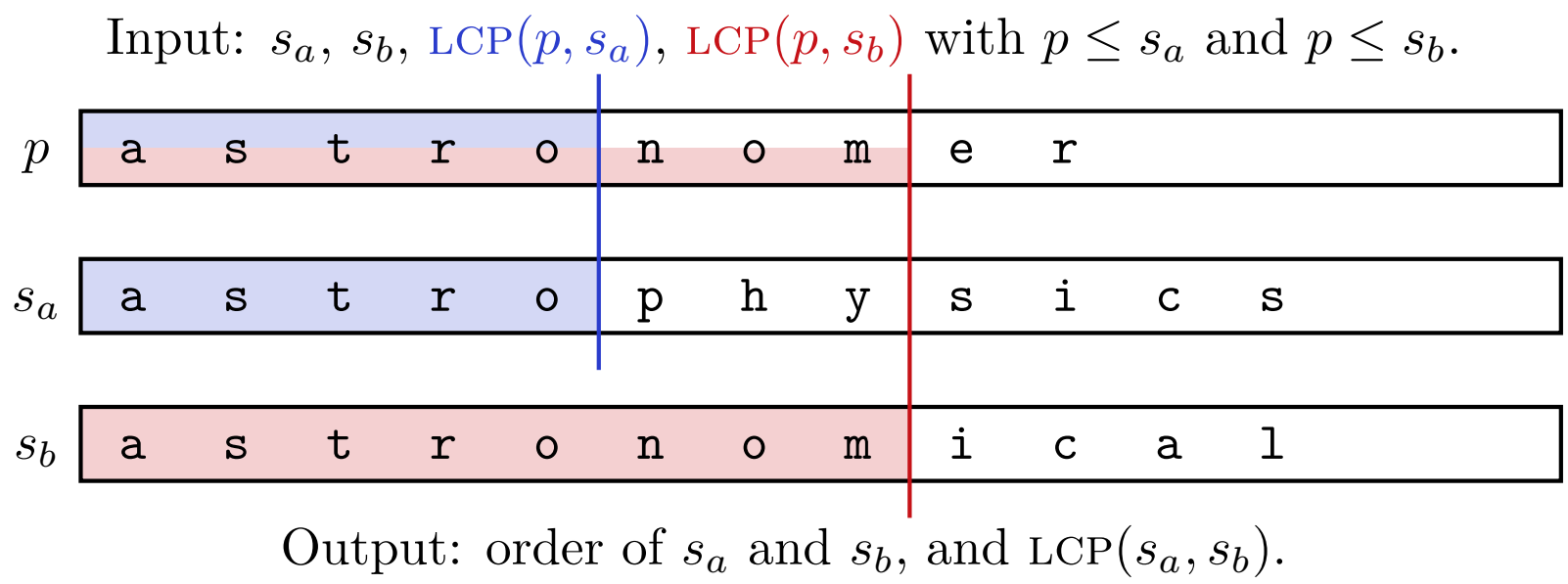

**Figure 4.8:** Visualization of LCP-Compare: $s_a > s_b$, since $p \leq s_a$, $p \leq s_b$, and $\text{LCP}(p, s_a) < \text{LCP}(p, s_b)$ (symbolized by the blue and green bars).

In the next section, we first review binary LCP-aware merging. On this foundation we then propose and analyze (parallel) $K$-way LCP-merging with tournament trees in sections 4.3.2 to 4.3.4. For node-local LCP calculations, we extended pS$^5$ appropriately, and describe the necessary base case sorter, LCP-insertion sort, in section 4.3.5. Further information on the results of this section is available in the bachelor thesis of Andreas Eberle [Ebe14], which we supervised.

Another way to perform a $K$-way LCP-merge is to use the LCP-aware string heap described by Kent, Lewenstein, and Sheinwald [KLS07; KLS12] as an on-demand string sorter for the smallest strings of the $K$ sorted sequences. This solution yields the same asymptotic time bounds as our approach, but as with merging of atomic objects, tournament trees promise smaller constant factors and better practical runtime than heaps. Our extension of tournament trees and insertion sort to use LCP-aware comparisons can be seen as an application of the theoretical “black-box” framework by Amir, Franceschini, Grossi, et al. [AFG+14], which takes any comparison-driven data structure and augments it with LCP-awareness. However, this theoretical framework is complex and does not yield practical algorithms, as the authors themselves note.

### 4.3.1 Binary LCP-Compare and LCP-Mergesort

Here we reformulate the binary LCP-merge and -mergesort presented by Ng and Kakehi [NK08] in a different way. Our exposition is somewhat more verbose than necessary, but this is intentional and lays the groundwork for a simpler description of $K$-way LCP-merge in the following subsection.

Consider the basic comparison of two strings $s_a$ and $s_b$. If there is no additional LCP information, the strings must be compared characterwise until a mismatch is found. However, if we have additionally the LCP of $s_a$ and $s_b$ with a preceding string $p$, then we can first compare these LCP values.

**Algorithm 4.2 :** Binary LCP-Compare

**Function** LCP-Compare$((a, s_a, h_a), (b, s_b, h_b))$
  **Input :** $(a, s_a, h_a)$ and $(b, s_b, h_b)$ where $s_a$ and $s_b$ are two strings together with LCPs $h_a = \textsc{lcp}(p, s_a)$ and $h_b = \textsc{lcp}(p, s_b)$, where $p$ is another string with $p \leq s_a$ and $p \leq s_b$.
  **if** $h_a = h_b$ **then** *// case 1: LCPs are equal ⇒ compare more characters,*
    $h' := h_a$ *// starting at $h' = h_a = h_b$.*
    **while** $s_a[h'] \neq 0$ **and** $s_a[h'] = s_b[h']$ **do** *// Compare characters and*
      $h'$++ *// increase total LCP.*
    **if** $s_a[h'] \leq s_b[h']$ **then return** $(a, h_a, b, h')$
    **else return** $(b, h_b, a, h')$
  **else if** $h_a < h_b$ **then return** $(b, h_b, a, h_a)$ *// case 2: $s_b[h_a{+}1] < s_a[h_a{+}1]$.*
  **else return** $(a, h_a, b, h_b)$ *// case 3: $s_a[h_b{+}1] < s_b[h_b{+}1]$.*
  **Output :** $(x, h_x, y, h')$ with $s_x \leq s_y$, $\{x, y\} = \{a, b\}$, and $h' = \textsc{lcp}(s_a, s_b)$.

**Lemma 4.5 (LCP-Aware String Comparison)**
*Given two strings $s_a$ and $s_b$, and their longest common prefixes $\textsc{lcp}(p, s_a)$ and $\textsc{lcp}(p, s_b)$ with a third string $p$ with $p \leq s_a$ and $p \leq s_b$, then*

*(i) $s_a > s_b$ if $\textsc{lcp}(p, s_a) < \textsc{lcp}(p, s_b)$, and*

*(ii) $s_a < s_b$ if $\textsc{lcp}(p, s_a) > \textsc{lcp}(p, s_b)$.*

*Proof.* Since both reference $p$, we know that $s_a$ and $s_b$ share a common prefix $\min\{\textsc{lcp}(p, s_a), \textsc{lcp}(p, s_b)\}$, and that this common prefix is maximal (i.e. longest). Thus if $\textsc{lcp}(p, s_a) < \textsc{lcp}(p, s_b)$, then the two strings $s_a$ and $s_b$ differ at position $\ell := \textsc{lcp}(p, s_a) + 1$: $s_a[\ell] \neq s_b[\ell]$ (compare also figure 4.8). If we now furthermore consider $\textsc{lcp}(p, s_b) > \ell$, then we immediately see $p[\ell] = s_b[\ell]$. Together with $p \leq s_a$, we have $p[\ell] = s_b[\ell] < s_a[\ell]$ and $s_a > s_b$. The argument can be applied symmetrically if $\textsc{lcp}(p, s_a) > \textsc{lcp}(p, s_b)$. □

There remains the case $\textsc{lcp}(p, s_a) = \textsc{lcp}(p, s_b)$. Here, the LCP information only reveals that both have a common prefix $\textsc{lcp}(p, s_a)$, and additional character comparisons starting at the common prefix are necessary to order the strings.

The pseudocode in algorithm 4.2 implements all three cases. In preparation for $K$-way LCP-merge, the function LCP-Compare additionally takes variables $a$ and $b$, which are corresponding indices and returns these instead of $s_a$ or $s_b$. It also calculates more information than just the order of $s_a$ and $s_b$, since future LCP-aware comparisons also require $\textsc{lcp}(s_a, s_b)$.

In the cases where $\textsc{lcp}(p, s_a) \neq \textsc{lcp}(p, s_b)$, the $\textsc{lcp}(s_a, s_b)$ is easily inferred since the character after the smaller LCP differs in $s_a$ and $s_b$. From this follows $\textsc{lcp}(s_a, s_b) = \min\{\textsc{lcp}(p, s_a), \textsc{lcp}(p, s_b)\}$ as already stated above. For $\textsc{lcp}(p, s_a) = \textsc{lcp}(p, s_b)$ each

**Algorithm 4.3 :** Binary LCP-Merge

**Input :** $\mathcal{S}_1$ and $\mathcal{S}_2$ two sorted sequences of strings with LCP arrays $H_1$ and $H_2$. Assume sentinels $\mathcal{S}_k[|\mathcal{S}_k|] = \infty$ for $k = 1, 2$, and $\mathcal{S}_0[-1] = \varepsilon$.

1 $i_1 := 0, \quad i_2 := 0, \quad j := 0$ — *// Indices for $\mathcal{S}_1$, $\mathcal{S}_2$ and $\mathcal{S}_0$.*

2 $h_1 := 0, \quad h_2 := 0$ — *// Invariants: $h_k = \text{LCP}(\mathcal{S}_0[j-1], \mathcal{S}_k[i_k])$ for $k = 1, 2$*

3 **while** $j < |\mathcal{S}_1| + |\mathcal{S}_2|$ **do** — *// and $j = i_1 + i_2$.*

4     $s_1 := \mathcal{S}_1[i_1], \quad s_2 := \mathcal{S}_2[i_2]$ — *// Fetch strings $s_1$ and $s_2$,*

5     $(x, \bot, y, h') :=$ LCP-Compare$((1, s_1, h_1), (2, s_2, h_2))$ — *// compare them,*

6     $(\mathcal{S}_0[j], H_0[j]) := (s_x, h_x), \quad j{+}{+}$ — *// put smaller into output*

7     $i_x{+}{+}, \quad (h_x, h_y) := (H_x[i_x], h')$ — *// and advance to next.*

**Output :** $\mathcal{S}_0$ contains sorted $\mathcal{S}_1$ and $\mathcal{S}_2$, and $\mathcal{S}_0$ has the LCP array $H_0$.

additionally compared equal character is common to both $s_a$ and $s_b$, and the comparison loop in line 4 of algorithm 4.2 breaks at the first mismatch or zero termination. Thus afterwards $h' = \text{LCP}(s_a, s_b)$, and can be returned as such.

Using LCP-Compare we can now build a binary LCP-aware merging method which merges two sorted string sequences with associated LCP arrays. One only needs to take $s_a$ and $s_b$, compare them using LCP-Compare, write the smaller of them, say $s_a$, to the output and fetch its successor $s'_a$ from the sorted sequence. The written string $s_a$ then plays the role of $p$ in the discussion above, and the next two candidate strings $s'_a$ and $s_b$ can be compared, since $\text{LCP}(p, s_b) = \text{LCP}(s_a, s_b)$ is returned by LCP-Compare and $\text{LCP}(p, s'_a) = \text{LCP}(s_a, s'_a)$ is known from the corresponding LCP array. This procedure is detailed in algorithm 4.3. For binary merging, we can ignore the $h_x$ returned by LCP-Compare. Notice that using the indices $x$ and $y$, the LCP invariant can be restored using just one assignment in line 7. Since algorithm 4.3 is somewhat more complex than necessary due to LCP-Compare, we offer a more straight-forward version of binary LCP-Merge in algorithm 4.4 for reference.

**Theorem 4.6 (Complexity of Binary LCP-Mergesort)**

*Using algorithm 4.3, one can implement a binary LCP-mergesort algorithm which requires at most $L + n\lceil\log_2 n\rceil$ character comparisons and runs in time $\mathcal{O}(D + n\log n)$, where $\mathcal{S}$ is a set of $n$ strings, $L = \sum_{i=1}^{n-1} \text{LCP}_{\mathcal{S}}(i)$, and $D$ the distinguishing prefix of $\mathcal{S}$.*

*Proof.* We assume the divide step of binary LCP-mergesort to do straight-forward halving as in non-LCP mergesort [Knu98], which is why we omitted its pseudocode. Likewise, the recursive division steps have at most depth $\lceil\log_2 n\rceil$ when reaching the base case. If we briefly ignore the character comparison loop in LCP-Compare, line 4, and regard it as a single comparison, then the standard divide-and-conquer recurrence $T(n) \leq T(\lfloor\frac{n}{2}\rfloor) + T(\lceil\frac{n}{2}\rceil) + n$ of non-LCP mergesort still holds. Regarding the character comparison loop, we can establish that each increment of $h'$ ultimately increases the overall LCP sum by exactly one, since in all other statements LCPs are only moved, swapped or stored, but never decreased or discarded. Another way to see this is that the character comparison loop is the only place where characters are compared, thus to be able to establish the correctly sorted order, all distinguishing characters must be compared here.

We regard the three different comparison expressions in lines 4 to 6 as one ternary comparison, as the same values are checked again and zero-terminators can be handled using flag tests. To count the total number of comparisons, we can thus account for all *true*-outcomes of the while loop condition in LCP-Compare (line 4) using $L$, and all *false*-outcomes using $n\lceil\log_2 n\rceil$, since this is the highest number of times case 1 can occur in the mergesort recursion. This is an upper bound, and for most string sets, cases 2 and 3 reduce the number of comparisons in the second term. Since $L \leq D$, the time complexity $\mathcal{O}(D + n \log n)$ follows immediately. □

Ng and Kakehi [NK08] do not give an explicit worst case analysis. Their average case analysis shows that the total number of character comparisons of binary LCP-mergesort is about $n(\mu_a - 1) + P_\omega n \log_2 n$, where $\mu_a$ is the average length of distinguishing prefixes and $P_\omega$ the probability of a "breakdown", which corresponds to case 1 in LCP-Compare. Taking $P_\omega = 1$ and $\mu_a = \frac{D}{n}$, their equation matches our worst-case result, except for the minor difference between $D$ and $L$.

### 4.3.2 *K*-way LCP-Merge

To accelerate LCP-merge for current NUMA systems with four or even eight NUMA nodes, we extended the binary LCP-merge approach to *multiway* LCP-merge using tournament trees [Knu98; San99; San00]. We could not find any reference to multiway LCP-merge with tournament trees in the literature, even though the idea to store and reuse LCP information inside the tree is very intuitive. The algorithmic details, however, require precise elaboration. Compared to the LCP-aware string heap [KLS12], our $K$-way LCP-aware tournament tree is only better by constant factors, but these are very important for practical implementations.

As commonly done in multiway mergesort, to perform $K$-way merging one regards selection of the next item as a tournament with $K$ players (see figure 4.9). Players compete against each other using binary comparisons, and these games are organized in a binary tree. Each node in the tree corresponds to one game, and we label the nodes of the tree with the "losers" of that particular game. The "winner" continues upward and plays further games, until the overall winner is determined. The winner is commonly placed on the top, in an additional node, and with this node, the tournament tree contains each player exactly once. Thus the tree has exactly $K$ nodes, since we do not consider the input, output, or players part of the tree. For sorting strings into ascending sequences, the "overall winner" of the tournament is the lexicographically smallest string.

The first winner is determined by playing an initial round on all $K$ nodes from the bottom up. This winner can then be sent to the output, and the next item from the

**Algorithm 4.4 :** Binary LCP-Merge with Integrated LCP-Compare

**Input :** $\mathcal{S}_1$ and $\mathcal{S}_2$ two sorted sequences of strings with LCP arrays $H_1$ and $H_2$. Assume sentinels $\mathcal{S}_k[|\mathcal{S}_k|] = \infty$ for $k = 1, 2$, and $\mathcal{S}_0[-1] = \varepsilon$.

```
i_1 := 0,  i_2 := 0,  j := 0                     // Indices into S_1, S_2 and S_0.
h_1 := 0,  h_2 := 0          // Invariants: h_k = LCP(S_0[j-1], S_k[i_k]) for k = 1,2
while j < |S_1| + |S_2| do                                     // and j = i_1 + i_2
    s_1 := S_1[i_1],  s_2 := S_2[i_2]  // Fetch strings s_1 and s_2, then first compare LCPs.
    if h_1 = h_2 then          // case 1: LCPs are equal ⇒ compare more characters.
        h' := h_1                                           // Start with base LCP,
        while s_1[h'] ≠ 0 and s_1[h'] = s_2[h'] do         // compare characters and
            h'++                                            // increase total LCP.
        if s_1[h'] ≤ s_2[h'] then        // If s_1 is smaller, place into S_0 with associated
            (S_0[j], H_0[j]) := (s_1, h_1)          // LCP h_1 = LCP(S_0[j-1], s_1), and
            i_1++,  (h_1, h_2) := (H_1[i_1], h')  // advance to H_1[i_1] and h' = LCP(s_1, s_2).
        else                             // If s_2 is smaller, place into S_0 with associated
            (S_0[j], H_0[j]) := (s_2, h_2)          // LCP h_2 = LCP(S_0[j-1], s_2), and
            i_2++,  (h_1, h_2) := (h', H_2[i_2])  // advance to h' = LCP(s_1, s_2) and H_2[i_2].
    else if h_1 > h_2 then         // case 2: LCP(p, s_1) > LCP(p, s_2) with p = S_0[j-1],
        (S_0[j], H_0[j]) := (s_1, h_1)      // so s_1 < s_2. Place s_1 into S_0, and advance to
        i_1++,  h_1 := H_1[i_1]             // H_1[i_1], keeping h_2 = LCP(p, s_2) = LCP(s_1, s_2).
    else                           // case 3: LCP(p, s_1) < LCP(p, s_2) with p = S_0[j-1],
        (S_0[j], H_0[j]) := (s_2, h_2)      // so s_1 > s_2. Place s_2 into S_0, and advance to
        i_2++,  h_2 := H_2[i_2]             // H_2[i_2], keeping h_1 = LCP(p, s_1) = LCP(s_1, s_2).
    j++
```

**Output :** $\mathcal{S}_0$ contains $\mathcal{S}_1$ and $\mathcal{S}_2$ sorted, and $H_0$ is its LCP array.

corresponding input sequence takes its place. Thereafter, only $\log_2 K$ games must be replayed per output item, since the previous winner only took part in those games along the path from the corresponding input to the root of the tournament tree. This can be repeated until all streams are empty. By using sentinels for empty inputs, special cases can be avoided, and we can assume $K$ to be a power of two, filling up with empty inputs as needed. Thus the tournament tree can be assumed to be a perfect binary tree, and can be stored implicitly in an array. Navigating upward in the tree corresponds to division by two: $\lceil \frac{i}{2} \rceil$ is the parent of child $i$, unless $i = 1$ (note that we use a one-based array here). Thus finding the path from input leaf to root when replaying the game can be implemented very efficiently. Inside the tree nodes, we save the loser *input index* $y_i$, or winner index $w$ (renamed from $y_1$), instead of storing the string $s_i$ or a reference thereto.

We now discuss how to make the tournament tree LCP-aware. The binary comparisons between players are done using LCP-Compare (algorithm 4.2), which may perform explicit character comparisons in case 1. Since we want to avoid comparing characters already found equal, we store an LCP value $h_i$ alongside the loser input index $y_i$ in the tree node. The LCP $h_i$ represents the LCP of the stored losing string $s_{y_i}$ with the particular game's winner string, which passes upward to play further comparisons. If we call the corresponding winner $x_i$, even though it's not explicitly stored, then $h_i = \textsc{lcp}(s_{x_i}, s_{y_i})$.

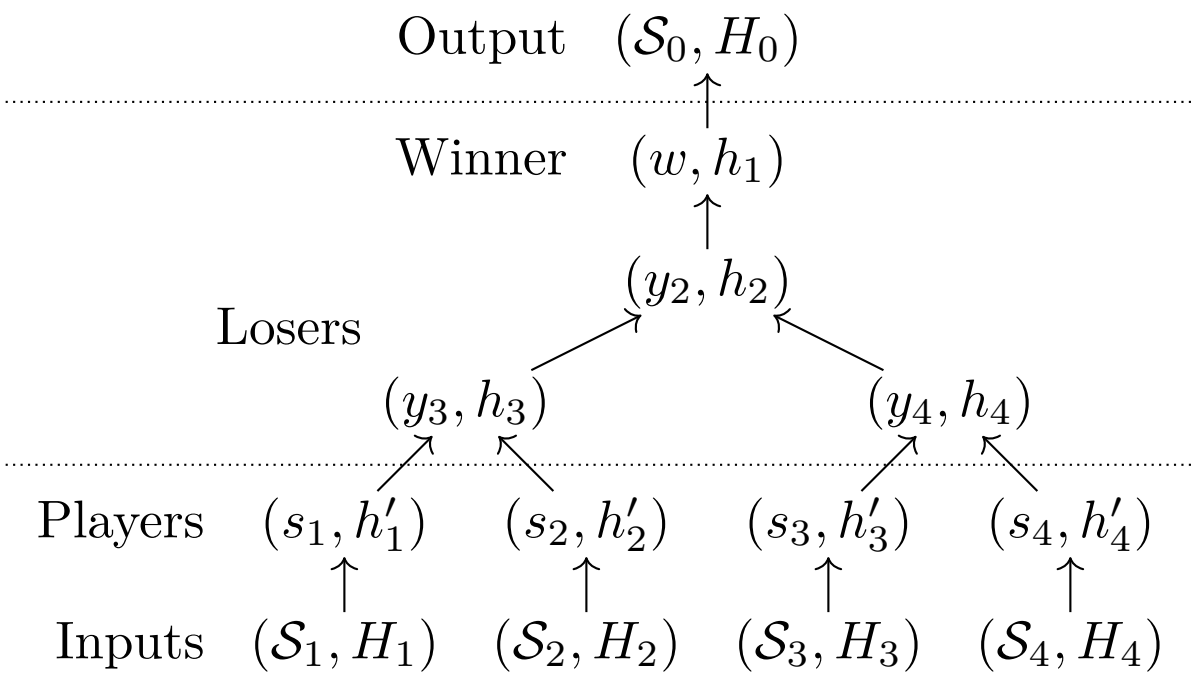

**Figure 4.9:** LCP-aware tournament tree with $K = 4$ showing input and output streams, their front items as players, the winner node $(w, h_1)$, and loser nodes $(y_i, h_i)$, where $y_i$ is the index of the losing player of the particular game and $h_i$ is the LCP of $s_{y_i}$ and the winner of the comparison at node $i$.

After the initial overall winner $w$ is determined, we have to check that all requirements of LCP-Compare are fulfilled when replaying the games on the path from input $w$ to the root. The key argument is that the overall winner $w$ was also the immediate winner of all individual games on the path, while the losers on this path are themselves winners of the subtree leading to the path but not *on* the path. These subtree winners are exactly all players against which $w$ won along the way to being overall winner. The LCP outcome of this game remained unchanged at the loser position, since any winner prior to $w$ cannot have originated below this game.

As a consequence, all games $i$ on that path have $h_i = \textsc{lcp}(s_w, s_{y_i})$. Thus after writing $s_w$ to the output and advancing to the next item $(s'_w, h''_w)$ from the input $(\mathcal{S}_w, H_w)$, we have $p = s_w$ as the common, smaller predecessor string. The previous discussion about the path to the overall winner $w$ is also valid for the path to the individual winner $x_i$ of any node $i$ in the tree, since it is the winner of all games leading from input $x_i$ to node $i$.

The function signature $(x, h_x, y, h_y) :=$ LCP-Compare$((a, s_a, h_a), (b, s_b, h_b))$ was designed to be played by two nodes $(a, h_a)$ and $(b, h_b)$ of the LCP-aware tournament tree. When replaying a path, we can picture a node $(a, h_a)$ moving "upward" along the edges. LCP-Compare is called with this moving node and the loser information $(b, h_b) := (y_i, h_i)$ saved in the encountered node $i$. After performing the comparisons,

**Algorithm 4.5 :** $K$-way LCP-Merge

**Input :** $\mathcal{S}_1, \ldots, \mathcal{S}_K$ sorted sequences of strings with LCP arrays $H_1, \ldots, H_K$ and common prefix $\overline{h}$. Assume sentinels $\mathcal{S}_k[|\mathcal{S}_k|] = \infty$ for $k = 1, \ldots, K$, and $K$ a power of two.

1 $i_k := 0 \; \forall \, k = 1, \ldots, K, \quad j := 0$ *// Initialize indices for* $\mathcal{S}_1, \ldots, \mathcal{S}_K$ *and* $\mathcal{S}_0$.
2 **for** $k := 1, \ldots, K$ **do** *// Initialize loser tree, building*
3     $s_k := \mathcal{S}_k[i_k]$ *// perfect subtrees left-to-right.*
4     $(x, h') := (k, \overline{h}), \quad v := K + k$ *// Play from input node* $v$*, upward till the root*
5     **while** $v$ is even **and** $v > 2$ **do** *// of a perfect odd-based subtree is reached.*
6         $v := \frac{v}{2}, \quad (x, h', y_v, h_v) :=$ LCP-Compare$((x, s_x, h'), (y_v, s_{y_v}, h_v))$
7     $v := \lceil \frac{v}{2} \rceil, \; (y_v, h_v) := (x, h')$ *// Save winner node at top of odd-based subtree.*
8 $w := y_1$ *// Initial winner after all games (rename* $y_1 \to w$*).*
9 **while** $j < \sum_{k=1}^{K} |\mathcal{S}_k|$ **do** *// Loop until output is done.*
10     $(\mathcal{S}_0[j], H_0[j]) := (s_w, h_1), \quad j{+}{+}$ *// Output winner string* $s_w$ *with LCP* $h_1$.
11     $i_w{+}{+}, \quad s_w := \mathcal{S}_w[i_w]$ *// Replace winner with next item from input.*
12     $(x, h') := (w, H_w[i_w]), \quad v := K + w$ *// Play from input node* $v$*, all games*
13     **while** $v > 2$ **do** *// upward to root (unrollable loop).*
14         $v := \lceil \frac{v}{2} \rceil, \quad (x, h', y_v, h_v) :=$ LCP-Compare$((x, s_x, h'), (y_v, s_{y_v}, h_v))$
15     $(w, h_1) := (x, h')$ *// Save next winner at top.*

**Output :** $\mathcal{S}_0$ contains sorted $\mathcal{S}_1, \ldots, \mathcal{S}_K$ and has the LCP array $H_0$

the pair $(x, h_x)$ is the winner node, which passes upwards, and $(y, h_y)$ is the loser information, which is saved in the node $i$. Thus LCP-Compare effectively selects the winner of each game, and computes the loser information for future LCP-aware comparisons. Due to the recursive property discussed in the previous paragraph, the requirements of LCP-Compare remain valid along all paths and LCP-Compare can switch between them.

This LCP-aware $K$-way merging procedure is shown in pseudocode in algorithm 4.5. We build the initial tournament tree incrementally from left to right, playing all games only on the right-most path of every odd-based perfect subtree. This right-most side contains only nodes with even index.

The pseudocode works with one-based arrays, but our implementation uses a zero-based implicit tree, which reduces the number of operations slightly due to rounding up or down. The pseudocode also contains one superfluous run of lines 11 to 15, which occurs at the end and only calculates a sentinel element when all streams are empty. We keep the current representation for better exhibition, as it separates initialization from output iterations. The extra run can be removed by moving the output instruction (line 10) to the end of the second loop, and replacing the last run ($k = K$) of the first loop with a run of second loop by setting $w = K$ and $i_w = 0$.

The following theorem considers only a single execution of $K$-way LCP-merging, since this is what is needed in our NUMA scenario:

**Theorem 4.7 (Complexity of Multiway LCP-Merge)**
*Algorithm 4.5 requires at most $\Delta L + n \log_2 K + K$ character comparisons, where $n = |\mathcal{S}_0|$ is the total number of strings and $\Delta L = L(H_0) - \sum_{k=1}^{K} L(H_k)$ is the sum of increments to LCP array entries.*

*Proof.* We need only consider the character comparisons in the subfunction LCP-Compare, since algorithm 4.5 itself does not contain any character comparisons. As in the proof of theorem 4.6, we can account for all *true*-outcomes of the while loop condition in LCP-Compare (line 4) using $\Delta L$ since it increments the overall LCP. The number of *false*-outcomes can be bounded by considering the number of calls to LCP-Compare, which occur exactly $K$ times when building the tournament tree, and then $\log_2 K$ times for each of the $n$ output string (excluding the superfluous run in the pseudocode). As before, this upper bound, $\Delta L + n \log_2 K + K$, is only attained in pathological cases, and for most inputs, cases 2 and 3 in LCP-Compare reduce the overall number of character comparisons significantly. □

**Theorem 4.8 (Complexity of Multiway LCP-Mergesort)**
*Using algorithm 4.5 one can implement a $K$-way LCP-mergesort algorithm which requires at most $L + n\lceil \log_K n \rceil \log_2 K + (n-1)\frac{K}{K-1}$ character comparisons and runs in time $\mathcal{O}(D + n \log n + \frac{n}{K})$.*

*Proof.* We assume the divide step of $K$-way LCP-mergesort to split the string set into $K$ subproblems of nearly equal size. Applying theorem 4.7 yields the recurrence $T(n) = K \cdot T(\frac{n}{K}) + n \log_2 K + K$ with $T(1) = 0$ if one omits the character comparisons loop. Assuming $n = K^d$ for some integer $d$, the recurrence can be solved elementarily using induction, yielding $T(n) = n \log_K n \cdot \log_2 K + \frac{K(n-1)}{K-1} = n \log_2 n + (n-1) + \frac{n-1}{K-1}$. For $n \neq K^d$, the input cannot be split evenly into recursive subproblems. However, to keep this analysis simple, we use $K$-way mergesort even when $n < K$, and thus incur the cost of theorem 4.7 also at the base level. So, we have $\lceil \log_K n \rceil$ levels of recursion. As in previous proofs, all matching character comparisons are accounted for with $L$, and all others with the highest number of occurrences of case 1 in LCP-Compare in the whole recursion, which is $T(n)$. Since $L \leq D$, the running time follows. □

The proof assumes the use of $K$-way LCP-merge even in the base level. In an implementation, one would rather choose a different merger when $n < K$. By selecting 2-way LCP-merge, the number of comparisons in the lowest level of recursion is reduced, and we can get a bound of $L + n \log_2 n + \mathcal{O}(\frac{n}{K})$.

### 4.3.3 Practical Parallelization of $K$-way LCP-Merge

We now discuss how to parallelize $K$-way LCP-merge of $K$ sorted input streams. The problem is that merging itself cannot be parallelized without significant overhead [Col88], as opposed to the classification and distribution in $pS^5$. Instead, we want to split the problem into *disjoint areas* of independent work, as is commonly done in practical parallel multiway mergesort algorithms and implementations [AS87; SSP07].

In contrast to atomic merging, a perfect split with respect to the number of elements in the subproblems *by no means* guarantees good load balance for string merging. Rather, the amount of work in each piece depends on the unknown lengths of the common prefixes. Therefore, *dynamic load balancing* is necessary in any case and one can settle for a simple and fast routine for splitting the input into pieces that are small enough to allow good load balancing. This also means that it is vital to split the workloads into many more portions than $p$ (a technique we will call *overpartitioning*), again because there is no way of predicting the amount of work per area.

We developed three approaches: naive binary splitting, multiway splitting, and a new LCP splitting approach.

**Binary and Multiway Splitting.** Multiway splitting is the standard approach for parallel multiway mergesort on multi-core architectures: each of the $p$ processors first sorts $\frac{n}{p}$ items, then the $p$ sorted sequences are split into $p$ areas each, such that each processor multiway merges $p$ subsequences of the sorted sequences to generate $\frac{n}{p}$ items of the output.

This approach is adapted for our parallel LCP-merging by randomly sampling $fp$ splitter strings from the $K$ inputs (with overpartitioning factor $f$) and performing a full-depth string sort of the splitters using multikey quicksort. The boundaries determined by these splitters are then located using binary search in each of the $K$ sorted inputs (figure 4.10). This *sampling* approach was faster than Varman, Scheufler, Iyer, and Ricard's [VSIR91] *perfect* multiway splitting technique, probably due to the high cost of comparing strings.

We simplified this approach to *binary* splitting due to the observation that the multiway approach takes a considerable amount of time, especially for larger numbers of processors. During this time, only one processor performs the splitting, as we did not parallelize it. This caused a large start-up delay for the sorting algorithm due to the *first split*, which also operates on the largest string set. To alleviate this problem, we decided to try splitting the input into only *two* subareas (see also figure 4.11), which are then recursively split by other processors, thereby parallelizing the splitting operations.

**LCP Splitting.** The previous two splitting methods ignored the fact that we are processing strings, and string sorting is notable in that access to string characters incurs costly cache faults. We thus focused on using the information in the LCP array to help split the input streams. We developed the following heuristic, which basically merges the top of the LCP interval trees [AKO04] of the $K$ input streams to find independent areas.

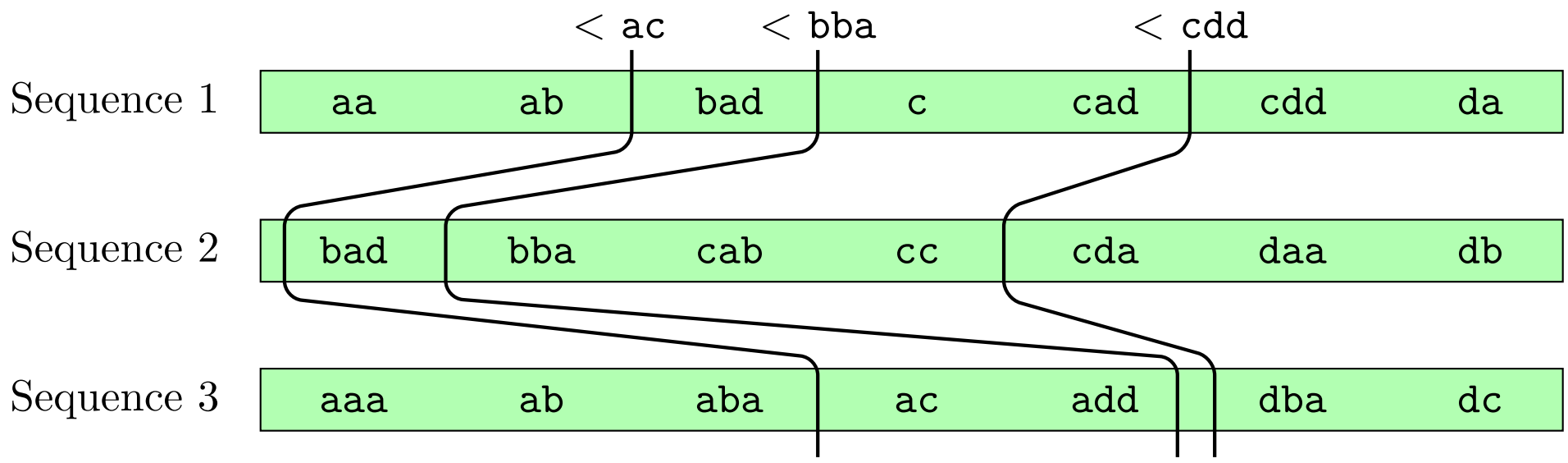

**Figure 4.10:** Multiway splitting technique, adapted from [Ebe14].

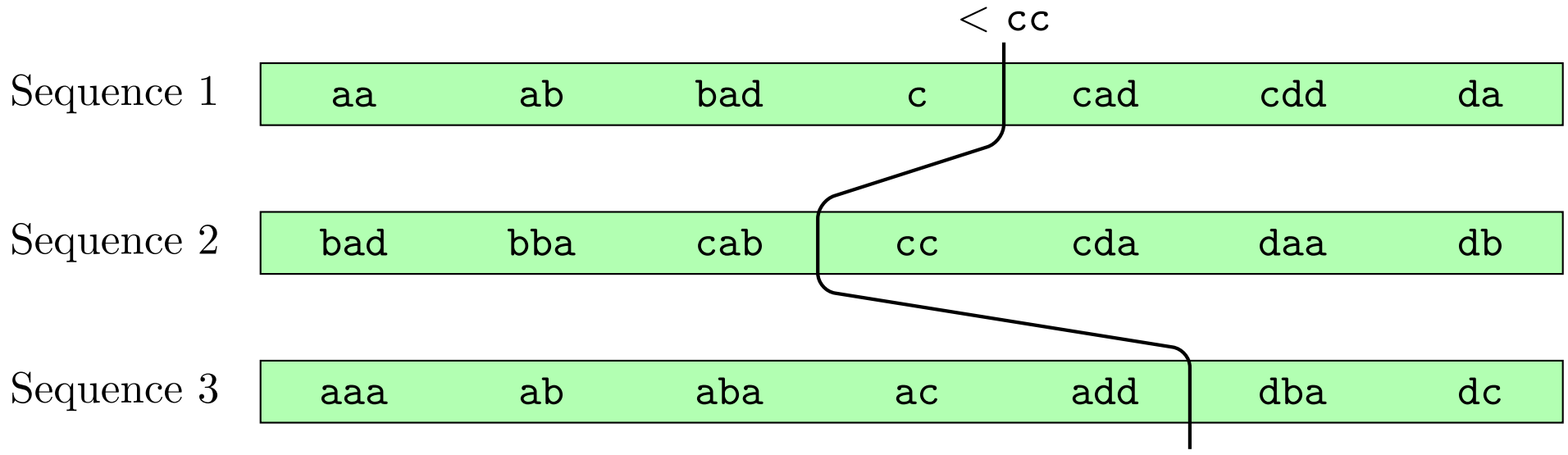

**Figure 4.11:** Binary splitting technique, adapted from [Ebe14].

The essential fact used for LCP splitting is that the input string set is split into disjoint areas starting with *distinct* prefixes by the occurrences of the global minimum in an LCP array. The only remaining challenge is to match equal prefixes from the $K$ input streams, and for this matching we need to inspect the first distinguishing characters of any string in the area. Matching areas can then be merged independently.

Depending on the input, considering only the global LCP minima may not yield enough independent work. However, we can apply the same splitting method again to matching subareas, within which all strings have a longer common prefix and the global minimum of the subarea is larger.

These ideas are combined in a splitting heuristic, which scans the $K$ input LCP arrays sequentially once and creates merge jobs while scanning. We start by reading $w$ characters from the first string of all $K$ input streams and select those inputs with the smallest character block $\bar{c}$. In each of these selected inputs, the LCP array is scanned forward, skipping over all entries $> w$ and checking entries $= w$ for equal character blocks until either an entry $< w$ or a mismatching character block is found. This forward scan encompasses all strings with prefix $\bar{c}$, and an independent merge job can be started. The process is then repeated with the next strings on all $K$ inputs.

The heuristic is started with $w = 8$ (loading a 64-bit register full of characters), but reduces $w$ depending on how many jobs are started, as otherwise the heuristic may create too many splits, e.g. for random input strings. Therefore, an *expected* number of independent jobs is calculated and $w$ is adapted depending on how much input is left and how many jobs were already created. This adaptive procedure keeps $w$ high for inputs with high average common prefix and low otherwise.

Figure 4.12 shows a sorted sequence of strings with its corresponding LCP array visualized as red lines at the appropriate height. In the example, the minimum LCP is $h = 2$ and can be found at the four positions 4, 6, 11, and 17, dividing the sequence into the five disjoint areas $[0..4)$, $[4..6)$, $[6..11)$, $[11..17)$, and $[17..20]$. As described before, the minimum LCP in these areas is of height at least $h + 1 = 3$ and all strings in an area have a common character at index $h + 1 = 3$.

Depending on the input data and alphabet, splitting only at positions of global LCP minimum $h$ might not yield enough independent merge jobs. However, the same approach can be applied to subareas of already splitted regions, since they can be considered to be independent sequences of their own. Due to the fact that the independent subregions created in the first run have a minimum LCP of at least $h+1$, the minimum LCP in these areas will also be at least $h + 1$.

The implementation of parallelized $K$-way LCP-merge uses the same load balancing framework as $\mathrm{pS}^5$ (see section 4.2.8). During merge jobs, the global unsynchronized counter variable is checked to determine whether other threads are idle. To reduce balancing overhead, the counter is read only once every 4 096 processed strings. If idle threads are detected, a $K$-way merge job is further split up into independent jobs using the same splitting heuristic, except that a common prefix of all strings may be known and is used to offset the character blocks of size $w$.

### 4.3.4 Implementation Details

Our experimental platforms have $m \in \{4, 8\}$ NUMA nodes, and we use parallel $K$-way LCP-merge only as a top-level merger on $m$ input streams (see figure 4.13). Thus we assume the $N$ inputs characters to be divided evenly onto the $m$ memory nodes. On the individual NUMA memory nodes, we pin $\frac{p}{m}$ threads (rounded suitably) and run $\mathrm{pS}^5$ on the string subset. We do not rebalance the string sets or thread associations when $\mathrm{pS}^5$ finishes early on one of the NUMA nodes, as the synchronization overhead incurred was larger than the gain. For skewed inputs, however, this problem remains to be solved.

Since $K$-way LCP-mergesort requires the LCP arrays of the sorted sequences, we extended $\mathrm{pS}^5$ to optionally save the LCP value while sorting. The string pointers and LCP arrays are kept separately, as opposed to interleaving them as "annotated" strings [NK08]. This was done, because $pS^5$ already requires an additional pointer array during out-of-place redistribution. The additional string array and the original string array are alternated between in recursive calls. When a subset is finally sorted, the correctly ordered pointers are copied back to the original array, if necessary. This allows us to place the LCP values in the additional array.

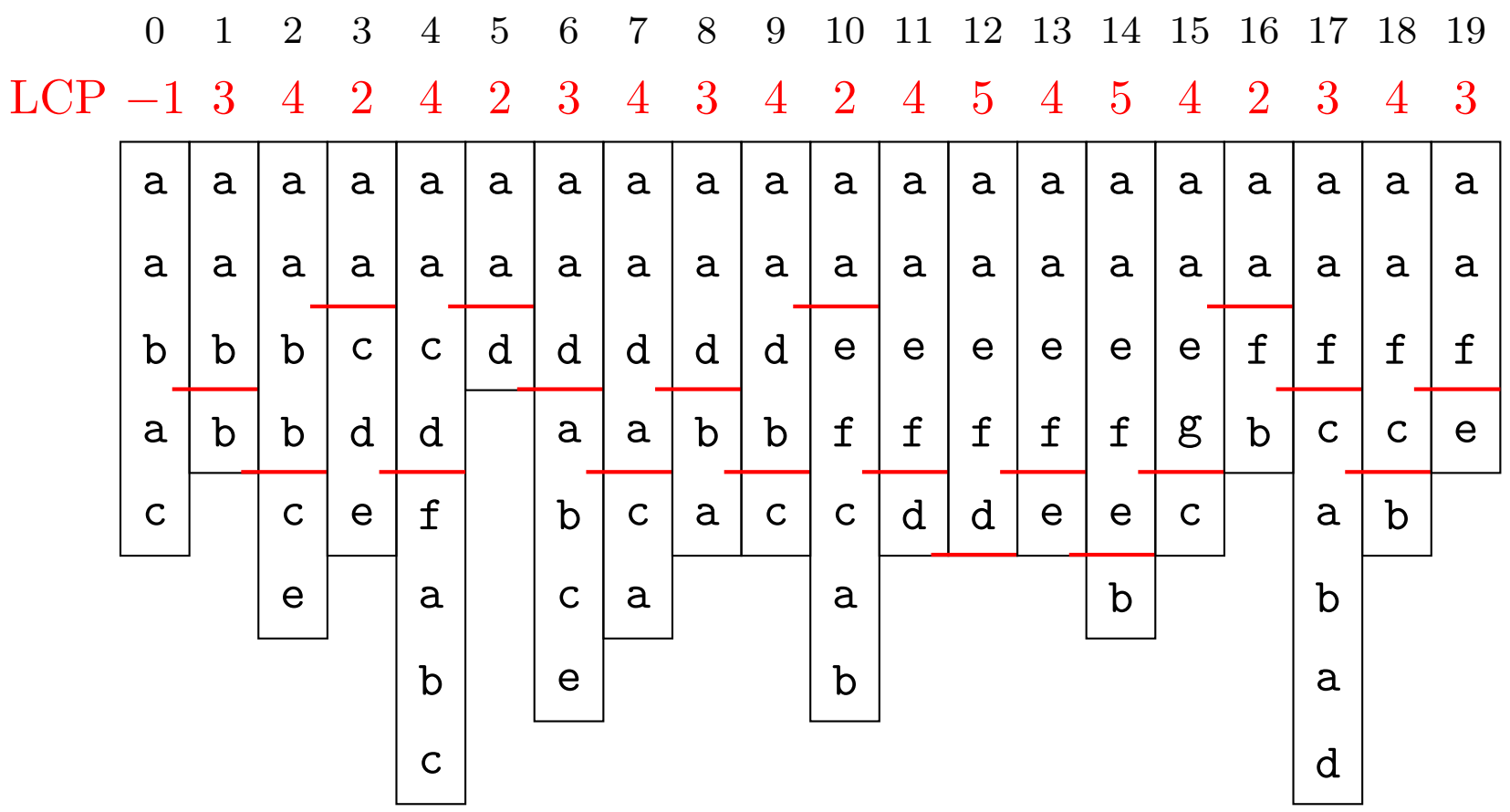

**Figure 4.12:** LCP splitting technique, adapted from [Ebe14].

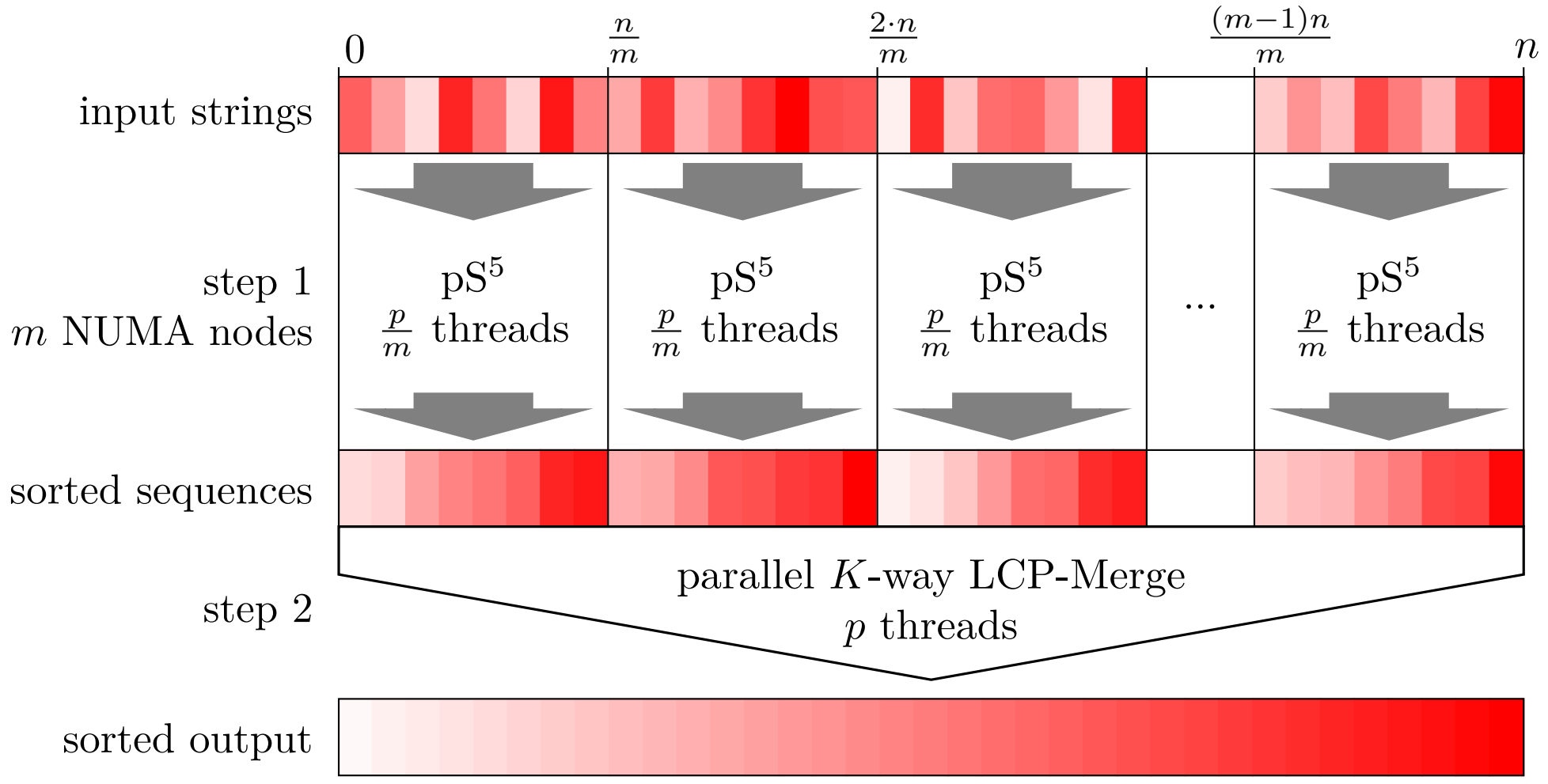

**Figure 4.13:** On NUMA architectures we run $pS^5$ independently on each NUMA node, and then merge the result using parallel $K$-way merge. Figure adapted from [Ebe14].

The additional work and space needed by $pS^5$ to save the LCP values is negligible. Most LCPs are calculated in the base case sorter of $pS^5$, and hence we describe LCP-aware insertion sort in the next section. All other LCPs are located at the boundaries of buckets separated by either multikey quicksort or string sample sort. We calculate these boundary LCPs after recursive levels are finished, and use the saved splitters or pivot elements whenever possible.

The splitting heuristic of parallel $K$-way LCP-merge creates jobs with varying $K$, and we created special implementations for the 1-way (plain copying) and 2-way (binary merging) cases, while all other $K$-way merges are performed using the LCP-aware tournament tree.

To make parallel $K$-way LCP-merge more cache- and NUMA transfer-efficient, we devised a *caching variant*. In LCP-Compare the first character needed for additional character comparisons during the merge can be predicted (if comparisons occur at all). This character is the distinguishing character between two strings, which we label $\hat{c}_i = s_i[h_i]$, where $h_i = \textsc{lcp}_{\mathcal{S}}(i)$. Caching this character while sorting is easy, since it is loaded in a register when the final, distinguishing character comparison is made. We extended $pS^5$ to save $\hat{c}_i$ in an additional array and employ it in a modified variant of LCP-Compare to save random accesses across NUMA nodes. Using this caching variant all character comparisons accounted for in the $n \log_2 K + K$ term in theorem 4.7 can be done using the cached $\hat{c}_i$, thus only $\Delta L$ random accesses to strings are needed for a $K$-way merge.

### 4.3.5 LCP-Insertion Sort

As mentioned in the preceding sections, $pS^5$ was extended to save LCP values. Thus its base-case string sorter, insertion sort, also needed to be extended. Again, saving and reusing LCPs during insertion sort is a very intuitive idea, but we found no reference or analysis in the literature.

Assuming the array $\mathcal{S} = [\, s_0, \dots, s_{j-1}\,]$ is already sorted and the associated LCP array $H$ is known, insertion sort locates a position $i$ at which a new string $x = s_j$ can be inserted while keeping the array sorted. After insertion, at most the two LCP values $h_i$ and $h_{i+1}$ may need to be updated. While scanning for the correct position $i$, customarily from the right, values of both $\mathcal{S}$ and $H$ can already be shifted to allocate a free position.

**Algorithm 4.6 :** LCP-InsertionSort

**Input :** $\mathcal{S} = [\, s_0, \ldots, s_{n-1} \,]$ is an array of $n$ strings with common prefix length $\overline{h}$

```
1  for j = 0, ..., n − 1 do                 // Insert x = s_j into sorted sequence [ s_0, ..., s_{j−1} ].
2      i := j,  x := move(s_j),  h' := h̄   // Start candidate LCP h' with common prefix h̄.
3      while i > 0 do
4          if h_i < h' then  break          // case 1: LCP decreases ⇒ insert after s_{i−1}.
5          else if h_i = h' then            // case 2: LCP equal ⇒ compare more characters.
6              p := h'                                                // Save LCP of x and s_i.
7              while x[h'] ≠ 0 and x[h'] = s_{i−1}[h'] do             // Compare characters.
8                  h'++
9              if x[h'] ≥ s_{i−1}[h'] then     // If x is larger or equal, insert x after s_{i−1},
10                 h_i := h',   h' := p             // set h_i, the LCP of s_{i−1} and x, but
11                 break                                 // set h_{i+1} after loop in line 14.
12         s_i := move(s_{i−1}),  h_{i+1} := h_i // case 3: LCP larger ⇒ no comparison needed.
13         i−−
14     s_i := move(x),  h_{i+1} := h'  // Insert x at correct position, update h_{i+1} with LCP.
```

**Output :** $\mathcal{S} = [\, s_0, \ldots, s_{n-1} \,]$ is sorted and has the LCP array $[\, \bot, h_1, \ldots, h_{n-1} \,]$

Beyond just calculating the LCP array while sorting, we can actually use the information in the preliminary LCP array to accelerate the algorithm. The scan for $i$ can skip over certain areas in which the LCP values attest a mismatch. In a sense, the accelerated scan corresponds to walking down the LCP interval tree, testing only one item of each child node, and descending down if it matches. In plainer words, areas of strings with a common prefix can be identified using the LCP array (as already mentioned in section 4.3.3), and it suffices to *check once* if the candidate matches this common prefix. If not, then the entire area can be skipped.

Algorithm 4.6 presents rather intricate pseudocode of LCP-InsertionSort. The common prefix of the candidate $x$ is kept in $h'$ and increased while characters match. When a mismatch occurs, the scan is continued to the left and all strings can be skipped if the LCP reveals that the mismatch position remains unchanged (case 3). If the LCP goes below $h'$, then a smaller string precedes and therefore the insertion point $i$ is found (case 1). At positions with equal LCP more characters need to be compared (case 2). In the pseudocode these three cases are fused with a copy-loop moving items to right. Note that the pseudocode sets $h_{n+1} := \overline{h}$ in the last iteration when $i = j = n$, which requires a sentinel array position or an out-of-bounds check in the last iteration. In the C++ implementation, all iterations but the last do not have out-of-bounds checks; the last iteration then uses specialized code with checks.

Due to the complex nature of algorithm 4.6, a detailed correctness proof is presented, followed by a time complexity analysis:

**Theorem 4.9 (Correctness of LCP-Aware Insertion Sort)**
*LCP-aware insertion sort (algorithm 4.6) returns the sorted string array $\mathcal{S}$ and associated LCP array $H$ for any given input string set.*

*Proof.* Our first goal is to show that the loop lines 3–13 always finds the correct insertion point $i$ of $x$ into the sorted sequence $[\, s_0, \ldots, s_{j-1}\,]$. During the search for the correct position $i$, the existing LCP array $[\,\bot, h_1, \ldots, h_{j-1}\,]$ and sentinels $h_0 := -1$ and $h_j := \overline{h}$ are used. Consider all remaining candidate insertion points as the range $[a\,..\,b]$. Initially this is $[0\,..\,j]$, and once $[a\,..\,b]$ is a single position with $a = b$, we are done and can insert the string. In the pseudocode $i$ starts at $b$ and $a$ is implicit. To accelerate the search, the common prefix of $x$ with all strings $s_a, \ldots, s_b$ is maintained in $h'$, and thus always requires $h_k \geq h'$ for all $k \in\, ]a\,..\,b[$, and $h_a < h'$, $h_b \leq h'$.

To understand what happens, consider all $k \in\, ]a\,..\,b[$ with $h_k = h'$. These positions are boundaries of string sets which differ starting at character $h'$. The loop 3–13 compares each of these sets with $x$ in case 2, starting with the rightmost set and using the rightmost string. If $x[h'] = s_k[h']$, the search range $[a\,..\,b]$ can be narrowed down to $[a'\,..\,k]$, where $a' = \min\{a' \in [a\,..\,k] \mid h_j > h' \;\forall j \in [a'\,..\,k]\}$. This new range encompasses all strings which have larger common prefix $h' + 1$ with $x$. The search is repeated using the new, narrower range. This corresponds to walking down the LCP interval tree [AKO04].

In the loop 3–13 multiple operations are fused, the left boundary $a$ is handled implicitly and instead of a fixed right boundary $b$, $i$ iterates over strings. Narrowing of $[a'\,..\,k]$ happens in lines 7–8, and the common prefix $h'$ may increase by more than one letter. By using the sentinel $h_j = \overline{h}$, case 2 always occurs for the rightmost string $s_{j-1}$. Instead of a variable $b$ and explicit search for $a'$, the variable $i$ iterates over $h_i > h'$. Case 3 is essentially a fused copy loop over the subset with $x[h'] \neq s_i[h']$, which was checked at the rightmost string of this set. The left boundary $a$ can be identified by $h_i < h'$ or $i = 0$, in which case $x[h'] < s_k[h']$ for all $k \in\, ]a\,..\,b[$. When this occurs as case 1 in line 4, $x$ is inserted between $s_i$ and $s_a$. The only other occurrence of $a = b$ is in line 9, when $x[h'] \geq s_{i-1}[h']$ (where '=' only occurs for $x[h'] = 0$). In this case, $x$ is inserted into $[i\,..\,i]$.

As mentioned above, when inserting $x$ into the sorted sequence $[\, s_0, \ldots, s_{j-1}\,]$ at $i$, at most two LCP values $h_i$ and $h_{i+1}$ need to be modified. In case 1, the LCP value $h_i = \textsc{lcp}(s_{i-1}, s_i) < h'$ need not be modified, as $\textsc{lcp}(x, s_i) = h'$, which means that the replaced string starts with the same $h'$ characters. In case 3 however, both LCP values need to be updated, $h_i = \textsc{lcp}(s_{i-1}, x)$ is set in line 10 as calculated in line 8, whereas $h_{i+1}$ is set when $\textsc{lcp}(x, s_i)$ as saved in line 12.

By showing that each new string $s_j$ is inserted into the correct position, and that the LCP array is updated accordingly, we determined that the algorithm sorts correctly and outputs the correct LCP array. ☐

**Theorem 4.10 (Complexity of LCP-Aware Insertion Sort)**
*LCP-aware insertion sort (algorithm 4.6) requires at most* $L + \frac{n(n-1)}{2}$ *character comparisons and runs in time* $\mathcal{O}(D + n^2)$.

*Proof.* The only lines containing character comparisons in algorithm 4.6 are lines 7 and 9. If the while loop condition is *true*, then $h'$ is incremented. In the remaining algorithm the value of $h'$ is only shifted, never discarded or decreased. Thus we can count the number of comparisons yielding a while-loop repetition with $L$. The while loop is encountered at most $\frac{n(n-1)}{2}$ times, as this is the maximum number of times the inner loop in lines lines 4 to 13 is executed. We can regard the exiting comparison of line 7 and the following comparison in line 9 as one ternary comparison, as the same values are checked again. This ternary comparison occurs at most once for each run of the inner loop, which is at most $\frac{n(n-1)}{2}$ times. With $L \leq D$, the running time follows from the number of iterations of the for loop (lines 1 to 14) and the while loop (lines 3 to 13). □

We close with the remark that non-LCP insertion sort requires $\mathcal{O}(nD)$ steps in the worst case, when all strings are equal except for the last character. Hence, for strings LCP-InsertionSort will often have a better practical running time than plain insertion sort.

## 4.4 More Shared-Memory Parallel String Sorting

### 4.4.1 Parallel Radix Sort

Radix sort is very similar to sample sort, except that classification is much faster and easier. Hence, we can use the same parallelization toolkit as with $S^5$. Again, we use three subalgorithms for differently sized subproblems: fully parallel radix sort for the original string set and large subsets, a sequential radix sort for medium-sized subsets and insertion sort for base cases. Fully parallel radix sort consists of a counting phase, global prefix sum, and a redistribution step. Like in $S^5$, the redistribution is done out-of-place by copying pointers into a shadow array.

We experimented with 8-bit and 16-bit radixes for the fully parallel step. Smaller recursive subproblems are processed independently by sequential radix sort with in-place permuting, and here we found 8-bit radixes to be faster than 16-bit sorting. Our parallel radix sort implementation uses the same load balancing method as parallel $S^5$, freeing the largest subproblems when other threads are idle.

### 4.4.2 Parallel Caching Multikey Quicksort

Our preliminary experiments with sequential string sorting algorithms (see section 2.3) showed a surprise winner: an enhanced variant of multikey quicksort by Rantala [Ran07] often outperformed more complex algorithms.

This variant employs both caching of characters and uses a super-alphabet of $w = 8$ characters, which is exactly the number that fit into a machine word. The string pointer array is augmented with $w$ cache bytes for each string, and a string subset is *partitioned using a whole machine word* as splitter. Thereafter, the cached characters are reused for the recursive subproblems $\mathcal{S}_<$ and $\mathcal{S}_>$, and access to strings is needed only for sorting $\mathcal{S}_=$ unless the pivot contains a zero-terminator. In this chapter "caching" means *copying* of characters into another array, not necessarily into the processor's cache. Key to the algorithm's good performance is the following observation:

**Theorem 4.11 (Character Accesses of Caching Multikey Quicksort)**
*Caching multikey quicksort needs at most $\lfloor \frac{D}{w} \rfloor + n$ (random) accesses to string characters in total, where $w$ is the number of characters cached per access.*

*Proof.* Per string access $w$ characters are loaded into the cache, and these $w$ characters are never fetched again. We can thus account for all accesses to distinguishing characters using $\lfloor \frac{D}{w} \rfloor$, since the characters are fetched in blocks of size $w$. Beyond these, at most one access per string can occur, which accounts for fetching the last $w$ characters in a string of which not all are need for sorting. □

In light of this variant's good performance, we designed a parallelized version. We use three subalgorithms: *fully parallel caching multikey quicksort*, the original sequential caching variant (with an explicit recursion stack) for medium and small subproblems, and insertion sort as base case. For the fully parallel subalgorithm, we generalized a block-wise processing technique from (two-way) parallel atomic quicksort [TZ03] to three-way partitioning.

The input array is viewed as a sequence of blocks containing $B$ string pointers together with their $w$ cache characters (see figure 4.14). Each thread holds exactly three blocks and performs ternary partitioning around a globally selected pivot. When all items in a block are classified as $<$, $=$ or $>$, then the block is added to the corresponding output set $\mathcal{S}_<$, $\mathcal{S}_=$, or $\mathcal{S}_>$. This continues as long as unpartitioned blocks are available. If no more input blocks are available, an extra empty memory block is allocated and a second phase starts. The second partitioning phase ends with fully classified blocks which might be only partially full. Per fully parallel partitioning step there can be at most $3p$ partially filled blocks. The output sets $\mathcal{S}_<$, $\mathcal{S}_=$, and $\mathcal{S}_>$ are processed recursively with threads divided as evenly among them as possible. The cached characters are updated only for the $\mathcal{S}_=$ set.

In our implementation we use atomic compare-and-swap operations for block-wise processing of the initial string pointer array and Intel TBB's lock-free queue for sets of blocks, both as output sets and input sets for recursive steps. When a partition reaches the threshold for sequential processing, a continuous array of string pointers plus cache characters is allocated and the block set is copied into it. On this continuous array, the usual ternary partitioning scheme of multikey quicksort is applied sequentially. Like in the other parallelized algorithms, we use dynamic load balancing and free the largest level when re-balancing is required. We empirically determined $B = 10^{17} = 128\,\text{Ki}$ as a good block size.

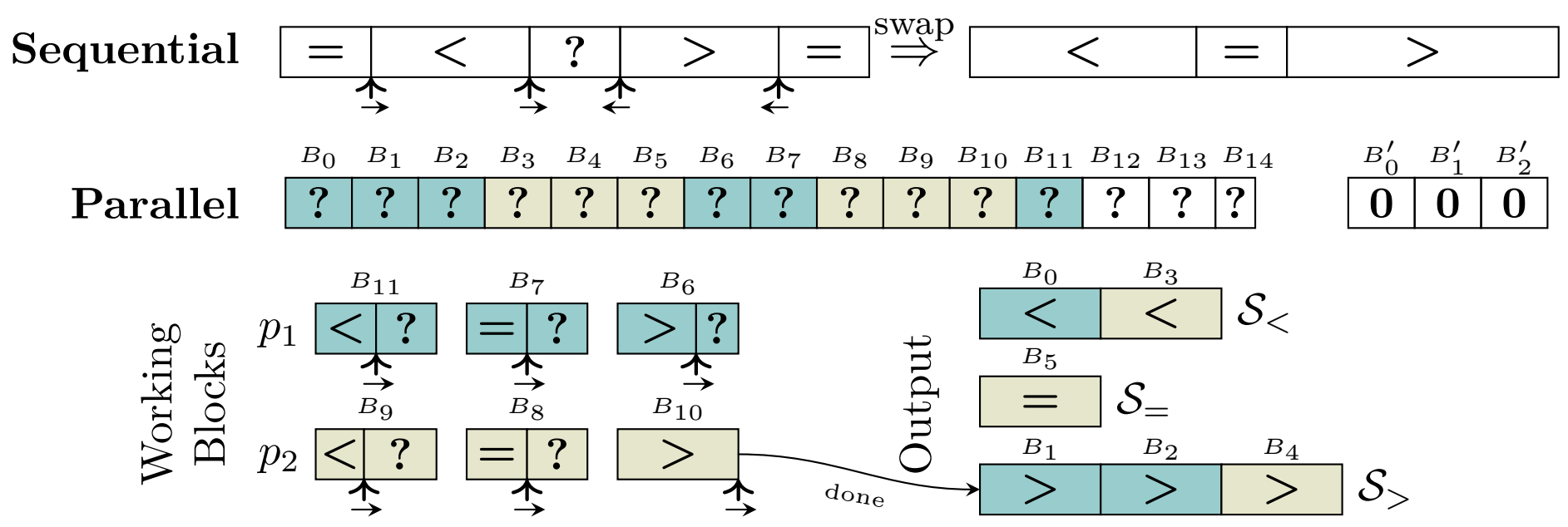

**Figure 4.14:** Block schema of sequential and parallel multikey quicksort's ternary partitioning process.

### 4.4.3 Burstsort

Burstsort [SZ03a; SZ03b; SZ04a; SZ04b; SZ05; SZR07; SW08; SW10] is one of the fastest string sorting algorithms and cache-efficient for many inputs (see also section 2.2.3), but it appears difficult to parallelize. Keeping a common burst trie would require a prohibitive amount of synchronized operations, while building independent burst tries on each processor would lead to the question of how to merge multiple tries of different structure. This problem of merging tries is related to parallel $K$-way LCP-merge, and future work may find a way to combine these approaches.

## 4.5 Empirical Performance of Parallel Algorithms

To evaluate the practical performance of our parallel string sorting algorithms, we implemented parallel versions of $S^5$, $K$-way LCP-merge, multikey quicksort, and radix sort in C++, and compare them with the few parallel string sorting implementations we could find in an extensive online and literature search. In section 2.3 we already discussed and evaluated the performance of *sequential* string sorting implementations. All of these evaluations were performed using our test framework, which contains a large number of both sequential and parallel string sorting implementations. The test framework and most input sets are available from `http://panthema.net/2013/parallel-string-sorting`.

For the performance evaluation in this section, we used the same six platforms and seven inputs as for evaluating sequential string sorting in section 2.3. The experimental platforms in table 2.1 (page 40) are composed of a wide variety of multi-core machines of different age. We used the same seven inputs, **URLs**, **Random**, **GOV2**, **Wikip**, **Sinha URLs**, **Sinha DNA**, and **Sinha NoDup**, which are described in section 2.3.2 (page 41).

Our parallel experiments cover all algorithms described in this chapter: **B.pS$^5$-UI** is a variant of pS$^5$ from section 4.2 which interleaves four unrolled descents of the classification tree, while **B.pS$^5$-E** unrolls only a single descent, but tests equality at each splitter node. **B.pS$^5$-UIC** is the variant of pS$^5$ with the additional bit-trick discussed in section 4.2.3. As in our preliminary evaluation in section 4.2.4, we selected a fixed splitter tree size of $2^{10} - 1$.

From section 4.3, we included a *pure* parallelized $K$-way LCP-mergesort with multiway splitting, **BE.pMS-ms**, and two variants of the NUMA-aware algorithm which first run pS$^5$-UI independently on each NUMA node for separate parts of the input and then merge the presorted parts using our parallel $K$-way LCP-merge algorithm. As predicted in section 4.3.4, we saw huge speed improvements due to *caching* of just the distinguishing character $\hat{c}$ in BE.pMS-ms, and do not consider the non-caching variant in our results. For the evaluation we included a version with multiway splitting, **BE.pS$^5$+M-ms**, and a version with LCP splitting, **BE.pS$^5$+M-ls**. For the large input sets in the experiments, binary splitting was always slower than these two variants.

We also tried to *rebalance* threads to other NUMA nodes once work on a node is finished, and called this feature *assisting* (other nodes). Compared to work stealing, our assistance technique processes voluntarily freed work from the job queues on other NUMA nodes (see also figure 4.5). However, this did not work well since the additional inter-node synchronization was too costly. We thus have to leave the question of how to rebalance sorting work on NUMA systems for highly skewed inputs to future research.

From the additional parallel algorithms in section 4.4, we draw our parallel multikey quicksort implementations, **B.pMKQS**, and radix sorts with 8-bit and 16-bit fully parallel steps, **B.pRS-8bit** and **B.pRS-16bit**. The 16-bit version only uses 16-bit for the fully parallel steps, the sequential subproblems then use 8-bit radix sort.

As discussed in the introduction to this chapter, only very few parallelized string sorting implementations by other authors exist. For the experiments, we included the parallel radix sort implemented by Akiba [Aki11] as **A.pRS**, and a very basic parallelized 2-way LCP-merge sort by Shamsundar [Sha09] as **S.pMS-2way**, which is based on Ng and Kakehi's LCP-mergesort [NK08].

**Table 4.3:** Description of parallel string sorting algorithms in experiment.

| Name | Description and Author |
|---|---|
| B.pS$^5$-E | Our parallel Super Scalar String Sample Sort (section 4.2) with *equality checking* at each splitter node. |
| B.pS$^5$-UI | Our parallel Super Scalar String Sample Sort (section 4.2) with *unrolled* and *interleaved* tree descents. |
| B.pS$^5$-UIC | Our parallel Super Scalar String Sample Sort (section 4.2) with unrolled and interleaved tree descents, and the *bit-transformation* from pre-order to level-order from lemma 4.1. |
| B.pMKQS | Our parallel multikey quicksort (section 4.4.2) with caching of $w = 8$ characters. |
| B.pRS-8bit | Our parallel radix sort (section 4.4.1) with *8-bit* alphabet |
| B.pRS-16bit | Our parallel radix sort (section 4.4.1) with *16-bit* alphabet at the fully parallel levels, and 8-bit alphabets for sequentially processed subproblems. |
| A.pRS | Akiba's [Aki11] parallel radix sort. |
| S.pMS-2way | Parallel 2-way LCP-mergesort by Shamsundar [Sha09], which is based on Ng and Kakehi's LCP-mergesort [NK08]. |
| BE.pMS-ms | Our parallel multiway LCP-merge with multiway LCP-mergesort on each NUMA node, caching the distinguishing character, and multiway splitting. |
| BE.pS$^5$+M-ms | Our parallel multiway LCP-merge with pS$^5$ on each NUMA node and multiway splitting. |
| BE.pS$^5$+M-ls | Our parallel multiway LCP-merge with pS$^5$ on each NUMA node and LCP splitting. |

The full list and a description of each of the parallel string sorting variants we selected for our experiments is shown in table 4.3.

The results plotted in figures 4.15 to 4.20 show the speedup of each parallel algorithm over the best sequential one for increasing thread count across all six experimental platforms. Note that the speedup values for one thread are often less than one because highly tuned sequential algorithm can be considerably faster than parallelized ones. Each experiment instance was run three times, and the graphs show the median of the runs.

Tables 4.4 to 4.15 show absolute running times of our experiments, with the fastest algorithm's time printed in bold text. These lists also contain the fastest sequential algorithm for each instance, which is usually the baseline for all speedup calculations. The tables containing absolute running times are included for reference and to show that our parallel implementations scale well both for very large instances on multi-core platforms and also for small inputs on machines with fewer cores.

Overall, our parallel string sorting implementations yield high speedups, which are much higher than those of all previously existing parallel string sorters. Each individual parallel algorithm's speedup depends highly on hardware characteristics like processor speed, RAM and cache performance (see chapter 3 for parallel memory bandwidth and latency experiments), the interconnection between sockets, and the input's characteristics. In general, the speedup of string sorting for high thread counts is bounded by memory bandwidth, not processing power. This is the reason why no algorithm scales perfectly with the number of cores. Instead, they tend to scale up well to the number of memory channels, after which the speedup only increases marginally.

*A.Intel-1×8* (figure 4.15, tables 4.4 to 4.5) and *F.AMD-1×16* (figure 4.20, tables 4.14 to 4.15) are *consumer-grade* single-socket machines from Intel and AMD with fast RAM and cache hierarchy. The latter AMD machine is almost ten years younger than the first. Both are not NUMA architectures, which is why we did not run our NUMA-aware variants on them. They are more classic architectures and exhibit most of the effects targeted in our algorithms to gain good speedups. Our $pS^5$ variants are fastest on all inputs, except very random ones (Random and NoDup), and curiously on URLs on F.AMD-1×16. For the high-entropy inputs, our radix sorts outperform $pS^5$ slightly because classification is simpler and faster.

For all test instances except URLs, the fully parallel subalgorithm of $pS^5$ was run between one and four times. Thereafter, the input was split up into sufficiently many subsets, and most of the additional speedup is gained by load-balancing the sequential subalgorithms well. Overall, which particular $pS^5$ variant is fastest seems to be a coin flip: the B.$pS^5$-E variant handles URL instances better, as many equal matches occur here, but for all other inputs, B.$pS^5$-UI with interleaved tree descents fares better, even though it has higher theoretical running time.

Our B.pMKQS also shows good overall speedups, but is never particularly fast. This is due to the high memory bandwidth required by caching multikey quicksort, as it reads and re-reads an array of string pointers to partition around just one pivot. Machine F.AMD-1×16 is special in the sense that it has the fewest memory channels per core of our platforms, which explains why pMKQS is relatively slow on it. For URLs, our parallelized multiway LCP-mergesort clearly outperforms on F.AMD-1×16, but not on A.Intel-1×8, probably because the newer machine is able to better optimize sequential memory scanning.

*D.Intel-4×8* (figure 4.18, tables 4.10 to 4.11) and *E.Intel-2×16* (figure 4.19, tables 4.12 to 4.13) are *server-grade* 2- and 4-socket multi-core NUMA machines with Intel Xeon CPUs, whereas *B.AMD-4×4* (figure 4.16, tables 4.6 to 4.7) and *C.AMD-4×12* (figure 4.17, tables 4.8 to 4.9) are *server-grade* 4- and 2-socket multi-core NUMA machines with AMD Opteron CPUs. Due to their internal make-up, the AMD Opterons have twice the number of NUMA nodes per socket.

While the NUMA multi-core platforms also exhibit very good speedups across the board, the individual algorithms' performance characteristics are more complex on these architectures. Relative results seem to mostly depend on how well the inner loops and memory transfers are optimized by the hardware on each particular system.

Platform *E.Intel-2×16* (figure 4.19, tables 4.12 to 4.13) is from 2016, the newest NUMA machine in our experiment, and exhibits the cleanest results for large inputs: our $pS^5$ scales virtually linearly up to the number of real cores (excluding simultaneous multithreading (SMT)), and then the speedup continues slower using SMT cores. For the small inputs, Sinha URLs, DNA, and NoDup, the speedup factor is up to ten, but they are unstable, probably simply because the input is too small to parallelize over so many processors.

For our large URLs and GOV2 inputs, BE.$pS^5$+M-ms outperforms all other algorithms with a speedup factor of up to 24, closely followed by B.$pS^5$-UI with a speedup of up to 19. As before, on our Random input, parallel radix sort is faster due to the simpler classification, and on URLs our parallel LCP-mergesort is a strong contender. On Wikip we observe the largest speedups of around 32, probably because this is the largest input and this allows algorithms to hide overhead. BE.$pS^5$+M-ms however exhibits very slow speedups for Wikip due to the long suffixes: During multiway splitting the sample strings are sorted in their entirety, which can take very long time for suffixes. Overall, the parallelized NUMA step in BE.$pS^5$+M-ms seems to accelerate the sorting process somewhat, but the larger improvement come from $pS^5$.

Platform *D.Intel-4×8* (figure 4.18, tables 4.10 to 4.11) is also an Intel CPU, but of older date. Comparing the results to E.Intel-2×16, one can see the advances in CPU technology, faster memory channels, and how the newer CPUs appear to better hide NUMA effects. Relative to E.Intel-2×16, our NUMA-optimized sorting algorithms using LCP-merge (BE.pMS-ms and BE.$pS^5$+M-ms) accelerate the sorting process by a much larger factor: on URLs, BE.$pS^5$+M-ms is nearly twice as fast as B.$pS^5$-UI, and the effect is also visible on other inputs, though less pronounced. This effect is clearly due to $pS^5$ ignoring the NUMA architecture and thus incurring a relatively large penalty for expensive inter-node random string accesses.

Platform *C.AMD-4×12* (figure 4.17, tables 4.8 to 4.9) is a somewhat older AMD multi-core machine with high core count, but relatively slow RAM and a slower interconnect. On this platform with eight NUMA nodes, random access is even more costly and the inter-node connections are easily congested, which is why $pS^5$ fares relatively well against algorithms with LCP-merge compared to D.Intel-4×8. On the same note, radix sort is still very fast on all NUMA machines for random inputs.

*B.AMD-4×4* (figure 4.16, tables 4.6 to 4.7) is an earlier NUMA architecture with four NUMA nodes, and the slowest RAM speed and interconnect in our experiment. However, on this machine random access, memory bandwidth, and processing power (in cache) seems to be more balanced for $pS^5$ than on the newer NUMA machines.

Considering other algorithms across all machines, we notice that our parallel multikey quicksort (B.pMKQS) is a very strong contender, and achieves excellent speedups, often on par with $pS^5$. We analyzed the number of string accesses of pMKQS in theorem 4.11, after which the characters are saved and accessed in a scanning pattern. This scanning apparently works well on the NUMA machines, as it is very cache-efficient, can be easily predicted by the processor's memory prefetcher, and costly cache line transfers between nodes contain characters from eight strings. However, compared to $pS^5$, the larger memory bandwidth requirement clearly limits the achievable speedup.

Of the algorithms by other authors, only Akiba's radix sort scales fairly well on single-socket consumer machines. By regarding the difference in performance on Random and URL inputs, we can already identify the implementation's main problems: it does not parallelize recursive sorting steps (only the top level is parallelized) and only performs simple load balancing. This is most visible on URLs and GOV2. Shamsundar's 2-way LCP-mergesort does not show good speedups on any input or machine, partly because it is already pretty slow sequentially, and partly because it is not fully parallelized.

## 4.6 Conclusions and Future Work

We have demonstrated that string sorting can be parallelized successfully on modern multi-core shared-memory and NUMA machines. In particular, our new string sample sort algorithm $pS^5$ combines favorable features of some of the best sequential algorithms – robust multiway divide-and-conquer from burstsort, efficient data distribution from radix sort, asymptotic guarantees similar to multikey quicksort, and word parallelism from caching multikey quicksort. For NUMA machines we developed parallel $K$-way LCP-merge to further decrease costly inter-node random accesses.

Both algorithms are practical for many applications, and our implementations are available as templates for further customization. For general use, we recommend our $pS^5$ implementation as its performance is most reliable across all platforms.

We want to highlight that using our $pS^5$ (which can save LCPs) and $K$-way LCP-merge implementations it is straight-forward to construct a fast parallel *external memory* string sorter for short strings ($\leq B$) using shared-memory parallelism. The sorting throughput of our string sorters is probably higher than the available I/O bandwidth.

Implementing some of the refinements discussed in the next subsection is likely to yield further improvements for string sample sort and $K$-way LCP-merge.

As most important vectors of future work, we see the splitting heuristic of LCP-Merging and how to rebalance work for skewed inputs on NUMA machines.

### 4.6.1 Future Practical Refinements

**Memory conservation.** For use of our algorithms in applications like database systems or MapReduce libraries, hard guarantees on the amount of memory required by the implementations are paramount. Our experiments clearly show, that caching of characters accelerates string sorting, but this speed comes at the cost of memory usage. A future challenge is thus further explore the Pareto front on how to sort quickly with limited memory. In this respect, $pS^5$ is a very promising candidate, as it can be restricted to use only the classification tree and a recursion stack if little additional memory is available. But if more memory is available, caching, saving oracle values, and out-of-place redistribution can be enabled adaptively.

**Multipass data distribution.** There are two constraints on the maximum sensible value for the number of splitters $v$: the cache size of the classification data structure and the resources needed for data distribution. Already in the plain external memory model these two constraints differ ($v = \mathcal{O}(M)$ versus $v = \mathcal{O}(M/B)$). In practice, things are even more complicated since multiple cache levels, cache replacement policy, TLBs, etc. play a role as well. In any case, we can increase $v$ by performing the data distribution in multiple passes (usually two in practice). Note that this fits very well with the approach to compute oracles even for single pass data distribution. This approach can be viewed as LSD radix sort using the oracles as keys. Initial experiments indicate that this could indeed lead to some performance improvements.

**Alphabet compression.** If we know that only $\sigma' < \sigma$ different characters from $\Sigma$ appear in the input, we can compress characters into $\lceil \log \sigma' \rceil$ bits. For $S^5$, this allows us to pack more characters into a single machine word. For example, for DNA input, we might pack 32 characters into a single 64 bit machine word. Note that this compression can be done on the fly without changing the input/output format and the compression overhead is amortized over $\log v$ key comparisons.

**Jump tables.** In $S^5$, the $k$ most significant bits of a key are often already sufficient to define a path in the classification tree of length up to $k$. We can exploit this by precomputing a jump table of size $2^k$ storing a pointer to the end of this path. During element classification, a lookup in this jump table can replace the traversal of the path. This might reduce the gap to radix sort for easy instances.

**Using tries in practice.** The success of burstsort indicates that traversing tries can be made efficient. Thus, we might also be able to use a tuned trie-based implementation of $S^5$ in practice. One ingredient to such an implementation could be the word parallelism used in the pragmatic solution – we define the trie over an enlarged alphabet. This reduces the number of required hash table accesses by a factor of $w$. Experiential results with the tuned van Emde Boas trees [DKMS04] suggest that this data structure might work in practice.

**Adaptivity.** By inspecting the sample, we can adaptively tune the algorithm. For example, if the sample suggests that a lot of information* can be gained from a few

*The entropy $\frac{1}{n}\sum_i \log \frac{n}{|b_i|}$ can be used to define the amount of information gained by a set of splitters. The bucket sizes $b_i$ can be estimated using their size within the sample.

most significant bits in the sample keys, the algorithm might decide to switch to radix sort. On the other hand, when even the $w$ most significant characters do not give a lot of information, a trie based implementation could be used. In turn, this trie can be adapted to the input, for example using hash tables for low degree trie nodes and arrays for high degree nodes.

**Acknowledgements.** We would like the thank the anonymous reviewer of our journal paper [BES17] for extraordinarily thoroughly checking our algorithms and proofs, and for kind suggestions on how to improve the paper.

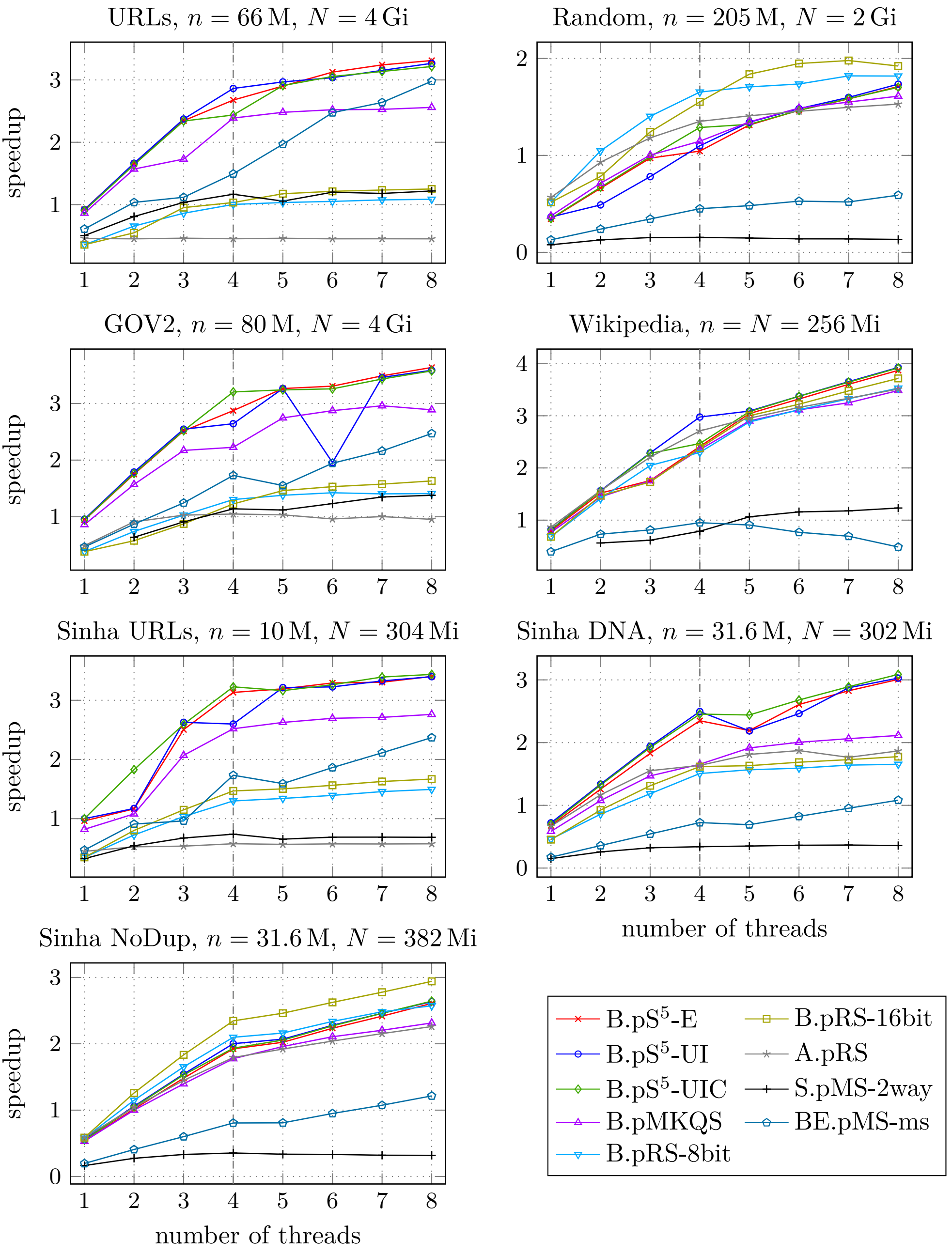

**Figure 4.15:** Speedup of parallel algorithm implementations on A.Intel-1×8.

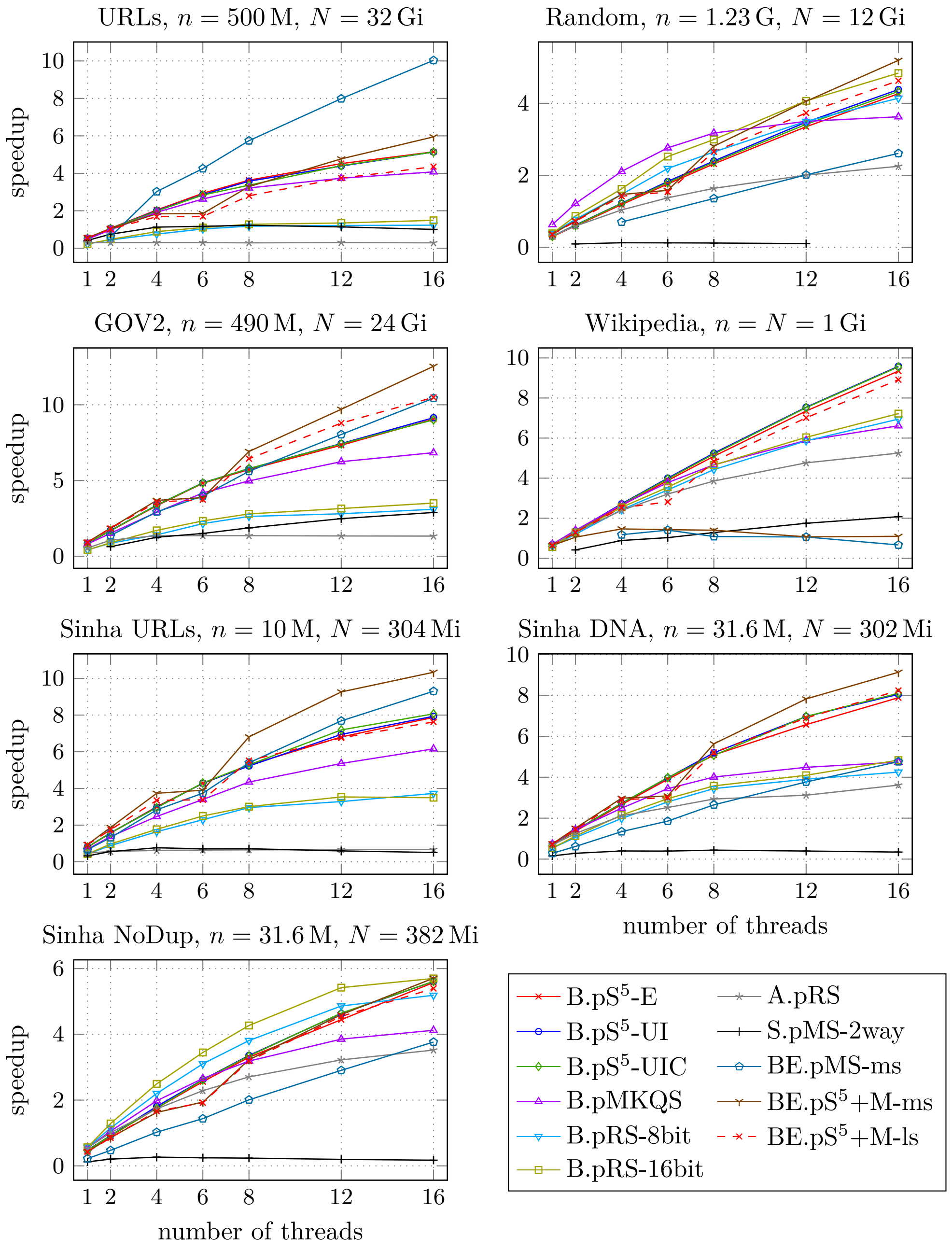

**Figure 4.16:** Speedup of parallel algorithm implementations on B.AMD-4×4.

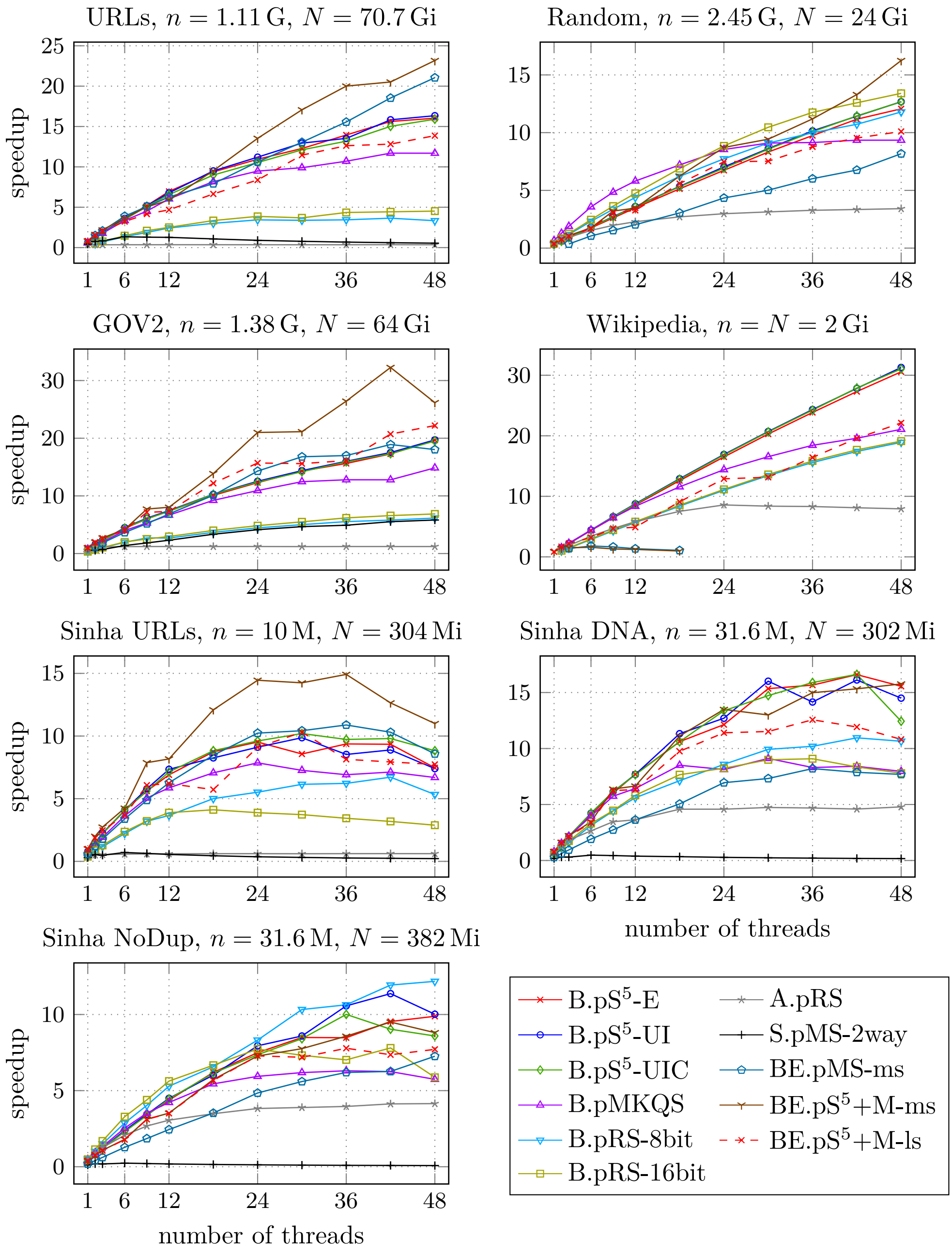

**Figure 4.17:** Speedup of parallel algorithm implementations on C.AMD-4×12.

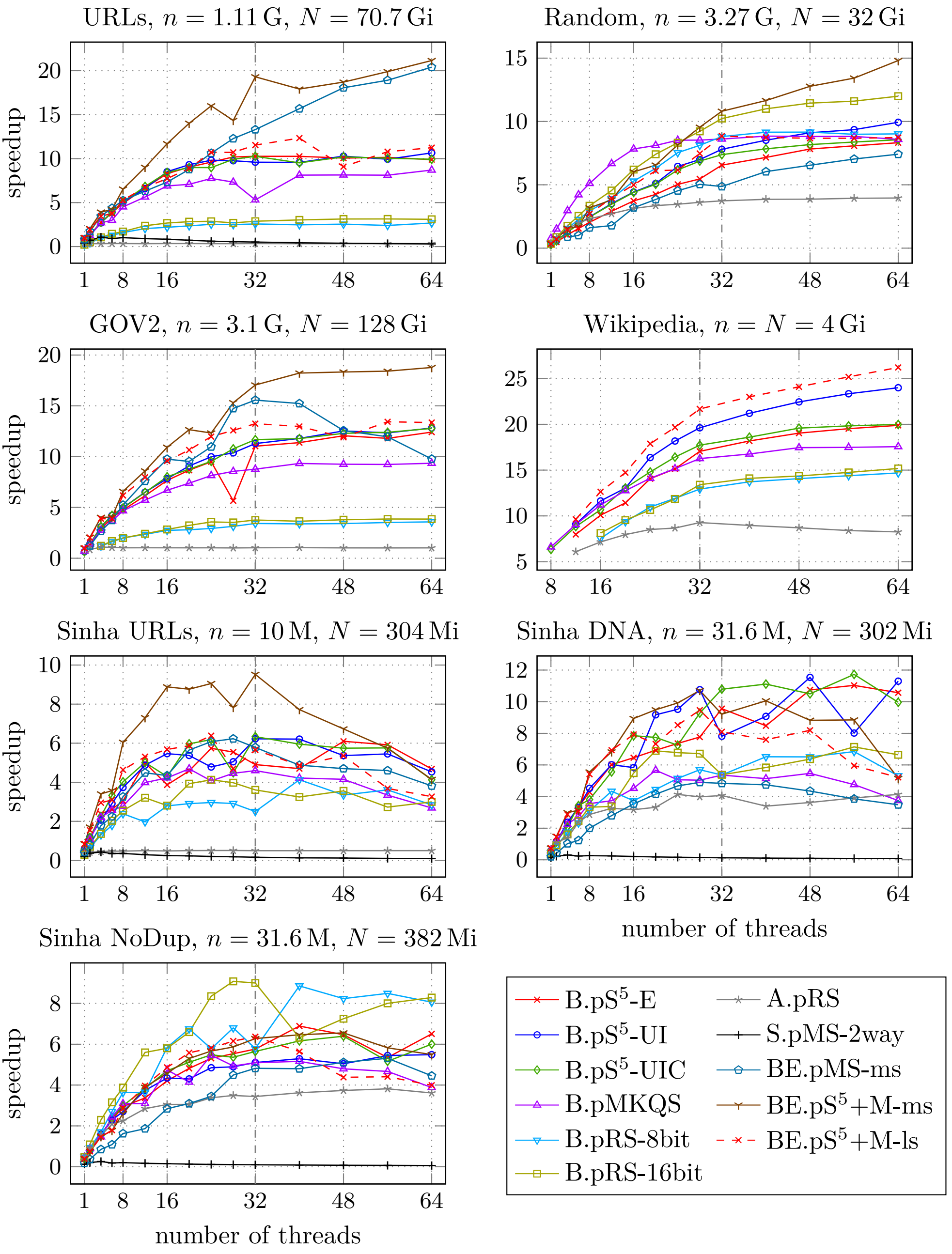

**Figure 4.18:** Speedup of parallel algorithm implementations on D.Intel-4×8.

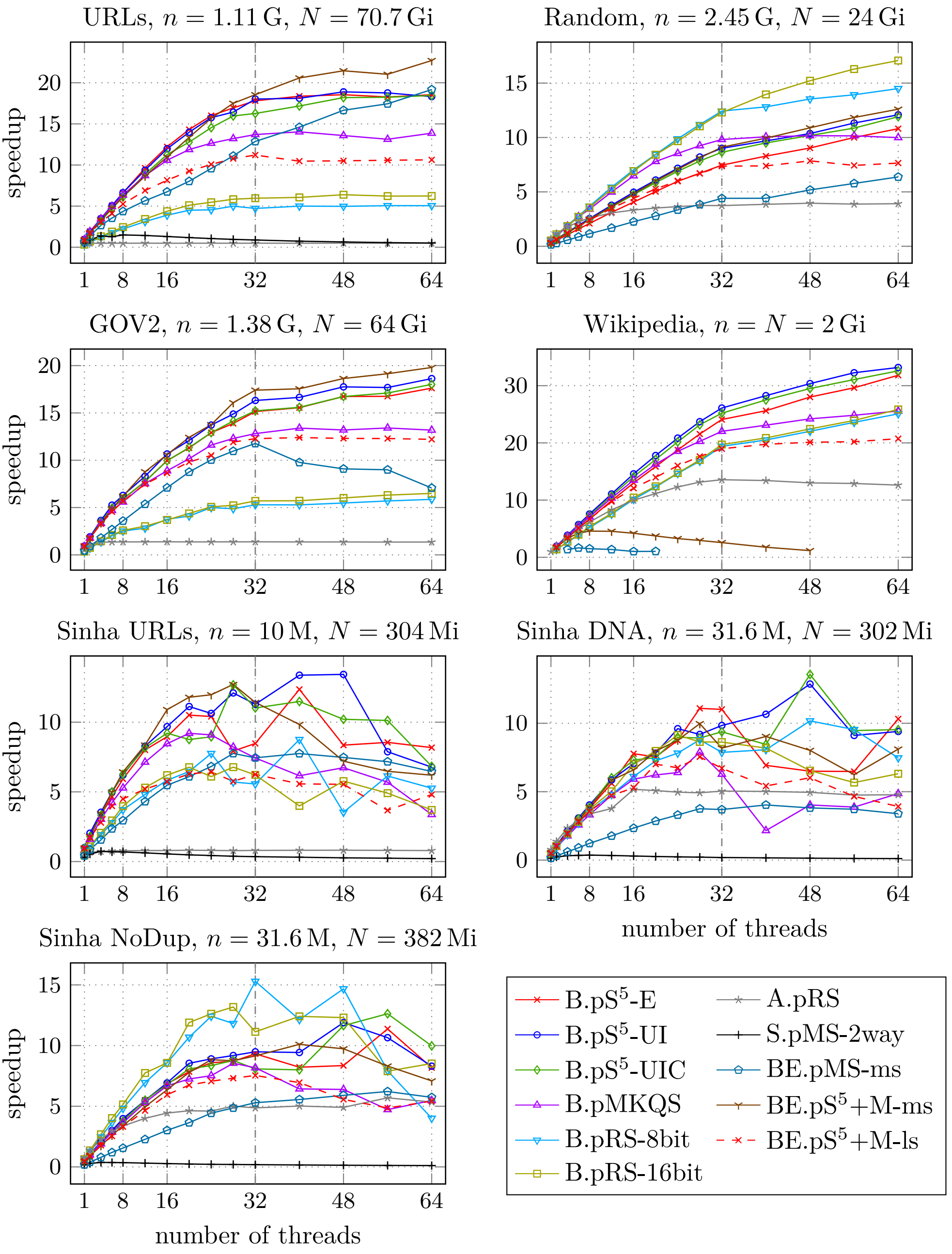

**Figure 4.19:** Speedup of parallel algorithm implementations on E.Intel-2×16.

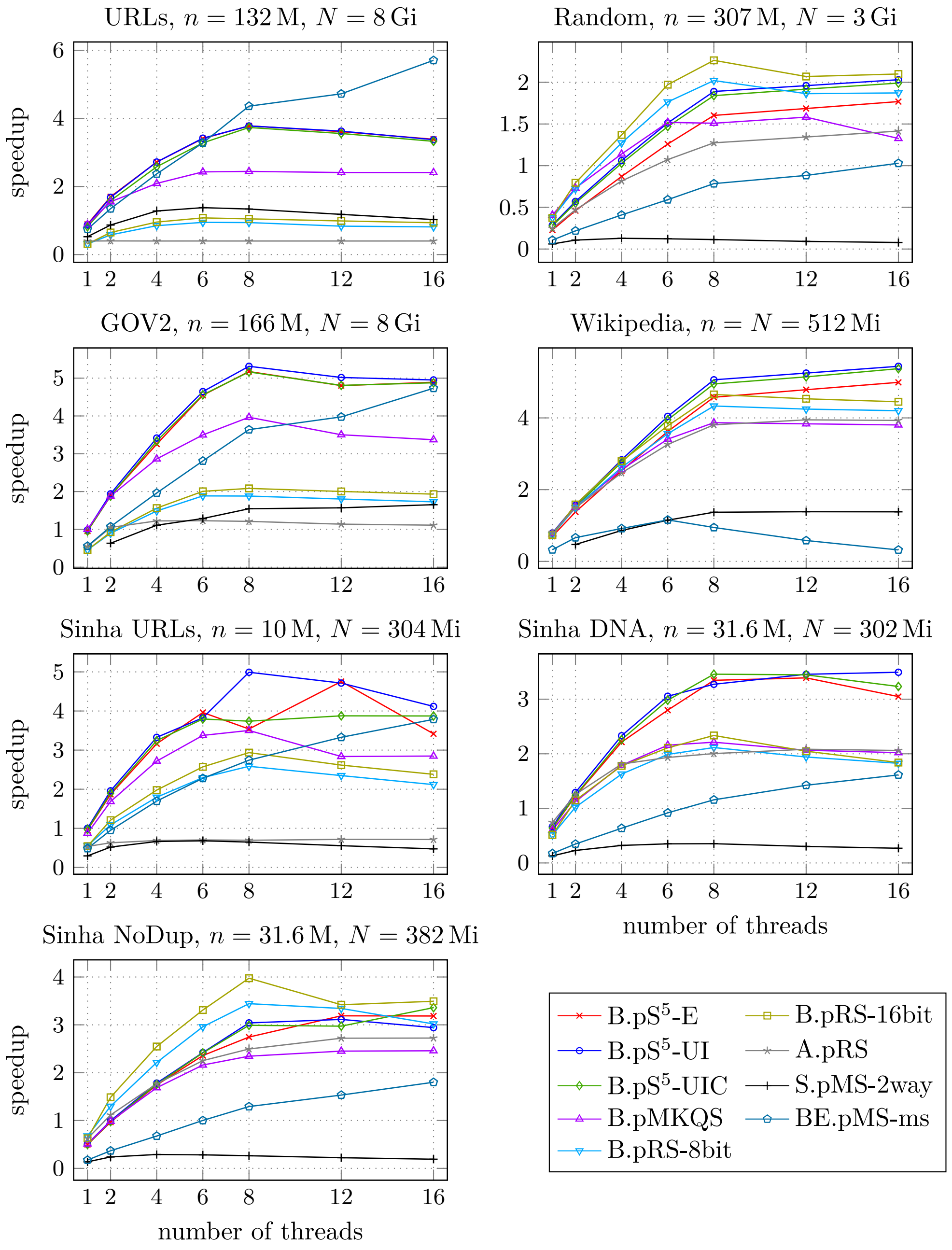

**Figure 4.20:** Speedup of parallel algorithm implementations on F.AMD-1×16.

**Table 4.4:** Absolute running time of parallel and best sequential algorithms on A.Intel-1×8 in seconds, median of three runs, larger test instances. See table 4.3 for a description of each.

| PEs | 1 | 2 | 3 | 4 | 5 | 6 | 7 | 8 |
|---|---|---|---|---|---|---|---|---|
| | **URLs**, $n = 65.7\,\mathrm{M}$, $N = 4\,\mathrm{Gi}$, $\frac{D}{N} = 92.7\,\%$ | | | | | | | |
| R.mkqs-cache8 | **14.8** | | | | | | | |
| B.pS$^5$-E | 16.0 | 8.9 | 6.3 | 5.5 | 5.09 | **4.73** | **4.56** | **4.46** |
| B.pS$^5$-UI | 16.1 | **8.9** | **6.2** | **5.2** | **4.98** | 4.86 | 4.68 | 4.52 |
| B.pS$^5$-UIC | 16.3 | 9.0 | 6.3 | 6.1 | 5.08 | 4.83 | 4.71 | 4.59 |
| B.pMKQS | 17.1 | 9.4 | 8.5 | 6.2 | 5.95 | 5.86 | 5.84 | 5.77 |
| B.pRS-8bit | 41.5 | 22.5 | 17.2 | 14.7 | 14.29 | 14.05 | 13.74 | 13.61 |
| B.pRS-16bit | 41.4 | 27.0 | 15.5 | 14.3 | 12.58 | 12.17 | 11.97 | 11.81 |
| A.pRS | 32.3 | 32.8 | 32.1 | 32.7 | 32.14 | 32.73 | 32.66 | 32.76 |
| S.pMS-2way | 29.4 | 18.3 | 14.3 | 12.7 | 13.99 | 12.32 | 12.53 | 12.12 |
| BE.pMS-ms | 24.3 | 14.3 | 13.2 | 9.9 | 7.50 | 5.97 | 5.60 | 4.96 |
| | **Random**, $n = 205\,\mathrm{M}$, $N = 2\,\mathrm{Gi}$, $\frac{D}{N} = 42.1\,\%$ | | | | | | | |
| KR.radixsort-DB | **12.3** | | | | | | | |
| B.pS$^5$-E | 35.6 | 18.6 | 12.6 | 11.7 | 9.34 | 8.37 | 7.75 | 7.17 |
| B.pS$^5$-UI | 33.6 | 25.1 | 15.7 | 11.2 | 9.14 | 8.27 | 7.67 | 7.07 |
| B.pS$^5$-UIC | 35.2 | 18.3 | 12.4 | 9.5 | 9.29 | 8.38 | 7.74 | 7.21 |
| B.pMKQS | 32.6 | 17.2 | 12.2 | 10.7 | 9.12 | 8.24 | 7.91 | 7.61 |
| B.pRS-8bit | 23.7 | **11.7** | **8.7** | **7.4** | 7.19 | 7.06 | 6.74 | 6.74 |
| B.pRS-16bit | 23.8 | 15.7 | 9.9 | 7.9 | **6.67** | **6.30** | **6.20** | **6.38** |
| A.pRS | 21.6 | 13.2 | 10.4 | 9.1 | 8.71 | 8.43 | 8.20 | 8.02 |
| S.pMS-2way | 157.0 | 95.8 | 80.4 | 79.5 | 83.16 | 88.14 | 88.45 | 92.13 |
| BE.pMS-ms | 95.1 | 51.2 | 35.7 | 27.2 | 25.48 | 23.23 | 23.61 | 20.81 |
| | **GOV2**, $n = 80\,\mathrm{M}$, $N = 4\,\mathrm{Gi}$, $\frac{D}{N} = 69.8\,\%$ | | | | | | | |
| R.mkqs-cache8 | **13.9** | | | | | | | |
| B.pS$^5$-E | 14.8 | 8.0 | 5.5 | 4.8 | **4.27** | **4.22** | **4.00** | **3.83** |
| B.pS$^5$-UI | 14.6 | **7.8** | **5.5** | 5.3 | 4.27 | 7.13 | 4.03 | 3.89 |
| B.pS$^5$-UIC | 14.8 | 7.9 | 5.5 | **4.3** | 4.30 | 4.28 | 4.06 | 3.90 |
| B.pMKQS | 16.2 | 8.9 | 6.4 | 6.3 | 5.08 | 4.85 | 4.71 | 4.83 |
| B.pRS-8bit | 36.5 | 18.9 | 13.6 | 10.7 | 10.11 | 9.80 | 9.93 | 9.89 |
| B.pRS-16bit | 36.5 | 24.2 | 16.0 | 11.4 | 9.54 | 9.10 | 8.85 | 8.54 |
| A.pRS | 28.5 | 15.3 | 13.6 | 13.3 | 13.50 | 14.52 | 13.90 | 14.63 |
| S.pMS-2way | | 21.9 | 15.3 | 12.2 | 12.46 | 11.32 | 10.34 | 10.11 |
| BE.pMS-ms | 30.0 | 16.0 | 11.2 | 8.1 | 8.99 | 7.17 | 6.45 | 5.64 |
| | **Wikipedia**, $n = N = 256\,\mathrm{Mi}$, $D = 13.8\,\mathrm{G}$ | | | | | | | |
| KR.radixsort-CE7 | **55.8** | | | | | | | |
| B.pS$^5$-E | 70.3 | 36.7 | 31.7 | 23.1 | 18.3 | 16.8 | 15.48 | 14.41 |
| B.pS$^5$-UI | 68.0 | 35.7 | **24.4** | **18.7** | **18.1** | **16.5** | **15.27** | **14.21** |
| B.pS$^5$-UIC | 68.5 | 35.9 | 24.5 | 22.6 | 18.1 | 16.5 | 15.31 | 14.24 |
| B.pMKQS | 73.0 | 38.6 | 31.8 | 23.7 | 19.2 | 17.9 | 17.17 | 16.01 |
| B.pRS-8bit | 81.8 | 39.5 | 27.2 | 24.3 | 19.4 | 17.9 | 16.77 | 15.80 |
| B.pRS-16bit | 81.9 | 37.9 | 32.2 | 23.4 | 18.6 | 17.3 | 16.04 | 15.01 |
| A.pRS | 64.4 | **35.6** | 25.3 | 20.6 | 18.9 | 17.7 | 16.72 | 15.88 |
| S.pMS-2way | | 99.2 | 90.7 | 70.9 | 52.4 | 48.2 | 47.37 | 45.27 |
| BE.pMS-ms | 141.1 | 76.3 | 68.6 | 58.6 | 61.5 | 72.7 | 80.63 | 115.01 |

**Table 4.5:** Absolute running time of parallel and best sequential algorithms on A.Intel-1×8 in seconds, median of three runs, smaller test instances. See table 4.3 for a description of each.

| PEs | 1 | 2 | 3 | 4 | 5 | 6 | 7 | 8 |
|---|---|---|---|---|---|---|---|---|
| | **Sinha URLs** (complete), $n = 10\,\mathrm{M}$, $N = 304\,\mathrm{Mi}$, $\frac{D}{N} = 97.5\,\%$ | | | | | | | |
| R.mkqs-cache8 | 1.84 | | | | | | | |
| B.pS$^5$-E | 1.87 | 1.55 | 0.72 | 0.58 | 0.565 | **0.548** | 0.545 | 0.530 |
| B.pS$^5$-UI | 1.80 | 1.54 | **0.69** | 0.69 | **0.561** | 0.559 | 0.542 | 0.530 |
| B.pS$^5$-UIC | **1.80** | **0.99** | 0.69 | **0.56** | 0.570 | 0.553 | **0.531** | **0.525** |
| B.pMKQS | 2.20 | 1.67 | 0.87 | 0.72 | 0.687 | 0.669 | 0.665 | 0.653 |
| B.pRS-8bit | 5.22 | 2.49 | 1.74 | 1.39 | 1.344 | 1.294 | 1.237 | 1.206 |
| B.pRS-16bit | 5.21 | 2.25 | 1.57 | 1.23 | 1.200 | 1.153 | 1.108 | 1.081 |
| A.pRS | 4.03 | 3.44 | 3.37 | 3.13 | 3.189 | 3.144 | 3.152 | 3.137 |
| S.pMS-2way | 5.56 | 3.34 | 2.68 | 2.44 | 2.758 | 2.623 | 2.621 | 2.633 |
| BE.pMS-ms | 3.84 | 1.98 | 1.88 | 1.04 | 1.133 | 0.968 | 0.853 | 0.761 |
| | **Sinha DNA** (complete), $n = 31.6\,\mathrm{M}$, $N = 302\,\mathrm{Mi}$, $\frac{D}{N} = 100\,\%$ | | | | | | | |
| KR.radixsort-CE7 | **2.53** | | | | | | | |
| B.pS$^5$-E | 3.74 | 2.01 | 1.38 | 1.08 | 1.152 | 0.969 | 0.894 | 0.840 |
| B.pS$^5$-UI | 3.50 | **1.89** | **1.30** | **1.01** | 1.155 | 1.026 | 0.879 | 0.834 |
| B.pS$^5$-UIC | 3.63 | 1.91 | 1.31 | 1.03 | **1.035** | **0.943** | **0.874** | **0.818** |
| B.pMKQS | 4.32 | 2.35 | 1.72 | 1.53 | 1.319 | 1.260 | 1.225 | 1.195 |
| B.pRS-8bit | 5.48 | 2.94 | 2.13 | 1.68 | 1.612 | 1.586 | 1.539 | 1.527 |
| B.pRS-16bit | 5.57 | 2.73 | 1.93 | 1.56 | 1.548 | 1.497 | 1.462 | 1.422 |
| A.pRS | 3.78 | 2.16 | 1.63 | 1.54 | 1.394 | 1.349 | 1.431 | 1.353 |
| S.pMS-2way | 16.81 | 9.82 | 7.81 | 7.42 | 7.199 | 6.953 | 6.883 | 7.043 |
| BE.pMS-ms | 14.57 | 7.05 | 4.65 | 3.48 | 3.656 | 3.065 | 2.649 | 2.334 |
| | **Sinha NoDup** (complete), $n = 31.6\,\mathrm{M}$, $N = 382\,\mathrm{Mi}$, $\frac{D}{N} = 73.4\,\%$ | | | | | | | |
| KR.radixsort-CE7 | **3.05** | | | | | | | |
| B.pS$^5$-E | 5.71 | 3.00 | 2.06 | 1.58 | 1.507 | 1.365 | 1.262 | 1.169 |
| B.pS$^5$-UI | 5.45 | 2.88 | 1.97 | 1.52 | 1.473 | 1.338 | 1.240 | 1.157 |
| B.pS$^5$-UIC | 5.55 | 2.91 | 1.99 | 1.58 | 1.483 | 1.343 | 1.240 | 1.153 |
| B.pMKQS | 5.74 | 3.05 | 2.18 | 1.72 | 1.556 | 1.447 | 1.383 | 1.318 |
| B.pRS-8bit | 5.28 | 2.65 | 1.85 | 1.45 | 1.410 | 1.305 | 1.228 | 1.185 |
| B.pRS-16bit | 5.20 | **2.42** | **1.66** | **1.30** | **1.239** | **1.162** | **1.098** | **1.037** |
| A.pRS | 5.19 | 2.92 | 2.10 | 1.70 | 1.587 | 1.494 | 1.416 | 1.349 |
| S.pMS-2way | 18.51 | 11.12 | 9.18 | 8.61 | 9.078 | 9.167 | 9.513 | 9.568 |
| BE.pMS-ms | 15.51 | 7.49 | 5.07 | 3.78 | 3.770 | 3.210 | 2.838 | 2.510 |

**Table 4.6:** Absolute running time of parallel and best sequential algorithms on B.AMD-4×4 in seconds, median of three runs. See table 4.3 for a description of each.

| PEs | 1 | 2 | 4 | 6 | 8 | 12 | 16 |
|---|---|---|---|---|---|---|---|
| | **URLs**, $n = 500\,\text{M}$, $N = 32\,\text{Gi}$, $\frac{D}{N} = 95.4\,\%$ | | | | | | |
| B.LCP-MS-2way | **288** | | | | | | |
| B.pS$^5$-E | 531 | **270** | 141 | 98 | 79 | 63.6 | 55.9 |
| B.pS$^5$-UI | 538 | 274 | 143 | 101 | 80 | 65.5 | 56.0 |
| B.pS$^5$-UIC | 543 | 274 | 145 | 101 | 85 | 65.2 | 56.0 |
| B.pMKQS | 592 | 293 | 150 | 110 | 89 | 76.9 | 70.5 |
| B.pRS-8bit | 1 264 | 650 | 381 | 280 | 244 | 237.4 | 233.7 |
| B.pRS-16bit | 1 264 | 609 | 324 | 261 | 225 | 213.5 | 193.1 |
| A.pRS | 966 | 953 | 950 | 951 | 1 000 | 947.1 | 996.6 |
| S.pMS-2way | 685 | 384 | 255 | 246 | 234 | 252.5 | 286.3 |
| BE.pMS-ms | | 464 | **95** | **68** | **50** | **36.0** | **28.7** |
| BE.pS$^5$+M-ms | 516 | 275 | 156 | 155 | 87 | 60.3 | 48.4 |
| BE.pS$^5$+M-ls | 519 | 282 | 171 | 170 | 103 | 76.8 | 66.0 |
| | **Random**, $n = 1.23\,\text{G}$, $N = 12\,\text{Gi}$, $\frac{D}{N} = 43.7\,\%$ | | | | | | |
| KR.radixsort-CE2 | **236** | | | | | | |
| B.pS$^5$-E | 793 | 396 | 199 | 134 | 102 | 70.4 | 55.3 |
| B.pS$^5$-UI | 761 | 381 | 192 | 129 | 98 | 67.9 | 53.9 |
| B.pS$^5$-UIC | 772 | 386 | 194 | 131 | 100 | 69.0 | 54.5 |
| B.pMKQS | 373 | **194** | **112** | **85** | **74** | 67.4 | 65.1 |
| B.pRS-8bit | 607 | 303 | 160 | 108 | 89 | 67.5 | 57.0 |
| B.pRS-16bit | 607 | 271 | 145 | 94 | 79 | **58.0** | 48.8 |
| A.pRS | 747 | 399 | 227 | 172 | 144 | 117.7 | 104.8 |
| S.pMS-2way | | 2 495 | 1 832 | 1 875 | 1 952 | 2 275.6 | |
| BE.pMS-ms | | | 336 | | 174 | 117.2 | 90.4 |
| BE.pS$^5$+M-ms | 638 | 319 | 161 | 149 | 84 | 58.2 | **45.4** |
| BE.pS$^5$+M-ls | 654 | 326 | 166 | 153 | 89 | 63.2 | 51.0 |
| | **GOV2**, $n = 490\,\text{M}$, $N = 24\,\text{Gi}$, $\frac{D}{N} = 72.4\,\%$ | | | | | | |
| R.mkqs-cache8 | **293** | | | | | | |
| B.pS$^5$-E | 346 | 173 | 88 | 61 | 51 | 39.9 | 32.3 |
| B.pS$^5$-UI | 341 | 172 | 87 | **60** | 51 | 39.3 | 32.0 |
| B.pS$^5$-UIC | 343 | 172 | 87 | 60 | 51 | 39.5 | 32.5 |
| B.pMKQS | 362 | 193 | 101 | 70 | 59 | 46.9 | 42.8 |
| B.pRS-8bit | 699 | 338 | 208 | 136 | 111 | 104.4 | 94.4 |
| B.pRS-16bit | 698 | 327 | 171 | 125 | 105 | 92.9 | 83.4 |
| A.pRS | 536 | 276 | 215 | 215 | 215 | 218.3 | 220.0 |
| S.pMS-2way | | 462 | 233 | 194 | 156 | 117.9 | 101.2 |
| BE.pMS-ms | | 213 | 100 | 74 | 52 | 36.4 | 28.1 |
| BE.pS$^5$+M-ms | 314 | **157** | **79** | 76 | **42** | **30.2** | **23.3** |
| BE.pS$^5$+M-ls | 319 | 160 | 82 | 78 | 45 | 33.3 | 27.9 |
| | **Wikipedia**, $n = N = 1\,\text{Gi}$, $D = 40\,\text{G}$ | | | | | | |
| KR.radixsort-CE7 | **533** | | | | | | |
| B.pS$^5$-E | 803 | 404 | 203 | 137 | 105 | 72.6 | 57.1 |
| B.pS$^5$-UI | 776 | 390 | **196** | **133** | **102** | **70.7** | **55.7** |
| B.pS$^5$-UIC | 785 | 394 | 198 | 135 | 103 | 70.9 | 55.8 |
| B.pMKQS | 755 | **381** | 200 | 141 | 114 | 90.7 | 80.6 |
| B.pRS-8bit | 899 | 444 | 220 | 156 | 120 | 91.0 | 76.8 |
| B.pRS-16bit | 946 | 413 | 210 | 150 | 115 | 88.3 | 73.9 |
| A.pRS | 777 | 408 | 223 | 166 | 138 | 111.9 | 101.6 |
| S.pMS-2way | | 1 273 | 600 | 517 | 415 | 305.1 | 256.7 |
| BE.pMS-ms | | | 453 | 379 | 490 | 501.0 | 799.9 |
| BE.pS$^5$+M-ms | 825 | 506 | 364 | 373 | 380 | 497.6 | 486.9 |
| BE.pS$^5$+M-ls | 823 | 415 | 211 | 190 | 110 | 76.0 | 59.8 |

**Table 4.7:** Absolute running time of parallel and best sequential algorithms on B.AMD-4×4 in seconds, median of three runs. See table 4.3 for a description of each.

| PEs | 1 | 2 | 4 | 6 | 8 | 12 | 16 |
|---|---|---|---|---|---|---|---|
| | **Sinha URLs** (complete), $n = 10\,\text{M}$, $N = 304\,\text{Mi}$, $\frac{D}{N} = 97.5\,\%$ | | | | | | |
| R.mkqs-cache8 | **5.02** | | | | | | |
| B.pS$^5$-E | 6.34 | 3.20 | 1.67 | 1.17 | 0.95 | 0.74 | 0.64 |
| B.pS$^5$-UI | 6.25 | 3.18 | 1.69 | 1.17 | 0.96 | 0.72 | 0.63 |
| B.pS$^5$-UIC | 6.29 | 3.18 | 1.70 | **1.17** | 0.95 | 0.70 | 0.62 |
| B.pMKQS | 7.28 | 3.64 | 2.03 | 1.48 | 1.15 | 0.94 | 0.82 |
| B.pRS-8bit | 11.82 | 5.68 | 3.06 | 2.19 | 1.70 | 1.53 | 1.35 |
| B.pRS-16bit | 11.85 | 5.17 | 2.84 | 2.01 | 1.67 | 1.42 | 1.44 |
| A.pRS | 10.52 | 8.88 | 8.08 | 7.83 | 7.71 | 7.54 | 7.53 |
| S.pMS-2way | 15.53 | 8.96 | 6.56 | 7.10 | 7.04 | 8.45 | 9.88 |
| BE.pMS-ms | 7.54 | 3.82 | 1.77 | 1.34 | 0.93 | 0.65 | 0.54 |
| BE.pS$^5$+M-ms | 5.33 | **2.68** | **1.34** | 1.29 | **0.74** | **0.54** | **0.49** |
| BE.pS$^5$+M-ls | 5.40 | 2.87 | 1.50 | 1.48 | 0.91 | 0.74 | 0.66 |
| | **Sinha DNA** (complete), $n = 31.6\,\text{M}$, $N = 302\,\text{Mi}$, $\frac{D}{N} = 100\,\%$ | | | | | | |
| KRB.radixsort-CE3s | **7.59** | | | | | | |
| B.pS$^5$-E | 10.91 | 5.56 | 2.84 | 1.95 | 1.49 | 1.15 | 0.96 |
| B.pS$^5$-UI | 10.47 | 5.39 | 2.78 | 1.91 | 1.46 | 1.09 | 0.94 |
| B.pS$^5$-UIC | 10.67 | 5.42 | 2.78 | **1.90** | 1.50 | 1.09 | 0.94 |
| B.pMKQS | 10.23 | 5.28 | 3.05 | 2.20 | 1.89 | 1.69 | 1.60 |
| B.pRS-8bit | 14.04 | 7.17 | 3.82 | 2.71 | 2.20 | 1.94 | 1.78 |
| B.pRS-16bit | 14.10 | 6.74 | 3.53 | 2.57 | 2.12 | 1.85 | 1.57 |
| A.pRS | 11.01 | 6.07 | 3.58 | 3.01 | 2.58 | 2.43 | 2.10 |
| S.pMS-2way | 48.14 | 26.67 | 19.08 | 19.40 | 17.32 | 19.27 | 22.08 |
| BE.pMS-ms | 26.73 | 12.29 | 5.65 | 4.08 | 2.86 | 2.01 | 1.59 |
| BE.pS$^5$+M-ms | 10.18 | **5.02** | **2.55** | 2.46 | **1.35** | **0.97** | **0.83** |
| BE.pS$^5$+M-ls | 10.45 | 5.23 | 2.61 | 2.53 | 1.46 | 1.10 | 0.92 |
| | **Sinha NoDup** (complete), $n = 31.6\,\text{M}$, $N = 382\,\text{Mi}$, $\frac{D}{N} = 73.4\,\%$ | | | | | | |
| KR.radixsort-CE7 | **6.61** | | | | | | |
| B.pS$^5$-E | 14.38 | 7.29 | 3.76 | 2.58 | 2.00 | 1.49 | 1.19 |
| B.pS$^5$-UI | 13.86 | 7.15 | 3.66 | 2.52 | 1.97 | 1.43 | 1.18 |
| B.pS$^5$-UIC | 14.13 | 7.16 | 3.78 | 2.53 | 1.98 | 1.42 | 1.18 |
| B.pMKQS | 12.17 | 6.27 | 3.34 | 2.49 | 2.07 | 1.72 | 1.60 |
| B.pRS-8bit | 11.72 | 5.76 | 3.00 | 2.13 | 1.73 | 1.36 | 1.27 |
| B.pRS-16bit | 11.72 | **5.13** | **2.65** | **1.92** | **1.55** | **1.22** | 1.16 |
| A.pRS | 12.09 | 6.55 | 3.82 | 2.89 | 2.44 | 2.05 | 1.88 |
| S.pMS-2way | 54.02 | 32.22 | 24.90 | 26.91 | 28.10 | 33.57 | 38.48 |
| BE.pMS-ms | 29.68 | 14.02 | 6.45 | 4.61 | 3.29 | 2.28 | 1.76 |
| BE.pS$^5$+M-ms | 15.25 | 7.67 | 4.07 | 3.41 | 2.05 | 1.44 | **1.16** |
| BE.pS$^5$+M-ls | 15.43 | 7.77 | 3.98 | 3.45 | 2.07 | 1.45 | 1.23 |

**Table 4.8:** Absolute running time of parallel and best sequential algorithms on C.AMD-4×12 in seconds, median of three runs. See table 4.3 for a description of each.

| PEs | 1 | 2 | 3 | 6 | 9 | 12 | 18 | 24 | 36 | 42 | 48 |
|---|---|---|---|---|---|---|---|---|---|---|---|
| | **URLs** (complete), $n = 1.11\,\mathrm{G}$, $N = 70.7\,\mathrm{Gi}$, $\frac{D}{N} = 93.5\,\%$ | | | | | | | | | | |
| B.LCP-MS-2way | **632** | | | | | | | | | | |
| B.pS$^5$-E | 987 | 512 | 341 | 177 | **122** | **91** | 67 | 57.8 | 45.3 | 40.5 | 39.4 |
| B.pS$^5$-UI | 1 012 | 513 | 343 | 175 | 122 | 93 | 67 | 56.4 | 46.7 | 39.9 | 38.7 |
| B.pS$^5$-UIC | 1 024 | 523 | 349 | 182 | 124 | 96 | 70 | 60.0 | 48.0 | 42.1 | 39.7 |
| B.pMKQS | 1 103 | 540 | 357 | 189 | 140 | 105 | 77 | 66.7 | 59.1 | 54.0 | 54.1 |
| B.pRS-8bit | | 1 369 | 904 | 469 | 338 | 260 | 210 | 184.1 | 183.8 | 173.5 | 189.8 |
| B.pRS-16bit | | 1 241 | 828 | 430 | 302 | 252 | 189 | 163.5 | 145.4 | 142.6 | 139.9 |
| A.pRS | 1 794 | 1 800 | 1 798 | 1 793 | 1 788 | 1 775 | 1 779 | 1 790.2 | 1 791.2 | 1 758.2 | 1 786.6 |
| S.pMS-2way | 1 478 | 821 | 757 | 472 | 491 | 498 | 589 | 704.2 | 930.3 | 1 041.3 | 1 155.5 |
| BE.pMS-ms | | 422 | 304 | **162** | 124 | 101 | 80 | 59.6 | 40.6 | 34.1 | 30.0 |
| BE.pS$^5$+M-ms | 769 | **389** | **288** | 174 | 126 | 107 | **66** | **46.7** | **31.6** | **30.8** | **27.3** |
| BE.pS$^5$+M-ls | 776 | 409 | 310 | 196 | 152 | 135 | 95 | 75.5 | 50.1 | 49.3 | 45.6 |
| | **Random**, $n = 2.45\,\mathrm{G}$, $N = 24\,\mathrm{Gi}$, $\frac{D}{N} = 44.5\,\%$ | | | | | | | | | | |
| KR.radixsort-CE2 | **469** | | | | | | | | | | |
| B.pS$^5$-E | 1 608 | 807 | 537 | 268 | 180 | 135 | 91 | 69.6 | 48.1 | 42.0 | 38.9 |
| B.pS$^5$-UI | 1 539 | 778 | 516 | 258 | 172 | 130 | 88 | 66.6 | 46.2 | 41.0 | 37.0 |
| B.pS$^5$-UIC | 1 560 | 783 | 520 | 260 | 174 | 131 | 88 | 67.7 | 46.5 | 40.9 | 37.0 |
| B.pMKQS | 704 | **363** | **248** | **131** | **97** | **81** | **65** | 54.9 | 51.3 | 50.2 | 50.2 |
| B.pRS-8bit | 1 206 | 585 | 399 | 202 | 139 | 107 | 75 | 60.6 | 46.9 | 43.7 | 39.8 |
| B.pRS-16bit | 1 203 | 550 | 373 | 191 | 129 | 99 | 68 | **53.0** | **39.9** | 37.2 | 35.0 |
| A.pRS | 1 294 | 699 | 502 | 305 | 238 | 205 | 173 | 157.3 | 143.1 | 140.2 | 137.2 |
| S.pMS-2way | | | | | | | | | | | |
| BE.pMS-ms | | | 1 313 | 444 | 306 | 231 | 154 | 108.0 | 77.9 | 69.3 | 57.3 |
| BE.pS$^5$+M-ms | 1 198 | 599 | 432 | 267 | 146 | 135 | 75 | 53.8 | 41.9 | **35.2** | **28.9** |
| BE.pS$^5$+M-ls | 1 214 | 612 | 444 | 275 | 154 | 143 | 84 | 62.7 | 53.4 | 49.2 | 46.4 |
| | **GOV2**, $n = 1.38\,\mathrm{G}$, $N = 64\,\mathrm{Gi}$, $\frac{D}{N} = 77.0\,\%$ | | | | | | | | | | |
| R.mkqs-cache8 | 758 | | | | | | | | | | |
| B.pS$^5$-E | 901 | 448 | 301 | 156 | 113 | 92 | 67.1 | 55.1 | 43.4 | 39.2 | 34.4 |
| B.pS$^5$-UI | 888 | 446 | 299 | **151** | 110 | 91 | 65.8 | 54.0 | 42.3 | 38.7 | 34.4 |
| B.pS$^5$-UIC | 891 | 449 | 299 | 154 | 112 | 92 | 65.9 | 54.6 | 42.6 | 39.1 | 34.7 |
| B.pMKQS | 946 | 481 | 341 | 170 | 128 | 102 | 73.4 | 62.0 | 52.9 | 53.0 | 45.6 |
| B.pRS-8bit | 2 093 | 1 019 | 674 | 347 | 250 | 252 | 184.5 | 152.1 | 122.3 | 116.8 | 110.1 |
| B.pRS-16bit | 2 095 | 986 | 650 | 337 | 268 | 229 | 169.1 | 140.2 | 109.5 | 102.9 | 98.6 |
| A.pRS | 1 492 | 759 | 588 | 565 | 554 | 554 | 554.4 | 553.0 | 554.0 | 555.3 | 556.5 |
| S.pMS-2way | | 1 183 | 965 | 487 | 368 | 293 | 202.2 | 163.8 | 138.0 | 122.6 | 116.4 |
| BE.pMS-ms | | 564 | 370 | 185 | 131 | 99 | 67.0 | 47.4 | 39.9 | 35.8 | 37.6 |
| BE.pS$^5$+M-ms | **677** | **339** | **250** | 162 | **88** | **84** | **48.9** | **32.3** | **25.6** | **21.0** | **25.9** |
| BE.pS$^5$+M-ls | 686 | 344 | 255 | 168 | 95 | 93 | 55.6 | 43.1 | 42.0 | 32.7 | 30.5 |
| | **Wikipedia**, $n = N = 2\,\mathrm{Gi}$, $D = 116\,\mathrm{G}$ | | | | | | | | | | |
| R.mkqs-cache8 | **1 407** | | | | | | | | | | |
| B.pS$^5$-E | | 967 | 646 | 323 | 218 | 164 | 112 | 85.3 | 59.0 | 51.5 | 46.0 |
| B.pS$^5$-UI | | 949 | 633 | 318 | **211** | **160** | **109** | **83.2** | **57.8** | 50.6 | **45.0** |
| B.pS$^5$-UIC | | 955 | 635 | 320 | 213 | 162 | 110 | 83.7 | 58.1 | **50.4** | 45.3 |
| B.pMKQS | | 934 | **627** | **318** | 218 | 169 | 122 | 97.7 | 76.4 | 71.8 | 66.8 |
| B.pRS-8bit | | 1 455 | 982 | 491 | 328 | 249 | 168 | 128.0 | 90.3 | 80.8 | 74.4 |
| B.pRS-16bit | | 1 388 | 940 | 473 | 318 | 242 | 165 | 126.2 | 88.8 | 79.6 | 73.6 |
| A.pRS | | 1 097 | 754 | 412 | 300 | 242 | 187 | 164.1 | 169.2 | 173.5 | 177.4 |
| S.pMS-2way | | | | | | | | | | | |
| BE.pMS-ms | | | 1 027 | 783 | 851 | 1 046 | 1 315 | | | | |
| BE.pS$^5$+M-ms | 1 646 | 1 006 | 923 | 893 | 1 077 | 1 148 | 1 451 | | | | |
| BE.pS$^5$+M-ls | 1 640 | **834** | 638 | 443 | 300 | 288 | 155 | 109.1 | 85.7 | 71.9 | 63.6 |

**Table 4.9:** Absolute running time of parallel and best sequential algorithms on C.AMD-4×12 in seconds, median of three runs. See table 4.3 for a description of each.

| PEs | 1 | 2 | 3 | 6 | 9 | 12 | 18 | 24 | 36 | 42 | 48 |
|---|---|---|---|---|---|---|---|---|---|---|---|
| | **Sinha URLs** (complete), $n = 10\,\text{M}$, $N = 304\,\text{Mi}$, $\frac{D}{N} = 97.5\,\%$ | | | | | | | | | | |
| R.burstsort-vecblk | 4.91 | | | | | | | | | | |
| B.pS$^5$-E | 5.39 | 2.71 | 1.83 | 1.00 | 0.72 | 0.59 | 0.46 | 0.42 | 0.43 | 0.43 | 0.54 |
| B.pS$^5$-UI | 5.30 | 2.69 | 1.80 | 0.99 | 0.70 | 0.55 | 0.49 | 0.44 | 0.47 | 0.46 | 0.55 |
| B.pS$^5$-UIC | 5.35 | 2.70 | 1.83 | 0.99 | 0.72 | 0.57 | 0.46 | 0.42 | 0.42 | 0.41 | 0.46 |
| B.pMKQS | 6.09 | 3.10 | 2.10 | 1.12 | 0.80 | 0.69 | 0.57 | 0.51 | 0.59 | 0.57 | 0.60 |
| B.pRS-8bit | 10.86 | 5.11 | 3.47 | 1.83 | 1.27 | 1.10 | 0.81 | 0.73 | 0.65 | 0.60 | 0.76 |
| B.pRS-16bit | 10.87 | 4.58 | 3.17 | 1.71 | 1.26 | 1.04 | 0.98 | 1.04 | 1.18 | 1.27 | 1.40 |
| A.pRS | 9.42 | 7.96 | 7.49 | 7.02 | 6.82 | 6.72 | 6.62 | 6.59 | 6.57 | 6.55 | 6.59 |
| S.pMS-2way | 13.29 | 7.61 | 8.28 | 5.70 | 6.38 | 7.34 | 9.05 | 11.10 | 15.09 | 16.54 | 18.61 |
| BE.pMS-ms | 6.98 | 3.38 | 2.28 | 1.20 | 0.83 | 0.65 | 0.48 | 0.40 | 0.37 | 0.39 | 0.47 |
| BE.pS$^5$+M-ms | **4.05** | **2.02** | **1.48** | **0.95** | **0.51** | **0.50** | **0.33** | **0.28** | **0.27** | **0.32** | **0.37** |
| BE.pS$^5$+M-ls | 4.10 | 2.17 | 1.64 | 1.10 | 0.67 | 0.65 | 0.71 | 0.44 | 0.50 | 0.51 | 0.52 |
| | **Sinha DNA** (complete), $n = 31.6\,\text{M}$, $N = 302\,\text{Mi}$, $\frac{D}{N} = 100\,\%$ | | | | | | | | | | |
| KRB.radixsort-CE3s | **7.02** | | | | | | | | | | |
| B.pS$^5$-E | 9.73 | 4.95 | 3.33 | 1.71 | 1.17 | **0.91** | 0.66 | 0.58 | 0.45 | 0.42 | 0.45 |
| B.pS$^5$-UI | 9.59 | 4.84 | 3.25 | 1.66 | 1.14 | 0.91 | **0.62** | 0.55 | 0.50 | 0.44 | 0.48 |
| B.pS$^5$-UIC | 9.39 | 4.81 | 3.23 | **1.66** | 1.14 | 0.92 | 0.66 | 0.53 | **0.44** | **0.42** | 0.56 |
| B.pMKQS | 8.95 | 4.57 | 3.28 | 1.77 | 1.22 | 1.09 | 0.83 | 0.86 | 0.85 | 0.83 | 0.88 |
| B.pRS-8bit | 12.56 | 6.42 | 4.47 | 2.25 | 1.59 | 1.25 | 0.99 | 0.82 | 0.69 | 0.64 | 0.66 |
| B.pRS-16bit | 12.58 | 6.03 | 4.04 | 2.13 | 1.58 | 1.21 | 0.92 | 0.85 | 0.77 | 0.84 | 0.90 |
| A.pRS | 9.88 | 5.44 | 3.87 | 2.66 | 2.03 | 1.93 | 1.53 | 1.53 | 1.50 | 1.53 | 1.47 |
| S.pMS-2way | 41.18 | 23.23 | 23.17 | 14.75 | 15.95 | 18.11 | 20.68 | 25.00 | 33.43 | 37.82 | 42.21 |
| BE.pMS-ms | 24.65 | 11.26 | 7.42 | 3.67 | 2.57 | 1.94 | 1.39 | 1.01 | 0.86 | 0.89 | 0.91 |
| BE.pS$^5$+M-ms | 8.75 | **4.32** | **3.17** | 2.04 | **1.10** | 1.06 | 0.63 | **0.52** | 0.47 | 0.46 | **0.45** |
| BE.pS$^5$+M-ls | 8.93 | 4.44 | 3.26 | 2.08 | 1.13 | 1.11 | 0.72 | 0.62 | 0.56 | 0.59 | 0.65 |
| | **Sinha NoDup** (complete), $n = 31.6\,\text{M}$, $N = 382\,\text{Mi}$, $\frac{D}{N} = 73.4\,\%$ | | | | | | | | | | |
| KR.radixsort-CE7 | **5.24** | | | | | | | | | | |
| B.pS$^5$-E | 12.65 | 6.43 | 4.34 | 2.26 | 1.53 | 1.20 | 0.85 | 0.70 | 0.62 | 0.55 | 0.53 |
| B.pS$^5$-UI | 12.51 | 6.35 | 4.29 | 2.20 | 1.52 | 1.17 | 0.87 | 0.66 | 0.50 | 0.46 | 0.52 |
| B.pS$^5$-UIC | 12.37 | 6.47 | 4.36 | 2.27 | 1.56 | 1.18 | 0.85 | 0.71 | 0.52 | 0.58 | 0.61 |
| B.pMKQS | 11.08 | 5.66 | 3.96 | 2.06 | 1.49 | 1.24 | 0.96 | 0.88 | 0.83 | 0.84 | 0.91 |
| B.pRS-8bit | 10.68 | 5.22 | 3.50 | 1.82 | 1.30 | 0.99 | 0.80 | **0.63** | **0.49** | **0.44** | **0.43** |
| B.pRS-16bit | 10.72 | **4.58** | **3.10** | **1.59** | **1.19** | **0.93** | **0.79** | 0.68 | 0.75 | 0.67 | 0.89 |
| A.pRS | 10.72 | 5.82 | 4.16 | 2.53 | 1.96 | 1.71 | 1.50 | 1.37 | 1.32 | 1.27 | 1.26 |
| S.pMS-2way | 46.15 | 27.09 | 28.44 | 22.18 | 24.71 | 28.51 | 34.80 | 41.43 | 56.45 | 60.59 | 66.96 |
| BE.pMS-ms | 27.92 | 12.90 | 8.49 | 4.11 | 2.83 | 2.14 | 1.49 | 1.08 | 0.85 | 0.84 | 0.72 |
| BE.pS$^5$+M-ms | 13.51 | 6.76 | 4.84 | 2.91 | 1.68 | 1.49 | 0.91 | 0.72 | 0.61 | 0.55 | 0.59 |
| BE.pS$^5$+M-ls | 13.67 | 6.85 | 4.92 | 2.94 | 1.68 | 1.49 | 0.92 | 0.72 | 0.67 | 0.71 | 0.68 |

**Table 4.10:** Absolute running time of parallel and best sequential algorithms on D.Intel-4×8 in seconds, median of three runs. See table 4.3 for a description of each.

| PEs | 1 | 2 | 4 | 8 | 12 | 16 | 24 | 32 | 48 | 64 |
|---|---|---|---|---|---|---|---|---|---|---|
| | **URLs** (complete), $n = 1.11\,\text{G}$, $N = 70.7\,\text{Gi}$, $\frac{D}{N} = 93.5\,\%$ | | | | | | | | | |
| B.LCP-MS-2way | 460 | | | | | | | | | |
| B.pS$^5$-E | 624 | 305 | 144 | 80 | 59 | 47.4 | 40.9 | 38.3 | 39.1 | 39.7 |
| B.pS$^5$-UI | 614 | 296 | 146 | 78 | 60 | 46.2 | 40.0 | 41.0 | 38.2 | 36.9 |
| B.pS$^5$-UIC | 616 | 297 | 142 | 78 | 57 | 46.6 | 43.8 | 38.3 | 38.5 | 39.6 |
| B.pMKQS | 684 | 318 | 145 | 87 | 70 | 57.0 | 50.7 | 74.0 | 48.3 | 45.2 |
| B.pRS-8bit | 1 916 | 921 | 409 | 246 | 192 | 179.2 | 155.2 | 152.8 | 154.1 | 146.9 |
| B.pRS-16bit | 1 905 | 811 | 375 | 230 | 166 | 146.8 | 136.8 | 136.4 | 125.4 | 126.8 |
| A.pRS | 1 182 | 1 201 | 1 194 | 1 153 | 1 199 | 1 148.8 | 1 190.0 | 1 147.9 | 1 201.2 | 1 149.0 |
| S.pMS-2way | 1 033 | 544 | 369 | 388 | 435 | 473.4 | 643.4 | 767.5 | 1 036.9 | 1 310.2 |
| BE.pMS-ms | 492 | 226 | 114 | 75 | 62 | 53.6 | 37.0 | 29.5 | 21.8 | 19.3 |
| BE.pS$^5$+M-ms | **393** | **193** | **100** | **60** | **44** | **33.6** | **24.6** | **20.4** | **21.0** | **18.6** |
| BE.pS$^5$+M-ls | 398 | 203 | 114 | 74 | 58 | 50.9 | 36.7 | 34.0 | 43.2 | 35.0 |
| | **Random**, $n = 3.27\,\text{G}$, $N = 32\,\text{Gi}$, $\frac{D}{N} = 44.9\,\%$ | | | | | | | | | |
| KR.radixsort-DB | **512** | | | | | | | | | |
| B.pS$^5$-E | | 1 076 | 485 | 251 | 176 | 137.1 | 101.7 | 78.1 | 65.4 | 61.5 |
| B.pS$^5$-UI | | 908 | 398 | 208 | 146 | 115.5 | 79.3 | 65.6 | 56.2 | 51.5 |
| B.pS$^5$-UIC | 1 917 | 887 | 400 | 209 | 148 | 116.3 | 82.8 | 69.3 | 62.6 | 59.8 |
| B.pMKQS | 655 | **339** | **173** | **100** | **77** | **65.4** | **60.0** | 59.4 | 58.1 | 59.6 |
| B.pRS-8bit | 1 510 | 698 | 328 | 174 | 131 | 96.7 | 67.9 | 58.1 | 55.9 | 56.7 |
| B.pRS-16bit | 1 513 | 590 | 290 | 153 | 112 | 82.5 | 61.5 | 50.0 | 44.7 | 42.7 |
| A.pRS | 1 254 | 621 | 333 | 221 | 188 | 163.3 | 147.6 | 137.0 | 132.4 | 129.3 |
| S.pMS-2way | | | | | | | | | | |
| BE.pMS-ms | | | 582 | 318 | 288 | 160.3 | 113.0 | 105.2 | 78.2 | 69.0 |
| BE.pS$^5$+M-ms | 1 461 | 689 | 328 | 161 | 134 | 85.0 | 62.7 | **47.3** | **40.0** | **34.5** |
| BE.pS$^5$+M-ls | 1 487 | 720 | 352 | 188 | 130 | 102.5 | 83.0 | 57.8 | 59.0 | 58.6 |
| | **GOV2**, $n = 3.1\,\text{G}$, $N = 128\,\text{Gi}$, $\frac{D}{N} = 82.7\,\%$ | | | | | | | | | |
| R.mkqs-cache8 | 1 083 | | | | | | | | | |
| B.pS$^5$-E | | 777 | 346 | 213 | 165 | 132 | 106.6 | 91.8 | 84.0 | 81.6 |
| B.pS$^5$-UI | 1 573 | 725 | 328 | 204 | 156 | 129 | 101.4 | 89.7 | 80.8 | 79.0 |
| B.pS$^5$-UIC | 1 552 | 726 | 325 | 205 | 156 | 126 | 106.2 | 86.7 | 82.3 | 79.3 |
| B.pMKQS | 1 425 | 718 | 365 | 217 | 177 | 151 | 124.4 | 115.5 | 109.4 | 108.3 |
| B.pRS-8bit | | | 847 | 507 | 431 | 369 | 345.5 | 296.0 | 296.9 | 282.7 |
| B.pRS-16bit | | | 826 | 508 | 420 | 358 | 283.4 | 269.2 | 267.3 | 262.7 |
| A.pRS | | 1 214 | 913 | 987 | 980 | 984 | 984.2 | 981.2 | 993.4 | 992.5 |
| S.pMS-2way | | | | | | | | | | |
| BE.pMS-ms | | 820 | 380 | 193 | 133 | 104 | 92.3 | 65.1 | 80.5 | 103.6 |
| BE.pS$^5$+M-ms | **1 014** | **491** | **252** | **154** | **118** | **93** | **82.1** | **59.4** | **55.2** | **53.9** |
| BE.pS$^5$+M-ls | 1 027 | 501 | 260 | 164 | 127 | 106 | 84.6 | 76.4 | 85.1 | 75.8 |
| | **Wikipedia**, $n = N = 4\,\text{Gi}$, $D = 249\,\text{G}$ | | | | | | | | | |
| KR.radixsort-CE2 | **2 084** | | | | | | | | | |
| B.pS$^5$-E | | | | | 261 | 207 | 148.2 | 122.2 | 109.5 | 104.8 |
| B.pS$^5$-UI | | | | | 229 | 179 | 127.3 | 106.2 | 92.8 | 86.8 |
| B.pS$^5$-UIC | | | | 329 | 236 | 195 | 140.8 | 117.5 | 106.4 | 104.3 |
| B.pMKQS | | | | **315** | 230 | 186 | 147.7 | 128.2 | 119.4 | 118.7 |
| B.pRS-8bit | | | | | | 276 | 190.0 | 161.2 | 148.1 | 141.8 |
| B.pRS-16bit | | | | | | 256 | 195.7 | 155.4 | 145.1 | 137.3 |
| A.pRS | | | | | 342 | 291 | 244.6 | 224.7 | 239.4 | 252.4 |
| S.pMS-2way | | | | | | | | | | |
| BE.pMS-ms | | | | | | | | | | |
| BE.pS$^5$+M-ms | | | | | | | | | | |
| BE.pS$^5$+M-ls | | | | | **216** | **165** | **116.6** | **96.1** | **86.5** | **79.5** |

**Table 4.11:** Absolute running time of parallel and best sequential algorithms on D.Intel-4×8 in seconds, median of three runs. See table 4.3 for a description of each.

| PEs | 1 | 2 | 4 | 8 | 12 | 16 | 24 | 32 | 48 | 64 |
|---|---|---|---|---|---|---|---|---|---|---|
| | **Sinha URLs** (complete), $n = 10\,\mathrm{M}$, $N = 304\,\mathrm{Mi}$, $\frac{D}{N} = 97.5\,\%$ | | | | | | | | | |
| R.mkqs-cache8 | **1.92** | | | | | | | | | |
| B.pS$^5$-E | 3.50 | 1.72 | 0.81 | 0.63 | 0.38 | 0.50 | 0.33 | 0.39 | 0.31 | **0.41** |
| B.pS$^5$-UI | 3.31 | 1.63 | 0.93 | 0.51 | 0.39 | 0.35 | 0.40 | 0.31 | 0.36 | 0.42 |
| B.pS$^5$-UIC | 3.41 | 1.63 | 0.85 | 0.48 | 0.38 | 0.45 | 0.31 | 0.30 | 0.33 | 0.46 |
| B.pMKQS | 3.68 | 1.79 | 0.87 | 0.69 | 0.48 | 0.45 | 0.47 | 0.42 | 0.46 | 0.71 |
| B.pRS-8bit | 6.57 | 2.99 | 1.47 | 0.80 | 0.97 | 0.69 | 0.65 | 0.77 | 0.57 | 0.68 |
| B.pRS-16bit | 6.65 | 2.73 | 1.41 | 0.75 | 0.60 | 0.68 | 0.47 | 0.53 | 0.54 | 0.64 |
| A.pRS | 5.53 | 4.61 | 4.23 | 4.04 | 3.83 | 3.97 | 3.79 | 3.86 | 3.85 | 3.83 |
| S.pMS-2way | 9.24 | 5.17 | 4.48 | 5.37 | 6.60 | 7.76 | 9.51 | 11.95 | 15.84 | 21.02 |
| BE.pMS-ms | 4.65 | 2.39 | 1.08 | 0.59 | 0.43 | 0.44 | 0.32 | 0.33 | 0.41 | 0.50 |
| BE.pS$^5$+M-ms | 2.25 | **1.13** | **0.56** | **0.32** | **0.26** | **0.22** | **0.21** | **0.20** | **0.29** | 0.46 |
| BE.pS$^5$+M-ls | 2.27 | 1.24 | 0.65 | 0.41 | 0.36 | 0.34 | 0.30 | 0.34 | 0.36 | 0.59 |
| | **Sinha DNA** (complete), $n = 31.6\,\mathrm{M}$, $N = 302\,\mathrm{Mi}$, $\frac{D}{N} = 100\,\%$ | | | | | | | | | |
| KR.radixsort-CE7 | **3.55** | | | | | | | | | |
| B.pS$^5$-E | 6.68 | 3.24 | 1.57 | 0.83 | 0.59 | 0.55 | 0.48 | 0.37 | 0.33 | 0.34 |
| B.pS$^5$-UI | 6.42 | 3.05 | 1.49 | 0.79 | 0.59 | 0.61 | 0.37 | 0.45 | **0.31** | **0.31** |
| B.pS$^5$-UIC | 6.38 | 3.11 | 1.70 | 0.93 | 0.64 | 0.45 | 0.49 | **0.33** | 0.34 | 0.36 |
| B.pMKQS | 6.10 | 3.04 | 1.59 | 1.00 | 0.95 | 0.79 | 0.70 | 0.66 | 0.65 | 0.95 |
| B.pRS-8bit | 8.03 | 4.10 | 2.03 | 1.15 | 0.82 | 0.95 | 0.69 | 0.66 | 0.55 | 0.67 |
| B.pRS-16bit | 8.00 | 3.88 | 2.21 | 1.07 | 1.05 | 0.65 | 0.52 | 0.66 | 0.56 | 0.53 |
| A.pRS | 6.25 | 3.20 | 2.56 | 1.24 | 1.09 | 1.12 | 0.86 | 0.87 | 0.98 | 0.86 |
| S.pMS-2way | 27.38 | 15.93 | 11.52 | 13.45 | 14.40 | 16.85 | 22.18 | 27.33 | 36.88 | 46.95 |
| BE.pMS-ms | 17.57 | 8.65 | 3.47 | 1.78 | 1.27 | 1.00 | 0.76 | 0.73 | 0.82 | 1.02 |
| BE.pS$^5$+M-ms | 4.76 | **2.39** | **1.19** | **0.64** | 0.52 | **0.40** | **0.36** | 0.39 | 0.40 | 0.69 |
| BE.pS$^5$+M-ls | 4.87 | 2.44 | 1.24 | 0.66 | **0.51** | 0.45 | 0.42 | 0.44 | 0.43 | 0.68 |
| | **Sinha NoDup** (complete), $n = 31.6\,\mathrm{M}$, $N = 382\,\mathrm{Mi}$, $\frac{D}{N} = 73.4\,\%$ | | | | | | | | | |
| KR.radixsort-CE7 | **3.56** | | | | | | | | | |
| B.pS$^5$-E | 9.32 | 4.54 | 2.21 | 1.21 | 1.06 | 0.85 | 0.67 | 0.62 | 0.55 | 0.55 |
| B.pS$^5$-UI | 8.97 | 4.43 | 2.24 | 1.32 | 0.95 | 0.82 | 0.73 | 0.70 | 0.71 | 0.65 |
| B.pS$^5$-UIC | 9.03 | 4.41 | 2.33 | 1.19 | 0.93 | 0.76 | 0.65 | 0.63 | 0.56 | 0.59 |
| B.pMKQS | 7.84 | 3.94 | 2.43 | 1.15 | 1.15 | 0.75 | 0.65 | 0.70 | 0.74 | 0.92 |
| B.pRS-8bit | 7.63 | 3.81 | 2.09 | 0.98 | 0.98 | **0.61** | 0.62 | 0.61 | **0.43** | 0.44 |
| B.pRS-16bit | 7.53 | **3.26** | **1.55** | **0.92** | **0.64** | 0.61 | **0.43** | **0.40** | 0.49 | **0.43** |
| A.pRS | 7.27 | 3.83 | 2.33 | 1.58 | 1.25 | 1.17 | 1.05 | 1.04 | 0.95 | 0.99 |
| S.pMS-2way | 28.84 | 17.39 | 13.96 | 17.76 | 21.43 | 23.92 | 31.69 | 37.15 | 50.18 | 62.27 |
| BE.pMS-ms | 20.61 | 9.81 | 4.19 | 2.19 | 1.91 | 1.25 | 1.03 | 0.74 | 0.70 | 0.80 |
| BE.pS$^5$+M-ms | 9.06 | 4.78 | 2.42 | 1.37 | 0.91 | 0.78 | 0.63 | 0.57 | 0.54 | 0.65 |
| BE.pS$^5$+M-ls | 9.20 | 4.71 | 2.43 | 1.24 | 0.90 | 0.73 | 0.61 | 0.56 | 0.81 | 0.89 |

**Table 4.12:** Absolute running time of parallel and best sequential algorithms on E.Intel-2×16 in seconds, median of three runs, larger test instances. See table 4.3 for a description of each.

| PEs | 1 | 2 | 4 | 8 | 12 | 16 | 24 | 32 | 48 | 64 |
|---|---|---|---|---|---|---|---|---|---|---|
| | **URLs**, $n = 132\,\text{M}$, $N = 8\,\text{Gi}$, $\frac{D}{N} = 92.6\,\%$ | | | | | | | | | |
| R.mkqs-cache8 | **318** | | | | | | | | | |
| B.pS$^5$-E | 322 | **166** | 91.2 | 48.0 | **33.0** | **26.0** | **19.9** | 17.8 | 17.1 | 17.1 |
| B.pS$^5$-UI | 336 | 170 | **91.0** | **47.9** | 33.9 | 26.6 | 20.2 | 17.6 | 16.8 | 17.4 |
| B.pS$^5$-UIC | 369 | 186 | 99.3 | 52.0 | 35.8 | 28.2 | 21.9 | 19.5 | 17.5 | 17.2 |
| B.pMKQS | 337 | 167 | 92.1 | 50.0 | 36.5 | 30.0 | 25.1 | 23.2 | 23.4 | 22.9 |
| B.pRS-8bit | 1 047 | 495 | 266.7 | 142.9 | 101.9 | 81.5 | 69.8 | 67.3 | 63.9 | 62.9 |
| B.pRS-16bit | 1 016 | 452 | 240.4 | 129.7 | 92.5 | 72.6 | 58.1 | 53.3 | 49.8 | 51.0 |
| A.pRS | 660 | 651 | 651.6 | 651.1 | 643.5 | 648.2 | 643.0 | 642.8 | 635.7 | 643.5 |
| S.pMS-2way | 650 | 348 | 233.3 | 212.1 | 222.0 | 243.4 | 301.2 | 360.5 | 492.7 | 623.8 |
| BE.pMS-ms | 470 | 231 | 119.2 | 72.8 | 56.2 | 47.1 | 33.2 | 24.7 | 19.1 | 16.6 |
| BE.pS$^5$+M-ms | 361 | 182 | 95.0 | 51.8 | 36.8 | 29.0 | 20.3 | **17.1** | **14.8** | **14.0** |
| BE.pS$^5$+M-ls | 363 | 185 | 103.6 | 60.5 | 46.1 | 39.0 | 31.5 | 28.4 | 30.2 | 29.9 |
| | **Random**, $n = 307\,\text{M}$, $N = 3\,\text{Gi}$, $\frac{D}{N} = 42.8\,\%$ | | | | | | | | | |
| KR.radixsort-DB | **333** | | | | | | | | | |
| B.pS$^5$-E | 1 261 | 623 | 313 | 159 | 106.6 | 82.0 | 55.8 | 44.7 | 36.9 | 30.8 |
| B.pS$^5$-UI | 1 034 | 506 | 256 | 131 | 88.3 | 67.0 | 46.5 | 37.0 | 32.1 | 27.6 |
| B.pS$^5$-UIC | 1 069 | 531 | 269 | 137 | 92.3 | 70.2 | 48.3 | 38.5 | 32.7 | 28.0 |
| B.pMKQS | 689 | 352 | 180 | 98 | 66.8 | 51.4 | 39.0 | 33.9 | 32.8 | 33.3 |
| B.pRS-8bit | 641 | 315 | **173** | **91** | **61.3** | **47.4** | **33.7** | **26.8** | 24.6 | 23.0 |
| B.pRS-16bit | 620 | 302 | 179 | 93 | 63.5 | 48.2 | 34.5 | 27.1 | **21.9** | **19.5** |
| A.pRS | 495 | **287** | 187 | 127 | 108.4 | 100.0 | 91.4 | 89.1 | 84.1 | 85.1 |
| S.pMS-2way | | | | | | | | | | |
| BE.pMS-ms | 2 356 | 1 223 | 583 | 288 | 197.5 | 147.2 | 99.6 | 75.6 | 64.3 | 52.3 |
| BE.pS$^5$+M-ms | 1 066 | 524 | 267 | 135 | 90.7 | 69.0 | 47.1 | 36.6 | 30.6 | 26.4 |
| BE.pS$^5$+M-ls | 1 072 | 529 | 272 | 139 | 95.3 | 75.4 | 55.3 | 45.2 | 42.4 | 43.5 |
| | **GOV2**, $n = 166\,\text{M}$, $N = 8\,\text{Gi}$, $\frac{D}{N} = 70.6\,\%$ | | | | | | | | | |
| R.mkqs-cache8 | **724** | | | | | | | | | |
| B.pS$^5$-E | 829 | 409 | 212 | 123 | 93.4 | 72.7 | 56.1 | 47.8 | 43.3 | 41.1 |
| B.pS$^5$-UI | 775 | **379** | **199** | **115** | 87.2 | **67.9** | 52.8 | 44.4 | 40.8 | 38.9 |
| B.pS$^5$-UIC | 823 | 403 | 209 | 122 | 93.4 | 72.8 | 56.1 | 47.6 | 43.3 | 40.2 |
| B.pMKQS | 822 | 407 | 218 | 130 | 96.2 | 81.5 | 62.4 | 56.5 | 54.9 | 54.9 |
| B.pRS-8bit | 2 160 | 1 019 | 529 | 288 | 256.3 | 192.4 | 144.9 | 136.8 | 132.2 | 123.5 |
| B.pRS-16bit | 2 121 | 982 | 507 | 280 | 238.9 | 196.1 | 142.7 | 126.9 | 120.4 | 111.5 |
| A.pRS | 1 293 | 655 | 536 | 528 | 527.1 | 524.2 | 527.0 | 525.2 | 530.8 | 535.0 |
| S.pMS-2way | | | | | | | | | | |
| BE.pMS-ms | 1 674 | 794 | 406 | 202 | 135.1 | 102.1 | 72.1 | 61.5 | 79.7 | 102.5 |
| BE.pS$^5$+M-ms | 778 | 406 | 214 | 117 | **82.5** | 68.4 | **52.6** | **41.6** | **38.9** | **36.6** |
| BE.pS$^5$+M-ls | 785 | 409 | 221 | 127 | 96.9 | 83.7 | 68.9 | 59.0 | 58.8 | 59.3 |
| | **Wikipedia**, $n = N = 512\,\text{Mi}$, $D = 21.5\,\text{G}$ | | | | | | | | | |
| R.mkqs-cache8 | **1 663** | | | | | | | | | |
| B.pS$^5$-E | | 953 | 486 | 249 | 168 | 128 | 88.6 | 69.1 | 59.4 | 52.3 |
| B.pS$^5$-UI | | **852** | **431** | **220** | **150** | **114** | **79.8** | **63.7** | **54.8** | **50.1** |
| B.pS$^5$-UIC | | 882 | 446 | 228 | 155 | 118 | 82.5 | 66.0 | 56.4 | 51.0 |
| B.pMKQS | | 881 | 452 | 235 | 161 | 123 | 89.8 | 75.5 | 68.7 | 65.0 |
| B.pRS-8bit | | 1 225 | 624 | 320 | 221 | 164 | 113.5 | 85.8 | 75.5 | 66.3 |
| B.pRS-16bit | | 1 202 | 611 | 309 | 217 | 159 | 112.9 | 84.4 | 74.1 | 64.4 |
| A.pRS | 1 692 | 876 | 476 | 270 | 200 | 167 | 135.3 | 122.7 | 127.8 | 131.8 |
| S.pMS-2way | | | | | | | | | | |
| BE.pMS-ms | | | 1 206 | 1 103 | 1 216 | 1 626 | | | | |
| BE.pS$^5$+M-ms | | 955 | 497 | 363 | 366 | 396 | 508.1 | 647.1 | 1 420.8 | |
| BE.pS$^5$+M-ls | | 934 | 484 | 248 | 171 | 138 | 103.9 | 87.7 | 82.7 | 80.2 |

**Table 4.13:** Absolute running time of parallel and best sequential algorithms on E.Intel-2×16 in seconds, median of three runs, smaller test instances. See table 4.3 for a description of each.

| PEs | 1 | 2 | 4 | 8 | 12 | 16 | 24 | 32 | 48 | 64 |
|---|---|---|---|---|---|---|---|---|---|---|
| | **Sinha URLs** (complete), $n = 10\,\text{M}$, $N = 304\,\text{Mi}$, $\frac{D}{N} = 97.5\,\%$ | | | | | | | | | |
| R.mkqs-cache8 | 2.49 | | | | | | | | | |
| B.pS$^5$-E | 2.21 | 1.15 | 0.62 | 0.35 | 0.26 | 0.23 | 0.20 | 0.24 | 0.25 | **0.25** |
| B.pS$^5$-UI | **2.07** | **1.02** | **0.59** | 0.33 | 0.25 | 0.21 | 0.20 | 0.18 | **0.15** | 0.31 |
| B.pS$^5$-UIC | 2.18 | 1.16 | 0.62 | 0.33 | 0.25 | 0.23 | 0.23 | 0.19 | 0.20 | 0.30 |
| B.pMKQS | 2.53 | 1.29 | 0.68 | 0.39 | 0.29 | 0.25 | 0.23 | 0.28 | 0.31 | 0.61 |
| B.pRS-8bit | 4.58 | 2.04 | 1.08 | 0.56 | 0.43 | 0.36 | 0.27 | 0.37 | 0.59 | 0.39 |
| B.pRS-16bit | 4.57 | 1.91 | 1.00 | 0.52 | 0.39 | 0.33 | 0.34 | 0.33 | 0.36 | 0.56 |
| A.pRS | 3.63 | 3.11 | 2.85 | 2.68 | 2.63 | 2.62 | 2.59 | 2.60 | 2.54 | 2.63 |
| S.pMS-2way | 6.84 | 3.84 | 2.85 | 3.01 | 3.33 | 3.79 | 4.78 | 5.92 | 7.77 | 9.85 |
| BE.pMS-ms | 4.97 | 2.46 | 1.33 | 0.70 | 0.48 | 0.38 | 0.30 | 0.28 | 0.28 | 0.32 |
| BE.pS$^5$+M-ms | 2.16 | 1.16 | 0.60 | **0.32** | **0.25** | **0.19** | **0.17** | **0.18** | 0.29 | 0.33 |
| BE.pS$^5$+M-ls | 2.18 | 1.22 | 0.74 | 0.46 | 0.40 | 0.36 | 0.32 | 0.33 | 0.37 | 0.43 |
| | **Sinha DNA** (complete), $n = 31.6\,\text{M}$, $N = 302\,\text{Mi}$, $\frac{D}{N} = 100\,\%$ | | | | | | | | | |
| KRB.radixsort-CE3s | **2.61** | | | | | | | | | |
| B.pS$^5$-E | 4.98 | 2.50 | 1.33 | 0.69 | 0.46 | **0.34** | 0.30 | **0.24** | 0.40 | **0.25** |
| B.pS$^5$-UI | 4.75 | 2.40 | 1.22 | **0.65** | 0.44 | 0.37 | **0.27** | 0.26 | 0.20 | 0.28 |
| B.pS$^5$-UIC | 4.97 | 2.46 | 1.31 | 0.67 | **0.43** | 0.36 | 0.29 | 0.28 | **0.19** | 0.27 |
| B.pMKQS | 5.45 | 2.82 | 1.47 | 0.79 | 0.55 | 0.44 | 0.41 | 0.41 | 0.65 | 0.54 |
| B.pRS-8bit | 5.71 | 2.77 | 1.41 | 0.73 | 0.54 | 0.43 | 0.33 | 0.33 | 0.26 | 0.35 |
| B.pRS-16bit | 5.72 | 2.52 | 1.34 | 0.70 | 0.52 | 0.37 | 0.29 | 0.30 | 0.40 | 0.41 |
| A.pRS | 3.40 | **1.91** | **1.11** | 0.77 | 0.69 | 0.50 | 0.52 | 0.52 | 0.53 | 0.55 |
| S.pMS-2way | 20.37 | 11.38 | 8.10 | 7.17 | 7.73 | 8.71 | 11.07 | 13.81 | 18.73 | 24.24 |
| BE.pMS-ms | 17.69 | 8.46 | 4.28 | 2.13 | 1.48 | 1.11 | 0.79 | 0.71 | 0.68 | 0.77 |
| BE.pS$^5$+M-ms | 5.02 | 2.45 | 1.29 | 0.67 | 0.45 | 0.39 | 0.30 | 0.32 | 0.32 | 0.32 |
| BE.pS$^5$+M-ls | 5.13 | 2.51 | 1.35 | 0.73 | 0.55 | 0.49 | 0.39 | 0.39 | 0.43 | 0.66 |
| | **Sinha NoDup** (complete), $n = 31.6\,\text{M}$, $N = 382\,\text{Mi}$, $\frac{D}{N} = 73.4\,\%$ | | | | | | | | | |
| KR.radixsort-CE7 | **3.61** | | | | | | | | | |
| B.pS$^5$-E | 7.03 | 3.63 | 1.84 | 0.95 | 0.66 | 0.53 | 0.42 | 0.39 | 0.43 | 0.44 |
| B.pS$^5$-UI | 6.79 | 3.41 | 1.76 | 0.91 | 0.66 | 0.52 | 0.41 | 0.38 | 0.30 | 0.43 |
| B.pS$^5$-UIC | 7.03 | 3.62 | 1.84 | 0.95 | 0.66 | 0.53 | 0.43 | 0.45 | 0.31 | **0.36** |
| B.pMKQS | 7.23 | 3.62 | 1.90 | 0.99 | 0.68 | 0.54 | 0.48 | 0.44 | 0.57 | 0.66 |
| B.pRS-8bit | 5.64 | 2.81 | 1.46 | 0.75 | 0.52 | **0.42** | 0.29 | **0.24** | **0.25** | 0.90 |
| B.pRS-16bit | 5.64 | **2.65** | **1.34** | **0.70** | **0.47** | 0.42 | **0.29** | 0.32 | 0.29 | 0.42 |
| A.pRS | 4.77 | 2.69 | 1.61 | 1.07 | 0.90 | 0.81 | 0.79 | 0.74 | 0.74 | 0.68 |
| S.pMS-2way | 22.27 | 12.61 | 9.78 | 10.18 | 11.59 | 13.17 | 17.08 | 19.57 | 26.37 | 32.43 |
| BE.pMS-ms | 19.93 | 9.24 | 4.56 | 2.32 | 1.59 | 1.20 | 0.82 | 0.68 | 0.61 | 0.63 |
| BE.pS$^5$+M-ms | 7.76 | 3.98 | 2.07 | 1.07 | 0.73 | 0.57 | 0.41 | 0.39 | 0.37 | 0.51 |
| BE.pS$^5$+M-ls | 7.79 | 4.00 | 2.03 | 1.09 | 0.78 | 0.61 | 0.51 | 0.48 | 0.65 | 0.67 |

**Table 4.14:** Absolute running time of parallel and best sequential algorithms on F.AMD-1×16 in seconds, median of three runs, larger test instances. See table 4.3 for a description of each.

| PEs | 1 | 2 | 4 | 6 | 8 | 12 | 16 |
|---|---|---|---|---|---|---|---|
| | **URLs**, $n = 132\,\mathrm{M}$, $N = 8\,\mathrm{Gi}$, $\frac{D}{N} = 92.6\,\%$ | | | | | | |
| B.LCP-MS-2way | **82** | | | | | | |
| B.pS$^5$-E | 92 | **48** | 30.1 | 24.0 | 21.6 | 22.6 | 24.1 |
| B.pS$^5$-UI | 93 | 49 | **30.1** | **23.9** | 21.6 | 22.5 | 24.2 |
| B.pS$^5$-UIC | 100 | 52 | 31.7 | 24.9 | 21.9 | 23.0 | 24.6 |
| B.pMKQS | 93 | 53 | 39.1 | 33.6 | 33.5 | 33.9 | 33.9 |
| B.pRS-8bit | 263 | 142 | 96.3 | 86.4 | 87.0 | 98.0 | 100.5 |
| B.pRS-16bit | 262 | 126 | 85.6 | 75.9 | 77.9 | 82.7 | 87.4 |
| A.pRS | 205 | 204 | 205.2 | 205.1 | 205.3 | 205.3 | 205.0 |
| S.pMS-2way | 156 | 94 | 63.8 | 59.3 | 61.1 | 69.2 | 79.5 |
| BE.pMS-ms | 110 | 61 | 34.5 | 24.9 | **18.7** | **17.3** | **14.3** |
| | **Random**, $n = 307\,\mathrm{M}$, $N = 3\,\mathrm{Gi}$, $\frac{D}{N} = 42.8\,\%$ | | | | | | |
| KR.radixsort-CE2 | **59** | | | | | | |
| B.pS$^5$-E | 257 | 128 | 67.5 | 46.6 | 36.6 | 34.8 | 33.2 |
| B.pS$^5$-UI | 205 | 103 | 55.4 | 38.8 | 31.1 | 30.0 | 28.9 |
| B.pS$^5$-UIC | 212 | 107 | 57.1 | 40.0 | 31.9 | 30.6 | 29.5 |
| B.pMKQS | 144 | 80 | 51.5 | 38.7 | 38.9 | 37.1 | 44.3 |
| B.pRS-8bit | 157 | 81 | 46.1 | 33.3 | 29.1 | 31.5 | 31.4 |
| B.pRS-16bit | 157 | **74** | **42.9** | **29.8** | **25.9** | **28.4** | **28.0** |
| A.pRS | 237 | 125 | 72.0 | 54.7 | 46.1 | 43.7 | 41.5 |
| S.pMS-2way | 968 | 547 | 456.7 | 478.8 | 517.5 | 641.4 | 749.4 |
| BE.pMS-ms | 549 | 269 | 143.7 | 99.2 | 74.8 | 66.5 | 57.1 |
| | **GOV2**, $n = 166\,\mathrm{M}$, $N = 8\,\mathrm{Gi}$, $\frac{D}{N} = 70.6\,\%$ | | | | | | |
| R.mkqs-cache8 | 78.4 | | | | | | |
| B.pS$^5$-E | 80.5 | 40.6 | 23.5 | 16.9 | 14.8 | 16.0 | 15.7 |
| B.pS$^5$-UI | 77.8 | **39.5** | **22.4** | **16.5** | **14.4** | **15.3** | **15.5** |
| B.pS$^5$-UIC | 80.1 | 40.8 | 23.0 | 16.8 | 14.8 | 15.9 | 15.7 |
| B.pMKQS | **76.6** | 40.7 | 26.8 | 21.9 | 19.3 | 21.9 | 22.7 |
| B.pRS-8bit | 167.9 | 85.4 | 51.6 | 40.6 | 40.7 | 42.5 | 44.3 |
| B.pRS-16bit | 167.9 | 82.6 | 49.0 | 38.2 | 36.8 | 38.3 | 39.7 |
| A.pRS | 140.0 | 73.0 | 62.6 | 62.4 | 63.3 | 67.4 | 69.0 |
| S.pMS-2way | | 121.1 | 69.1 | 59.5 | 49.6 | 48.8 | 46.4 |
| BE.pMS-ms | 137.4 | 71.4 | 39.0 | 27.3 | 21.1 | 19.3 | 16.2 |
| | **Wikipedia**, $n = N = 512\,\mathrm{Mi}$, $D = 21.5\,\mathrm{G}$ | | | | | | |
| KR.radixsort-CE7 | **163** | | | | | | |
| B.pS$^5$-E | 233 | 118 | 64.4 | 45.0 | 35.6 | 34.0 | 32.6 |
| B.pS$^5$-UI | 206 | 104 | **57.5** | **40.3** | **32.2** | **31.0** | **29.9** |
| B.pS$^5$-UIC | 211 | 107 | 58.9 | 41.2 | 32.9 | 31.7 | 30.3 |
| B.pMKQS | 212 | 107 | 63.6 | 47.8 | 42.1 | 42.5 | 42.8 |
| B.pRS-8bit | 221 | 109 | 62.1 | 45.8 | 37.6 | 38.4 | 38.8 |
| B.pRS-16bit | 220 | **103** | 58.3 | 43.1 | 35.0 | 35.9 | 36.6 |
| A.pRS | 204 | 109 | 66.0 | 49.9 | 42.8 | 41.3 | 41.4 |
| S.pMS-2way | | 346 | 189.1 | 141.7 | 118.8 | 117.6 | 117.8 |
| BE.pMS-ms | 501 | 246 | 177.5 | 140.8 | 172.2 | 279.5 | 509.4 |

**Table 4.15:** Absolute running time of parallel and best sequential algorithms on F.AMD-1×16 in seconds, median of three runs, smaller test instances. See table 4.3 for a description of each.

| PEs | 1 | 2 | 4 | 6 | 8 | 12 | 16 |
|---|---|---|---|---|---|---|---|
| | **Sinha URLs** (complete), $n = 10\,\mathrm{M}$, $N = 304\,\mathrm{Mi}$, $\frac{D}{N} = 97\,5\,\%$ | | | | | | |
| R.burstsort-vecblk | 1.49 | | | | | | |
| B.pS$^5$-E | 1.38 | 0.71 | 0.42 | **0.33** | 0.37 | **0.28** | 0.39 |
| B.pS$^5$-UI | **1.32** | **0.67** | **0.40** | 0.34 | **0.26** | 0.28 | **0.32** |
| B.pS$^5$-UIC | 1.36 | 0.69 | 0.41 | 0.35 | 0.35 | 0.34 | 0.34 |
| B.pMKQS | 1.51 | 0.78 | 0.48 | 0.39 | 0.38 | 0.46 | 0.46 |
| B.pRS-8bit | 2.39 | 1.21 | 0.73 | 0.57 | 0.51 | 0.56 | 0.62 |
| B.pRS-16bit | 2.42 | 1.09 | 0.67 | 0.51 | 0.45 | 0.50 | 0.55 |
| A.pRS | 2.46 | 2.09 | 1.91 | 1.87 | 1.90 | 1.84 | 1.84 |
| S.pMS-2way | 4.46 | 2.53 | 1.99 | 1.94 | 2.03 | 2.37 | 2.78 |
| BE.pMS-ms | 2.72 | 1.39 | 0.78 | 0.58 | 0.48 | 0.40 | 0.35 |
| | **Sinha DNA** (complete), $n = 31.6\,\mathrm{M}$, $N = 302\,\mathrm{Mi}$, $\frac{D}{N} = 100\,\%$ | | | | | | |
| KR.radixsort-CE7 | **1.67** | | | | | | |
| B.pS$^5$-E | 2.64 | 1.36 | 0.75 | 0.60 | 0.50 | 0.49 | 0.55 |
| B.pS$^5$-UI | 2.51 | **1.29** | **0.72** | **0.55** | 0.51 | **0.48** | **0.48** |
| B.pS$^5$-UIC | 2.61 | 1.35 | 0.74 | 0.56 | **0.48** | 0.48 | 0.52 |
| B.pMKQS | 2.77 | 1.49 | 0.93 | 0.77 | 0.75 | 0.81 | 0.83 |
| B.pRS-8bit | 3.26 | 1.63 | 1.02 | 0.84 | 0.79 | 0.86 | 0.91 |
| B.pRS-16bit | 3.25 | 1.45 | 0.94 | 0.79 | 0.71 | 0.81 | 0.91 |
| A.pRS | 2.25 | 1.33 | 0.92 | 0.86 | 0.83 | 0.80 | 0.81 |
| S.pMS-2way | 12.88 | 7.28 | 5.20 | 4.76 | 4.73 | 5.51 | 6.19 |
| BE.pMS-ms | 9.59 | 4.81 | 2.62 | 1.82 | 1.44 | 1.17 | 1.03 |
| | **Sinha NoDup** (complete), $n = 31.6\,\mathrm{M}$, $N = 382\,\mathrm{Mi}$, $\frac{D}{N} = 73.4\,\%$ | | | | | | |
| KR.radixsort-CE7 | **1.98** | | | | | | |
| B.pS$^5$-E | 3.99 | 2.05 | 1.14 | 0.84 | 0.72 | 0.62 | 0.62 |
| B.pS$^5$-UI | 3.88 | 1.99 | 1.11 | 0.82 | 0.65 | 0.64 | 0.67 |
| B.pS$^5$-UIC | 3.97 | 2.03 | 1.13 | 0.82 | 0.66 | 0.67 | 0.59 |
| B.pMKQS | 3.88 | 2.01 | 1.18 | 0.92 | 0.84 | 0.81 | 0.81 |
| B.pRS-8bit | 2.93 | 1.53 | 0.89 | 0.67 | 0.58 | 0.59 | 0.66 |
| B.pRS-16bit | 3.15 | **1.33** | **0.78** | **0.60** | **0.50** | **0.58** | **0.57** |
| A.pRS | 3.17 | 1.78 | 1.12 | 0.88 | 0.79 | 0.73 | 0.73 |
| S.pMS-2way | 14.59 | 8.31 | 6.84 | 7.02 | 7.51 | 8.90 | 10.38 |
| BE.pMS-ms | 11.42 | 5.39 | 2.93 | 1.98 | 1.53 | 1.29 | 1.10 |

**Table 4.16:** Absolute running times in seconds and slowdown factor of sequential string sorters run independently in parallel on all eight cores of A.Intel-1×8 over running the sorter on only a single core.

| | Overall | | Our Datasets | | | | Sinha's | | |
|---|---|---|---|---|---|---|---|---|---|
| | Rank | AriM | URLs | Random | GOV2 | Wikip | URLs | DNA | NoDup |
| | single core | | | | | | | | |
| R.mkqs-cache8 | 3 | 1.99 | **1.51** | 3.36 | **1.50** | 6.37 | **0.19** | 0.49 | 0.54 |
| KRB.radixsort-CI3s | 4 | 2.18 | 3.28 | 2.01 | 3.10 | 5.87 | 0.28 | 0.38 | 0.37 |
| KR.radixsort-CE6 | 2 | 1.93 | 3.39 | **1.62** | 2.66 | 4.93 | 0.28 | 0.30 | **0.31** |
| KR.radixsort-CE7 | 1 | **1.92** | 3.33 | 1.64 | 2.65 | **4.93** | 0.29 | **0.30** | 0.31 |
| B.Seq-$S^5$-E | 7 | 2.68 | 2.89 | 3.58 | 3.36 | 7.74 | 0.28 | 0.43 | 0.47 |
| B.Seq-$S^5$-UI | 5 | 2.53 | 3.02 | 3.17 | 3.31 | 7.07 | 0.28 | 0.41 | 0.48 |
| B.Seq-$S^5$-UIC | 6 | 2.60 | 3.07 | 3.32 | 3.36 | 7.24 | 0.29 | 0.43 | 0.48 |
| | in parallel on eight cores | | | | | | | | |
| R.mkqs-cache8 | 1 | **4.77** | 5.64 | 6.31 | **3.58** | 15.40 | **0.47** | 1.00 | 1.02 |
| KRB.radixsort-CI3s | 2 | 5.09 | 8.84 | 4.38 | 5.77 | 14.52 | 0.58 | 0.84 | 0.68 |
| KR.radixsort-CE6 | 7 | 5.20 | 10.47 | **4.04** | 5.44 | 14.29 | 0.64 | 0.90 | 0.66 |
| KR.radixsort-CE7 | 3 | 5.10 | 9.80 | 4.04 | 5.45 | **14.20** | 0.64 | 0.90 | **0.66** |
| B.Seq-$S^5$-E | 5 | 5.13 | **5.06** | 5.68 | 5.94 | 17.29 | 0.48 | **0.68** | 0.75 |
| B.Seq-$S^5$-UI | 4 | 5.11 | 5.44 | 5.67 | 5.95 | 16.75 | 0.49 | 0.68 | 0.76 |
| B.Seq-$S^5$-UIC | 6 | 5.18 | 5.56 | 5.78 | 5.97 | 17.02 | 0.51 | 0.69 | 0.77 |
| | slowdown factor | | | | | | | | |
| R.mkqs-cache8 | 5 | 2.40 | 3.75 | 1.88 | 2.39 | 2.42 | 2.42 | 2.04 | 1.89 |
| KRB.radixsort-CI3s | 4 | 2.18 | 2.70 | 2.18 | 1.86 | 2.47 | 2.05 | 2.18 | 1.82 |
| KR.radixsort-CE6 | 7 | 2.56 | 3.09 | 2.49 | 2.04 | 2.90 | 2.27 | 2.98 | 2.14 |
| KR.radixsort-CE7 | 6 | 2.53 | 2.94 | 2.46 | 2.06 | 2.88 | 2.21 | 3.01 | 2.12 |
| B.Seq-$S^5$-E | 1 | **1.74** | **1.75** | **1.59** | **1.77** | **2.23** | **1.70** | **1.57** | 1.58 |
| B.Seq-$S^5$-UI | 3 | 1.82 | 1.80 | 1.79 | 1.80 | 2.37 | 1.77 | 1.65 | **1.58** |
| B.Seq-$S^5$-UIC | 2 | 1.80 | 1.81 | 1.74 | 1.78 | 2.35 | 1.73 | 1.60 | 1.59 |

**Table 4.17:** Absolute running times in seconds and slowdown factor of sequential string sorters run independently in parallel on all 16 cores of F.AMD-1×16 over running the sorter on only a single core.

| | Overall | | Our Datasets | | | | Sinha's | | |
|---|---|---|---|---|---|---|---|---|---|
| | Rank | AriM | URLs | Random | GOV2 | Wikip | URLs | DNA | NoDup |
| | single core | | | | | | | | |
| R.mkqs-cache8 | 4 | 0.88 | **0.59** | 2.21 | **0.56** | 2.30 | **0.08** | 0.20 | 0.19 |
| KRB.radixsort-CI3s | 3 | 0.72 | 1.07 | 1.20 | 0.75 | 1.68 | 0.09 | 0.12 | 0.12 |
| KR.radixsort-CE6 | 1 | **0.66** | 1.31 | 0.81 | 0.76 | **1.44** | 0.10 | **0.10** | 0.10 |
| KR.radixsort-CE7 | 2 | 0.66 | 1.26 | **0.81** | 0.73 | 1.53 | 0.11 | 0.10 | **0.10** |
| B.Seq-$S^5$-E | 7 | 1.05 | 0.98 | 2.91 | 0.86 | 2.18 | 0.11 | 0.15 | 0.16 |
| B.Seq-$S^5$-UI | 5 | 0.95 | 1.00 | 2.27 | 0.84 | 2.14 | 0.10 | 0.15 | 0.16 |
| B.Seq-$S^5$-UIC | 6 | 0.96 | 1.01 | 2.33 | 0.84 | 2.10 | 0.11 | 0.15 | 0.16 |
| | in parallel on 16 cores | | | | | | | | |
| R.mkqs-cache8 | 7 | 3.24 | 3.26 | 7.73 | 2.01 | 8.29 | 0.24 | 0.64 | 0.53 |
| KRB.radixsort-CI3s | 6 | 2.92 | 5.18 | 5.85 | 2.39 | 6.02 | 0.24 | 0.48 | 0.32 |
| KR.radixsort-CE6 | 5 | 2.57 | 6.06 | **3.54** | 2.33 | **5.03** | 0.25 | 0.48 | 0.30 |
| KR.radixsort-CE7 | 4 | 2.50 | 5.52 | 3.55 | 2.32 | 5.08 | 0.25 | 0.48 | **0.30** |
| B.Seq-$S^5$-E | 3 | 2.49 | **2.39** | 6.16 | 1.74 | 6.26 | **0.21** | 0.34 | 0.35 |
| B.Seq-$S^5$-UI | 2 | 2.35 | 2.42 | 5.47 | 1.72 | 5.88 | 0.24 | **0.34** | 0.36 |
| B.Seq-$S^5$-UIC | 1 | **2.35** | 2.40 | 5.46 | **1.71** | 5.94 | 0.22 | 0.35 | 0.34 |
| | slowdown factor | | | | | | | | |
| R.mkqs-cache8 | 5 | 3.59 | 5.53 | 3.49 | 3.56 | 3.61 | 2.96 | 3.18 | 2.80 |
| KRB.radixsort-CI3s | 6 | 3.65 | 4.83 | 4.88 | 3.20 | 3.57 | 2.53 | 3.96 | 2.56 |
| KR.radixsort-CE6 | 7 | 3.66 | 4.61 | 4.36 | 3.08 | 3.48 | 2.51 | 4.60 | 2.98 |
| KR.radixsort-CE7 | 4 | 3.59 | 4.38 | 4.39 | 3.19 | 3.32 | 2.27 | 4.58 | 2.98 |
| B.Seq-$S^5$-E | 1 | **2.25** | 2.45 | **2.12** | **2.02** | 2.87 | **1.92** | **2.26** | 2.13 |
| B.Seq-$S^5$-UI | 3 | 2.35 | 2.43 | 2.40 | 2.06 | **2.75** | 2.29 | 2.30 | 2.22 |
| B.Seq-$S^5$-UIC | 2 | 2.30 | **2.38** | 2.35 | 2.03 | 2.83 | 2.07 | 2.37 | **2.08** |

# II

# Sorting Suffixes in External Memory

*Suffix arrays [MM90; MM93] were created in 1990 as a space saving alternative to suffix trees, and earlier as an index for the Oxford English Dictionary under the name PAT array [GBYS92]. They list all suffixes of a text in lexicographical order, allow fast substring search in unstructured text, and are the starting point for many other stringology algorithms. However, they are computationally expensive to construct, especially for large inputs. To enable searching and processing of large texts using suffix arrays, we consider* scalable *suffix sorting in this part and in part III of this dissertation.*

*In chapter 5 we start with a survey of the history of suffix array construction algorithms, which is incredibly rich and is an interesting example of scientific progress.*

*Then we present our first contribution to this field in chapter 6: the first external memory algorithm based on the induced sorting principle, called* eSAIS*. This principle is used by all the fastest RAM-based suffix sorters, and using it we are able to accelerate external memory suffix sorting by a factor of two. We then continue and extend eSAIS to also construct the LCP array. Our implementation of this algorithm is the first fully external suffix and LCP array construction published in the literature. We demonstrate its scalability with experiments on real-world inputs of up to 80 GiB of Wikipedia text.*

Chapter 5

# 5 A Brief History of Suffix and LCP Array Construction

*Does scientific progress occur by large leaps and bounds or by small incremental steps? Do inventions appear when the time is ripe or when a genius has a sparking idea? Kuhn [Kuh62] made the case that science progresses in phases of small steps interrupted by paradigm shifts when the incremental progress enters a crisis period. The history of suffix array construction algorithms definitely supports his thesis that progress is both evolutionary and revolutionary. The suffix array [MM90; MM93] itself was invented simultaneously in the fields of string and DNA processing and for indexing the Oxford English Dictionary under the name* PAT arrays *[GBYS92]. From 1990 to 2017, more than 23 algorithms or implementations were proposed by no less than 39 authors. In this chapter we review the unique contributions of all these researchers and trace the lineage of their ideas as they evolved in ever more sophisticated suffix sorting algorithms.*

*Suffix arrays* are among the most popular data structures for text indexing and the basis for many string and compression algorithms. They simply list the indices of all suffixes of a static text in lexicographically ascending order. This array not only allows one to efficiently locate arbitrary patterns in unstructured texts (like DNA, East Asian languages, etc.) in time proportional to the *pattern* length (as opposed to *text* length), but also enables fast phrase searches (e.g., "to be or not to be"). In combination with additional auxiliary arrays like the LCP array or the $\Phi$ array, suffix arrays can emulate powerful text data structures like the suffix tree [AKO02; AKO04] and be used to solve a myriad of combinatorial string problems [Apo85; Gus97; CR03] but also real-world bioinformatics applications [Ohl13; MBCT15].

The first and often computationally most expensive step in using suffix arrays is the efficient construction of the index (also called "*suffix sorting*"), where the term "efficient" encompasses both time and space. In this dissertation, we focus on construction of suffix arrays for very large inputs, which is a huge challenge despite asymptotically fast algorithms. The goal is no less than to make progress towards establishing a more powerful indexing technology for searching all mankind's knowledge than the currently popular inverted indices [BCC16]. For these reasons, we are most interested in external memory, parallel, and distributed suffix sorting algorithms that are not limited by the amount of RAM on a machine. Additionally, the problem of suffix sorting can be seen as string sorting with an interesting algorithmic twist due to the suffix properties. As such it can be seen as a natural extension of our work on string sorting in part I.

The original suffix array inventors already proposed an $\mathcal{O}(n \log n)$ construction algorithm [MM93], and in the following years many more with the same time complexity and slower were designed. Up until 2003, the string algorithm community was in the dilemma that while suffix *tree* construction was elegantly possible in linear time [Wei73; McC76; Ukk95; GK97], *direct* suffix *array* construction was only possible in $\mathcal{O}(n \log n)$ time. This was unsatisfactory, since one can simply take the detour of first constructing the suffix tree and then deriving the suffix array from it in linear time via an Euler tour. Hence, a linear-time construction method for suffix arrays was always known, but it required as much space as suffix tree construction.

In 2003, four seminal papers appeared virtually simultaneously proposing direct linear-time suffix array construction algorithms [KS03; KA03; KSPP03; HSS03]. Maybe most surprising is that all four algorithms use different approaches to achieve the same goal. While the theoretical dilemma was solved, it turned out that in practice highly-tuned superlinear algorithms still outperformed the new linear-time algorithms [ARS+04].

This divide between theory and practice was narrowed in 2009 by Nong, Zhang, and Chan's [NZC09a] introduction of the *SA-IS* algorithm, which is linear and also very fast in practice. Yuta Mori, the author of *divsufsort* [Mor06], then picked up SA-IS and published an engineered version that greatly improved the academic version, but did not quite reach divsufsort's speed for real world inputs. Mori's divsufsort and sais [Mor08] are the fastest suffix sorting implementations to date.

The field of suffix sorting has a rich history of algorithm engineering and following the lineage of ideas is very interesting as authors pick them up from previous papers and reforge them into newer, better algorithms or implementations. This lineage is a compelling example of how scientific progress is formed by a series of incremental steps with rare large iconic leaps. Puglisi, Smyth, and Turpin [PST07] published a comprehensive survey of suffix array construction algorithms in 2007. We present our own updated and highly subjective map of the history of suffix array construction in figure 5.2. It is subjective in the sense that the relationship and positioning of the algorithms are our interpretation.

While designing novel sequential RAM-based suffix sorting algorithms is the "ultimate" discipline, there are many related problems, such as LCP array construction in RAM, direct specialized Burrows-Wheeler transform (BWT) construction in RAM, and semi-external or fully external memory construction algorithms for suffix array, LCP array, BWT, or any integrated combination of these. In the following sections, we will review the history of these fields in more details.

Section 5.1 gives a preliminary introduction to the basic definitions and algorithms related to suffix arrays, after which the remainder of this chapter is a guided tour of the history of suffix array construction algorithms. The tour starts with sequential RAM-based algorithms in section 5.2, followed by parallel algorithms in section 5.3, external memory algorithms in section 5.4, and distributed suffix array construction algorithms in section 5.5. The final section 5.6 of the history then switches focus to LCP array construction.

| $i$ | $\mathsf{SA}_T[i]$ | $\mathsf{LCP}_T[i]$ | $\mathsf{BWT}_T[i]$ | $T[\mathsf{SA}_T[i]..n) = [t_{\mathsf{SA}_T[i]}, \dots, t_{n-1}]$ | $t_i$ | $\mathsf{ISA}_T[i]$ |
|---|---|---|---|---|---|---|
| 0 | 14 | - | c | $ | d | 12 |
| 1 | 9 | 0 | b | a b d c c $ | b | 4 |
| 2 | 2 | 1 | b | a d c b c c b a b d c c $ | a | 2 |
| 3 | 8 | 0 | c | b a b d c c $ | d | 13 |
| 4 | 1 | 2 | d | b a d c b c c b a b d c c $ | c | 9 |
| 5 | 5 | 1 | c | b c c b a b d c c $ | b | 5 |
| 6 | 10 | 1 | a | b d c c $ | c | 11 |
| 7 | 13 | 0 | c | c $ | c | 8 |
| 8 | 7 | 1 | c | c b a b d c c $ | b | 3 |
| 9 | 4 | 2 | d | c b c c b a b d c c $ | a | 1 |
| 10 | 12 | 1 | d | c c $ | b | 6 |
| 11 | 6 | 2 | b | c c b a b d c c $ | d | 14 |
| 12 | 0 | 0 | $ | d b a d c b c c b a b d c c $ | c | 10 |
| 13 | 3 | 1 | a | d c b c c b a b d c c $ | c | 7 |
| 14 | 11 | 2 | b | d c c $ | $ | 0 |

**Figure 5.1:** Suffix array example for $T = [\overset{0}{\texttt{d}}, \overset{1}{\texttt{b}}, \overset{2}{\texttt{a}}, \overset{3}{\texttt{d}}, \overset{4}{\texttt{c}}, \overset{5}{\texttt{b}}, \overset{6}{\texttt{c}}, \overset{7}{\texttt{c}}, \overset{8}{\texttt{b}}, \overset{9}{\texttt{a}}, \overset{10}{\texttt{b}}, \overset{11}{\texttt{d}}, \overset{12}{\texttt{c}}, \overset{13}{\texttt{c}}, \overset{14}{\texttt{\$}}]$ with LCP array, Burrows-Wheeler transform, longest common prefixes marked, inverse suffix array, and 1-buckets marked.

In chapter 6 we then present *eSAIS*, which is our external memory adaptation of induced suffix sorting. eSAIS is faster than previous external suffix sorters by a factor of two, and can elegantly be extended to also calculate the LCP array.

This chapter was newly written for this dissertation to put our external memory (chapter 6) and distributed parallel memory algorithms (chapter 8) into perspective.

## 5.1 Definitions and Preliminaries

Let $[0\,..\,n] := \{0, \dots, n\}$ and $[0\,..\,n) := \{0, \dots, n-1\}$ be ranges of integers. For any array $A$, we write $A[\ell\,..\,r]$ to denote the subarray of $A$ ranging from index $\ell$ to $r$. Analogously, $A[\ell\,..\,r)$ denotes the subarray from $\ell$ to $r-1$.

Given a string $T = [t_0, \dots, t_{n-1}]$ of $n$ characters drawn from a totally ordered alphabet $\Sigma$, we call the substring $T[i..n) = [t_i, \dots, t_{n-1}]$ the *$i$-th suffix* of $T$. For a simpler exposition, we assume that $t_{n-1}$ is a unique character '$' that is lexicographically smallest. Notice that all suffixes $\{T[i..n) \mid i \in [0, n)\}$ of a string are distinct, as each has a different length.

The *suffix array* $\mathsf{SA}_T$ of $T$ is the permutation of the integers $[0\,..\,n)$, such that $T[\mathsf{SA}_T[i]..n) < T[\mathsf{SA}_T[j]..n)$ for all $0 \le i < j < n$ (lexicographic order is always intended when comparing strings by "$<$"). The unique inverse permutation of $\mathsf{SA}_T$ is called the *inverse suffix array* $\mathsf{ISA}_T$, and $\mathsf{ISA}_T[i]$ is the *lexicographic rank* of the $i$-th suffix. Figure 5.1 shows an example of a suffix array, the corresponding inverse suffix array, and more arrays discussed later in this section.

While the ranks of all suffixes are distinct, we will often use the notion of a *lexicographic name*. Lexicographic names are *representatives* of suffixes (or more generally, of any strings) which need not be distinct but do respect the lexicographic ordering. Formally, $n_i$ and $n_j$ are lexicographic names of two suffixes $T[i..n) < T[j..n)$ if and only if $n_i \le n_j$. The most common way of defining lexicographic names is to map the set of suffixes $\{T[i..n) \mid i \in [0\,..\,n)\}$ to integers $[0\,..\,m)$ such that the map respects the lexicographic order. For example, mapping each suffix to its first two letters defines lexicographic names.

The auxiliary array $\mathsf{LCP}_T$ of length $n$ is defined as

$$\mathsf{LCP}_T[i] := \textsc{lcp}(T[\mathsf{SA}_T[i-1]..n), T[\mathsf{SA}_T[i]..n))\,,$$

for $i = 1, \ldots, n-1$, but $\mathsf{LCP}_T[0]$ remains undefined. $\mathsf{LCP}_T[i]$ is the length of the longest common prefix (LCP) of the suffixes $T[\mathsf{SA}[i-1]..n)$ and $T[\mathsf{SA}[i]..n)$.

Furthermore, we label the *maximum* and *average* LCP value as

$$\operatorname{maxlcp}(T) := \max_{i=1,\ldots,n-1} \mathsf{LCP}_T[i] \qquad \text{and} \qquad \operatorname{avglcp}(T) := \frac{1}{n-1}\sum_{i=1}^{n-1} \mathsf{LCP}_T[i]\,.$$

Closely related to LCP array and string sorting, we can also define the *distinguishing prefix* of a suffix $i$ with its lexicographic predecessor

$$\operatorname{dps}_T(i) := 1 + \max\{\mathsf{LCP}_T[i-1], \mathsf{LCP}_T[i]\}\,,$$

assuming the sentinels $\mathsf{LCP}_T[0] := 0$ and $\mathsf{LCP}_T[n] := 0$. In the context of suffix sorting, the distinguishing prefix is mainly used to determine the *average logarithmic distinguishing prefix* of suffixes in a string $T$:

$$\operatorname{avglogdps}(T) := \frac{1}{n}\sum_{i=0}^{n-1} \log_2(\operatorname{dps}_T(i))\,,$$

which is a measure of difficulty for the sorting problem instance $T$. For the example string in figure 5.1 these values are $\operatorname{maxlcp}(T) = 2$, $\operatorname{avglcp}(T) = 1.0$, and $\operatorname{avglogdps}(T) \approx 1.334$.

Another companion array to the suffix array is the Burrows-Wheeler transform (BWT) [BW94], which is defined as $\mathsf{BWT}_T[i] := T[\mathsf{SA}_T[i]-1]$ if $\mathsf{SA}_T[i] > 1$, and $\mathsf{BWT}_T[i] = \$$ if $\mathsf{SA}_T[i] = 0$. In words, the Burrows-Wheeler transform is the character

**Algorithm 5.1 :** Simple $\mathcal{O}(|P| \log n)$ Suffix Array Search

```
Function SASearch(P, T, SA)
    Input : P a pattern to search for, T a text string, and SA the suffix array of T.
    n := |T|,   (L, R) := (0, n)                  // Initialize binary search
    while L < R do                                // boundaries [L .. R).
        M := (L + R)/2                            // Find middle suffix and perform
        if T[SA[M]..n) < P then                   // full string comparison.
            L := M + 1                            // Go right if pattern is larger
        else                                      // than the middle,
            R := M                                // or left if pattern is smaller.
    return L                                      // Return left boundary.
    Output : T[L..n) is the smallest suffix of T greater or equal to P.
```

preceding each suffix in lexicographic order, cyclically wrapping around the string if needed. The Burrows-Wheeler transform is popular in lightweight suffix array construction algorithms, text compression, and compressed suffix array representations.

Let us consider some basic pattern search queries using suffix arrays. The simplest search query asks “Is a pattern $P$ a substring of $T$?”, where $P$ is a string. Such a request is called a *decision query*, and can be solved with a plain suffix array in time $\mathcal{O}(|P| \log n)$ using a binary search (see algorithm 5.1). Likewise, the query “Where are all *occ* occurrences of $P$ in $T$?” is called a *enumeration query*, and can be solved with a plain suffix array in time $\mathcal{O}(|P| \log n + occ)$.

By adding more auxiliary arrays, the query time can be improved. The initial paper by Manber and Myers [MM90; MM93] already contained an algorithm answering decision queries in time $\mathcal{O}(|P| + \log n)$, which requires two additional LCP-like arrays to accelerate the search.

Abouelhoda, Kurtz, and Ohlebusch [AKO02; AKO04] proposed *enhanced* suffix arrays, which can fully emulate suffix tree operations. Depending on the specific operation, the suffix array is enhanced with the LCP array, the LCP interval tree, the BWT, the inverse suffix array, or the suffix link table. Using the LCP array and a new table called the *child* table, decision queries can be accelerated to $\mathcal{O}(|P|)$ time, and enumeration queries to $\mathcal{O}(|P| + occ)$ time, which are both optimal. They also show how to store the LCP array and the child table in $n$ bytes each, instead of the trivial $n$ integers.

For processing decision queries in external memory, the binary search algorithm on plain suffix arrays would require the obvious $\mathcal{O}(\lceil \frac{|P|}{B} \rceil \log n)$ I/Os. As this is comparably slow, many authors have proposed different array layouts and compression schemes to accelerate the search.

One approach is based on the idea of building a B-Tree over the suffix array. The resulting data structures is called a *String B-Tree* [FG95; FG96; FG99; Sch02; NP04] and contains Patricia tries [Mor68; FS86] in each tree node. A *blind search* in the String B-Tree requires $\mathcal{O}((|P| + occ)/B + \log_B n)$ I/Os in worst case.

While the String B-Tree has interesting theoretical properties, the most popular and practical solution is a *two-level* index: one small data structure in memory, and a larger one on disk. For example, the *LOF-SA* data structure [SPMT08] consists of a small LOF-trie in memory and a larger external array combining suffix array, LCP array, and fringe characters. Queries are accelerated by first searching the LOF-trie and then the fringe characters in external memory. However, the LOF-SA requires $8n + f$ bytes for 4 byte indices and $f$ fringe characters (the authors suggest $f = 4$), which is close to the space requirements of a suffix tree. By packing, reordering, and compressing the entries in the LOF-SA, the space requirements can be brought down to around $7n$ [MPS09].

Following efforts on the LOF-SA, Gog and Moffat [GM13] proposed the *RoSA* data structure, which is also a two-level index but replaces the in-memory trie with a condensed BWT-based search data structure employing backward search. The on-disk data structures are also reduced by finding duplicate subintervals in the suffix array and replacing them with references. The space requirements of RoSA is around $3n$. The space can again be brought down further using an in-memory dictionary (phrase-book) [GMC+14] which is used to compress both the in-memory BWT data structure and the external memory data store.

## 5.2 Sequential RAM-based Suffix Sorting Algorithms

Despite the large number of suffix array construction algorithms, they all can be viewed as founded on only *three* basic sorting principles: *prefix doubling*, *induced sorting*, originally called *induced copying*, and *divide-and-conquer* aka *recursion*. In figure 5.2 we show a very subjective map of the lineage of the underlying ideas and attempt to assign each algorithm to its prevailing principle.

Informally speaking we describe prefix doubling as "*forward sorting*" of prefixes of the suffixes using a modified string sorting algorithm. Since the suffixes are not independent, the sorting can be accelerated, doubling the sorted prefix in each step. The decisive characteristic is that suffixes are sorted from the front, similar to strings.

Induced sorting or induced *copying*, however, we prefer to describe as "*sorting backwards*", because it induces the order of suffixes from a subset of already ordered suffixes by prepending characters. We consider this backwards, because characters are added to the front of suffixes in an ordered working set. The trick of induced sorting algorithms is how to organize prepending characters and maintain order while doing so.

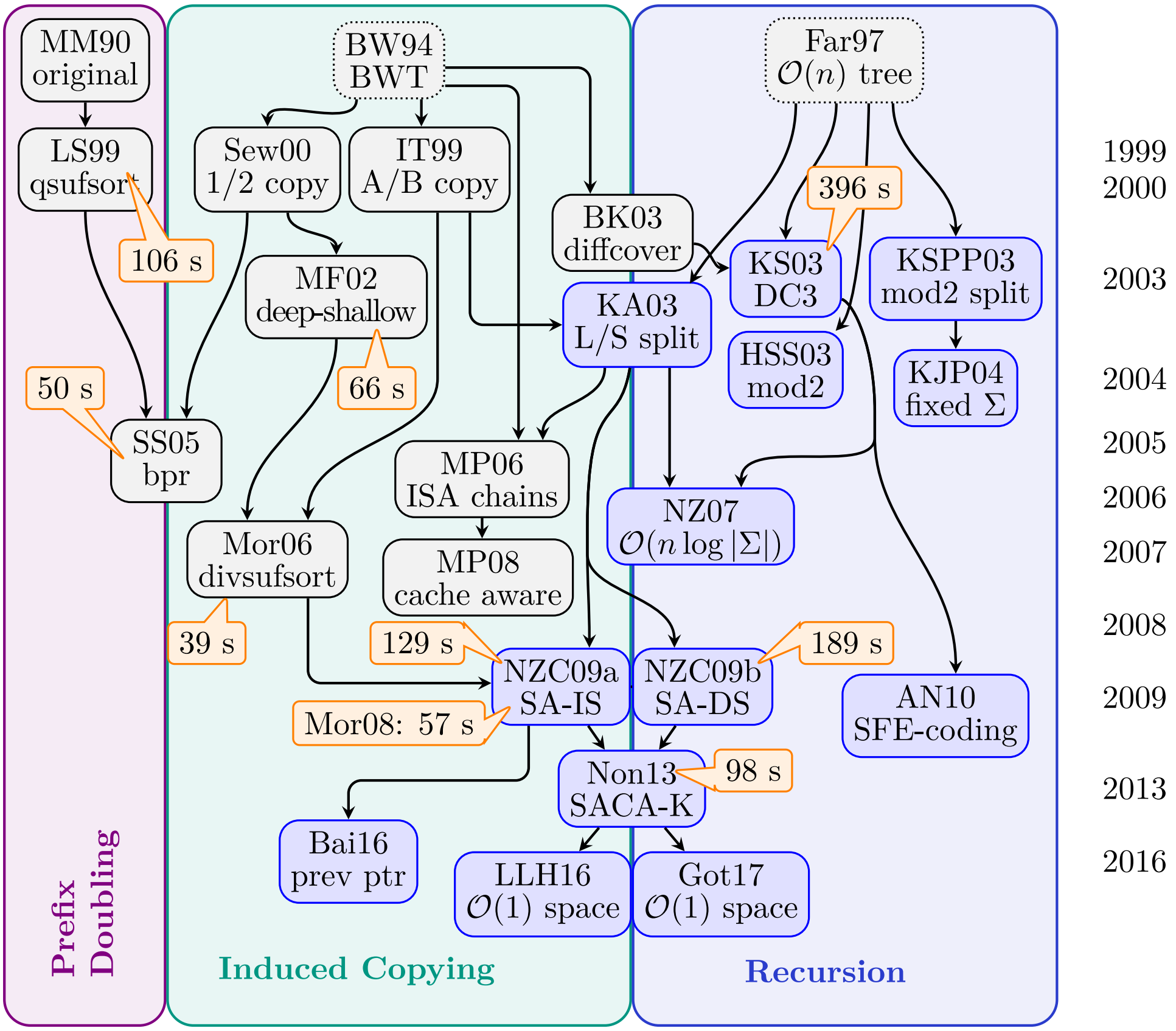

**Figure 5.2:** Our view of the historic lineage of many suffix array construction algorithms, based on [PST07], but heavily modified and updated. Shaded blue algorithms are linear time (for a fixed alphabet size), while gray algorithms are superlinear. The running times is the call out bubbles are for suffix sorting 256 MiB of English text from Project Gutenberg in RAM on an Intel i7 2.67 GHz (A.Intel-1×8 in table 2.1).

And finally, recursion is "*grouping*" of characters of the string to sort them as *larger entities*. Recursion turned out to be the key to the invention of linear-time suffix sorting algorithms. However, each recursive suffix sorting algorithm has a different approach to dividing the input string, grouping characters and creating one or more strings of representatives, and to using the information gained from recursively suffix sorting those strings.

We will review all suffix array construction algorithms in the literature in chronological order in the following paragraphs and compare their running times. Much confusion about the asymptotic running time is due to the alphabet $\Sigma$: some authors only consider an alphabet of *constant size* (e.g. 256 ASCII characters) and hence the size is hidden by the $\mathcal{O}$-notation. Other authors consider alphabets of *fixed size* and introduce a parameter $\sigma = |\Sigma|$ into the running time. Finally, the third alphabet model is purely *comparison-based*, which immediately implies an $\Omega(n \log n)$ lower bound on suffix sorting.

More confusion is created when denoting bounds on the amount of space used by a suffix sorting algorithm. Some authors include the input text and output suffix array in the tally, others count bits and include $n \log_2 n$ for the suffix array space. Furthermore, nearly all authors assume 32-bit integers in their practical programs, such that text and suffix array together need $5n$ bytes. However, 32-bit integers limit the algorithm to input sizes of up to 4 GiB, which is considered relatively small today. So, more recent authors use 40-bit, 48-bit, or even 64-bit integers, hence the text and suffix array account for $6n$, $7n$, or $9n$ bytes. Due to this ambiguity, in this dissertation we distinguish *bytes* (characters are assumed to be bytes) and *integers* when counting space. However, to compare with previous work we often denote *total* space assuming 4 byte integers.

The original suffix array papers by Manber and Myers [MM90; MM93] present a suffix array construction algorithm with $\mathcal{O}(n \log n)$ running time based on prefix doubling [KMR72]. Basically, the suffix array is built by first sorting all suffixes by their first character and by creating a preliminary suffix array ordered to the depth $h = 1$. Then in each iteration, instead of straight-forward string sorting, the sorting depth is doubled to $2h$ by using the order information known in depth $h$. We will review prefix doubling in more depth in section 8.2. The challenge is to efficiently organize bookkeeping of the buckets in the preliminary suffix array. In this context, a *$k$-bucket* is a subrange of the (preliminary) suffix array which contains all suffixes starting with the same $k$ characters (see also the 1-buckets in figure 5.1).

While Manber and Myers' [MM90; MM93] algorithm scans over the entire preliminary suffix array in each iteration, the improved prefix doubling algorithm by Larsson and Sadakane [Sad98; LS99; LS07], called *qsufsort*, uses a more clever technique which allows it to skip over all buckets that contain only one suffix, and hence are fully sorted. Their algorithm also has an $\mathcal{O}(n \log n)$ running time, but is much faster in practice. The authors of both prefix doubling algorithms also mention an engineering technique such that the algorithms require only one array of $n$ integers ($4n$ bytes for 32-bit integers) in addition to the input string and the output suffix array.

In 1994, four years after Manber and Myers [MM90; MM93] invented the suffix array, Burrows and Wheeler [BW94] proposed their famous lossless compression scheme based on a cyclic rotation of lexicographically ordered suffixes. Constructing the BWT is closely related to suffix array construction and algorithms from one problem can often be directly adapted to other. In the original BWT paper [BW94], the authors describe a suffix sorting algorithm based on naive quicksort with some ad-hoc optimizations.

They do not give a performance analysis and we suspect it has worst-case $\mathcal{O}(n^2 \log n)$ running time.

As BWT-based compression was popularized by the *bzip2* tool [Sew97], Seward performed an empirical experiment on suffix sorting approaches [Sew00]. While most of these were naive $\mathcal{O}(n^2 \log n)$ algorithms, they nevertheless outperformed qsufsort on the real-world inputs used in the evaluation. Seward's paper is important because it is one of the first appearances of *induced copying*, which later is more often called *induced sorting*. He writes "once we know the sorted order for all strings beginning with $c_2$, we also know the ordering for all strings starting $c_1c_2$, for all 256 possible $c_1$ values" [Sew00, p. 4]. In his *copy* algorithm, this technique is used after an initial radix sort of all suffixes up to depth two. In the subsequent loops which finish the sorting of each bucket, the algorithm copies the order of a 1-bucket $c_2$ to all 2-buckets $c_1c_2$, and marks them as complete.

Seward notes that his method may synthesize buckets which are already complete, because it sorts buckets in lexicographic order. Itoh and Tanaka [IT99] invented *induced copying* independently from Seward for suffix array construction and avoid superfluously calculating buckets twice in a different manner: They classify each suffix $T[i..n)$ as either type `A`, if $t_i > t_{i+1}$, or type `B`, if $t_i \leq t_{i+1}$. Their key insight is that once all type `B` suffixes are fully sorted, the order of all type `A` suffixes can be induced from them by prepending one character. While they simply use a string sorting algorithm for the first phase, their classification of suffixes into two distinct types was seminal for future developments. Both Seward's [Sew00] and Itoh and Tanaka's [IT99] algorithms run in $\mathcal{O}(n^2 \log n)$ time due to the string sorting and require at least one array of $n$ integers of additional space.

At this time, $\mathcal{O}(n \log n)$ algorithms [Sad98; LS99; LS07] were considerably slower on real-world texts than asymptotically slower algorithms using the *copying* methods. Hence, authors focused on engineering string sorting and copying. Manzini and Ferragina developed the *deep-shallow* suffix sorter [MF02; MF04], which uses many sophisticated string sorting implementations, such as multikey quicksort [BS97] and a method derived from the String B-Tree [FG99] called *blind sorting*. Their implementation uses less than $0.03n$ bytes of additional space, which is *strictly less* than the usual one array of $n$ integers and opens the field of what they call "*lightweight* suffix sorting". Their *deep-shallow* algorithm, however, still has the $\mathcal{O}(n^2 \log n)$ worst-case running time bound of all known suffix sorters relying on forward sorting.

The new field of lightweight suffix sorting was tightened by Burkhardt and Kärkkäinen [BK03], who proposed an algorithm with $\mathcal{O}(n \log n)$ running time bound that uses $\mathcal{O}(n/\sqrt{\log n})$ extra space. The basic idea behind it was to calculate lexicographic ranks of only a specific *sample* of the suffixes. This sample should be constructed in such a way that when comparing two arbitrary suffixes $T[i..n)$ and $T[j..n)$, one can find an *anchor pair* $T[i + k..n)$ and $T[j + k..n)$ in the sample such that their relative order is already known. When this is possible, at most the first $k$ characters plus the known ranks of the anchor pair must be compared. Burkhardt and Kärkkäinen [BK03] proposed to use a *difference cover* [Sin38] to determine the sample suffixes: a difference cover $D$ is a subset of the integers $X = \{0, \ldots, v-1\}$ such that for all $i \in X$ there is a pair $x, y \in D$ with $i = (x - y) \bmod v$. Furthermore, for each $v$ a difference cover of size $\Theta(\sqrt{v})$ exists [CL00]. In their algorithm, they select a small power of two for $v$ and sort the $\mathcal{O}(n/\sqrt{v})$ sample suffixes using the suffix sorter by Larsson and Sadakane [Sad98; LS99; LS07]. For $v > 2^4$, the space needed by the recursive suffix sorter is small enough such that the overall algorithm can be called "lightweight".

In 2003, there was then a break-through in suffix array construction: four direct *linear-time* algorithms were invented [KS03; KA03; KSPP03; HSS03]. All four proposed algorithms incorporate one new algorithmic ingredient: *recursion*. The fundamental idea to use recursion stems from Farach-Colton, Ferragina, and Muthukrishnan [Far97; FCFM00], who proposed to construct the suffix *tree* for *arbitrary* alphabets using recursion and sorting. They classify all suffixes into odd and even positions and construct the suffix tree for only the odd positions recursively by generating a new string. They group pairs of adjacent characters into a single character in the new string by assigning lexicographic names, hence, reducing the string length by two. The even tree is derived after recursion from the odd tree, and then merged with it to gain the complete suffix tree. Their algorithm achieves linear time in the RAM model for alphabets of polynomial size, and sorting complexity in the external memory model (see section 1.1.2), both matching the intuitive lower bound.

Kim, Sim, Park, and Park's [KSPP03; KSPP05] linear-time suffix array construction algorithm is based directly on the split-recurse-merge approach of odd and even suffixes proposed by Farach-Colton, Ferragina, and Muthukrishnan [Far97; FCFM00]. However, their merge routine uses only the suffix and LCP arrays instead of building actual suffix trees. To achieve linear time they have to pre-process the LCP array for fast RMQs, which enables them to emulate most suffix tree operations. While they do not discuss space usage in their paper, we can assume the algorithm needs at least $2n$ additional integer arrays for the LCP array, the RMQ data structures, and the recursion.

Independently, Hon, Sadakane, and Sung [HSS03; HSS09] proposed a similar adaptation of Farach-Colton, Ferragina, and Muthukrishnan's [Far97; FCFM00] recursive odd-even construction method for suffix arrays. However, their goal was mainly to design a suffix sorting algorithm with $\mathcal{O}(n)$ time *and* $\mathcal{O}(n)$ space in the worst-case, which breaks the previously best bounds of $\mathcal{O}(n \log n)$ time with $\mathcal{O}(n)$ space (e.g. [Sad98; LS99; LS07]), or $\mathcal{O}(n)$ time with $\mathcal{O}(n \log n)$ space (e.g. via the suffix tree). Their space-economical solution requires a plethora of advanced techniques from the compressed suffix array [GV00; GV05] and FM-index [FM00; FM05] areas to meet the $\mathcal{O}(n)$ bound. To this day, their proposal has never been implement and remains a purely theoretical consideration.

Kärkkäinen, Sanders, and Burkhardt's [KS03; KSB06] linear-time suffix sorting algorithm can be seen as a synthesis of Farach-Colton, Ferragina, and Muthukrishnan's [Far97; FCFM00] recursive construction method with Burkhardt and Kärkkäi-nen's [BK03] difference cover sampling approach. Due to its elegant simplicity, it has quickly become a showcase string algorithm and is now being taught in many computer science classes around the world. Instead of naively sorting the sample, a new string of representatives is constructed from the sample positions and this string is recursively suffix sorted. By this process, the lexicographic ranks of the difference cover samples are calculated. The other suffixes then only need to be sorted up to the next sample position, and in a final merge all sample and non-sample suffixes are ordered with a constant-time comparison operation. Kärkkäinen, Sanders, and Burkhardt [KS03; KSB06] propose to use a small difference cover, e.g. $|X| = 3$ or 7. In the first paper [KS03] they presented *skew* with $|X| = 3$, later called *skew3*, and suggest using larger difference covers. The journal paper [KSB06] renames the algorithm to *DC3*, and proposes the more general version *DC* for any difference cover. We will discuss this difference cover algorithm in more detail in section 8.3, and present our distributed external implementation using Thrill in section 8.3.1.

Ko and Aluru [KA03; KA05] on the other hand pick up ideas from Itoh and Tanaka [IT99] and add recursion to create a linear-time algorithm. They classify all suffixes into two types: type `S`, if $T[i..n) < T[i+1..n)$, or type `L`, if $T[i..n) > T[i+1..n)$ (the case "=" cannot occur). They then create a new string by splitting the input using `S` or `L`, depending on which occurs fewer times. We will assume `S`s are fewer. Each letter in the new string represents a sequence of `SL`...`LS`-type characters, which they call a type `S` substring. The new string representing these type `S` substrings is suffix sorted recursively. Thereafter, the order of all `S`-type suffixes is calculated from the order of all `S` substrings using string sorting and a rather complex process involving distance lists to the preceding `S`-type position. Once the ranks of all `S` suffixes are known, the final suffix array can be induced with an elegant sweep and one pointer per bucket: iterate from right to left to induce all `L`-type suffixes from the known `S`-type. As the recursive problem is less than half in size and only linear work is performed in each recursive step, the whole algorithm is linear.

While the break-through invention of direct linear-time construction algorithms was a big success for the theoretical string algorithm community, their practical impact was very muted due to their disappointing performance on real-world inputs [LP04; ARS+04; PST05]. Among the four linear-time suffix sorters, two were more of theoretical interest [HSS03; HSS09; KSPP03; KSPP05], leaving only two implementable algorithms. Lee and Park [LP04] compared two engineered versions of Ko and Aluru's [KA03; KA05] and Kärkkäinen, Sanders, and Burkhardt's [KS03; KSB06] algorithms with a version by Manber and Myers [MM90; MM93]. Their results clearly show that Ko and Aluru's [KA03; KA05] algorithm outperforms its competitors ("contrary to the fact that Ko's own implementation of Ko-Aluru is worse than Kärkkäinen-Sanders" [LP04]). However, Puglisi, Smyth, and Turpin [PST05] compared Lee and Park's [LP04] implementations with the main superlinear algorithms [Sad98; LS99; LS07; MF02; MF04; BK03], and found them to be "even though linear in the worst case, [...] in practice *not* as fast as several other supralinear suffix array construction algorithms" [PST05]. The reasons for the dissatisfying performance of linear-time suffix sorters are large constant factors hidden by the $\mathcal{O}(n)$ asymptotic and larger space requirements than more straight-forward superlinear sorters.

Hence, even after 2003, researchers and practitioners continued to engineer superlinear algorithms for better performance. Kim, Jo, and Park [KJP04] developed a recursive suffix sorting algorithm which is almost linear and exploits having a *fixed size* alphabet $\Sigma$. Their approach picks up Kim, Sim, Park, and Park's [KSPP03; KSPP05] linear-time suffix array construction method and uses backward search on uncompressed suffix arrays [SKPP03; SKPP05] to accelerate the expensive merging step. Their algorithm runs in worst case in $\mathcal{O}(n \log\log n)$ time.

Schürmann and Stoye [SS05; SS07; Sch07] presented the *bucket pointer* algorithm as an "incomplex" suffix array construction method. It is based on Larsson and Sadakane's [Sad98; LS99; LS07] qsufsort combined with induced copying after sorting only a subset of the buckets. Hence, this combination uses prefix doubling to sort some buckets, and uses those to induce the rest. Which buckets to sort first and which are induced from them is selected by heuristics. While their space requirements are large and while they could not show a tighter bound than $\mathcal{O}(n^2 \log n)$, the algorithm adapts well to the input's properties and is quite fast in practice.

Maniscalco and Puglisi [MP06; MP08] proposed a wholly different approach to suffix array construction by first calculating the inverse suffix array, hence, the lexicographic ranks of each suffix, and then inverting it. This twist allows them to use the output suffix array space for auxiliary information and to form a linked list inside the array which describes both the order of the currently sorted buckets and their boundaries. Their algorithm, *msufsort*, is definitely superlinear but still very lightweight, requiring only about $n$ bytes beyond the input and output.

Due to the importance of suffix sorting for compression, practitioners outside the academic space also engineered highly efficient implementations around 2006 [Ale06; Mor06]. The efforts by Mori resulted in *divsufsort* [Mor06], which is still the fastest practical in-memory suffix sorter to date and used in a large variety of applications [GBMP14]. While divsufsort *is* open-source, there is little primary documentation [Mor05] about how it works and the source code itself is highly engineered and difficult to comprehend. From our own interpretation of the source code, we find that it is based on a combination of `A`/`B`-type copy/induced sorting [IT99; KA03; KA05] with engineered string sorting algorithms. Stepping beyond Ko and Aluru's [KA03; KA05] complex `S`-distance sorting, Mori makes the important observation that one only needs to sort the *last* representative in a sequence of `B`-type suffixes, which he calls `B`*-suffixes. The order of these special suffixes is sufficient to then first induce all `B`-type suffix, followed by all `A`-type suffixes. In divsufsort, the `B`*-suffixes are sorted using highly engineered string sorting implementations such as multikey introsort (multikey quicksort [BS97] with introsort-like worst-case guarantee [Mus97]) and in-place mergesort for memory-constrained environments. Hence, contrary to some other authors' claims, divsufsort is not linear, but very fast in practice since it avoids the overhead of recursion and fully utilizes induced copying. Furthermore, divsufsort only needs very little space beyond the input and output array, mainly for bucket pointer arrays of size $\Sigma$ and for the recursion stack during string sorting. In summary, economical space usage and fast string sorters with worst-case avoidance heuristics make divsufsort the fastest suffix sorter to date. Mori's [Mor06] experimental results show that divsufsort is a factor 3.1 faster than the bucket pointer algorithm [SS05; SS07; Sch07], a factor 4.1 faster than the first difference cover algorithm [BK03], a factor 2.1 faster than deep-shallow, a factor 2.6 faster than Ko and Aluru's [KA03; KA05] induced sorting algorithm, a factor 5.2 faster than DC3, a factor 1.7 faster than msufsort, and a factor 2.7 faster than qsufsort across 98 different input files of various size. Recently, Fischer and Kurpicz [FK17] published a paper describing how the divsufsort algorithm works and extending it to also output the LCP array.

At the end of 2006, research on suffix sorting had maneuvered into a stalemate between two linear-time algorithms [KA03; KA05; KS03; KSB06], which had provably good performance for *all inputs*, and divsufsort, which was used in all practical applications and far outperformed the linear algorithms. Hence, authors turned to reducing the space requirements of algorithms [DPT12]. Franceschini and Muthukrishnan [FM07] presented a tricky suffix sorter implementation which runs in $\mathcal{O}(n \log n)$ time and uses only $\mathcal{O}(1)$ space in addition to the input and output arrays. Nong and Zhang [NZ07] designed an algorithm which achieves $\mathcal{O}(n \log |\Sigma|)$ time and $|\Sigma| \lceil \log n \rceil + \mathcal{O}(1)$ working space, which translates to $\mathcal{O}(n)$ time and $\mathcal{O}(1)$ additional space for constant size alphabets. Their algorithm uses the difference cover approach [KS03; KSB06] for deeper recursion levels when enough memory is available, and induced sorting for the top levels when memory is restrained.

The stalemate between theoretically and practically fast algorithms was resolved in 2009 by Nong, Zhang, and Chan [NZC09a; NZC09b; NZC11], who presented two linear-time algorithms which were *also fast in practice*. The first, called *SA-IS* [NZC09a], is an elegant formulation of Ko and Aluru [KA03; KA05] combined with the improvements by Mori [Mor06]. After classifying all suffixes into `L` or `S` types, it partitions the string at all *left-most* `S`*-type* (LMS) positions and constructs a recursive string representing each *LMS-substring*, which only extends from one LMS position to the next. SA-IS is composed of four main steps: a) scan the text, classify each suffix into `L`/`S`-type, and find all LMS positions b) sort all LMS-substrings using two induced sorting sweeps and construct a new string from them, c) recursively suffix sort the new string to determine the full lexicographic order of all LMS suffixes, and d) place only the LMS positions in correct lexicographic order into the output suffix array, and generate the remaining suffix array by performing two induced sorting sweeps. Due to their construction there are at most $\frac{n}{2}$ LMS-strings and on average only $\frac{n}{4}$. Since all other steps are mainly linear sweeps over the suffix array, the algorithm is linear. And maybe most surprising, the algorithm relies mostly on induced sorting of suffixes and no longer performs string sorting in the performance critical parts. The authors claim the algorithm requires at most $n \log n + n + \mathcal{O}(1)$ bits of space, including the output suffix array, when the alphabet is integer. Okanohara and Sadakane [OS09] quickly modified SA-IS for lightweight BWT construction to use only $\mathcal{O}(n \log |\Sigma| \log\log_{|\Sigma|} n)$ space.

The second algorithm by Nong, Zhang, and Chan [NZC09b; NZC11], called *SA-DS*, is another elegant reformulation of the previous induced sorting algorithms. SA-DS picks up the idea of LMS-substrings from SA-IS, but limits their depth to a constant $d$. These substrings are then called $d$-critical and can be sorted using a string sorter with limited depth, e.g. with radix sort. Once all $d$-critical substrings are sorted, a new string is constructed from lexicographic names of the $d$-critical substrings and suffix sorted recursively. The remaining induction steps to gain the suffix array are identical with SA-IS. As with SA-IS, there are at most $\frac{n}{2}$ $d$-critical substrings and only linear work is performed in each recursive step, hence, SA-DS is a linear-time algorithm. While SA-DS was usually slightly slower than SA-IS in the authors experiments [NZC11], the advantage of SA-DS is that string sorting the $d$-critical substrings can be parallelized (e.g. see part I of this dissertation), as opposed to the pure induced sorting of SA-IS, which is much harder to accelerate.

While the original authors' implementation of SA-IS was already faster than all previous linear suffix sorting algorithms, Mori [Mor08] soon provided an independent, highly engineered implementation of SA-IS that outperformed the algorithms by Burkhardt and Kärkkäinen [BK03] (difference cover), Manzini and Ferragina [MF02; MF04] (deep-shallow), Kärkkäinen, Sanders, and Burkhardt's [KS03; KSB06] (DC aka skew), Ko and Aluru [KA03; KA05], and Larsson and Sadakane [Sad98; LS99; LS07] (qsufsort, prefix doubling) on *almost all* instances in a large experiment test set with many real-world inputs [Mor08]. Mori's sais was a factor 5.7 faster than the first difference cover algorithm [BK03] with $v = 32$, a factor 12 faster than deep-shallow, a factor 2.8 faster than Ko and Aluru's [KA03; KA05] induced sorting algorithm, and a factor 4.6 faster than qsufsort. This was definitely a break-through in suffix sorting, because the algorithm was both fast in practice and had provably good worst-case performance guarantees. Compared with divsufsort, sais was only a factor 1.58 slower.

In 2013, Nong [Non11; Non13] followed up their research with another SA-IS variation, called SACA-K, which only needs $n \log |\Sigma| + n \log n + |\Sigma| \log n$ bits of space, including the input string and the suffix array. If one considers the alphabet size $|\Sigma|$ to be constant, then the algorithm requires only $\mathcal{O}(1)$ space and it still has $\mathcal{O}(n)$ worst-case running time. This approaches the lower limit of what is possible. Louza, Gog, and Telles [LGT16; LGT17a] adapted SA-IS, SACA-K, and Fischer [Fis11]'s LCP extension of SA-IS [LGT17b] to generalized suffix arrays. In terms of practical speed, the current implementation of SACA-K is about 33% faster than SA-IS in their experiments, but is not as good as Mori's engineered suffix sorters, which still lead in terms of performance to this day.

Since the development and engineering of divsufsort and SA-IS, competing in the field of suffix sorting in terms of performance has become very difficult. Nevertheless, some experimental suffix sorting algorithms have been proposed. These explore new algorithmics ideas and generally avoid targeting performance. Adjeroh and Nan [AN08; AN10] proposed to use Shannon-Fano-Elias coding and recursion to construct a suffix sorter with smaller recursive problems. Baier [Bai15; Bai16] proposed the first linear-time suffix sorter which does not use recursion. Instead he applies prefix doubling to induced sorting, which, in a sense, doubles the "backwards" context during induced sorting in each step via pointers.

Li, Li, and Huo [LLH16] continued to improve Nong's [Non11; Non13] efforts at space saving. While SACA-K requires $|\Sigma| \log n$ additional bits, the new space saving tricks by Li, Li, and Huo [LLH16] bring this down to an actual $\mathcal{O}(1)$, independent of $|\Sigma|$. They propose different implementations depending on whether the input string is read-only or may be written and restored by the algorithm. In a sense these algorithms are truly optimal in terms of space and asymptotically optimal in time. However, they have not yet compared them with the currently fastest practical suffix sorters on real-world data.

Goto [Got17] also proposed an optimal linear-time, $\mathcal{O}(1)$ extra space suffix array construction algorithm for integer alphabets based on the induced sorting framework. Compared to Li, Li, and Huo [LLH16], his algorithm is somewhat simpler and he also describes how to construct the LCP array in-place. However, he too does not report any experimental results comparing his algorithm with divsufsort or sais.

## 5.3 Parallel Suffix Sorting Algorithms

Many researchers have ventured into areas beyond basic sequential suffix array construction algorithms for the RAM model. In this section, we give an overview of work on parallel algorithms for shared-memory machines, for GPUs, and then conduct a short excursion into the area of suffix *tree* construction. The next sections 5.4 to 5.6 then cover external memory suffix sorting, distributed suffix sorting, and LCP array construction.

**Parallel Shared-Memory Suffix Sorting.** As suffix sorting is a compute intensive task, parallelization is important to enable future speedups in performance. CPU technology increasingly delivers more cores with relatively low clock rates, since these are cheaper to produce and run. Parallelization of the basic suffix sorting algorithms, however, turns out to be an extremely difficult area. The reason is that of the three sorting principles discussed in the last section, only prefix doubling readily allows parallelization. Parallelizing induced sorting and recursion turns out to be difficult or impossible.

The earliest reference we could find in the literature to a shared-memory parallelization of Larsson and Sadakane's qsufsort algorithm [Sad98; LS99; LS07] is as one entry of many algorithms in a collection by Shun, Blelloch, Fineman, et al. called the Problem-Based Benchmark Suite [SBF+12; LSB16; LSB17]. This approach seems to have been too straight-forward for an earlier independent publication: one can simply sort the remaining $h$-groups in parallel, as has been done in many other contexts.

The most popular and fastest practical suffix sorter, Mori's divsufsort [Mor06], actually already contains an OpenMP-based parallelization of its first phase: the $\mathtt{B}^*$-suffixes are sorted using parallelized string sorting algorithms. The second phase, inducing the suffix array from the ranks of the $\mathtt{B}^*$-suffixes, however, is not parallelized. So in a sense, the most popular suffix sorter is already partially parallel.

Homann, Fleer, Giegerich, and Rehmsmeier [HFGR09] created the mkESA suffix array construction tool for the bioinformatics field. It is based on parallelizing the deep-shallow suffix sorter. However, genomes are a special case, since they can be expected to be more random than usual text. Random text on the other hand is an input instance, where deep-shallow's fast string sorters dominate more complex suffix sorting methods.

Mohamed and Abouelhoda [MA10] presented an attempt to parallelize the "incomplex" bucket pointer algorithm by Schürmann and Stoye [SS05; SS07; Sch07]. This approach is quite promising, because the bucket pointer algorithm uses heuristics to determine which buckets to sort using prefix doubling, which can be parallelized, and uses these to then induce other buckets. However, the author's experiments showed that their implementation does not scale well with the number of cores.

Singler [Sin10, chapter 4.6] implemented a parallelized version of Kärkkäinen, Sanders, and Burkhardt's skew algorithm [KS03; KSB06] as a case study for applying parallel integer sorting implementations from the MCSTL [PSS07; SSP07]. On an AMD Opteron with 8 cores they achieve speedups of about 4.5, but do not compare with faster algorithms such as divsufsort.

Shun, Blelloch, Fineman, et al. [SBF+12] also evaluated a parallelized implementation of the skew algorithm [KS03; KSB06] as part of their larger collection of parallelized algorithms. They showed good speedups, but also only relative to the same sequential implementation, not to faster suffix sorters such as SA-IS or divsufsort.

Labeit, Shun, and Blelloch [LSB16; LSB17] looked into parallelizing many string problems, such as lightweight wavelet tree, FM-index, and also suffix array construction. They proposed a parallelization approach to divsufsort, which also includes parallelizing the induced sorting part. They can show a worst-case bound of $\mathcal{O}(|\Sigma| \log n \sum_{\alpha \in \Sigma} R_\alpha)$ depth (the algorithm's critical path) and $\mathcal{O}(n)$ work, where $R_\alpha$ is the longest run of the character $\alpha$ in the input string. Note that for a unary string $\mathtt{a}^n$, $R_\mathtt{a} = n$, hence *no* work can be parallelized in the worst-case with this approach. We believe that this is due to the inherent sequential dependencies that steps in induced sorting have, and that parallelization with induced sorting cannot work beyond a certain point. In the future, other approaches are needed.

These six approaches are the only practical shared-memory parallelization attempts for CPU which we could find in the literature. Given the importance of parallelizing suffix sorting this is remarkably little work on the subject compared to the plethora of sequential suffix sorters in section 5.2. It turns out that parallelizing suffix sorting is a *very* difficult challenge, because SA-IS and divsufsort are extremely efficient in terms of CPU operations per suffix. Parallelized algorithms often require more CPU operations than sequential ones, and in the case of suffix sorting, this ratio is still very high because the sequential algorithms are near absolute optimal efficiency.

**Parallel Suffix Sorting on GPUs.** On GPUs, which are massively parallel shared-memory machines, researchers have implemented many parallelized suffix sorters. Osipov [Osi12; Osi14] proposed a GPU version of qsufsort performing prefix doubling with a highly parallel radix sort. Deo and Keely [DK13] analyzed the three suffix sorting principles and concluded that induced sorting is not suited for GPU parallelization. They therefore decided to implement a parallelized skew version. Wang, Baxter, and Owens [WBO15; WBO16] picked up the work of previous authors and implemented an optimized version of skew on GPU, which achieves a speedup of 1.45x over Deo and Keely [DK13]. They then continued to fuse prefix doubling and skew into a highly optimized, GPU friendly algorithm, which is 12.76x faster than Deo and Keely [DK13], 4.4x faster than Osipov [Osi12; Osi14], and 7.9x times faster than their own skew-only implementation. While other authors also implemented skew/DC3 on GPU [LMZT15], Wang, Baxter, and Owens's [WBO15; WBO16] implementation is currently the fastest and most sophisticated.

**Suffix Tree Construction for PRAM and in External Memory.** While we could not find any early work on suffix array construction for PRAM, there are some important precursor papers for parallelizing construction of the *suffix tree* on PRAMs [LSV87; AIL+88; Har94], which tackle the main problem of distributing work onto processors. Due to the demand from bioinformatics applications, there is a lot more work on external memory and parallel construction of large suffix trees.

Building large suffix trees in external memory was long seen as impractical due to the $\mathcal{O}(n^2)$ random disk accesses required when applying traditional RAM-based suffix tree construction algorithms [GK97]. However, the rise in processing of genome data in bioinformatics has lead many authors to proposed heuristic solutions to external memory suffix tree construction nonetheless. A survey paper [BST10] from 2010 summarizes these approaches very well.

Hunt, Atkinson, and Irving [HAI01] were the first to discuss an algorithm for building a large suffix tree in external memory. Bedathur and Haritsa [BH04] presented *TOP-Q*, which implements intelligent buffering methods adapted to Ukkonen's suffix tree construction algorithm [Ukk95]. Cheung, Yu, and Lu [CYL05] proposed *DynaCluster*, which replaces static pre-partitioning with dynamic clustering of suffix prefixes onto disk blocks. Clifford and Sergot [CS03; Cli05] presented the *PST* (paged suffix tree) algorithm for external memory and the *DST* (distributed suffix tree) algorithm for distributed memory, which are both based on sparse suffix links between subtree clusters, which can then be built in linear time. Tian, Tata, Hankins, and Patel [TTHP05] propose both the *TDD* (top down disk-based) algorithm, which adapts the lazy suffix tree construction method *wotd* [GKS99; GKS03] to external memory, and the *ST-Merge* algorithm, which is the first to build small suffix trees in main memory and then merge them together in external memory. Phoophakdee and Zaki [PZ07; PZ08] presented the *TRELLIS* suffix tree construction algorithm, which performs heuristic clustering of the output suffix tree prior to applying the partition-merge steps pioneered by ST-Merge. The dynamic clustering helps avoid data skew in the suffix tree due to characteristics of the input string. Barsky, Stege, Thomo, and Upton [BSTU08] proposed *DiGeST*, which is based on an external memory multiway merge sort similar to ST-Merge. Barsky, Stege, Thomo, and Upton [BSTU09; Bar10; BST11] then presented $B^2ST$, which first constructs the suffix and LCP arrays for partitions of the input and then merges these together using a two-phase multiway mergesort.

As the genome data pool and disk capacity grew, researchers started to look into constructing external suffix trees on parallel machines. Ghoting and Makarychev [GM09a; GM09b; GM10] presented the *Wavefront* algorithm, which consists of a prefix set creation step which partitions the resulting suffix tree, a parallelizable subtree construction step applied to each partition, and an optional suffix link recovery step. Wavefront was parallelized both for shared-memory architectures and for massively parallel IBM BlueGene supercomputers. Mansour, Allam, Skiadopoulos, and Kalnis [MASK11] proposed the *ERA* (elastic range) algorithm, which divides the suffix tree construction both vertically into independent subtrees, and horizontally into partitions, while elastically balancing work load among all tasks. While ERA is very popular and also runs on massively parallel supercomputers, a tight theoretical analysis is difficult [Jek13; JB16]. Tsirogiannis and Koudas [TK10] present two suffix tree construction algorithms parallelized for shared-memory machines with a common cache. The first, *CMPUTree*, is derived from Ukkonen's algorithm [Ukk95], and the second from Giegerich et al.'s *wotd* [GKS99; GKS03]. Comin and Farreras [CF13; CF14] proposed the *PCF* (parallel continuous flow) algorithm, which constructs the suffix and LCP arrays of small partitions of the input string, then builds order arrays of pairs of the suffix arrays, and finally merges all order arrays to gain the final suffix tree. Flick and Aluru [FA17] assume the suffix and LCP array of the input is already calculated on a distributed machine, and derive the suffix tree from them using a batch of all-nearest smaller-value (ANSV) queries.

## 5.4 External Memory Suffix Sorting Algorithms

External memory suffix tree algorithms require a lot of disk space for construction and storage of the tree. Hence, direct suffix array construction in external memory is attractive for a number of important applications such as genome analysis and text search. However, as is the case with internal memory, while the final suffix array is $n$ integers is size, the working space required by an algorithm can be much larger. Even though disk memory is 1–2 orders of magnitude cheaper than RAM (see figure 1.2), the intermediate working space and long running times limit the applicability of external memory suffix array construction.

**Table 5.1:** History of external memory suffix array construction algorithms with I/O and space requirements. The space column assumes 4 byte integers, and loglcp is short for $\lceil \log_2(1 + \text{maxlcp}) \rceil$, which is the number of iterations needed by prefix doubling algorithms.

| Algorithm | Upper Bound on Total I/Os | Space |
|---|---|---|
| Manber-Myers [CF02] | $\mathcal{O}(n \log_2 n)$ | $8n$ |
| BGS [CF02] | $\mathcal{O}((n^3 \log_2 M)/(MB))$ | $8n$ |
| New BGS [CF02] | $\mathcal{O}((n/M) \cdot \text{scan}(n))$ | $8n$ |
| doubling [CF02] | $\mathcal{O}(\text{sort}(n) \cdot \text{loglcp})$ | $24n$ |
| doubling+discard [CF02] | $\mathcal{O}(\text{sort}(n) \cdot \text{loglcp})$ | $24n$ |
| doubling+discard+radix [CF02] | $\mathcal{O}((n/B)(\log_{M/(B \log n)} n) \cdot \text{loglcp})$ | $12n$ |
| doubling [DKMS05] | $\text{sort}(5n) \cdot \text{loglcp} + \mathcal{O}(\text{sort}(n))$ | ? |
| doubling+discard [DKMS05] | $\text{sort}(5n) \cdot \text{avglogdps} + \mathcal{O}(\text{sort}(n))$ | ? |
| $a$-tupling [DKMS05] | $\text{sort}(\frac{a+3}{\log_2 a} n) \log_2(\text{maxlcp}) + \mathcal{O}(\text{sort}(n))$ | ? |
| $a$-tupling+discard [DKMS05] | $\text{sort}(\frac{a+3}{\log_2 a} n)\, \text{avglogdps} + \mathcal{O}(\text{sort}(n))$ | ? |
| 4-tupling [DKMS05] | $\text{sort}(3.5n) \log_2(\text{maxlcp}) + \mathcal{O}(\text{sort}(n))$ | ? |
| DC3 [DKMS05] | $\text{sort}(30n) + \text{scan}(6n)$ | $46n^*$ |
| DC7 [DKMS05] | $\text{sort}(24.75n) + \text{scan}(3.5n)$ | ? |
| bwt-disk [FGM10] | $\mathcal{O}((n/M) \cdot \text{scan}(n)/ \log n)$ | $\frac{1}{8}n$ |
| eSAIS [BFO13] (chapter 6) | $\text{sort}(17n) + \text{scan}(9n)$ | $25n^*$ |
| EM-SA-DS [NCZG14] | $\mathcal{O}(d\, \text{scan}(n))^\dagger$ | $25.5n^*$ |
| DSA-IS [NCHW15] | $\mathcal{O}(n/W)^\dagger$ | $40.1n^\ddagger$ |
| SAIS-PQ [LNCW15] | $\mathcal{O}(\text{sort}(n))$ | $23n^*$ |
| SAIS-PQ+ [LNCW15] | $\mathcal{O}(\text{sort}(n))$ | $15n^*$ |
| SAscan [KK14a] | $\mathcal{O}(\text{scan}(n)(1 + \frac{n}{M \log_\sigma n}))$ | $6.5n$–$11.5n$ |
| pSAscan [KKP15b] | $\mathcal{O}(\text{scan}(n)(1 + \frac{n}{M \log_\sigma n}) + \text{sort}(n))$ | $7.5n^\ddagger$ |
| fSAIS [KKPZ17] | $\mathcal{O}(\text{scan}(n) \cdot \log^2_{M/B}(n/B))$ | $8.2n^\ddagger$ |

$^*$ Space usage measured in experiments, not by theoretical analysis.
$^\dagger$ EM-SA-DS requires $M = \Omega(\sqrt{nB})$ space, and DSA-IS requires $M = \Omega(\sqrt{nW})$ to achieve these I/O bounds.
$^\ddagger$ Space measured in experiments by Kärkkäinen, Kempa, Puglisi, and Zhukova [KKPZ17].

Table 5.1 shows a tabular overview of all external memory suffix array construction algorithms we could find in the literature. While the original PAT-array paper [GBYS92] was already concerned with external memory, it was not their main focus and contained no theoretical considerations for I/O complexity.

The first paper explicitly on construction of suffix arrays in external memory was published by Crauser and Ferragina [CF02] as a comprehensive theoretical and experimental study of the following six algorithms.

The first is simply Manber and Myers' $\mathcal{O}(n \log n)$ RAM algorithm [MM90; MM93] run with automatically paged external memory. This is of course an overly naive approach, since it performs $\Theta(n \log_2 n)$ random I/Os, which yields unacceptable running times even for moderate $n$.

The original algorithm by Baeza-Yates-Gonnet-Snider (BGS) [GBYS92] incrementally builds an external suffix array $\mathsf{SA}_{\text{ext}}$ in $m = \Theta(n/M)$ steps by suffix sorting $m$ parts of the input text in main memory, and then merging them with $\mathsf{SA}_{\text{ext}}$. The problem is that during merging, many disk accesses may be required to compare suffixes outside of the currently loaded text part. The worst case occurs for an unary string $\mathtt{a}^n$: in total BGS needs up to $\mathcal{O}((n^3 \log_2 M)/(MB))$ I/Os, which again is unacceptable running time for even moderate $n$. However, Crauser and Ferragina were able to improve the algorithm (then called "New BGS") by processing the string back-to-front, which guarantees the better running time $\mathcal{O}(n/M \operatorname{scan}(n))$. Highlight of the BGS family of algorithms is the low space requirement of only $2n$ additional integers.

The fourth algorithm is a variant of prefix *doubling*, as used by Manber and Myers [MM90; MM93] and in Larsson and Sadakane's qsufsort [Sad98; LS99; LS07], but expressed in an external memory friendly way using tuples. Each tuple represents one suffix and they are sorted and reordered such that in round $h$ the lexicographic names of suffix $i$ and $i + 2^{h-1}$ are contained in one tuple, which then allows the algorithm to infer lexicographic names of depth $2^h$ for round $h + 1$. Seen on a high level, the doubling algorithm emulates qsufsort's batched random access into the lexicographic names array via two sorts [CGG+95]. This leads to $\mathcal{O}(\operatorname{sort}(n)\lceil\log_2(1 + \text{maxlcp})\rceil)$ I/Os since the algorithm can stop after $\lceil\log_2(1 + \text{maxlcp})\rceil$ rounds. In section 8.2.1 we review this algorithm in detail and present a distributed implementation of it.

As fifth and sixth algorithms, the authors then propose two new optimized versions of doubling: the first incorporates a *discarding* technique which allows finished suffixes to drop out of the set sorted in each doubling step. While this does not reduce the I/O complexity in the worst-case, for many practical instances the optimization substantially reduces the I/O volume (see also section 8.2.5, page 279). The second optimization extends doubling+discarding with a monotone integer *radix priority queue* for external memory. With this optimization the authors target the maximum space usage and bring it down to $14n$, which is a large improvements over the hefty $24n$ working space of the other doubling algorithms.

Crauser and Ferragina [CF02] collected all six algorithms and evaluated the implementations using the external memory library LEDA-SM [CM99], except for Manber and Myers' [MM90; MM93] which they tested on paged virtual memory. Their conclusion was that BGS is a good choice for "reasonably large" text collections with long repeated substrings, but for "very large" text collections or short repeated substrings, doubling+discarding should be preferred.

The second systematic evaluation of suffix array construction algorithms for external memory was performed by Dementiev, Kärkkäinen, Mehnert, and Sanders [Meh04; DKMS05; DKMS08][Dem06, chapter 5]. They implemented five algorithms using the then new external memory library STXXL [DKS05; DKS08]: *doubling*, *doubling with discarding*, *a-tupling*, *a-tupling with discarding*, and *DC3*, or more generally *DC*. The main engineering improvement is that STXXL supports *pipelining*, which avoids unnecessary data rounds trips to disk between stages in an algorithm. Practically their paper is interesting, because they present the first implementation of an asymptotically optimal suffix sorting algorithm in external memory: DC3, which runs in $\mathcal{O}(\mathrm{sort}(n))$. Furthermore, they attempt to determine the exact constant factors for the number of I/Os of the implemented algorithms, which crucially determine real-world performance. They however do not determine the maximum disk space used.

Their implementation of the *doubling* algorithm requires at most $\mathrm{sort}(5n)$ I/Os in each iteration, because it sorts two sets: one of tuples containing 3 integers and one containing 2 integers. The algorithm stops when all suffixes are fully ordered, which is after $\lceil \log_2(1 + \mathrm{maxlcp}) \rceil$ iterations. They then account with $\mathcal{O}(\mathrm{sort}(n))$ for all operations outside of the iterations, like reading the input and sorting the final output, which yield a total I/O bound of $\mathrm{sort}(5n) \cdot \lceil \log_2(1 + \mathrm{maxlcp}) \rceil + \mathcal{O}(\mathrm{sort}(n))$ for doubling.

Using the pipelining features of STXXL, they then improve on Crauser and Ferragina's doubling+discarding algorithm by saving superfluous scans. Their doubling+discarding requires only $\mathrm{sort}(5n\,\mathrm{avglogdps}) + \mathcal{O}(\mathrm{sort}(n)) = \mathrm{sort}(5 \sum_{i=1}^{n} \log_2(\mathrm{dps}_T(i))) + \mathcal{O}(\mathrm{sort}(n))$ I/Os because it *fully discards* suffixes which are no longer needed, completely removing them from the I/O working set. Crauser and Ferragina only classify suffixes as unfinished and partially discarded, hence their doubling+discarding has an additional $\mathcal{O}(\mathrm{scan}(n \lceil \log_2(1 + \mathrm{maxlcp}) \rceil))$ term because suffixes are kept in the working set.

Dementiev et al. then generalize doubling to *a-tupling*, which sorts all suffixes up to depth $a^k$ in iteration $k$. This requires larger tuples to be sorted, hence more I/O volume per iteration but only $\log_a(\mathrm{maxlcp})$ iterations are needed for plain doubling until all suffixes are fully sorted. In total, their $a$-tupling approach need at most $\mathrm{sort}((a + 3)n) \log_a(\mathrm{maxlcp}) + \mathcal{O}(\mathrm{sort}(n))$ I/Os. They then combine $a$-tupling with discarding and obtain an algorithm with $\mathrm{sort}(\frac{a+3}{\log_2 a} n)\,\mathrm{avglogdps} + \mathcal{O}(\mathrm{sort}(n)) = \mathrm{sort}((a + 3) \sum_{i=1}^{n} \log_a(\mathrm{dps}_T(i))) + \mathcal{O}(\mathrm{sort}(n))$ I/Os. A simple theoretical analysis shows that $a = 4$ or $5$ are optimal values.

Maybe the most important contribution of Dementiev et al. is the implementation of an asymptotically I/O optimal algorithm for external memory. They proposed a tuple-based external memory formulation of *DC3* aka *skew*, and more generally *DC*, which is one of the four linear-time algorithms proposed for the RAM model in the breakthrough of 2003. DC is composed of only scanning, sorting, and merging — all algorithmic primitives that can be implemented in asymptotically optimal time in external memory. Hence, DC in external memory also has sorting complexity, which is an obvious lower bound for suffix sorting. Their analysis shows that DC3 requires at most $\text{sort}(30n) + \text{scan}(6n)$ I/Os, while DC7 needs only $\text{sort}(24.75n) + \text{scan}(3.5n)$. For difference cover sizes larger than 7, the I/O requirements increases again. We describe DC3 and DC7 in detail in section 8.3, in which we present our distributed external memory adaptation using Thrill.

Dementiev et al. perform extensive experiments with doubling, doubling+discarding, 4-tupling, 4-tupling+discarding, and DC3. They clearly show that DC3 outperforms the other implementations on five real-world inputs. DC3 and DC7 was later also implemented by Weese for the SeqAN bioinformatics library [Wee06; DWRR08; Wee13].

Ferragina, Gagie, and Manzini [FGM10; FGM12] proposed an external memory BWT construction algorithm called *bwt-disk*, which only uses *n bits* of working space on disk. They term this a *lightweight* BWT construction algorithm and achieve it by avoiding to build the suffix array explicitly. The algorithm itself is derived from the original PAT-array algorithm [GBYS92] with the improvements by Crauser and Ferragina [CF02]: it partitions the input into $m = \Theta(n/M)$ parts and constructs the external BWT in $m$ steps by creating the BWT for the parts in RAM and merging them with the external BWT array. As this is done in $m$ rounds, and each round consists of a scan of $n/\log_2 n$ words of external memory, the bwt-disk algorithm needs in the worst case $\mathcal{O}((n/M) \cdot \text{scan}(n/\log n))$ I/Os. Despite constructing the BWT, we included bwt-disk in our experimental evaluation of external memory suffix array construction in section 6.3.

For almost a decade no work on external suffix array construction was published, then we presented the first external memory algorithm, *eSAIS*, which is based on *induced sorting* [BFO13; BFO16]. The key new ingredient is to use an external memory *priority queue* in addition to sorting, scanning, and merging. And since priority queue implementations exist with sorting complexity, eSAIS is an I/O optimal algorithm with $\mathcal{O}(\text{sort}(n))$ I/Os. In chapter 6 (page 193) we present eSAIS in detail and establish an upper bound of $\text{sort}(17n) + \text{scan}(9n)$ for a specific parameter set. Furthermore, we extended eSAIS to also construct the LCP array in external memory, which resulted in the *first practical implementation* of an algorithm that constructs both suffix and LCP array for very large inputs in external memory. For the experiments in section 6.3 we also implemented DC3 with LCP construction together with our student Daniel Feist [Fei13].

Soon after our paper on external memory suffix and LCP array construction multiple papers appeared. Louza, Telles, and Aguiar Ciferri [LTAC13] proposed *eGSA*, which is an algorithm for *generalized* suffix array construction based on induced sorting. It is useful for large collections of independent small strings, which is an easier problem than suffix sorting one long string. However, they do not provide any theoretical I/O analysis in their paper.

Nong, Chan, Zhang, and Guan [NCZG14] then presented two algorithms for external memory based on induced sorting. *EM-SA-DS* is based on SA-DS, which uses $d$-critical characters and LMS-substrings to recursively suffix sort a sample from which the whole suffix array can be induced. While EM-SA-DS is proposed as an external memory algorithm with $\mathcal{O}(d\,\mathrm{scan}(n))$ I/Os, they assume a RAM capacity of $M = \Omega(\sqrt{nB})$. This is reasonable in a practical scenario but at the same time limits its theoretical scalability. The authors do not compare EM-SA-DS with eSAIS, because they were published simultaneously and because the analysis of EM-SA-DS does not regard constant factors. Furthermore, the constant factor hides the recursive work, so one cannot assume the constant to be small.

The second algorithm by Nong, Chan, Hu, and Wu [NCHW15], *DSA-IS*, emulates SA-IS in an external memory setting. To avoid lots of random accesses, DSA-IS uses many I/O buffers of size $W$ to amortize the cost. Under the assumption that $M = \Omega(\sqrt{nW})$, which allows for enough I/O buffers, the I/O complexity of DSA-IS is given as $\mathcal{O}(n/W)$. But this again hides the constant factors and the constraint on $M$ makes any comparison with other external memory algorithms difficult. In their experiments, "the best experimental times are achieved by DSA-IS and eSAIS. [...] Although they have similar speed, eSAIS uses around 20% more disk space than DSA-IS" [NCHW15].

Liu, Nong, Chan, and Wu [LNCW15] then criticize eSAIS, EM-SA-DS, and DSA-IS for being overly complex and to consist of thousands of lines of code, which are "challenging to be revised for a specific application" ([LNCW15]). They pick up eSAIS and improve it by proposing *SAIS-PQ* and *SAIS-PQ+*, which are "coded in around 800 and 1600 lines in C++" ([LNCW15]). While building on the same algorithm structure, they attempt to simplify eSAIS by using preceding cache items instead of reinsertions into the induction PQ. In their experiments, SAIS-PQ and SAIS-PQ+ are usually significantly slower than eSAIS, on some instances up to a factor of two, though instances exist where SAIS-PQ+ outperforms eSAIS. The reason is surmised to be that SAIS-PQ and SAIS-PQ+'s I/O volumes are about twice the volume of eSAIS, which significantly hinders performance. On the other hand, in those cases were SAIS-PQ+ outperforms, eSAIS appears to be CPU bound instead of I/O bound. While the main selling point of SAIS-PQ and SAIS-PQ+ is simplicity of the code, the experimental evaluation of SAIS-PQ+ shows a much lower peak disk space usage of $15n$, compared to the $25n$ of eSAIS.

Another line of research in external memory algorithms was performed by Kempa [Kem15] and his coauthors. The first algorithm proposed by Kärkkäinen and Kempa [KK14a; KK17a] was *SAscan*, which picks up the idea of constructing the suffix array in parts in internal memory and merging the parts into the large final suffix array, which is stored in external memory. This idea goes back to Gonnet, Baeza-Yates, and Snider [GBYS92] and was improved several times [CF02; FGM10; FGM12]. Kärkkäinen and Kempa reduce the I/O volume by a factor $\mathcal{O}(\log_\sigma n)$ by using backwards search in the merge phase. This yields a worst-case running time of $\mathcal{O}(\mathrm{scan}(n)(1 + n/(M \log_\sigma n)))$. Due to the remaining $\frac{n}{M}$ factor the authors surmise, "SAscan is a true external memory algorithm [...], but its time complexity $\Omega(n^2/M)$ makes it hopelessly slow when the text is much larger than the RAM. However, when the text is too large for an internal memory SACA, i.e., larger than about *one fifth of the RAM size*, but not too much bigger than the RAM, SAscan is probably the fastest SACA in practice.” ([KK14a], emphasis added by us).

Kärkkäinen, Kempa, and Puglisi [KKP15b] then parallelized SAscan and created *pSAscan*. In pSAscan, both the internal sorting of parts of the suffix array in RAM and the merging process are performed in parallel. The backwards search to find the insertion gaps and the actual merging are performed with dynamic load balancing. While they did not publish any speedup experiments, their implementation is faster than eSAIS on their real-world inputs, but scales less well for increasing input sizes. However, the scalability is sufficient to construct “the suffix array of a 1 TiB text in a little over 8 days” [KKP15b].

The most recent improvement for external memory was proposed by Kärkkäinen, Kempa, Puglisi, and Zhukova [KKPZ17] and named *fSAIS*. It is derived from eSAIS and DSA-IS: from eSAIS they inherit induction using a priority queue and from DSA-IS they lift the idea of “pre-inducing” blocks to gain an array containing the previous characters (needed for induced sorting) in exactly the order necessary. They combine the two algorithms and further engineer the implementation to use a stable monotone radix heap instead of a general-purpose priority queue and to work with large alphabets. Their algorithm needs only $\mathcal{O}((n/B)\log^2_{M/B}(n/B))$ I/Os, and the experimental study in their paper shows that fSAIS is about two times faster than eSAIS while retaining its stability for larger inputs. Maybe more important, it also drastically reduces eSAIS's space requirements to only $8.2n$ for a worst-case input. This makes fSAIS the fastest and most advanced approach for external memory suffix sorting to date.

## 5.5 Distributed Suffix Sorting Algorithms

Distributed algorithms on clusters of “shared-nothing” machines have always been interesting to utilize scalable computing power for suffix array construction.

Early research in this direction ignored the interrelationship of suffixes and focused on optimizing distributed quicksort and merge sort adaptations [KRZ96; KRRNZ97; NKRNZ97]. The text and suffix array are distributed evenly onto $p$ processors and each first sorts its local suffixes. Then the local suffix array is transformed into an intermediate global array using an all-to-all data exchange. To complete the sort, however, the suffix array parts need to be merged, and for this characters from the text on remote nodes may be necessary. This may incur high additional communication cost and the authors propose to transmit “pruned suffixes” or “sistrings” with additional characters to alleviate the problem.

The first paper to focus on special purpose suffix sorting algorithms was published by Kitajima and Navarro [KN99]. They take Manber and Myers' [MM90; MM93] prefix doubling algorithm and *emulate* it on a distributed system. The large arrays are stored as slices across all nodes, loops are parallelized, and remote accesses are performed in batch via slow inter-node queries. In their analysis they show that the distributed algorithms take $\mathcal{O}(\frac{n}{p}\log\log n)$ time on average and $\mathcal{O}(\frac{n}{p}\log n)$ in the worst case.

Futamura, Aluru, and Kurtz [FAK01] also discuss parallel suffix sorting and propose an elaborate load balancing method for radix sorting. However, they assume that the full input string is available on each processor, which is not a practical assumption once the input grows large.

In 2006, Kulla and Sanders [Kul06; KS06b; KS07] presented an adaption of DC3 to a distributed system. They implemented the algorithm using C++ and MPI and ran experiments on up to 512 PEs using a cluster machine. They were able to demonstrate that *pDC3* is a practicable and scalable algorithm for building large suffix arrays. In chapter 8 we pick up their work and implement DC3 and DC7 as distributed external memory algorithms using Thrill.

Menon, Bhat, and Schatz [MBS11] took up suffix array construction using the MapReduce framework Hadoop. They present two algorithms: the first is a native string sorting approach using MapReduce as sorter, the second attempts to improve on the first by detecting long character repetitions. In their experiments, the authors conclude that the "end-to-end speedup was modest compared to the number of cores used in the experiments." Furthermore, they state the pDC3 was three times faster, though on different hardware.

In 2014, Abdelhadi, Kandil, and Abouelhoda [AKA14] presented an independent MPI implementation of Futamura, Aluru, and Kurtz's [FAK01]'s algorithm under the name *cloudSACA*, which is specialized for running on Amazon Web Service's cloud infrastructure. Later, Metwally, Kandil, and Abouelhoda [MKA16] also implemented Kulla and Sanders's [KS07] algorithm in the same package using C++ and OpenMPI. In their experiments, they show that the prefix doubling algorithm performs much better on their input set which however is composed of DNA with short LCPs.

Maybe the largest-scale suffix array construction to date was performed by Flick and Aluru [FA15] in 2015. They adapted the prefix doubling approach of Larsson and Sadakane's qsufsort [Sad98; LS99; LS07] to a distributed system and implemented both suffix and LCP array construction using C++ and MPI for cluster computers with two techniques. Their first technique uses global sorting (the common approach for prefix doubling), while the second technique sorts only remaining, non-singleton $h$-groups once most suffix groups are sorted (this is similar to discarding, but not identical). The second technique accelerates inputs with uneven LCPs. They demonstrate the scalability of their implementation by sorting the human genome in less than 8 seconds on 1 024 cores (64 nodes) of a compute cluster, which is a 110-fold speed-up over divsufsort on a single host. While these results are very impressive, we consider the cost efficiency of their results rather low: the amount of utilized hardware is immense considered that one can suffix sort the human genome (around 3 GiB) on a simple laptop in a few minutes. With 1 024 cores, each core only has a *mere 3 MB* of the input but 8 GiB of available RAM. But for suffix sorting with high throughput, where absolute running time is the only criterion, their highly parallelized distributed implementation definitely established new records.

Our own paper [BGK16], the most recent paper on distributed suffix array construction, uses Thrill to advance suffix sorting to more cost efficient *distributed external* memory and is the topic of chapter 8.

## 5.6 LCP Array Construction

In applications, the suffix array is usually used in conjunction with additional auxiliary arrays, among which the most important is the LCP array. Combinations of such arrays created by additional preprocessing are generically called "enhanced" suffix arrays [AKO02; AKO04] and these emulate the suffix tree in a variety of applications from stringology and bioinformatics [Gus97; CR03; Ohl13; MBCT15].

Construction of the LCP array is usually possible *during* suffix sorting, and most algorithms can be adapted accordingly. However, this is not always an easy task and in the following we focus only on papers where the adaption is discussed *explicitly.*

Besides suffix sorters with integrated LCP array construction, there are also a number of "standalone" LCP construction algorithms taking the text, the suffix array, and sometimes other arrays to construct the LCP array using these as input. Such algorithms are usually simpler and faster since suffix sorting is considered to be computationally more expensive than LCP array construction. In table 5.2 we present a tabular history of LCP array construction algorithms.

In the remainder of this section, we review most algorithms explicitly designed to construct the LCP array, both integrated into suffix sorters and "standalone" algorithms. In chapter 6 we present the first implementation of a suffix sorting algorithm for external memory which can also construct the LCP array in near-optimal sorting complexity.

The very first suffix sorting algorithm by Manber and Myers [MM90; MM93] already contained an integrated LCP array construction technique. The classical RAM algorithm runs in $\mathcal{O}(n \log n)$ time and requires $n$ integers of additional space, besides the input, the output suffix array, and the output LCP array.

Kasai, Lee, Arimura, et al. [KLA+01] then presented an algorithm, called *KLAAP*, which takes the text and suffix array and calculates the LCP array in *linear time* (in the RAM model). Their algorithm was optimal in theory, thus relieved all future theoretical RAM-based suffix array construction algorithms from concerning themselves with LCP array construction. It was also fast in practice, compared to any suffix array construction algorithm of the time, and the additional space required by the algorithm are only $n$ integers for the inverse suffix array. However, as suffix array construction

**Table 5.2:** History of LCP construction algorithms containing both integrated into suffix sorters and standalone algorithms. The space column assumes 4 byte integers.

| Algorithm | Input | → Output | Time / I/Os | Space | Model |
|---|---|---|---|---|---|
| [MM90] | T | → SA,LCP | $\mathcal{O}(n \log n)$ | $4n$ | RAM |
| [KLA+01] KLAAP | T,SA | → LCP | $\mathcal{O}(n)$ | $4n$ | RAM |
| [KS03] DC-LCP | T | → SA,LCP | $\mathcal{O}(n)$ / $\mathcal{O}(\text{sort}(n))$ | $\mathcal{O}(n)$ | RAM/EM |
| [Man04] Lcp9 | T,SA | → LCP | $\mathcal{O}(n)$ | $\mathcal{O}(1)$ | RAM |
| [Man04] Lcp5 | T,SA | → LCP | $\mathcal{O}(n)$ | $n + 4H_k(T)^{\dagger}$ | RAM |
| [PT08] | T,SA | → $v$-LCP | $\mathcal{O}(nv)$ | $6n+\mathcal{O}(\frac{n}{\sqrt{v}}+v)$ | RAM |
| [KMP09] Φ | T,SA | → PLCP | $\mathcal{O}(n)$ | $4n$ | RAM |
| [KMP09] I | T,SA | → PLCP | $\mathcal{O}(n \log n)$ | $5n + \frac{3}{8}n$ | RAM |
| [GO11] | T,SA,BWT,LF | → LCP | $\mathcal{O}(n^2)$ | $2n$ | Semi-EM |
| [Fis11] (with SA-IS) | T | → SA,LCP | $\mathcal{O}(n)$ | $\mathcal{O}(1)$ | RAM |
| [BGOS11] | T,WT(BWT) | → LCP | $\mathcal{O}(n \log \sigma)$ | $2.2n^*$ | RAM |
| [BFO13] (chapter 6) | T | → SA,LCP | $\mathcal{O}(\text{sort}(n))^{\ddagger}$ | $\mathcal{O}(n)$ | EM |
| [KK14b] LCPscan | T,SA | → LCP | $\mathcal{O}(\frac{n^2}{MB(\log_\sigma n)^2} + \text{sort}(n))$ | $\mathcal{O}(n)$ | EM |
| [KK16c] | T,SA | → LCP | $\mathcal{O}(\text{sort}(n))$ | $\mathcal{O}(n)$ | EM |
| [KK16a] EM-SΦ | T,SA | → LCP | $\mathcal{O}(\frac{n^2}{MB(\log_\sigma n)^2} + \text{sort}(n))$ | $4n$ | EM |
| [KK16a] EM-SI | T,SA | → LCP | $\mathcal{O}(\frac{n^2}{MB(\log_\sigma n)^2} + \frac{n \log \sigma}{B} + \text{sort}(n))$ | $1.25n^*$ | EM |
| [KK17b] EM-SΦ | T,SA | → LCP | $\mathcal{O}(\frac{n^2}{MB(\log_\sigma n)^2} + \text{sort}(n))$ | $\mathcal{O}(1)$ | EM |
| [KK17b] EM-SI | T,SA | → LCP | $\mathcal{O}(\frac{n^2}{MB(\log_\sigma n)^2} + \frac{n \log \sigma}{B} + \text{sort}(n))$ | $\mathcal{O}(1)$ | EM |
| [Got17] | T | → SA,LCP | $\mathcal{O}(n)$ | $\mathcal{O}(1)$ | RAM |
| [FK17] (with divsufsort) | T | → SA,LCP | $\mathcal{O}(n \log n)$ | $\mathcal{O}(1)$ | RAM |

* Space usage measured in experiments, not by theoretical analysis.

† [Man04]: additional space over the text and suffix array, which is transformed in-place.

‡ eSAIS-LCP is sorting complexity provided $n \leq C \cdot M^2$ for a small constant $C$.

algorithms accelerated in later years, it became clear that the highly random memory access pattern performed by KLAAP is inferior to integrating LCP array construction with suffix sorting.

Furthermore, Kasai, Lee, Arimura, et al. [KLA+01] is only linear in the RAM model. The first external memory algorithm to also construct the LCP array in sorting complexity was DC3/DC by Kärkkäinen, Sanders, and Burkhardt [KS03; KSB06]. Their solution, called *DC-LCP*, however remained a theoretical side remark at the end of the paper. In the preparation of this dissertation we supervised Feist's [Fei13] bachelor thesis on an exact description and implementation of DC3-LCP, which we greatly recommend for details on the topic.

Manzini [Man04] engineered two optimized "standalone" LCP construction algorithms which take the text and suffix array as given. Both remain linear in the RAM model and improve on the additional space requirements. The first algorithm, *Lcp9*, reuses the space of the suffix and output LCP arrays such that only $\mathcal{O}(1)$ *additional* space is needed besides the text, suffix, and LCP arrays. The second algorithm, *Lcp6*, then even considers how to construct the LCP array *in-place*, given the text and suffix array. The suffix array is transformed into the LCP array using only $kn$ additional space. The parameter $k$ depends on the empirical entropy $H_k(T)$ of the text and ranges from 1 (highly compressible) to 5 (random). In their experiments, Lcp9 is only 10% slower than the original KLAAP algorithm and Lcp6 is about two times slower than the original.

To reduce the space usage even further, Puglisi and Turpin [PT08] introduced the $v$-LCP array which is a sampled LCP array at indices of a difference cover of size $v$. Their construction algorithm for a $v$-LCP array of size $\Theta(n/\sqrt{v})$ runs in $\mathcal{O}(n\sqrt{v})$ time. They continue to give an algorithm which constructs the ordinary LCP arrays and runs in $\mathcal{O}(nv)$ time with $n + \mathcal{O}(\frac{n}{\sqrt{v}} + v)$ bytes of working space. They call the parameter $v$ a *space-time* trade-off and suggest $v \geq 32$, such that the $v$-LCP is less than $n$ bytes in size.

A wholly different approach to the LCP array was proposed by Kärkkäinen, Manzini, and Puglisi [KMP09] with the *permuted* LCP array, called PLCP, which is in text order (position order) rather than suffix array order (lexicographic order). This permuted array has interesting properties such as $\mathsf{PLCP}[i] \geq \mathsf{PLCP}[i-1] - 1$ for all $i$. In their paper they propose a succinct representation with at most $3n$ bits, a sparse PLCP array containing only every $k$-th entry, and two fundamentally different construction techniques. The first technique uses the $\Phi$ array, defined as $\Phi[\mathsf{SA}[i]] = \mathsf{SA}[i-1]$, and the second so called *irreducible* LCP values. An entry $\mathsf{LCP}[i]$ is called *reducible* if $\mathsf{BWT}[i] = \mathsf{BWT}[i-1]$ and *irreducible* otherwise. The first method is not very space efficient but outperforms even KLAAP, while the second needs only $3n$ bits of space but $\mathcal{O}(n \log n)$ time for construction.

Gog and Ohlebusch [GO11] presented an algorithm that takes as input the text, the suffix array, the Burrows-Wheeler transform, and the last-to-first mapping (LF) to calculate the LCP array. The distinguishing property is that all arrays, except the text, are read sequentially, which makes the algorithm *semi-external* as it is I/O friendly. The algorithm needs only a string of $n$ bytes in RAM and a reduced LCP array of $n$ bytes. Larger values are stored externally as integers. While the practical performance of their algorithm is better than KLAAP due to fewer random accesses and cache misses, the worst-case running time is $\mathcal{O}(n^2)$. However, in the addendum they sketch an idea how to make it linear.

Since KLAAP, in theory RAM-based suffix sorting algorithms no longer needed to be able to generate the LCP array. However, experimentally due to the advances in suffix sorting algorithm it was now the case that divsufsort and SA-IS were *faster* in constructing the suffix array than KLAAP was for calculating the LCP array afterwards. This was a strange situation because suffix sorting is considered a computationally more difficult problem than LCP array calculation. Fischer [Fis11] therefore extended SA-IS to also *induce* the LCP array while inducing the suffix array. We will review his approach in section 6.1.1, which is based on RMQs over the LCP arrays while it is being built. In experiments their new algorithm increases the running time of SA-IS only by about 40-60% and this overhead is always less than running KLAAP after SA-IS. Integration of LCP array construction into algorithms based on the induced sorting framework was later also done by Goto [Got17] for an optimal $\mathcal{O}(1)$ extra space algorithm and by Fischer and Kurpicz [FK17] for divsufsort.

Beller, Gog, Ohlebusch, and Schnattinger [BGOS11; BGOS13] presented an algorithm which does not use the suffix array as input as the authors consider it to use too much space. Instead they take a wavelet tree of the Burrows-Wheeler transform as input and construct the LCP array front-to-back in a streaming manner to disk. The algorithm runs in $\mathcal{O}(n \log \sigma)$ time, which is linear for a constant alphabet, and needs only $2.2n$ bytes of internal memory in their experiments.

All previous LCP construction algorithms were sequential. The first author to consider parallelizing LCP construction on shared memory was Shun [Shu14], who presented six parallelized algorithms: *skew-LCP* based on skew3/DC3 [KS03; KSB06], *par-LCP* based on KLAAP [KLA+01], *par-PLCP* based on the first algorithm and *par-iLCP* based on the second algorithm by Kärkkäinen, Manzini, and Puglisi [KMP09], *dk-LCP* based on the GPU parallelization by Deo and Keely [DK13], and *dk-PLCP* which is dk-LCP modified to generate the PLCP array. Among all these implementations, par-PLCP performs best on a variety of real-world and artificial inputs, reaching a speedup of 14.4–21.8 with 40 cores.

Building on sequential, shared-memory parallelized, and semi-external LCP algorithms, Kärkkäinen and Kempa [Kem15] formed a line of research into *fully* external memory LCP array construction. They proposed the first such algorithm for *standalone* LCP construction, called *LCPscan* [KK14b; KK16b]. All previous algorithms were either semi-external (required the text in RAM) or were integrated into a suffix sorter (only two explicit implementations were known [KS03; KSB06; BFO13; BFO16]). LCPscan is an agglomeration of different LCP algorithms: KLAAP and the two algorithms for PLCP construction by Kärkkäinen, Manzini, and Puglisi [KMP09]. In LCPscan they exploit the fact that if given all irreducible LCP values, then the other, reducible LCP values are easy to compute via scanning. Hence, the goal is to identify and calculate the irreducible LCPs, which they do by loading segments into RAM and handling rare overflows when needed. While each individual irreducible LCP value can be up to $n$, Kärkkäinen, Kempa, and Piątkowski [KKP15a; KKP16] showed that *the sum* of all irreducible LCP values is at most $n \log n$. This property allows them to bound the worst-case I/O volume of LCPscan with $\mathcal{O}(\frac{n^2}{MB(\log_\sigma n)^2} + \mathrm{sort}(n))$, which is still much higher than the $\mathcal{O}(\mathrm{sort}(n))$ of the best LCP array construction algorithms integrated with a suffix sorter. However, they argue that for current machines with a *balanced* RAM/EM ratio, LCPscan nevertheless must outperform eSAIS (the LCP construction overhead only) and back this theoretical claim with experimental results. The combination of eSAIS (suffix array only) + LCPscan achieves a 35% speedup over eSAIS with integrated LCP construction.

In their second paper, Kärkkäinen and Kempa [KK16c] proposed a fully external standalone LCP array construction algorithm with at most $\mathcal{O}(\mathrm{sort}(n))$ I/Os. This was a breakthrough since previously only LCP construction algorithms integrated into suffix sorting algorithms were able to achieve these bounds. Their algorithm however is quite complex and guarantees this bound using *sampling* both of irreducible LCPs (via a sparse PLCP array) and of LCPs in a difference cover. They construct a recursive hierarchy of sparse PLCP arrays which are used to attain approximate LCP values. These approximate values are then refined via the difference cover and finally the true LCP is determined using $\mathcal{O}(1)$ character comparisons. This complex algorithm breaks a theoretical bound but was not implemented by the authors.

Kärkkäinen and Kempa [KK16a] then propose two practical external memory LCP array construction algorithms in their third paper: *EM-S*$\Phi$ and *EM-SI*. They are in one sense external memory adaptations of $\Phi$ and I by Kärkkäinen, Manzini, and Puglisi [KMP09] and in another sense use the sparsification and recursion ideas from their previous paper [KK16c] to reduce the I/O complexity. In a follow-up paper [KK17b] they then even parallelized the approach and bring down the external memory requirement to $\mathcal{O}(1)$ by reusing the LCP array space for its construction. While the algorithms are the fastest available standalone external memory LCP construction methods in practice, their known theoretical worst-case I/O complexity are $\mathcal{O}(\frac{n^2}{MB(\log_\sigma n)^2} + \mathrm{sort}(n))$ for EM-S$\Phi$ and $\mathcal{O}(\frac{n^2}{MB(\log_\sigma n)^2} + \frac{n \log \sigma}{B} + \mathrm{sort}(n))$ for EM-SI. Both make heavy use of irreducible LCP values, which are about 25% of all LCP values in natural texts and DNA, but can be up to 100% in pathological cases.

Chapter 6

# 6 Inducing Suffix and LCP Arrays in External Memory

*Motivated by the superior performance of the SA-IS algorithm over other suffix array construction algorithms in internal memory, we investigate in this chapter how the induced sorting principle can be exploited in the external memory model. We have two goals in mind: first to engineer an external memory suffix sorting algorithm that outperforms the previously best one, DCX [DKMS05; DKMS08], while keeping it within sorting complexity, and second to implement the first external memory LCP array construction algorithm that is faster than a DCX-based approach. Both of our algorithms are based on a reformulation of SA-IS [NZC11; Fis11]. In the experimental section, we report on the first comparative study of suffix sorting in external memory that includes algorithms based on the induced sorting principle and demonstrate that our implementations scale up to inputs of 80 GiB in size.*

In this chapter we consider full text index construction in external memory for large inputs on a single machine. The following contents is based on two papers [BFO13; BFO16], which are joint work with Johannes Fischer and Vitaly Osipov and were first published in 2013.

Our first contribution is a new external memory suffix array construction algorithm, called *eSAIS* (External Suffix Array construction by Induced Sorting), which is based on induced sorting and runs in external memory sorting complexity. It is the first external algorithm utilizing this suffix sorting principle and the second external algorithm in the literature with sorting complexity. Practical tests and theoretical analysis show that our new algorithm outperforms the previously best external memory suffix sorter [DKMS05; DKMS08] by a factor of about two in time and I/O-volume. Our analysis shows that eSAIS requires at most $\textsc{Sort}(17n) + \textsc{Scan}(9n)$ I/O volume. This is an improvement over the previously best external memory suffix sorter DC7, which requires at most $\textsc{Sort}(24.75n) + \textsc{Scan}(3.5n)$ I/O volume [DKMS08].

In many applications (e.g. for fast string matching), the suffix array needs to be augmented with the LCP array, which holds the lengths of the longest common prefixes (LCPs) of lexicographically consecutive suffixes (see also figure 5.1 and section 5.6). Our second contribution is to augment the first algorithm to also construct the LCP array. Historically, the only prior algorithm extension for fully external memory LCP construction was proposed for DC$X$ in a one paragraph side remark by the authors. We not only present the second fully external memory LCP array construction algorithm by modifying eSAIS, but also report on experiments with the first implementation of DC3-LCP. The overhead in time and I/O volume for eSAIS-LCP over plain suffix array construction is roughly a factor of two. Our implementations scale far beyond problem sizes previously considered in the literature at that time: a text of 80 GiB size using only 4 GiB of RAM.

We would like to point out that at the time of our first paper's publication in 2013, a truly scalable external LCP array construction algorithm was the only missing piece for fast practical external memory suffix *tree* construction, because, as Barsky, Stege, and Thomo [BST10, p. 986] state in their survey on external memory suffix *trees*: "The conversion of a suffix array into a suffix tree turned out to be disk-friendly, since reads of the suffix array and writes of the suffix tree can be performed sequentially. However, the suffix array needs to be augmented with the LCP information in order to be converted into a suffix tree." They also comment on the possibility of adapting external DC3 to LCP arrays: "It is currently not clear how efficient the presented algorithm for the LCP computation would be in a practical implementation." And finally they say: "It may be only one step that divides us from a scalable solution for constructing suffix trees on disk for inputs of any type and size. Once this is done, a whole world of new possibilities will be opened, especially in the field of biological sequence analysis." Our paper closed this gap, as outlined in the remaining chapter.

Please refer to section 5.1 (page 165) in the previous chapter for some basic definitions on suffix arrays and our notation for strings, suffixes, etc. In the following section 6.1 we recapitulate the SA-IS algorithm in internal memory. This algorithm is quite sophisticated and our explanation is different from previous authors, because it better fits our understanding of the underlying principles. Section 6.1.1 contains joint work with Johannes Fischer, who showed how SA-IS can be augmented to also construct the LCP array in internal memory [Fis11].

Then, in section 6.2, we show that SA-IS and more generally induced sorting itself is suitable for the external memory model by reformulating the original algorithm so that it uses only scanning, sorting, merging, and *priority queues*. The former three operations are certainly doable in external memory, and there are also external memory priority queues achieving amortized sorting complexity in theory [Arg95; Arg03]. In practice the most efficient priority queues are those of Sanders [San99; San00] and Dementiev, Kettner, and Sanders [DKS05; DKS08]. We make some careful implementation decisions in order to keep the I/O-volume low. As a result, our new algorithm, called eSAIS, is about two times faster than the external memory implementation of DC3 [DKMS05; DKMS08]. The I/O volume is reduced by a similar factor. In section 6.2.4 we engineer the first fully external memory algorithm for LCP array construction, building on the internal memory algorithm from section 6.1.1. It is 3–4 times faster than our own implementation of LCP construction using DC3 (recall there was no such implementation before). The increase in both time and I/O volume of eSAIS with LCP array construction compared to pure suffix array construction is only a factor of around two.

Our experimental results are given in section 6.3. There we apply our algorithms on very large instances. At the extreme end we could build the suffix array for an 80 GiB XML dump of the English Wikipedia in 2.5 µsec per character using only 4 GiB of main memory with a total of about 18 TiB of generated I/O-volume. In total, all experiments reported in this chapter took 34 computing days and 200 TiB I/O volume.

In section 6.4 we conclude and discuss algorithms and implementations presented *after* our papers were published. We believe to have reignited interest in external memory suffix and LCP construction, mainly because our papers were followed by six papers presenting seven distinct algorithms (see also table 5.1 on page 181). The currently fastest and most scalable algorithm, fSAIS is a direct descendant of eSAIS.

This chapter is based on joint work with Johannes Fischer and Vitaly Osipov [BFO13; BFO16]. The original idea of reformulating induced sorting using a priority queue came from the author of this dissertation, who also wrote the majority of the articles. Large parts of this chapter were copied verbatim from the articles. Vitaly Osipov worked with us in the early stages of eSAIS's development on the initial external memory construction and analysis. Johannes Fischer previously worked on in-memory LCP array construction using SA-IS [Fis11], so integrating LCP construction was a natural extension. While implementing external LCP array construction in eSAIS, we found a small oversight in the conference version [Fis11] which was corrected in the joint journal article [BFO16]. Johannes Fischer worked together with us on the theoretical questions of how to perform the necessary range minimum queries in internal and external memory. The parts of the text describing inducing in internal memory were collaboratively written with Johannes Fischer. The theoretical I/O analysis in section 6.2.3 was joint work written together with Vitaly Osipov. In the experimental section eSAIS is compared with DC3-LCP, which was implemented by Daniel Feist as part of his bachelor thesis [Fei13]. This thesis was supervised by us and contains a full description of (external) DC3-LCP, which was previously only a side remark in the original paper. The implementation of eSAIS, eSAIS-LCP, and all experiments were done by us using STXXL.

## 6.1 Induced Sorting in Main Memory

In this section we present our own reformulation of Nong, Zhang, and Chan's [NZC09a; NZC11] SA-IS algorithm in main memory. This suffix sorting algorithm is a challenge to present well, because induced sorting is a rather unintuitive approach.

The first step is to classify all suffixes into two *types*: `S` and `L`. For suffix $T[i..n)$ the type($i$) is `S` if $T[i..n) < T[i+1..n)$, and `L` if $T[i..n) > T[i+1..n)$. The last suffix $T[n-1..n)$ is fixed as type `S`. Furthermore, we distinguish the first suffix in a run of consecutive suffixes of the same type as $\mathtt{S}^*$ or $\mathtt{L}^*$; more precisely, $T[i..n)$ is $\mathtt{S}^*$ if $T[i..n)$ is `S`-type and $T[i-1..n)$ is `L`-type. Symmetrically, $T[i..n)$ is $\mathtt{L}^*$-type if $T[i..n)$ is L-type and $T[i-1..n)$ is S-type. The last suffix $T[n-1..n) = [\$]$ is always S*, while the first suffix is never S* nor L*. Sometimes we also say the character $t_i$ is of type type($i$), even though this is actually a property of the suffix. Figure 6.1 illustrates S- and L-suffixes as transitions in height when drawing a string as a segmented line in a plot graph. Readers familiar with SA-IS will recognize that our S*-suffixes are identical with "LMS-suffixes" in the description of SA-IS, and that L*-suffixes are the symmetric equivalent.

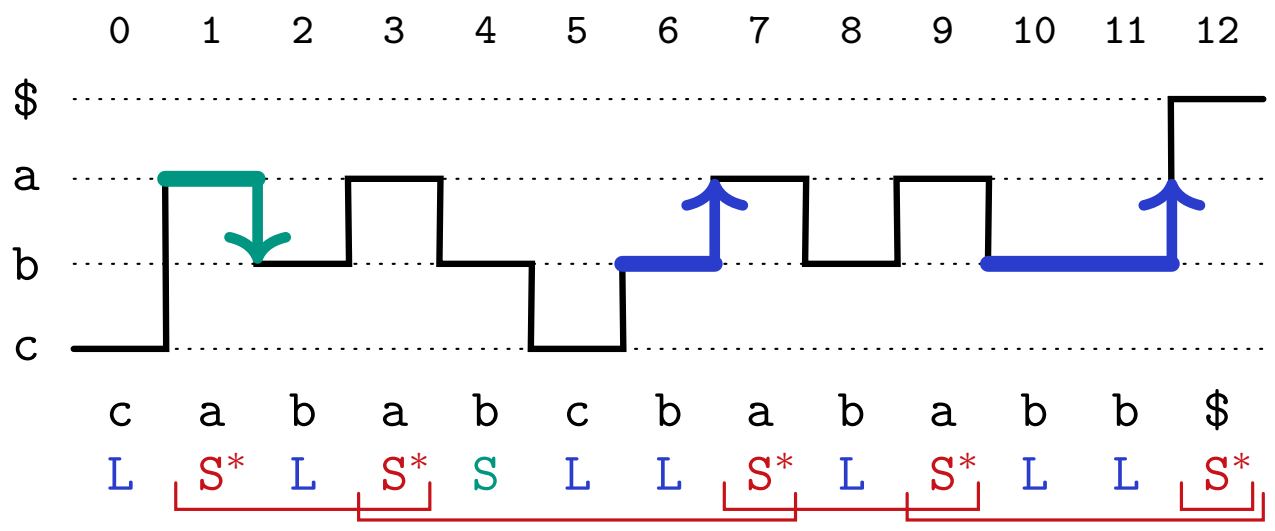

**Figure 6.1:** Pictorial representation of L- and S-suffixes in a height diagram: each letter of a string is shown as a horizontal segment in a continuous line. S-suffixes are letters where the next vertical transition goes "downwards" in the alphabet, and L-suffixes where the next transition goes "upwards".

Using these classifications, one can identify *subsequences* within the suffix array. The range of suffixes starting with the same character $c$ is called the *c-bucket*, which itself is composed of a sequence of L-suffixes followed by S-suffixes. We call these subsequences the $c$-L- and $c$-S-*subbuckets* or just L/S-subbuckets if the character is implied by the context. We also define the *repetition count* for a suffix $T[i..n)$ as $\mathrm{rep}(i) := \max_{k \in \mathbb{N}_0}\{t_i = t_{i+1} = \cdots = t_{i+k}\}$. Then the L/S-subbuckets can further be decomposed into ranges of equal repetition counts, which we call *repetition buckets*. Figure 6.2 shows these subsequences in the suffix array of an example string. The division between L-subbucket and S-subbucket is called the L/S-*seam*.

The principle behind *induced sorting* is to deduce the lexicographic order of unsorted suffixes from a set of already ordered suffixes. As described in our history of suffix sorting (chapter 5), many fast suffix sorting algorithms incorporate this principle in one way or another. SA-IS [NZC09a] and Ko and Aluru's [KA05] algorithm are built on the following basic idea:

*If the lexicographic order of all* S*-*suffixes is known, then the lexicographic order of all* L-*suffixes can be induced iteratively smallest to largest.*

We start with $\mathcal{L} := S^*$ as the lexicographically ordered list of S*-suffixes, and $\mathcal{U}$ as the set of L-suffixes. After running the following *inducing procedure*, the list $\mathcal{L}$ contains all L- and S*-suffixes in lexicographic order: Iteratively, choose the unsorted L-suffix $i \in \mathcal{U}$ such that

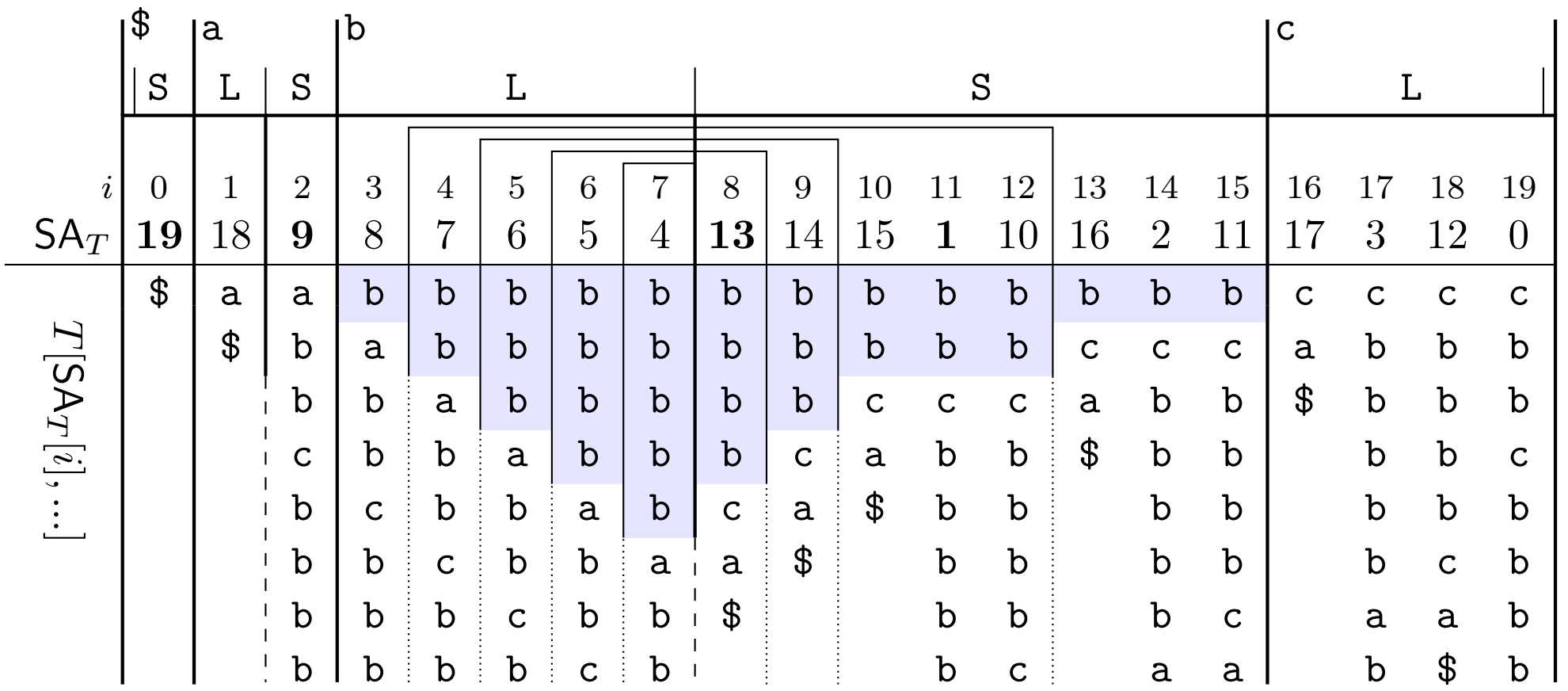

**Figure 6.2:** L/S-subbuckets in the suffix array of the string $T = [\texttt{c}, \texttt{b}, \texttt{b}, \texttt{c}, \texttt{b}, \texttt{b}, \texttt{b}, \texttt{b}, \texttt{b}, \texttt{a}, \texttt{b}, \texttt{b}, \texttt{c}, \texttt{b}, \texttt{b}, \texttt{b}, \texttt{b}, \texttt{c}, \texttt{a}, \$]$. In total there are four buckets as $|\Sigma| = 4$, of which the b-bucket being the largest. The b-bucket can be decomposed in the b-L-subbucket containing five suffixes, and the b-S-subbucket containing eight. Both can then be further decomposed into repetition buckets: there are five different repetition buckets within the b-L-subbucket and four in the b-S-subbucket.

(i) $T[i+1..n)$ is already ordered in $\mathcal{L}$,

(ii) $t_i$ is the smallest first character among those suffixes, and

(iii) $T[i+1..n)$ has smallest rank within $\mathcal{L}$ of those suffixes starting with the same first character $t_i$.

Insert $i$ into $\mathcal{L}$ as the next larger L-suffix among all suffixes that start with $t_i$. Remove $i$ from $\mathcal{U}$ and repeat until $\mathcal{U}$ is empty.

**Lemma 6.1 (Inducing L-Suffixes from S*-Suffixes)**

*The inducing procedure correctly establishes the lexicographic order of all* S*- *and* L-*suffixes in* $\mathcal{L}$.

*Proof.* The set $\mathcal{U}$ is already partially ordered with respect to the lexicographic order, since sequences of L-suffixes $T[i..n) > T[i+1..n) > \cdots > T[i+k..n)$ form >-chains. By combining these chains with the lexicographic order of S*-suffixes, which terminate them, the total order of all L-suffixes can be induced.

From the definition of the inducing procedure for a chosen suffix $T[i..n)$, we have that suffix $T[i+1..n)$ must already be in $\mathcal{L}$, that $t_i$ is the smallest first character, and in case of ties $T[i+1..n)$ is the smallest among those. From these properties, $T[i..n) < T[j..n)$ follows for all L-suffixes $T[j..n) \in \mathcal{U} \setminus \{T[i..n)\}$, because the L-type property forms descending chains of unsorted suffixes in $\mathcal{U}$. Due to the transitive ordering of L-suffix >-chains in $\mathcal{U}$ it suffices to consider only the "tails" of these chains, which are those suffixes $T[i..n) \in \mathcal{U}$ with $T[i+1..n) \in \mathcal{L}$. Among all the tails, the suffix with smallest first letter and in case of ties smallest remaining suffix $T[i+1..n)$ is always the overall smallest remaining suffix in $\mathcal{U}$.

Thus the iterative procedure always picks the smallest remaining suffix and places it as the next larger one in the $t_i$-L-subbucket. This procedure ultimately sorts all L-suffixes, because each has a (possibly empty) sequence of L suffixes terminated by an S*-suffix to its right. □

This rather technical proof can be motivated by the following intuition: picture a string as one long metal chain were each link represents a character. Sequences of L-suffixes correspond to pieces of that chain containing one or more chain links. The chain pieces have a direction, so we can picture them hanging downwards such that larger links hang lower. In each piece, however, one chain link (the smallest L-suffix) is at the top. The selection algorithm in the lemma then looks at all the top chain links and figures out which among these is smallest. This is sufficient to find the smallest link, because all chain links hanging downwards must be larger. We then cut off that smallest link and repeat the process. The L-type property guarantees that we only have to look at the top-most chain links, for all below it are larger suffixes. This induction process was first described by Ko and Aluru [KA05], but the challenge remained on how to cut the whole chain into pieces and how to sort the ends.

Analogously to lemma 6.1, the order of all S-suffixes can be induced iteratively largest to smallest, if the relative order of all L*-suffixes is known. This results in the following high-level four step algorithm SA-IS [NZC11]:

(1) Sort the S*-suffixes. This step will be explained in more detail below.

(2) Put the sorted S*-suffixes into their corresponding S-subbuckets, without changing their order. All other entries remain undefined. Prepare head and tail pointers for all L-subbuckets in SA.

(3) Induce the order of the L-suffixes by scanning SA from *left to right* (skipping undefined entries): for every position $i$ in SA, if $T[\mathsf{SA}[i]-1..n)$ is L-type, write index $\mathsf{SA}[i]-1$ to the current head of the $c$-L-subbucket, where $c = T[\mathsf{SA}[i]-1]$ (the preceding character), and increase the current head of that bucket by one. Note that this step can only induce "to the right" (the current head of the $c$-L-subbucket is larger than $i$).

(4) Induce the order of the S-suffixes by scanning SA from *right to left*: for every position $i$ in SA, if $T[\mathsf{SA}[i]-1..n)$ is S-type, write $\mathsf{SA}[i]-1$ to the current *tail* of the $c$-S-subbucket, where $c = T[\mathsf{SA}[i]-1]$, and *de*crease the current tail of that bucket by one. Note that this step can only induce "to the left," and might intermingle arbitrary S-suffixes with S*-suffixes.

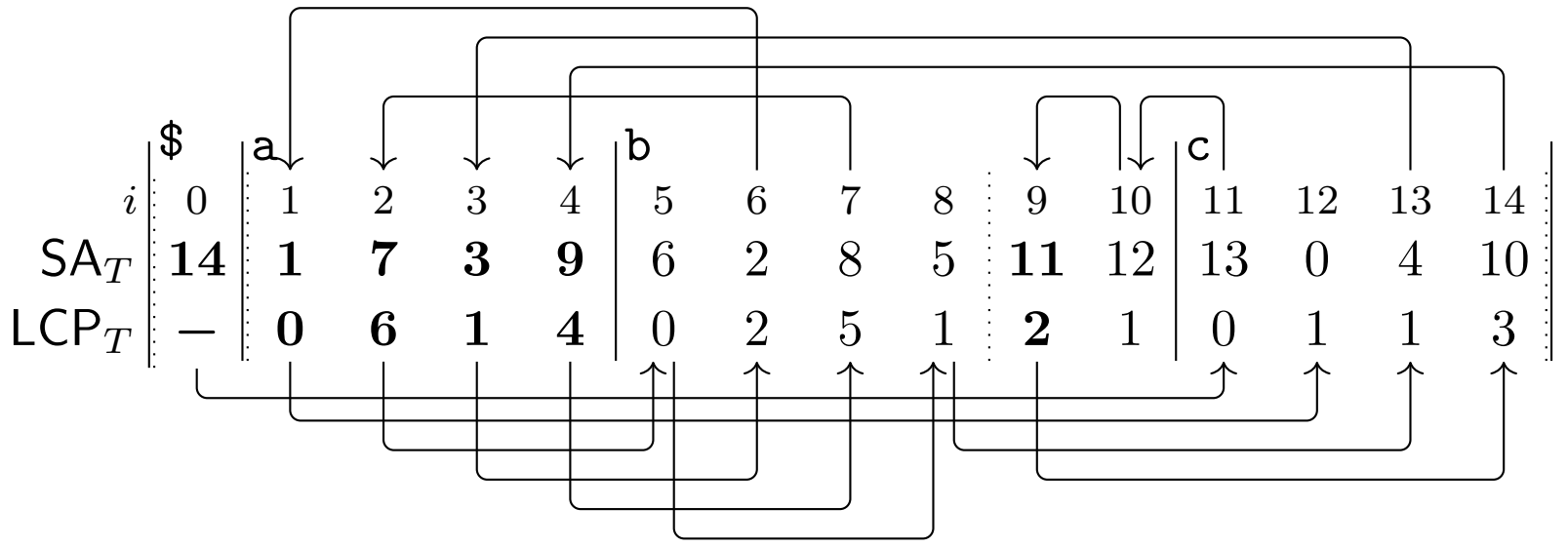

**Figure 6.3:** Example of the inducing steps on the string $T = [\mathtt{c}, \mathtt{a}, \mathtt{b}, \mathtt{a}, \mathtt{c}, \mathtt{b}, \mathtt{b}, \mathtt{a}, \mathtt{b}, \mathtt{a}, \mathtt{c}, \mathtt{b}, \mathtt{b}, \mathtt{c}, \$]$. Assume that the relative order of $\mathtt{S}^*$-suffixes (bold font) is known (see figure 6.5 for the recursion). They are placed into their corresponding S-subbuckets as described in step (2). In step (3), suffix and LCP values of the L-suffixes (normal font) are induced from the LCPs of $\mathtt{S}^*$-suffixes. Afterwards, in step (4), the reverse process (shown above the array) induces all S-suffixes from the $\mathtt{L}^*$-suffixes.

Figure 6.3 illustrates the inducing steps (3) and (4) with arrows, step (3) as arrows below the suffix array and step (4) above it (ignore for now the row labeled "$\mathsf{LCP}_T$").

It remains to explain how to find the relative order of $\mathtt{S}^*$-suffixes. To solve this problem, we break the string into smaller parts. For each $\mathtt{S}^*$-suffix $T[i..n)$, we define the $\mathtt{S}^*$-*substring* $[t_i, \dots, t_j]$, where $T[j..n)$ is the next $\mathtt{S}^*$-suffix in the string. The last $\mathtt{S}^*$-suffix [$] is fixed to be a sentinel $\mathtt{S}^*$-substring by itself. We call the last character $T[j]$ of each $\mathtt{S}^*$-substring the *overlapping character* since it is also the first character of the next $\mathtt{S}^*$-substring. Figure 6.1 shows the $\mathtt{S}^*$-substrings in the example as overlapping brackets below the $\mathtt{S}^*$ positions.

$\mathtt{S}^*$-substrings are ordered lexicographically, with each letter compared first by character and then by type, L-characters being smaller than S-characters in case of ties. This partial order allows one to apply *lexicographic naming* to $\mathtt{S}^*$-substrings. Lexicographic naming is a mapping $\nu$ of strings $s_i$ (or other objects) from the word-alphabet $\Sigma^*$ to another alphabet $\Sigma'$ such that strings $s_i$ retain their order, hence, if $s_i < s_j$ then $\nu(s_i) < \nu(s_j)$ for all strings $s_i$ and $s_j$. The value $\nu(s_i)$ is then called the lexicographic name of $s_i$. This mapping is often used in suffix sorting to map long ($\mathtt{S}^*$-)substrings to an integer alphabet $\Sigma' = \{1, 2, 3, \dots, m\}$ for recursion.

Due to how $\mathtt{S}^*$-suffixes are defined, there are at most $\frac{n}{2}$ $\mathtt{S}^*$-substrings in any text, and on average there are between $\frac{n}{3}$ and $\frac{n}{4}$ depending on $|\Sigma$. Hence, one can efficiently solve the problem of finding the relative order of all $\mathtt{S}^*$-suffixes by *recursively suffix sorting* the reduced string of lexicographic names of $\mathtt{S}^*$-substrings [NZC11]. For the remainder of this chapter, we denote the reduced string consisting of lexicographic names by $R$ and the recursively computed suffix array by $\mathsf{SA}_R$.

While the order on S*-substrings defined above first compares characters and then types at each index, this is in almost all cases equivalent to naive string comparisons. The rare exceptions occur only when the type contains information about characters beyond the end of the S*-substring. Consider the example [c, |a, c, b, |a, c, |b, c, |$], where | indicates the beginning of an S*-suffix. The substring a, c, b occurs twice, and the last character b is once of type L and once of type S*. The trick used to break S*-suffixes into S*-substrings is to determine when a naive string comparison can be terminated and the remaining order be calculated recursively. This is done by splitting S*-suffixes at S*-positions. The overlapping character is required to distinguish S*-positions from ordinary L-positions. In the previous example, the character b requires this distinction. If one splits the string into S*-substrings without the overlapping character the S*-suffix starting with |a, c, b, |a, c, ... and |a, c, |b, c, ... are given incorrect lexicographic names.

### 6.1.1 Inducing LCP Arrays in Main Memory

In this section we explain how the induced sorting algorithm can be modified to also compute the LCP array *in main memory*. This will form the foundation of our *external memory* adaption in section 6.2.4. The idea of inducing the LCP array in main memory was first proposed by Fischer [Fis11] in a conference proceeding. Later, we coauthored a journal article [BFO16] with Fischer and Osipov, which contains a corrected, longer explanation of LCP array construction using induced sorting. This journal article is the basis for this chapter.

The basic idea is that whenever we place two S- or L-suffixes $T[x..n)$ and $T[y..n)$ at adjacent places $k-1$ and $k$ in the same $c$-bucket of the final suffix array (see figure 6.4 and steps (3)–(4) of the algorithm in section 6.1), the length of their longest common prefix can be induced from the longest common prefix of the suffixes $T[x+1..n)$ and $T[y+1..n)$. As the latter suffixes are exactly those that caused the inducing of $T[x..n)$ and $T[y..n)$, we already know their LCP value $\ell$ (by the order in which we fill SA), and can hence set $\mathsf{LCP}_T[k]$ to $\ell+1$.

The details are described next. We augment the steps (1) of the induced sorting algorithm with (1′) as follows:

(1′) Compute $\mathsf{LCP}_{S^*}$, the array of LCP values of the S*-suffixes, as we will discuss in the following section 6.1.2.

(2′) Whenever we place an S*-suffix into its S-subbucket, we also store its LCP value (as computed in step (1′)) at the corresponding position in $\mathsf{LCP}_T$.

(3′) Suppose that the $j$-th inducing iteration in the left-to-right sweep just put suffix $T[y..n)$ with $y = \mathsf{SA}[k]$ into its $c$-L-subbucket ($c = t_y$) at position $k$. If $T[y..n)$ is the first suffix in its L-subbucket, we set $\mathsf{LCP}_T[k]$ to 0. Otherwise, suppose further that in a previous iteration $i < j$ the inducing step placed suffix $T[x..n)$ at $k-1$ in the same $c$-L-subbucket, with $x = \mathsf{SA}[k-1]$. Then if $i$ and $j$ are in different buckets, the corresponding suffixes $T[x+1..n)$ and $T[y+1..n)$ start with different characters, so we set $\mathsf{LCP}_T[k]$ to 1, as the suffixes $T[x..n)$ and $T[y..n)$ share only the common character $c$ at their beginnings. Otherwise, $x+1$ and $y+1$ are in the same $c'$-bucket, with $t_{x+1} = c' = t_{y+1}$. Then the length $h$ of the longest common prefix of the suffixes $T[x+1..n)$ and $T[y+1..n)$ is given by the *minimum* value in $\mathsf{LCP}_T[i+1\,..\,j]$, all of which are in the same $c'$-bucket and have therefore already been computed in previous iterations. We can hence set $\mathsf{LCP}_T[k]$ to $h+1$. We address the problem of how to compute these minima in section 6.1.3.

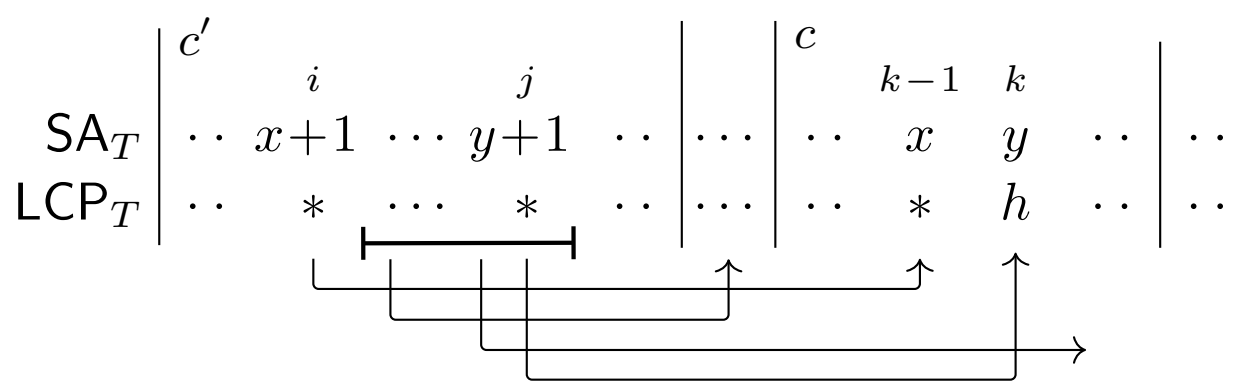

**Figure 6.4:** General scheme of the inducing step. When inducing index $k$ in the suffix array, the LCP value $h = \text{RMQ}_{\mathsf{LCP}_T}(i+1, j) + 1$ can be derived using a range minimum query (RMQ) between the previous and current relative ranks of the sources of the induction.

(4′) We modify the right-to-left sweep in this step symmetrically to step (3′).

We will resolve the problem of computing the LCP value at the L/S-seam between the last L-suffix and the first S-suffix in a bucket in section 6.1.4.

For an example, consider the inducing of suffixes $T[2..n)$ and $T[8..n)$ in figure 6.3 at the suffix array indices 6 and 7. Both suffixes start with character b. The suffixes that caused their inducing are $T[3..n)$ and $T[9..n)$ at positions 3 and 4 of $\mathsf{SA}_T$, respectively, both starting with a. Their LCP is 4, which is (trivially) determined by finding the minimum in $\mathsf{LCP}_T[3{+}1\,..\,4]$. Therefore, we set $\mathsf{LCP}_T[7]$ to 5.

### 6.1.2 Computing LCP Values of S*-suffixes

Here, we give the details of step (1′) above. From the recursion, we can assume that the LCP array $\mathsf{LCP}_R$ of the reduced string $R$ is calculated together with $\mathsf{SA}_R$, while in the base case with unique lexicographic names $\mathsf{LCP}_R$ is simply filled with zeros.

Let $s_1^*, \ldots, s_K^*$ be the $K$ positions of S*-substrings in $T$, ordered as in the input string. Given the recursively calculated LCP array $\mathsf{LCP}_R$ and $\mathsf{SA}_R$, we now show how to calculate $\mathsf{LCP}_{S^*}[k] := \text{LCP}_T(s^*_{\mathsf{SA}_R[k-1]}, s^*_{\mathsf{SA}_R[k]})$, which is the maximum number of equal characters (in $T$, not in $R$!) at the beginning of two lexicographically consecutive $\mathtt{S}^*$-suffixes $s^*_{\mathsf{SA}_R[k-1]}$ and $s^*_{\mathsf{SA}_R[k]}$. See also figure 6.5, which gives an example of all concepts presented in this section.

There are two main issues to be dealt with: firstly, a reduced character in $R$ is composed of several characters in $T$. Apart from the obvious need for *scaling* the values in $\mathsf{LCP}_R$ by the lengths of the corresponding $\mathtt{S}^*$-substrings, we note that even *different* characters in $R$ can have a common prefix in $T$ and thus contribute to the total LCP. For example, in figure 6.5 the first two $\mathtt{S}^*$-substrings $[\mathtt{a},\mathtt{b},\mathtt{a}]$ and $[\mathtt{a},\mathtt{c},\mathtt{b},\mathtt{b},\mathtt{a}]$ both start with an 'a', although they are different characters in $R$. The second issue is that lexicographically consecutive $\mathtt{S}^*$-suffixes with unequal first $\mathtt{S}^*$-substring can still have LCPs encompassing more than one $\mathtt{S}^*$-substring in one suffix, but not in the other. For example, the $\mathtt{S}^*$-suffixes $T[3..n) = [\mathtt{a},\mathtt{c},\mathtt{b},\mathtt{b},\mathtt{a},\mathtt{b},\mathtt{a},\mathtt{c},\mathtt{b},\mathtt{b},\mathtt{c\$}]$ and $T[9..n) = [\mathtt{a},\mathtt{c},\mathtt{b},\mathtt{b},\mathtt{c},\mathtt{\$}]$ have an LCP of 4 that spans two $\mathtt{S}^*$-substrings of the latter suffix.

To handle the first issue, we store the length of each $\mathtt{S}^*$-substring, minus the one overlapping character, in an array called $\mathsf{Size}_{S^*} := [\, s^*_{k+1} - s^*_k \mid k \in [0\,..\,K)\,]$ in string order, with $K$ being the number of $\mathtt{S}^*$-substrings and $s^*_K = n-1$. The second issue is more difficult: during the lexicographic naming process (which sorts the $\mathtt{S}^*$-substrings), we compute the LCPs of lexicographically consecutive $\mathtt{S}^*$-*suffixes* in an array $\mathsf{LCP}_N$ up to the length of the *longer* $\mathtt{S}^*$-substring of each pair. For $\mathtt{S}^*$-suffixes with equal first $\mathtt{S}^*$-substring this is just the length of the $\mathtt{S}^*$-substring. However, if the first $\mathtt{S}^*$-substrings differ, the required LCP calculation seems to require characters outside the shorter $\mathtt{S}^*$-substring. This can be solved by augmenting $\mathtt{S}^*$-substrings with the *repetition count* of the overlapping character, and we will describe this modification in detail later.

The resulting array $\mathsf{LCP}_N$ is then prepared for constant-time range minimum queries (RMQs) [FH11]; such queries return the minimum among all array entries in a given range: $\text{RMQ}_A(\ell, r) = \min_{\ell \le i \le r} A[i]$ for an array $A$. This allows us to find the common characters of arbitrary $\mathtt{S}^*$-suffixes, as shown in the next lemma.

**Lemma 6.2 (Calculating the LCP Values of $\mathtt{S}^*$-Suffixes)**
*Given* $\mathsf{SA}_R$, $\mathsf{ISA}_R$, $\mathsf{LCP}_R$, $\mathsf{Size}_{S^*}$, *and* $\mathsf{LCP}_N$, *the array* $\mathsf{LCP}_{S^*}[k]$ *can be calculated by*

$$\mathsf{LCP}_{S^*}[k] = \sum_{i=\mathsf{SA}_R[k]}^{\mathsf{SA}_R[k]+\mathsf{LCP}_R[k]-1} \mathsf{Size}_{S^*}[i] + \text{RMQ}_{\mathsf{LCP}_N}(\ell[k]+1, r[k]) \tag{6.1}$$

$$\textit{with} \quad \ell[k] = \mathsf{ISA}_R\big[\mathsf{SA}_R[k-1] + \mathsf{LCP}_R[k]\big]$$

$$\textit{and} \quad r[k] = \mathsf{ISA}_R\big[\mathsf{SA}_R[k] + \mathsf{LCP}_R[k]\big]\,,$$

*Proof.* We must show that this expression counts the maximum number of equal characters starting at the $\mathtt{S}^*$-suffixes $s^*_{\mathsf{SA}_R[k-1]}$ and $s^*_{\mathsf{SA}_R[k]}$, which are lexicographically consecutive. Because they are consecutive, $\mathsf{LCP}_R[k]$ was calculated recursively as the number of equal *complete* $\mathtt{S}^*$-substrings starting at these positions. Thus summing over the sizes of those equal S*-substring entries from $\mathsf{SA}_R[k]$ to $\mathsf{SA}_R[k] + \mathsf{LCP}_R[k] - 1$ (or, equivalently, $\mathsf{SA}_R[k-1]$ to $\mathsf{SA}_R[k-1] + \mathsf{LCP}_R[k] - 1$) yields the total number of equal characters in whole S*-substrings.

| $i$ | 0 | 1 | 2 | 3 | 4 | 5 | 6 | 7 | 8 | 9 | 10 | 11 | 12 | 13 | 14 |
|---|---|---|---|---|---|---|---|---|---|---|---|---|---|---|---|
| $T$ | c | a | b | a | c | b | b | a | b | a | c | b | b | c | $ |
| type($i$) | L | S* | L* | S* | L* | L | L | S* | L* | S* | L* | S* | S | L* | S* |
| $R$ | | | 1 | | | 2 | | | 1 | | 3 | | 4 | | 0 |
| Size$_{S^*}$ | | | 2 | | | 4 | | | 2 | | 2 | | 3 | | 0 |

| $k$ | 0 | 1 | 2 | 3 | 4 | 5 |
|---|---|---|---|---|---|---|
| SA$_R$ | 5 | 0 | 2 | 1 | 3 | 4 |
| LCP$_R$ | – | 0 | 1 | 0 | 0 | 0 |
| LCP$_{S^*}$ | – | 0 | 6 | 1 | 4 | 0 |

| $[t_i, \ldots, t_j]$ | $i$ | rep($j$) | LCP$_N$ |
|---|---|---|---|
| [$] | 14 | 0 | – |
| [a, b, a] | 1 | 0 | 0 |
| [a, b, a] | 7 | 0 | 3 |
| [a, c, b, b, a] | 3 | 0 | 1 |
| [a, c, b] | 9 | 1 | 4 |
| [b, b, c, $] | 11 | 0 | 0 |

**Figure 6.5:** Example of the structures before and after the recursive call of the induced sorting algorithm. Left: the top part shows the text, the classification of suffixes, and the reduced string $R$ on which the algorithm is run recursively. The resulting suffix and LCP arrays for $R$ are shown in the lower part (SA$_R$ and LCP$_R$). Whereas the former has a direct correspondence to the S*-suffixes in $T$, the latter needs to be expanded to LCP$_{S^*}$ to account for the different alphabets in $T$ and $R$. Right: additional information needed to expand LCP$_R$ to LCP$_{S^*}$, consisting of the sorted S*-substrings and associated information. The last column LCP$_N$ shows the LCPs of lexicographically consecutive S*-suffixes up to the length of the longer S*-substring.

It thus remains to determine the longest common prefix of the first pair of unequal S*-substrings contained in both S*-suffixes. This first pair of unequal S*-substrings begins at $\mathsf{SA}_R[k-1] + \mathsf{LCP}_R[k]$ and $\mathsf{SA}_R[k] + \mathsf{LCP}_R[k]$ in $R$. However, instead of calculating the LCP of the suffixes $SA_R[k-1] + \mathsf{LCP}_R[k]$ and $\mathsf{SA}_R[k] + \mathsf{LCP}_R[k]$ in $T$ directly (after mapping from $R$ to $T$), we resort to looking up the lexicographic ranks of these positions in $\mathsf{ISA}_R$. So, $\mathsf{ISA}_R[\mathsf{SA}_R[k-1] + \mathsf{LCP}_R[k]]$ and $\mathsf{ISA}_R[\mathsf{SA}_R[k] + \mathsf{LCP}_R[k]]$ are the lexicographic ranks of the pair of unequal S*-substrings. These ranks need not be adjacent in $\mathsf{SA}_R$, therefore instead of a direct lookup in $\mathsf{LCP}_N$, an RMQ between these ranks becomes necessary. Notice that $\mathsf{LCP}_N$ is constructed from names, while the queries boundaries are ranks. This is however still correct, as the range in $\mathsf{LCP}_N$ corresponding to the same lexicographic name is filled with the length of the name, except for the first entry. Because the LCP to both predecessor or successor name is no longer than the length, for RMQ calculation it suffices to take any rank of the same lexicographic name.

If $\mathsf{LCP}_R[k] = 0$, then the whole expression reduces to $\mathsf{LCP}_N[k]$, as one would expect. □

Let us return to the calculation of $\mathsf{LCP}_N$: in SA-IS, letters of S*-substrings are compared first by character and then by type. For LCP construction, however, we must count equal characters even though they may have *different* types.

Consider any two suffixes that start with an equal character but different types. We can calculate their LCP by considering only the number of *repetitions* of the equal character. This is sufficient since if the same character occurs with different types, then these differing types are defined by the next differing character of each suffix, where one suffix is L and the other S, and obviously that character must be different. Thus the LCP of the two considered suffixes is the minimum of the two repetition counts. For occurrences within an $S^*$-substring the repeating letters can be counted directly. But, for cases where the equal sequence extends beyond the end of an $S^*$-substring, we need the repetition count of the overlapping character. Hence, we augment each $S^*$-substring $[T[i], \ldots, T[j]]$ with the repetition count $\text{rep}(j)$, which enables direct calculation of $\mathsf{LCP}_N$: first compare the characters $[T[i], \ldots, T[j]]$ in the $S^*$-substring and if they are equal, add the minimum of $\text{rep}(j)$. This procedure also works correctly, if $T[j]$ has the same type.

For example, regard the penultimate row on the right side of figure 6.5. Even though there are only 3 common characters in $[\mathtt{a}, \mathtt{c}, \mathtt{b}]$ and its preceding $S^*$-substring $[\mathtt{a}, \mathtt{c}, \mathtt{b}, \mathtt{b}, \mathtt{a}]$, for the calculation in lemma 6.2 to be correct, there must be a '4' in $\mathsf{LCP}_N$. This LCP value can be deduced from the repetition count '1' of the shorter string $[\mathtt{a}, \mathtt{c}, \mathtt{b}]$, which matches the second 'b' of the longer string $[\mathtt{a}, \mathtt{c}, \mathtt{b}, \mathtt{b}, \mathtt{a}]$.

Due to the augmentation with the repetition count, the $S^*$-substring sort order must be extended to encompass $\text{rep}(j)$. As before, overlapping L characters are smaller than S characters. Of two overlapping L characters, the one with *lower* repetition count is considered as smaller. Symmetrically, of two S characters, the one with *higher* repetition count is smaller. This distinction is only necessary when comparing the overlapping characters of two $S^*$-substrings of equal length; arbitrary characters and types can be compared as before, since further comparison of characters delivers the same result as comparing repetition counts.

### 6.1.3 Finding Minima

To find the minimum value in $\mathsf{LCP}[i+1\mathinner{.\,.} j]$ or $\mathsf{LCP}[j+1\mathinner{.\,.} i]$ (steps (3′) and (4′) above), we have several alternatives. Let us focus on the left-to-right scan (step (3′)); the right-to-left scan (step (4′)) is symmetric. The simplest idea is to naively scan the whole interval in LCP, however, this results in overall $O(n^2)$ running time.

A better alternative would be to keep an array $M$ of size $|\Sigma|$, such that the minimum is always given by $M[c]$ if we induce an LCP value in bucket $c$. More formally, we define the array $M[1\mathinner{.\,.}|\Sigma|]$ by $M[c'] := \min \mathsf{LCP}[i_{c'}+1\mathinner{.\,.} j]$, where $c' \in \Sigma$ and $i_{c'} \le j$ is the last position from where we induced to the $c'$-bucket. To keep $M$ up to date, before retrieving $h = \text{RMQ}_{\mathsf{LCP}_T}(i+1, j) + 1$ from $M[c]$, we update all entries in $M$ that are larger than $\mathsf{LCP}[j]$ by $\mathsf{LCP}[j]$, since their corresponding range minimum queries overlap with position $j$. Finally, we set $M[c] = +\infty$. This ensures that in the next iteration $j+1$ the value $M[c]$ will be set correctly. In total, this approach has $O(n|\Sigma|)$ running time. A further refinement of this technique stores the values in $M$ in sorted order and uses binary search on $M$ to find the minima, similar to the stack used by [GO11]. This results in overall $O(n \lg |\Sigma|)$ running time.

Yet, we can also update the minima in $O(1)$ amortized running time. Recall that the queries lie within a single bucket (called $c'$), and every bucket is subdivided into an L- and an S-subbucket. The idea is to also subdivide the query into an L- and an S-query, and return the minimum of the two. The S-queries are simple to handle: in step (3′), only S*-suffixes will be scanned, and these are static. Hence, we can preprocess every S*-subbucket (consisting of S*-suffixes starting with the same character) with a static data structure for constant-time range minima, using overall linear space [Fis10, Thm. 1].

The L-queries are more difficult, as elements keep being written to them during the scan. However, these updates occur in a very regular fashion, namely in a left-to-right manner. This makes the problem simpler: we maintain a *LRM-tree* [BFN11; BFN12, Def. 1] $\mathcal{M}_{c'}$ for each bucket $c'$, which is initially empty (no L-suffixes written so far). When a new L-suffix along with LCP value $\ell+1$ is written into its $c$-bucket, we climb up the rightmost path of $\mathcal{M}_c$ until we find an element $x$ whose corresponding array-entry is strictly smaller than $\ell + 1$ ($\mathcal{M}_c$ has an artificial root holding LCP value $-\infty$, which guarantees that such an element always exists). The new element is then added as $x$'s new rightmost leaf. An easy amortized argument shows that this results in overall linear time. Further, $\mathcal{M}_c$ is stored along with a data structure for constant-time *lowest common ancestor queries* (LCAs) which supports dynamic leaf additions in $O(1)$ worst-case time [CH99; CH05]. Then the minimum in any range in the processed portion of the L-subbucket can be found in $O(1)$ time [Fis10, Lemma 2].

What we have described in the preceding paragraph is actually more general than what we really need: a solution to the *semi-dynamic range minimum query problem* with constant $O(1)$ query- and amortized $O(1)$ insertion-time, with the restriction that new elements can only be appended at the end (or beginning, respectively) of the array. This solution might also have interesting applications in other problems. In our setting, though, the problem is slightly more specific: the sizes of the arrays to be prepared for RMQs are known in advance (namely the sizes of the L- or S-subbuckets).

### 6.1.4 Computing LCP Values at the L/S-Seam

There is one subtlety in the above inducing algorithm we have withheld so far, namely that of computing the LCP values between the last L-suffix and the first S-suffix in a given $c$-bucket (the L/S-seam). More precisely, when reaching an L/S-seam in step (3′), we have to re-compute the LCP value between the first S*-suffix in the $c$-bucket (if it exists) and the last L-suffix in the same $c$-bucket (the one that we just induced), in order to induce correct LCP values when stepping through the S*-suffixes in subsequent iterations. Likewise, when placing the very first S-suffix in its $c$-bucket in step (4′), we need to compute the LCP value between this induced S-suffix and the largest L-suffix in the same $c$-bucket. (Note that step (4) might place an S-suffix before all S*-suffixes, so we cannot necessarily re-use the LCP value computed at the L/S-seam in step (3′).)

Despite these complications, the following lemma shows that the LCP computation at L/S-seams is actually particularly easy:

**Lemma 6.3 (Common Prefix of Suffixes in the Same $c$-Bucket)**
*Let $T[i..n)$ be an L-suffix, $T[j..n)$ an S-suffix, and $T[i] = c = T[j]$ (the suffixes are in the same $c$-bucket in SA). Further, let $\ell \geq 1$ denote the length of the longest common prefix of $T[i..n)$ and $T[j..n)$. Then*

$$[\,T[i] \ldots T[i+\ell-1]\,] = c^{\ell} = [\,T[j] \ldots T[j+\ell-1]\,] \;.$$

*Proof.* Assume that $t_{i+k} = c' = t_{j+k}$ for some $2 \leq k < \ell$ and $c' \neq c$. In case $c' < c$, both $T[i..n)$ and $T[j..n)$ are of type L, and in case $c' > c$, they are both of type S. In both cases, this is a contradiction to the assumption that $T[i..n)$ is of type L, and $T[j..n)$ of type S. □

In words, the above lemma states that the longest common prefix at the L/S-seam can only consist of one identical repeated character. Therefore, a *naive computation* of the LCP values at the L/S-seam is sufficient to achieve overall linear running time in main memory: every character $t_i$ contributes at most to the computation at the L/S-seam in the $t_i$-bucket, and not in any other $c$-bucket for $c \neq t_i$.

## 6.2 Induced Sorting in External Memory

We now design an external memory algorithm based on the induced sorting principle that runs in sorting complexity and has a lower constant factor than DC$X$ [DKMS05; DKMS08]. The basis for this algorithm is an efficient external memory priority-queue (PQ) [DKS05; DKS08], as suggested by the proof of lemma 6.1. Since it is derived from RAM-based SA-IS, we call our new algorithm *eSAIS* (External Suffix Array construction by Induced Sorting). We first comment on details of the pseudocode shown as algorithm 6.1, which is a simplified variant of eSAIS. Section 6.2.1 is then devoted to complications that arise due to large S*-substrings.

Our analysis in section 6.2.3 shows that eSAIS requires at most $\textsc{Sort}(17n) + \textsc{Scan}(9n)$ I/O volume. This is an improvement over external memory DC3 and DC7 which need at most $\textsc{Sort}(30n) + \textsc{Scan}(5n)$ and $\textsc{Sort}(24.75n) + \textsc{Scan}(3.5n)$ I/O volume (see also table 5.1 on page 181). DC7 was the previous best known external memory suffix sorting algorithm, as the I/O volume of DC13 rises back to $\textsc{Sort}(30.1n) + \textsc{Scan}(2.9n)$ and the coefficients increase further for larger difference covers [Meh04; DKMS05; Wee06; DKMS08].

Let $R$ denote the reduced string consisting of lexicographic names of S*-suffixes. The objective of lines 2 to 9 is to create the inverse suffix array $\mathsf{ISA}_R$, containing the ranks of all $\mathtt{S}^*$-suffixes in $T$ (corresponding to step (1) of the high-level algorithm in section 6.1). In line 2, the input is scanned back-to-front and the type of each suffix $i$ is determined from $t_i$, $t_{i+1}$, and type$(i+1)$. Thereby, $\mathtt{S}^*$-suffixes are identified and we assume there are $K$ $\mathtt{S}^*$-suffixes with $K-1$ $\mathtt{S}^*$-substrings between them, plus the sentinel $\mathtt{S}^*$-substring. For each $\mathtt{S}^*$-substring, the scan creates one tuple. These tuples are then sorted as described at the end of section 6.1. (Note that the type of each character inside the tuple can be deduced from the characters and the type of the overlapping character. These overlapping characters are currently all $\mathtt{S}^*$, but this will change in the next section.) After sorting, in line 3 the $\mathtt{S}^*$-substring tuples are lexicographically named with respect to the $\mathtt{S}^*$-substring ordering and the output tuple array $N$ is naturally ordered by names $n_k \in [0 \mathinner{..} K)$. The names must be sorted back to string order in line 4. This yields the reduced string $R$, wherein each character represents one $\mathtt{S}^*$-substring. If the lexicographic names are unique, the lexicographic ranks of $\mathtt{S}^*$-substrings are simply the names in $R$ (lines 8 to 9). Otherwise the ranks are calculated recursively by calling eSAIS and inverting $\mathsf{SA}_R$ (lines 5 to 7).

With $\mathsf{ISA}_R$ containing the ranks of $\mathtt{S}^*$-suffixes, we apply lemma 6.1 in lines 10 to 18. The PQ contains quintuples $(t_i, y, r, [t_{i-1}, \dots, t_{i-\ell}], i)$ with $(t_i, y, r)$ being the sort key, which is composed of character $t_i$, indicator $y = \text{type}(i)$ with $\mathtt{L} < \mathtt{S}$ and relative rank $r$ of suffix $T[i+1..n)$. To efficiently implement lemma 6.1, instead of checking *all* unsorted $\mathtt{L}$-suffixes, we design the PQ to create the relative order of $\mathtt{S}^*$- and $\mathtt{L}$-suffixes as described in the proof. Extraction from the PQ always yields the smallest unsorted $\mathtt{L}$-suffix, or, if all $\mathtt{L}$-suffixes within a $c$-bucket are sorted, the smallest $\mathtt{S}^*$-suffix $i$ with unsorted preceding $\mathtt{L}$-suffix at position $i-1$ (hence $t_{i-1} > c$). Thus diverging slightly from the proof, the PQ only contains $\mathtt{L}$-suffixes $T[i..n)$ where $T[i+1..n)$ is already ordered, plus all $\mathtt{S}^*$-suffixes where $T[i-1..n)$ has not been ordered; so at any time the PQ contains at most $K$ items. In line 11, the PQ is initialized with the array $S^*$, which is built in line 10 by reading the input back-to-front again, re-identifying $\mathtt{S}^*$-suffixes and merging them with $\mathsf{ISA}_R$ to get the rank for each tuple. Notice that the characters of $\mathtt{S}^*$-substrings are saved in *reverse* order. The while loop in lines 12 to 18 then repeatedly removes the minimum item and assigns it the next relative rank as enumerated by $\rho_L$. This is the *inducing* process. If the extracted tuple represents an $\mathtt{L}$-suffix, the suffix position $i$ is saved in $A_L$ as the next $\mathtt{L}$-suffix in the $t_i$-bucket (line 13). Extracted $\mathtt{S}^*$-suffixes do not have an output. If the preceding suffix $T[i-1..n)$ is $\mathtt{L}$-type, then we shorten the tuple by one character to represent this suffix, and reinsert the tuple with its relative rank (line 16). However, if the preceding suffix $T[i-1..n)$ is $\mathtt{S}$-type, then the suffix $T[i..n)$ is $\mathtt{L}^*$-type, and it must be saved for the inducing of $\mathtt{S}$-suffixes (line 18). When the PQ is empty, all $\mathtt{L}$-suffixes are sorted in $A_L$, and $L^*$ contains all $\mathtt{L}^*$-suffixes ranked by their lexicographic order.

With the array $L^*$ the while loop is repeated to sort all $\mathtt{S}$-suffixes (line 19). This process is symmetric with the PQ order being reversed and using $\rho_S$`--` instead of incrementing. If $t_{i-1} > t_i$ occurs, the tuple can be dropped, because there is no need to recreate the array $S^*$ (as all $\mathtt{L}$-suffixes are already sorted). When both $A_L$ and $A_S$ are computed, the suffix array can be constructed by merging together the $\mathtt{L}$-

**Algorithm 6.1 :** eSAIS Description in Tuple Pseudocode

1 **Function** eSAIS($T = [t_0 \dots t_{n-1}]$)<br>
2 Scan $T$ back to front, create $[\,(s_k^*) \mid k \in [0\,..\,K)\,]$ for $K$ $\mathtt{S}^*$-suffixes,<br>
and sort $\mathtt{S}^*$-substrings (with $s_K^* := n-1$ as sentinel):<br>
$P := \mathrm{Sort}_{S^*}([\,(([t_i \dots t_j], i, \mathrm{type}(j)) \mid (i,j) = (s_k^*, s_{k+1}^*),\ k \in [0\,..\,K)\,])$<br>
3 $N = [\,(n_k, i)\,] := \mathrm{Lexname}_{S^*}(P)$ // *choose lexnames* $n_k \in [0\,..\,K)$ *for* $\mathtt{S}^*$*-substrings*<br>
4 $R := [\,n_k \mid (n_k, i) \in \mathrm{Sort}(N$ by second comp.$)\,]$ // *sort names back to string order*<br>
5 **if** the lexnames in $N$ are not unique **then**<br>
6 $\mathsf{SA}_R := \mathrm{eSAIS}(R)$ // *recursion with* $|R| \le \frac{|T|}{2}$<br>
7 $\mathsf{ISA}_R := [\,r_k \mid (k, r_k) \in \mathrm{Sort}([\,(\mathsf{SA}_R[k], k) \mid k \in [0\,..\,K)\,])$ // *invert permutation*<br>
8 **else** // (Sort *sorts lexicographically unless stated otherwise.*)<br>
9 $\mathsf{ISA}_R := R$ // $\mathsf{ISA}_R$ *has been generated directly*<br>
10 $S^* := [\,(t_j, \mathtt{S}, \mathsf{ISA}_R[k], [t_{j-1} \dots t_i], j) \mid (i,j) = (s_{k-1}^*, s_k^*),\ k \in [0\,..\,K)\,]$ // $(s_{-1}^* := 0)$<br>
11 $\rho_L := 0,\quad Q_L := \mathrm{CreatePQ}(S^*$ by $(t_i, y, r, [t_{i-1} \dots t_{i-\ell}], i))$ // *use PQ to induce*<br>
12 **while** $(t_i, y, r, [t_{i-1} \dots t_{i-\ell}], i) = Q_L.\mathrm{extractMin}()$ **do** // *from next* $\mathtt{S}^*$*- or* $\mathtt{L}$*-suffix*<br>
13 **if** $y = \mathtt{L}$ **then** $A_L.\mathrm{append}((t_i, i))$ // *save* $i$ *as next* $\mathtt{L}$*-type in* $\mathsf{SA}$<br>
14 **if** $i = 0$ **then** do nothing // *No reinsertion of first suffix.*<br>
15 **else if** $t_{i-1} \ge t_i$ **then**<br>
16 $Q_L.\mathrm{insert}(t_{i-1}, \mathtt{L}, \rho_L{+}{+}, [t_{i-2} \dots t_{i-\ell}], i-1)$ // $T[i{-}1..n)$ *is* $\mathtt{L}$*-type?*<br>
17 **else**<br>
18 $L^*.\mathrm{append}(\,(t_i, \mathtt{L}, \rho_L{+}{+}, [t_{i-1} \dots t_{i-\ell}], i)\,)$ // $T[i{-}1..n)$ *is* $\mathtt{S}$*-type*<br>
19 Repeat lines 11 to 18 and construct $A_S$ from $L^*$ array with inverted PQ order,<br>
counting down by setting $\rho_S := n$ and decrementing $\rho_S$--.<br>
20 **return** $[\,i \mid (t, i) \in \mathrm{Merge}($ // *Combine* $A_L$ *and* $A_S$ *into final suffix array*<br>
$(t_i, i) \in A_L$ and $(t_j, j) \in A_S.\mathrm{reverse}()$ by first component, prefer $A_L)\,]$

and $\mathtt{S}$-subsequences bucket-wise, with $\mathtt{L}$ preceding $\mathtt{S}$ (line 20). $A_S$ has to be reversed first, because the $\mathtt{S}$-suffix order is generated largest to smallest. Note that in this formulation the alphabet $\Sigma$ is only used for comparison.

### 6.2.1 Splitting Large Tuples

After the detailed description of algorithm 6.1, we must tackle two challenges that occur in the external memory setting. While $\mathtt{S}^*$-substrings are usually very short, at least three characters long and between four and five on average, in pathological cases they can encompass nearly the whole string. Thus in lines 2 to 3 of algorithm 6.1, the tuples would grow larger than an I/O block $B$, and one would have to resort to long string sorting [AFGV97]. More importantly, in the special case of [$] being the only $\mathtt{S}^*$-suffix, the while-loop in lines 12 to 18 inserts $\frac{n(n+1)}{2}$ characters, which leads

to quadratic I/O volume. Both issues arise due to long $\mathtt{S}^*$-substrings, but we will deal with them differently, once splitting $\mathtt{S}^*$-substrings from their *beginning* and the second time from their *end*.

Long string sorting in external memory can be dealt with using lexicographic naming and doubling [AFGV97, Section 4]. However, instead of explicitly sorting long strings, we integrate the doubling procedure into the suffix sorting recursion and ultimately only need to sort short strings in line 2 of algorithm 6.1. This is done by dividing the $\mathtt{S}^*$-substrings into *split substrings* of length at most $B$, starting at their *beginning*, and lexicographically naming them along with all other substrings. Thereby, a long $\mathtt{S}^*$-substring is represented by a sequence of lexicographic names in the reduced string. The corresponding split tuples are formed in the same way as $\mathtt{S}^*$-substring tuples in $P$: they also overlap by one character, except that the overlapping character need not be $\mathtt{S}^*$-type. Thus split tuples are distinct from ordinary $\mathtt{S}^*$-substrings and the recursive super-alphabet $\Sigma' = (\Sigma \times \{\mathtt{L}, \mathtt{S}\})^*$ (each character of the reduced string corresponds to a split substring, within which each character has a letter and a type). After the recursive call, long $\mathtt{S}^*$-substrings are correctly ordered among all other $\mathtt{S}^*$-substrings due to suffix sorting, and split tuples can easily be discarded in line 10 as they do not correspond to any $\mathtt{S}^*$-suffix. The use of *d-critical* characters in SA-DS [NZC09b; NZC11] (SA-IS' sibling algorithm) is a similar approach.

The second issue arises due to repeated re-insertions of payload characters into the PQ in line 16, possibly incurring quadratic I/O volume. Our solution is to place a limit on the number of characters stored in the PQ, and fetch additional characters when needed. Since the characters in the PQ tuples are ordered in reverse, we must again split $\mathtt{S}^*$-substrings, but this time from their *end*. We call the items containing the last $D_0$ characters of an $\mathtt{S}^*$-substring the *seed tuples*, and all items containing additional (up to $D$) characters *continuation tuples*. When the currently processed PQ tuple requires additional characters, we say it *underruns*.

The challenge in external memory is to have the additional characters readily available when needed, since we cannot spend an I/O to fetch each continuation tuple. We solve this by noting that we can predict *when* a continuation tuple is required. The additional characters are needed exactly at the *boundaries between repetition buckets* (see section 6.1 for the definition of repetition buckets). To understand this, consider what happens when a tuple underruns. The point is that we need not fetch the missing characters immediately, since the earliest output position which may change due to the additional characters lies in the next *repetition* bucket. This occurs when the characters in the continuation tuple themselves induce into the current bucket. Thus we can postpone matching of continuation tuples with underrun tuples to the boundaries between repetition buckets. We have thus established time points when underrun tuples must be matched, however, this also implicitly determines *which* tuples are matched at these boundaries. We can thus presort the set of continuation tuples by repetition bucket (and text position) and have them readily available for merging with underrun tuples.

**Algorithm 6.2 :** Inducing Step with $\mathtt{S}^*$-substrings split by $D_0$ and $D$, replacing lines 10 to 18 of Algorithm 6.1

$\mathcal{D} := \{\, s^*_k - D_0 - \nu{\cdot}D \mid \nu \in \mathbb{N},\, s^*_k - D_0 - \nu{\cdot}D > s^*_{k-1},\, k \in [0\,..\,K) \,\}$ // *calculate all split positions, with* $s^*_{-1} = 0$

$S^* := \text{Sort}[\,(t_j, \mathsf{ISA}_R[k], [t_{j-1}\ldots t_i], j, \mathbb{1}_{i\in\mathcal{D}}) \mid j = s^*_k,\ i = \max(s^*_{k-1}, j - D_0),\ k \in [0\,..\,K)\,]$

$L := \text{Sort}[\,(t_j, \text{rep}(j), j, [t_{j-1}\ldots t_i], \mathbb{1}_{i\in\mathcal{D}}) \mid j \in \mathcal{D},\ i = \max(s^*_{k-1}, j - D),\ t_j$ is $\mathtt{L}$-type$\,]$

$S := \text{Sort}[\,(t_j, \text{rep}(j), j, [t_{j-1}\ldots t_i], \mathbb{1}_{i\in\mathcal{D}}) \mid j \in \mathcal{D},\ i = \max(s^*_{k-1}, j - D),\ t_j$ is $\mathtt{S}$-type$\,]$

$\rho_L := 0, \quad a := \bot, \quad r_a = 0, \quad S^* := \text{Stack}(S^*)$

$Q_L := \text{CreatePQ}(\emptyset$ by $(t_i, r, [t_{i-1}\ldots t_{i-\ell}], i, c))$

**while** $Q_L$.NotEmpty() **or** $S^*$.NotEmpty() **do**

  **while** $Q_L$.Empty() **or** $t < Q_L$.TopChar() **with** $(t, \ldots) = S^*$.Top() **do**

    $(t, r, [t_{i-1}\ldots t_{i-\ell}], i, c) = S^*$.Top(), $S^*$.Pop() // *induce from*

    $Q_L$.insert$(t_{i-1}, \rho_L{+}{+}, [t_{i-2}\ldots t_{i-\ell}], i-1, c)$ // $\mathtt{S}^*$*-suffixes*

  $a' := a, \quad a := Q_L$.TopChar(), $\quad r_a := (r_a + 1)\mathbb{1}_{a'=a}$ // *next a-repetition bucket*

  $m := \rho_L, \quad M := \emptyset$

  **while** $Q_L$.TopChar() $= a$ **and** $Q_L$.TopRank() $< m$ **do** // *induce from* $\mathtt{L}$*-suffixes*

    $(t_i, r, [t_{i-1}\ldots t_{i-\ell}], i, c) = Q_L$.extractMin()

    $A_L$.append$((t_i, i))$ // *save i as next* $\mathtt{L}$*-type*

    **if** $\ell > 0$ **then**

      **if** $t_{i-1} \geq t_i$ **then** // $T[i-1..n)$ *is* $\mathtt{L}$*-type*

        $Q_L$.insert$(t_{i-1}, \rho_L{+}{+}, [t_{i-2}\ldots t_{i-\ell}], i-1, c)$

      **else** // $T[i-1..n)$ *is* $\mathtt{S}$*-type*

        $L^*$.append$(\,(t_i, \rho_L{+}{+}, [t_{i-1}\ldots t_{i-\ell}], i, c)\,)$

    **else if** $\ell = 0$ **and** $c = 1$ **then** // *need continuation?*

      $M$.append$(i, \rho_L{+}{+})$

  **foreach** $((a, r_a, i, r), (a, r_a, i, [t_{i-1}, \ldots, t_{i-\ell}], c)) \in$
  Merge$([\,(a, r_a, i, r) \mid (i, r) \in \text{Sort}(M)\,]$ with $(a, r_a, i, [t_{i-1}, \ldots, t_{i-\ell}], c) \in L)$ **do**

    **if** $t_{i-1} \geq t_i$ **then** // $T[i-1..n)$ *is* $\mathtt{L}$*-type*

      $Q_L$.insert$(t_{i-1}, r, [t_{i-2}\ldots t_{i-\ell}], i-1, c)$

    **else** // $T[i-1..n)$ *is* $\mathtt{S}$*-type*

      $L^*$.append$(\,(a, r, [t_{i-1}\ldots t_{i-\ell}], i, c)\,)$

This procedure is the key idea of algorithm 6.2, which replaces lines 10 to 18 in algorithm 6.1 and which we describe in the following. Let $\mathcal{D}$ be the set of splitting positions, counting first $D_0$ and then $D$ characters backwards starting at each $\mathtt{S}^*$-suffix until the preceding $\mathtt{S}^*$-suffix is met ($D_0 \geq D$ indicates when to split at all, and $D \geq 1$ being the split length of continuation tuples). As before, for each $\mathtt{S}^*$-substring, a seed tuple is stored in the $S^*$ array, except that only the initial $D_0$ payload characters are copied. If an $\mathtt{S}^*$-substring consists of more than $D_0$ characters, a continuation tuple is stored in one of the two new arrays $L$ or $S$ in lines 3 to 4, depending on the type of its overlapping character. This overlapping character $t_i$ will later be used together with its *repetition count* $\mathrm{rep}(i)$ to efficiently match continuation tuples with preceding tuples at repetition bucket boundaries; $\mathrm{rep}(i)$ is easily calculated while reading the text back-to-front. Along with both seed and continuation tuples we save a flag $\mathbb{1}_{i\in\mathcal{D}}$ marking whether a continuation exists.

With these different sources of characters pre-computed, we have to break up the elegant while loop of algorithm 6.1 into three separate phases: (1) inducing from $\mathtt{S}^*$-suffixes in lines 8 to 10, (2) inducing from $\mathtt{L}$-suffixes in lines 11 to 22, and (3) finding continuation tuples for underrun PQ items in lines 23 to 27. Since we must match continuation tuples at each repetition bucket boundary, one iteration of the large while loop (lines 7 to 27) is designed to induce all items of one repetition bucket. An additional difference from algorithm 6.1 is that in line 6 the PQ is initialized as empty and $S^*$ will be processed as a stack.

More details of algorithm 6.2 are described next. The two induction sources, the $S^*$ and $L$ arrays, are alternated between, with precedence depending on their top character: $Q_L.\mathrm{TopChar}() := t_i$ and $Q_L.\mathrm{TopRank}() := r$ with $(t_i, r, \tau, i, c) = Q_L.\mathrm{Top}()$. Since $\mathtt{L}$-suffixes are smaller than $\mathtt{S}^*$-suffixes if they start with the same character, the while loop in lines 8 to 10 may only induce from $\mathtt{S}^*$-suffixes with the first character being smaller than $Q_L.\mathrm{TopChar}()$; otherwise, the while loop in lines 11 to 22 has precedence. When line 11 is reached, the loop in lines 13 to 22 extracts all suffixes from the PQ starting with $a$, after which the $S^*$ stack must be checked again. In lines 17 to 20 the extracted tuple is handled as in algorithm 6.1, however, when there is no preceding character $t_{i-1}$ in the tuple and the continuation flag $c$ is set, the tuple *underruns* and the matching continuation must be found. For each underrun tuple, the required position $i$ and its assigned rank $\rho_L$ is saved in the buffer $M$, which will be sorted and merged with the $L$ array in line 23. Matching of the continuation tuple can be postponed up to the smallest rank at which a continued tuple may be reinserted into the PQ. This earliest rank is $m = \rho_L$, as set in line 12, because any reinsertion will have $r \geq \rho_L$, and thus the while loop lines 11 to 22 extracts exactly the $r_a$-th repetition bucket of $a$. Because continuation tuples must only be matched exactly once per repetition bucket, the continuation tuples are sorted by $(t_j, \mathrm{rep}(j), j)$, whereby $L$ can be sequentially merged with $M$ if $M$ is kept sorted by the first component and $L$ is scanned as a stack.

In section 6.2.3 we compute the optimal values for $D_0$ and $D$, and analyze the resulting I/O volume.

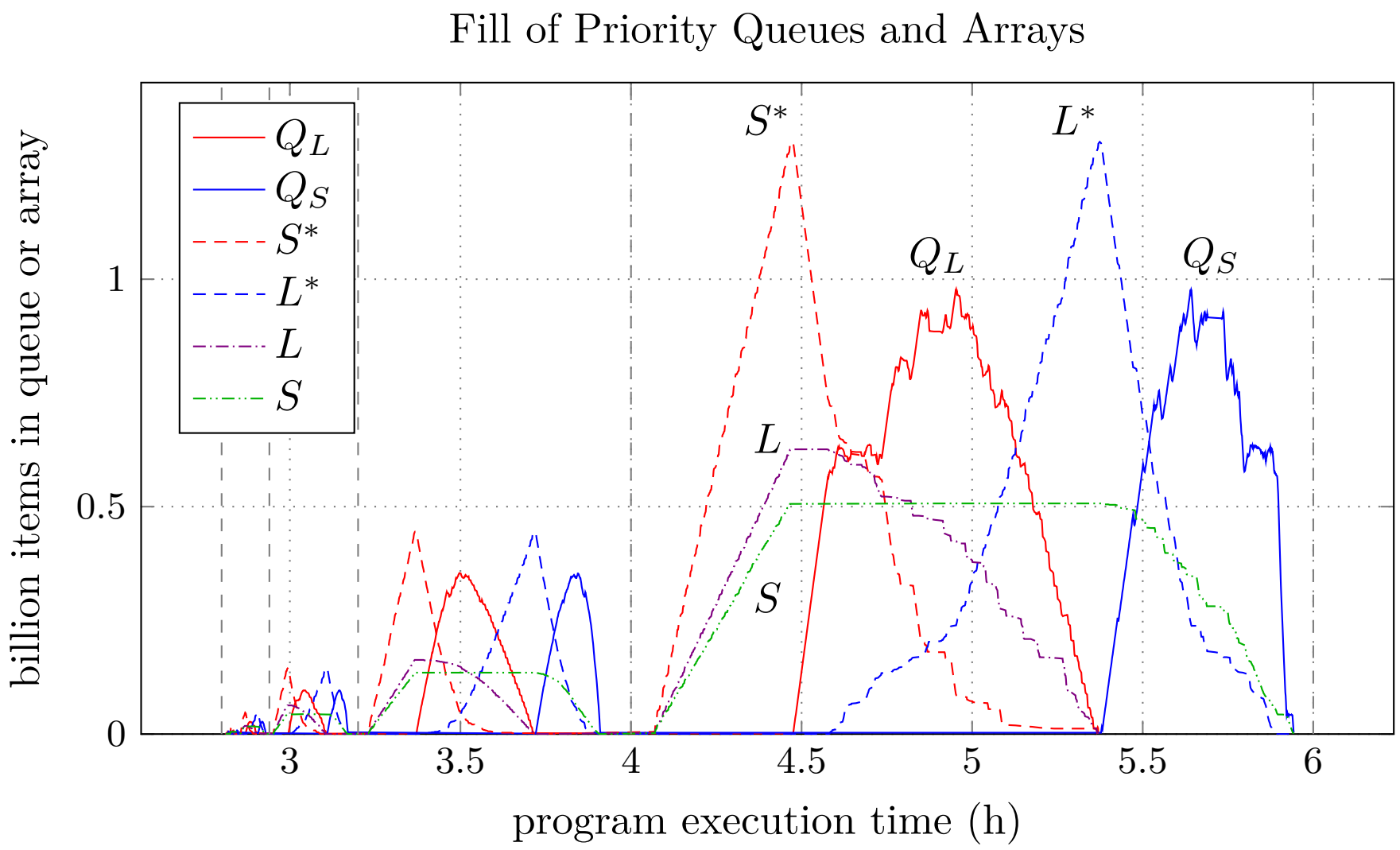

**Figure 6.6:** The graph shows the number of items in the six main data structures used in algorithm 6.1, plotted over the program execution time of one run of our eSAIS implementation on 4 GiB of Wikipedia input.

### 6.2.2 Fill of Priority Queues and Arrays

In this section we give a visual insight into the eSAIS algorithm using the example of the plot in figure 6.6. The graph shows the number of items contained in the two PQs and the most important four arrays for an example run of our eSAIS implementation on 4 GiB of Wikipedia XML (see section 6.3 for details on the implementation, input, and experimental setup).

One can see the unwinding of four recursive levels, each composed of the inducing process described in algorithm 6.1, lines 12 to 18, and augmented by algorithm 6.2. In the first phase (lines 1 to 4 of algorithm 6.2), the arrays $S^*$, $L$ and $S$ are simultaneously constructed by reading the input and the recursively calculated $\mathsf{ISA}_R$. Thereafter, the while-loop in lines 7 to 27 runs until both $Q_L$ and $S^*$ are empty. In this phase, all L-suffixes are ordered. The array $L$ contains continuation tuples for tuples that underrun when processing $Q_L$, thus $L$ is fully consumed when $Q_L$ is empty. While processing L-suffixes, the while-loop outputs the array $L^*$ in line 20. These tuples are the seeds for the symmetric while-loop, which orders all S-suffixes using $Q_S$ and consuming $S$.

Note that the peak fill of arrays $S^*$ and $L^*$ is the same. This corresponds to the number of $\mathtt{S}^*$-substrings, as each substring contains exactly one $\mathtt{S}^*$- and one $\mathtt{L}^*$-character, except for the first and last. The irregular fill of $Q_L$ and $Q_S$ is due to the particular real-world input. It shows an uneven distribution of the ASCII characters in the English text: the short plateaus in $Q_L$ and $Q_S$ are probably due to the large number of spaces.

### 6.2.3 I/O Analysis of eSAIS with Split Tuples

We now analyze the overall I/O performance of our algorithm and find the best splitting parameters $D_0$ and $D$ under practical assumptions. We will focus on calculating the I/O volume processed by Sort in lines 2 to 4 and 23 of algorithm 6.2, and by the PQs.

For simplicity, we assume that there is only one elementary data type, disregarding the fact that characters can be smaller than indices, for instance. Thus a tuple is composed of multiple elements of equal size. This assumption is also used by previous authors [DKMS05; DKMS08] and makes our results better comparable. Let $\textsc{Sort}(n)$ or $\textsc{Scan}(n)$ be the number of I/Os needed to sort or scan an array of $n$ elements. We also assume that the PQ has amortized I/O complexity $\textsc{Sort}(n)/n$ for insertion and extraction ("sorting") of one element; an assumption that is supported by preliminary experiments. In the proofs we count the number of *elements* sorted with $\textsc{Sort}(\cdot)$ and the number of elements scanned with $\textsc{Scan}(\cdot)$. This is the sorting or scanning *volume*, not the number of items sorted. Small numbers of elements can be sorted efficiently in main memory, and all scanning in our algorithms runs over large arrays. Since all processing in the algorithms occurs in large batches, the resulting number of I/O operations for sorting or scanning $k$ elements is $\textsc{Sort}(k)$ or $\textsc{Scan}(k)$ I/Os.

For our practical experiments we assume $M < n \leq \frac{M^2}{B}$, and thus can relate $\textsc{Sort}(n) = 2\,\textsc{Scan}(n)$ for *pipelined* sorting, which is equivalent to saying that $n$ elements can be sorted with one merge step if the input is read as a stream from prior stages and the output is written on-the-fly to subsequent stages. With parameters $M = 2^{30}$ (1 GiB) and $B = 2^{10}$ (1 MiB), as used in our experiments, up to $2^{50}$ (1 PiB) elements can be sorted under this assumption. This assumption will be used in the following analysis only once, when a relation between $\textsc{Sort}$ and $\textsc{Scan}$ is required to calculate a practical value for $D$ and $D_0$.

In the analysis the length of $\mathtt{S}^*$-substrings is denoted *excluding* the overlapping character, thus the sum of their lengths is the string length. The overlapping character is counted separately. For further simplicity, we assume that line 22 of algorithm 6.2 always stores continuation requests in $M$, and unmatched requests are later discarded. Thus our analysis can ignore the boolean continuation variables.

For a broader view of the algorithm, algorithm 6.1 (including algorithm 6.2) is visualized as an abstract pipelined data-flow graph [DKS08; DKMS08] in figure 6.7.

**Lemma 6.4 (Optimal Tuning Parameters $D$ and $D_0$ for eSAIS)**

*To minimize I/O cost algorithm 6.2 should use $D = 3$ and $D_0 = 8$ for splitting $\mathtt{S}^*$-strings, when $n \leq \frac{M^2}{B}$.*

*Proof.* We first focus on the number of elements sorted and scanned by the algorithm for one long $\mathtt{S}^*$-substring of length $\ell = kD$ for $k \in \mathbb{N}_1$ when splitting by period $D$ and assume for now $D_0 := D$.

For one $\mathtt{S}^*$-substring the algorithm incurs $\textsc{Sort}(D+3)$ for sorting $\mathtt{S}^*$ (line 2) and $\textsc{Sort}((\frac{\ell}{D}-1)\cdot(D+3))$ for sorting $L$ and $S$ (lines 3–4). In $Q_L$ and $Q_S$ a total of $\textsc{Sort}(\frac{\ell}{D}(\frac{1}{2}D(D+1)) + \ell \cdot 3)$ occurs due to repeated reinsertions into the PQs with decreasing lengths. The buffer $M$ (line 23) requires at most $\textsc{Sort}((\frac{\ell}{D}-1)\cdot 2)$, while reading from $L$ and $S$ is already accounted for. Additionally, at most $\textsc{Scan}((D-1)+3)$ occurs when switching from $Q_L$ to $Q_S$ via $L^*$, as at least the first $\mathtt{S}$-character was removed. Overall, this is $\textsc{Sort}(\frac{\ell}{D}(\frac{1}{2}D^2 + \frac{9}{2}D + 5) - 2) + \textsc{Scan}(D+2)$, which is minimized for $D = \sqrt{10} \approx 3.16$, when assuming $\textsc{Sort} = 2\,\textsc{Scan}$. Taking $D = 3$, we get at most $\textsc{Sort}(\frac{23}{3}\ell - 3) + \textsc{Scan}(5)$ per $\mathtt{S}^*$-substring.

Next, we determine the value of $D_0$ (as the length at when to start splitting by $D$). This offset is due to the base overhead of using continuations over just reinserting into the PQ. Given an $\mathtt{S}^*$-substring of length $\ell$, repeated reinsertions without continuations would incur $\textsc{Sort}(\frac{1}{2}\ell(\ell+1) + \ell \cdot 3)$. By putting this quadratic cost in relation to the one with splitting by $D = 3$ and finding a minimum by differentiation, we get that at length $\ell \approx 7.7$ the cost in both approaches is balanced. With this parameter the first iteration with $D_0$ characters takes the same I/O volume as each further iteration, which additionally requires insert into the merge buffer and matching with continuation tuples. Therefore, we choose to start splitting at $D_0 = 8$. □

**Theorem 6.5 (Sorting and Scanning I/O Volume of eSAIS)**

*For a string of length $n$ the I/O volume of algorithm 6.1 is bounded by* $\textsc{Sort}(17n) + \textsc{Scan}(9n)$*, when splitting with $D = 3$ and $D_0 = 8$ in algorithm 6.2.*

*Proof.* To bound the I/O volume, we consider a string that consists of $\frac{n}{\ell}$ $\mathtt{S}^*$-substrings of length $\ell$, and determine the maximum volume over all $2 \le \ell \le n$, where $\ell = 2$ is the smallest possible length of $\mathtt{S}^*$-substrings, due to exclusion of the overlapping character. Algorithm 6.1 needs $\textsc{Scan}(2n)$ to read $T$ twice (in lines 2 and 10) and $\textsc{Sort}(n + \frac{n}{\ell}\cdot 2)$ to construct $P$ in line 3, counting the overlapping character and excluding the boolean type, which can be encoded into $i$. In this $\textsc{Sort}$ the I/O volume of $\text{Lexname}_{S^*}$ is already accounted for. Creating the reduced string $R$ requires sorting of $N$, and thus $\textsc{Sort}(2 \cdot \frac{n}{\ell})$ I/Os. Then the suffix array of the reduced string $R$ with $|R| \le \frac{n}{\ell}$ is computed recursively and inverted using $\textsc{Sort}(2 \cdot \frac{n}{\ell})$, or the names are already unique. After creating $\mathsf{ISA}_R$, algorithm 6.2 is used with the parameters derived in lemma 6.4, incurring the total I/O cost calculated there for all $\frac{n}{\ell}$ $\mathtt{S}^*$-substrings. The final merging of $A_L$ and $A_S$ (line 20) needs $\textsc{Scan}(2n)$. In sum this is

$$\begin{aligned} V(n) \le{} & \textsc{Scan}(2n) + \textsc{Sort}(n + \tfrac{n}{\ell}\cdot 2) + \textsc{Sort}(\tfrac{n}{\ell}\cdot 2) + V(\tfrac{n}{\ell}) \\ & + \textsc{Sort}(\tfrac{n}{\ell}\cdot 2) + \textsc{Scan}(2n) + \tfrac{n}{\ell}\cdot \min\{\textsc{Sort}(\tfrac{23}{3}\ell - 3) + \textsc{Scan}(5)\,, \\ & \qquad \textsc{Sort}(\tfrac{1}{2}\ell(\ell+1) + \ell\cdot 3) + \textsc{Scan}(\tfrac{\ell}{2})\}\,. \end{aligned}$$

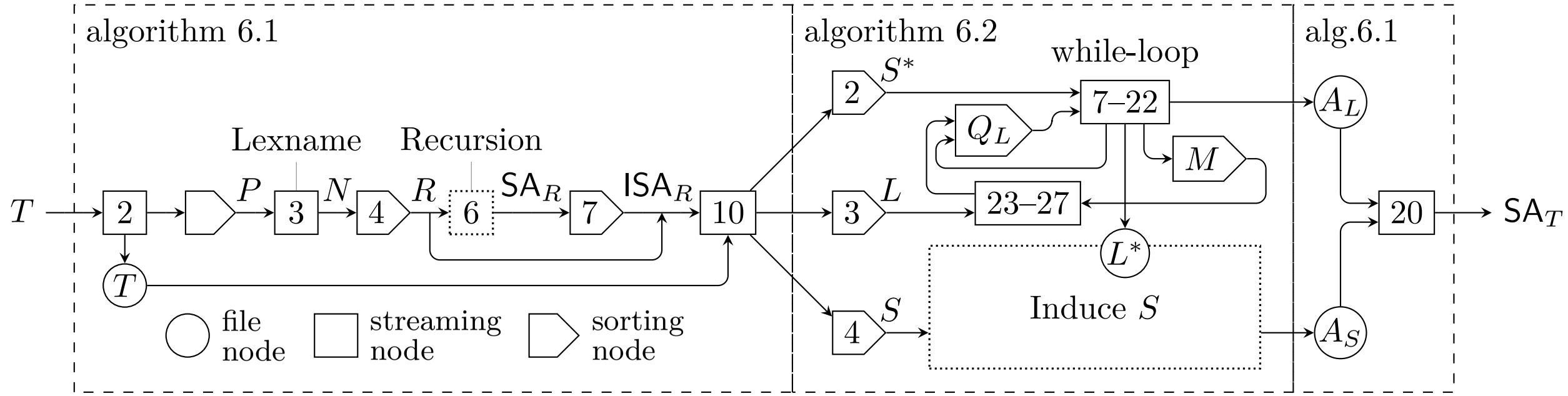

**Figure 6.7:** Data flow graph of the eSAIS algorithm. Numbers refer to the line numbers of algorithm 6.1 and algorithm 6.2, respectively. The input $T$ is read and saved to a file (2), while creating tuples. Sorting these tuples yields $P$, whose entries are lexicographically named in $N$ (3) and sorted again by string index, resulting in $R$ (4). If names are not unique in $R$, the algorithm calls itself recursively (6) to calculate $\mathsf{SA}_R$. The suffix array is inverted into $\mathsf{ISA}_R$ (7) and resulting ranks are merged with $T$ to create seed and continuation tuples (10), which are distributed into sorters (2,3,4) in algorithm 6.2. The main while-loop (7–22) reads from array $S^*$ and priority-queue $Q_L$. Depending on the calculation, the while-loop outputs final L-suffix order information into $A_L$, stores merge requests to $M$ when tuples underrun, reinserts a shortened tuple in $Q_L$, or saves L*-tuples. Merge requests are handled by matching tuples from $M$ and $L$ (23–27), and reinserting into $Q_L$. When the while-loop for inducing L-suffixes finishes, the process is repeated with seed tuples from $L^*$ and continuation tuples from $S$, yielding the final S-suffix order values in $A_S$. The output suffix array is constructed by merging $A_L$ and $A_S$ (20).

**Algorithm 6.3 :** External memory calculation of the LCP array of $\mathtt{S}^*$-suffixes.

$\mathsf{SALCP}_R :=$ eSAIS-LCP$(R)$ and $\mathsf{ISA}_R$ calculated from $\mathsf{SA}_R$

$Q_1 :=$ Sort$([\,(\mathsf{SA}_R[k]-1,k),(\mathsf{SA}_R[k]+\mathsf{LCP}_R[k]-1,k) \mid$ *// range sum over* $\mathsf{Size}_{S^*}$
$(\mathsf{SA}_R[k],\mathsf{LCP}_R[k]) \in \mathsf{SALCP}_R$ with $\mathsf{LCP}_R[k] > 0\,])$

$A_1 :=$ Sort$([\,(k,\sum_{i=0}^{s}\mathsf{Size}_{S^*}[i]) \mid$
Merge$((s,k) \in Q_1$ and $(s,\sum_{i=0}^{s}\mathsf{Size}_{S^*}[i]) \in$ PrefixSum$(\mathsf{Size}_{S^*}))\,])$

$Q_2 :=$ Sort$([\,(\mathsf{SA}_R[k-1]+\mathsf{LCP}_R[k],k),(\mathsf{SA}_R[k]+\mathsf{LCP}_R[k],k) \mid$
$((\mathsf{SA}_R[k-1],\mathsf{LCP}_R[k-1]),(\mathsf{SA}_R[k],\mathsf{LCP}_R[k])) = (\mathsf{SALCP}_R[k-1],\mathsf{SALCP}_R[k])\,])$
*// batched random access on* $\mathsf{ISA}_R$

$A_2 :=$ Sort$([\,(k,\mathsf{ISA}_R[p]) \mid$ Merge$((p,k) \in Q_2$ and $(p,\mathsf{ISA}_R[p]) \in \mathsf{ISA}_R)\,])$

$Q_3 := [\,\mathrm{RMQ}(\ell+1,r,k) \mid ((k,\ell),(k,r)) = (A_2[i],A_2[i+1])\,]$ *// RMQs on* $\mathsf{LCP}_N$

$A_3 := [(k,\textsc{rmq}_{\mathsf{LCP}_N}(\ell,r)] =$ AnswerRMQ$(Q_3,\mathsf{LCP}_N)$

$\mathsf{LCP}_{S^*} := [\,(s_2-s_1)+m \mid$
Merge$((k,s_1)=A_1[i],\ (k,s_2)=A_1[i+1]$ and $(k,m)=A_3[j])\,]$

Maximizing $V(n,\ell)$ for $2 \le \ell \le n$ by $\ell = 2$, we get $V(n,\ell) \le V(n,2) \le \textsc{Sort}(8.5n) + \textsc{Scan}(4.5n) + V(\frac{n}{2})$ and, solving the recurrence, $V(n,\ell) \le \textsc{Sort}(17n) + \textsc{Scan}(9n)$. In section 6.3 a worst-case string called Skyline is constructed with $\mathtt{S}^*$-substrings of length $\ell = 2$ on every recursive level. □

The proof of theorem 6.5 does not need to assume $n \le \frac{M^2}{B}$, since we take $D$ and $D_0$ as fixed parameters independent of $n$. These $D$ and $D_0$ give minimal I/O cost under our practical assumptions, yet theorem 6.5 holds whether of not these parameters are optimal.

eSAIS an improvement over the previously best external memory suffix sorting algorithms, DC3 and DC7, which need at most $\textsc{Sort}(30n)+\textsc{Scan}(5n)$ and $\textsc{Sort}(24.75n)+\textsc{Scan}(3.5n)$ I/O volume (see also table 5.1 on page 181). For larger difference cover sizes the I/O volume rises again: DC13 requires at most $\textsc{Sort}(30.1n) + \textsc{Scan}(2.9n)$ and the coefficients increase further [Meh04; DKMS05; Wee06; DKMS08].

## 6.2.4 Inducing the LCP Array in External Memory

In this section we describe the first practical algorithm that calculates the LCP array in external memory. The general method of integrating LCP construction into SA-IS has already been described in section 6.1.1; here, we adapt it to work in external memory.

### Calculating $\mathsf{LCP}_{S^*}$

We first show how to compute the LCP values of $\mathtt{S}^*$-suffixes in $\mathsf{LCP}_{\mathtt{S}^*}$ as described in section 6.1.2 (step $(1')$ on page 200). The pseudocode of our external memory algorithm is shown in algorithm 6.3, where $Q_1, Q_2, Q_3$ are sets of queries, and $A_1, A_2, A_3$ their respective answers. Line 1 recursively calculates $\mathsf{SA}_R$ and $\mathsf{LCP}_R$. According to lemma 6.2 (page 202), two subproblems must be solved efficiently in external memory: range sums over $\mathsf{Size}_{S^*}$ (lines 2 to 3), and range minimum queries over $\mathsf{LCP}_N$ (lines 4 to 7). The first is solved by preparing query tuples for the sum boundaries and then performing a prefix-sum scan on $\mathsf{Size}_{S^*}$. In more detail, from two consecutive entries, prepare two range sum query tuples $(\mathsf{SA}_R[k]-1, k)$, $(\mathsf{SA}_R[k]+\mathsf{LCP}_R[k]-1, k)$, sort these by first component, and perform a prefix-sum scan on $\mathsf{Size}_{S^*}$, which delivers $\sum_{k=0}^{\mathsf{SA}_R[k]-1} \mathsf{Size}_{S^*}[k]$ and $\sum_{k=0}^{\mathsf{SA}_R[k]+\mathsf{LCP}_R[k]-1} \mathsf{Size}_{S^*}[k]$, from which the range sum is easily calculated.

For the static range minimum queries in $\mathsf{LCP}_N$, we follow a common RAM-technique [FH11]: we precompute $\mathcal{O}(n)$ potential subqueries by a scan of $\mathsf{LCP}_N$, and store them on disk. The actual queries are divided into three subqueries, sorted, and merged with the precomputed queries (first by left, then by right query end). A final sort by query IDs brings the answers to subqueries back together. This technique was already sketched in the DC3 algorithm [KSB06].

### Computing LCPs by Finding Minima

The RMQs from section 6.1.3 delivering the LCP values are created in batch while inducing $\mathsf{SA}$ and answered afterwards, forming the LCP array. This is possible, as the indices $i$ and $j$ in $\textsc{rmq}_{\mathsf{LCP}_T}(i+1, j)+1$ are the relative ranks '$\rho_L$' of two consecutively extracted tuples from the PQ $Q_L$ (and symmetrically for the second phase). Notice that the first while-loop in algorithm 6.1 orders only the $\mathsf{L}$-suffixes in $\mathsf{SA}$. Likewise, the batch process computes only all LCP values of $\mathsf{L}$-type suffixes. The corresponding RMQs are calculated on a virtual array, denoted by $\mathsf{LCP}_T|_{Q_L}$, which interleaves the entries of $\mathsf{LCP}_{S^*}$ with LCP values of $\mathsf{L}$-suffixes bucket-wise, and is indexed by the relative rank $\rho_L$.

As we saw in section 6.1.3, solving the RMQs on $\mathsf{LCP}_T|_{Q_L}$ is in fact a semi-dynamic problem. To solve it, we decide not to explore which of the multi-purpose external memory data structures such as buffer trees [Arg95; Arg03] are suitable for solving this task within sorting complexity. Instead, we make the highly realistic assumption that the main memory size $M$ is large enough such that $\frac{n}{M} = \mathcal{O}(M)$; or, more precisely, $n \le C \cdot M^2$ for some small constant $C$. This means we can handle problems of size $n \le 2^{58}$ (almost one Exabyte) with only one GiB of main memory and $C = \frac{1}{4}$. This assumption is more lax than the one used in section 6.2.3.

Under this assumption we can split the array $\mathsf{LCP}_T|_{Q_L}$ into *blocks* of size $s := C \cdot M$ and keep the $\mathsf{LCP}_T|_{Q_L}$-values of the current block in main memory. Further, we can keep the *minima* of all $\mathcal{O}(n/M) = \mathcal{O}(M)$ previous blocks in RAM. We build succinct semi-dynamic RAM-based RMQ-structures over both arrays, as in section 6.1.3. Then every range minimum query can be split into three subqueries: the first and last subquery being contained in a block of size $s$, and the middle (possibly large) subquery perfectly aligning with block boundaries on both ends. The former two subqueries are answered when the block is held in RAM, while the latter subquery is answered when the last block it contains has been processed. This takes overall $\mathcal{O}(n)$ time and $\mathcal{O}(n/B)$ I/Os.

We make some additional optimizations for cases where $\mathsf{LCP}_T|_{Q_L}$-values can be induced without range minimum queries. One interesting case is related to the repetition counts: consider among all L-suffixes in a $c$-bucket ($c \in \Sigma$) the first suffixes starting with $c$, $cc$, $ccc$, etc. Their LCP values are 0,1,2, etc., which is exactly their repetition count. The current repetition count, however, is the readily-available variable '$r_a$' when extracting from the PQ, and thus the LCP can be set immediately without any RMQ. This optimization turned out to be very effective for highly repetitive texts.

Finally, we note that we have also implemented a completely in-memory version of RMQs that relies on the fact that only the right-to-left minima (looking left from the current position $i$) are candidates for the minima. Except for pathological inputs there are only $\mathcal{O}(M)$ such right-to-left minima, because the minimum at each bucket boundary is zero. Therefore they all fit in RAM and can be searched in a binary manner or using more complex heuristics (see section 6.1.3).

As already discussed in section 6.1.4, the LCP value at the L/S-seam requires special consideration. For handling the seam in external memory we reapply lemma 6.3 in a different manner: for each $c$-bucket we save the maximum repetition count in the L-subbucket during the first while-loop. Then, when inducing S-suffixes in the symmetric while-loop, the L/S-seam LCP value can be determined from the maximum repetition count in L- and S-subbucket. As suggested by lemma 6.3, the true value is the smaller of both repetition counts.

## 6.3 Experimental Evaluation

We implemented the eSAIS algorithm with integrated LCP construction in C++ using the external memory library STXXL [DKS05; DKS08]. This library provides efficient external memory sorting and a priority queue that is modeled after Sanders' design for cached memory [San99; San00]. Note that in STXXL all I/O operations bypass the operating system cache; therefore the experimental results are not influenced by system cache behavior. Our implementation and selected input files are available from `http://panthema.net/2012/esais/`.

Before describing the experiments, we highlight some details of the implementation. Most notably, STXXL does not support variable length structures, nor are we aware of a library with a PQ that does. Therefore, in the implementation the tuples in the PQ and the associated arrays are of fixed length, and superfluous I/O transfer volume occurs. Due to fixed length structures, the results from the I/O analysis for the tuning parameter $D$ does not directly apply. We found that $D = D_0 = 3$ are good splitting values in practice, which match the theoretical average `S`*-substring length. All results of the algorithms were verified using a suffix array checker [DKMS08, Section 8] and a semi-external version of Kasai's LCP algorithm [KLA+01] when possible. We designed the implementation to use an implicit sentinel instead of '`$`,' so that input containing zero bytes can be suffix sorted as well. Since our goal is to sort large inputs, the implementation can use different data types for array positions: regular 32-bit integers, a special 40-bit data type stored in five bytes, or 64-bit integers. The input data type is also a template parameter: while our real-world inputs are all composed of regular 8-bit ASCII characters, the recursive levels internally work with 32- or 40-bit data types. When sorting ASCII strings in memory, an efficient in-place radix sort [KR08] is used. Strings of larger data types are sorted in RAM using `gcc-4.4` STL's version of introsort. The initial sort of short strings into $P$ is implemented using a variable length tuple sorter.

We chose a wide variety of large inputs, both artificial and from real-world applications:

**Wikipedia** is an XML dump of the most recent version of all pages in the English Wikipedia, which was obtained from `http://dumps.wikimedia.org/`; our dump is dated `enwiki-20120601`.

**Gutenberg** is a concatenation of all ASCII text documents from Project Gutenberg by document id as available in September 2012 from `http://www.gutenberg.org/robot/harvest`. The Gutenberg data contains a version of the human genome as a substring.

**Human Genome** consists of all DNA files from the UCSC human genome assembly "hg19" downloadable from `http://genome.ucsc.edu/`. The files were normalized to upper-case and stripped of all characters but $\{\mathtt{A}, \mathtt{G}, \mathtt{C}, \mathtt{T}, \mathtt{N}\}$. Note that this input contains very long sequences of unknown `N` placeholders, which influences the LCPs.

**Pi** are the decimals of $\pi$, written as ASCII digits and starting with "3.1415."

**Skyline** is an artificial string for which SA-IS And eSAIS have maximum recursion depth. To achieve this, the string's suffixes must have type sequence `LSLS`...`LS` at each level of recursion. Such a string can be constructed for a length $n = 2^p$, $p \geq 1$, using the alphabet $\Sigma = [\, \$, \sigma_1, \dots, \sigma_p \,]$ and the grammar $\{S \to T_1\$,\ T_i \to T_{i+1}\sigma_i T_{i+1}$ for $i = 1, \dots, p-1$ and $T_p \to \sigma_p\}$. For $p = 4$ and $\Sigma = [\, \$, \mathtt{a}, \mathtt{b}, \mathtt{c}, \mathtt{d} \,]$, we get $[\, \mathtt{d}, \mathtt{c}, \mathtt{d}, \mathtt{b}, \mathtt{d}, \mathtt{c}, \mathtt{d}, \mathtt{a}, \mathtt{d}, \mathtt{c}, \mathtt{d}, \mathtt{b}, \mathtt{d}, \mathtt{c}, \mathtt{d}, \$ \,]$; for the test runs we replaced `$` with $\sigma_0$. The name "Skyline" comes from the corresponding height diagram, which looks like a metropolitan skyline, when drawn with smallest character on top as in figure 6.1.

The input Skyline is generated depending on the experiment size, all other inputs are cut to size. The inputs are available from the same URL as our implementation's source code.

Our main experimental **platform A** is a cluster computer, with one node exclusively allocated when running a test instance. The nodes have an Intel Xeon X5355 processor clocked with 2.66 GHz and 4 MiB of level 2 cache. In all tests only one core of the processor is used. Each node has 850 GiB of available disk space striped with RAID 0 across four local "Seagate Barracuda 7200.10 ST3250820AS" disks of size 250 GiB; the rest is reserved by the system. A single disk's write and read throughput ranges between 80 MiB/s on the outside and 72 MiB/s on the inside. Parallel I/O speed to the four disks ranges between 320 MiB/s and 240 MiB/s, and was measured using STXXL's `benchmark_disks` tool. We limit the main memory usage of the algorithms to 1 GiB of RAM, and use a block size of 1 MiB. The block size was optimized in preliminary experiments.

Due to the limited local disk space in the cluster computer, we chose to run some additional, larger experiments on **platform B**: an Intel Xeon X5550 processor clocked with 2.66 GHz and 8 MiB of level 2 cache. The main memory usage is limited to 4 GiB RAM, we keep the block size at 1 MiB, and up to seven local SATA disk with 1 TB of local space are available. The disks are labeled "Seagate SV35.5 ST31000525SV" and an individual disk's throughput ranged from 110 MiB/s to 90 MiB/s. All disks together reach at most 520 MiB/s and on average 450.0 MiB/s when writing 4 TiB of data.

Programs on both platforms were compiled using `g++` 4.4.6 with `-O3` and native architecture optimization.

### 6.3.1 Plain Suffix Array Construction

As noted in the introduction, the previously fastest external memory suffix sorting implementation is DC3 [DKMS05; DKMS08]. We adapted and optimized the original source code*, which was already implemented using STXXL, to our current setup and larger data types. An implementation of DC7 exists that is reported to be about 20% faster in the special case of human DNA [Wee06], but we did not include it in our experiments. We also report on some results of bwt-disk [FGM10; FGM12], even though it generates the BWT instead of the suffix array. These two suffix sorting implementations were state-of-the-art in 2013, when we first proposed eSAIS in our conference paper [BFO13].

Figure 6.8 shows the construction time and I/O volume of eSAIS, DC3, and bwt-disk on platform A using 32-bit keys. The three algorithms eSAIS (red, solid lines), DC3 (blue, dashed lines), and bwt-disk (green, dotted lines) were run on prefixes $T[0, 2^k)$ of all five inputs, with only Skyline being generated specifically for each size. In total the plots of eSAIS and DC3 took 3.2 computing days and over 16.8 TiB of I/O volume, which is why only one run was performed for each of the 90 test instances. The bwt-disk experiments were run only once.

For all real-world inputs eSAIS's construction time is about half of DC3's. The I/O volume required by eSAIS is also only about 60% of the volume of DC3. The two

*`http://algo2.iti.kit.edu/dementiev/esuffix/docu/`

(a) Construction Time

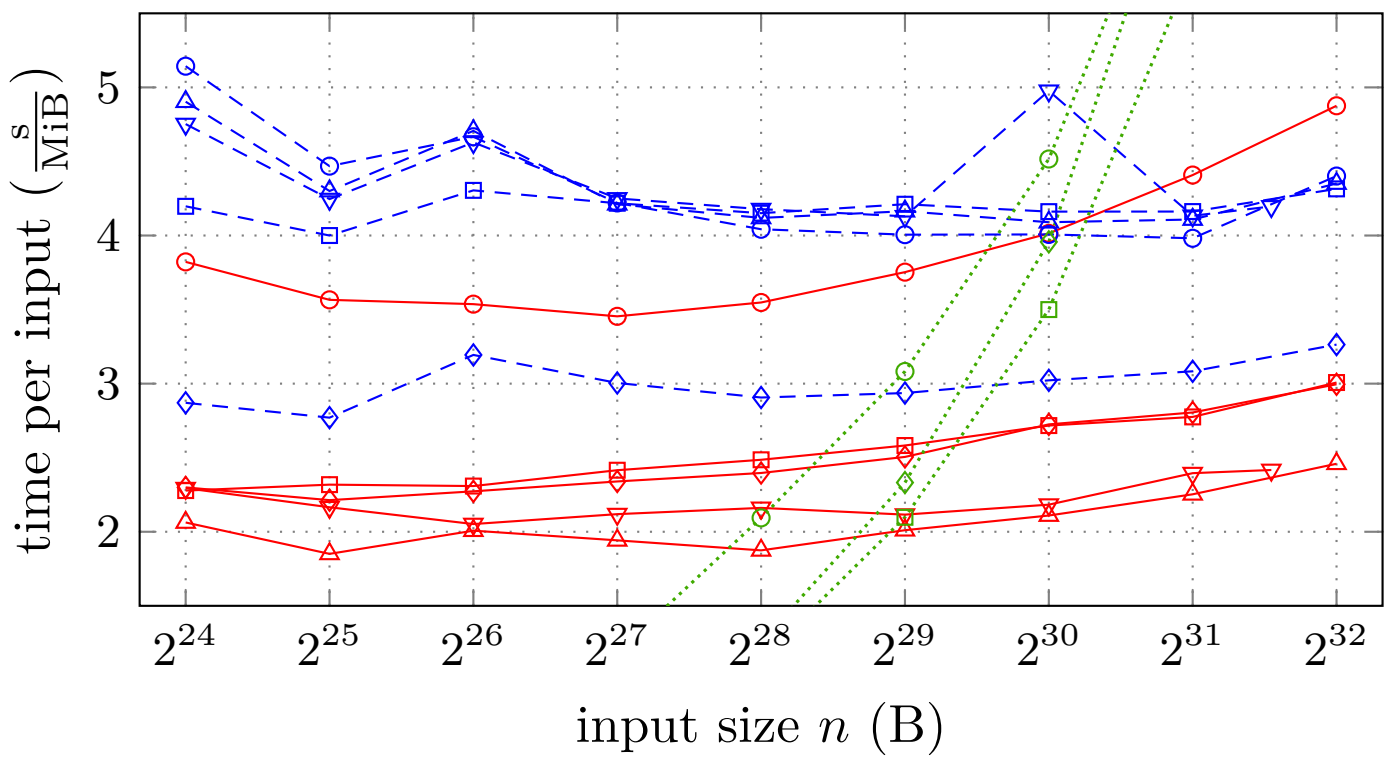

(b) I/O Volume

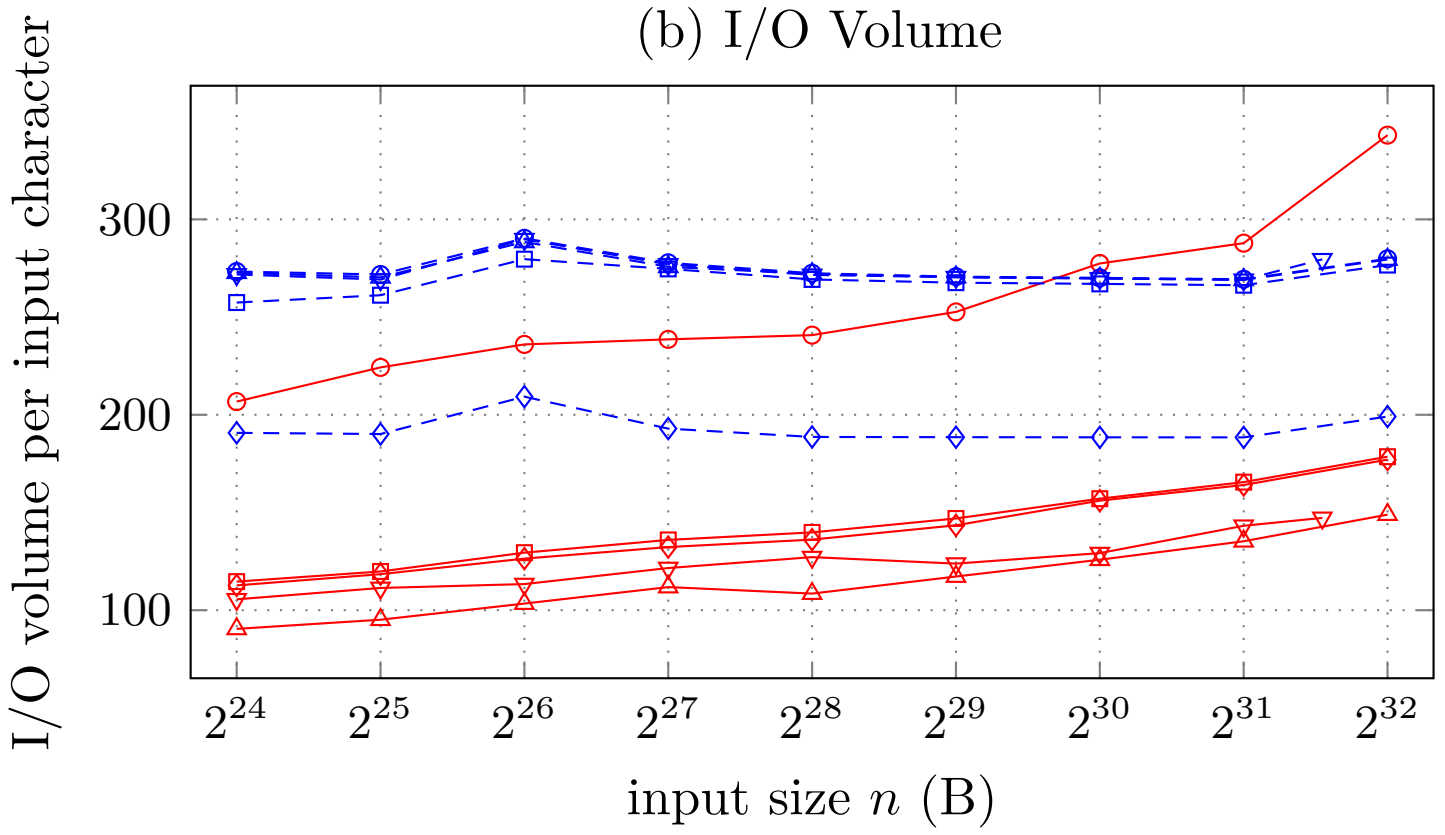

(c) Input Characteristics

| Plot | Input | $\lvert T\rvert$ | $\lvert\Sigma\rvert$ |
|---|---|---|---|
| —□— | Wikipedia XML | 79 479 MiB | 213 |
| —△— | Gutenberg Text | 22 975 MiB | 256 |
| —▽— | Human Genome | 2 992 MiB | 5 |
| —◇— | Decimals of $\pi$ | $\infty$ | 11 |
| —○— | Skyline (worst-case) | $\infty$ | $\lfloor \log_2 n \rfloor$ |

—○— eSAIS - -○- - DC3 ···○··· bwt-disk
(all inputs)

**Figure 6.8:** The two plots show (a) construction time and (b) I/O volume of eSAIS (red, solid lines), DC3 (blue, dashed lines), and bwt-disk (green, dotted lines) on experimental platform A. The table (c) shows selected characteristics of the input strings.

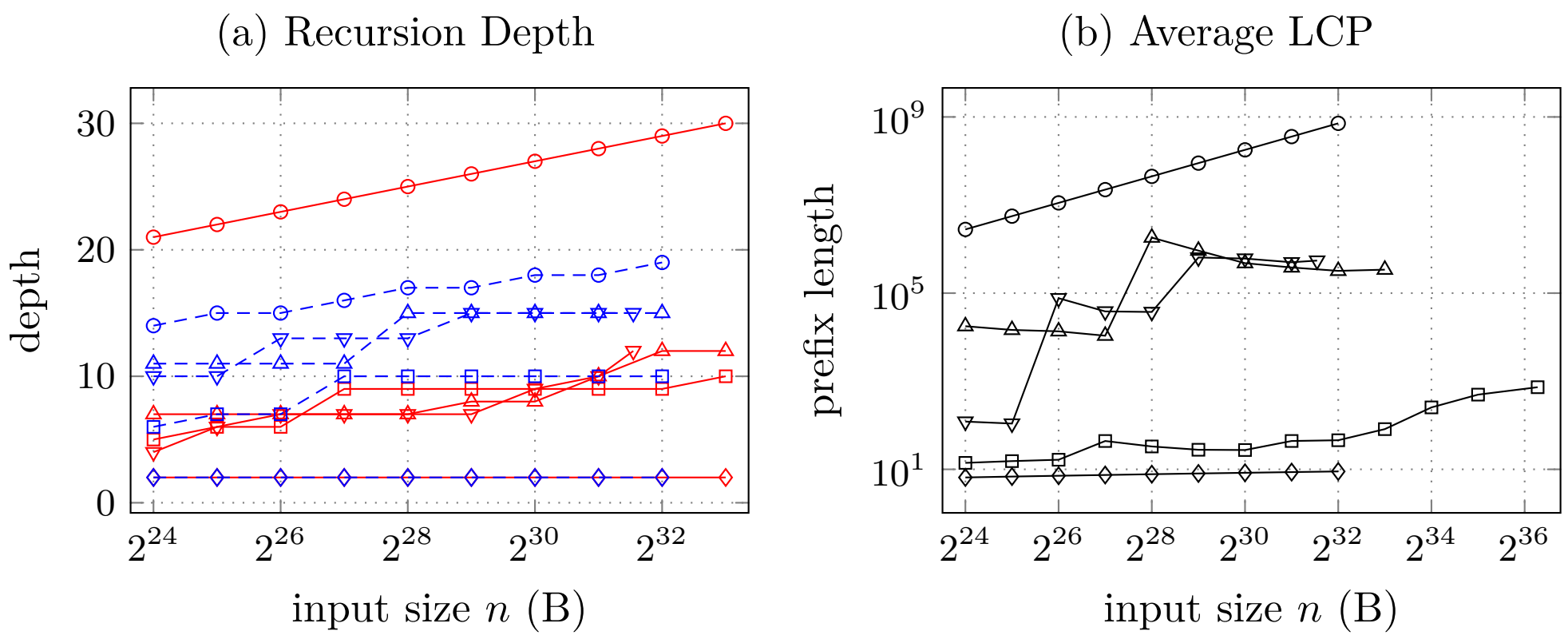

**Figure 6.9:** Panel (a) shows the maximum recursion depth reached by eSAIS and DC3 during the experiments on platform A (running with 40-bit positions). Subfigure (b) contains the average LCP of increasing $2^k$ slices of the inputs, calculated using eSAIS-LCP. The plots have the same legend as in figure 6.8.

artificial inputs exhibit the extreme results they were designed to provoke: Pi is random input with short LCPs, which is an easy case for DC3. Nevertheless, eSAIS is still faster, but not twice as fast. The results from eSAIS's worst-case Skyline show another extreme: eSAIS has highest construction time on its worst input, whereas DC3 is moderately fast because Skyline can efficiently be sorted by triples. The high I/O volume of eSAIS for Skyline is due to its maximum recursion depth, reducing the string only by $\frac{1}{2}$ and filling the PQ with $\frac{n}{2}$ items on each level (see figure 6.9 (a)). The PQ implementation requires more I/O volume than sorting, because it recursively combines short runs to keep the arity of mergers in main memory small. Even though DC3 reduces by $\frac{2}{3}$, the recursion depth is limited by $\log_3 n$ and sorting is more straightforward.

We configured bwt-disk to also use 1 GiB of main memory on platform A. Thus bwt-disk can suffix sort quite large chunks in internal memory, and behaves like its in-memory suffix sorter (divsufsort) for small input sizes. But once the input does not fit into memory, multiple chunks are merged and this merging causes the high increase in construction time seen in figure 6.8. This is probably due to bwt-disk's theoretical I/O complexity, $O(n^2/(MB))$, and quadratic CPU time, $O(n^2/M)$ [FGM12]. We could not measure the required I/O volume of bwt-disk, and the program does not output such statistics. The main feature of bwt-disk is that it needs only very little additional disk space, which is why the authors call it "lightweight".

Besides the basic eSAIS algorithm, we also implemented a variant which "discards" sequences of multiple unique names from the reduced string prior to recursion, similar to [DKMS05; DKMS08] and [PST05]. However, we discovered that this optimization has much smaller effect in eSAIS than in other suffix sorters (see figure 6.10 (a)-(d)).

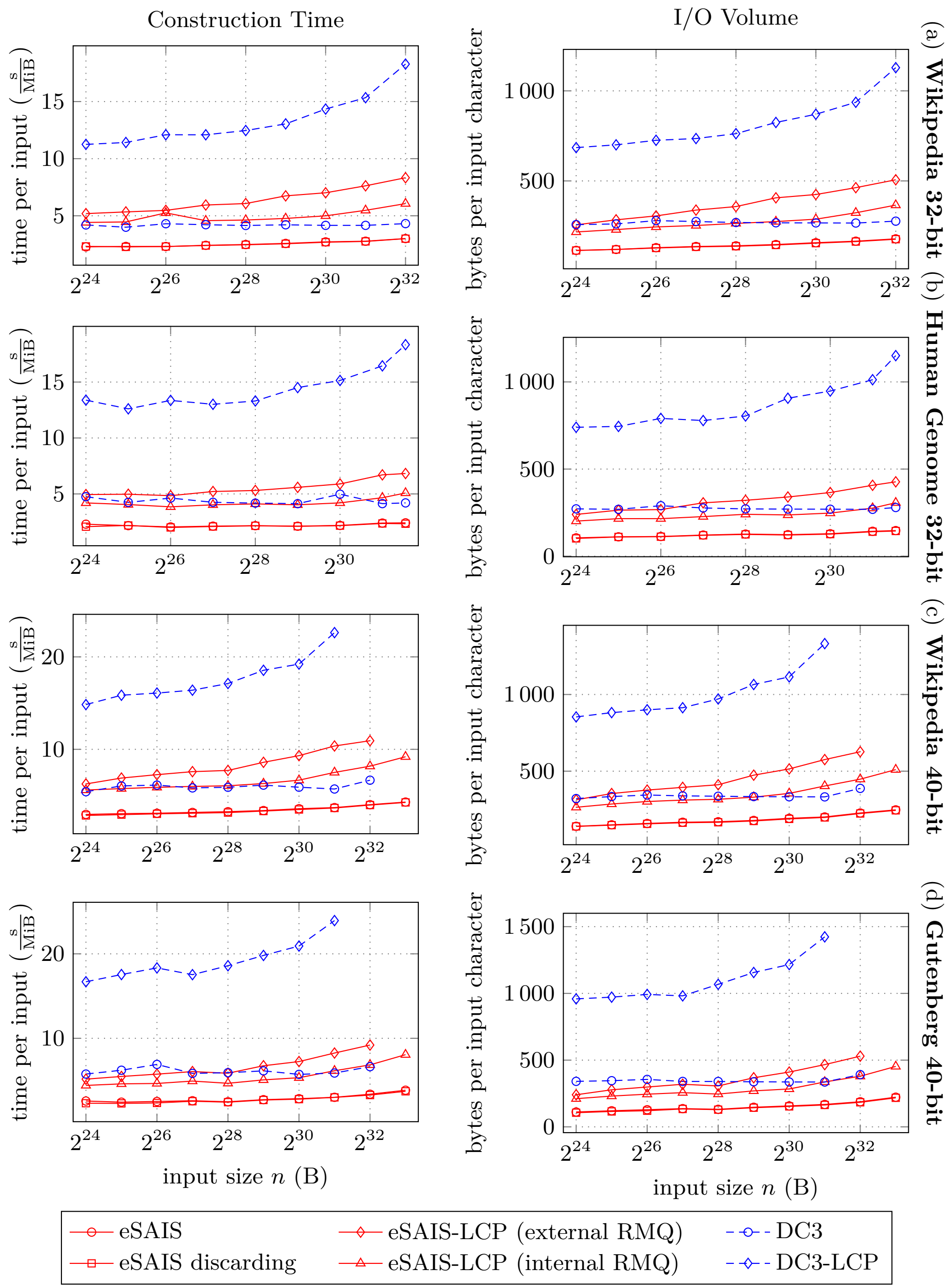

**Figure 6.10:** Subfigures (a)-(d) show construction time and I/O volume of all six implementations run on platform A for three different inputs. Subfigures (a)-(b) use 32-bit positions, while (c)-(d) runs with 40-bit.

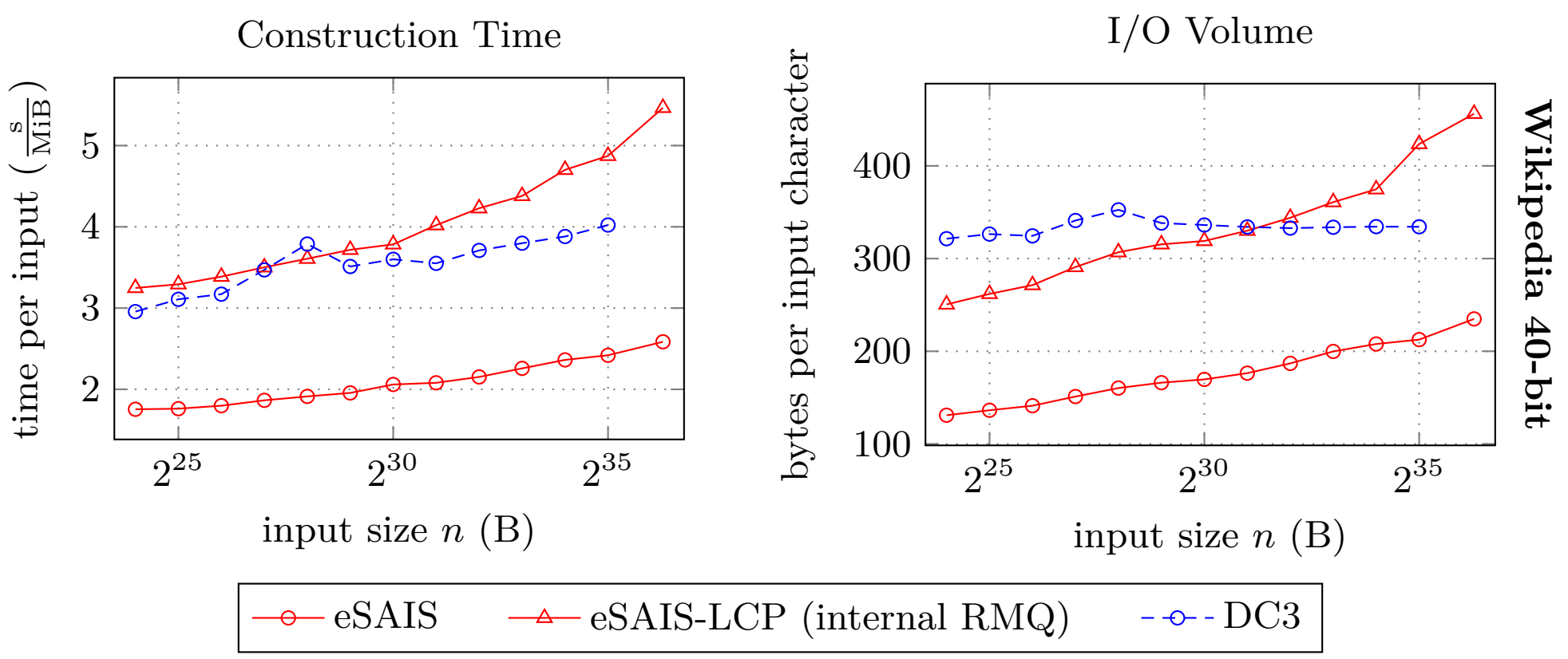

**Figure 6.11:** Measured construction time and I/O volume of three implementations is shown for the largest test instance Wikipedia run on platform B using 40-bit positions.

This is probably due to the induced sorting algorithm already adapting very efficiently to the input string's characteristics.

## 6.3.2 Suffix and LCP Array Construction

We implemented two variants of LCP construction: one solving RMQs in external memory (LCPext), and the other entirely in RAM (LCPint). The external memory solution saves RMQs to disk during the inducing process, and constructs the LCP array from these queries after the SA was completed. In contrast, the RAM solution precalculates the LCP for each induced position from an in-memory structure and saves the LCP in the PQ. Thus the LCP array is constructed at the same time as the SA (when extracting from the PQ). The size of the in-memory RMQ structure is related to the maximum LCP and the number of different inducing targets within one bucket, and grows up to 300 MiB for the Human Genome. The in-memory RMQ construction also requires the preceding character $t_{i-1}$ to be available when processing the while loop, a restriction that requires an overlap of two characters in continuation tuples and thus leads to a larger I/O volume.

Since no external memory variant of DC3 with LCP construction in STXXL is available, we extended the original implementation to also calculate the LCP array recursively [Fei13], as suggested in [KS03]. Similar to section 6.1.2, one must save an array $\mathsf{LCP}_N$ during the lexicographic naming phase. Each entry in the output $\mathsf{LCP}_T$ is composed of three parts: the number of equal characters found when merging sample and non-sample tuples, the expanded value from $\mathsf{LCP}_R$, and the result of an RMQ on $\mathsf{LCP}_N$. The second and third occur if the suffixes are ordered depending on the ranks of sub-suffixes, which is usually the case. Part one can be counted easily during merging. The second component requires processing of batched RMQs on $\mathsf{LCP}_R$ with the distinguishing ranks of sub-suffixes as boundaries. The result is multiplied by three for DC3. To determine the third summand, the previously calculated value from $\mathsf{LCP}_R$ is used for a batched random lookup on $\mathsf{SA}_R$ and $\mathsf{ISA}_R$ (if the recursive LCP was not zero) yielding the ranks of the first pair of mismatching reduced characters. The third component represents the LCP of these character triples and is computed using an RMQ on $\mathsf{LCP}_N$. These steps are similar to those needed in eSAIS-LCP (see algorithm 6.3), however, DC3-LCP generally requires two batched random lookups and two generally unpredictable RMQs per output value. In eSAIS-LCP on the other hand, the lexicographic names encompass variable length substrings, thus requiring the prefix-sum, followed by the same batched random access and an RMQ on $\mathsf{LCP}_N$. But due to the structure of the inducing process, fewer operations are required after calculating $\mathsf{LCP}_{S^*}$ and the RMQ ranges are "local" to the currently induced bucket.

**Table 6.1:** Maximum disk allocation in bytes required by the algorithms in our experiments, averaged and rounded over all our inputs.

| | eSAIS | -LCPint | -LCPext | DC3 | -LCP |
|---|---|---|---|---|---|
| 32-bit | $25n$ | $44n$ | $52n$ | $46n$ | $88n$ |
| 40-bit | $28n$ | $54n$ | $63n$ | $58n$ | $109n$ |

Figure 6.10 (a)-(d) shows the results of all six variants of the algorithms on the real-world inputs run on platform A. We observe that eSAIS-LCP internal or external are the first viable methods to calculate suffix array and LCP array in external memory; our version of DC3-LCP finishes in justifiable time only for very small instances. On all real-world inputs the construction time of eSAIS-LCP is never more than twice the time of DC3 *without* LCP construction. As expected, in-memory RMQs are consistently faster than external memory RMQs and also require fewer I/Os, even though the PQ tuples are larger.

To exhibit experiments with building large suffix arrays, we configured the algorithms to use 40-bit positions on platform A. Figure 6.10 (c)-(d) show results for the Wikipedia and Gutenberg input only up to $2^{33}$, because larger instances require more local disk space than available at the node of the cluster computer. On average over all tests instances of Wikipedia, calculation using 40-bit positions take about 33% more construction time and the expected 25% more I/O volume.

The size of suffix arrays that can be built on platform A was limited by the local disk space, we therefore determined the maximum disk allocation required. Table 6.1 shows the average maximum disk allocation measured empirically over our test inputs for 32-bit and 40-bit offset data types.

On platform B we had the necessary 4 TiB disk space required to process the full Wikipedia instance, and these results are shown in figure 6.11. The maximum size of the in-memory RMQ structure was only about 12 MiB. Sorting of the whole Wikipedia input with eSAIS took 2.4 days and 18 TiB I/O volume, and eSAIS with LCP construction (internal memory RMQs) took 5.0 days and 35 TiB I/O volume.

## 6.4 Conclusions and Future Work

We presented a better external memory suffix sorter based on induced sorting that can also construct the LCP array. Both theoretical analysis and our experiments show that it is about twice as fast and needs about half as many I/Os than the previously best external memory suffix sorter DC3/DC7. Adding LCP construction however cuts the speed again by one half. eSAIS with LCP is the first implementation of an LCP array construction algorithm fully in external memory.

Although our implementations are already very practical, we point out some optimizations that could yield an even better performance in the future. Because eSAIS is largely compute bound, a more efficient internal memory priority queue implementation, e.g. a radix heap, may improve suffix array construction time significantly. Another fact that could lead to significantly better performance is that any reinsertion into the PQ is always after the last tuple of the current repetition bucket. Thus the PQ's main-memory merge buffer could be bypassed in many cases. Performance on inputs relying heavily on sorting (like Pi and Skyline) could also be improved by sorting $\mathtt{S}^*$-substrings deeper than only three characters if they are very short.

As a whole, the potential of further speed improvements by optimization of eSAIS is higher than for DC3. We note that the final recursive stage can also output the Burrows-Wheeler transform [BW94] directly from the extracted PQ tuple, instead of the suffix array. Obviously, for real-world applications one should stop sorting in external memory when the reduced string can be suffix sorted internally. This is currently not implemented. Finally, it is possible to combine the two variants of eSAIS-LCP (internal and external RMQs) into one algorithm with a bounded in-memory RMQ structure, where unanswered RMQs are saved to external memory and solved later.

### 6.4.1 New External Memory Algorithms Succeeding Our Work

We believe our papers on eSAIS [BFO13; BFO16], which are the foundation of this chapter, rekindled interested in external memory suffix and LCP array construction. Sections 5.4 and 5.6 already reviewed the history of these topics in depth, here we give more details on the algorithms presented after eSAIS was published.

Virtually at the same time as our presentation of eSAIS, Nong, Chan, Zhang, and Guan [NCZG14] proposed *EM-SA-DS*, which is their externalization of SA-DS, the $d$-critical induced suffix sorting algorithm related to SA-IS. They do not use an external memory library such as STXXL, hence, it is unclear how well buffering and pipelining between algorithms stages is implemented. Instead of rewriting induced sorting as a priority queue problem, EM-SA-DS more closely follows the RAM algorithm and performs induction by writing into buffers for each bucket in the output suffix array. Intuitively this approach can be described as “pushing” the next suffix index into the right bucket buffer, while eSAIS “pulls” the next smaller suffix from the PQ. In eSAIS all buffer handling is offloaded into the PQ algorithm, while EM-SA-DS must perform it manually.

Nong, Chan, Hu, and Wu [NCHW15] then proposed another external memory algorithm, *DSA-IS*, which closely emulates SA-IS. They also did not use an external memory library and implemented external merge sort and read/write functions from scratch. As in EM-SA-DS, the algorithm also “pushes” suffixes into the buffers corresponding to output buckets. From the information in the paper, it is unclear to us how they manage the large number of I/O buffers needed during the induction process. Their experimental evaluation shows that “the best experimental times are achieved by DSA-IS and eSAIS. [...] Although they have similar speed, eSAIS uses around 20% more disk space than DSA-IS” [NCHW15].

Two more adaptations of SA-IS to external memory were presented by Liu, Nong, Chan, and Wu [LNCW15]: *SAIS-PQ* and *SAIS-PQ+*. Both attempt to simplify the ideas behind eSAIS and use STXXL [DKS05; DKS08] to implement the core induction algorithm in 800 and 1 600 lines of C++ code. While eSAIS sorts $D_0$ and $D$ characters in a tuple, SAIS-PQ stores only one and searches for the preceding character in a “preceding cache item” array prepared for this purpose. One can see this as setting $D_0 = D = 1$ in eSAIS, instead of our default of 3. In their experiments, they show that SAIS-PQ has an I/O volume about twice as that of eSAIS and that SAIS-PQ and SAIS-PQ+ are half as fast as eSAIS, except for two input instances were eSAIS becomes CPU bound. However, SAIS-PQ and SAIS-PQ+ are more-space efficient and need only $23n$ and $15n$ peak disk usage, respectively.

Kärkkäinen and Kempa [KK14a; KK17a] then proposed *SAscan* and later parallelized it together with Puglisi [KKP15b] resulting in *pSAscan*. These external memory algorithms are not based on induced sorting. Instead, they build on ideas from bwt-disk and even earlier papers [CF02; GBYS92], which generate the suffix array of parts of the input in internal memory and then merge all parts externally. The SAscan algorithms suffer from a large worst-case I/O complexity because merging suffix arrays can require comparing long common prefixes of suffixes spanning almost the entire string. The authors can show a worst-case bound of $\mathcal{O}(\text{scan}(n)(1 + \frac{n}{M \log_\sigma n}))$, but their experiments document that SAscan is faster than eSAIS by a factor of three on real-world data. While it is obvious that SAscan can be faster for small inputs, asymptotically eSAIS has better worst-case bounds. Kärkkäinen and Kempa [KK17a] then calculate the break-even point of eSAIS by extrapolation from their experiment results and surmise that SAscan “is faster than eSAIS when the text size is less than about five times the available RAM, at which point the disk space usage of eSAIS is already well over 125 times the available RAM" ([KK17a]). Another highlight of these scan-based algorithms is their low peak disk space usage: pSAscan requires only $7.5n$ bytes in external memory. The authors demonstrate the scalability of their algorithms by constructing the suffix array of a 1 TiB text in a little over eight days.

Kärkkäinen, Kempa, Puglisi, and Zhukova [KKPZ17] presented with *fSAIS* the most recent advancement in external suffix array construction. As its name suggests, fSAIS is a faster, greatly improved version of eSAIS. In their experiments the new algorithm is twice as fast as eSAIS, and, just as significant, uses only a third of its peak disk space ($8.1n$ for the Skyline input). fSAIS brings induced sorting's performance back in line with pSAscan, but gives better asymptotic guarantees. While the main induced sorting idea remains the same, instead of storing preceding character in large tuples in a PQ (like eSAIS), fSAIS performs block-wise *preinducing*. The result of this preprocessing phase is an array of preceding characters in exactly the order in which they will be needed during induced sorting (using the PQ as in eSAIS). The preinducing phase obviously takes time to compute, but enables use of smaller fixed length tuples in the PQ. Furthermore, they implemented our idea for using a monotone stable radix heap [BCFM99; BCFM00], instead of a general purpose external memory PQ. All these improvements make fSAIS the fastest external memory suffix array construction algorithm for inputs that are about 75 times larger than the available RAM. For smaller inputs, pSAscan is faster (fSAIS dominates pSAscan for larger inputs, despite pSAscan being parallelized). fSAIS has not yet been extended to also construct the LCP array in external memory, however, today there are many stand-alone fully external LCP array construction algorithms available (see section 5.6).

Table 5.1 on page 181 in our history section gives an overview of the running time and space requirements of these external memory algorithms.

# III

# Distributed Suffix Sorting with Thrill

*To scale beyond a single multi-core or external memory machine one needs to not only cope with parallelization but also communication and synchronization. In chapter 7 we present* Thrill*, our new C++ framework for distributed computation. Thrill provides a convenient modern C++ template meta-programming interface composed of scalable primitives such as* Map, ReduceByKey, Sort, Zip, *and* Window, *which operate on virtual distributed immutable arrays called DIAs. One programs in Thrill by parameterizing and composing the primitives into large complex distributed algorithms. In our experimental section 7.4, Thrill is compared against the popular frameworks Apache Spark and Apache Flink and we show that it outperforms them on five benchmark kernels.*
*In chapter 8 Thrill is then applied to distributed suffix sorting. We focus on two classes of algorithms: prefix doubling and difference cover algorithms. For both we discuss in detail how they can be implemented using only the few primitive operations in Thrill and then run them on up to 32 hosts with fast external NVMe SSDs in the AWS Elastic Compute Cloud.*

Chapter 7

# 7

# Thrill: An Algorithmic Distributed Big Data Batch Processing Framework in C++

*We present the design and a first performance evaluation of Thrill – a prototype of a general purpose big data processing framework with a convenient data-flow style programming interface. Thrill is somewhat similar to Apache Spark and Apache Flink with at least two main differences. First, Thrill is based on C++ which enables performance advantages due to direct native code compilation, a more cache-friendly memory layout, and explicit memory management. In particular, Thrill uses template meta-programming to compile chains of subsequent local operations into a single binary routine without intermediate buffering and with minimal indirections. Second, Thrill uses arrays rather than multisets as its primary data structure which enables additional operations like sorting, prefix sums, window scans, or combining corresponding fields of several arrays (zipping).*

*We compare Thrill with Apache Spark and Apache Flink using five kernels from the HiBench suite. Thrill is consistently faster and often several times faster than the other frameworks. At the same time, the source codes have a similar level of simplicity and abstraction.*

In this chapter we present Thrill, our new open-source C++ framework for algorithmic distributed batch data processing. The need for parallel and distributed algorithms cannot be ignored anymore, since individual processor cores' clock speeds have stagnated in recent years. At the same time, we have experienced an explosion in data volume so that scalable distributed data analysis has become a bottleneck in an ever-increasing range of applications. With Thrill we want to make a step at bridging the gap between two traditional scenarios of "big data" processing.

On the one hand, in academia and high-performance computing (HPC), distributed algorithms are often handcrafted in C/C++ and use MPI for explicit communication. This can achieve high efficiency at the price of difficult implementation and costly developer time. On the other hand, global players in the software industry created their own ecosystem to cope with their data analysis needs. Google popularized the MapReduce [DG08] model in 2004 and described their in-house implementation. Apache Hadoop and more recently Apache Spark [ZCF+10] and Apache Flink [ABE+14] have gained attention as open-source Scala/Java-based solutions for heavy duty data processing. These frameworks provide a simple programming interface and promise *automatic* work parallelization and scheduling, *automatic* data distribution, and *automatic* fault tolerance. While most benchmarks highlight the scalability of these frameworks, the bottom line efficiency has been shown to be lacking [MIM15], surprisingly with the CPU often being the bottleneck [ORR+15].

Thrill's approach to bridging the gap between HPC and data science frameworks is a library of *scalable algorithmic primitives* such as *Map*, *ReduceByKey*, *Sort*, and *Window*, which can be combined efficiently to construct large complex algorithms and applications using pipelined data-flow style programming. Thrill is written in modern C++14 from the ground up, has minimal external dependencies, and compiles cross-platform on Linux, Mac OS, and Windows. By using C++, Thrill is able to exploit compile-time optimization, template meta-programming, and explicit memory management. Thrill enables efficient processing of fixed-length items like single characters or fixed-dimensional vectors without object overhead due to the zero overhead abstractions of C++. It treats data types of operations as opaque and utilizes template programming to instantiate operations with user-defined functions (UDFs). For example, the comparison function of the sorting operation is compiled into the actual internal sorting and merging algorithms (similar to `std::sort`). At the same time, Thrill makes no attempts to optimize the execution order of operations, as this would require introspection into the data and how UDFs manipulate it.

Thrill programs run in a collective bulk-synchronous manner similar to most MPI programs. Thrill primarily focuses on fast in-memory computation, but transparently uses external memory when needed. The functional programming style used by Thrill enables easy parallelization, which also works remarkably well for shared-memory parallelism. Hence, due to the restriction to scalable primitives, Thrill programs run on a wide range of homogeneous parallel systems.

By using C++, Thrill aims for high performance distributed algorithms. Java virtual machine (JVM)-based frameworks are often slow due to the overhead of the interpreted bytecode, even though just-in-time (JIT) compilation has leveled the field somewhat. Nevertheless, due to object indirections and garbage collection, Java/Scala must remain less cache-efficient. While efficient CPU usage should be a matter of course, especially when processing massive amounts of data, the ultimate bottleneck for scalable distributed application is the (bisection) bandwidth of the network. But by using more tuned implementations, more CPU time is left for compression, deduplication [SSM13], and other algorithms to reduce communication. Nevertheless, in smaller networks the CPU is often the bottleneck [ORR+15], and for most applications a small cluster is sufficient.

A consequence of using C++ is that memory management has to be done explicitly. While this is desirable for more predictable and higher performance than garbage collected memory, it does make programming more difficult. However, with modern

C++11 this has been considerably alleviated, and Thrill uses reference counting extensively outside of inner loops.

While scalable algorithms promise eventually higher performance with more hardware, the performance hit going from parallel shared memory to a distributed scenario is large. This is due to the communication latency and bandwidth bottlenecks. This network overhead and the additional management overheads of big data frameworks often make speedups attainable only with unjustifiable hardware costs [MIM15]. Thrill cannot claim zero overhead, as network costs are unavoidable. But by overlapping computation and communication, and by employing binary optimized machine code, we keep the overhead small.

Development of Thrill started in winter 2014 with a one year practical student lab course with seven master-level students — Robert Hangu, Emanuel Jöbstl, Sebastian Lamm, Huyen Chau Nguyen, Alexander Noe, Matthias Stumpp, and Tobias Sturm — three PhD students — Michael Axtmann, Sebastian Schlag, and the author of this dissertation — and Peter Sanders who initiated and supervised the project. In the lab course the design and base implementation of the current prototype were developed. The author of this dissertation supervised and managed the lab course, including contributing most of the lower layers of Thrill, guiding the design decisions that were made in the group, and setting up code quality control measures such that the master students would produce stable code. After the lab course ended, we continued further development with some volunteer support and finally published a paper in a high-ranking big data conference [BAJ+16a].

Thrill is open-source under the BSD 2-clause license and available as a community project on GitHub: `http://github.com/thrill/thrill`. It currently has more than 61 K lines of C++ code and approximately a dozen developers have contributed. The author of this dissertation is the primary developer and we roughly estimate that 70–80 % of the code base were directly contributed by us.

This chapter is based on our joint conference paper [BAJ+16a] and technical report [BAJ+16b], which the author of this dissertation wrote almost entirely. Parts of these publications were copied verbatim for this chapter and extended with a much more detailed description of DIA operations in section 7.2.3, more details in section 7.3 on the underlying data, network, and I/O layers, and of the reduce operations in section 7.3.3, and a complete report of the experiments in section 7.4. Thrill itself is the collaborative work of many authors, as outlined in the history paragraph above.

**Overview.** The following section discusses related work with an emphasis on Spark and Flink. Section 7.2 discusses the design of Thrill, in particular its API and the rationale behind the chosen concept. We present a complete WordCount example in section 7.2.2, followed by an overview of the current portfolio of operations. Details of their implementation and of pipelining are discussed in section 7.3. In section 7.4, results of an experimental comparison of Thrill, Spark, and Flink based on six micro benchmarks including PageRank and KMeans are shown. Section 7.5 concludes with an outlook on future work.

**Our Contributions.** Thrill demonstrates that with the advent of C++11 lambda expressions, it has become feasible to use C++ for big data processing using an abstract and convenient API comparable to currently popular frameworks like Spark or Flink. This not only harvests the usual performance advantages of C++, but allows us moreover to transparently compile sequences of local operations into a single binary code via sophisticated template meta-programming. By using arrays as the primary data type, we enable additional basic operations that have to be emulated by more complicated and more costly operations in traditional multiset-based systems. Our experimental evaluation demonstrates that even the current prototypical implementation already offers a considerable performance advantage over Spark and Flink.

## 7.1 Related Work

Due to the importance and hype of the "big data" topic, a myriad of distributed data processing frameworks have been proposed in recent years [CZ14]. These cover many different aspects of this challenge like data warehousing and batch processing, stream aggregation [TTS+14], interactive queries [MGL+10], and specialized graph [MAB+10; LBG+12] and machine learning frameworks [ABC+16].

**MapReduce/Hadoop.** In 2004, Google established the MapReduce [DG08] paradigm as an easy-to-use interface for scalable data analysis. Their paper spawned a whole research area on how to express distributed algorithms using just *map* and *reduce* in as few rounds as possible. Soon, Apache Hadoop was created as an open-source MapReduce framework written in Java for commodity hardware clusters. Most notable from this collection of programs was the Hadoop distributed file system (HDFS) [SKRC10], which is key for fault tolerant data management for MapReduce. Subsequently, a large body of academic work was done optimizing various aspects of Hadoop like scheduling and data shuffling [LLC+12].

MapReduce and Hadoop are very successful due to their simple programming interface, which at the same time is a severe limitation. For example, iterative computations are reported to be very slow due to the high number of MapReduce rounds, each of which may need a complete data exchange and round-trip to disks. More recent frameworks such as Apache Spark and Apache Flink offer a more general interface to increase usability and performance.

**Apache Spark** operates on an abstraction called *resilient distributed datasets* (RDDs) [ZCF+10]. This abstraction provides the user with an easy-to-use interface which consists of a number of deterministic coarse-grained operations. Each operation can be classified either as *transformation* or *action.* A transformation is a lazy operation that defines a new RDD given another one, e.g. *map* or *join.* An action returns computed results to the user program, e.g. *count*, *collect*, or reads/writes data from/to external storage. When an action triggers computation, Spark examines the sequence of previously called transformations and identifies so-called execution *stages*. Spark runs in a master-worker architecture. While the driver program runs on the master, the actual computation occurs on the workers with a *block-based* work-partitioning and scheduling system. Spark can maintain already computed RDDs in main memory to be reusable by future operations, in order to speed-up iterative computations [ZCD+12].

In more recent versions, Spark added two more APIs: DataFrames [AXL+15] and Datasets. Both offer domain specific languages for higher level declarative programming similar to SQL, which allows Spark to optimize the query execution plan. Even further, it enables Spark to generate optimized query bytecode online, aside of the original Scala/Java program. The optimized bytecode can use more efficient direct access methods to the data, which no longer needs to be stored as JVM objects, and hence garbage collection can be avoided. The DataFrame engine is built on top of the original RDD processing interface.

**Apache Flink** originated from the Stratosphere research project [ABE+14] and is progressing from an academic project to industry. While Flink shares many ideas with Spark such as the master-worker model, lazy operations, and iterative computations, it tightly integrates concepts known from parallel database systems. Flink's core interface is a domain-specific declarative language. Furthermore, Flink's focus has turned to streaming rather than batch processing.

In Flink, the optimizer takes a user program and produces a graph of logical operators. The framework then performs rule- and cost-based optimizations, such as reordering of operations, pipelining of local operations, selection of algorithms, and evaluation of different data exchange patterns to find an execution plan Flink believes is best for a given user program and cluster configuration. Flink is based on a pipelined execution engine comparable to parallel database systems, which is extended to integrate streaming operations with rich windowing semantics. Iterative computations are sped up by performing delta-iterations on changing data only, as well as placing computation on the same worker across iterations. Fault tolerance is achieved by continuously taking snapshots of the distributed data streams and operator states [CFE+15], a concept inspired by Chandy-Lamport snapshots [CL85]. Flink also has an own memory management system separate from the JVM's garbage collector for higher performance and better control of memory peaks.

**Comparing Spark and Flink.** The interfaces and designs of Spark and Flink differ in some very important ways [MCAPH16]. Flink's optimizer requires introspection into the components of data objects and how the UDFs operate on them. This requires many Scala/Java annotations to the UDFs and incurs an indirection for access to the values of components. In contrast to Spark's RDD interface, where users can make use of *host language control-flow*, Flink provides custom iteration operations. Hence, Flink programs are in this respect more similar to declarative SQL statements than to an imperative language. The newer DataFrame and Dataset interfaces introduce similar concepts to Spark, but extend them further with a custom code generation engine. At its core, Spark is an in-memory batch engine that executes streaming jobs as a series of mini-batches. In contrast, Flink is based on a pipelined execution engine used in database systems, allowing Flink to process streaming operations in a pipelined way with lower latency than in the micro-batch model. In addition, Flink supports external memory algorithms whereas Spark is mainly an in-memory system with spilling to external memory.

Overall the JVM is currently the dominant platform for open-source big data frameworks. This is understandable from the point of view of programmer productivity but surprising when considering that C++ is the predominant language for performance critical systems – and big data processing is inherently performance critical. Spark [ADD+15] (with Project Tungsten) and Flink (with MemorySegments) therefore put great efforts into overcoming performance penalties of the JVM, for example by using explicit *Unsafe* memory operations and generating optimized bytecode to avoid object overhead and garbage collection. With Thrill we present a C++ framework as an alternative that does not incur these overheads in the first place.

**Other Frameworks and Approaches.** All three frameworks, Thrill, Apache Spark, and Apache Flink are similar in the sense that they provide the user with a data-flow processing methodology composed of virtual distributed datasets and methods to lazily manipulate them.

However, this is by far not the only approach to programming compute clusters. Maybe most well-known is the message passing interface (MPI) standard [MPI95; MPI97], and its popular implementations [GLDS96; GFB+04]. This interface provides little more than explicit communication functions and collective operations, but these are implemented well with full hardware acceleration on the largest supercomputers. A lot of scientific software is written using MPI to scale on these machines.

A newer approach are partitioned global address space (PGAS) languages [YBC+07] such as Unified Parallel C (UPC) and Titanium. PGAS offers a programming abstraction similar to shared-memory systems, but with the underlying memory distributed onto a cluster of machines. This abstraction is implemented by augmenting known languages (like C, Java, and Fortran) and intercepting operations on variables to execute them via a communication layer called GASNet. Operations on variables result in *one-sided* communication operations: no implicit synchronization is performed. This simple protocol can naturally be implemented using remote direct memory address (RDMA) hardware, which bypasses the CPU and reads/writes directly via the RAM bus. This avoids much of the overhead of explicit communication, such as with MPI.

## 7.2 Design of Thrill

Thrill programs are written in C++ and compile into binary programs. The execution model of this binary code is similar to MPI programs: one identical program is run collectively on $h$ machines (see figure 7.1). Our prototype of Thrill currently expects all machines to have nearly identical hardware, since it balances work and data equally between the machines. The binary program is started simultaneously on all machines, and connects to the others via a network protocol. Thrill currently supports TCP sockets and MPI as network backends. The startup procedures depend on the specific backend and cluster environment.

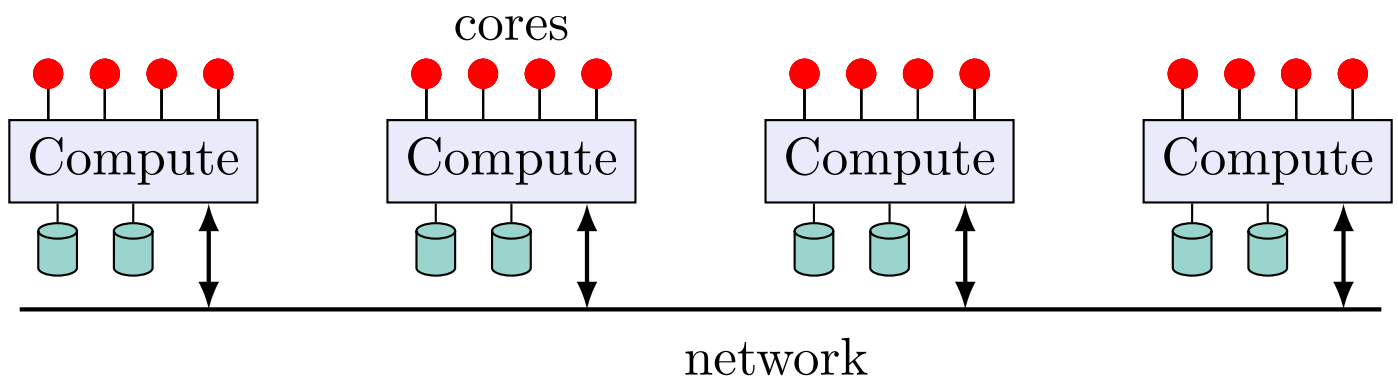

**Figure 7.1:** Thrill binaries run collectively on $h$ compute hosts, with $c$ cores each, use all local disks and communicate via a TCP or MPI network.

Each machine is called a *host*, and each work thread on a host is called a *worker*. Currently, our prototype requires all hosts to have the same number of cores $c$, hence, in total there are $p = h \cdot c$ worker threads. Additionally, each host has one thread for network/data handling and one for asynchronous disk I/O. Each of the $h$ hosts have $h - 1$ reliable network connections to the other hosts, and the hosts and workers are enumerated $0 \dots h - 1$ and $0 \dots p - 1$. Thrill does not have a designated master or driver host, as all communication and computation is done collectively.

Thrill currently provides no fault tolerance. While our data-flow API permits smooth integration of fault tolerance using asynchronous checkpoints [CFE+15; CL85], the execution model of exactly $h$ machines may have to be changed.

### 7.2.1 Distributed Immutable Arrays

The central concept in Thrill's high-level data-flow API is the *distributed immutable array* (DIA). A DIA is an array of items which is distributed over the cluster in some way. No direct array access is permitted. Instead, the programmer can apply so-called *DIA operations* to the array as a whole. These operations are a set of *scalable primitives*, listed in table 7.1, which can be composed into complex distributed algorithms. DIA operations can create DIAs by reading files, transform existing DIAs by applying user functions, or calculate scalar values collectively, used to determine the further program control flow. In a Thrill program, these operations are used to lazily construct a DIA *data-flow* graph in C++ (see figure 7.2). The data-flow graph is only executed when an *action* operation is encountered. How DIA items are actually stored and in what way the operations are executed on the distributed system remains transparent to the user.

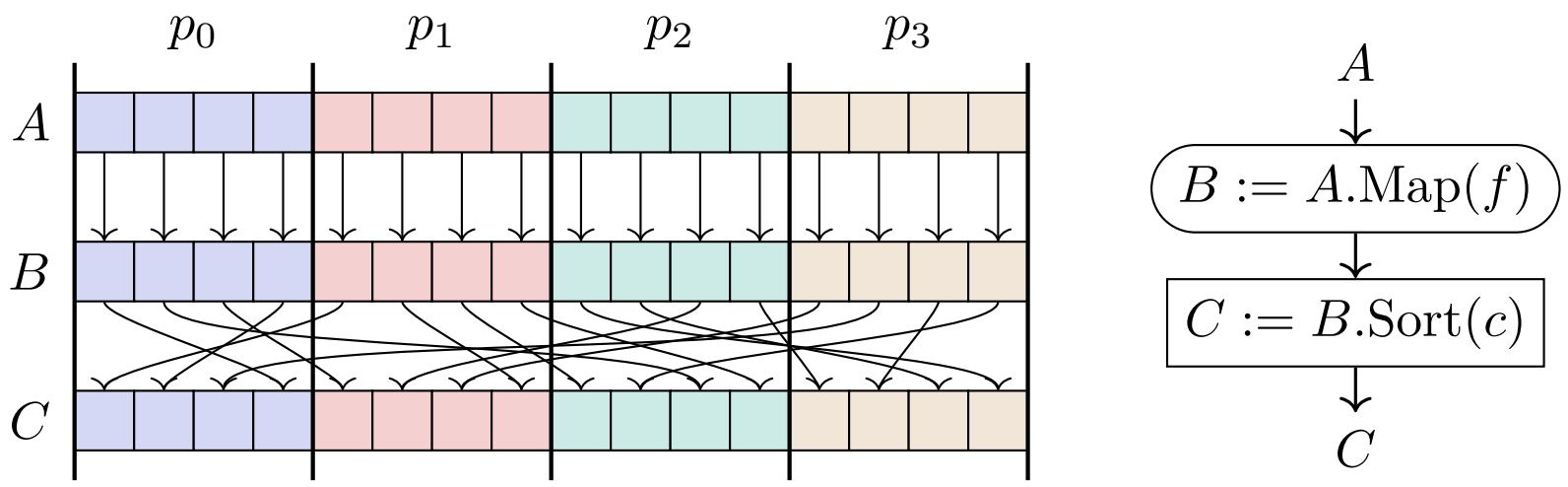

**Figure 7.2:** Mental model of the distribution of a DIA among processors (left) and a data-flow graph (right).

In the current prototype of Thrill, the array is usually distributed evenly between the $p$ workers in order. DIAs can contain any C++ data type, provided serialization methods are available (more in section 7.3.2). Thrill contains built-in serialization methods for all primitive types and most STL types; only custom non-trivial classes require additional methods. Each DIA operation in table 7.1 is implemented as a C++ template class, which can be instantiated with appropriate UDFs. Instead of diving directly into the details of each DIA operation, we present WordCount in the next section as a introductory example.

### 7.2.2 Example: WordCount

We now present a complete source code example of the popular WordCount benchmark in algorithm 7.1 to demonstrate how easy it is to program in Thrill. The program counts the number of occurrences of each unique word in a text. WordCount in Thrill, including file I/O, consists of only five DIA operations.

*ReadLines* (line 4) and *WriteLines* (line 21) are used to read the text and write the result from/to the file system. Thrill currently uses standard POSIX filesystem methods to read and write to disk, and it requires a distributed parallel file system such as NFS, Lustre, or Ceph to provide a common view to all compute hosts. ReadLines takes a `thrill::Context` object, which is only required for source DIA operations, and a set of files. The result of ReadLines is a `DIA⟨std::string⟩` which contains each line of the files as an item. The set of files is ordered lexicographically and the set of lines is partitioned equally among the workers.

However, this DIA is not assigned to a variable name. Instead, we immediately append a *FlatMap* operation (line 5) which splits each text line into words and emits one `std::pair⟨std::string, size_t⟩` (aliased as `Pair`) containing $(word, 1)$ per word. In the example, we use a custom `Split` function and `std::string_view` to reference characters in the text line, and copy them into word strings (lines 7 to 10). The `emit auto` parameter of the *FlatMap* lambda function (line 7) enables Thrill to pipeline the `FlatMap` with the following *ReduceByKey* operation. Details on pipelining are

```
1 void WordCount(thrill::Context& ctx,
2   std::string input, std::string output) {
3   using Pair = std::pair<std::string, size_t>;
4   auto word_pairs = ReadLines(ctx, input)
5   .template FlatMap<Pair>(
6     // flatmap lambda: split and emit each word
7     [](const std::string& line, auto emit) {
8       Split(line, '␣', [&](std::string_view sv) {
9         emit(Pair(sv.to_string(), 1));
10      });
11    });
12  word_pairs.ReduceByKey(
13    // key extractor: the word string
14    [](const Pair& p) { return p.first; },
15    // commutative reduction: add counters
16    [](const Pair& a, const Pair& b) {
17      return Pair(a.first, a.second + b.second);
18    })
19  .Map([](const Pair& p) {
20    return p.first + ":␣" + std::to_string(p.second); })
21  .WriteLines(output);
22 }
```

**Algorithm 7.1:** Complete WordCount example C++ source code in Thrill using five DIA operations.

discussed in section 7.3.1. The result of *FlatMap* is a `DIA`⟨`Pair`⟩, which is assigned to the variable `word_pairs`. Note that the keyword `auto` makes C++ infer the appropriate type for `word_pairs` automatically.

The operation *ReduceByKey* is then used to reduce $(word, 1)$ pairs by *word*. This DIA operation must be parameterized with a key extractor (take *word* out of the pair, line 14) and a reduction function (sum two pairs with the same key together, line 17). Thrill currently implements *ReduceByKey* using hash tables, as described in section 7.3.3. Notice that C++ will infer most types during instantiation of *ReduceByKey*, both input and output are implicit; only with *FlatMap* it is necessary to specify what type gets emitted.

The output of *ReduceByKey* is again a `DIA`⟨`Pair`⟩. We need to use a *Map* to transform the `Pair`s into printable strings (lines 19 to 20), which can then be written to disk using the *WriteLines* action. Again, the return type of the *Map* (`std::string`) is inferred automatically, and hence the result of the *Map* operation is implicitly a `DIA`⟨`std::string`⟩.

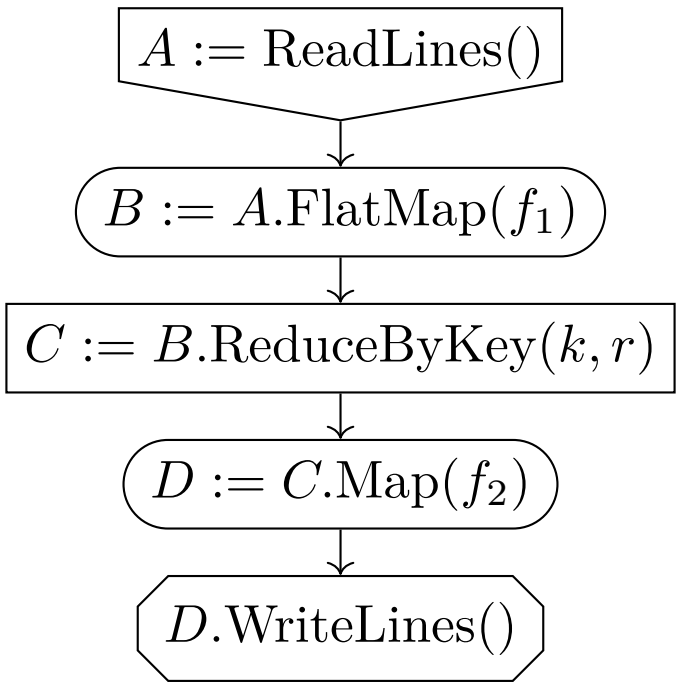

**Figure 7.3:** DIA data-flow graph of WordCount example.

Notice that it is not obvious that the code in algorithm 7.1 describes a parallel and distributed algorithm. It is the *implementation* of the DIA operations in the lazily built data-flow graph which perform the actual distributed execution. The code instructs the C++ compiler to instantiate and optimize these template classes with the UDFs provided. At runtime, objects of these template classes are procedurally created and evaluated when actions are encountered in the DIA data-flow graph.

### 7.2.3 Overview of DIA Operations

Table 7.1 gives an overview of the DIA-operations currently supported by Thrill. The immutability of a DIA enables functional-style data-flow programming. As DIA operations can depend on other DIAs as inputs, these form a directed acyclic graph (DAG), which is called the *DIA data-flow graph.* We denote DIA operations as vertices in this graph, and directed edges represent a dependency. Intuitively, one can picture a directed edge as the *values* of a DIA as they flow from one operation into the next. While the data-flow graph's structure is thus well-defined, it is unclear how much more details an illustration should contain to be useful.

Figure 7.3 shows the data-flow graph of the WordCount example with each vertex labeled alphabetically. For the sake of clarity, this graph representation does not contain details of the UDFs parameterizing the primitives, as these can be very verbose as can be seen in the full source code. In the representation in figure 7.3 we therefore only label them, and could discuss them in more detail in accompanying prose. In figure 7.4 we show another example of a data-flow graph, this time of the PageRank implementation used in section 7.4 for benchmarking. In this illustration we replaced the UDFs with very short descriptions of what they are to accomplish in each step. For example, Map(*parse*) should parse the text lines into pairs, which are then further mapped and counted to calculate the number of vertices (pages) $n_v$ and the number of edges $n_e$. While these brief descriptions allow one to better grasp the structure of the whole algorithm, the description themselves are not precise and again require more explanation in prose. The following chapter 8 will consider more data-flow graphs of suffix sorting algorithms.

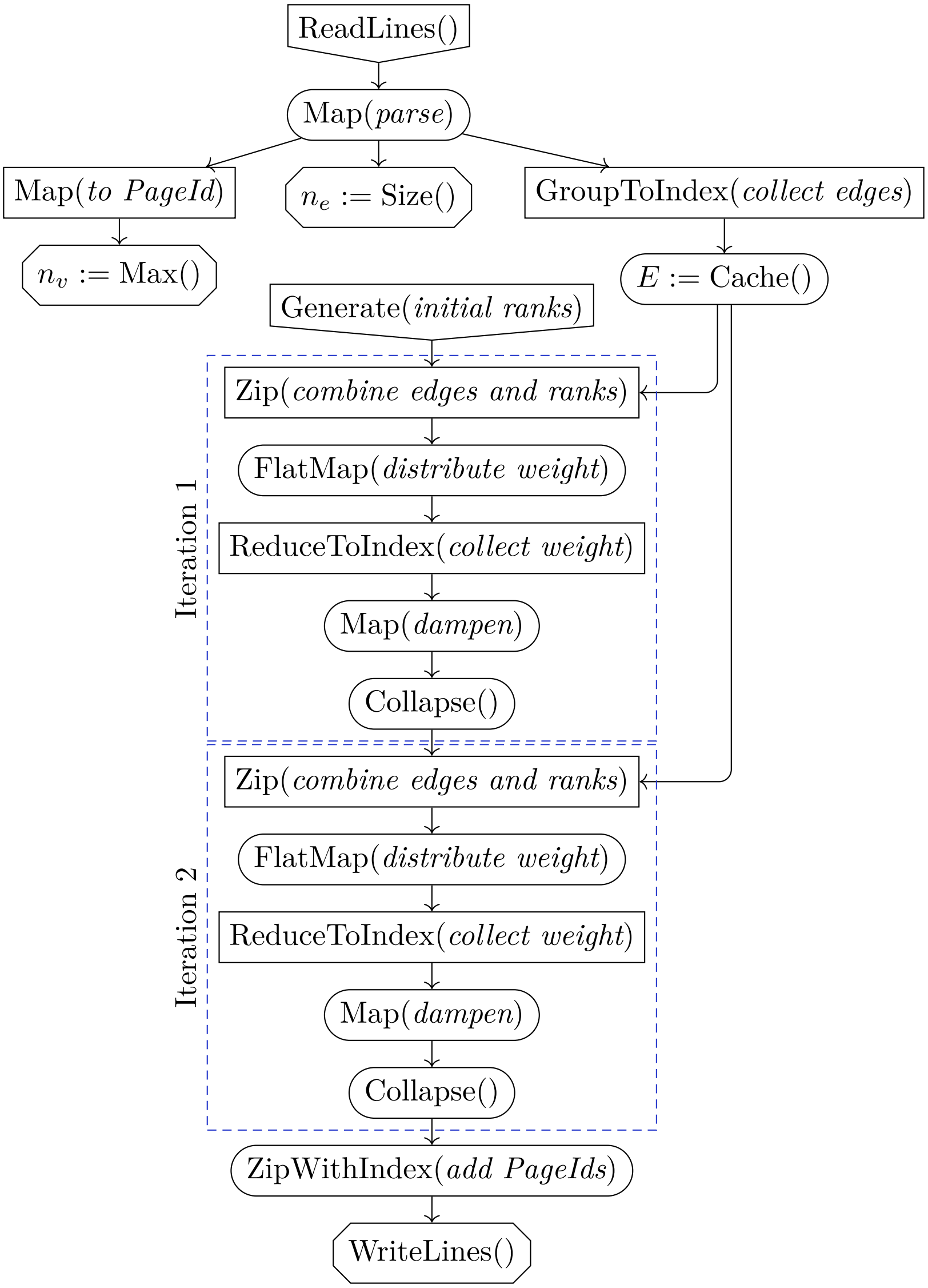

**Figure 7.4:** DIA data-flow graph of PageRank with two iterations.

We classify all DIA operations into four categories. *Source* operations have no incoming edges and generate a DIA from external sources like files, database queries, or simply by generating the integers $0 \ldots n-1$. Operations which have one or more incoming edges and return a DIA are classified further as *local* (LOps) and *distributed* operations (DOps). Examples of LOps are *Map* or *Filter*, which apply a function to every item of the DIA independently. LOps can be performed locally and in parallel, without any communication between workers. On the other hand, DOps such as *ReduceByKey* or *Sort* may require communication and a full data round-trip to disks.

The fourth category are *actions*, which do not return a DIA and hence have no outgoing edges. The DIA data-flow graph is built lazily, i.e., DIA operations are not immediately executed when created. Actions trigger evaluation of the graph and return a value to the user program. For example, writing a DIA to disk or calculating the sum of all values are actions. By inspecting the results of actions, a user program can determine the future program flow, e.g. to iterate a loop until a condition is met. Hence, control flow decisions are performed *collectively* in C++ with imperative loops or recursion (host language control-flow).

The first DIAs in a Thrill program are generated using *source* operations:

**Generate($n, g$)** creates a `DIA`$\langle A \rangle$ of size $n$ by mapping each integer $[0 \mathinner{..} n)$ to an item using a generator function $g : [0, n) \to A$. The generator $g$ can also be omitted, in which case Generate($n$) simply returns a DIA containing $[0 \mathinner{..} n)$.

**ReadLines(*file-pattern*)** creates a `DIA`$\langle$`std::string`$\rangle$ which contains the lines of the text files in order of appearance. If multiple files are given, the file names are sorted lexicographically.

**ReadBinary$\langle A \rangle$(*file-pattern*)** reads binary data from the file system and creates a `DIA`$\langle A \rangle$ containing this data. The on-disk binary objects are read with the same serialization methods of class $A$ used in Thrill for storing and transmitting C++ objects. Due to Thrill's internal Block management system, the external data is actually only virtually mapped into the DIA and I/Os are performed lazily when needed.

Source operations are performed in parallel on all workers: e.g. *ReadLines* and *ReadBinary* determine the overall size and equally split the size such that all workers read portions of the files.

Local operations (LOps) are operations without communication. This classification restricts their capabilities, but allows them to be efficiently chained with preceding operations.

**FlatMap($f$)** corresponds to the *map* step in the MapReduce paradigm. Each item of the input DIA is mapped to zero, one, or more output items by a function $f$ (see also figure 7.5). In C++ this is done by calling a specially crafted *emit* function for each item, as shown in the WordCount example. All items are concatenated in order and form a new DIA.

**Table 7.1:** DIA Operations of Thrill.

| Operation | User Defined Functions |
|---|---|
| Sources | |
| **Generate**$(n)$ : $\texttt{DIA}[0,\dots,n-1]$<br>**Generate**$(n,g)$ : $\texttt{DIA}\langle A\rangle$ | $n$ : DIA size<br>$g : [0\mathinner{..}n) \to A$ |
| **ReadLines**(files) : $\texttt{DIA}\langle\textbf{string}\rangle$<br>**ReadBinary**$\langle A\rangle$(files) : $\texttt{DIA}\langle A\rangle$ | <br>$A$ : data type |
| Local Operations (no communication) | |
| **FlatMap**$(f)$ : $\texttt{DIA}\langle A\rangle \to \texttt{DIA}\langle B\rangle$ | $f : A \to \textbf{list}(B)$ |
| **Map**$(f)$ : $\texttt{DIA}\langle A\rangle \to \texttt{DIA}\langle B\rangle$ | $f : A \to B$ |
| **Filter**$(f)$ : $\texttt{DIA}\langle A\rangle \to \texttt{DIA}\langle A\rangle$ | $f : A \to \textbf{bool}$ |
| **BernoulliSample**$(p)$ : $\texttt{DIA}\langle A\rangle \to \texttt{DIA}\langle A\rangle$ | $p$ : success probability |
| **Union**$()$ : $\texttt{DIA}\langle A\rangle \times \texttt{DIA}\langle A\rangle \cdots \to \texttt{DIA}\langle A\rangle$ | |
| **Cache**$()$ : $\texttt{DIA}\langle A\rangle \to \texttt{DIA}\langle A\rangle$ | |
| **Collapse**$()$ : $\texttt{DIA}\langle A\rangle \to \texttt{DIA}\langle A\rangle$ | |
| Distributed Operations (communication between workers) | |
| **ReduceByKey**$(k,r)$ : $\texttt{DIA}\langle A\rangle \to \texttt{DIA}\langle A\rangle$<br>**ReduceToIndex**$(i,r,n)$ : $\texttt{DIA}\langle A\rangle \to \texttt{DIA}\langle A\rangle$<br><br>**GroupByKey**$(k,g)$ : $\texttt{DIA}\langle A\rangle \to \texttt{DIA}\langle B\rangle$<br>**GroupToIndex**$(i,g,n)$ : $\texttt{DIA}\langle A\rangle \to \texttt{DIA}\langle B\rangle$ | $k : A \to K$<br>$i : A \to [0\mathinner{..}n)$<br>$r : A \times A \to A$<br>$g : \textbf{iterable}(A) \to B$<br>$n$ : result size |
| **Sort**$(c)$ : $\texttt{DIA}\langle A\rangle \to \texttt{DIA}\langle A\rangle$ | $c : A \times A \to \textbf{bool}$ |
| **Merge**$(c)$ : $\texttt{DIA}\langle A\rangle \times \texttt{DIA}\langle A\rangle \cdots \to \texttt{DIA}\langle A\rangle$ | $c : A \times A \to \textbf{bool}$ |
| **Concat**$()$ : $\texttt{DIA}\langle A\rangle \times \texttt{DIA}\langle A\rangle \cdots \to \texttt{DIA}\langle A\rangle$ | |
| **PrefixSum**$(s,a)$ : $\texttt{DIA}\langle A\rangle \to \texttt{DIA}\langle A\rangle$<br>**ExPrefixSum**$(s,a)$ : $\texttt{DIA}\langle A\rangle \to \texttt{DIA}\langle A\rangle$ | $s : A \times A \to A$<br>$a$ : initial value |
| **Sample**$(n)$ : $\texttt{DIA}\langle A\rangle \to \texttt{DIA}\langle A\rangle$ | $n$ : result size |
| **Zip**$(z)$ : $\texttt{DIA}\langle A\rangle \times \texttt{DIA}\langle B\rangle \cdots \to \texttt{DIA}\langle C\rangle$ | $z : A \times B \cdots \to C$ |
| **ZipWithIndex**$(z)$ : $\texttt{DIA}\langle A\rangle \to \texttt{DIA}\langle B\rangle$ | $z : A \times \mathbb{N}_0 \to B$ |
| $\textbf{Window}_k(w)$ : $\texttt{DIA}\langle A\rangle \to \texttt{DIA}\langle B\rangle$<br>$\textbf{FlatWindow}_k(f)$ : $\texttt{DIA}\langle A\rangle \to \texttt{DIA}\langle B\rangle$ | $k$ : window size<br>$w : \mathbb{N}_0 \times A^k \to B$<br>$f : \mathbb{N}_0 \times A^k \to \textbf{list}(B)$ |
| $\textbf{ZipWindow}_{[k_1,k_2,\dots]}(z)$ :<br>$\texttt{DIA}\langle A_1\rangle \times \texttt{DIA}\langle A_2\rangle \cdots \to \texttt{DIA}\langle B\rangle$ | $k_1, k_2, \dots : \mathbb{N}_0$<br>$z : \mathbb{N}_0 \times A_1^{k_1} \times A_2^{k_2} \cdots \to B$ |
| Actions | |
| **Size**$()$ : $\texttt{DIA}\langle A\rangle \to \textbf{unsigned}$ | |
| **AllGather**$()$ : $\texttt{DIA}\langle A\rangle \to \textbf{vector}(A)$ | |
| **Sum**$(s,a)$ : $\texttt{DIA}\langle A\rangle \to A$<br>**Min**$()$ : $\texttt{DIA}\langle A\rangle \to A$<br>**Max**$()$ : $\texttt{DIA}\langle A\rangle \to A$ | $s : A \times A \to A$<br>$a$ : initial value |
| **WriteLines**$()$ : $\texttt{DIA}\langle\textbf{string}\rangle \to$ files<br>**WriteBinary**$()$ : $\texttt{DIA}\langle A\rangle \to$ files | |
| **Execute**$()$ | |

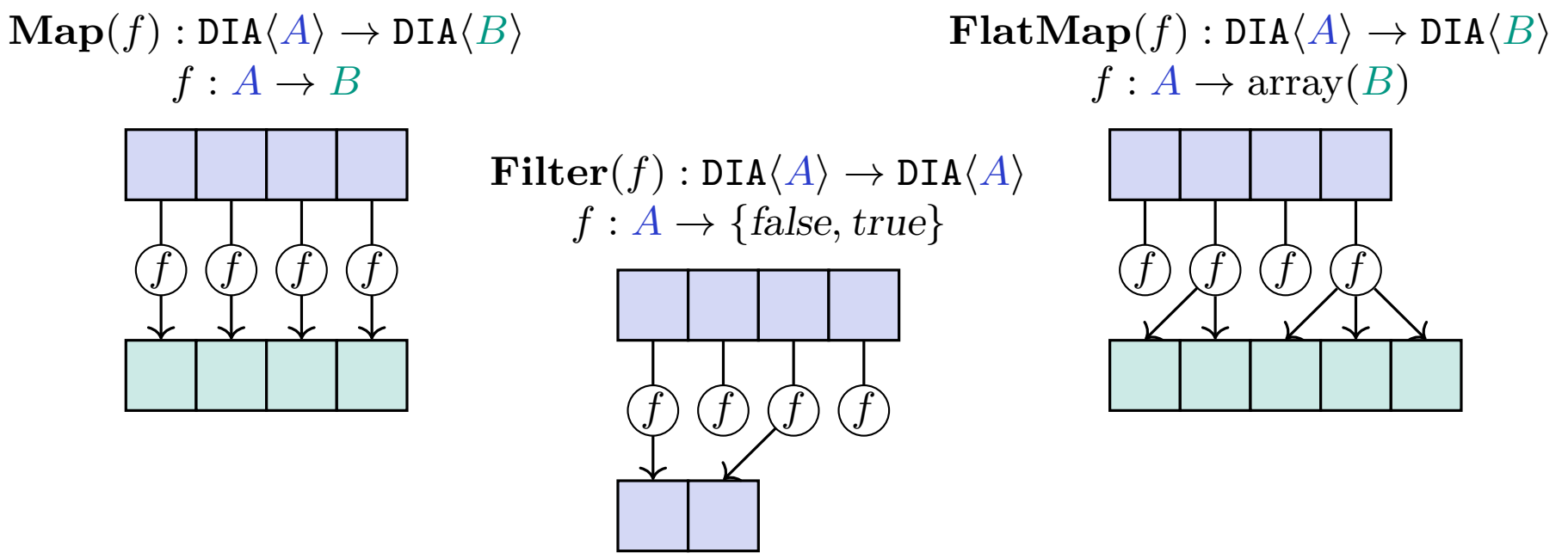

**Figure 7.5:** Illustrations of how *Map*, *Filter*, and *FlatMap* operate on a DIA.

**Map($\boldsymbol{f}$)** applies the function $f : A \to B$ to each item in an input $\texttt{DIA}\langle A\rangle$ $X$, and returns a $\texttt{DIA}\langle B\rangle$ $Y$ with $Y[i] = f(X[i])$ for all $i = 0, \dots, |X| - 1$ (see figure 7.5 again).

**Filter($\boldsymbol{f}$)** takes a $\texttt{DIA}\langle A\rangle$ $X$ and a function $f : A \to \texttt{bool}$, and returns the $\texttt{DIA}\langle A\rangle$ containing $[\, x \in X \mid f(x) \,]$ within which the order of items is maintained (see also figure 7.5). Both *Map* and *Filter* are special cases of *FlatMap*.

**BernoulliSample($\boldsymbol{p}$)** takes a $\texttt{DIA}\langle A\rangle$ $X$ and samples each item independently with constant probability $p$.

**Union($\boldsymbol{X_1, \dots, X_n}$)** : Given a set of $\texttt{DIA}\langle A\rangle$s $X_1, \dots, X_n$, Union returns $\texttt{DIA}\langle A\rangle$ $Y = \bigcup_{i=1}^{n} X_i$ containing all items of the input in an arbitrary order.

**Cache()** explicitly materializes and caches the result of a DIA operation for later use.

**Collapse()** on the other hand only folds a pipeline of functions, as described in more detail in section 7.3.3.

Distributed operations (DOps) are more costly and powerful than LOps. DOps contain at least one BSP barrier at which all workers synchronize, and can contain zero or more BSP additional supersteps.

**ReduceByKey($\boldsymbol{k, r}$)** and **GroupByKey($\boldsymbol{k, g}$)** : The *reduce* step from the MapReduce paradigm is represented by Thrill's *ReduceByKey* and *GroupByKey* DOps, which are illustrated in figure 7.6. In both operations, input items are grouped by a key. Keys are extracted from items using the *key extractor* function $k : A \to K$, and then mapped to workers using a hash function $h$. In *ReduceByKey*, the associative reduction function $r : A \times A \to A$ specifies how two items are combined into one. In *GroupByKey*, all items with a certain key are collected on one worker and processed by the group function $g$ : iterable($A$) $\to B$. When possible, *ReduceByKey* should be preferred as it allows local reduction and thus lowers communication volume and running time.

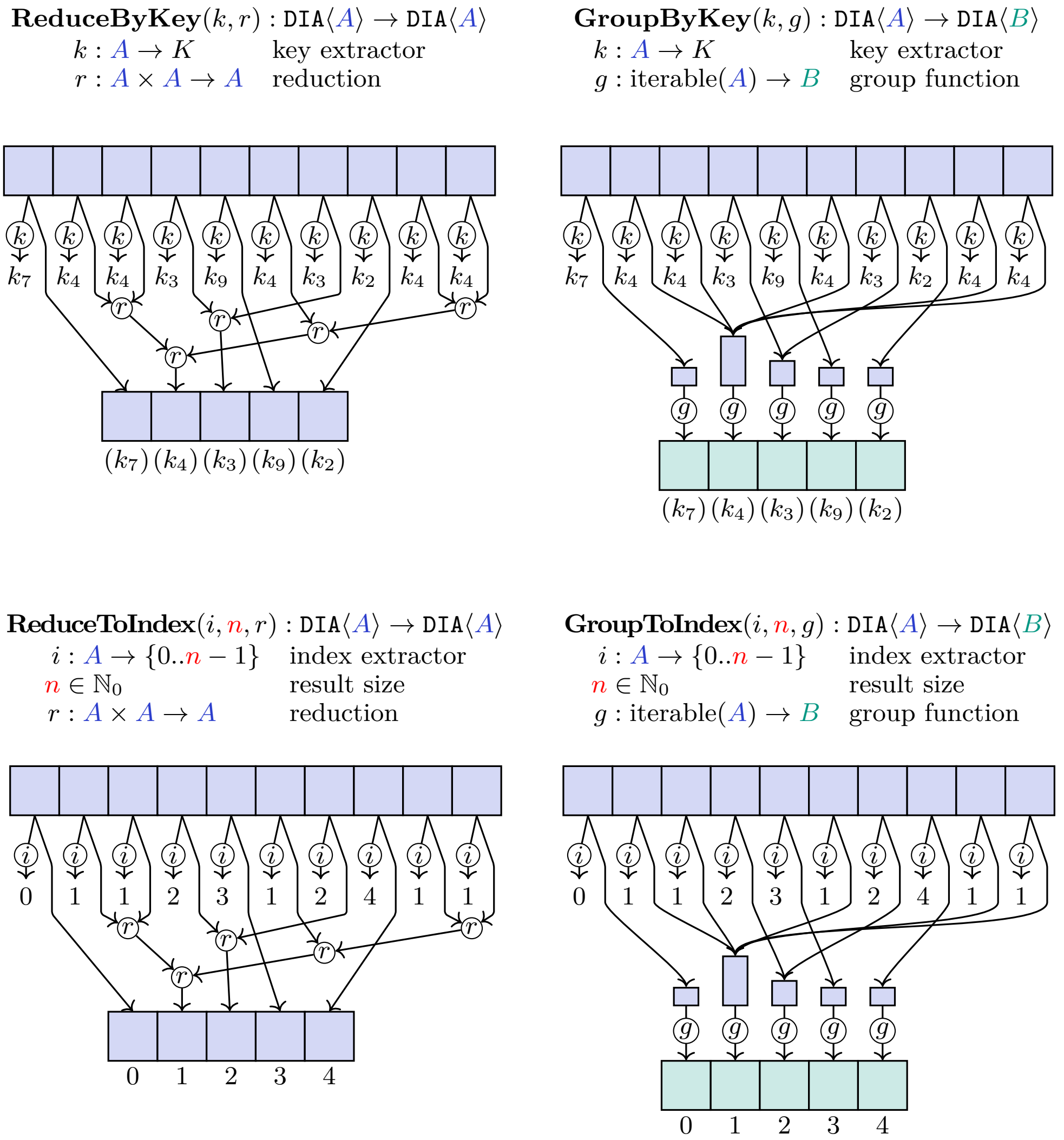

**Figure 7.6:** Illustrations of how *ReduceByKey*, *GroupByKey*, *ReduceToIndex*, and *GroupToIndex* operate on a DIA.

**ReduceToIndex($i, r, n$)** and **GroupToIndex($i, g, n$)** : Both *ReduceByKey* and *GroupByKey* also offer a *ToIndex* variant (see figure 7.6 again), wherein each item of the input DIA is mapped by a function $i : A \to [0 \mathinner{..} n)$ to an index in the output DIA. The size of the resulting DIA must be given as $n$. Items which map to the same index are either reduced using an associative reduction function $r : A \times A \to A$, or processed by a group function $g$ : iterable($A$) $\to B$. Empty slots in the resulting DIA are filled with a neutral item.

**Sort($c$)** sorts an input `DIA`$\langle A \rangle$ $X$ with respect to a less-comparison function $c : A \times A \to$ bool (see figure 7.7). If *Sort* is denoted without a comparison function, we assume the tuples are compared component-wise with the first component being most significant, the second component the second most significant, and so on.

**Merge($c$)** : Given a set of sorted `DIA`$\langle A \rangle$s $X_1, \dots, X_n$ and a less-comparison function $c : A \times A \to$ bool, *Merge* returns `DIA`$\langle A \rangle$ $Y$ that contains all tuples of $X_1, \dots, X_n$ and is sorted with respect to $c$ (see figure 7.7). If *Merge* is denoted without a comparison function we assume the tuples are compared component-wise as in *Sort*.

**Concat()** takes a list of `DIA`$\langle A \rangle$s $X_1, \dots, X_n$ and returns all items concatenated in order. Thrill's current implementation performs an all-to-all data shuffle to rearrange the items, which makes this an expensive operation. In nearly all cases, however, *Concat* can be avoided by changing the order in preceding or subsequent operations.

**PrefixSum($s, a$)** and **ExPrefixSum($s, a$)** : Given an input `DIA`$\langle A \rangle$ $X$, an associative operation $s : A \times A \to A$ (by default $s = +$), and an initial value $a$, *PrefixSum* returns a `DIA`$\langle A \rangle$ $Y$ such that $Y[0] = s(a, X[0])$ and $Y[i] = s(Y[i-1], X[i])$ for all $i = 1, \dots, |X| - 1$. *ExPrefixSum* returns a `DIA`$\langle A \rangle$ $Y$ such that $Y[0] = a$ and $Y[i] = s(Y[i-1], X[i-1])$ for all $i = 1, \dots, |X| - 1$.

**Sample($n$)** selects $n$ items from a DIA $X$ uniform at random, but *without replacement* [SLHS+17]. The selected items are returned as a new DIA.

**Zip($z$)** combines two or more DIAs index-wise using a zip function $z$ similar to functional programming languages. The function $z$ is applied to all items with index $i$ to deliver the new item at index $i$ (see also figure 7.7).
Given a set of DIAs $X_1, \dots, X_n$ of type $A_1, \dots, A_n$ of equal size ($|X_1| = \dots = |X_n|$) and a function $z : A_1 \times \dots \times A_n \to B$, *Zip* returns `DIA`$\langle B \rangle$ $Y$ with $Y[i] = z(X_1[i], \dots, X_n[i])$ for all $i = 0, \dots, |X_1| - 1$. The regular *Zip* function requires all DIAs to have equal length, but Thrill also provides variants which cut the DIAs to the shortest or pad them to the longest.

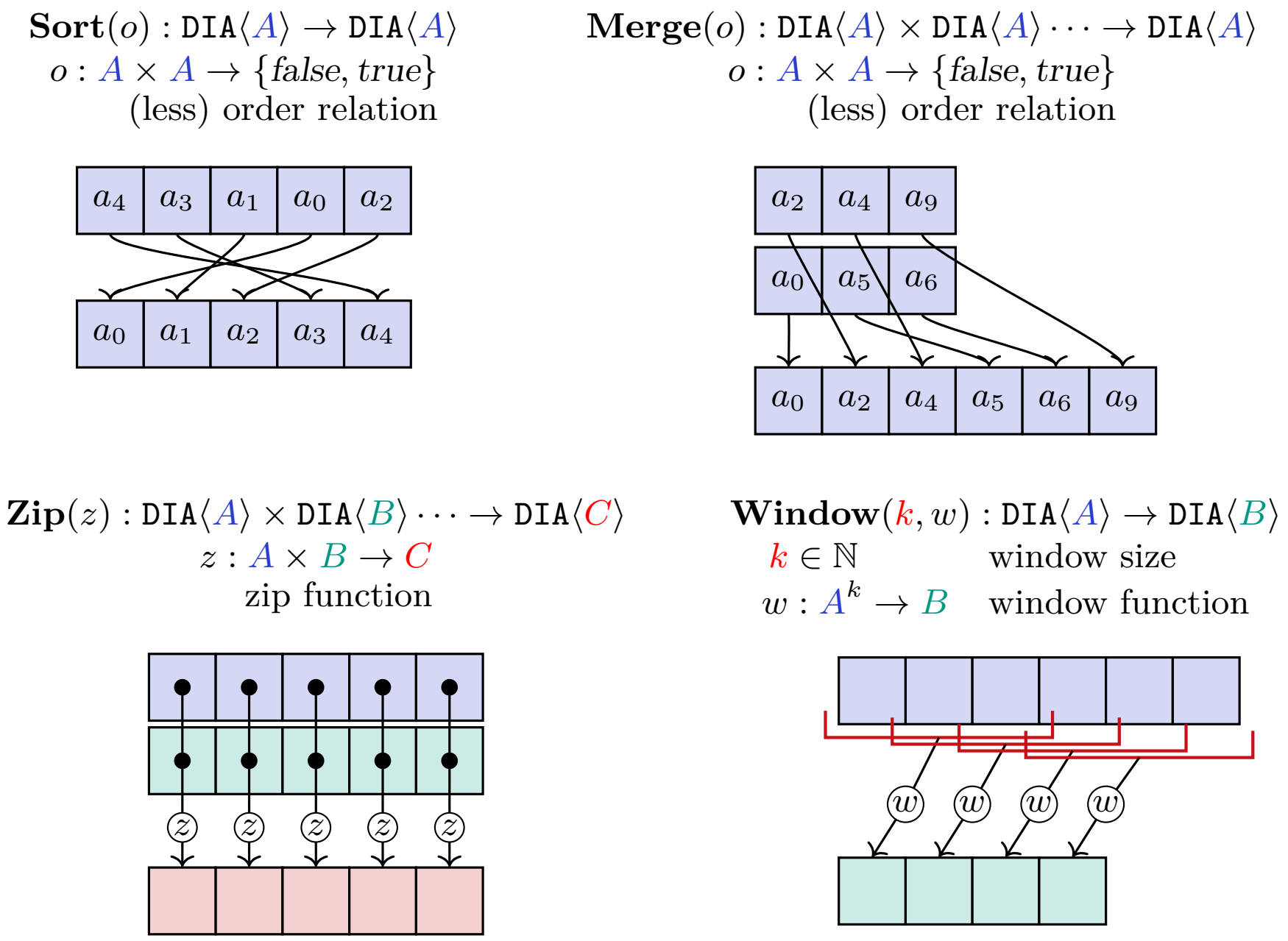

**Figure 7.7:** Illustrations of how the operations *Sort*, *Merge*, *Zip*, and *Window* work on a DIA.

**ZipWithIndex($z$)** zips each DIA item with its global index. While *ZipWithIndex* can be emulated using *Generate* and *Zip*, the combined variant requires less communication, because only the number of items needs to be exchanged. Given an input $\texttt{DIA}\langle A\rangle$ $X$ and a function $z : (A, \mathbb{N}_0) \to B$, *ZipWithIndex* returns $\texttt{DIA}\langle B\rangle$ $Y$ with $Y[i] = z(X[i], i)$ for all $i = 0, \dots, |X| - 1$

**Window$_k$($w$)** and **FlatWindow$_k$($f$)** take an input $\texttt{DIA}\langle A\rangle$ $X$ and a window function $w : \mathbb{N}_0 \times A^k \to B$. The operation scans over $X$ with a window of size $k$ and applies $w$ once to each set of $k$ consecutive items from $X$ and their index in $X$ (see also figure 7.7). The final $k - 1$ indices with less than $k$ consecutive items are delivered to $w$ as partial windows padded with sentinel values. The result of all invocations of $w$ is returned as a $\texttt{DIA}\langle B\rangle$ containing $|X|$ items in the order.

FlatWindow is a variant of Window which takes an input $\texttt{DIA}\langle A\rangle$ $X$ and a window function $f : \mathbb{N}_0 \times A^k \to \text{list}(B)$. The only difference compared to Window is, that $f$ can *emit* zero or more items that are concatenated in the resulting $\texttt{DIA}\langle B\rangle$ in the order they are emitted. Thrill also provides specializations which delivers all disjoint windows of $k$ consecutive items.

**ZipWindow$_{[k_1,k_2,\dots]}$($z$)** is a faster combination of *Zip* and *Window* applied to a set of DIAs $X_1, \dots, X_n$. In each step, the *Zip* function $z$ is delivered a window of items from each DIA: from DIA $X_i$ at most $k_i$ items are extracted into a vector and delivered to the zip lambda function.

The list of distributed operation above is long, still not complete, and we expect future authors to add even more operations. Recently, Alexander Noe implemented a *Join* operation [Noe17], and Moritz Kiefer and Tino Fuhrmann contributed a *HyperLogLog* method [FFGM07; HNH13]. Furthermore, additional helper operations such as *ConcatToDIA*, *EqualToDIA*, *Distribute*, *Gather*, and so on are also available. Instead of detailing more DOps, let us focus on action operations:

**Size()** : Given an input `DIA`$\langle A \rangle$ $X$, *Size* returns the number of items in $X$, i.e. $|X|$.

**AllGather()** returns a whole DIA as `std::vector`$\langle T \rangle$ on *each* worker. This obviously only works for small DIAs.

**Sum($s, a$)** computes an associative function $s$ over all items in a DIA and returns the result on every worker. By default *Sum* uses $+$, and $a$ is an optional initial value.

**Min()** and **Max()** are specializations of *Sum* with other operators to calculate the minimum and maximum values of a DIA.

**WriteLines(*path-pattern*)** writes a `DIA`$\langle$`std::string`$\rangle$ to multiple text files. As with *ReadLines*, all workers write files in *parallel.* Each worker writes a separate file to enable parallel writing on all file systems.

**WriteBinary$\langle A \rangle$(*path-pattern*)** writes a `DIA`$\langle A \rangle$ into binary files using Thrill's (built-in) serialization. Primitive types, such as characters and integers, are simply written without any preamble.

**Execute()** is a special action, which can be used to explicitly trigger evaluation of DIA operations. It does not insert a vertex into the DIA data-flow graph.

Actions trigger evaluation of all DIA operations needed to calculate the result. Besides regular actions which trigger evaluation, Thrill also provides *action futures*, called *SumFuture*, *MinFuture*, *AllGatherFuture*, etc, which only insert an action vertex into the DIA data-flow graph, but *do not* trigger evaluation. Using action futures one can calculate multiple results (e.g. the minimum *and* maximum item) with just one data round trip.

The current set of scalable primitive DIA operations listed in table 7.1 is definitely not final, and more distributed algorithmic primitives may be added in the future as necessary and prudent. In section 7.3.3 we describe the implementations of some of the operations in more detail. We also envision future work on how to accelerate scalable primitives, which can then be use as drop-in replacement to our current straight-forward implementations.

### 7.2.4 Why Arrays?

Thrill's DIA API is obviously similar to Spark and Flink's data-flow languages, which themselves are similar to many functional programming languages [MDAT16; MDAT17]. However, we explicitly define the items in DIAs to be ordered. This order may be arbitrary after operations like *ReduceByKey*, which hash items to indices in the array, but they do have an order. Many of our operations like *PrefixSum*, *Sort*, *Merge*, *Zip*, and especially *Window* only make sense with an ordered data type.

Having an order on the distributed array opens up new opportunities in how to exploit this order in algorithms. Essentially, the order reintroduces the concept of *locality* into distributed data-flow programming. While one cannot access DIA items directly, such as in a imperative for loop over an array, one *can* iterate over them using a *Window* function *in parallel* with adjacent items as context. A common design pattern in Thrill programs is to use *Sort* or *ReduceToIndex* to bring items into a desired order, and then to process them using a *Window*. Furthermore, if the computation in a *Window* needs context from more than one DIA, these can be *Zip*-ped together first.

We are looking forward to future work on how this order paradigm can be exploited. Furthermore, extending Thrill beyond one-dimensional arrays to higher dimensional arrays, (sparse) matrices, or graphs is not only useful but also conceptually interesting since these data types have a more complex concept of locality.

## 7.3 Implementation of Thrill

### 7.3.1 Data-Flow Graph Implementation

Contrary to the picture of DIAs we have drawn for application programmers in the preceding sections, the distributed array of items usually does not exist explicitly. Instead, a DIA remains purely a conceptual data-flow between two concrete DIA operations. This data-flow abstraction allows us to apply an optimization called *pipelining* or *chaining*. Chaining in general describes the process of combining the logic of one or more functions into a single one (called *pipeline*). In Thrill we chain together all independently parallelizable local operations (e.g. *FlatMap*, *Map*, *Filter*, and *BernoulliSample*), and the first local computation step of the next distributed DIA operation into one block of optimized binary code. Via this chaining, we reduce both the overhead of the data flow between them, as well as the total number of operations, and obviate the need to store intermediate explicit arrays. Additionally, we leverage the C++ compiler to combine the local computations *on the assembly level* with full optimization, thus reducing the number of indirections to a minimum, which additionally improves cache efficiency. In essence, we combine all local computation of one bulk-synchronous parallel (BSP) superstep [GV94] using chaining into one block of assembly code.

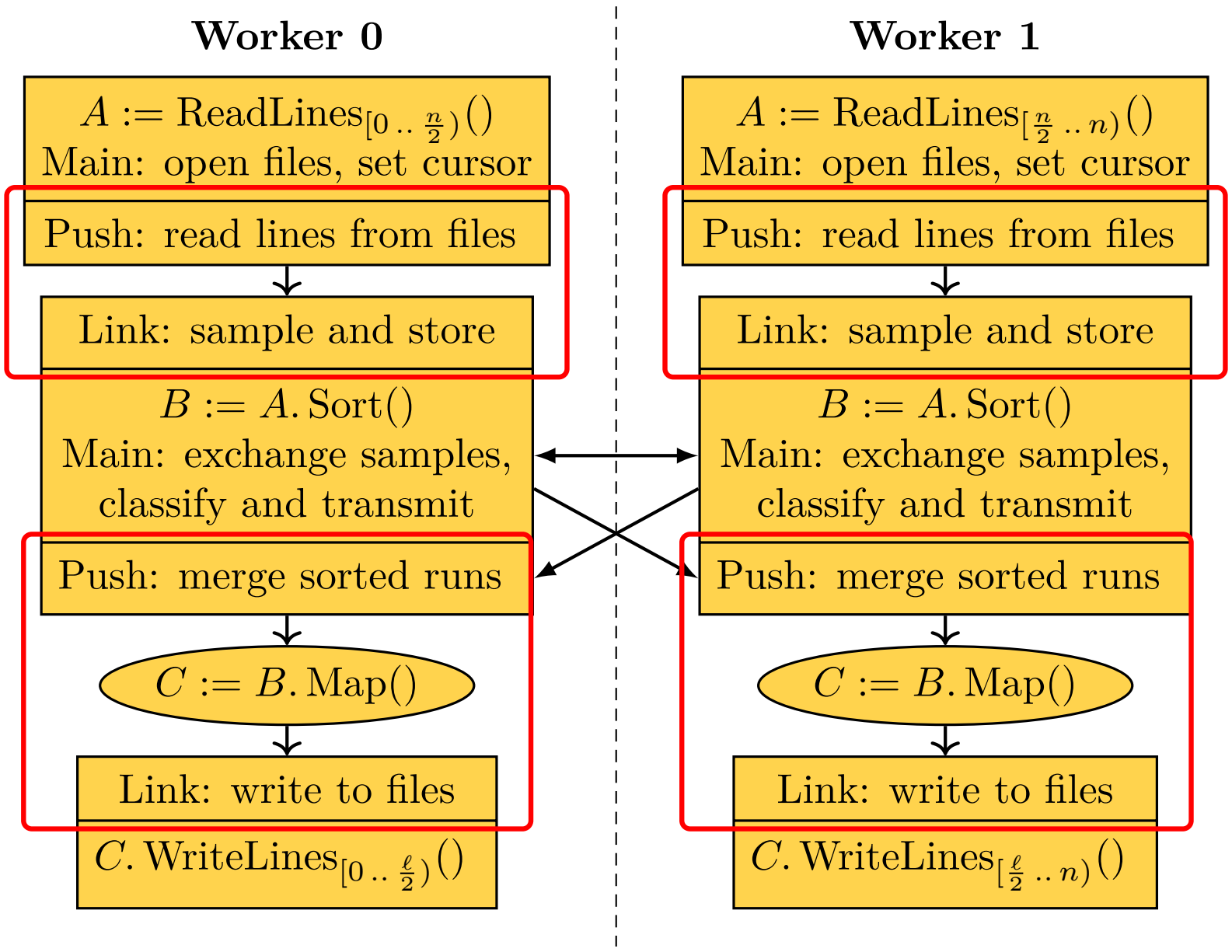

**Figure 7.8:** Subdivisions of DOps and chained Push, LOps, and Link parts.

To integrate the implementations of distributed DIA operations into the pipelining framework we subdivide them into three parts: *Link*, *Main* and *Push* (see figure 7.8 for an example). The Link part (sometimes also called *PreOp*) handles incoming items by performing some *finalizing local work* like storing or transmitting them. This process closes the pipeline and results in a single executable code block containing its logic. The Main part contains the actual DIA operation logic like sorting, synchronous communication, etc. And finally, the Push part (also called *PostOp*) represents the start of a new pipeline by *emitting* items for further processing. Depending on the type of a DIA operation, subdivisions can also be empty or trivial.

We explain these subdivisions using *PrefixSum* as an example. In the Link part, *PrefixSum* receives a stream of items from a preceding operation and stores them in sequence. While storing them, each worker keeps a local sum over all items. In the Main part, the workers perform a global synchronous exclusive prefix sum on the local sums to calculate the initial value of their items. This local initial value is then added to items while they are being read and Push-ed into the next operation.

Chaining also affects how data dependencies between DIA operations are represented in Thrill's data-flow graph. Due to pipelining of local operations into one assembly block, all LOp are fused with the succeeding DOp vertices. Hence only vertices representing distributed operations remain in the DAG. This optimized data-flow DAG corresponds to a set of BSP supersteps and their data dependencies, and is executed lazily when an action is encountered.

Execution is done by Thrill's *StageBuilder*, which performs a reverse breadth-first *stage* search in the optimized DAG to determine which DIA operations need to be calculated. The gathered vertices are then executed in topological order such that their data dependencies are resolved prior to execution. Unnecessary recomputations are avoided by maintaining the state and data in each vertex. The vertices in the DIA graph are reference counted objects, such that both child vertices and $\texttt{DIA}\langle T\rangle$ references in user programs can keep a handle to the object. The data in DIA operations is automatically disposed when all children are calculated and no further children can be added.

We implemented chaining and our execution model by making heavy use of C++ template programming. More precisely, we compose a pipeline by chaining together the underlying (lambda) functions using their static functor types. Since these types can be deduced by static analysis, chaining can take place during compile time, and hence chained operations can be optimized into single pipelined functions on the assembly code level. In the end all trivially-parallel local operation like *Map*, *FlatMap*, etc. introduce zero overhead during runtime, and are combined with the following DIA operation's *Link* part.

The caveat of Thrill's chaining mechanism is that the preceding LOp and DOp's (lambda) functions $f_1, f_2, \ldots$ become part of the DIA operation's template instantiation types as $\texttt{DIA}\langle T, f_1, f_2, \ldots\rangle$. This is usually not a problem, since with C++11 we can encourage liberal use of the `auto` keyword instead of using concrete $\texttt{DIA}\langle T\rangle$ types. However, in iterative or recursive algorithms $\texttt{DIA}\langle T\rangle$ variables have to be updated. These variables are only references to the actual DIA operations, which are immutable, but the references must point to the same underlying DIA operation template type. We address this issues by adding a special operation named *Collapse* which constructs a $\texttt{DIA}\langle T\rangle$ from $\texttt{DIA}\langle T, f_1, f_2, \ldots\rangle$. This operation creates an additional vertex in the data-flow DAG that closes the current pipeline, stores it, and creates a new (empty) one. The framework will issue compilation errors when *Collapse* is required.

In Thrill we took pipelining of data processing one step further by enabling *consumption* of source DIA storage *while* pushing data to the next operation. DIA operations transform huge data sets, but a naive implementation would read all items from one DIA, push them all into the pipeline for processing, and then deallocate the data storage. Assuming the next operation also stores all items, this requires twice the amount of storage. However, with *consume* enabled, the preceding DIA operation's storage is deallocated while processing the items, hence the storage for all items is needed only once, plus a small overlapping buffer.

### 7.3.2 Data, Network, and I/O Layers

Below the convenient high-level DIA API of Thrill lie several software layers which do the actual data handling (see the layer diagram in figure 7.9). DIA operations are

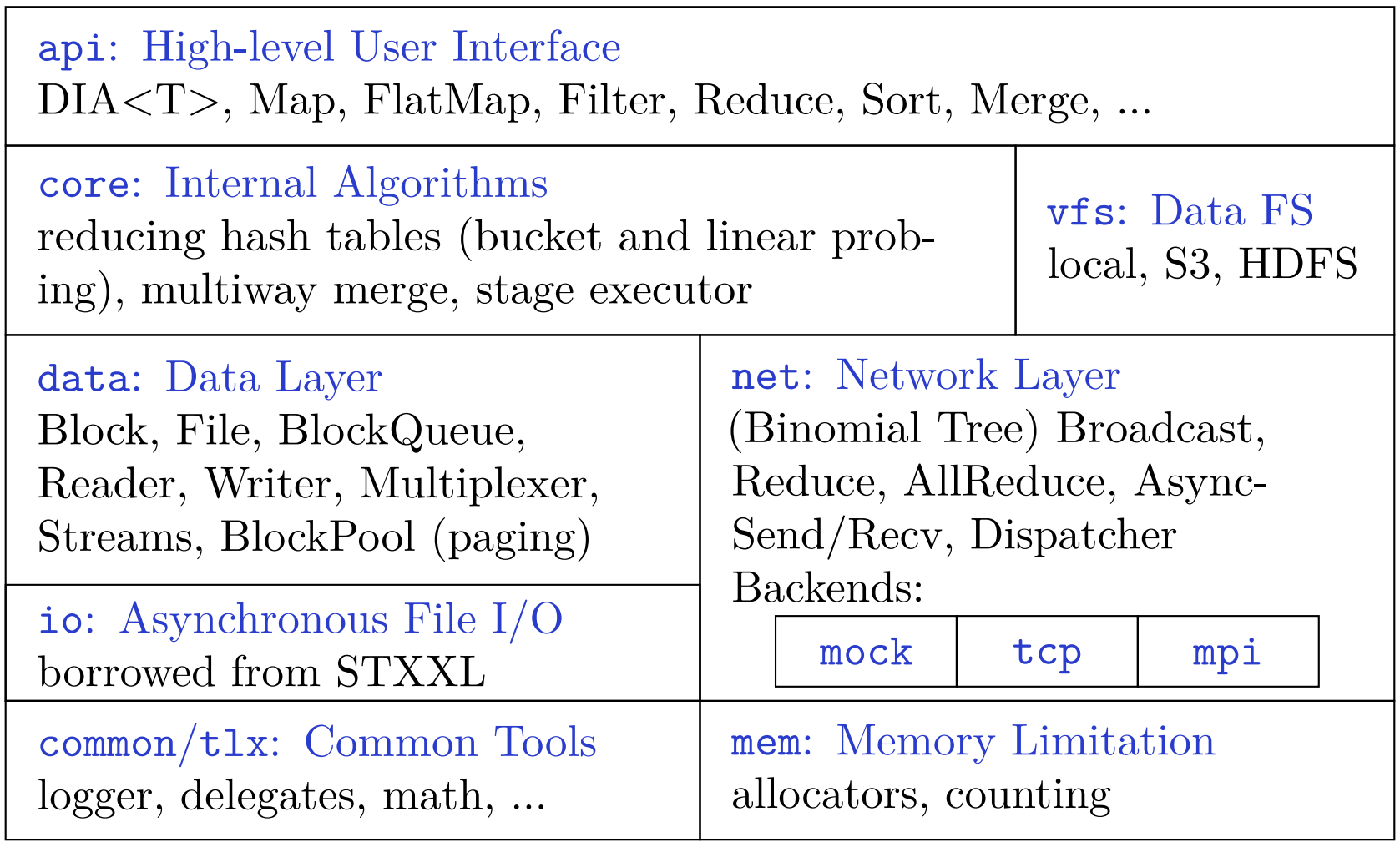

**Figure 7.9:** Layer diagram of Thrill.

C++ template classes which are chained together as described in section 7.3.1. These operations store and transmit the items using the *data*, *net*, and *io* layers.

Items have to be serialized to byte-sequences for transmission via the network or for storage on disk. Thrill contains a custom C++ serialization framework which aims to deliver high performance and low to zero overhead. This is possible because neither signatures nor versioning are needed. In general, fixed-length trivial items like integers and fixed-size numerical vectors are stored with zero overhead. Variable length items like strings and variable-length vectors are prepended with their length. Compound objects are stored as a sequence of their components.

DIA operations process a stream of items, which need to be transmitted or stored, and then read. Such a stream of items is serialized directly into the memory buffer of a *Block*, which is by default 2 MiB in size. Items in a Block are stored without separators or other per item overhead, as illustrated in figure 7.10. This is possible because Thrill's serialization methods correctly advance a cursor to the next item. Hence, currently only four integers are required as overhead per Block and zero per item. This efficient Block storage format is important for working with small items like plain integers or characters, but Thrill can also process large blobs spanning multiple Blocks.

A sequence of Blocks is called a *File*, even though it is usually stored in main memory. DIA operations read/write items sequentially to/from Files using template *BlockReader* and *BlockWriter* classes.

To transmit items to other workers, DIA operations have two choices. One is a set of efficient *synchronous* collective communication primitives similar to MPI, such as

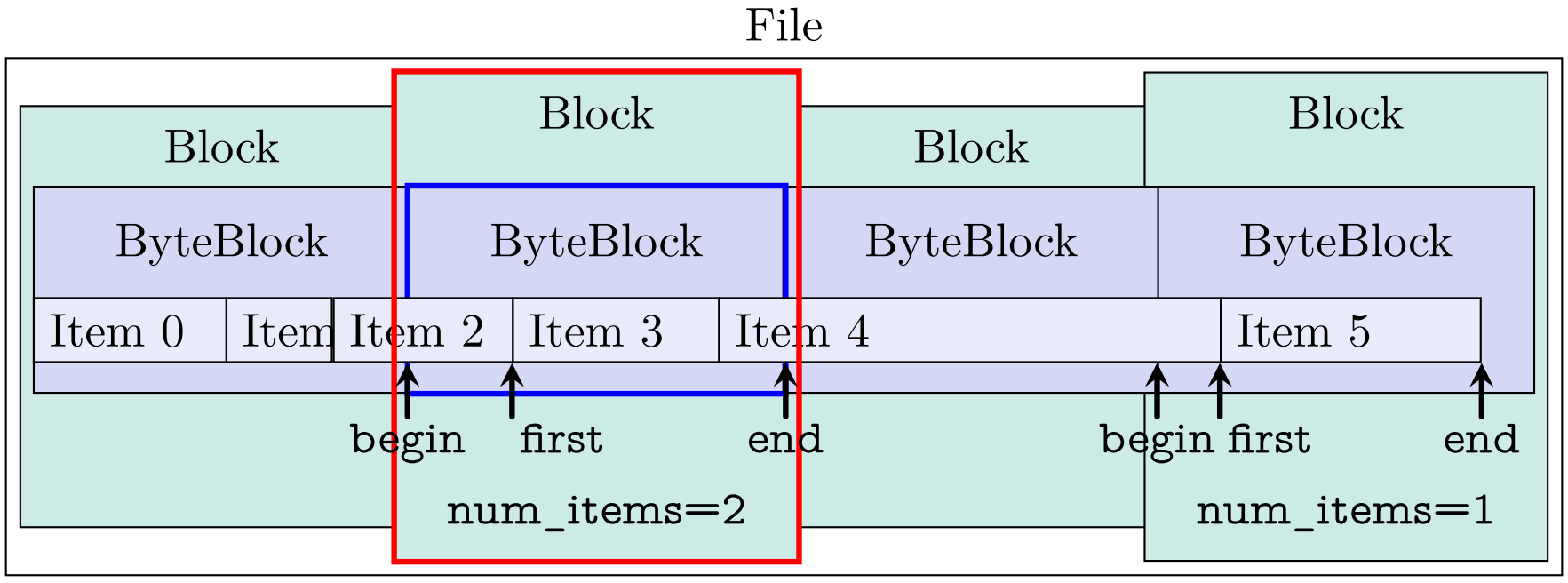

**Figure 7.10:** Thrill's File structure: a File contains an array of Blocks, which reference portions of ByteBlocks containing the actual items with zero overhead.

*AllReduce*, *Broadcast*, and *PrefixSum*. These utilize the same serialization framework and are mostly used for blocking communication of small data items, e.g. an integer AllReduce is often used to calculate the total number of items in a DIA.

The second choice are *Streams* for transmitting large amounts of items *asynchronously*. Streams enable bulk all-to-all communication between all workers (see figure 7.11). Thrill contains two subtypes of Streams which differ in the order items are received from other workers: *CatStreams* deliver items strictly in worker rank order, while *MixStreams* deliver items in the arbitrary order in which Blocks are received from the network. Besides transmitting items in Blocks using the BlockReader and BlockWriter classes, Streams can also scatter whole ranges of a File to other workers without an additional deep copy of the Block's data in the network layer. Items in Blocks scattered via Streams to workers on the same host are "transmitted" via reference counting and not deeply copied. All communication with workers on the same host is done via shared memory within the same process space.

All Blocks in a Thrill program are managed by the *BlockPool*. Blocks are reference counted and automatically deleted once they are no longer in any File or used by the network system. The BlockPool also keeps track of the total amount of memory used in Blocks. Once a user-defined limit is exceeded, the BlockPool asynchronously swaps out the least recently used Blocks to a local disk. To distinguish which Blocks may be evicted and which are being used by the data system, Blocks have to be *pinned* to access their data. Pins can be requested asynchronously to enable prefetching from external memory. However, all the complexity of pinning Blocks is hidden in the BlockReader/Writer such as to make implementation of DIA operations easy.

Thrill divides available system memory into three parts (by default equally): BlockPool memory, DIA operations memory, and free floating heap memory for user objects like `std::string`. All memory is tracked in Thrill by hooking `malloc()`, hence the user application needs no special allocators. Memory limits for DIA operations' internal data structures are negotiated and defined when executed during evaluation. The StageBuilder determines which DIA operations participate in a stage and divides the allotted memory fairly between them. It is important for external memory support that the operations adhere to these internal memory limits, e.g. by correctly sizing their hash tables and sort buffers.

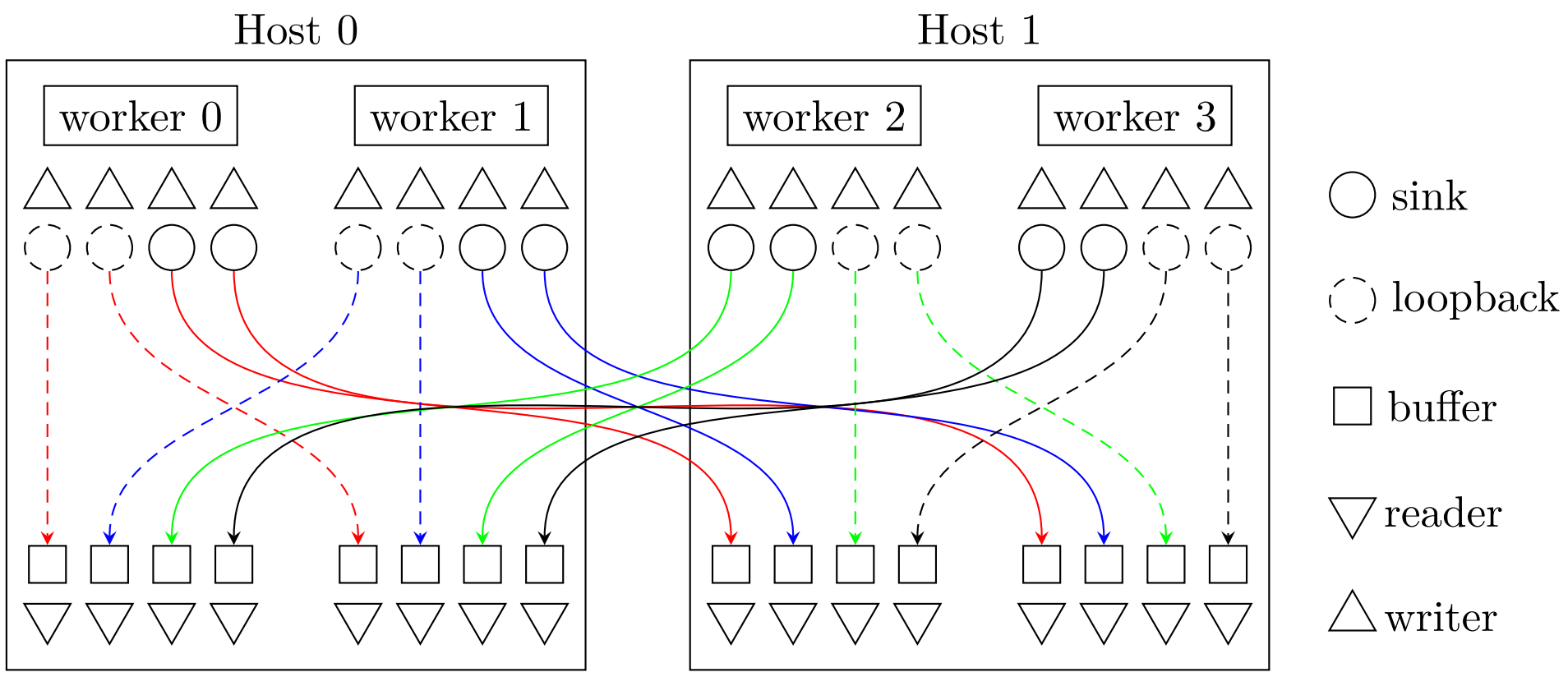

**Figure 7.11:** Thrill's Streams: asynchronous big data all-to-all transfers.

### 7.3.3 Details on the Reduce, Group, and Sort Implementations

Besides pipelining DIA operations, careful implementations of the core algorithms in the operations themselves are important for performance. Most operations are currently implemented rather straight-forwardly, and future work may focus on more sophisticated versions of specific DIA operations. Due to the generic DIA operation interface, these future implementations can then be easily plugged into existing applications.

#### Reduce Operations

*ReduceByKey* and *ReduceToIndex* are implemented using multiple levels of hash tables, because items can be immediately reduced due to associative or even commutative reduction operations $r : A \times A \to A$.

Thrill distinguishes two reduction phases: the pre-phase prior to transmission and the post-phase receiving items from other workers. Items which are pushed into the Reduce-DOp are first processed by the key extractor $k : A \to K$ or index function $i : A \to [0\mathinner{..}n)$ (see table 7.1). The key space $K$ or index space $[0\mathinner{..}n)$ is divided equally onto the range of workers $[0\mathinner{..}p)$. During the pre-phase, each worker hashes and inserts items into one of $p$ separate hash tables, each destined for one worker. If a hash table exceeds its fill-factor, its content is transmitted. If two items with matching keys are found, they are combined locally using $r$ *prior* to transmission.

Items that are received from other workers in the post-phase are inserted into a second level of hash tables. Again, matching items are immediately reduced using $r$. To enable truly massive data processing, Thrill may spill items into external memory during the post-phase. The second level of hash tables are again partitioned into $m$ separate tables. An incoming item is put into one of the $m$ tables and possibly reduced. If any of the $m$ tables exceeds its fill-factor, its whole content is spilled into a File. When all items have been received by the post-phase, the spilled Files are recursively reduced by choosing a new hash function and reusing the $m$ hash tables. The recursive step rereduces the smaller set of items using a new hash space, and if any new Files are spilled, then they are added to the list of Files to process further.

The pre- and post-phases use custom linear probing hash tables with built-in reduction on collisions. One large memory segment is used for $m$ separate hash tables. Initially, only a small area of each partition is filled and used to save allocation time. When a hash table is flushed or spilled, its allocated size is doubled until the memory limit prescribed by the StageBuilder is reached.

### Group Operations

*GroupByKey* and *GroupToIndex* are based on sorting and multiway merging of sorted runs. Items pushed into the Group-DOp are first processed by the key extractor $k : A \to K$ or index function $i : A \to [0\,..\,n)$, the result space $K$ or $[0\,..\,n)$ is distributed evenly onto all $p$ workers. After determining the destination worker, items are immediately transmitted to the appropriate worker via a Stream. Each worker stores all received items in an in-memory vector. Once the vector is full or heap memory is exhausted, the vector is sorted by key, and serialized into a File which may be swapped to external memory. Once all items have been received, the sorted runs are merged using an efficient multiway merger. The stream of sorted items is separated into subsequences with equal keys, and these sequences are delivered to the group function $g : \text{iterable}(A) \to B$ as a multiway merge iterator.

### Distributed Sorting

The operation *Sort* rearranges all DIA items into a global order as defined by a comparison function. In the Link step on each worker, all local incoming items are written to a File. Simultaneously, a random sample is drawn using reservoir sampling [Vit85; Li94] and sent to worker 0 once all items have been seen. In the Main part, Thrill uses Super Scalar Sample Sort [SW04] to redistribute items between workers: worker 0 receives all sample items, sorts them locally, chooses $p-1$ equidistant splitters, and broadcasts the splitters back to all workers. These build a balanced binary tree with $p$ buckets to determine the target worker for each item in $\lceil \log p \rceil$ comparisons. As Super Scalar Sample Sort requires the number of buckets to be a power of two, the tree is filled with sentinels as necessary. Items are then read from the File, classified using the splitter tree, and transmitted via a Stream to the appropriate worker. When a worker reaches its memory limit while receiving items, the items are sorted and written to a File. If multiple sorted Files are created, these are merged during the Push part.

Datasets with many duplicated items can lead to load balance problems if sorting is implemented naively. To mitigate skew, Thrill uses the global array position of the item to break ties and determine its recipient. When an item is equal to a splitter, it will be sent to the lower rank worker if and only if its global array position is lower than the corresponding quantile of workers.

## 7.4 Experimental Results

We compared Apache Spark 2.0.1, Apache Flink 1.0.3, and Thrill using six micro benchmarks on the Amazon Web Services (AWS) Elastic Compute Cloud (EC2). Our benchmark and input set is based on HiBench [HHD+10], which we extended* with implementations for Flink and Thrill.

We selected five micro benchmark kernels: *WordCount*, *PageRank*, *TeraSort*, *KMeans*, and *Sleep*. WordCount is run on two different inputs. To focus on the performance of the frameworks themselves, we attempted to implement the benchmarks equally well using each of the frameworks, and made sure that the same basic algorithms were used. Spark and Flink can be programmed in Java or Scala, and we include implementations of both whenever possible. The code for Spark and Flink benchmarks was taken from different sources, all implementations for Thrill were written by us and are included with the Thrill C++ source code as examples. While we tried to configure Spark and Flink best possible, the complexity and magnitude of configuration options these frameworks provide make it possible that we may have missed some tuning parameters. For the most part we kept the parameters from HiBench. The experiments are run with weak scaling of the input, which means that the input size increases linearly with the number of hosts $h$, where each AWS host has 32 cores.

### 7.4.1 The Micro Benchmarks

Implementations of WordCount were available in Java and Scala from the examples accompanying Spark and Flink. The **WordCount1000** benchmark processes $h \cdot 32$ GiB of text generated by a C++ version of Hadoop's RandomTextWriter. There are only 1 000 distinct words in this random text, which we do not consider a good benchmark for reduce, since only very little data needs to be communicated, but this input seems to be an accepted standard.

*`http://github.com/thrill/fst-bench`

Additionally, we ran WordCount on text data extracts from the CommonCrawl[†] corpus (September 2016). In **WordCountCC**, $h \cdot 128$ gzip-compressed "WET" archives were processed. Each WET archive is about 155 MiB compressed and extracts to approximately 392 MiB of plain text. This sums up to about $h \cdot 19.3$ GiB gzip-compressed text and $h \cdot 49$ GiB uncompressed text. Decompression of the archives is performed on-the-fly by the frameworks. Contrary to the synthetic WordCount1000 benchmark, this real-world text contains a huge number of words with few occurrences.

For **PageRank** we used only implementations which perform ten iterations of the naive algorithm involving a join of the current ranks with all outgoing edges and a reduction to collect all contributions to the new ranks. We took the implementation from Spark's examples and modified it to use integers instead of strings as page keys. We adapted Flink's example to calculate PageRank without normalization and to perform a fixed number of iterations. Thrill emulates a join operation using *ReduceToIndex* and *Zip* with the page id as the index into the DIA. The input graph for the experiments contained $h \cdot 4$ M vertices with an average of 39.5 edges per vertex, totaling $\approx h \cdot 2.7$ GiB in size, and generated using the PagerankData generator in HiBench.

**TeraSort** requires sorting 100 byte records, and we used the standard sort method in each framework. HiBench provided a Java implementation for Spark, and we used an unofficial Scala implementation[‡] [DW15] for Flink. Hadoop's *teragen* was used to generate $h \cdot 16$ GiB as input.

For **KMeans** we used the implementations from Spark and Flink's examples. Spark calls its machine learning package, while Flink's example is a whole algorithm. We made sure that both essentially perform ten iterations of Lloyd's algorithm [Llo82] using random initial centroids, and we implemented this algorithm in Thrill. We fixed the number of dimensions to three, because Flink's implementation required a fixed number of dimensions, and the number of clusters to ten. Following HiBench's settings, Apache Mahout's GenKMeansDataset was used to generate $h \cdot 16$ M sample points, and the binary Mahout format was converted to text for reading with Flink and Thrill ($\approx h \cdot 8.8$ GiB in size).

The **Sleep** benchmark is used to measure framework startup overhead time. It launches one map task per core which sleeps for 60 seconds.

### 7.4.2 The Platform

We performed our micro benchmarks on AWS using $h$ r3.8xlarge EC2 instances. Each instance contains 32 vCPU cores of an Intel Xeon E5-2670 v2 with 2.5 GHz, 244 GiB

[†] `http://commoncrawl.org`

[‡] `https://github.com/eastcirclek/terasort`

**Table 7.2:** Input Size and Resource Utilization of Frameworks during Benchmarks.

| Benchmark | Input Size | | Spark (Scala) | Flink (Scala) | Thrill |
|---|---|---|---|---|---|
| WordCount1000 | $h \cdot 32$ GiB | CPU | 50 s (65 %) | 251 s (73 %) | 43 s (66 %) |
| | | Net | 931 MiB/s | 195 MiB/s | 1 016 MiB/s |
| WordCountCC | $h \cdot 19.3/49$ GiB | CPU | 367 s (60 %) | 776 s (81 %) | 146 s (61 %) |
| | | Net | 107 MiB/s | 70 MiB/s | 284 MiB/s |
| PageRank | $h \cdot 2.7$ GiB | CPU | 383 s (28 %) | 284 s (62 %) | 37 s (17 %) |
| | | Net | 36 MiB/s | 154 MiB/s | 260 MiB/s |
| TeraSort | $h \cdot 16$ GiB | CPU | 73 s (20 %) | 65 s (22 %) | 42 s (21 %) |
| | | Net | 425 MiB/s | 396 MiB/s | 425 MiB/s |
| KMean | $h \cdot 8.8$ GiB | CPU | 81 s (22 %) | 190 s (4.3 %) | 35 s (50 %) |
| | | Net | 64 MiB/s | 27 MiB/s | 250 MiB/s |

The table shows the CPU utilization as seconds and percentage of total running time, and the average network bandwidth in MiB/s, both averaged over all hosts during the run with 16 hosts. TeraSort shows Spark (Java), as we have no Scala implementation.

RAM, and two local 320 GiB SSD disks. We measured 86 GiB/s single-core/L1-cache, 11.6 GiB/s single-core/RAM, and 74 GiB/s 32-core/RAM memory bandwidth using our parallel memory bandwidth benchmark tool *pmbw* (see chapter 3). The SSDs reached 460 MiB/s when reading 8 MiB blocks, and 397 MiB/s when writing.

The $h$ instances were allocated in one AWS availability zone and were connected with a 10 gigabit network. Our network measurements showed $\approx 100\,\mu$s ping latency, and up to 1 GiB/s sustained point-to-point bandwidth. All frameworks used TCP sockets for transmitting data.

We experimented with AWS S3, EBS, and EFS as data storage for the benchmark inputs, but ultimately chose to run a separate CephFS cluster on the EC2 instances. Ceph provided reliable, repeatable performance and minimized external factors in our experiments. Each EC2 instance carried one Ceph ODS on a local SSD, and we configured the Ceph cluster to keep only one replication block to minimize bandwidth due to data transfer. We did not use HDFS because a POSIX-based distributed file system (DFS) provides a standard view for all frameworks. The other SSD was used for temporary files created by the frameworks.

All Spark implementations use the RDD interface. Support for fault tolerance in Spark and Flink incurred no additional overhead, because no checkpoints were written. By default checkpointing is deactivated and must be explicitly configured. All run-time compression was deactivated, and Spark was configured to use Kyro serialization.

We used Ubuntu 16.04 LTS (Xenial Xerus) with Linux kernel 4.4.0-31, Ceph 10.2.2 (jewel), Oracle Java 1.8.0_101, Apache Spark 2.0.1, Apache Flink 1.0.3, and compiled Thrill using gcc 5.4.0 with cmake in Release mode.

### 7.4.3 The Results

Figure 7.12 shows the median result of three benchmark runs for $h = 1, 2, 4, 8, 16$ hosts. We plotted the throughput per host in MiB/s with the input size in raw bytes, which is roughly proportional to the number of items. The issue here is that the benchmark inputs use different units. For example, WordCount's input is counted in words or characters, while PageRank's input is measured in number of edges, and KMeans's input in number of points and clusters.

Figure 7.13 shows the same results as Figure 7.12, except plotted as the *slowdown* in running time of each framework over the fastest. Additionally, we measured a performance profile of the CPU, network, and disk I/O utilization during the benchmarks using information from the Linux kernel. The profiles are plotted in figures 7.14 to 7.16 for $h = 16$ over the execution time of each benchmark run and summarized results are shown in table 7.2.

Thrill consistently outperforms Spark and Flink in all benchmarks on all numbers of hosts, and is often several times faster than the other frameworks. The speedup of Thrill over Spark and Flink is often highest on a single host, and grows smaller as network and disk I/O become bottlenecks.

In WordCount1000, the text is read from the DFS, split into words, and the word pairs are reduced locally. As only 1 000 unique words occur, the overall result is small and communication thereof is negligible. Thrill maximizes network utilization with 1 016 MiB/s via the DFS and uses 66% of the available CPU time for splitting and reducing. Spark also nearly maximizes the network with 931 MiB/s, and utilizes the CPU 65% of the running time. Flink is a factor 5.2 slower than Thrill in WordCount with 16 hosts, uses the CPU 73% of the time, and is not network bound. Thrill's reduction via hash tables are very fast, Spark is about the same, while Flink requires considerably more CPU time for the same task. With 16 hosts Thrill is network bound due to the DFS, and Spark (Scala) is only a factor 1.2 slower.

In WordCountCC, the DFS is no longer the bottleneck since the text archives are gzip-compressed. Surprisingly, the on-the-fly decompression was less overhead than expected, hence, this benchmark focuses on the reduction itself. As the reduced entries contain a dynamically allocated string, we believe that heap memory allocation strategies play the most important role in this benchmark. With 16 hosts Thrill is a factor 2.5 faster than Spark, and a factor of almost 4 faster than Flink.

In PageRank, the current rank values are joined with the adjacency lists of the graph and transmitted via the network to sum all rank contributions for the next iteration in a reduction. Hence, the PageRank benchmark switches back and forth ten times between high CPU load while joining, and high network load while reducing. These cycles are clearly visible in the execution profile (figure 7.15). Spark (Java) is a factor 4.5 slower than Thrill on 16 hosts, while Flink (Java) is a factor 1.9 slower. Flink's pipelined execution engine works well in this benchmark, and reaches 62% CPU and 15% network utilization. From the internal profile of Spark one can see that it does not balance work well between the hosts due to stragglers. Hence, each iteration takes longer than necessary. We believe Thrill's performance could be increased even further by using a *Join* algorithm.

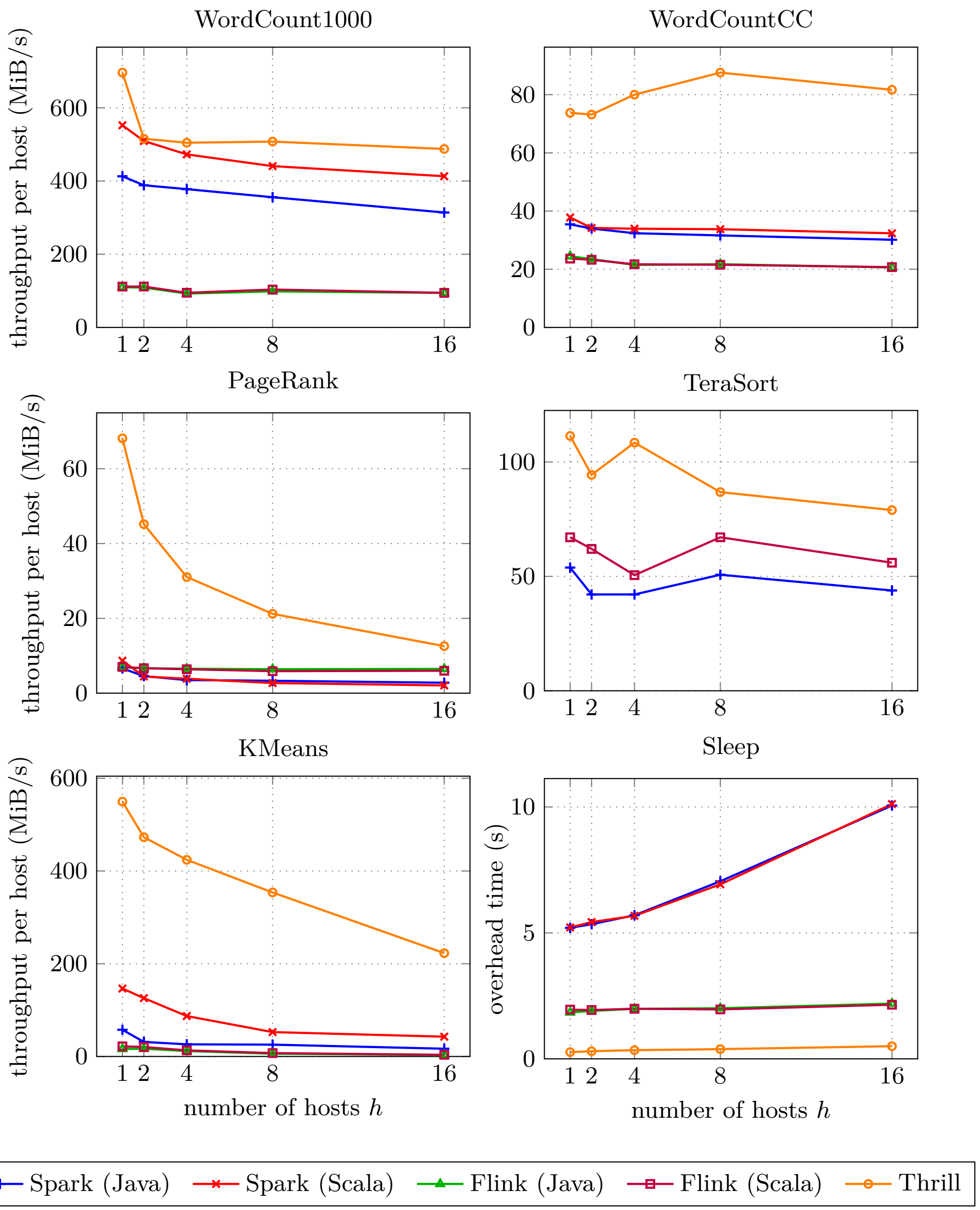

**Figure 7.12:** Experimental results of Apache Spark 2.0.1, Apache Flink 1.0.3, and Thrill on $h$ AWS r3.8xlarge hosts

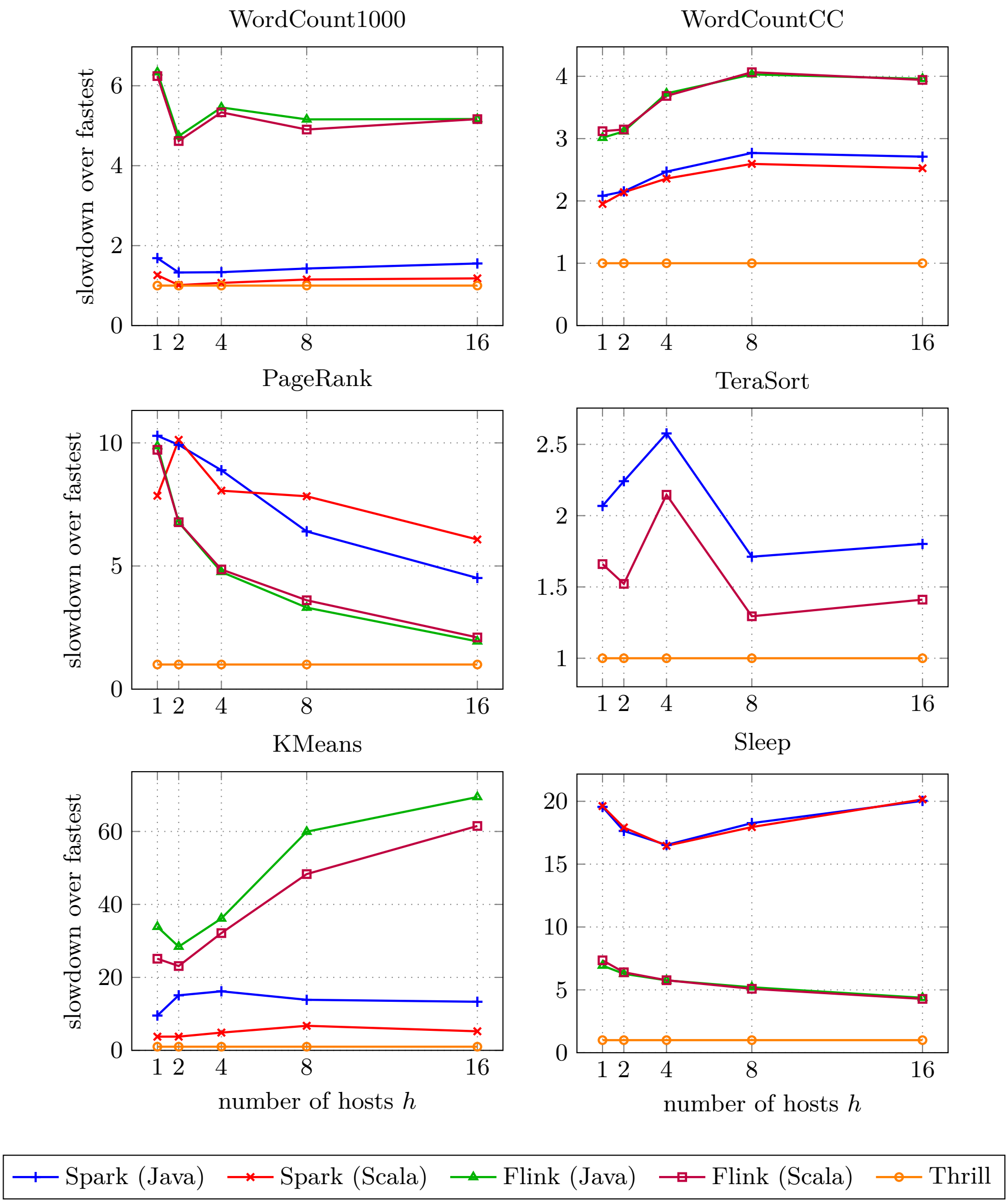

**Figure 7.13:** Slowdown of Apache Spark 2.0.1, Apache Flink 1.0.3, and Thrill on $h$ AWS r3.8xlarge hosts over the fastest framework

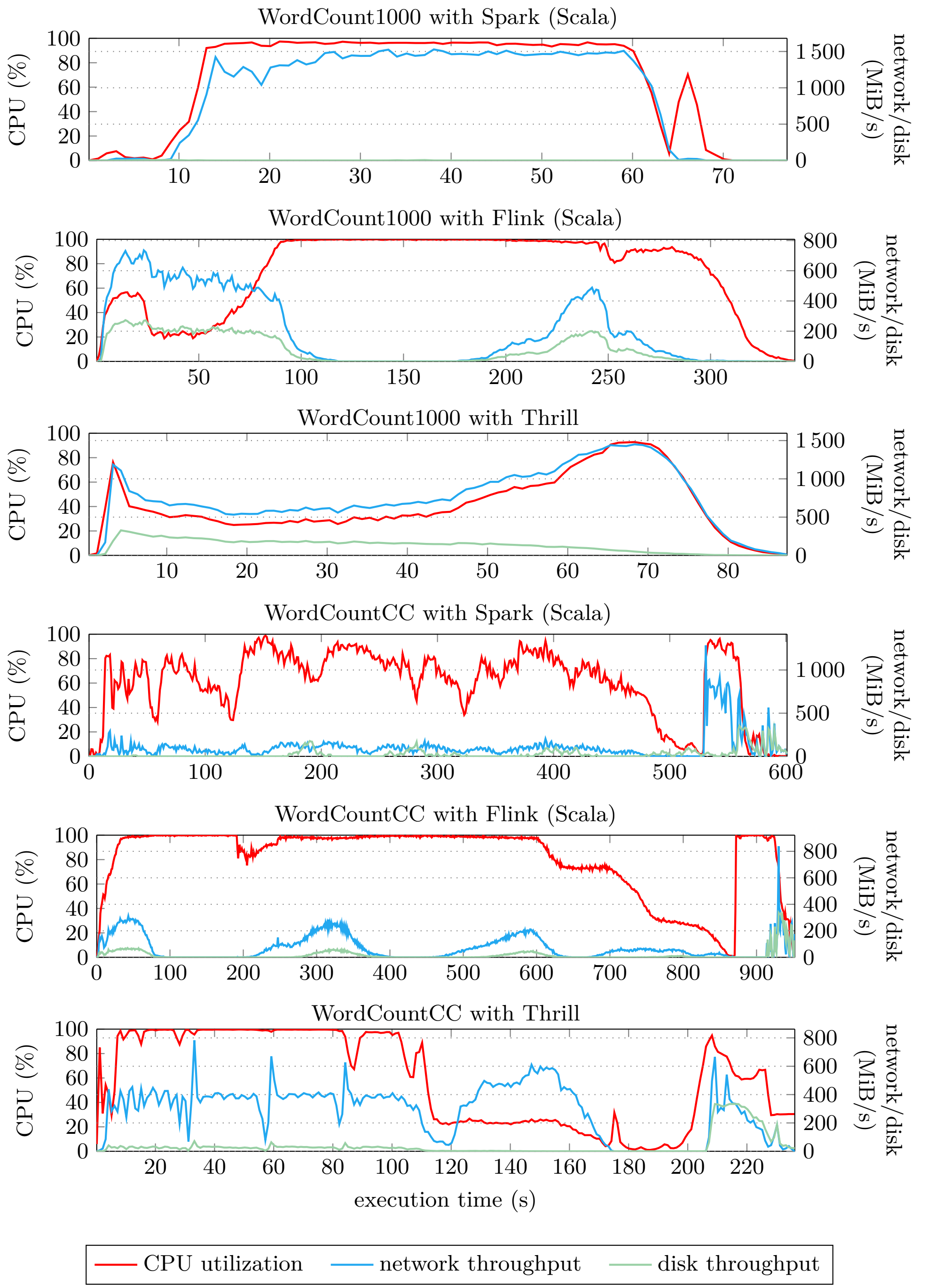

**Figure 7.14:** CPU utilization, and network and disk throughput averaged over all hosts during the median WordCount1000 and WordCountCC benchmark runs with 16 hosts.

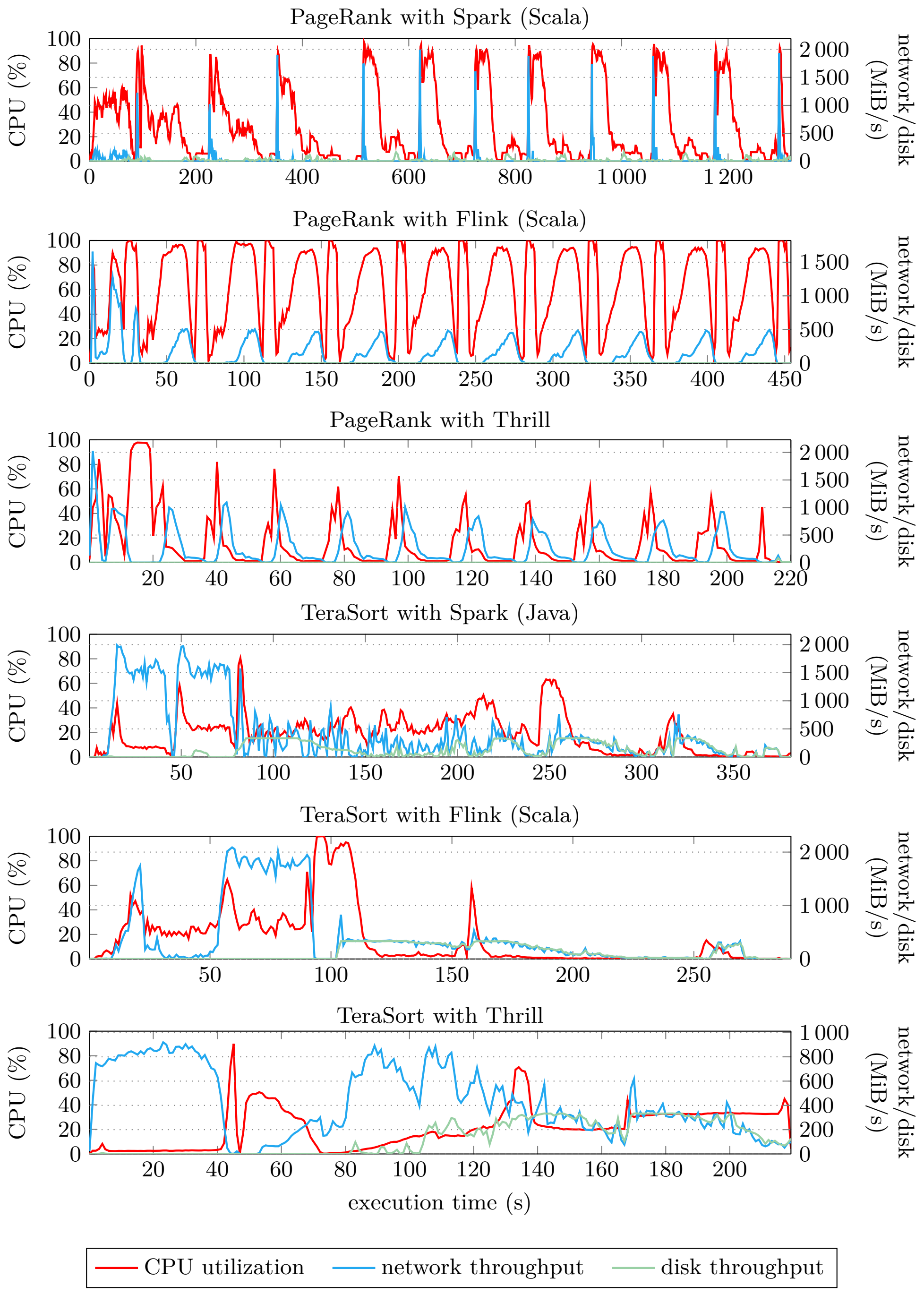

**Figure 7.15:** CPU utilization, and network and disk throughput averaged over all hosts during the median PageRank and TeraSort benchmark runs with 16 hosts.

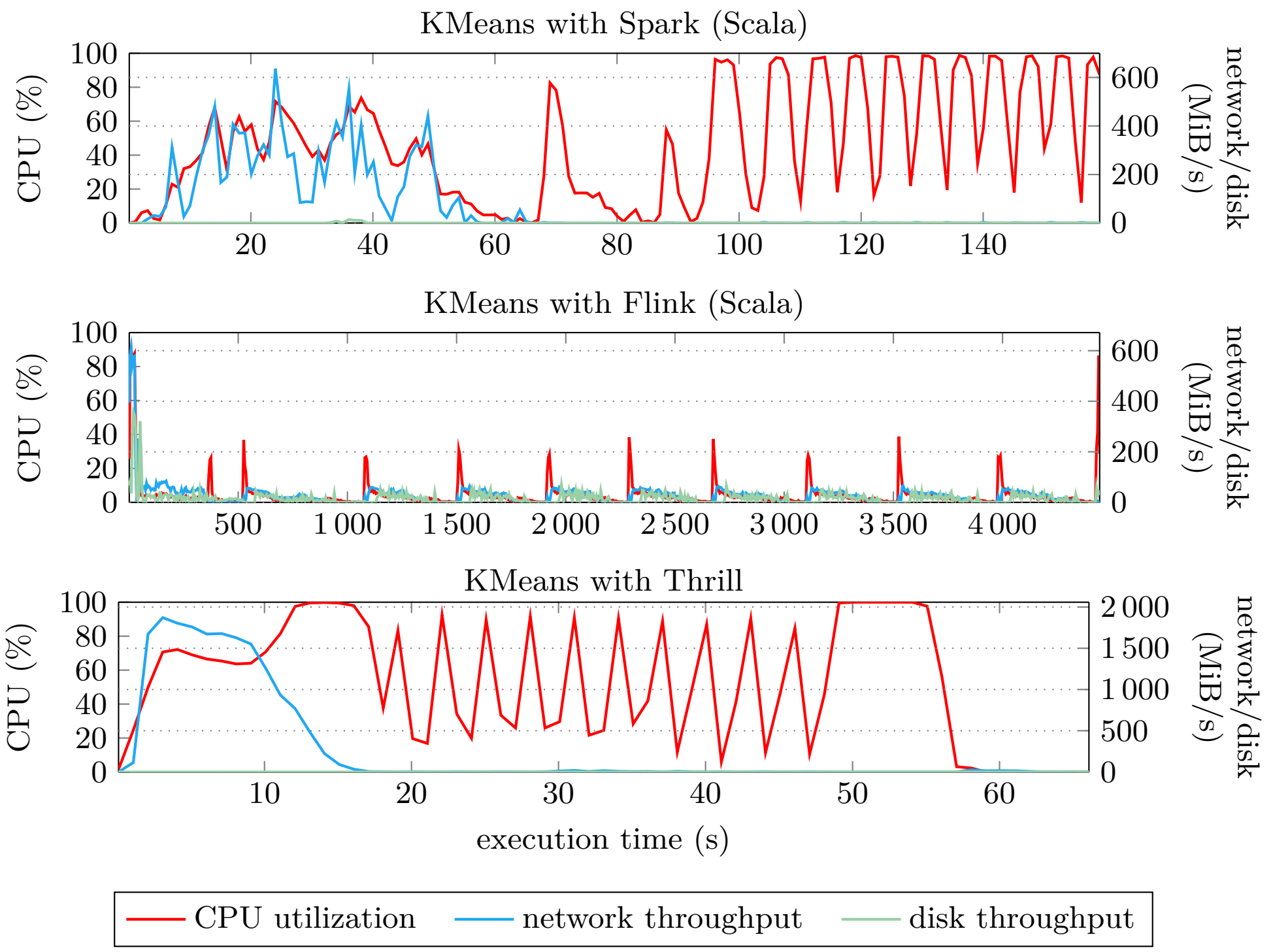

**Figure 7.16:** CPU utilization, and network and disk throughput averaged over all hosts during the median KMeans benchmark runs with 16 hosts.

In TeraSort, Spark is only a factor 1.8 slower and Flink a factor 1.4 than Thrill on 16 hosts. Spark reaches only 20% CPU and 42% network utilization on average, Flink 22% and 40%, and Thrill 21% and 43%, respectively. Flink's pipelined execution outperforms Spark in TeraSort, as was previously shown by another author [DW15]. The different phases of distributed sorting — local run formation, partitioning, exchange, and local merging — can be identified in the execution profiles in (figure 7.15). The implementations appear well tuned, however, due to the low CPU and network utilization, we believe all can be improved.

In the KMeans algorithm, the set of centroids are broadcast. Then all points are reclassified to the closest centroid, after which new centroids are determined from all points via a reduction. Like PageRank, the KMeans algorithm interleaves high local work and high network load (a reduction and a broadcast). Spark (Scala) is a factor 5.2 slower than Thrill on 16 hosts, Spark (Java) a factor 13, and Flink more than 60. We believe this is due to the JVM object overhead for vectors, and to inefficiencies in the way Spark and Flink broadcast the centroids. Flink's query optimizer does not seem to work well for the KMeans example accompanying their source package. Thrill utilizes the CPU 50% and the network 25% of the running time, while Spark reach 22% CPU and only 6% network utilization.

The Sleep benchmark highlights the startup time of the frameworks. We plotted the running time excluding the slept time in figure 7.12. Spark requires remarkably close to $5 + h \cdot 0.4$ seconds to start up. Apparently, hosts are not started in parallel. Flink's start up time was much lower, and Thrill's less than one second.

## 7.5 Conclusion and Future Work

With Thrill we have demonstrated that a C++ library can be used as a distributed data processing framework reaching a similarly high level of abstraction as the currently most popular systems based on Java and Scala while gaining considerable performance advantages. In the future, we want to use Thrill on the one hand for implementing scalable parallel algorithms (e.g. for construction of succinct text indices) that are both advanced and high level. Thrill has already been used for more than five suffix sorting algorithms, on which we will focus in the next chapter. Additionally, many smaller examples such as logistic regression, graph generators, triangle counting, stochastic gradient descent, and more have been implemented by contributors. On the other hand, at a much lower level, we want to use Thrill as a platform for developing algorithmic primitives for big data tools that enable massively scalable load balancing, communication efficiency, and fault tolerance.

While Thrill is so far a prototype and research platform, the results of this dissertation are sufficiently encouraging to see a possible development into a main stream big data processing tool. Of course, a lot of work remains in that direction such as implementing interfaces for other popular tools like Hadoop and the AWS stack, and creating frontends in scripting languages like Python for faster algorithm prototyping. To achieve practical scalability and robustness for large clusters, we also need significant improvements in issues like load balancing, fault tolerance and native support for high performance networks like InfiniBand or Omni-Path.

Furthermore, we view it as useful to introduce additional operations and data types like graphs and multidimensional arrays in Thrill (see also section 7.2.4). But, we are not sure whether automatic query plan optimization as in Flink should be a focus of Thrill, because that makes it more difficult to implement complex algorithms with a sufficient amount of control over the computation. Rather it may be better to use Thrill as an intermediate language for a yet higher level tool that would no longer be a plain library but a true compiler with a query optimizer.

## Acknowledgment

We would like thank the AWS Cloud Credits for Research program for making the experiments in section 7.4 possible. Our research was supported by the Gottfried Wilhelm Leibniz Prize 2012, and the Large-Scale Data Management and Analysis (LSDMA) project in the Helmholtz Association.

**Chapter 8**

# 8 Suffix Array Construction with Thrill

*The suffix array is the key to efficient solutions for myriads of string processing problems in different applications domains, like data compression, data mining, or bioinformatics. With the rapid growth of available data, suffix array construction algorithms have to be adapted to advanced computational models such as external memory and distributed computing. In this chapter, we present five suffix array construction algorithms utilizing the new algorithmic big data batch processing framework Thrill from chapter 7, which allows us to conveniently write efficient distributed external memory algorithms.*

Suffix arrays [MM90; MM93; GBYS92] are the basis for many text indices and string algorithms. Suffix array construction is theoretically linear work, but practical suffix sorting is computationally intensive and often limits the applicability of advanced text data structures on large datasets. While fast sequential algorithms exist in the RAM model [Mor06; NZC09a], these are limited by the CPU power and RAM size of a single machine. External memory algorithms on a single machine are limited by disk space [DKMS08; BFO13; KKP15b], and often have long running times due to mostly sequential computation or limited I/O bandwidth.

Most suffix array construction algorithms focus only on sequential computation models. However, while the volume of data is increasing, the speed of individual CPU cores is not, as discussed in the introduction (chapter 1). This leaves us no choice but to consider shared-memory parallelism and distributed cluster computation to gain considerable speedups in the future. For this reason, we propose to use our new big data processing framework *Thrill* to implement *distributed external memory* suffix sorting algorithms for large inputs.

As already reviewed in the brief history of suffix array construction in chapter 5, most suffix array construction algorithms use a subset of three basic suffix sorting principles: *prefix doubling*, *recursion*, and *inducing* [PST07]. Our first three algorithms in Thrill are based on prefix doubling. In section 8.2 we first review this technique in a RAM setting, and then show how to implement it using only the scalable primitives provided by Thrill. The result are prefix doubling *using the inverse suffix array* (section 8.2.3), prefix doubling *using sorting* (section 8.2.4), and prefix doubling *with discarding* (section 8.2.5).

Our last two algorithms are distributed formulations of the linear-time difference cover algorithm *DC* presented by Kärkkäinen, Sanders, and Burkhardt [KS03; KSB06].

These employ sorting, recursion, prefix sums, and merging of arrays, which are all scalable primitives in Thrill. In section 8.3 we recall the DC3 algorithm in abstract form, and then present DC3 and DC7 in section 8.3.1 using concrete Thrill pseudocode.

Concerning pseudocode, we would like to warn the reader: the amount in this chapter is daunting. However, we try to mitigate the highly technical tuple code with easy to read comments alongside it, and with additional explanations in the text. In concert, Thrill pseudocode and comments yield a concrete and formalized expression of complex algorithms.

In section 8.4 we run the five Thrill implementations on up to 32 hosts with fast NVMe SSDs and limited RAM in the AWS Elastic Compute Cloud (EC2). We compare them against two independent MPI implementations and the two fastest non-distributed sequential suffix sorters as a baseline. Using 32 hosts, we can suffix sort 16 GiB of Wikipedia text in 30 min, or 16 GiB of digits of $\pi$ in 15 min. Our Thrill implementations scale higher than the MPI implementations, which are constrained by RAM and fail after 2 GiB. Comparing to the fastest sequential suffix sorters, our best Thrill implementations outperform on digits of $\pi$ when run with 2 hosts (32 cores), and on Wikipedia when run with 4 hosts (64 cores). While these results are impressive, we believe future work on Thrill's core algorithms can further improve the performance.

This chapter is based on a technical report [BGK16] written together with Florian Kurpicz, who implemented most of the prefix doubling algorithms, while we focused on the difference cover algorithms. All text was rewritten for this dissertation to improve clarity, but most of the algorithm pseudocode remained identical to our previous report except for minor corrections. The experiments in section 8.4 were newly created and may lead to a future conference publication.

## 8.1 Related Work

While there exist numerous papers and articles on sequential suffix array construction, as considered in our history chapter 5, there is much less work on distributed suffix sorting and no publications using distributed external memory. In this section we only review the publications most relevant for our implementations in Thrill, please refer to section 5.4 (page 180) for a longer discussion of previous external, and section 5.5 (page 186) for previous distributed suffix sorting algorithms.

In 2003, Kärkkäinen, Sanders, and Burkhardt [KS03; KSB06] presented a linear-time suffix sorting algorithm, the so called DC3 algorithm, that works well in multiple advanced models of computation such as external memory and also parallel and distributed environments. Kulla and Sanders later demonstrated the scalability of the DC3 algorithm [KS06b; Kul06; KS07] by presenting a MPI implementation.

More recently, Flick and Aluru presented an implementation of a prefix doubling algorithm in MPI that can also compute the longest common prefix array [FA15]. Suffix array construction has also been considered in external memory, where in theory the DC3 algorithm is optimal. Dementiev, Kärkkäinen, Mehnert, and Sanders [DKMS05; DKMS08] compared multiple variants of prefix doubling and DC3 for external memory in practice. Our prefix doubling and difference cover algorithm implementations in Thrill are based on these preceding publications.

As we successfully redesigned induced sorting for external memory in chapter 6, the question naturally arises why we did not consider it in a distributed environment. The problem is that induced sorting appears very difficult to parallelize. Up to now, it yields only well to parallelization for specific inputs on small alphabets [LSB16; LSB17], and hence does not appear to be a promising approach for general inputs in a highly parallel and distributed setting. However, more future work in this direction is needed, as important applications such as bioinformatics have small alphabets, and parallelization of induced sorting is possibly easier when considering *generalized* suffix array construction.

## 8.2 Prefix Doubling Algorithms

In this section, we start by reviewing *prefix doubling* for suffix sorting. But first, let us take a step back and consider that the suffix array can just as well be constructed using naive string sorting: just treat all suffixes as individual strings, and apply any string sorting algorithm from part I. This approach however completely ignores the special property that *suffixes of suffixes of $T$ are themselves suffixes of $T$*. If a comparison-based string sorting algorithm takes $\Omega(n \log n)$ time for $n$ characters (disregarding the distinguishing prefix), then suffix sorting a string of length $n$ may take $\Omega(n^2 \log n)$ time, because the corresponding naive string sorting problem contains $\frac{n(n+1)}{2}$ characters. Likewise, a naive radix sort approach to suffix sorting may take $\Omega(n^2)$ time. These complexities are prohibitive for large inputs.

In their seminal paper introducing the suffix array, Manber and Myers [MM90; MM93] already presented a suffix sorting algorithm with $\mathcal{O}(n \log n)$ time. Like naive string sorting, they too consider suffix sorting as sorting an array $[0 \mathinner{..} n)$ by the suffix strings. However, they *reuse* the information gained from previous sorting steps to further sort suffixes of suffixes of $T$ and advance the depth by doubling the common prefix in each step [KMR72] to achieve logarithmic depth. This is called *prefix doubling*.

For better exposition of these ideas, we define the *lexicographic $h$-order* or just *$h$-order* $\leq_h$ on strings as their lexicographic order limited to depth $h$: $a \leq_h b$ if and only if $a|_h \leq b|_h$, where $a|_h$ or $b|_h$ is the string $a$ or $b$ truncated to $h$ characters. Other comparison operators like $a =_h b$ and $a <_h b$ are defined accordingly. For $h < n$ the $h$-order of suffixes of a string $T$ may not be unique, e.g. with respect to $\leq_2$ two suffixes starting with the same 2 characters are considered equal and hence their order is not fixed. A set of suffixes equal under $=_h$ is called an *$h$-group* and they all start with the same $h$ characters.

The $\mathcal{O}(n \log n)$ suffix sorting algorithm by Manber and Myers [MM90; MM93] is based on the following fact:

**Observation 8.1 (Gaining a 2$h$-Order of Suffixes from an $h$-Order [MM90])** *Given a lexicographic $h$-order $\leq_h$ for the suffixes of a string $T$, one can gain a lexicographic $2h$-order $\leq_{2h}$ by sorting all suffixes $T[i..n)$ by their rank in $\leq_h$ as primary key and by the rank of $T[i+h..n)$ in $\leq_h$ as secondary key.*

Instead of Manber and Myers's [MM93] rather complex suffix sorting algorithm, we want to review the improved *qsufsort* prefix doubling algorithm by Larsson and Sadakane [Sad98; LS99; LS07]. The main improvements over the previous algorithm [MM93] are that it is simpler, faster, and introduces an array $L$ for better bookkeeping of already sorted groups.

### 8.2.1 Prefix Doubling using the Inverse Suffix Array in RAM

We express the idea behind the qsufsort algorithm in a slightly different form than the original literature: we call an index a *lexicographic $h$-name* or just *$h$-name* for $T[i..n)$ if it is *a* rank with respect to $\leq_h$. Since these ranks are not unique, we will settle on using the lowest rank as lexicographic $h$-name. For example, for the string $[\,\mathtt{a},\mathtt{b},\mathtt{a},\mathtt{b},\$\,]$, the suffix $[\,\$\,]$ has 2-name 0, $[\,\mathtt{a},\mathtt{b},\mathtt{a},\mathtt{b},\$\,]$ and $[\,\mathtt{a},\mathtt{b},\$\,]$ have 2-name 1, and $[\,\mathtt{b},\mathtt{a},\mathtt{b},\$\,]$ and $[\,\mathtt{b},\$\,]$ have 2-name 3.

Using the previous definitions, we can now define the $h$-suffix array $\mathsf{SA}^h$ and the inverse $h$-suffix array $\mathsf{ISA}^h$. The *$h$-suffix array* contains a permutation of the indices $[0\mathinner{..}n)$ such that $T[\mathsf{SA}^h[i]..n) \leq_h T[\mathsf{SA}^h[j]..n)$ for all $i < j$, which is the same condition as the suffix array, except limited to the first $h$ characters. Because the $h$-order is not necessarily unique, a string may have different $\mathsf{SA}^h$. For example, for the string $[\,\mathtt{a},\mathtt{b},\mathtt{a},\mathtt{b},\$\,]$ both $[\,5,0,2,1,3\,]$ and $[\,5,2,0,3,1\,]$ are possible $\mathsf{SA}^2$.

Likewise, the *inverse $h$-suffix array* shall contain at $\mathsf{ISA}^h[i]$ the *smallest* lexicographic $h$-name of $T[i..n)$ with respect to $\leq_h$. This implies $\mathsf{ISA}^h[i] < \mathsf{ISA}^h[j]$ if $T[i..n) <_h T[j..n)$ and $\mathsf{ISA}^h[i] = \mathsf{ISA}^h[j]$ if $T[i..n) =_h T[j..n)$. While $\mathsf{ISA}^h$ is not necessarily a permutation, as $h$-order ranks are not unique, $\mathsf{ISA}^h$ itself *is* unique for a string due to our additional requirement of choosing the smallest rank. For the example string $[\,\mathtt{a},\mathtt{b},\mathtt{a},\mathtt{b},\$\,]$, only $[\,1,3,1,3,0\,]$ is a valid $\mathsf{ISA}^2$. Obviously, for $h \geq n$ the new arrays $\mathsf{SA}^h$ and $\mathsf{ISA}^h$ are equivalent to the suffix and inverse suffix arrays.

Prefix doubling algorithms are based on calculating $\mathsf{SA}^{2h}$ and $\mathsf{ISA}^{2h}$ from $T$, $\mathsf{SA}^h$, and $\mathsf{ISA}^h$, the difficulty is how to organize sorting to exploit observation 8.1. Larsson and Sadakane's [Sad98; LS99; LS07] qsufsort requires an additional array $L$ which keeps track of $h$-groups in the arrays. The following steps give a outline of the algorithm and figure 8.1 shows an example execution.

(i) Initialize $\mathsf{SA}^0$ as $\mathsf{SA}^0[i] = i$ such that $[\,0,1,\ldots,n-1\,]$ enumerates all suffixes of $T$ with regard to $\leq_0$.

| $i$ | 0 | 1 | 2 | 3 | 4 | 5 | 6 | 7 | 8 | 9 | 10 | 11 | 12 | 13 |
|---|---|---|---|---|---|---|---|---|---|---|---|---|---|---|
| $T[i]$ | t | o | b | e | o | r | n | o | t | t | o | b | e | $ |
| $h = 1$, $\mathsf{SA}^1[i]$ | 13 | 2 | 11 | 3 | 12 | 6 | 1 | 4 | 7 | 10 | 5 | 0 | 8 | 9 |
| $L^1[i]$ | -1 | 2 | | 2 | | -1 | 4 | | | | -1 | 3 | | |
| $\mathsf{ISA}^1[i]$ | 11 | 6 | 1 | 3 | 6 | 10 | 5 | 6 | 11 | 11 | 6 | 1 | 3 | 0 |
| $\mathsf{ISA}^1[\mathsf{SA}^1[i]+1]$ | | 3 | 3 | 6 | 0 | | 1 | 10 | 11 | 1 | | 6 | 11 | 6 |
| $h = 2$, $\mathsf{SA}^2[i]$ | 13 | 2 | 11 | 12 | 3 | 6 | 1 | 10 | 4 | 7 | 5 | 0 | 9 | 8 |
| $L^2[i]$ | -1 | 2 | | -3 | | | 2 | | -3 | | | 2 | | -1 |
| $\mathsf{ISA}^2[i]$ | 11 | 6 | 1 | 4 | 8 | 10 | 5 | 9 | 13 | 11 | 6 | 1 | 3 | 0 |
| $\mathsf{ISA}^2[\mathsf{SA}^2[i]+2]$ | | 8 | 0 | | | | 4 | 3 | | | | 1 | 1 | |
| $h = 4$, $\mathsf{SA}^4[i]$ | 13 | 11 | 2 | 12 | 3 | 6 | 10 | 1 | 4 | 7 | 5 | 0 | 9 | 8 |
| $L^4[i]$ | -11 | | | | | | | | | | | 2 | | -1 |
| $\mathsf{ISA}^4[i]$ | 11 | 7 | 2 | 4 | 8 | 10 | 5 | 9 | 13 | 11 | 6 | 1 | 3 | 0 |
| $\mathsf{ISA}^4[\mathsf{SA}^4[i]+4]$ | | | | | | | | | | | | 8 | 0 | |
| $h = 8$, $\mathsf{SA}^8[i]$ | 13 | 11 | 2 | 12 | 3 | 6 | 10 | 1 | 4 | 7 | 5 | 9 | 0 | 8 |
| $L^8[i]$ | -14 | | | | | | | | | | | | | |

**Figure 8.1:** Example run of Larsson and Sadakane [LS99; LS07] prefix doubling suffix sorting algorithm. Example adapted from their paper with our notation.

(ii) Sort array $\mathsf{SA}^0$ by $T[i]$ for index $i$ to get $\mathsf{SA}^1$. Set $h = 1$, and mark in $L$ the length of each unfinished 1-group (buckets of equal characters). Each finished singleton bucket (containing only one suffix) is marked in $L$ with a $-1$, add adjacent $-1$ entries to accelerate later skips.

(iii) Scan over $\mathsf{SA}^1$ and set $\mathsf{ISA}^1[i]$ to the current lexicographic 1-name of suffix $T[i..n)$ by setting $\mathsf{ISA}^1[\mathsf{SA}^1[i]] = i'$, but only update $i'$ when the 1-group (bucket) changes.

(iv) Sort each unfinished $h$-group in $\mathsf{SA}^h$ (with ternary-quicksort or any other integer sorting algorithm) using $\mathsf{ISA}^h[\mathsf{SA}^h[i] + h]$ as the sort key for index $i$. This generates $\mathsf{SA}^{2h}$ in-place. Mark in $L$ the length of each unfinished $2h$-group (regions of equal keys), flip the sign bit for finished singletons and consolidate adjacent negative numbers.

(v) Scan over $\mathsf{SA}^h$ and update $\mathsf{ISA}^h$ as in 8.1 (iii) for all processed groups, finished groups can be skipped. Set $h = 2h$.

(vi) If $\mathsf{SA}^h$ still contains any unsorted groups (check via $L[0]$), go to step 8.1 (iv).

Since $h$ doubles in each round, $h \geq n$ after $\lceil \log_2 n \rceil$ rounds and the algorithm above has worst case running time $\mathcal{O}(n \log n)$. To be more precise, the algorithm terminates when $\mathsf{SA}^h$ has no more unsorted groups and becomes the suffix array. This already happens after $\lceil \log_2(\text{maxlcp}(T)) \rceil$ iterations, yielding $\mathcal{O}(n \log(\text{maxlcp}(T)))$ running time.

**Algorithm 8.1 :** Generic Prefix Doubling Algorithm.

```
function PrefixDoubling(T ∈ DIA⟨Σ⟩)
    S := T.Window_2((i, [t_0, t_1]) ↦ (i, t_0, t_1))        // Initial triples (i, T[i], T[i+1]).
    for k := 1 to ⌈log_2 |T|⌉ − 1 do
        S := S.Sort((i, n_0, n_1) by (n_0, n_1))            // Sort triples by name pair.
        N := S.FlatWindow_2((j, [a, b]) ↦ CmpName(j, a, b))  // Outputs 0 or j.
        if N.Filter((i, n) ↦ (n = 0)).Size() = 1 then       // If all names distinct, then
            return N.Map((i, n) ↦ i)                        // return names as suffix array,
        N := N.PrefixSum((i, n), (i′, n′) ↦ (i′, max(n, n′)))  // else make new names
        S := Generate new name pairs using N                // and run next iteration.
```

The space requirements of Larsson and Sadakane's [Sad98; LS99; LS07] prefix doubling algorithm are only $T$, $\mathsf{SA}^h$, $\mathsf{ISA}^h$ and $L$. In their papers [LS99; LS07] they detail many refinements to the base algorithm above, including how to eliminate the extra array $L$ by placing the group length information into $\mathsf{SA}^h$. So beyond the input text the algorithm requires only $2n \log_2 n$ bits if indices are stored as $\log_2 n$ data types, or $8n$ bytes if indices are 4 bytes.

A worst-case input for a prefix doubling algorithm is simply a unary string, like $[\,\mathsf{a}, \mathsf{a}, \ldots, \mathsf{a}, \$\,]$, because these have the highest LCP value sum possible.

### 8.2.2 Distributed External Prefix Doubling with Thrill

Dementiev, Kärkkäinen, Mehnert, and Sanders [DKMS05; DKMS08] adapted the prefix doubling idea to external memory comparing six different external suffix sorting implementations. Based on their work, in this section we will describe how to adapt prefix doubling to distributed external memory using Thrill.

The essential goal of a prefix doubling algorithm is to give each suffix of $T$ a lexicographic $2^k$-name in iteration $k$ using information from iteration $k-1$. More precisely, observation 8.1 states that one can compute a $2^k$-name for the prefix $T[i..i+2^k)$ of suffix $T[i..n)$ using already computed $2^{k-1}$-names of the prefixes $T[i..i+2^{k-1})$ and $T[i+2^{k-1}..i+2^k)$. The main idea to bringing this to external and distributed memory is to store and sort tuples $(i, n_i)$ containing names $n_i$ in such a way that we gain triples $(i, n_i, n_{i+2^{k-1}})$ in iteration $k$. Using two such triples one can take the step from $2^{k-1}$-names to $2^k$-names: consider $(i, n_i, n_{i+2^{k-1}})$ and $(j, n_j, n_{j+2^{k-1}})$ with $n_i = n_j$. This means that suffixes $i$ and $j$ start with the same $2^{k-1}$ characters, $T[i..n) =_{(2^{k-1})} T[j..n)$. By comparing $n_{i+2^{k-1}}$ and $n_{j+2^{k-1}}$, we can determine the lexicographic order of the next $2^{k-1}$ characters, and hence compute new lexicographic names. As the depth of the lexicographic names doubles in each iteration, the algorithm runs at most $\lceil \log_2 n \rceil$ iterations.

**Algorithm 8.2 :** Identifications of Suffix Array Intervals.

```
1 function CmpName(j ∈ ℕ₀, (i, n₀, n₁), (i', n'₀, n'₁) ∈ N)
2     if j = 0 then emit (i, 0)                    // First DIA item has no offset.
3     if (n₀, n₁) ≠ (n'₀, n'₁) then emit (i', j)   // Set name if name pair differs,
4     else emit (i', 0)    // otherwise T[i..n) and T[i'..n) get the previous name.
```

Algorithm 8.1 describes the basic structure of the prefix doubling algorithms presented in this section. The whole algorithm requires one `DIA` $N$ storing pairs and one `DIA` $S$ storing triples.

For the first iteration, $S$ contains the triples $(i, T[i], T[i+1])$ for all $i = 0, \ldots, n-1$ (line 2). These triples contain a text position and the *name pair* for that position, i.e, the two names that are required to compute the new name for the suffix starting at the text position. For bootstrapping the first iteration $k = 1$, we can simply use the characters as lexicographic 1-names.

In our actual implementation, we accelerated the first iteration with *alphabet compression.* Quite often the input string does not use the whole alphabet range. Hence, one can reduce the alphabet size by first counting how many distinct characters occur in the input, and then monotonously mapping them to a compressed range $[0 .. |\Sigma|)$. The input string and the mapped string have the same suffix array, as the lexicographic order of suffixes does not change. In our implementation, we then *pack* as many mapped characters as possible into an integer index, and hence sort to a higher depth in the first iteration. This optimization is particularly important for DNA input.

For subsequent iterations, we continue on line 4 and sort $S$ with respect to the name pair, which brings *equal* $2^{k-1}$-names together. These entries with equal $2^{k-1}$-name need to be extended to prefix depth $2^k$. These new $2^k$-names are calculated using a $\text{FlatWindow}_2$ on $S$ (line 5) and the function CmpName(), which takes the current position $i$ in $S$ and the items $S[i]$ and $S[i+1]$ as input and emits a tuple consisting of a text position and a new name (algorithm 8.2). We know that the suffixes are sorted with respect to their name pairs. Therefore, we can scan $S$ and mark every position where the name pair differs from its predecessor. CmpName() marks these non-unique name pairs by giving them the name 0. All unique name pairs get a name equal to their current position in $S$. If there is only one suffix with name 0, then we know that all names differ and that we have finished the computation, see line 6. Otherwise, we can use the DIA operation PrefixSum() with a *max* operator to set the name of each tuple to the largest preceding name (line 8). The sequence of names is initialized by emitting an arbitrary first name (zero in line 2 of algorithm 8.2) as first item in the DIA. With this extra item, the name array $N$ always contains $|T|$ items. Now each suffix has a new, more refined name.

The next step (line 9) is to identify the ranks of the suffixes required for the next doubling step. During the $k$-th doubling step, we fill $S$ with one triple for each index

**Algorithm 8.3 :** Prefix Doubling using the Inverse Suffix Array.

```
1 function PrefixDoublingISAWindow(k ∈ ℕ_0)
2     N := N.Sort((i, n) by i)                                  // Compute ISA^{2^k}.
3     S := N.Window_{2^k+1}( (j, [(i, n), ..., (i', n')]) ↦ { (i, n, n')  if j + 2^k < |T|,
                                                              (i, n, 0)   otherwise. } )
```

$i = 0, \ldots, |T| - 1$ that contains the current name of the suffix at position $i$ and the current name of the suffix at position $i + 2^{k-1}$. For this step, we propose two different approaches: one using the inverse suffix array and a Window operation (section 8.2.3), and one using sorting (section 8.2.4). These are alternative implementations for line 9.

### 8.2.3 Distributed External Prefix Doubling using the Inverse Suffix Array

We can obtain the $h$-names of the required suffixes using the inverse $h$-suffix array. This approach is based on the qsufsort algorithm [Sad98; LS99; LS07] reviewed in section 8.2.1 and was pioneered in a distributed setting using MPI by Flick and Aluru [FA15].

Algorithm 8.3 shows in line 2 how we can sort pairs in $N$ based on their position in the text, such that we get the inverse $2^k$-suffix array $\mathsf{ISA}^{2^k}$ in iteration $k$. This inverse $2^k$-suffix array contains the current $2^k$-name of each suffix. For each position $i$, we need the name of the $(i + 2^k)$-th suffix. To get this name, we can simply scan over the `DIA` $N$ with a Window() operation of width $2^k + 1$, i.e., the same as shifting the inverse $2^k$-suffix array by $2^k$ positions and appending 0s until its length is $|T|$ again.

While our experiments show that this approach is faster than prefix doubling using sorting (described in the next subsection), it is obvious that it only works as long as the Window() size $2^k + 1$ fits into the RAM of each worker. The second solution using sorting does not suffer from this limitation and can be used as a fallback method.

We give an example run of the algorithm for the text $T = [\,\mathtt{b}, \mathtt{d}, \mathtt{a}, \mathtt{c}, \mathtt{b}, \mathtt{d}, \mathtt{a}, \mathtt{c}, \mathtt{b}, \$\,]$ in example 8.4. The comment at the end of each line refers to the line of code responsible for the change from the previous line where $x.y$ denotes line $y$ in Algorithm $x$.

### 8.2.4 Distributed External Prefix Doubling using Sorting

In the original suffix array paper by Manber and Myers [MM90; MM93], the presented construction algorithm was a prefix doubling algorithm using sorting. This idea was

**Example 8.4 :** Example of Prefix Doubling using the Inverse Suffix Array in Thrill.

$T = [\,\texttt{b},\texttt{d},\texttt{a},\texttt{c},\texttt{b},\texttt{d},\texttt{a},\texttt{c},\texttt{b}(,\$)\,]$
$S = [\,(0,\texttt{b},\texttt{d}),(1,\texttt{d},\texttt{a}),(2,\texttt{a},\texttt{c}),(3,\texttt{c},\texttt{b}),(4,\texttt{b},\texttt{d}),(5,\texttt{d},\texttt{a}),(6,\texttt{a},\texttt{c}),(7,\texttt{c},\texttt{b}),(8,\texttt{b},\$)\,]$ *// 8.1.2*
$\boldsymbol{k = 1}$ *// 8.1.3*
$S = [\,(2,\texttt{a},\texttt{c}),(6,\texttt{a},\texttt{c}),(8,\texttt{b},\$),(0,\texttt{b},\texttt{d}),(4,\texttt{b},\texttt{d}),(3,\texttt{c},\texttt{b}),(7,\texttt{c},\texttt{b}),(1,\texttt{d},\texttt{a}),(5,\texttt{d},\texttt{a})\,]$ *// 8.1.4*
$N = [\,(2,0),(6,0),(8,2),(0,3),(4,0),(3,5),(7,0),(1,7),(5,0)\,]$ *// 8.1.5*
**4 items with name 0** *// 8.1.6*
$N = [\,(2,0),(6,0),(8,2),(0,3),(4,3),(3,5),(7,5),(1,7),(5,7)\,]$ *// 8.1.8*
$N = [\,(0,3),(1,7),(2,0),(3,5),(4,3),(5,7),(6,0),(7,5),(8,2)\,]$ *// 8.3.2*
$S = [\,(0,3,0),(1,7,5),(2,0,3),(3,5,7),(4,3,0),(5,7,5),(6,0,2),(7,5,0),(8,2,0)\,]$ *// 8.3.3*
$\boldsymbol{k = 2}$ *// 8.1.3*
$S = [\,(6,0,2),(2,0,3),(8,2,0),(0,3,0),(4,3,0),(7,5,0),(3,5,7),(1,7,5),(5,7,5)\,]$ *// 8.1.4*
$N = [\,(6,0),(2,1),(8,2),(0,3),(4,0),(7,5),(3,6),(1,7),(5,0)\,]$ *// 8.1.5*
**2 items with name 0** *// 8.1.6*
$N = [\,(6,0),(2,1),(8,2),(0,3),(4,3),(7,5),(3,6),(1,7),(5,7)\,]$ *// 8.1.8*
$N = [\,(0,3),(1,7),(2,1),(3,6),(4,3),(5,7),(6,0),(7,5),(8,2)\,]$ *// 8.3.2*
$S = [\,(0,3,3),(1,7,7),(2,1,0),(3,6,5),(4,3,2),(5,7,0),(6,0,0),(7,5,0),(8,2,0)\,]$ *// 8.3.3*
$\boldsymbol{k = 3}$ *// 8.1.3*
$S = [\,(6,0,0),(2,1,0),(8,2,0),(4,3,2),(0,3,3),(7,5,0),(3,6,5),(5,7,0),(1,7,7)\,]$ *// 8.1.4*
$N = [\,(6,0),(2,1),(8,2),(4,3),(0,4),(7,5),(3,6),(5,7),(1,8)\,]$ *// 8.1.5*
**1 item with name 0** *// 8.1.6*
**Result:** $[\,6,2,8,4,0,7,3,5,1\,]$ *// 8.1.7*

refined by Dementiev, Kärkkäinen, Mehnert, and Sanders [DKMS05; DKMS08] who presented an external memory algorithm that we adapted to Thrill.

The idea is to compute the new name pairs by sorting the old names with respect to the starting position of the suffix, as shown in algorithm 8.5. We make use of the fact that during each iteration we know for each suffix the suffix index whose current name is required to compute the next refined name. Hence, we can sort the tuples containing the starting positions of the suffixes and their current name in such a way that if there is another name required for a name pair, then it is the name of the succeeding tuple (line 2). To do so, we use the following comparison operator: $<^k_{\mathrm{op}}\colon (\mathbb{N}_0,\mathbb{N}_0) \times (\mathbb{N}_0,\mathbb{N}_0) \to$ `bool` (see equation (8.1)) in algorithm 8.5:

$$(i,n) <^k_{\mathrm{op}} (i',n') = \begin{cases} i \ \mathbf{div}\ 2^k < i' \ \mathbf{div}\ 2^k & \text{if } i \equiv i' \pmod{2^k}, \\ i \ \mathbf{mod}\ 2^k < i' \ \mathbf{mod}\ 2^k & \text{otherwise.} \end{cases} \tag{8.1}$$

This relation orders pairs $(i,n)$ first by the $w-k$ most significant bits of $i$ and then by the $k$ least significant bits, where $w$ is the number of bits used to store $i$. For example $<^2_{\mathrm{op}}$ reorders $[0\mathinner{..}8)$ to $[\,0,4,1,5,2,6,3,7\,]$.

After sorting using the $<^k_{\mathrm{op}}$-comparator, we need to ensure that two consecutive names are the ones required to compute the new name, since the required name may not

**Algorithm 8.5 :** Prefix Doubling using Sorting.

1 **function** PrefixDoublingSorting($N, k \in \mathbb{N}_0$)

2 $\quad N := N.\mathrm{Sort}(<^k_{\mathrm{op}})$ *// Sort such that names of $i$ and $i + 2^k$ are consecutive.*

3 $\quad S := N.\mathrm{Window}_2\left((j, [\,(i, n_0, n_1), (i', n'_0, n'_1)\,]) \mapsto \begin{cases} (i, n_0, n'_0) & \text{if } i + 2^k = i'\,, \\ (i, n_0, 0) & \text{otherwise.} \end{cases}\right)$

**Example 8.6 :** Example of Prefix Doubling using Sorting in Thrill.

$T = [\,\mathtt{b}, \mathtt{d}, \mathtt{a}, \mathtt{c}, \mathtt{b}, \mathtt{d}, \mathtt{a}, \mathtt{c}, \mathtt{b}(, \$)\,]$

$S = [\,(0, \mathtt{b}, \mathtt{d}), (1, \mathtt{d}, \mathtt{a}), (2, \mathtt{a}, \mathtt{c}), (3, \mathtt{c}, \mathtt{b}), (4, \mathtt{b}, \mathtt{d}), (5, \mathtt{d}, \mathtt{a}), (6, \mathtt{a}, \mathtt{c}), (7, \mathtt{c}, \mathtt{b}), (8, \mathtt{b}, \$)\,]$ *// 8.1.2*

$\boldsymbol{k = 1}$ *// 8.1.3*

$S = [\,(2, \mathtt{a}, \mathtt{c}), (6, \mathtt{a}, \mathtt{c}), (8, \mathtt{b}, \$), (0, \mathtt{b}, \mathtt{d}), (4, \mathtt{b}, \mathtt{d}), (3, \mathtt{c}, \mathtt{b}), (7, \mathtt{c}, \mathtt{b}), (1, \mathtt{d}, \mathtt{a}), (5, \mathtt{d}, \mathtt{a})\,]$ *// 8.1.4*

$N = [\,(2,0), (6,0), (8,2), (0,3), (4,0), (3,5), (7,0), (1,7), (5,0)\,]$ *// 8.1.5*

**4 items with name 0** *// 8.1.6*

$N = [\,(2,0), (6,0), (8,2), (0,3), (4,3), (3,5), (7,5), (1,7), (5,7)\,]$ *// 8.1.8*

$N = [\,(0,3), (2,0), (4,3), (6,0), (8,2), (1,7), (3,5), (5,7), (7,5)\,]$ *// 8.5.2*

$S = [\,(0,3,0), (2,0,3), (4,3,0), (6,0,2), (8,2,0), (1,7,5), (3,5,7), (5,7,5), (7,5,0)\,]$ *// 8.5.3*

$\boldsymbol{k = 2}$ *// 8.1.3*

$S = [\,(6,0,2), (2,0,3), (8,2,0), (0,3,0), (4,3,0), (7,5,0), (3,5,7), (1,7,5), (5,7,5)\,]$ *// 8.1.4*

$N = [\,(6,0), (2,1), (8,2), (0,3), (4,0), (7,5), (3,6), (1,7), (5,0)\,]$ *// 8.1.5*

**2 items with name 0** *// 8.1.6*

$N = [\,(6,0), (2,1), (8,2), (0,3), (4,3), (7,5), (3,6), (1,7), (5,7)\,]$ *// 8.1.8*

$N = [\,(0,3), (4,3), (8,2), (1,7), (5,7), (2,1), (6,0), (3,6), (7,5)\,]$ *// 8.5.2*

$S = [\,(0,3,3), (4,3,2), (8,2,0), (1,7,7), (5,7,0), (2,1,0), (6,0,0), (3,6,5), (7,5,0)\,]$ *// 8.5.3*

$\boldsymbol{k = 3}$ *// 8.1.3*

$S = [\,(6,0,0), (2,1,0), (8,2,0), (4,3,2), (0,3,3), (7,5,0), (3,6,5), (5,7,0), (1,7,7)\,]$ *// 8.1.4*

$N = [\,(6,0), (2,1), (8,2), (4,3), (0,4), (7,5), (3,6), (5,7), (1,8)\,]$ *// 8.1.5*

**1 item with name 0** *// 8.1.6*

**Result:** $[\,6, 2, 8, 4, 0, 7, 3, 5, 1\,]$ *// 8.1.7*

exist due to the length of the text. This occurs during the $k$-th iteration for each suffix beginning at a text position greater than $n - 2^k$. In this case we use the sentinel name 0, which compares smaller than any valid name (line 3). In both cases, we return one triple for each position, consisting of a text position, the current name of the suffix beginning at that position and the name of the suffix $2^k$ positions to the right (if it exists and 0 otherwise).

Again, to better illustrate prefix double using sorting in Thrill, we present the corresponding data-flow graph in figure 8.2 and a sample execution for the same text as before in example 8.6. As in the previous example, the comments refer to the algorithm and line responsible for the change.

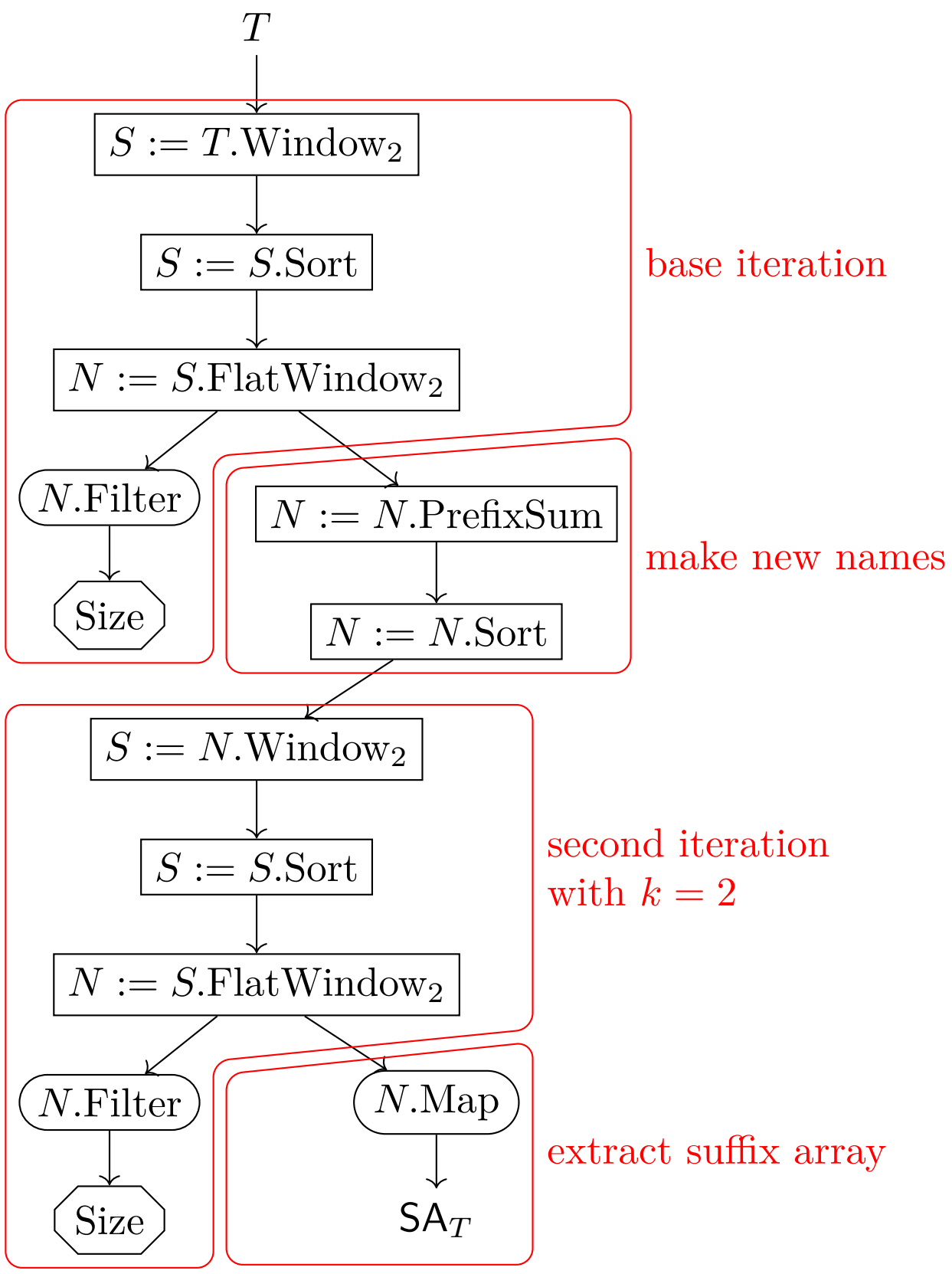

**Figure 8.2:** DIA data-flow graph of two iterations of prefix doubling using sorting.

### 8.2.5 Distributed External Prefix Doubling with Discarding

Both prefix doubling variants presented in the previous two sections incur large I/O costs from repeatedly re-ranking suffixes whose final rank is already known. These are included in each distributed shuffling operation and incur needless overhead. Crauser and Ferragina [CF02] and Dementiev, Kärkkäinen, Mehnert, and Sanders [DKMS05; DKMS08] presented a method called *discarding* to alleviate this by omitting all suffixes no longer needed from sorting operations. Crauser and Ferragina [CF02] keep two lists: one for unfinished items, and one for finished suffixes. Dementiev, Kärkkäinen, Mehnert, and Sanders [DKMS05; DKMS08] improved this by splitting the finished list into two: one for partially discarded entries still needed for sorting others and one for fully discarded suffixes no longer needed. This technique was later transferred to DC3 by Puglisi, Smyth, and Turpin [PST05]. We also implemented discarding in eSAIS (see also chapter 6), however, the performance benefit is largest for prefix doubling.

For discarding it is necessary to classify suffixes into three categories:

(i) Suffixes that do not yet have a unique name are called *not unique* (N) or *undecided*, which is also the initial state,

(ii) suffixes that have an unique name, but are required to compute another name pair for a suffix that does not yet have a unique name, are called *unique* (U), and finally

(iii) suffixes that have a unique name and are no longer needed for any other can be *discarded* (D).

Using this classification in a prefix doubling algorithm we can exclude all discarded suffixes from expensive sorting operations. Our implementation of prefix doubling with discarding in Thrill, shown in algorithm 8.7 (see figure 8.3 for the data-flow graph), initially behaves like the generic prefix doubling algorithm. As before, we present a sample execution for the text $T = [\mathtt{b}, \mathtt{d}, \mathtt{a}, \mathtt{c}, \mathtt{b}, \mathtt{d}, \mathtt{a}, \mathtt{c}, \mathtt{b}, \$]$ in example 8.9. The comments in the example refer to the algorithm and line responsible for the change.

Given a list of pairs $(i, n_i)$ sorted by names $n_i$, it is easy to identify *unique* suffixes if given *three* consecutive pairs $(i, n)$, $(i', n')$, and $(i'', n'')$: the middle suffix $i'$ can be marked as *unique* if $n \neq n'$ and $n' \neq n''$.

As before, for the initial round we compute character pairs for suffixes (line 2) and compute the 2-names for all suffixes (lines 3 to 5) in the same way as in the generic algorithm. Given a list of $2^k$-names $N = \{(i, n_i)\}$, the function Unique() is applied to *three* consecutive pairs in line 8 of algorithm 8.7 to identify unique names (see algorithm 8.8 for the function along with two edge cases).

The previous step results in triples $(i, n_i, s_i)$ with $s_i \in \{\mathtt{N}, \mathtt{U}\}$. We can then apply to this list the same steps as in prefix doubling with sorting (section 8.2.4): reorder them using $<^k_{\text{op}}$ such that suffixes $i$ and $i + 2^k$ are consecutive in $P$ (line 9).

For the actual discarding step, we created an enhanced version of CmpName() called NPairs() (algorithm 8.8), which again requires *three* tuples $a = (i, n_i, s_i)$, $b = (i', n'_i, s'_i)$, and $c = (i'', n''_i, s''_i)$ from the list $P$. If $s''_i$ is *unique* and either $s_i$ or $s'_i$ is also *unique*, then $c$ can be *discarded* because both $a$ and $b$ will get a unique name pair during this iteration (line 18, algorithm 8.8). Otherwise, $s''_i$ is *unique* but $s$ and $s'$ are *not unique*, then $c$ cannot be *discarded* as $a$ will not get a unique name pair during this iteration and we require the name of $c$ during the next iteration to compute the name pair (line 18, algorithm 8.8).

The function NPairs() is applied to the list $P$ of suffixes in interleaved order (line 10), and returns a stream of 4-tuples $(i, n_0, n_1, s)$ for calculating the next names. This stream is split up by state $s$ into three DIAs: $D'$ contains all newly discarded suffixes (line 11), $U'$ all unique suffixes which no longer need to be named (line 13), and $I'$

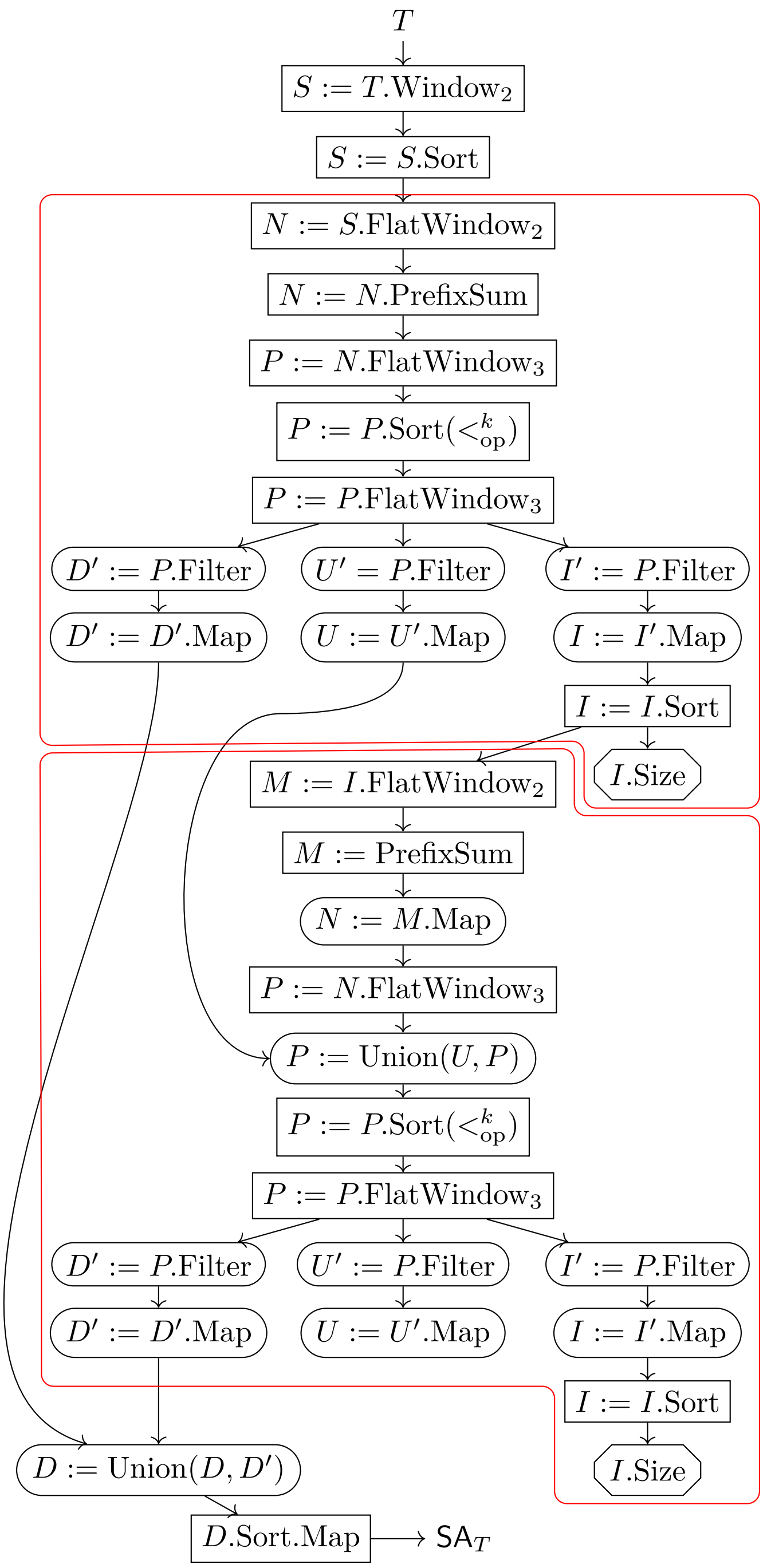

**Figure 8.3:** DIA data-flow graph of two iterations of prefix doubling with discarding.

**Algorithm 8.7 :** Prefix Doubling with Discarding.

```
1  function PrefixDoublingDiscarding(T ∈ DIA⟨Σ⟩)
2      S := T.Window_2((i, [t_0, t_1]) ↦ (i, t_0, t_1)) // Create initial triples (i, T[i], T[i+1]).
3      S := S.Sort((i, n_0, n_1) by (n_0, n_1))              // Sort triples by name pairs.
4      N := S.FlatWindow_2((i, [a, b]) ↦ CmpName(i, a, b))   // Map names to 0 or i.
5      N := N.PrefixSum(((i, n), (i', n')) ↦ (i', max(n, n'))) // Calculate initial names.
6      D := [ ],  U := [ ]                   // D will contain discarded, and U unique items.
7      for k := 1 to ⌈log_2 |T|⌉ do
8          P := N.FlatWindow_3((i, [a, b, c]) ↦ Unique(i, a, b, c))        // Compute states.
9          P := Union(U, P).Sort(<^k_op) // Concatenate undiscarded items and sort them.
10         P := P.FlatWindow_3((i, [a, b, c]) ↦ NPairs(i, a, b, c, k)) // Compute new name
11         D' := P.Filter((i, n_0, n_1, s)) ↦ (s = D)   // pairs and update state. Then filter
12         D := Union(D, D'.Map((i, n_0, n_1, s) ↦ (i, n_0)))  // out newly discarded items.
13         U' := P.Filter((i, n_0, n_1, s) ↦ (s = U))        // Separate already unique items
14         U := U'.Map((i, n_0, n_1, s) ↦ (i, n_0, s))        // into U, and not unique items,
15         I' := P.Filter((i, n_0, n_1, s) ↦ (s = N))       // which still need to be computed,
16         I := I'.Map((i, n_0, n_1, s) ↦ (i, n_0, n_1)).Sort((i, n_0, n_1) by (n_0, n_1))  // into I.
17         if I.Size() = 0 then                                   // If all items are unique,
18             return D.Sort((i, n) by n).Map((i, n) ↦ i)   // return SA from discarded.
19         M := I.FlatWindow_2((i, [a, b]) ↦ NameDiscarding(i, a, b))        // Form names
20         M := M.PrefixSum(((i, n_0, c_0, c_1), (i', n'_0, c'_0, c'_1)) ↦          // that comply
                   (i', n'_0, max(c_0, c'_0), max(c_1, c'_1)))
21         N := M.Map((i, n_0, c_0, c_1) ↦ (i, n_0 + (c_1 − c_0)))        // with the old names.
```

the remaining not unique suffixes (line 15). Following each Filter() we immediately reduce the 4-tuples by extracting only the necessary information using a Map(). Due to the way Thrill chains LOps, the combination of Filter() followed by a Map() has no run-time overhead.

*Discarded* suffixes already have their final names, and we can collect all in $D$ by repeatedly performing a Union() operation on $D'$ and $D$ (line 12). When all items are unique, $D$ can be used to output the final suffix array by sorting pairs $(i, n)$ by final rank (line 18).

*Unique* suffixes in $U$ no longer need to be ranked, but they are needed for naming other suffixes. To avoid sorting unique suffixes, we filter them out in line 13 and create the Union() of the set of undecided suffixes and the remaining required unique items in line 9.

Remaining *undecided* suffixes are stored in $I$ as triples $(i, n_0, n_1)$, and in principle dealt with in the same manner as in our generic prefix doubling algorithm (algorithm 8.1). However, other than in the generic algorithm the `DIA` $I$ no longer contains *all* suffixes and we cannot use the same naming schema as before, because entries in $I$ will be

**Algorithm 8.8 :** Prefix Doubling with Discarding – Additional Functions

**function** Unique($j \in \mathbb{N}_0,\ (i,n),(i',n'),(i'',n'') \in N$)

  **if** $j = 0$ **then**

    **emit** $\begin{cases} (i,n,\mathtt{U}) & \text{if } n \neq n', \\ (i,n,\mathtt{N}) & \text{otherwise.} \end{cases}$  *// First item is unique // if its name differs from its successor.*

  **else if** $j+2 = |T|$ **then**

    **emit** $\begin{cases} (i'',n'',\mathtt{U}) & \text{if } n' \neq n'', \\ (i'',n'',\mathtt{N}) & \text{otherwise.} \end{cases}$  *// Final item is unique // if its name differs from its precursor.*

  **emit** $\begin{cases} (i',n',\mathtt{U}) & \text{if } n \neq n' \text{ and } n' \neq n'', \\ (i',n',\mathtt{N}) & \text{otherwise.} \end{cases}$  *// An item is unique // if its name is unique.*

**function** NPairs($j \in \mathbb{N}_0,\ (i,n,s),(i',n',s'),(i'',n'',s'') \in P,\ k \in \mathbb{N}_0$)

  **if** $j = 0$ **then**

    **emit** $\begin{cases} (i,n,0,\mathtt{D}) & \text{if } s = \mathtt{U}, \\ (i',n',0,\mathtt{D}) & \text{if } s' = \mathtt{U}. \end{cases}$  *// The first two items can be discarded // if they are unique. Emit $\leq 2$ items.*

  **else if** $j+2 = |T|$ **then**

    **if** $s' = \mathtt{N}$ **then**

      **emit** $\begin{cases} (i',n',n'',\mathtt{N}) & \text{if } i' + 2^k = i'', \\ (i',n',0,\mathtt{N}) & \text{otherwise.} \end{cases}$  *// If the last two items of the // DIA are undecided, then we need*

    **if** $s'' = \mathtt{N}$ **then**

      **emit** $(i'',n'',0,\mathtt{N})$  *// to fuse the names required for renaming.*

  **if** $s = \mathtt{N}$ **then**

    **emit** $\begin{cases} (i,n,n',\mathtt{N}) & \text{if } i + 2^k = i', \\ (i,n,0,\mathtt{N}) & \text{otherwise.} \end{cases}$  *// The names for renaming are // consecutive and fused accordingly.*

  **if** $s'' = \mathtt{U}$ **then**

    **emit** $\begin{cases} (i'',n'',0,\mathtt{D}) & \text{if } s = \mathtt{U} \text{ or } s' = \mathtt{U}, \\ (i'',n'',0,\mathtt{U}) & \text{otherwise.} \end{cases}$  *// Unique items are discarded // if not needed in future renaming.*

**function** NameDiscarding($j \in \mathbb{N}_0,\ (i,n_0,n_1),(i',n_0',n_1') \in I$)

  **if** $j = 0$ **then**

    **emit** $(i,n_0,0,0)$  *// The new names must comply with the old ones.*

  **emit** $\begin{cases} (i',n_0',j+1,j+1) & \text{if } n_0 \neq n_0', \\ (i',n_0',0,j+1) & \text{else if } n_1 \neq n_1', \\ (i',n_0',0,0) & \text{otherwise.} \end{cases}$  *// The first name determines // the group and new names // are consistent within groups.*

put into $D$ and later sorted to generate the final suffix array. Hence, names in $P$ and $N$ must be kept such that they are $2^k$-names as defined in section 8.2.1 (the *smallest* name within a $2^k$-group of suffixes starting with the same $2^k$ characters). Each triple $(i, n_0, n_1)$ contains the current name of $T[i..n)$ as $n_0$ and of $T[i+2^k..n)$ as $n_1$ (or zero if it is beyond the end of the string). The function NameDiscarding() together with the following PrefixSum in line 20 is used to calculate the preceding indices in $I$ where both $n_0$ and $n_1$, or only $n_1$ changes. Suffixes that do not have a unique name form *intervals* in $\mathsf{SA}^{2^k}$ because all suffixes in an unfinished $2^k$-group are given the same name. To get $2^{k+1}$-names in $N$, each name entry has to be increased by $(c_1 - c_0)$ where $c_0$ is the starting index of the $2^k$-group with name $n_0$, and $c_1$ is the starting index of the subgroup with $T[i+2^k..n)$ having name $n_1$. Hence, $(c_1 - c_0)$ is the *relative shift* in name due to the current prefix doubling step. Another way to view this, is that lines 19 to 21 perform a mapping from the two component space $(n_0, n_1)$ to a relative scalar shift such that $n_0 + (c_1 - c_0)$ form new $2^{k+1}$-names.

For example, if $I = [\dots, (x_1, 7, 0), (x_2, 7, 0), (x_3, 7, 1), (x_4, 7, 1), (x_5, 7, 1), (x_6, 7, 2), \dots]$, so if $I$ contains a $2^k$-group of suffixes $x_i$ with names $n_0 = 7$, then NameDiscarding() and PrefixSum determine $M = [\,(x_1, 7, 10, 10), (x_2, 7, 10, 10), (x_3, 7, 10, 12), (x_4, 7, 10, 12), (x_5, 7, 10, 12), (x_6, 7, 10, 15)\,]$, where 10 is the first index of $n_0 = 7$ in $I$. Using these $(i, n_0, c_0, c_1)$ tuples, the new $2^{k+1}$-names can be calculated as $N = [\dots, (x_1, 7), (x_2, 7), (x_3, 9), (x_4, 9), (x_5, 9), (x_6, 12), \dots]$.

At the beginning of the next iteration we add all unique names to the new names and check if they can be discarded. As soon as all names are unique (line 17) we know that all names have been discarded and can compute $\mathsf{SA}$ by sorting the discarded tuples by their names (line 18).

All prefix doubling algorithms need at most $\lceil \log_2(\text{maxlcp}(T)) \rceil$ iterations, each containing one or more sorting steps. While this approach works well for inputs with low maximum LCP (such as DNA and random inputs), on natural English text the algorithms are rather slow. In the next section, we will focus on difference cover algorithms, which run in $\mathcal{O}(\text{sort}(|T|))$, albeit with a relatively large hidden constant.

## 8.3 Difference Cover Algorithms

In 2003, the *skew* aka *DC3* suffix sorting algorithm was proposed by Kärkkäinen and Sanders [KS03], and later generalized to *DC* by Kärkkäinen, Sanders, and Burkhardt [KSB06]. They employ recursion on a subset of the suffixes to reach linear running time in the sequential RAM model. Furthermore, the approach can be transferred to the external memory model, resulting in an algorithm with *sorting* complexity, and to the EREW-PRAM model, resulting in $\mathcal{O}(\log^2 n)$ time and $\mathcal{O}(n \log n)$ work. While the reference implementation by the authors is in the sequential RAM model, the algorithms were later implemented for external memory [DKMS05; DKMS08;

**Example 8.9 :** Example of Prefix Doubling with Discarding in Thrill.

1 $T = [\,\texttt{b},\texttt{d},\texttt{a},\texttt{c},\texttt{b},\texttt{d},\texttt{a},\texttt{c},\texttt{b}(,\texttt{\$})\,]$

2 $S = [\,(0,\texttt{b},\texttt{d}),(1,\texttt{d},\texttt{a}),(2,\texttt{a},\texttt{c}),(3,\texttt{c},\texttt{b}),(4,\texttt{b},\texttt{d}),(5,\texttt{d},\texttt{a}),(6,\texttt{a},\texttt{c}),(7,\texttt{c},\texttt{b}),(8,\texttt{b},\texttt{\$})\,]$ *// 8.7.2*

3 $S = [\,(2,\texttt{a},\texttt{c}),(6,\texttt{a},\texttt{c}),(8,\texttt{b},\texttt{\$}),(0,\texttt{b},\texttt{d}),(4,\texttt{b},\texttt{d}),(3,\texttt{c},\texttt{b}),(7,\texttt{c},\texttt{b}),(1,\texttt{d},\texttt{a}),(5,\texttt{d},\texttt{a})\,]$ *// 8.7.3*

4 $N = [\,(2,0),(6,0),(8,2),(0,3),(4,0),(3,5),(7,0),(1,7),(5,0)\,]$ *// 8.7.4*

5 $N = [\,(2,0),(6,0),(8,2),(0,3),(4,3),(3,5),(7,5),(1,7),(5,7)\,]$ *// 8.7.5*

6 $\boldsymbol{k=1}$: $D = [\ ]$, $U = [\ ]$ *// 8.7.7*

7 $P = [\,(2,0,\texttt{N}),(6,0,\texttt{N}),(8,2,\texttt{U}),(0,3,\texttt{N}),(4,3,\texttt{N}),(7,5,\texttt{N}),(3,5,\texttt{N}),(1,7,\texttt{N}),(5,7,\texttt{N})\,]$ *// 8.7.8*

8 $P = [\,(0,3,\texttt{N}),(2,0,\texttt{N}),(4,3,\texttt{N}),(6,0,\texttt{N}),(8,2,\texttt{U}),(1,7,\texttt{N}),(3,5,\texttt{N}),(5,7,\texttt{N}),(7,5,\texttt{N})\,]$ *// 8.7.9*

9 $P = [\,(0,3,0,\texttt{N}),(2,0,3,\texttt{N}),(4,3,0,\texttt{N}),(6,0,2,\texttt{N}),(8,2,0,\texttt{U}),$
$(1,7,5,\texttt{N}),(3,5,7,\texttt{N}),(5,7,5,\texttt{N}),(7,5,0,\texttt{N})\,]$ *// 8.7.10*

10 $D = [\ ]$ *// 8.7.12*

11 $U = [\,(8,2,\texttt{U})\,]$ *// 8.7.14*

12 $I = [\,(6,0,2),(2,0,3),(4,3,0),(0,3,0),(7,5,0),(3,5,7),(5,7,5),(1,7,5)\,]$ *// 8.7.16*

13 **condition:** $I$.Size() $= 8$ *// 8.7.17*

14 $M = [\,(6,0,0,0),(2,0,0,1),(4,3,2,2),(0,3,0,0),$
$(7,5,4,4),(3,5,0,5),(5,7,6,6),(1,7,0,0)\,]$ *// 8.7.19*

15 $M = [\,(6,0,0,0),(2,0,0,1),(4,3,2,2),(0,3,2,2),$
$(7,5,4,4),(3,5,4,5),(5,7,6,6),(1,7,6,6)\,]$ *// 8.7.20*

16 $N = [\,(6,0),(2,1),(4,3),(0,3),(7,5),(3,6),(5,7),(1,7)\,]$ *// 8.7.21*

17 $\boldsymbol{k=2}$: $D = [\ ]$, $U = [\,(8,2,\texttt{U})\,]$ *// 8.7.7*

18 $P = [\,(6,0,\texttt{U}),(2,1,\texttt{U}),(4,3,\texttt{N}),(0,3,\texttt{N}),(7,5,\texttt{U}),(3,6,\texttt{U}),(5,7,\texttt{N}),(1,7,\texttt{N})\,]$ *// 8.7.8*

19 $P = [\,(0,3,\texttt{N}),(4,3,\texttt{N}),(8,2,\texttt{U}),(1,7,\texttt{N}),(5,7,\texttt{N}),(2,1,\texttt{U}),(6,0,\texttt{U}),(3,6,\texttt{U}),(7,5,\texttt{U})\,]$ *// 8.7.9*

20 $P = [\,(0,3,3,\texttt{N}),(4,3,2,\texttt{N}),(8,2,0,\texttt{U}),(1,7,7,\texttt{N}),(5,7,0,\texttt{N}),$
$(2,1,0,\texttt{U}),(6,0,0,\texttt{D}),(3,6,0,\texttt{D}),(7,5,0,\texttt{D})\,]$ *// 8.7.10*

21 $D = [\,(6,0),(3,6),(7,5)\,]$ *// 8.7.12*

22 $U = [\,(8,2,\texttt{U}),(2,1,\texttt{U})\,]$ *// 8.7.14*

23 $I = [\,(4,3,2),(0,3,3),(5,7,0),(1,7,7)\,]$ *// 8.7.16*

24 **condition:** $I$.Size() $= 4$ *// 8.7.17*

25 $M = [\,(4,3,0,0),(0,3,0,1),(5,7,2,2),(1,7,0,3)\,]$ *// 8.7.19*

26 $M = [\,(4,3,0,0),(0,3,0,1),(5,7,2,2),(1,7,2,3)\,]$ *// 8.7.20*

27 $N = [\,(4,3),(0,4),(5,7),(1,8)\,]$ *// 8.7.21*

28 $\boldsymbol{k=3}$: $D = [\,(6,0),(3,6),(7,5)\,]$, $U = [\,(8,2,\texttt{U}),(2,1,\texttt{U})\,]$ *// 8.7.7*

29 $P = [\,(4,3,\texttt{U}),(0,4,\texttt{U}),(5,7,\texttt{U}),(1,8,\texttt{U})\,]$ *// 8.7.8*

30 $P = [\,(0,4,\texttt{U}),(8,2,\texttt{U}),(1,8,\texttt{U}),(2,1,\texttt{U}),(4,3,\texttt{U}),(5,7,\texttt{U})\,]$ *// 8.7.9*

31 $P = [\,(0,4,0,\texttt{D}),(8,2,0,\texttt{D}),(1,8,0,\texttt{D}),(2,1,0,\texttt{D}),(4,3,0,\texttt{D}),(5,7,0,\texttt{D})\,]$ *// 8.7.10*

32 $D = [\,(6,0),(3,6),(7,5)\,] \cup [\,(0,4),(8,2),(1,8),(2,1),(4,3),(5,7)\,]$ *// 8.7.12*

33 $U = [\ ]$ *// 8.7.14*

34 $I = [\ ]$ *// 8.7.16*

35 **condition:** $I$.Size() $= 0$ *// 8.7.17*

36 **Result:** $[\,6,2,8,4,0,7,3,5,1\,]$ *// 8.7.18*

Wee06], and DC3 was implemented for distributed memory using MPI [KS06b; Kul06; KS07].

The DC algorithms are based on scanning, sorting, and merging, and hence are asymptotically optimal in many models, provided optimal theoretical base algorithms. As Thrill supplies all of these base algorithms as scalable distributed external algorithmic building blocks, implementing DC is a natural choice.

The key notion of DC is to recursively calculate the ranks of suffixes in only a *difference cover* [Sin38] of the original text. A set $D \subseteq \mathbb{N}_0$ is a difference cover for $v \in \mathbb{N}_0$, if $\{(i - j) \bmod v \mid i, j \in D\} = \{0, \ldots, v - 1\}$. Examples of difference covers are $D_3 = \{1, 2\}$ for $v = 3$, $D_7 = \{0, 1, 3\}$ for $v = 7$, and $D_{13} = \{0, 1, 3, 9\}$ for $v = 13$. In general, a difference cover of size $\mathcal{O}(\sqrt{v})$ can be calculated for any $v$ in $\mathcal{O}(\sqrt{v})$ time [KSB06]. With respect to suffix sorting, the difference cover has the interesting property that it *samples* suffixes for recursive sorting and given the rank of all samples allows one to order the non-sample suffixes using a *constant-time* comparison operation. This is possible because for any two indices $i, j \in [0 \mathinner{..} v)$ there is an $\ell < v$ such that $((i + \ell) \bmod v) \in D$ and $((j + \ell) \bmod v) \in D$.

The basic steps of the DC3 algorithm are the following:

(D.1) Calculate ranks for all suffixes starting at positions in the difference cover $D_3 = \{1, 2\}$ modulo 3. This is done by sorting the triples $(T[i], T[i+1], T[i+2])$ for $(i \bmod 3) \in D_3$, calculating lexicographic names, sorting the names back to string order, and recursively calling a suffix sorting algorithm on a reduced string $T_R$ of size $\frac{2}{3}n$ if necessary. This reduced string represents *two* concatenated copies of the input string using the lexicographic names: the first copy are all names for suffixes with $i = 1 \bmod 3$ followed by a second copy for all suffixes with $i = 2 \bmod 3$. Hence, each character in $T_R$ embodies three characters. The result of step (D.1) are two arrays, $R_1$ and $R_2$, containing the ranks of suffixes $i = 1 \bmod 3$ and $i = 2 \bmod 3$, which can be calculated by inverting the recursively constructed suffix array of $T_R$.

(D.2) Scan the text $T$ and rank arrays $R_1$ and $R_2$ to generate three arrays: $S_0$, $S_1$, and $S_2$, where array $S_j$ contains one tuple for each suffix $i$ with $i = j \bmod 3$. For each suffix $i$, the arrays store one tuple containing the two *following* ranks from $R_1$ and $R_2$ and all characters from $T$ *up to* the next ranks. This is exactly the information required such that the following merge step is able to deduce the suffix array correctly. Due to the difference cover property the following rank for each suffix $i$ is among the three elements $R_1[i]$, $R_1[i+1]$, and $R_1[i+2]$ for $R_1$, and analogously for $R_2$.

(D.3) Sort $S_0$, $S_1$, and $S_2$ and merge them using a custom comparison function which compares the suffixes represented in the tuples using characters and ranks. Only a constant number of characters and ranks need to be accessed in each comparison. Output the suffix array using the indices stored in tuples.

The first two steps of the difference cover suffix sorting algorithms can be seen as preparation for the final merge in step (D.3). Step (D.1) delivers ranks for all suffixes $(i \bmod 3) \in D_3$ in $R_1$ and $R_2$. In step (D.2) tuples are created in $S_0$, $S_1$, and $S_2$ which are constructed from the recursively calculated ranks and characters from the text. The tuples are designed such that the comparison function can fully determine the final suffix array.

The difference cover algorithm DC3 generalizes to *DC* using a difference cover $D$ for any ground set size $v \geq 3$. DC constructs a recursive subproblem of size $\lceil (|T|/v)|D| \rceil$, which, considering $|D| = \mathcal{O}(\sqrt{v})$, is of size $\Theta(\frac{1}{\sqrt{v}})$. The algorithm has at most $\log_v |T|$ recursion levels and only one recursion branch.

At every level of the recursion, only work with sorting complexity is needed, and a straight-forward application of the Master theorem [CLRS01] to the recurrence $T(n) = T(\Theta(\frac{n}{\sqrt{v}})) + \mathcal{O}(\text{sort}(n))$ shows that the whole algorithm has sorting complexity due to the small recursive subproblem.

In the RAM model and with integer alphabets one can use radix sort for lexicographic naming in each level *and* for the merging step, and thus DC has linear running time. For our distributed scenario, DC has the same complexity as sorting and merging.

Due to the subproblem size $\lceil (|T|/v)|D| \rceil$ it is best to use the largest $v$ for a specific difference cover size. For $|D| = 2$ this is $v = 3$, aka DC3. For difference covers of size three, the largest $v = 7$ which yields DC7 with $D_7 = \{0, 1, 3\}$. And for difference covers of size four, the largest $v = 13$. Dementiev, Kärkkäinen, Mehnert, and Sanders [Meh04; DKMS05; DKMS08][Dem06, chapter 5] showed that DC7 is optimal regarding the number of I/Os in the external memory model assuming one elemental data type. They however only implemented DC3. Weese [Wee06] verified their result under with the more detailed assumption that characters are one byte in the first level of recursion.

### 8.3.1 Distributed Difference Cover Algorithms with Thrill

The complete DC3 implementation in Thrill algorithm code is shown as algorithm 8.10, and example 8.11 shows the transcript of a run with the text $T = [\,\mathtt{d}, \mathtt{b}, \mathtt{a}, \mathtt{c}, \mathtt{b}, \mathtt{a}, \mathtt{c}, \mathtt{b}, \mathtt{d}\,]$. Figure 8.5 shows the dataflow graph of *DC7* instead of DC3, which is slightly more complex but shows the algorithmic structure better. In the algorithm pseudocode we omitted some details on padding and sentinels needed for inputs that are not a multiple of the difference cover size, but our actual implementation in Thrill covers all these edge cases.

Goal of lines 2 to 20 is to calculate $R_1$ and $R_2$ as an interleaved array $I_R$ (corresponding to step (D.1)). This is done by performing the following steps:

(i) Scan the text $T$ using a FlatWindow operation and create triples $(i, c_0, c_1, c_2)$ for all indices $(i \bmod 3) \in D_3 = \{1, 2\}$ (lines 2 to 4).

**Algorithm 8.10 :** DC3 Algorithm in Thrill.

**function** DC3($T \in \texttt{DIA}\langle\Sigma\rangle$)
 $T_3 := T.\text{FlatWindow}_3((i, [\,c_0, c_1, c_2\,]) \mapsto \text{MakeTriples}(i, c_0, c_1, c_2))$
 **with function** MakeTriples($i \in \mathbb{N}_0,\ c_0, c_1, c_2 \in \Sigma$)
  **if** $i \neq 0 \bmod 3$ **then emit** $(i, c_0, c_1, c_2)$ *// Make triples $i \in D_3$.*
 $S := T_3.\text{Sort}((i, c_0, c_1, c_2)$ by $(c_0, c_1, c_2))$ *// Sort triples lexicographically.*
 $I_S := S.\text{Map}((i, c_0, c_1, c_2) \mapsto i)$ *// Extract sorted indices.*
 $N' := S.\text{FlatWindow}_2((i, [\,p_0, p_1\,]) \mapsto \text{CmpTriple}(i, p_0, p_1))$ *// Compare triples.*
 **with function** CmpTriple($i \in \mathbb{N}_0,\ p_0 = (c_0, c_1, c_2),\ p_1 = (c'_0, c'_1, c'_2)$) *// Emit one*
  **if** $i = 0$ **then emit** 0 *// sentinel for index 0, and 0 or 1*
  **emit** (**if** $(c_0, c_1, c_2) = (c'_0, c'_1, c'_2)$ **then** 0 **else** 1) *// depending on previous tuple.*
 $N := N'.\text{PrefixSum}()$ *// Use prefix sum to calculate names.*
 $n_{\text{sub}} = \lceil 2|T|/3 \rceil,\quad n_{\text{mod1}} = \lceil |T|/3 \rceil$ *// Size of recursive problem and mod 1 part of $T_R$*
 **if** $N.\text{Max}() + 1 \neq n_{\text{sub}}$ **then** *// If duplicate names exist, sort names back to*
  $T'_R := \text{Zip}([\,I_S, N\,], (i, n) \mapsto (i, n)).\text{Sort}((i, n)$ by $(i \bmod 3, i \text{ div } 3))$ *// string order*
  $\mathsf{SA}_R := \text{DC3}(T'_R.\text{Map}((i, n) \mapsto n))$ *// interleave as $T_1 \oplus T_2$ and call suffix sorter.*
  $I'_R := \mathsf{SA}_R.\text{ZipWithIndex}((r, i) \mapsto (r, i))$ *// Invert resulting suffix array, but*
  $I_R := I'_R.\text{Sort}((r, i)$ by $(r \bmod n_{\text{mod1}}, r))$ *// interleave mod 1/2 ranks in ISA.*
 **else** *// Else, if all names/triples are unique, then $I_S$ is already the suffix array.*
  $R := I_S.\text{ZipWithIndex}((r, i) \mapsto (r, i))$ *// Invert it to get ISA, but*
  $I_R := R.\text{Sort}((r, i)$ by $(r \text{ div } 3, r))$ *// interleave mod 1/2 ranks in ISA.*
 $I_R := I_R.\text{Map}((r, i) \mapsto (i + 1))$ *// Extract ranks from $I_R$, free rank zero.*
 $Z' := \text{ZipWindow}_{[3,2]}([T, I_R], (i, [\,c_0, c_1, c_2\,], [\,r_1, r_2\,]) \mapsto (c_0, c_1, c_2, r_1, r_2))$ *// Combine*
 $Z := Z'.\text{Window}_2((i, [\,(z_1, z_2)\,]) \mapsto (i, z_1, z_2))$ *// characters and ranks*
 $S'_0 := Z.\text{Map}((i, (c_0, c_1, c_2, r_1, r_2), (\bar{c}_0, \bar{c}_1, \bar{c}_2, \bar{r}_1, \bar{r}_2)) \mapsto (3i + 0, c_0, c_1, r_1, r_2))$ *// to make*
 $S'_1 := Z.\text{Map}((i, (c_0, c_1, c_2, r_1, r_2), (\bar{c}_0, \bar{c}_1, \bar{c}_2, \bar{r}_1, \bar{r}_2)) \mapsto (3i + 1, c_1, r_1, r_2))$ *// arrays*
 $S'_2 := Z.\text{Map}((i, (c_0, c_1, c_2, r_1, r_2), (\bar{c}_0, \bar{c}_1, \bar{c}_2, \bar{r}_1, \bar{r}_2)) \mapsto (3i + 2, c_2, r_2, \bar{c}_0, \bar{r}_1))$ *// of*
 $S_0 := S'_0.\text{Sort}((i, c_0, c_1, r_1, r_2)$ by $(c_0, r_1))$ *// representatives for each*
 $S_1 := S'_1.\text{Sort}((i, c_1, r_1, r_2)$ by $(r_1))$ *// suffix class.*
 $S_2 := S'_2.\text{Sort}((i, c_2, r_2, \bar{c}_0, \bar{r}_1)$ by $(r_2))$
 **return** $\text{Merge}([\,S_0, S_1, S_2\,], \text{CompareDC3}).\text{Map}((i, \ldots) \mapsto i)$ *// Merge sorted*
 **with function** CompareDC3($z_1, z_2$) *// representatives to deliver final suffix array.*
  $(c_0, r_1) < (c'_1, r'_2)$ if $z_1 = (i, c_0, c_1, r_1, r_2) \in S_0,\ z_2 = (i', c'_1, r'_1, r'_2) \in S_1,$
  $(c_0, c_1, r_2) < (c'_2, \bar{c}'_0, \bar{r}'_1)$ if $z_1 = (i, c_0, c_1, r_1, r_2) \in S_0,\ z_2 = (i', c'_2, r'_2, \bar{c}'_0, \bar{r}'_1) \in S_2,$
  $(r_1) < (r'_2)$ if $z_1 = (i, c_1, r_1, r_2) \in S_1,\ z_2 = (i', c'_2, r'_2, \bar{c}'_0, \bar{r}'_1) \in S_2,$
  and symmetrically if $z_1 \in S_i, z_2 \in S_j$ with $i > j$ .

**Example 8.11 :** Example of DC3 Algorithm in Thrill.

1 $T = [\,\mathtt{d},\mathtt{b},\mathtt{a},\mathtt{c},\mathtt{b},\mathtt{a},\mathtt{c},\mathtt{b},\mathtt{d}\,]$ *// Example text $T$.*
2 $T_3 = [\,(1,\mathtt{b},\mathtt{a},\mathtt{c}),(2,\mathtt{a},\mathtt{c},\mathtt{b}),(4,\mathtt{b},\mathtt{a},\mathtt{c}),(5,\mathtt{a},\mathtt{c},\mathtt{b}),(7,\mathtt{b},\mathtt{d},\$),(8,\mathtt{d},\$,\$)\,]$ *// Triples ($i \bmod 3) \in D_3$.*
3 $S = [\,(2,\mathtt{a},\mathtt{c},\mathtt{b}),(5,\mathtt{a},\mathtt{c},\mathtt{b}),(1,\mathtt{b},\mathtt{a},\mathtt{c}),(4,\mathtt{b},\mathtt{a},\mathtt{c}),(7,\mathtt{b},\mathtt{d},\$),(8,\mathtt{d},\$,\$)\,]$ *// Sorted triples.*
4 $I_S = [\,2,5,1,4,7,8\,]$ *// Indices extracted from sorted triples.*
5 $N' = [\,0,0,1,0,1\,]$ *// 0/1 indicators depending if triples are unequal or equal.*
6 $N = [\,0,0,1,1,2,3\,]$ *// Prefix sum of 0/1 indicators delivers lexicographic names.*
7 $n_{\text{sub}} = 6$, $n_{\text{mod1}} = 3$ *// Calculate result size directly.*
8 Condition $(N.\text{Max}() + 1 = 4) \neq (6 = n_{\text{sub}})$, so follow recursion branch.
9 $T''_R = [\,(2,0),(5,0),(1,1),(4,1),(7,2),(8,3)\,]$ *// Zip lexicographic names and their string*
10 $T'_R = [\,(1,1),(4,1),(7,2),(2,0),(5,0),(8,3)\,]$ *// index, and sort them to string order*
11 $T_R = [\,1,1,2,0,0,3\,]$ *// to construct the recursive subproblem.*
12 $\mathsf{SA}_R = [\,3,4,0,1,2,5\,]$ *// Recursively calculate suffix array of $T_R$.*
13 $I'_R = [\,(3,0),(4,1),(0,2),(1,3),(2,4),(5,5)\,]$ *// Add index positions to suffix array*
14 $I_R = [\,(0,2),(3,0),(1,3),(4,1),(2,4),(5,5)\,]$ *// and sort into interleaved $R_1$ and $R_2$ ranks.*
15 $I_R = [\,3,1,4,2,5,6\,]$ *// Extract ranks from $I_R$, free rank zero.*
16 $Z' = [\,(\mathtt{d},\mathtt{b},\mathtt{a},3,1),(\mathtt{c},\mathtt{b},\mathtt{a},4,2),(\mathtt{c},\mathtt{b},\mathtt{d},5,6)\,]$ *// Zip $T$ and $I_R$ to make arrays $S_i$.*
17 $S'_0 = [\,(0,\mathtt{d},\mathtt{b},3,1),(3,\mathtt{c},\mathtt{b},4,2),(6,\mathtt{c},\mathtt{b},5,6)\,]$ *// Construct $(i,c_0,c_1,r_1,r_2) \in S_0$,*
18 $S'_1 = [\,(1,\mathtt{b},3,1),(4,\mathtt{b},4,2),(7,\mathtt{b},5,6)\,]$ *// $(i,c_1,r_1,r_2) \in S_1$, and*
19 $S'_2 = [\,(2,\mathtt{a},1,\mathtt{c},4),(5,\mathtt{a},2,\mathtt{c},5),(8,\mathtt{d},6,\$,0)\,]$ *// $(i,c_2,r_2,\bar{c}_0,\bar{r}_1) \in S_2$*
20 $S_0 = [\,(3,\mathtt{c},\mathtt{b},4,2),(6,\mathtt{c},\mathtt{b},5,6),(0,\mathtt{d},\mathtt{b},3,1)\,]$ *// as representatives of suffixes,*
21 $S_1 = [\,(1,3,\mathtt{b},1),(4,4,\mathtt{b},2),(7,5,\mathtt{b},6)\,]$ *// sort them among themselves*
22 $S_2 = [\,(2,1,\mathtt{a},\mathtt{c},4),(5,2,\mathtt{a},\mathtt{c},5),(8,6,\mathtt{d},\$,0)\,]$ *// such that merging delivers*
23 Result: $[\,2,5,1,4,7,3,6,8,0\,]$ *// the final suffix array.*

(ii) Sort the triples as $S$, scan $S$ and use a prefix sum to calculate lexicographic names $N$ (lines 5 to 11). The lexicographic names are constructed in the prefix sum from 0 and 1 indicators. The value 0 is used if two lexicographic consecutive triples are equal, which means they are assigned the same lexicographic name; the value 1 increments the name in the prefix sum and assigns the unequal triple a new name.

(iii) Check if all lexicographic names are different by comparing the highest lexicographic name against the maximum possible (lines 12 to 13).

(iv) If all lexicographic names are different, then $I_S$, which contains the indices of $S$, is already the suffix array of the suffixes in $D_3$ (lines 19 to 20). Hence, $R_1$ and $R_2$ can be created directly: the suffix array $I_S$ only needs to be inverted and split modulo 3. However, instead of constructing $R_1$ and $R_2$ as separate DIAs, we *interleave* them in $I_R$ using a Sort operation such that they are balanced on the distributed system, as we will be needing pairs of mod 1/2 ranks.

(v) Otherwise, prepare a recursive subproblem $T_R$ to calculate the ranks.

(a) Sort the lexicographic names back into string order such that $T_R = T_1 \oplus T_2$ where $\oplus$ is string concatenation (line 14). $T_1$ represents the complete text $T$

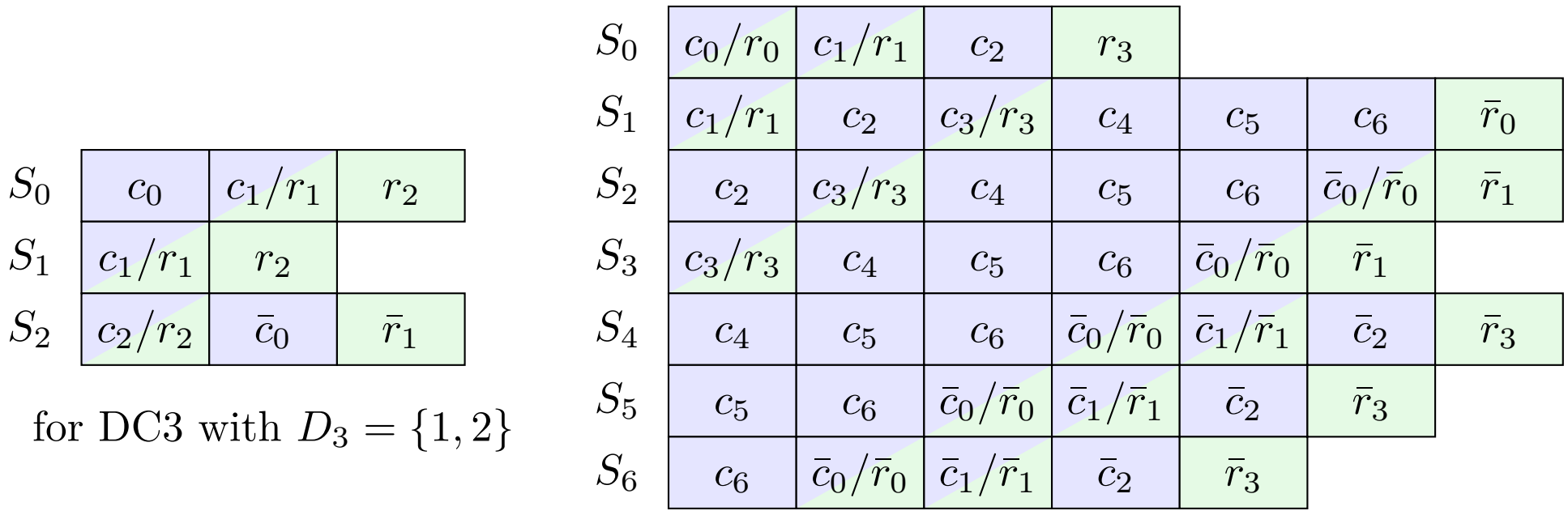

for DC3 with $D_3 = \{1, 2\}$

for DC7 with $D_7 = \{0, 1, 3\}$

**Figure 8.4:** Construction schema of tuples in arrays $S_i$ to represent suffixes in DC3 and DC7. The array $S_j$ contains a tuple for each suffix $i$ with $(i \bmod v) = j$. The diagram shows how the $v$ tuples in $S_0, \dots, S_{v-1}$ are constructed for each $(i \bmod v) = 0$ from the characters $[\, c_{i+0}, \dots, c_{i+v-1} \,]$ and $[\, c_{i+v+0}, \dots, c_{i+2v-1} \,] =:$ $[\, \bar{c}_{i+0}, \dots, \bar{c}_{i+v-1} \,]$, and ranks $[\, r_{i'+0}, \dots, r_{i'+|D|-1} \,]$ and $[\, r_{i'+|D|}, \dots, r_{i'+2|D|-1} \,] =:$ $[\, \bar{r}_{i'+0}, \dots, \bar{r}_{i'+|D|-1} \,]$ where $i' := \frac{i|D|}{v}$. For each suffix $i$ the next $|D|$ ranks $r_i$ or $\bar{r}_i$ stored in the tuple are marked green. All preceding characters $c_i$ or $\bar{c}_i$ up to but excluding the last rank are also stored in the tuple and marked blue. The comparison functions in algorithms 8.10 and 8.14 are constructed by considering pairs of rows in the diagram and comparing entries with matching colors componentwise up to a green rank. The difference cover property ensures that this is possible for every pair.

using the lexicographic names of all triples $i = 1 \bmod 3$, and $T_2$ is another complete copy of $T$ with triples $i = 2 \bmod 3$. By replacing the triples with lexicographic names, the original text is reduced by $\frac{2}{3}$.

(b) Recursively call any suffix sorting algorithm (e.g. DC3) on $T_R$ (line 15).

(c) Invert the permutation $\mathsf{SA}_R$ to gain ranks $R_1$ and $R_2$ of triples of $T$ in $D_3$, again interleave $I_R$ such that $R_1$ and $R_2$ are distributed on the workers after the Sort operation.

With $R_1$ and $R_2$ interleaved in $I_R$ from step (D.1) (lines 2 to 20), the objective of step (D.2) is to create $S_0$, $S_1$, and $S_2$ in lines 22 to 29. Each suffix $i$ has exactly one representative in the array $S_j$ where $j = i \bmod 3$. Its representative contains the recursively calculated ranks of the two following suffixes in the difference cover from $R_1$ and $R_2$ (two consecutive items from $I_R$), and the characters $T[i], T[i+1], T[i+2], \dots$ up to (but excluding) the next known rank.

For DC3 these are $T[i]$, $T[i+1]$, $I_R[\frac{2i}{3}]$, and $I_R[\frac{2i}{3}+1]$ for a suffix $i = 0 \bmod 3$ in $S_0$. $I_R[\frac{2i}{3}] = R_1[\frac{i}{3}]$ is the rank of the suffix $T[i+1..n)$ and $I_R[\frac{2i}{3}+1] = R_2[\frac{i}{3}]$ is the rank of suffix $T[i+2..n)$, which are both in the difference cover. We write the tuple as $(i, c_0, c_1, r_1, r_2)$ where the indices are interpreted relative to $i$ mod 3. Each suffix $i = 1$ mod 3 in $S_1$ stores $T[i]$, $R_1[\frac{i-1}{3}]$, and $R_2[\frac{i-1}{3}]$ and we write the tuples as $(i, c_1, r_1, r_2)$ where the indices again are relative to $i$ mod 3. And lastly, each suffix $i = 2$ mod 3 in $S_2$ stores $T[i]$, $T[i+1]$, $R_1[\frac{i-2}{3}+1]$, and $R_2[\frac{i-2}{3}]$ because $R_1[\frac{i-2}{3}+1]$ is the rank of suffix $T[i+2..n)$.

In the Thrill algorithm code we construct the tuples by zipping pairs from $I_R$, and triple groups from $T$ together. The ZipWindow $Z'$ (line 22) delivers $(c_0, c_1, c_2, r_1, r_2)$ for each index $i = 0$ mod 3. To construct the tuples in $S_i$ two adjacent tuples need to be used because $S_2$'s element are taken from the next tuple. This can be done in Thrill using a Window operation of size 2 (line 23). Thus to construct $S_0$, $S_1$, and $S_2$, we take $(c_0, c_1, c_2, r_1, r_2)$ for each index $i = 0$ mod 3 and $(\bar{c}_0, \bar{c}_1, \bar{c}_2, \bar{r}_1, \bar{r}_2)$ for the next index $i$ mod $3+3$, and output $(3i+0, c_0, c_1, r_1, r_2)$ for $S_0$, $(3i+1, c_0, c_1, r_1, c_2, r_2)$ for $S_1$, and $(3i+2, c_2, r_2, \bar{c}_0, \bar{r}_1)$ for $S_2$, as described above (lines 24 to 26). The three arrays are then sorted (lines 27 to 29) and merged, whereby the comparison function compares two representatives characterwise until a rank is found. The difference cover property guarantees that such a rank is found for every pair $S_i$, $S_j$ during the Merge (lines 30 to 35).

DC3 is definitely one of the most complex algorithm currently implemented with Thrill, but it also showcases the expressiveness of the data-flow processing approach with scalable primitives. Encouraged by this success, we ventured to implement DC7 in Thrill. Most of the previous discussion on DC3 can be extended to DC7 straightforwardly: Sort by seven characters instead of three, construct $T_R = T_0 \oplus T_1 \oplus T_3$ in case not all character tuples are unique, and have step (D.1) deliver $R_0$, $R_1$, and $R_3$ containing the ranks of all suffixes ($i$ mod 7) $\in D_7$. We included the Thrill algorithm code for DC7 in algorithms 8.12 to 8.14.

The key to implementing DC7 is the construction of the tuple contents of the seven arrays $S_0, \dots, S_6$ from $R_0$, $R_1$, $R_3$, and characters from $T$. Figure 8.4 shows a schematic to illustrate the underlying system. For each index $i$ there are three indices $((i+k_0)$ mod $7)$, $((i+k_1)$ mod $7)$, and $((i+k_3)$ mod $7)$ in the difference cover $D_7$. The offsets depend on $j = i$ mod 7 for each index $i$, which classifies the suffix into $S_j$. The tuples in the arrays must contain all characters from the text up to (but excluding) the last known rank, since this is the information needed for the comparison function to perform characterwise comparisons up to the next known rank. The components of the tuples in $S_0, \dots, S_6$ as visualized in figure 8.4 are selected in algorithm 8.13 from $Z$ using seven Map operations (lines 4 to 10). They are then sorted by characters up to the next known rank (lines 11 to 17) and then merged using CompareDC7 (algorithm 8.14), which compares tuples characterwise up to the next known rank for all possible $S_i/S_j$ pairs.

In our Thrill implementation, CompareDC7 is not rolled out as shown in the algorithm. Instead a lookup table is used to determine how many characters and which of the included ranks need to be compared. Surprisingly, this more complex code was

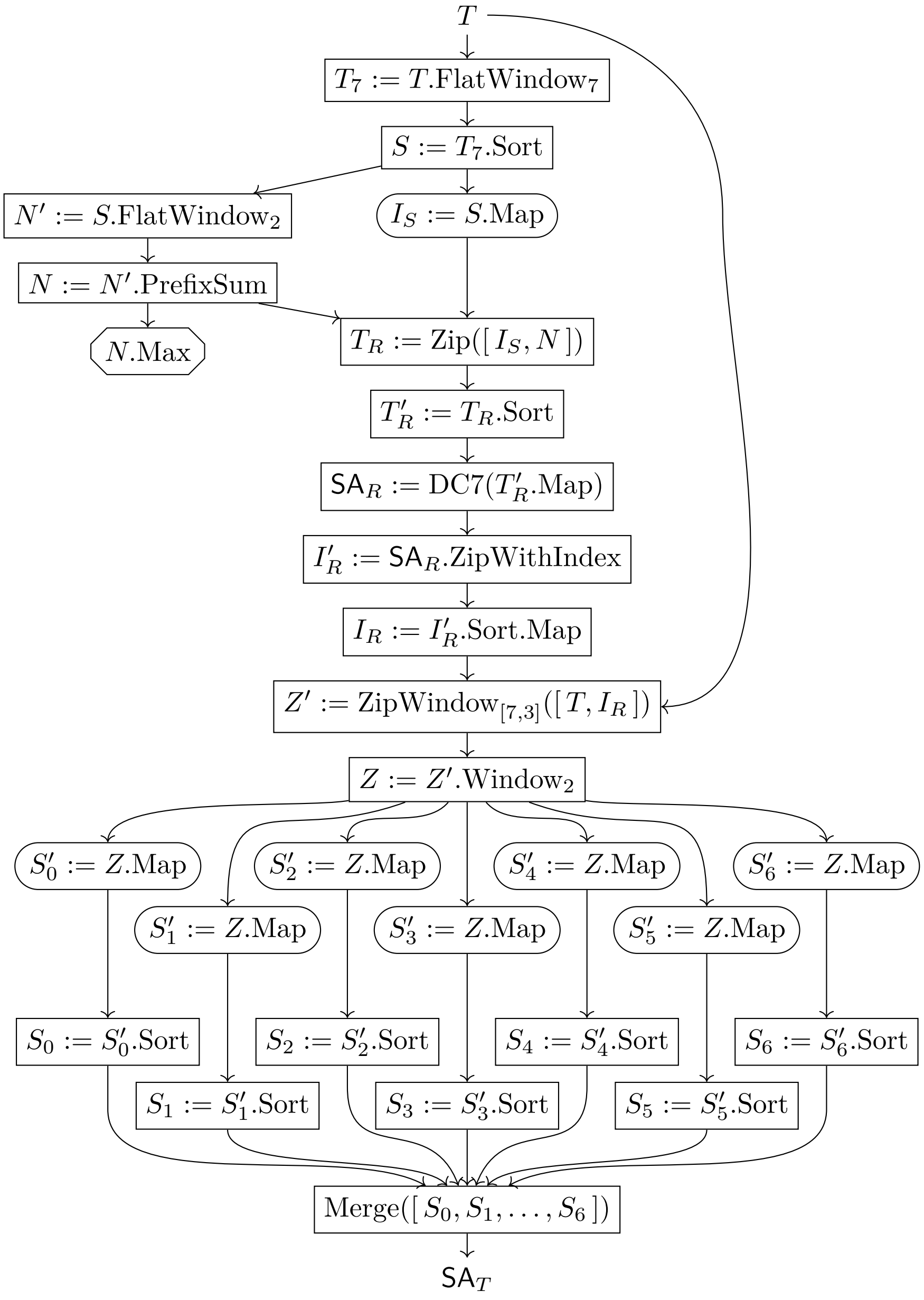

**Figure 8.5:** DIA data-flow graph of DC7 with recursion.

**Algorithm 8.12 :** DC7 Algorithm in Thrill – Part One.

**function** DC7PartOne($T \in \texttt{DIA}\langle\Sigma\rangle$)<br>
  $T_7 := T.\text{FlatWindow}_7((i, [\,c_0, c_1, \ldots, c_6\,]) \mapsto \text{MakeTuples}(i, c_0, c_1, \ldots, c_6))$<br>
  **with function** MakeTuples($i \in \mathbb{N}_0,\ c_0, c_1, \ldots, c_6 \in \Sigma$)<br>
    **if** $i \in D_7$ **then emit** $(i, c_0, c_1, \ldots, c_6)$ *// Make tuples in difference cover.*<br>
  $S := T_7.\text{Sort}((i, c_0, c_1, \ldots, c_6) \text{ by } (c_0, c_1, \ldots, c_6))$ *// Sort tuples lexicographically.*<br>
  $I_S := S.\text{Map}((i, c_0, c_1, \ldots, c_7) \mapsto i)$ *// Extract sorted indices.*<br>
  $N' := S.\text{FlatWindow}_2((i, [\,p_0, p_1\,]) \mapsto \text{CmpTuple}(i, p_0, p_1))$ *// Compare tuples.*<br>
  **with function** CmpTuple($i \in \mathbb{N}_0,\ p_0 = (c_0, c_1, \ldots, c_6),\ p_1 = (c_0', c_1', \ldots, c_6')$))<br>
    **if** $i = 0$ **then emit** 0 *// Emit one sentinel for index 0.*<br>
    **if** $(c_0, c_1, \ldots, c_6) = (c_0', c_1', \ldots, c_6')$ **then emit** 0 *// Emit 0 or 1 depending on*<br>
    **else emit** 1 *// whether the previous tuple is equal.*<br>
  $N := N'.\text{PrefixSum}()$ *// Use prefix sum to calculate names.*<br>
  $n_{\text{sub}} = \lceil 3|T|/7 \rceil, \quad n_{\text{mod0}} = \lceil |T|/7 \rceil$ *// Size of recursive problem and mod 0,*<br>
  $n_{\text{mod1}} = \lceil |T|/7 \rceil, \quad n_{\text{mod01}} = n_{\text{mod0}} + n_{\text{mod1}}$ *// mod 1 and both parts of $T_R$.*<br>
  **if** $N.\text{Max}() + 1 \neq n_{\text{sub}}$ **then** *// If duplicate names exist, sort names back to*<br>
    $T_R' := \text{Zip}([\,I_S, N\,], (i, n) \mapsto (i, n)).\text{Sort}((i, n) \text{ by } (i \bmod 7, i \text{ div } 7))$ *// string order*<br>
    $\mathsf{SA}_R := \text{DC7}(T_R'.\text{Map}((i, n) \mapsto n))$ *// as $T_0 \oplus T_1 \oplus T_3$ and call suffix sorter.*<br>
    $I_R' := \mathsf{SA}_R.\text{ZipWithIndex}((r, i) \mapsto (r, i))$ *// Invert resulting suffix array, but*<br>
    $I_R := I_R'.\text{Sort}((r, i) \text{ by } (\text{InterleavedRank}(r), r))$ *// interleave mod 0/1/3 ranks in* ISA.<br>
    **with function** InterleavedRank($r \in \mathbb{N}_0$)<br>
      **return** (**if** $r < n_{\text{mod0}}$ **then** $r$ **else if** $r < n_{\text{mod01}}$ **then** $r - n_{\text{mod0}}$ **else** $r - n_{\text{mod01}}$)<br>
  **else** *// Else, if all names/tuples are unique, then $I_S$ is already the suffix array.*<br>
    $R := I_S.\text{ZipWithIndex}((r, i) \mapsto (r, i))$ *// Invert it to get* ISA*, but*<br>
    $I_R := R.\text{Sort}((r, i) \text{ by } (r \text{ div } 7, r))$ *// interleave mod 0/1/3 ranks in* ISA.<br>
  $I_R := I_R.\text{Map}((r, i) \mapsto (i + 1))$ *// Extract ranks from $I_R$, free rank zero.*<br>
  **return** DC7PartTwo($T, I_R$)

faster in our preliminary experiments, possibly due to the larger cost of decoding the instructions or higher number of branch mispredictions in the large unrolled comparison function.

## 8.4 Distributed External Memory Experiments

As described in the previous sections of this chapter, we implemented five suffix sorting algorithms using Thrill in C++. These implementations are available as open-source among the examples included with Thrill at `https://github.com/thrill/thrill`, and are currently the most complex applications of the framework. In a sense, these suffix sorting algorithms were a main driver in the design and implementation of Thrill, but they are also a case study of how well one can express complex algorithms using a data-flow description in C++.

**Algorithm 8.13 :** DC7 Algorithm in Thrill – Part Two.

**function** DC7PartTwo($T \in \texttt{DIA}\langle\Sigma\rangle$, $I_R \in \texttt{DIA}\langle\mathbb{N}_0\rangle$)

$Z' := \text{ZipWindow}_{[7,3]}([\,T, I_R\,], (i, [\,c_0, \dots, c_6\,], [\,r_0, r_1, r_3\,]) \mapsto (c_0, \dots, c_6, r_0, r_1, r_3))$ *// Pull*

$Z := Z'.\text{Window}_2((i, [\,(z_1, z_2)\,]) \mapsto (i, z_1, z_2))$ *// chars and ranks using Zip from*

$S'_0 := Z.\text{Map}((i, (c_0, \dots, c_6, r_0, r_1, r_3), (\bar{c}_0, \dots, \bar{c}_6, \bar{r}_0, \bar{r}_1, \bar{r}_3)))$ *// T and $I_R$*
$\quad \mapsto (7i + 0, c_0, r_0, c_1, r_1, c_2, r_3))$

$S'_1 := Z.\text{Map}((i, (c_0, \dots, c_6, r_0, r_1, r_3), (\bar{c}_0, \dots, \bar{c}_6, \bar{r}_0, \bar{r}_1, \bar{r}_3)))$ *// to make*
$\quad \mapsto (7i + 1, c_1, r_1, c_2, c_3, r_3, c_4, c_5, c_6, \bar{r}_0)$

$S'_2 := Z.\text{Map}((i, (c_0, \dots, c_6, r_0, r_1, r_3), (\bar{c}_0, \dots, \bar{c}_6, \bar{r}_0, \bar{r}_1, \bar{r}_3)))$ *// arrays of*
$\quad \mapsto (7i + 2, c_2, c_3, r_3, c_4, c_5, c_6, \bar{c}_0, \bar{r}_0, \bar{r}_1)$

$S'_3 := Z.\text{Map}((i, (c_0, \dots, c_6, r_0, r_1, r_3), (\bar{c}_0, \dots, \bar{c}_6, \bar{r}_0, \bar{r}_1, \bar{r}_3)))$ *// representatives*
$\quad \mapsto (7i + 3, c_3, r_3, c_4, c_5, c_6, \bar{c}_0, \bar{r}_0, \bar{r}_1)$

$S'_4 := Z.\text{Map}((i, (c_0, \dots, c_6, r_0, r_1, r_3), (\bar{c}_0, \dots, \bar{c}_6, \bar{r}_0, \bar{r}_1, \bar{r}_3)))$ *// for each*
$\quad \mapsto (7i + 4, c_4, c_5, c_6, \bar{c}_0, \bar{r}_0, \bar{c}_1, \bar{r}_1, \bar{c}_2, \bar{r}_3)$

$S'_5 := Z.\text{Map}((i, (c_0, \dots, c_6, r_0, r_1, r_3), (\bar{c}_0, \dots, \bar{c}_6, \bar{r}_0, \bar{r}_1, \bar{r}_3)))$ *// suffix class.*
$\quad \mapsto (7i + 5, c_5, c_6, \bar{c}_0, \bar{r}_0, \bar{c}_1, \bar{r}_1, \bar{c}_2, \bar{r}_3)$

$S'_6 := Z.\text{Map}((i, (c_0, \dots, c_6, r_0, r_1, r_3), (\bar{c}_0, \dots, \bar{c}_6, \bar{r}_0, \bar{r}_1, \bar{r}_3)))$
$\quad \mapsto (7i + 6, c_6, \bar{c}_0, \bar{r}_0, \bar{c}_1, \bar{r}_1, \bar{c}_2, \bar{r}_3)$

$S_0 := S'_0.\text{Sort}((i, c_0, r_0, c_1, r_1, c_2, r_3) \text{ by } (r_0))$ *// Sort representatives*

$S_1 := S'_1.\text{Sort}((i, c_1, r_1, c_2, c_3, r_3, c_4, c_5, c_6, \bar{r}_0) \text{ by } (r_1))$ *// characterwise up to*

$S_2 := S'_2.\text{Sort}((i, c_2, c_3, r_3, c_4, c_5, c_6, \bar{c}_0, \bar{r}_0, \bar{r}_1) \text{ by } (c_2, r_3))$ *// next rank, and merge*

$S_3 := S'_3.\text{Sort}((i, c_3, r_3, c_4, c_5, c_6, \bar{c}_0, \bar{r}_0, \bar{r}_1) \text{ by } (r_3))$ *// sorted representatives*

$S_4 := S'_4.\text{Sort}((i, c_4, c_5, c_6, \bar{c}_0, \bar{r}_0, \bar{c}_1, \bar{r}_1, \bar{c}_2, \bar{r}_3) \text{ by } (c_4, c_5, c_6, \bar{r}_0))$ *// to deliver the*

$S_5 := S'_5.\text{Sort}((i, c_5, c_6, \bar{c}_0, \bar{r}_0, \bar{c}_1, \bar{r}_1, \bar{c}_2, \bar{r}_3) \text{ by } (c_5, c_6, \bar{r}_0))$ *// final suffix array.*

$S_6 := S'_6.\text{Sort}((i, c_6, \bar{c}_0, \bar{r}_0, \bar{c}_1, \bar{r}_1, \bar{c}_2, \bar{r}_3) \text{ by } (c_6, \bar{r}_0))$ *// See algorithm 8.14*

**return** $\text{Merge}([\,S_0, S_1, \dots, S_6\,], \text{CompareDC7}).\text{Map}((i, \dots) \mapsto i)$ *// for* CompareDC7.

**Algorithms.** We label prefix doubling with a Window on the inverse suffix array from section 8.2.3 as **T.PD-Window**, prefix doubling with sorting from section 8.2.4 as **T.PD-Sort**, and prefix doubling with discarding from section 8.2.5 as **T.PD-Discard**. From section 8.3.1 we include **T.DC3** and **T.DC7**. All Thrill implementations in the experiment are thus prefixed with a **T**, and they use 40-bit (5 byte) indices in the suffix array such as to support up to inputs of up to 1 TiB. The program can also be compiled to use 48-bit or 64-bit indices.

There are only two other distributed suffix sorting implementations available. The first is pDC3 implemented using MPI by Kulla and Sanders [KS06b; Kul06; KS07]. As a learning experience while preparing this dissertation, we took their implementation, generalized it, rewrote large parts, and extended it to **BKS.pDC3** and **BKS.pDC7**. Our improved pDCX version is available at `https://github.com/bingmann/pDCX`. These variants only support 32-bit (4 byte) indices, and thus are limited to inputs of up to 4 GiB.

The second implementation is **FA.psac** by Flick and Aluru [FA15] which is a highly engineered suffix sorter using MPI. It is based on prefix doubling with the inverse

**Algorithm 8.14 :** Full Comparison Function in DC7.

**function** CompareDC7($z_1, z_2$)

| | | | |
|---|---|---|---|
| $(r_0)$ | $<$ | $(r'_1)$ | if $z_1 \in S_0, z_2 \in S_1$, |
| $(c_0, r_1)$ | $<$ | $(c'_2, r'_3)$ | if $z_1 \in S_0, z_2 \in S_2$, |
| $(r_0)$ | $<$ | $(r'_3)$ | if $z_1 \in S_0, z_2 \in S_3$, |
| $(c_0, c_1, c_2, r_3)$ | $<$ | $(c'_4, c'_5, c'_6, \bar{r}'_0)$ | if $z_1 \in S_0, z_2 \in S_4$, |
| $(c_0, c_1, c_2, r_3)$ | $<$ | $(c'_5, c'_6, \bar{c}'_0, \bar{r}'_1)$ | if $z_1 \in S_0, z_2 \in S_5$, |
| $(c_0, r_1)$ | $<$ | $(c'_6, \bar{r}'_0)$ | if $z_1 \in S_0, z_2 \in S_6$, |
| $(c_1, c_2, c_3, c_4, c_5, c_6, \bar{r}_0)$ | $<$ | $(c'_2, c'_3, c'_4, c'_5, c'_6, \bar{c}'_0, \bar{r}'_1)$ | if $z_1 \in S_1, z_2 \in S_2$, |
| $(r_1)$ | $<$ | $(r'_3)$ | if $z_1 \in S_1, z_2 \in S_3$, |
| $(c_1, c_2, c_3, c_4, c_5, c_6, \bar{r}_0)$ | $<$ | $(c'_4, c'_5, c'_6, \bar{c}'_0, \bar{c}'_1, \bar{c}'_2, \bar{r}'_3)$ | if $z_1 \in S_1, z_2 \in S_4$, |
| $(c_1, c_2, r_3)$ | $<$ | $(c'_5, c'_6, \bar{r}'_0)$ | if $z_1 \in S_1, z_2 \in S_5$, |
| $(c_1, c_2, r_3)$ | $<$ | $(c'_6, \bar{c}'_0, \bar{r}'_1)$ | if $z_1 \in S_1, z_2 \in S_6$, |
| $(c_2, c_3, c_4, c_5, c_6, \bar{r}_0)$ | $<$ | $(c_3, c_4, c_5, c_6, \bar{c}_0, \bar{r}'_1)$ | if $z_1 \in S_2, z_2 \in S_3$, |
| $(c_2, c_3, c_4, c_5, c_6, \bar{c}_0, \bar{r}_1)$ | $<$ | $(c'_4, c'_5, c'_6, \bar{c}'_0, \bar{c}'_1, \bar{c}'_2, \bar{r}'_3)$ | if $z_1 \in S_2, z_2 \in S_4$, |
| $(c_2, c_3, c_4, c_5, c_6, \bar{r}_0)$ | $<$ | $(c'_5, c'_6, \bar{c}'_0, \bar{c}'_1, \bar{c}'_2, \bar{r}'_3)$ | if $z_1 \in S_2, z_2 \in S_5$, |
| $(c_2, r_3)$ | $<$ | $(c'_6, \bar{r}'_0)$ | if $z_1 \in S_2, z_2 \in S_6$, |
| $(c_3, c_4, c_5, c_6, \bar{r}_0)$ | $<$ | $(c'_4, c'_5, c'_6, \bar{c}'_0, \bar{r}'_1)$ | if $z_1 \in S_3, z_2 \in S_4$, |
| $(c_3, c_4, c_5, c_6, \bar{c}_0, \bar{r}_1)$ | $<$ | $(c'_5, c'_6, \bar{c}_0, \bar{c}_1, \bar{c}_2, \bar{r}'_3)$ | if $z_1 \in S_3, z_2 \in S_5$, |
| $(c_3, c_4, c_5, c_6, \bar{r}_0)$ | $<$ | $(c'_6, \bar{c}'_0, \bar{c}'_1, \bar{c}'_2, \bar{r}'_3)$ | if $z_1 \in S_3, z_2 \in S_6$, |
| $(c_4, c_5, c_6, \bar{r}_0)$ | $<$ | $(c'_5, c'_6, \bar{c}'_0, \bar{r}'_1)$ | if $z_1 \in S_4, z_2 \in S_5$, |
| $(c_4, c_5, c_6, \bar{c}_0, \bar{r}_0)$ | $<$ | $(c'_6, \bar{c}'_0, \bar{c}'_1, \bar{c}'_2, \bar{r}'_3)$ | if $z_1 \in S_4, z_2 \in S_6$, |
| $(c_6, \bar{c}_0, \bar{r}_1)$ | $<$ | $(c'_6, \bar{c}'_0, \bar{r}'_1)$ | if $z_1 \in S_5, z_2 \in S_6$, |

and symmetrically for $z_1 \in S_i, z_2 \in S_j$ if $i > j$ ,

with $z_1 = (i, c_0, r_0, c_1, r_1, c_2, r_3)$ if $z_1 \in S_0$,
$z_2 = (i', c'_0, r'_0, c'_1, r'_1, c'_2, r'_3)$ if $z_2 \in S_0$,
$z_1 = (i, c_1, r_1, c_2, c_3, r_3, c_4, c_5, c_6, \bar{r}_0)$ if $z_1 \in S_1$,
$z_2 = (i', c'_1, r'_1, c'_2, c'_3, r'_3, c'_4, c'_5, c'_6, \bar{r}'_0)$ if $z_2 \in S_1$,
$z_1 = (i, c_2, c_3, r_3, c_4, c_5, c_6, \bar{c}_0, \bar{r}_0, \bar{r}_1)$ if $z_1 \in S_2$,
$z_2 = (i', c'_2, c'_3, r'_3, c'_4, c'_5, c'_6, \bar{c}'_0, \bar{r}'_0, \bar{r}'_1)$ if $z_2 \in S_2$,
$z_1 = (i, c_3, r_3, c_4, c_5, c_6, \bar{c}_0, \bar{r}_0, \bar{r}_1)$ if $z_1 \in S_3$,
$z_2 = (i', c'_3, r'_3, c'_4, c'_5, c'_6, \bar{c}'_0, \bar{r}'_0, \bar{r}'_1)$ if $z_2 \in S_3$,
$z_1 = (i, c_4, c_5, c_6, \bar{c}_0, \bar{r}_0, \bar{c}_1, \bar{r}_1, \bar{c}_2, \bar{r}_3)$ if $z_1 \in S_4$,
$z_2 = (i, c'_4, c'_5, c'_6, \bar{c}'_0, \bar{r}'_0, \bar{c}'_1, \bar{r}'_1, \bar{c}'_2, \bar{r}'_3)$ if $z_2 \in S_4$,
$z_1 = (i, c_5, c_6, \bar{c}_0, \bar{r}_0, \bar{c}_1, \bar{r}_1, \bar{c}_2, \bar{r}_3)$ if $z_1 \in S_5$,
$z_2 = (i', c'_5, c'_6, \bar{c}'0, \bar{r}'_0, \bar{c}'_1, \bar{r}'_1, \bar{c}'_2, \bar{r}'_3)$ if $z_2 \in S_5$,
$z_1 = (i, c_6, \bar{c}_0, \bar{r}_0, \bar{c}_1, \bar{r}_1, \bar{c}_2, \bar{r}_3)$ if $z_1 \in S_6$,
$z_2 = (i', c'_6, \bar{c}'_0, \bar{r}'_0, \bar{c}'_1, \bar{r}'_1, \bar{c}'_2, \bar{r}'_3)$ if $z_2 \in S_6$.

suffix array, but they enhanced it with alphabet compression (see section 8.2.2) in the first iteration and by using list ranking instead of a full sort when only 1/10-th of all suffixes remain unordered. The psac implementation always works with 64-bit (8 byte) indices in the suffix array.

We also compare our distributed parallel implementations against the fastest non-distributed suffix sorters, Mori's divsufsort 2.0.2-1 [Mor06] and sais 2.4.1 [Mor08]. Divsufsort comes in two variants: **M.divsufsort** does not use any parallelization, and **M.divsufsort.par** uses OpenMP parallelization only in the first string sorting phase. Mori's sais is always only sequential, included as **M.sais**, and is an engineered version of SA-IS [NZC09a; NZC11]. We had to fix an error in sais 2.4.1 for 2 GiB inputs, and could not find a second bug which makes it crash for inputs larger than 4 GiB. We used the versions of divsufsort and sais with 64-bit (8 byte) indices in the suffix array.

**Inputs.** As inputs for our experiments, we reuse the **Gutenberg** and **Pi** inputs from section 6.3 (page 218). Their characteristics are shown in figure 6.8 (c) (page 221). We also include an English **Wikipedia** XML dump input, but in a more recent version than in the previous chapter: for these experiment we use `enwiki-201701`, which is 125.6 GiB in size, and probably has similar characteristics as the version used in section 6.3. Gutenberg and Wikipedia are diverse real-world inputs, while Pi is random. We ran some additional experiments with DNA data, but Pi delivered more consistent results for random inputs. For scalability tests, we take prefixes of size $[0\mathinner{..}2^k)$ of the inputs.

**Platform.** We ran our experiments on the Amazon Web Services (AWS) Elastic Compute Cloud (EC2) using **i3.4xlarge** instances. Each host had 16 Intel Xeon E5-2686 v4 Broadwell vCPUs with 2.30 GHz clock speed, 122 GB RAM, and $2 \times 1.9$ TB Non-Volatile Memory Express (NVMe) SSDs. We measured that these SSDs reach effective sustained sequential read speeds of 2.1 GiB/s and write speeds of 800 MiB/s each. The hosts were connected via the AWS network, and reached 1 144 MiB/s simultaneous pair-wise throughput bandwidth and 80 µs ping-pong round-trip latency in a four host test setup.

For performing distributed external memory experiments, we *limited* the available RAM on each host to 8 GB by setting the kernel option `mem=8G`. This actually only leaves about 7 GB for Thrill or other applications, because the kernel reserves itself a considerable portion. This limitation is extreme, but demonstrates that Thrill will efficiently utilize disk space when needed. For our non-distributed comparison experiments with divsufsort and sais we removed the memory limit.

All experiments were run with the Thrill master branch version from January 19th, 2018, compiled with `g++` 5.4.1 an Ubuntu Linux 16.04 "xenial" using Linux 4.4.0-1052-aws.

We would like thank the AWS Cloud Credits for Research program for making the experiments possible. Our suffix sorting inputs were stored on the AWS Elastic File

System (EFS) and transferred via NFS to the compute hosts. In total, the experiments reported in the next section took 4 125 compute instance-hours and cost $ 1 713.

### 8.4.1 The Results

We ran the algorithms on $h$ instances for all powers of two ranging from 1 to 32; as each host had 16 cores, the highest core/worker count in our experiment was 512. For each host configuration, we ran the algorithms on all input prefixes from 16 MiB to $h \cdot 1$ GiB, again doubling the size in each step. Due to the quantity and considerable cost of the experiments, we only ran each configuration once. All constructed suffix arrays were verified to be correct using a checking algorithm [DKMS08, Section 8].

First consider the results shown in figures 8.7 to 8.9. The graphs show the throughput of all suffix sorting configurations we ran in our experiment. As expected, not all algorithms succeeded in scaling to the large input sizes with limited available memory, including some of our implementations in Thrill.

The MPI implementations are clearly limited by RAM: considering that suffix sorting $n$ bytes with 8 byte indices requires at least $9n$ bytes of RAM, at most about 770 MiB can be sorted by a single host in this setting. For 5 byte indices, the constraint rises to about 1 150 MiB on a single host. However, the MPI implementations BKS.pDC3, BKS.pDC7, and FA.psac already stopped working much earlier than these hard limits: on one and two hosts they could process at most 128 MiB, on four hosts at most 256 MiB, on eight hosts at most 512 MiB, on 16 hosts at most 1 GiB, and on 32 hosts at most 2 GiB. FA.psac performed very well on the random Pi input, which was to be expected from a prefix doubling algorithm. It was also fast on Wikipedia, but was much slower on Gutenberg. Most remarkable, however, is that it did not scale well on any input: while it was very fast on Pi for a small number of hosts, on eight or more its performance degraded quickly. On Wikipedia the suffix sorting speed did not increase well when adding more hosts. A possible reason is that the inputs and available RAM size were too small for the algorithms to reach their full potential. BKS.pDC3 and BKS.pDC7 incur the same problems as FA.psac: their performance only increases slightly with more hosts. However, their overall performance is more stable across inputs due to the underlying difference cover algorithm. While BKS.pDCX may suffer from some less well-engineered implementation details, FA.psac is high quality code, which makes its unfavorable scalability in a memory constrained environment even more surprising and unlikely.

Let us now turn to the Thrill implementations. As discussed earlier, T.PD-Window is limited in the window size by the available RAM, and hence cannot sort inputs with long LCPs. Both T.PD-Window and T.PD-Sort are slow on Gutenberg and Wikipedia, but fast on Pi, again which is to be expected from prefix doubling. T.PD-Window and T.PD-Sort are also not able to sort large prefixes of Gutenberg and Wikipedia on many hosts.

The three best algorithms are T.PD-Discard, T.DC3, and T.DC7 which are able to process all inputs, except for the very largest instances. The implementations fail with 32 GiB on 32 hosts, probably because the amount of buffers and metadata in Thrill grows too large for the limited RAM available. Compared to T.PD-Window and T.PD-Sort, the discarding optimization in T.PD-Discarding really makes prefix doubling practical for larger inputs.

The throughput of all Thrill implementations increases first with the input size and also with the number of hosts, until the throughput starts dropping at 1–4 GiB size. The turning point is when external memory usage starts to impact sorting performance, as can be seen by comparing the throughput graphs with the resource utilization plots in figure 8.10. This obviously must slow down the Thrill implementations, however the NVMe SSDs are so fast that the throughput only drops down to levels that many MPI implementation reach with RAM.

On the input Pi, performance peaks at 1 GiB size, reaching more than 95 MiB/s throughput with 32 hosts. For larger Pi inputs, external memory usage increases, and throughput drops to 25 MiB/s. Pi is the only input for which our prefix doubling T.PD-Sort works well; the more complex T.PD-Discard is second best and gains no advantage over regular prefix doubling algorithms for this uniformly random input. On Wikipedia and Gutenberg the picture changes completely: unoptimized prefix doubling algorithms are much slower and often even fail to sort the input. Clearly the best implementation for these inputs is T.DC7, which outperforms on nearly all instance of Wikipedia and Gutenberg. T.DC3 appears to be slower by a constant factor, and T.PD-Discard also performs very well, but with a different characteristic than the difference cover algorithms.

To better compare the scalability properties, we present weak scaling plots in figure 8.6. These can be considered *slices* of the previous diagrams: the left panels shows all experiments with $h \cdot 256$ MiB input per host and the right panels all with $h \cdot 1$ GiB per host. Only Thrill implementations function properly with these parameters. Additionally, we include results from the non-distributed algorithms M.divsufsort, M.divsufsort.par, and M.sais. These run on only *one* host, but with the *same* amount of input data as the distributed experiments. As these three implementations are the fastest non-distributed suffix sorters, we can determine the number of hosts needed for distributed algorithms to outperform.

Despite the full 122 GB RAM available, M.divsufsort, M.divsufsort.par, and M.sais could not suffix sort 16 GiB inputs or larger, because the algorithms require at least $9n$ working space. M.sais failed even for 8 GiB due to an unknown error in the program. T.DC7 outperformed M.divsufsort on Gutenberg and Wikipedia with four hosts, and on Pi T.PD-Discard outperformed already for two host. Considering these numbers, one has to bear in mind, that Thrill uses all 16 cores on the hosts, while M.divsufsort only uses one. M.divsufsort.par uses OpenMP parallelism, but that does not have a large impact. M.sais is slightly faster than M.divsufsort on our inputs. Previous experiments on the performance of big data frameworks [MIM15] reported

"Configuration that Outperforms a Single Thread (COST)" ratios of 16 to 512 for PageRank, and 10 to 100 for graph connectivity. The COST ratio of our suffix sorters are thus 32–64, which is impressive considering that suffix sorters based on induced sorting have a significant algorithmic advantage (compare part II). However, we were unable to replicate the 110x speedup reported for FA.psac [FA15] over M.divsufsort, probably due to our cheaper commodity hardware.

Figures 8.11 to 8.13 show the *throughput ratio* of our Thrill implementations over the MPI implementation of the same algorithm. Using these plots we can determine how well Thrill executes the algorithms using the data-flow concepts. Overall, for small inputs and host numbers the throughput ratio is high, up to seven on Wikipedia but usually only 2–3 on other inputs, meaning MPI programs are much faster on small inputs. But once input sizes and host numbers increase Thrill's efficiency rises, and even beats the specialized BKS.pDC3 and BKS.pDC7 MPI implementations. FA.psac however is much better optimized, and T.PD-Window is considerably slower. But the plots in figures 8.11 to 8.13 only show the instances in which both algorithms correctly computed the suffix array, the Thrill implementations scale to larger instances and gain relative performance on these.

To gauge how well Thrill utilizes the computation resources, we include execution profiles of some of the largest instances in figures 8.14 to 8.17. Each profile has a top panel, containing the total amount of bytes in DIAs across all hosts, and two series distinguishing the amount of these bytes currently in RAM and on disk. The bottom panel of each profile shows the CPU utilization, and network and disk throughput averaged over all hosts.

The structure of the profiles correspond to how the algorithms operate. For example, one can clearly see the iterations in the simple prefix doubling algorithms T.PD-Window and T.PD-Sort in figure 8.14. T.PD-Discard also exhibits these iterations (they are more clearly seen in the CPU usage), but they are dampened in the data panel because fewer and fewer suffixes remain to be reordered, while their total number across all DIAs remains constant.

In the profiles of T.DC3 and T.DC7 in figures 8.15 to 8.17 one can identify the two phases of the recursion: the descent and the unwind. For T.DC3 in figure 8.15 the descent takes from 0–100 s and the unwind from 100–270 s. This is can identified by considering the sorting cycles in the data size series: As the suffix sorting problem grows smaller the cycles dampen, until the recursion is finished and the cycles grow larger again during the unwind. The remaining execution time corresponds to the suffix checker at the end. For T.DC7 in the same figure, the descent takes from 0–50 s and the unwind from 50–200 s. Both times the unwind is characteristic in the plateaus in which the DIAs $S_i$ are created and sorted (lines 4 to 17 of algorithm 8.13), and the drops in DIA data when they are merged. In figures 8.16 to 8.17 the same plateaus can be identified, but the recursion descent seems to be much quicker. Across all our DC runs, the recursion descent requires much less execution time compared to the unwind.

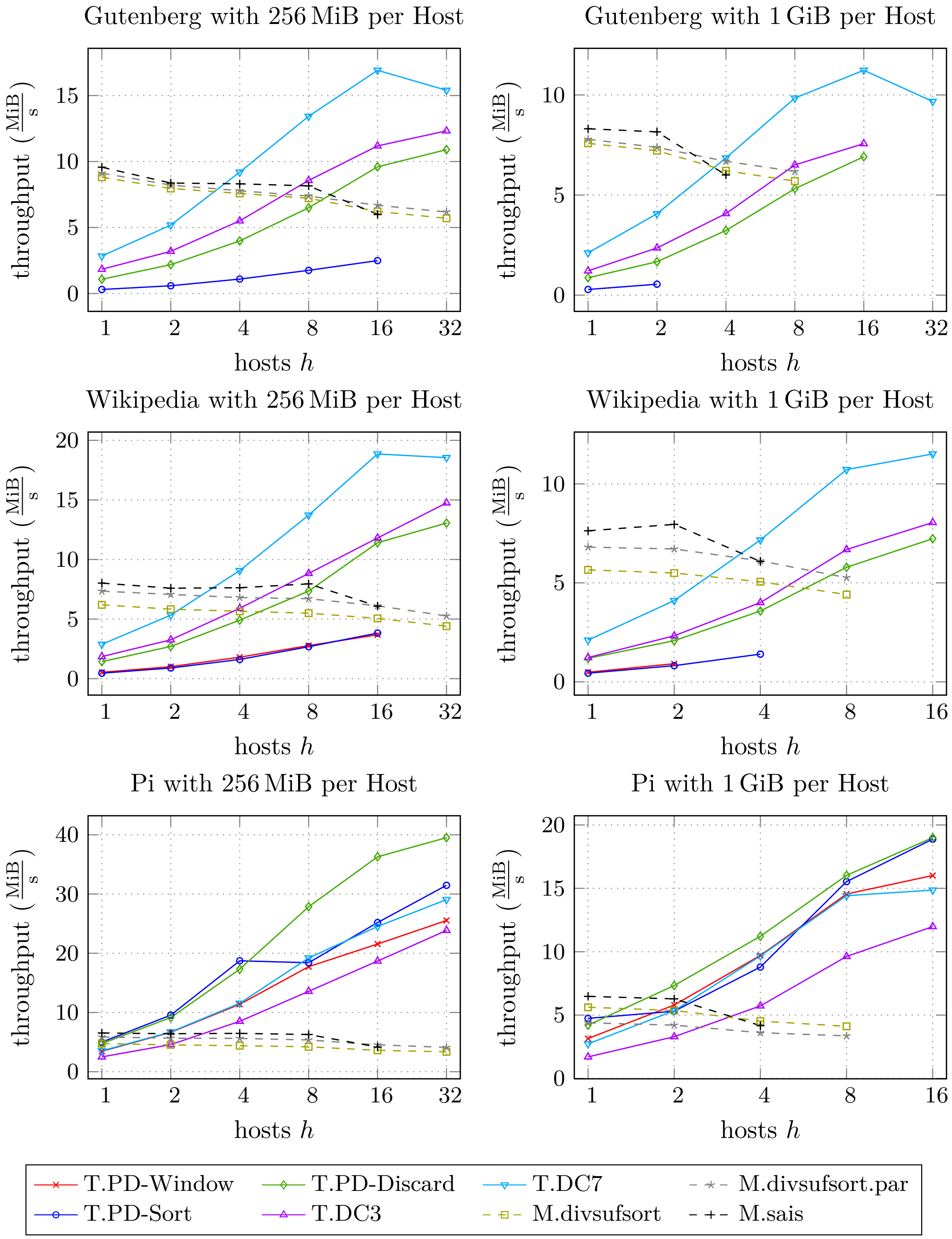

**Figure 8.6:** Weak scaling plots of distributed suffix sorting algorithms and of the fastest non-distributed suffix sorters run on one host with the same input size.

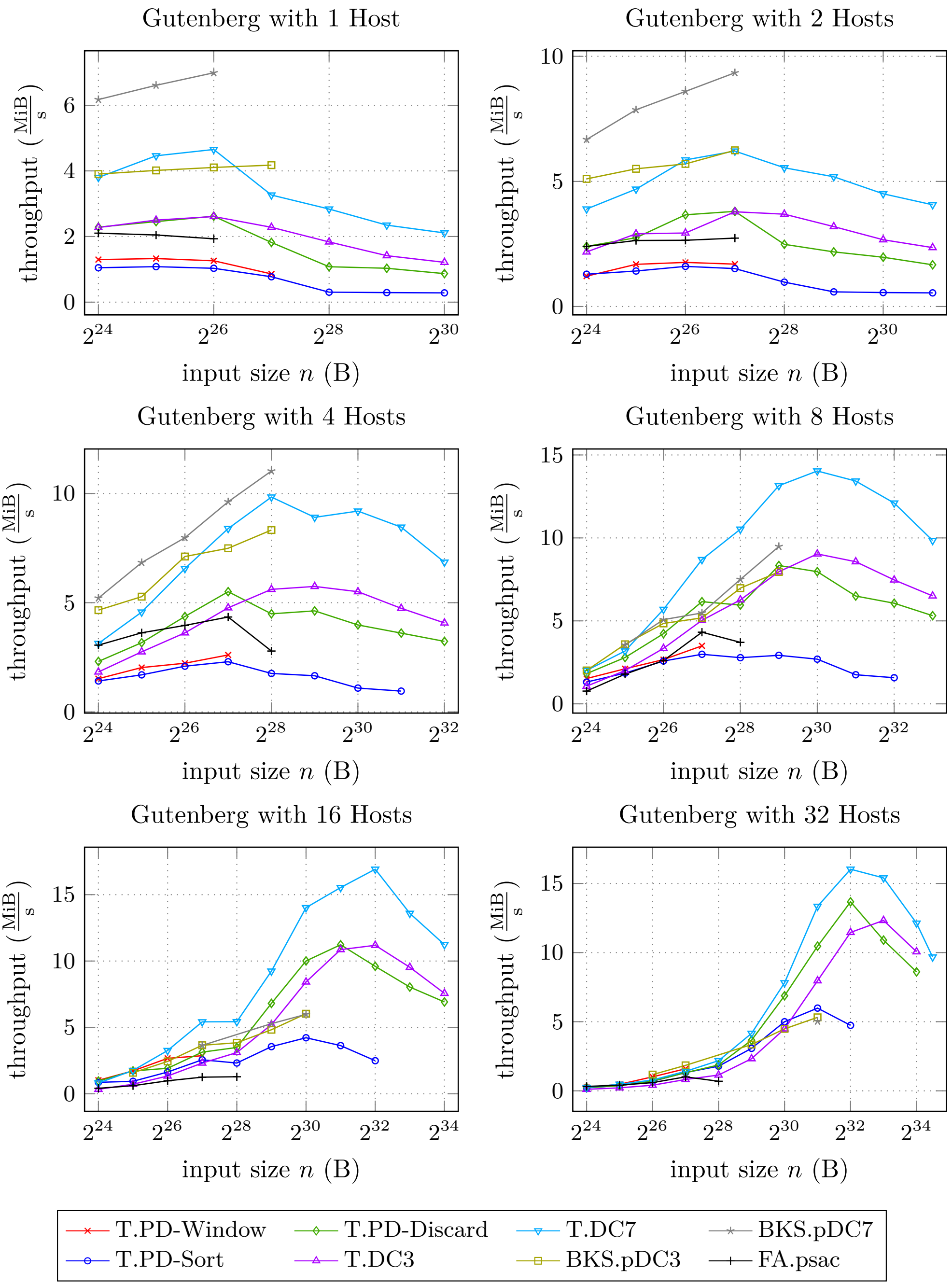

**Figure 8.7:** Throughput of algorithms with 1–32 hosts on Gutenberg input.

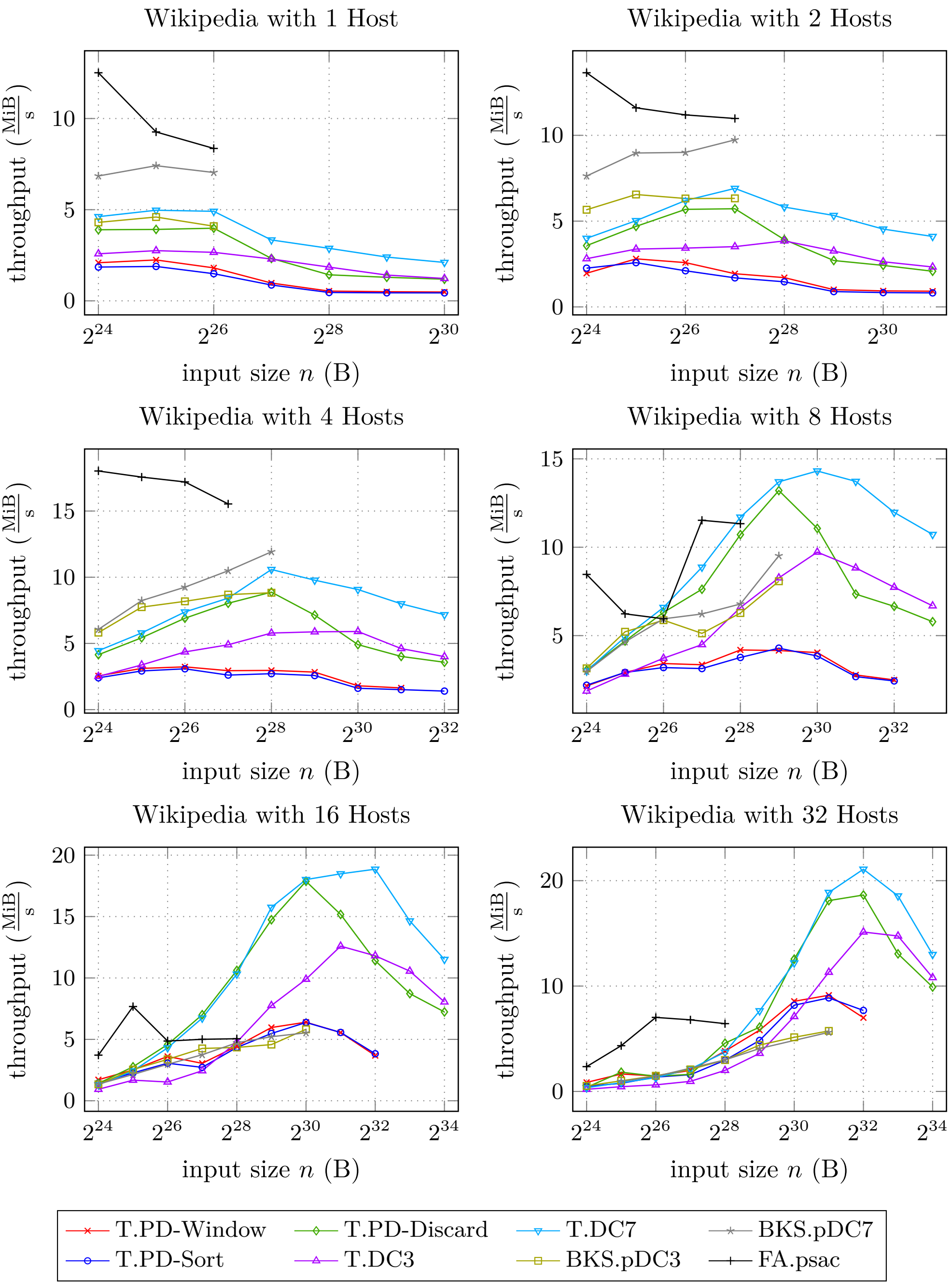

**Figure 8.8:** Throughput of algorithms with 1–32 hosts on Wikipedia input.

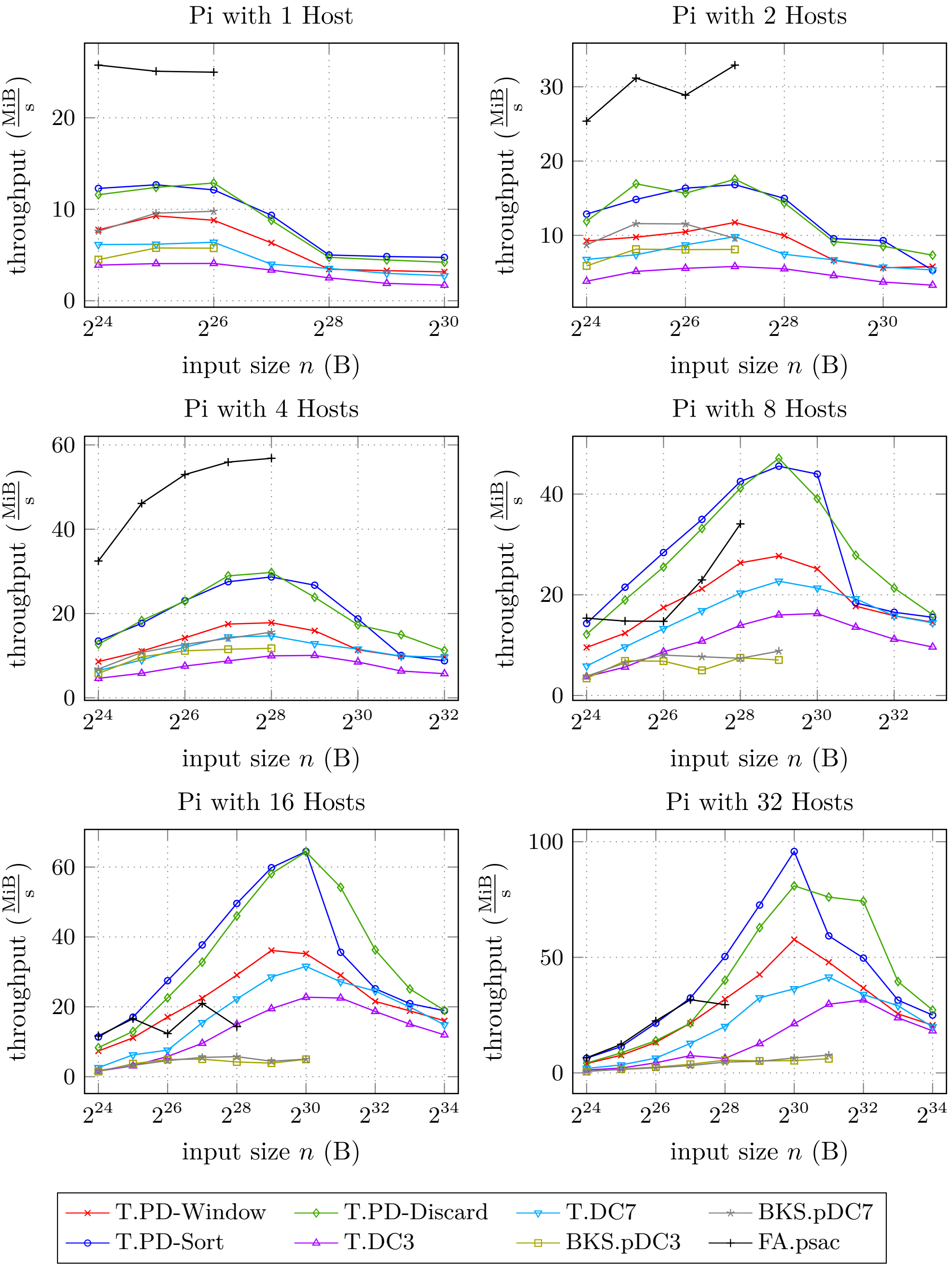

**Figure 8.9:** Throughput of algorithms with 1–32 hosts on Pi input.

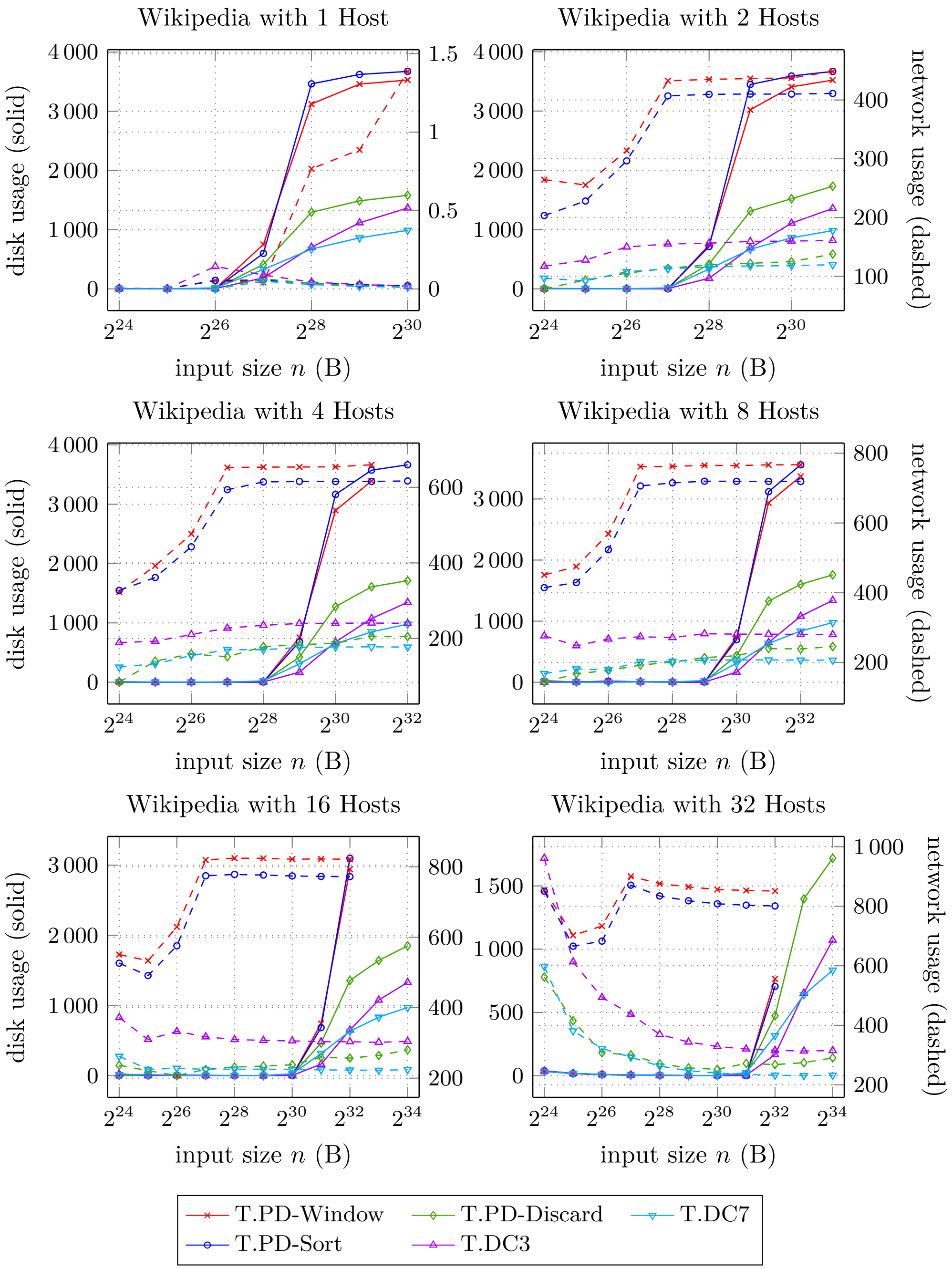

**Figure 8.10:** Disk and network resource usage as ratio of bytes transferred per input byte with 1–32 hosts on Wikipedia input.

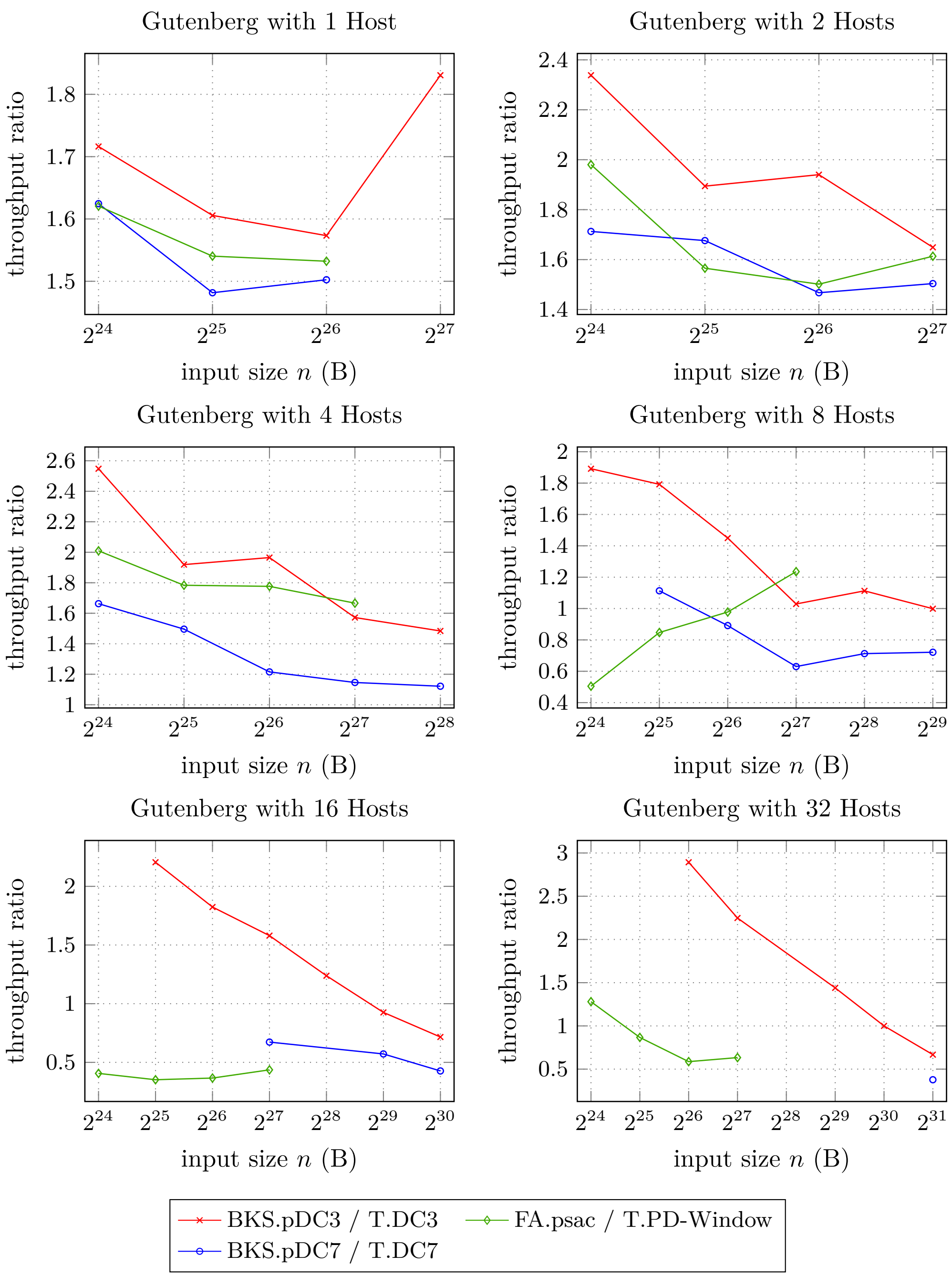

**Figure 8.11:** Throughput ratio of MPI implementation over same algorithm in Thrill on Gutenberg input.

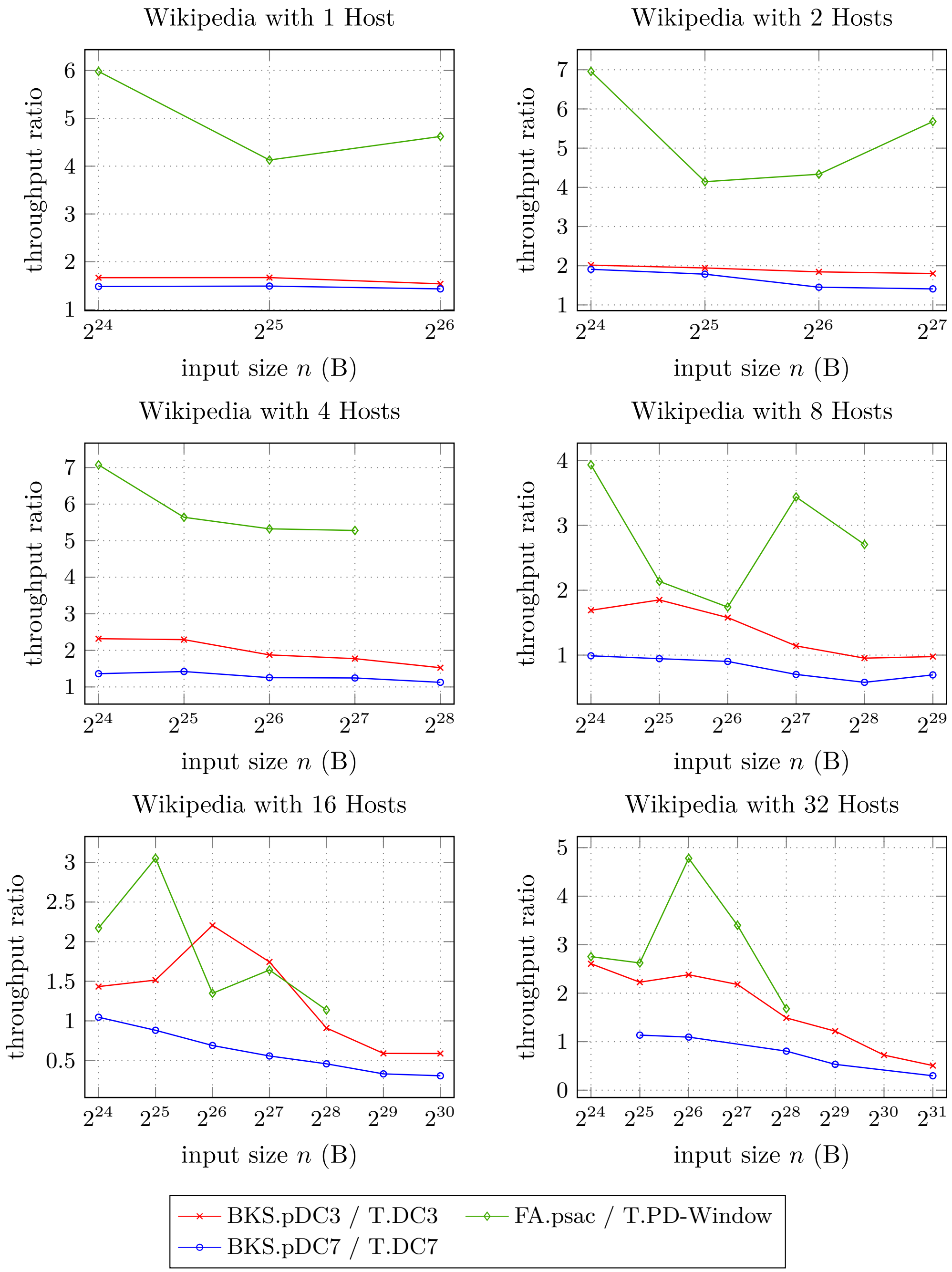

**Figure 8.12:** Throughput ratio of MPI implementation over same algorithm in Thrill on Wikipedia input.

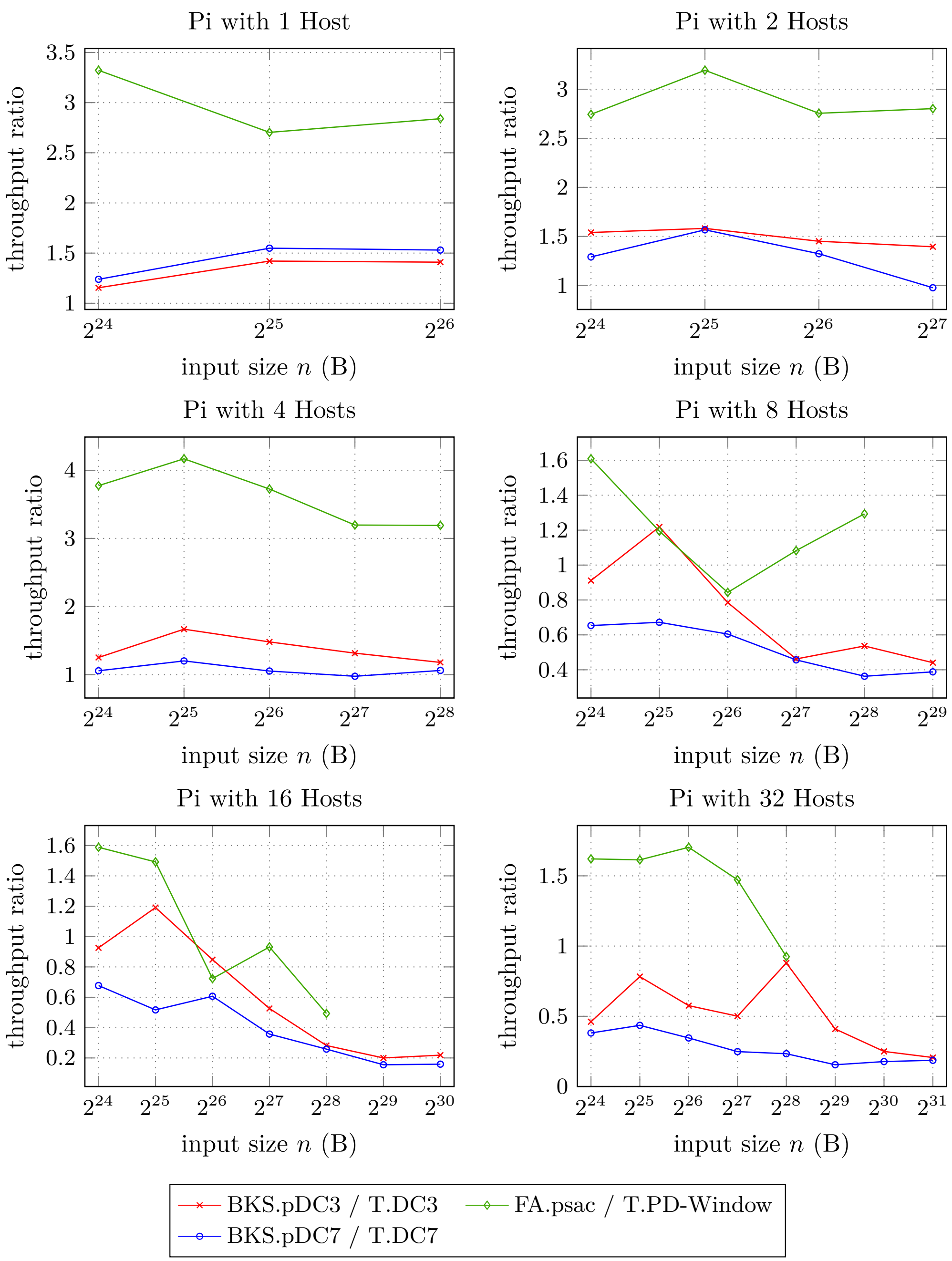

**Figure 8.13:** Throughput ratio of MPI implementation over same algorithm in Thrill on Pi input.

Considering the efficiency of the Thrill execution profiles, we believe a lot can still be improved by optimizing the underlying scalable primitives. For example, the resource utilization in the long plateaus of T.DC7 in figures 8.16 to 8.17 is very low. It is unclear what the bottleneck is during these phases. T.PD-Discard also has similar intervals of low resource usage. More work is needed in this direction, and by improving the Thrill constructs, all applications using them will benefit.

## 8.5 Conclusions and Future Work

We presented the implementation of five different suffix array construction algorithms in Thrill showing that the small set of algorithmic primitives provided by Thrill is sufficient to express the algorithms within the framework.

Our experimental results demonstrate that algorithms implemented in Thrill are competitive to hand-coded MPI implementations. By using the Thrill framework we gain additional benefits like future improvements of the algorithmic primitives in Thrill, and possibly even fault tolerance. Furthermore, Thrill already has automatic external memory support, hence our implementations are the first distributed external memory suffix array construction algorithms.

While our experimental results are already impressive, we believe that more future work should be directed at improving efficiency of the underlying sorting algorithm implementations in Thrill. The suffix sorting algorithms are the most complex algorithms currently implemented in the framework, and by improving their performance, all other applications will also gain.

But there are also other vectors for future work. One could extend the existing algorithms with LCP array construction and the difference cover algorithms with discarding tuples [PST05] similar to the technique we applied to the prefix doubling algorithms. The success of DC3 and DC7 suggests trying DC13 or an accelerated sampling strategy [PT13] where the difference cover size increases with the recursion depth. Furthermore, induced sorting should also be reconsidered, possibly in the context of *generalized* suffix array construction, as very large inputs are more commonly composed of many documents than of a long opaque string.

And then one can turn to post-processing the suffix array into other forms such as compressed indices, the FM-index [FM00; FM05], or specific on-disk suffix array representations such as RoSA [GM13]. Implementing these efficiently and scalable using the Thrill framework will open up new possibilities for applying advanced text algorithms to large datasets.

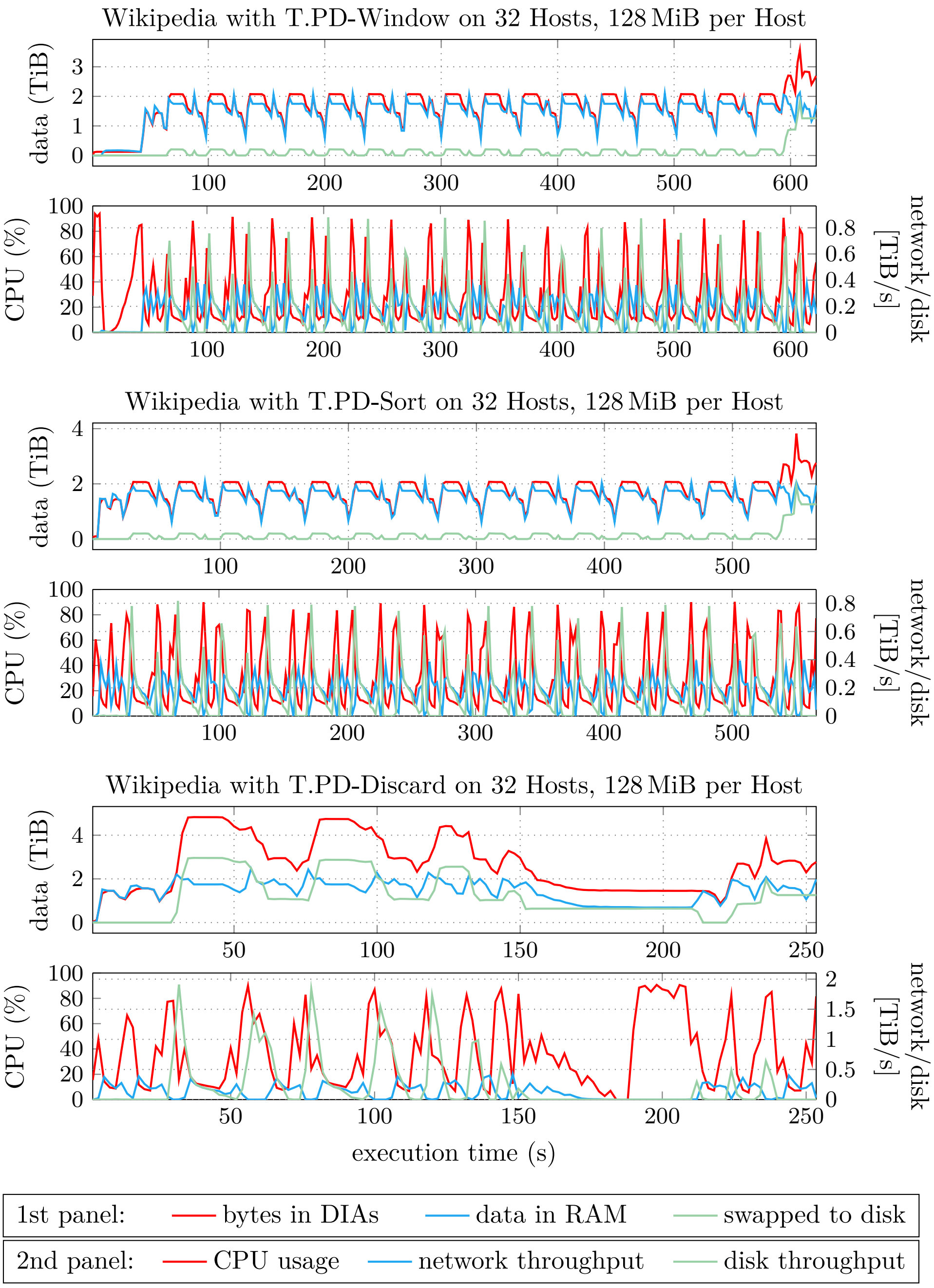

**Figure 8.14:** Bytes in DIAs, data in RAM, data swapped to disk over all hosts (first panel), and CPU utilization, and network and disk throughput averaged over all hosts (second panel) during T.PD-Sort, T.PD-Window, and T.PD-Discard running on Wikipedia with 32 hosts with 128 MiB per host.

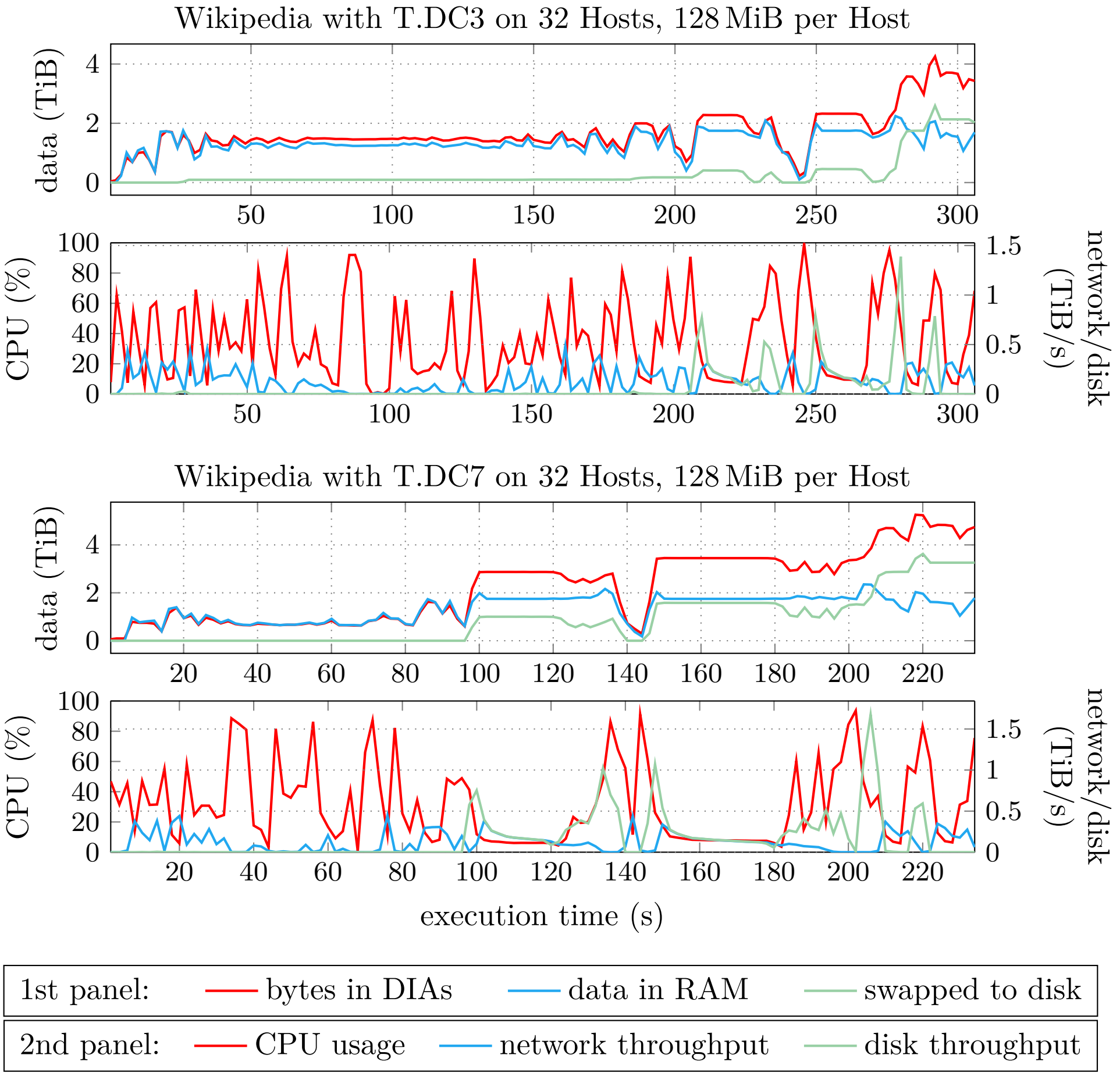

**Figure 8.15:** Bytes in DIAs, data in RAM, data swapped to disk over all hosts (first panel), and CPU utilization, and network and disk throughput averaged over all hosts (second panel) during T.DC3 and T.DC7 running on Wikipedia with 32 hosts with 128 MiB per host.

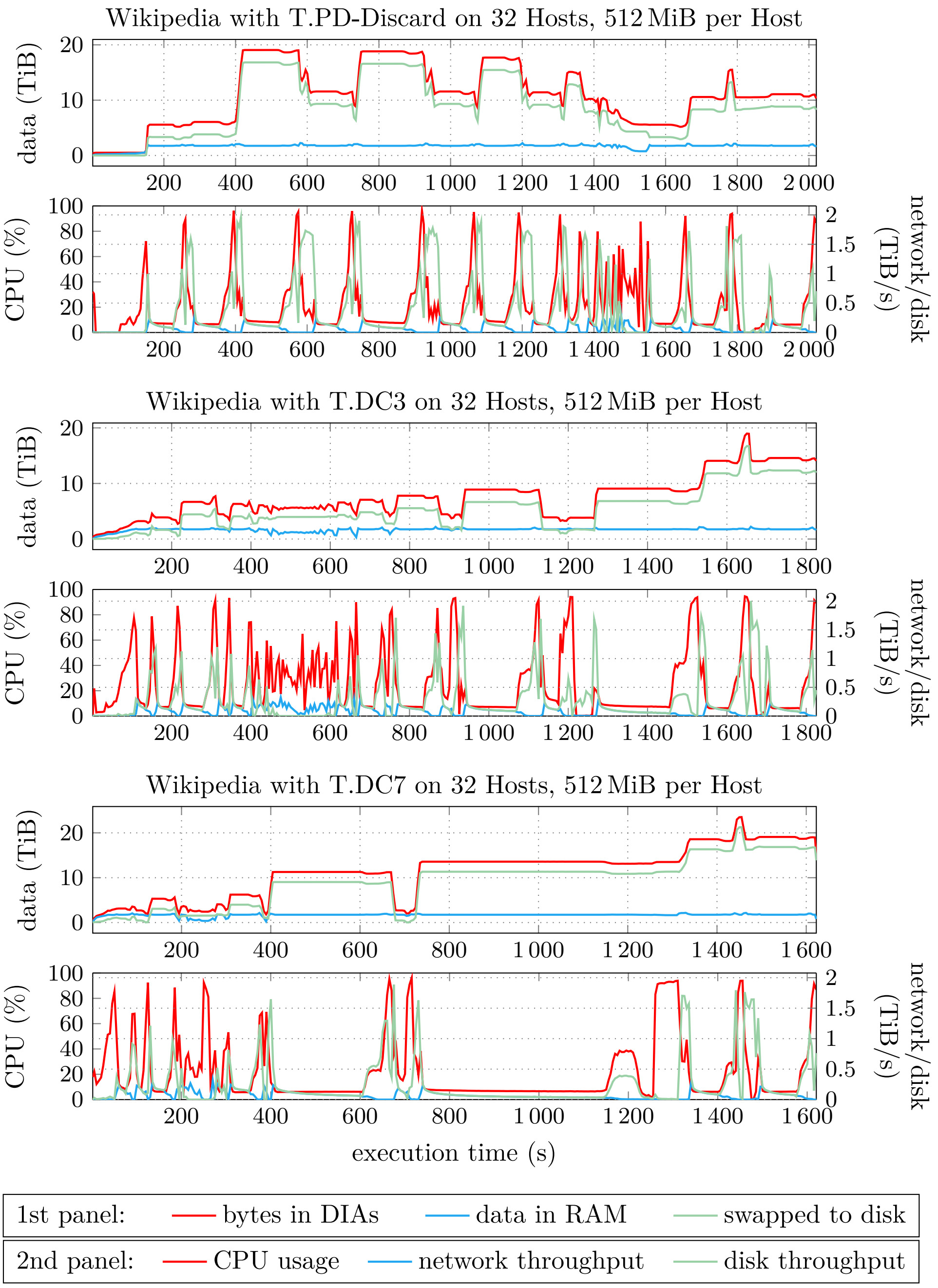

**Figure 8.16:** Bytes in DIAs, data in RAM, data swapped to disk over all hosts (first panel), and CPU utilization, and network and disk throughput averaged over all hosts (second panel) during T.PD-Discard, T.DC3, and T.DC7 running on Wikipedia with 32 hosts with 512 MiB per host.

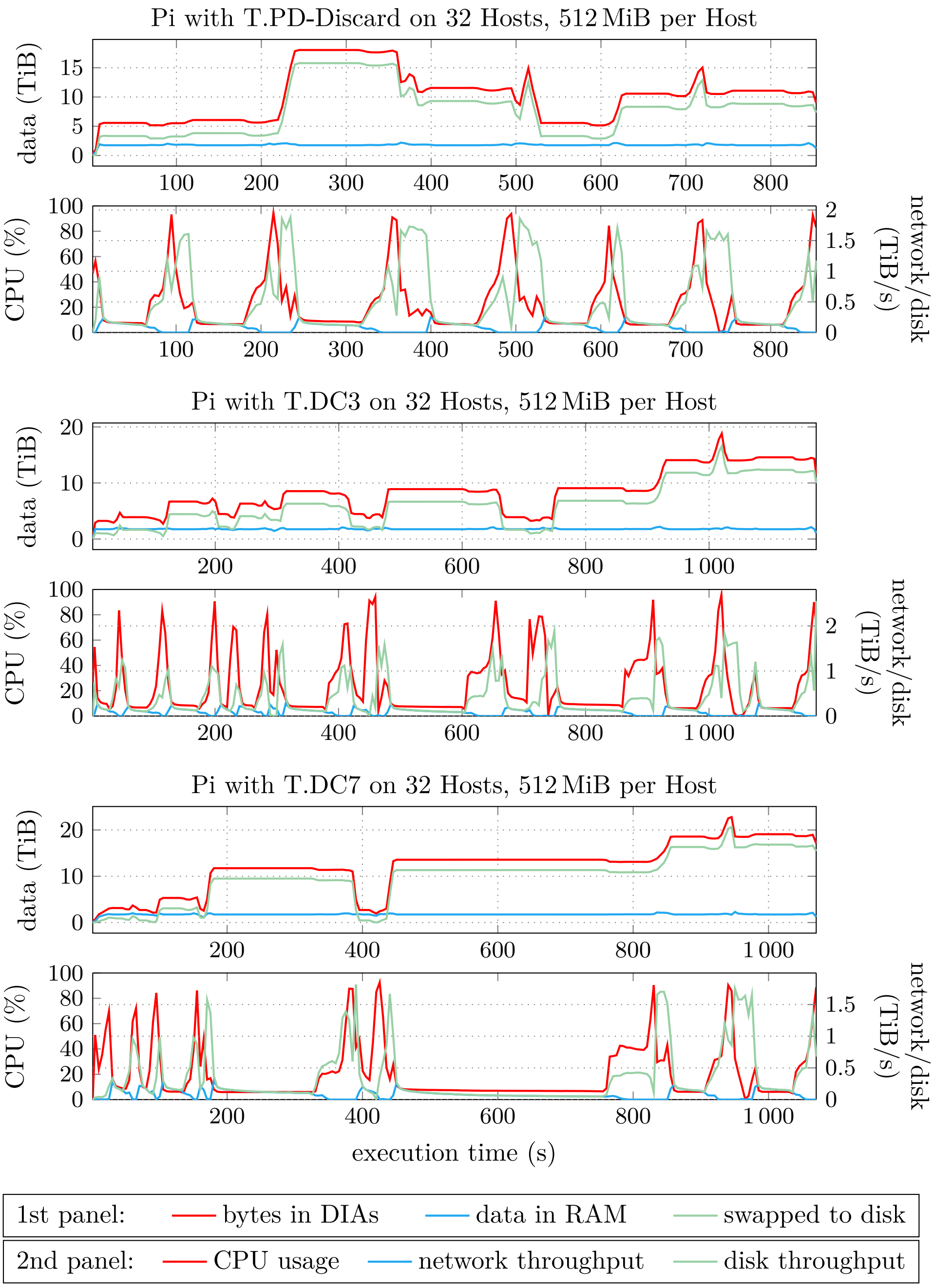

**Figure 8.17:** Bytes in DIAs, data in RAM, data swapped to disk over all hosts (first panel), and CPU utilization, and network and disk throughput averaged over all hosts (second panel) during T.PD-Discard, T.DC3, and T.DC7 running on Pi with 32 hosts with 512 MiB per host.

# Appendix

# List of Abbreviations

| | | |
|---|---|---|
| LMS | left-most S-type suffix | 175 |
| LOp | local DIA operation | 244 |
| MPI | message passing interface standard | 11 |
| MSD | most significant digit | 25, 28 |
| NUMA | non-uniform memory access | 113 |
| NVMe | non-volatile memory express | 10, 296 |
| PDM | parallel disk model | 8 |
| PE | processing elements (e.g. processors or cores) | 26 |
| PEM | parallel external memory model | 11 |
| PQ | priority queue | 206 |
| PRAM | parallel random access machine | 10 |
| RAM | random access machine | 6 |
| RDD | Apache Spark's resilient distributed datasets | 236 |
| RMQ | range minimum query | 202 |
| SMT | simultaneous multithreading (e.g. Intel's Hyperthreading) | 74 |
| SSD | solid-state drive or disk | 9 |
| TLB | translation look-aside buffer | 59 |
| UDF | user-defined function | 234 |

# List of Algorithms

# List of Figures

# List of Tables

# List of Theorems

# Publications and Supervised Theses

## In Conference Proceedings

## Journal Articles

## Technical Reports

## Theses

Timo Bingmann. “Visualisation of Very Large Graphs”. Study Thesis. Department of Computer Science, University of Karlsruhe (TH), Germany. Aug. 2006

Timo Bingmann. “Accuracy Enhancements of the 802.11 Model and EDCA QoS Extensions in ns-3”. Diploma Thesis. Department of Computer Science, University of Karlsruhe (TH), Germany. Apr. 2009

Timo Bingmann. “On the Structure of the Graph of Unique Symmetric Base Exchanges of Bispanning Graphs”. Diploma Thesis. Fakultät für Mathematik und Informatik, FernUniversität in Hagen, Germany. preprint arXiv:1601.03526. Jan. 2016

## Supervised Theses

Sascha Denis Knöpfle. „String Samplesort“. Bachelor Thesis. Karlsruhe Institute of Technology, Germany, in German. Nov. 2012

Daniel Feist. “External Batched Range Minimum Queries and LCP Construction”. Bachelor Thesis. Karlsruhe Institute of Technology, Germany. Apr. 2013

Thomas Keh. “Bulk-Parallel Priority Queue in External Memory”. Bachelor Thesis. Karlsruhe Institute of Technology, Germany. July 2014

Andreas Eberle. “Parallel Multiway LCP-Mergesort”. Bachelor Thesis. Karlsruhe Institute of Technology, Germany. Apr. 2014

Fellipe Bernardes Lima. “Implementation and Evaluation of an External Memory String B-Tree”. Bachelor Thesis. Karlsruhe Institute of Technology, Germany. Dec. 2014

Alexander Noe. “Bloom Filters for ReduceBy, GroupBy and Join in Thrill”. Master Thesis. Karlsruhe Institute of Technology, Germany. Jan. 2017

Florian Gauger. “Engineering a Compact Bit-Sliced Signature Index for Approximate Search on Genomic Data”. Master Thesis. Karlsruhe Institute of Technology, Germany. Feb. 2018